\documentclass[aps, rmp,onecolumn,showpacs,superscriptaddress,nofootinbib,longbibliography]{revtex4-2}

\usepackage[english]{babel}
\usepackage[utf8x]{inputenc}
\usepackage[T1]{fontenc}

\usepackage[a4paper,top=3cm,bottom=2cm,left=3cm,right=3cm,marginparwidth=1.75cm]{geometry}

\usepackage{amsmath}
\usepackage{amssymb}
\usepackage{graphicx}
\usepackage[colorinlistoftodos]{todonotes}
\usepackage[colorlinks=true, allcolors=blue]{hyperref}
\usepackage[inline]{enumitem}
\usepackage{ulem}
\usepackage{cleveref}
\usepackage{aasmacros}

\newcommand\be{\begin{equation}}
\newcommand\ee{\end{equation}}
\newcommand\bees{\begin{eqnarray}}
\newcommand\ees{\end{eqnarray}}
\newcommand{\dgw}{d_{\rm L}^{\,\rm gw}}
\newcommand{\dem}{d_{\rm L}^{\,\rm em}}
\newcommand{\vk}{{\bf k}}
\newcommand{\ra}{\rightarrow}
\newcommand\eq[1]{eq.~(\ref{#1})}

\newcommand{\dd}{\mathrm{d}}

\begin{document}

\title{New Horizons for Fundamental Physics with LISA}
\date{\today}

\author{K. G. Arun}
\affiliation{Chennai Mathematical Institute, Siruseri, 603103, India}

\author{Enis Belgacem}
\affiliation{Institute for Theoretical Physics, Utrecht University, Princetonplein 5, 3584 CC Utrecht, The Netherlands}

\author{Robert Benkel}
\affiliation{Max Planck Institute for Gravitational Physics (Albert Einstein Institute), Am M\"{u}hlenberg 1, Potsdam-Golm 14476, Germany}

\author{Laura Bernard}
\affiliation{LUTH, Observatoire de Paris, PSL Research University, CNRS, Universit\'e de Paris, 5 place Jules Janssen, 92195 Meudon, France}

\author{Emanuele Berti}
\affiliation{Department of Physics and Astronomy, Johns Hopkins University,
3400 N. Charles Street, Baltimore, Maryland 21218, USA}

\author{Gianfranco Bertone}
\affiliation{Gravitation Astroparticle Physics Amsterdam (GRAPPA),
Institute for Theoretical Physics Amsterdam and Delta Institute for Theoretical Physics, University of Amsterdam, Science Park 904, 1098 XH Amsterdam, The Netherlands}

\author{Marc Besancon}
\affiliation{CEA Paris-Saclay Irfu/DPhP,
Bat. 141, 91191 Gif sur Yvette, France}

\author{Diego Blas}
\affiliation{Theoretical Particle Physics and Cosmology Group, Physics Department, King's College London, University of London, Strand, London WC2R 2LS, UK}
\affiliation{Grup de F\'isica Te\`orica, Departament de F\'isica, Universitat Aut\`onoma de Barcelona, 08193 Bellaterra, Spain}
\affiliation{Institut de Fisica d'Altes Energies (IFAE), The Barcelona Institute of Science and Technology,\\ Campus UAB, 08193 Bellaterra, Spain}

\author{Christian G. B\"ohmer}
\affiliation{Department of Mathematics, University College London,
Gower Street, London, WC1E 6BT, United Kingdom}

\author{Richard Brito}
\affiliation{CENTRA, Departamento de F\'{\i}sica, Instituto Superior T\'ecnico -- IST, Universidade de Lisboa -- UL, Avenida Rovisco Pais 1, 1049 Lisboa, Portugal}

% ...

\begin{abstract}
\vspace{0.3cm}
\newpage
\begin{center}
{\bf Abstract}
\end{center}
The Laser Interferometer Space Antenna (LISA) has the potential to reveal wonders about the fundamental theory of nature at play in the extreme gravity regime, where the gravitational interaction is 
both strong and dynamical. 
\;~In this white paper, the Fundamental Physics Working Group of the LISA Consortium summarizes the current topics in fundamental physics where LISA observations of GWs can be expected to provide key input. We provide 
the briefest of reviews to then delineate avenues for future research directions 
and to discuss connections between this working group, other working groups and the consortium work package teams.
These connections must be developed for LISA to live up to its science potential in these areas.   
\end{abstract}

\maketitle

\tableofcontents

\section{Introduction}
\vspace{0.25cm}

Gravity is at the forefront of many of the deepest questions in fundamental physics. These questions include the classical dynamics and quantum nature of black holes (BHs), the matter and antimatter asymmetry of the observable universe, the processes at play during the expansion of the universe and during cosmological structure formation, and, of course, the intrinsic nature of dark matter and dark energy, and perhaps of spacetime itself. These questions are cross-generational and their answers are likely to require cross-disciplinary explorations, instead of being the purview of a particular sub-discipline. 


The recent observations of gravitational waves (GWs) are beginning to help us dig deeper into these questions, opening new research paths that enable the fruitful interaction of fundamental theory and observation. GWs have the potential to examine largely unexplored regions of the universe that are otherwise electromagnetically obscure. Examples of these regions include the vicinity of BH horizons, early phases in the formation of large-scale structure, and the hot big bang. Moreover, GWs can complement electromagnetic (EM) observations in astronomy and cosmology, thus enabling ``multi-messenger'' astrophysics and creating a path toward a better understanding of our universe. 


The Laser Interferometer Space Antenna (LISA) has the potential to contribute enormously in this quest, as this instrument is uniquely positioned to observe, for the first time, long-wavelength GWs \citep{LISA:2017pwj}, and therefore, new sources of GW radiation. Why is this? Because LISA can provide new information about gravitational anomalies, perhaps related to the quantum nature of gravity, about quantum-inspired effects in BH physics, and even about beyond the standard model, particle physics. Examples of the information that could be gained abound, but one concrete example is the following. LISA has the potential to constrain (or detect) the activation of scalar or vector degrees of freedom around black holes, which arise in certain modified gravity theories inspired by quantum gravity. This is because these fields typically carry energy-momentum away from BH binaries, forcing them to spiral faster into each other than predicted in GR. This faster rate of inspiral then modifies the time evolution of the GW frequency, and therefore, its phase, which LISA is particularly sensitive to. 


GW observations with LISA will complement the information we have gained (and will continue to gain) with ground-based GW interferometers, such as the Laser Interferometer Gravitational-wave Observatory (LIGO), Virgo and Kagra. This complementarity arises because ground-based instruments operate at higher frequencies than LISA, and therefore, they can hear GWs from a completely different type of sources. This, in turn, implies LISA is uniquely positioned to detect supermassive BH (SMBH) mergers, extreme mass-ratio inspirals (EMRIs), galactic binaries, and SGWB, none of which ground-based detectors are sensitive to. The GWs emitted by some of these sources (such as those emitted by SMBH mergers) will be extremely loud, generating signal-to-noise ratios (SNRs) in the thousands, which will allow us to prove fundamental physics deeply. Other GWs will be less loud, but they will be extremely complex, as is the case for waves generated by EMRIs; these waves will contain intricate amplitude and phase modulations that will encode the BH geometry in which the small compact object zooms and whirls. The observation of GWs with LISA will also complement observations with pulsar timing arrays and EM observations of the B-mode polarization in the cosmic microwave background (CMB), which can inform us about fundamental physics at even lower frequencies. 

\begin{figure}[t]
\begin{center}
\includegraphics[width=0.8\textwidth]{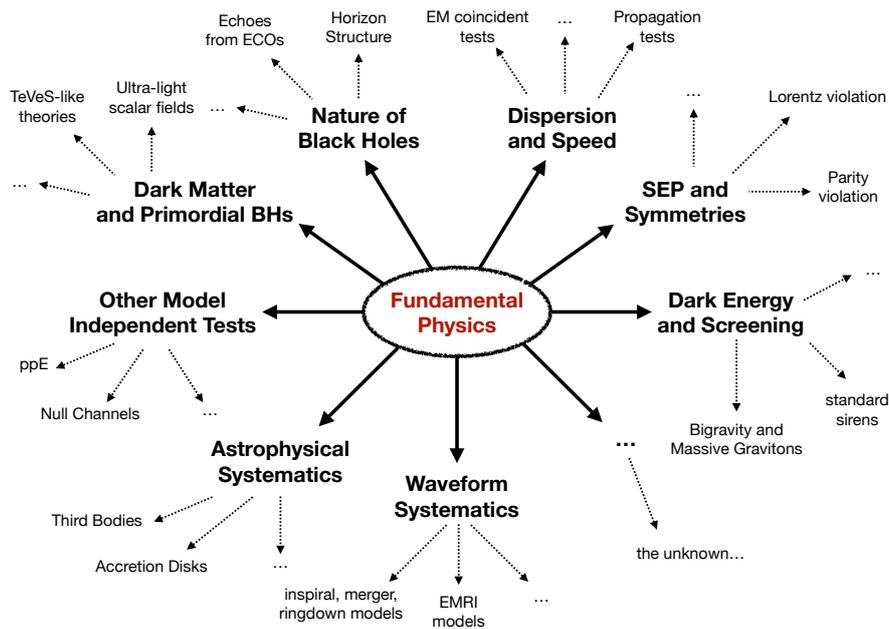}
\end{center}
\caption{Schematic diagrams of topics of relevance to the Fundamental Physics WG of the LISA Consortium. Some of these topics overlap naturally with other WGs, such as the Waveform modelling WG, the Astrophysics WG and the Cosmology WG.}
\label{branches}
\end{figure}

The goal of this white paper is threefold. First, we aim to identify those topics in fundamental physics beyond the current standard models of particle physics, gravity and cosmology that are particularly relevant for the scientific community, in order to delineate and sharpen LISA's potential in each of these areas. The organization of this article reflects this identification of topics, with each Section covering one area. The range of areas is illustrated in Figure \ref{branches}. Second, within each Section we summarize the state of the art of each of these areas, both from a theoretical and an observational point of view. In order to keep this article both manageable and useful, rather than presenting an extensive detailed review of each topic, we have opted for sufficiently concise ``reviews'', referring to excellent recent articles in Living Reviews and elsewhere for more details (see e.g., \citealt{Barack:2018yly}). Put differently, the philosophy behind this white paper is to strive towards completion in terms of what are relevant branches of fundamental physics for LISA that we think ought to be explored further, rather than providing extensive background and technical details on each of the topics. Third, each Section then assesses what must be done in order for LISA to live up to its scientific potential in these areas of fundamental physics. This is primarily where the living part of this article enters, which should be viewed in a LISA-specific context. Finally, we bring these different strands together and identify possible synergies in a Roadmap Section at the end.


In closing, let us note that this article is part of a series of LISA working group (WG) articles. Some of the topics discussed here are of interest to several WGs organized within the LISA Consortium. Strong synergies exist between the Fundamental Physics WG, the Cosmology, the Astrophysics, and the Waveform modelling WGs, each of which approaches the topics discussed in this article from different, complementary angles. One of the intended goals of this article is to promote these synergies and connect the WGs, specially the fundamental physics one to the LISA work package groups, so that the ideas presented here can be implemented and deployed in future LISA data analysis. 

\newpage
\subsection{Abbreviations}

We here list a set of abbreviations commonly used in this review article:
\begin{itemize}
\setlength\itemsep{0cm}
    \item BH = Black hole.
    \item BBH = Binary black hole.
    \item BMS = Bondi-Van der Burg-Metzner-Sachs. 
    \item CMB = Cosmic microwave background.
    \item DM = Dark matter.
    \item dCS = Dynamical Chern--Simons. 
    \item ECO = Exotic compact object.
    \item EdGB = Einstein dilaton Gauss Bonnet.
    \item EdM = Einstein dilaton Maxwell.
    \item EEP = Einstein equivalence principle.
    \item EM = Electromagnetic.
    \item EMRI = Extreme mass-ratio inspiral. 
    \item EsGB = Einstein scalar Gauss--Bonnet gravity.
    \item FLRW = Friedmann--Lema{\^{\i}}tre--Robertson--Walker.
    \item GR = General relativity. 
    \item GW = Gravitational waves.
    \item IMBH = Intermediate-mass black hole.
    \item IMR = Inspiral-Merger-Ringdown. 
    \item LIGO = Laser Interferometer Gravitational-Wave Observatory.
    \item LISA = Laser Interferometer Space Antenna.
    \item LLI = Local Lorentz invariance. 
    \item LPI = Local position invariance. 
    \item MCMC = Markov Chain Monte Carlo.
    \item MBH = Massive black hole. 
    \item NS = Neutron star. 
    \item QNM = Quasinormal model. 
    \item PBH = Primordial black hole.  
    \item PN = Post-Newtonian.
    \item ppE = Parameterized post-Einsteinian.
    \item ppN = Parametrized post-Newtonian.
    \item SCO = Stellar compact object.
    \item SEP = Strong equivalence principle. 
    \item SGWB = stochastic gravitational wave background.
    \item SMBH = Supermassive black hole.
    \item SNR = Signal-to-noise ratio. 
    \item TeVeS = Tensor-vector-scalar gravity. 
    \item WD = White dwarf. 
    \item WG = Working group.
    \item WEP = Weak equivalence principle.
\end{itemize}
Henceforth, we use geometric units in which $G = 1 = c$.

\newpage
\section{Tests of General Relativity}
\label{sec-testsGR}



\vspace{0.25cm}

\subsection{Tests of Gravity and its fundamental principles}

GWs and some of their major sources --- BHs and relativistic stars --- are among the most groundbreaking predictions of GR. It is hence rather intuitive that GW observations should offer unique insights into the workings of gravity and the fundamental principles that underpin GR.   

\subsubsection{The Equivalence Principle} 
\label{subsec:ep}

The various versions of the equivalence principle are commonly used as a theoretical foundation of GR and as guiding principles for tests of gravity \citep{Will:1993ns}.

The \textit{Weak Equivalence Principle} (WEP) postulates that the trajectory of a freely falling test body is independent of its structure and composition, providing the fact that there are no external forces acting on this body such as electromagnetism. The \textit{Einstein Equivalence Principle} (EEP) goes one step further and combines the WEP, Local Lorentz Invariance (LLI), and Local Position Invariance (LPI). The LLI is connected with the assumption that the outcome of any local non-gravitational test experiment is independent of the velocity of the freely-falling frame of reference where this experiment is performed. The LPI requires that the outcome of such an experiment is independent of where and when it is performed. The \textit{Strong Equivalence Principle} (SEP) is equivalent to the EEP with the WEP extended to self-gravitating bodies and the LLI and LPI to \textit{any} experiment.  

The EEP dictates the universality of couplings in the standard model because of LLI and LPI. The SEP takes this even further and implies that the gravitational coupling is fixed as well. Testing the SEP is often considered synonymous to testing GR itself \citep{Will:1993ns}. If the SEP is fulfilled, the only field that is a mediator of the gravitational interaction should be the spacetime metric. Conversely, the existence of new fields mediating gravity would generically lead to violations of the SEP (which might appear only in specific systems, depending on how elusive the fields are).

\subsubsection{Lovelock’s Theorem and GR uniqueness} \noindent 

The discussion above already suggests strongly that testing GR (classically) amounts to looking for new fields. An independent way to reach the same conclusion is to start from Lovelock's theorem \citep{Lovelock:1972vz}: Einstein's field equations are unique, assuming that we are working in four dimensions, diffeomorphism invariance is respected, the metric is the only field mediating gravity, and the equations are second-order differential equations. Violating any of these assumption can circumvent Lovelock's theorem and lead to distinct alternative theories of gravity. However, the vast majority of them\footnote{See \citet{Flanagan:2003rb,Barausse:2007pn,Pani:2012qd,Pani:2013qfa} for discussions regarding attempts to circumvent Lovelock's theorem without adding new fields.} will share one property: they will contain one or more additional fields (see e.g., \citealt{Sotiriou:2015lxa}). Demonstrating this mathematically  might require compactifying spacetime down to 4-dimensions, introducing additional fields to restore diffeomorphism invariance, and performing field redefinitions. 

The realization that testing the SEP and looking for deviations from GR largely amounts to looking for new fields allows one to understand intuitively how compact objects and GWs can be used to probe such deviations. When the new fields have a nontrivial configuration around a BH or compact star, the latter can be thought of as carrying a `charge' (this might not be a conserved charge associated to a gauge symmetry). Field theory intuition tells us that accelerating charges radiate. Hence, binaries beyond GR will exhibit additional GW polarizations. Emission in extra polarizations affects the rate of energy loss and, in turn, the orbital dynamics. So, the pattern of emission of conventional polarizations will also be affected.  In Sect.~\ref{sec2:BHbeyodGR} we discuss how the structure of BHs---primary sources for LISA---can be affected by new fields, whereas in the rest of Sect.~\ref{sec2:Testbinaries} we discuss how deviations from GR can be imprinted in different parts of the waveform or affect GW propagation. 

\subsection{Testing GR with compact objects}
\label{sec2:Testbinaries}
\subsubsection{BHs beyond GR and theories that predict deviations}
\label{sec2:BHbeyodGR}

There exist no-hair theorems for a wide range of alternative theories of gravity \citep{Hawking:1972qk,Bekenstein:1996pn,Sotiriou:2011dz,Hui:2012qt,Sotiriou:2015pka,Herdeiro:2015waa,Doneva:2020dji} stating that the quiescent, isolated BHs are indistinguishable from their GR counterpart. If the additional fields can be excited, quasinormal mode (QNM) ringing can still be distinct from GR \citep{Barausse:2008xv,Molina:2010fb,Tattersall:2018nve} and hence offer a possibility of test theories that are covered by no-hair theorems. Models that manage to circumvent no-hair theorems are expected to lead to more prominent deviations from GR in GW signals, as they can posses additional charges and exhibit additional interactions that can affect all parts of the waveform.

Scalar fields coupled to the Gauss--Bonnet invariant or the Pontryagin density (dynamical Chern--Simons gravity) are known to lead to BH hair \citep{Campbell:1991kz,Kanti:1995vq,Yunes:2011we,Yagi:2012ya,Sotiriou:2013qea,Sotiriou:2014pfa,Benkel:2016kcq,Benkel:2016rlz,Yunes:2009hc,Stein:2014xba,Delsate:2018ome,Delgado:2020rev} which can be observed with LISA \citep{Yagi:2012vf,Yagi:2016jml,Maselli:2020zgv}. Such couplings are expected in low-energy limits of  quantum gravity  (see e.g., \citealt{Metsaev:1986yb,Ashtekar:1988sw,Jackiw:2003pm}) and are part of the Horndeski class of scalar-tensor theories \citep{Kobayashi:2019hrl}.  A coupling to the Pontryagin density is the leading-order parity violating term in gravity in the presence of a (pseudo) scalar \citep{Jackiw:2003pm,Alexander:2009tp}, while the linear coupling with the Gauss--Bonnet invariant is the only term inducing hair for shift-symmetric (aka massless) scalars \citep{Sotiriou:2013qea,Saravani:2019xwx,Delgado:2020rev} and the leading correction for GW emission \citep{Witek:2018dmd}.  A nonlinear coupling with the Gauss--Bonnet invariant has recently been shown to give rise to BH \textit{spontaneous scalarization} triggered by curvature \citep{Doneva:2017bvd,Silva:2017uqg,Blazquez-Salcedo:2018jnn,Silva:2018qhn,Macedo:2019sem,Cunha:2019dwb,Collodel:2019kkx} or spin \citep{Dima:2020yac,Hod:2020jjy,Doneva:2020nbb,Herdeiro:2020wei,Berti:2020kgk,Doneva:2020kfv}. The full set of Horndeski theories that can give rise to BHs scalarization has been identified in \citet{Antoniou:2017acq}.

Attempts have also been made to circumvent no-hair theorems by relaxing their assumptions. For example, superradiance can support  long-lived scalar clouds for very light scalars \citep{Arvanitaki:2010sy,Brito:2015oca} or lead to hairy BHs for complex scalars  with a time dependent phase \citep{Herdeiro:2014goa,Kleihaus:2015iea,Herdeiro:2015tia,Herdeiro:2017oyt,Delgado:2019prc,Collodel:2020gyp};
long-lived scalar ``wigs'' can be formed around a Schwarzschild BH \citep{Barranco:2012qs,Barranco:2013rua};
configurations with time-dependent scalar fields can be supported by some non-trivial cosmological boundary conditions \citep{Jacobson:1999vr,Babichev:2013cya,Berti:2013gfa,Clough:2019jpm,Hui:2019aqm}; 
BHs immersed in an inhomogeneous scalar field were considered in~\citet{Healy:2011ef}; 
scalarization can be induced by matter surrounding a BH \citep{Cardoso:2013fwa,Cardoso:2013opa,Herdeiro:2018wub}.

BHs in Lorentz-violating theories, such as Einstein-aether theory \citep{Jacobson:2000xp,Jacobson:2008aj} and Ho{\vr}ava gravity \citep{Horava:2009uw,Blas:2009qj,Sotiriou:2010wn}, will generically carry hair, as the field that breaks Lorentz symmetry will have to be nontrivial and may backreact on the geometry (see, however, \citealt{Ramos:2018oku,Adam:2021vsk}).
In such theories, new fields can propagate superluminally or even instantaneously \citep{Blas:2011ni,Bhattacharyya:2015uxt}, even when all known constraints are satisfied \citep{Sotiriou:2017obf,Gumrukcuoglu:2017ijh}. Indeed, BHs in Lorentz-violating theories can have a nested structure of different horizons for different modes \citep{Eling:2006ec,Barausse:2011pu} and potentially a {\textit{universal}} horizon that traps all signals \citep{Barausse:2011pu,Blas:2011ni,Bhattacharyya:2015gwa}. GW observations can probe this richer causal structure and yield novel constraint on Lorentz symmetry breaking. 

Hairy BHs have been studied in a variety of other scenarios and theories, including generalized Proca theories \citep{Herdeiro:2016tmi,Minamitsuji:2017aan,Babichev:2017rti,Heisenberg:2017xda,Kase:2018owh,Rahman:2018fgy,Santos:2020pmh} and massive gravity theories  \citep{Berezhiani:2011mt,Comelli:2011wq,Rosen:2017dvn}.

\subsubsection{Tests with GW propagation}
\label{Sec:Tests_with_GW_propagation}

The propagation of GWs provides a clean test of their kinematics. In GR, gravitons are massless spin-2 particles, while in alternative gravity theories, they might be massive \citep{Will:1997bb,deRham:2010kj,Hassan:2011zd,TheLIGOScientific:2016src}, or even have a Lorentz-violating structure in the dispersion relation \citep{Jacobson:2000xp,Jacobson:2008aj,Horava:2009uw,Blas:2009qj,Sotiriou:2010wn,Mirshekari:2011yq, Kostelecky:2016kfm, Shao:2020shv}. 
Modifications to the dispersion relation could lead to frequency-dependent, polarization-dependent, direction-dependent propagating velocities of GWs.
Eventually, while these components travel at different velocities, the gravitational waveforms received on the Earth are distorted, with respect to their original chirping structure at generation \citep{Will:1997bb, Kostelecky:2016kfm}. Therefore, a matched-filter analysis allowing the possibility of the modified GW propagation in the waveform will reveal the nature of graviton kinematics (see e.g., \citealt{TheLIGOScientific:2016src}) and provide stringent constraints on the mass of the graviton and on Lorentz symmetry violations.
The binary neutron star (NS) merger GW170817 has already provided a strong, double-sided bound on the speed of GWs to a part in $10^{15}$ \citep{Monitor:2017mdv}. However, Lorentz-violating theories have multidimensional parameter spaces and generically exhibit additional polarizations \citep{Sotiriou:2017obf}. The speed of these other polarizations remains virtually unconstrained \citep{Gumrukcuoglu:2017ijh,Oost:2018tcv}. 
A combination of multiple GW events can also be used to simultaneously constrain a set of beyond-GR parameters \citep{Shao:2020shv}.

GWs travel over cosmological distances before they reach the detector.
On a Friedmann-Lemaître-Robertson-Walker (FLRW) background, we can describe GWs by $h_{ij}$, the transverse and traceless  perturbation of the spatial metric, $g_{ij} ({\bf x},t) = a^2(t) \left[ \delta_{ij} + h_{ij} ({\bf x},t) \right]$.
In Fourier space, the most general modification of the GW propagation equation can be written as  (assuming spatially flat models) \citep{Ezquiaga:2018gbw,Barausse:2020rsu,Ezquiaga:2021ayr} 
\begin{equation}
\ddot h_{ij}({\bf k},t) + \left[  3 H(t) +  \Gamma(k,t) \right] \dot h_{ij} ({\bf k},t) + \left[ c_{\rm T}^2(t) k^2 + D(k,t)   \right] h_{ij} ({\bf k},t) = 0 \;,
\label{eq:sec2-EOMwave}
\end{equation}
where $H \equiv \dot a /a$ is the Hubble rate. The parameters $c_{\rm T}$, $\Gamma$ and $D$ respectively describe the speed of  propagation of the wave, the damping of its amplitude and additional modifications of the dispersion relation. Scale-independent modifications, described by a $k$-independent $\Gamma$, are discussed in Sect.~\ref{sec:LambdaCDM}. One expects a single observation to exclude $\Gamma  \gtrsim L^{-1}$ and $D \gtrsim  f L^{-1} $, where $L$ is the distance to the source and $f$ the GW frequency, although the detailed constraints are also controlled by the frequency dependence of these quantities. $c_{\rm T}$, $\Gamma$ and $D$ will depend on the theory of gravity and on the cosmological background. Hence, one can translate bounds on these parameters into bounds on a given model. For example, the speed of GW bound has severely constrained generalized scalar-tensor theories \citep{Horndeski:1974wa,Deffayet:2011gz,Zumalacarregui:2013pma,Gleyzes:2014dya,Langlois:2015cwa,Crisostomi:2016czh,BenAchour:2016fzp} under the assumption that they account for dark energy \citep{Lombriser:2015sxa,McManus:2016kxu,Creminelli:2017sry,Ezquiaga:2017ekz,Sakstein:2017xjx,Baker:2017hug} (see also \citealt{deRham:2018red}).  Relaxing this assumption \citep{Franchini:2019npi,Noller:2019chl,Antoniou:2020nax} can lift the constraint. 

GW propagation tests can also reveal potential couplings and decays of gravitons into other particles (e.g., \citealt{Creminelli:2019kjy})
or oscillations between different states (analogous to neutrino oscillations). The latter are expected in bigravity models \citep{Hassan:2011zd}, where a massless and a massive spin-2 fields interact in a specific way to avoid ghost degrees of freedom.
The lightest tensor mode is the one that couples to matter and its speed can be constrained as above if there is a prompt EM counterpart. Its amplitude  determines the ratio between the luminosity distance of GWs and the one of EM radiation, which oscillates as a function of redshift.
Using several astrophysical population models for the population of massive black hole binaries \citep{Barausse:2012fy,Klein:2015hvg} and performing a $\chi^2$ analysis, it was found that oscillation effects can be observed for masses $m\gtrsim 2 \cdot 10^{-25}\,$eV \citep{Belgacem:2019pkk}. Thus LISA will provide a $\sim 3$ order of magnitude improvement in mass sensitivity over the current LIGO/Virgo limit, which probes $m \gtrsim 10^{-22}$eV \citep{Max:2017flc}, due to the larger oscillation baseline and the lower detection frequency. 

\subsubsection{Tests of GR with MBH coalescence } 
\label{sec:GR_test_MBH}

\paragraph{Inspiral} 

 LISA will  observe very long inspiral phases and it will therefore allow for high precision tests of gravity \citep{Berti:2004bd}. This stage can be modelled using approximate techniques such as the low-velocity, weak-field PN expansion \citep{poisson2014gravity}, or the parameterized post-Einsteinian approach (ppE) \citep{Yunes:2009ke}, that is better suited in certain cases for alternative theories of gravity. Even though tests of the GR nature of the waveforms can be  performed without linking to specific alternative theories of gravity within the PN or ppE approaches, connecting the predictions of the different generalizations of Einstein's theory with the possible deviations in the GR expectation values of the PN or ppE parameters is an inseparable part of testing the possible violations of the GR fundamental symmetries \citep{Berti:2018cxi}.

Using the inspiral observations, LISA will be able to improve the constraints on different non-GR predictions, such as the scalar dipole radiation, Lorenz symmetry, mass of the graviton, etc., by several orders of magnitude better compared to LIGO/Virgo \citep{Chamberlain:2017fjl}. Many alternative theories of gravity predict non-zero tidal deformation of BHs \citep{Cardoso:2017cfl} and that can be also tested using the inspiral waveforms.

Even more intriguing is the possibility for multiband GW observations of BBH mergers. The separation between the two members of a BBH determines the frequencies of the emitted GW signal. A binary that will be in the observational band of LISA at large separation can several years later enter the band of ground-based detectors. Using LISA observations and assuming GR, one can then obtain high-accuracy predictions of the time when the binary will become observable by ground-based detectors as well as its position on the sky \citep{Sesana:2016ljz}. Any deviation of the former, observed by a ground based detector, would imply a breakdown of GR. For example, joint observations of a LIGO/Virgo and LISA of a GW150914-like event could improve constraints on BBH dipole emission by 6 orders of magnitude \citep{Barausse:2016eii,Toubiana:2020vtf}. The possibility of observing the same system more than once for a prolonged period will offer the opportunity to test completely nonlinear predictions of some generalized scalar-tensor theories, such as dynamical BH scalarization \citep{Khalil:2019wyy}. 

\paragraph{Ringdown}

In GR, numerical simulations have shown that the end state of a BBH merger is a Kerr BH (e.g., \citealt{Jani:2016wkt,Boyle:2019kee,Healy:2020vre}) in agreement with the earlier analytical studies proving that the Kerr metric is the unique stationary, asytmptotically flat, axially symmetric BH spacetime \citep{Carter:1971zc,Robinson:1975bv,buntingthesis,Mazur:1982db} (see \citealt{heusler_1996,Chrusciel:2012jk} for a review).
Before reaching the final Kerr state, the BH emits GWs in QNMs, which are labeled by their overtone numbers, $n$, and angular numbers $(l,m)$ and are determined by these numbers $(l,m,n)$ plus the mass $M$ and spin parameter $a$ of the Kerr BH \citep{Berti:2009kk}.
Measuring several of these QNMs would allow for ``BH spectroscopy,'' in which a perturbed Kerr BH is identified by the spectrum of QNMs in its ringdown GWs \citep{Detweiler:1980gk,Dreyer:2003bv,Berti:2005ys}.
Specifically, for a given BBH merger, the number and amplitude of QNMs excited during the ringdown is determined by the individual BHs and their orbital parameters prior to merger \citep{Kamaretsos:2011um}.
If at least two of the QNMs can be measured from the ringdown of a BH, then by knowing which modes were excited, one can verify that the two QNM frequencies and two QNM damping times are both consistent with the same mass $M$ and spin $a$ parameters of the remnant Kerr BH, within the errors of the observation \citep{Dreyer:2003bv,Berti:2005ys}.
For a high SNR massive BBH event measured by LISA, this ``no hair'' (or ``final-state'') test can be performed to verify the remnant is consistent with a Kerr BH to high precision (e.g., \citealt{Gossan:2011ha}).

If the underlying gravity theory differs from GR, the ringdown of a BBH merger will also typically differ from the ringdown in GR.
Computing the QNMs in modified theories is typically challenging, as is performing NR simulations of BBH mergers in modified gravity theories (both of which have only been performed in a handful of cases, e.g., \citealt{Cardoso:2009pk,Blazquez-Salcedo:2016enn,Okounkova:2019zjf,Okounkova:2020rqw}).
This makes it difficult to look for deviations in a particular modified theory.
Parametrized tests are instead used to look for deviations.
In \citet{Tattersall:2017erk}, deviations from linear perturbations about a Schwarzschild BH in GR were parametrized at the level of a diffeomorphism-invariant action that encompassed a large range of possible theories that lead to second-order equations of motion.
A more phenomenological approach to the test for deviations from GR during ringdown is to perform a parametrized test in which the QNM frequencies are given by the Kerr values plus small deviations, in a ``post-Kerr'' expansion \citep{Glampedakis:2017dvb}.
There are also procedures to try to combine the deviation parameters from multiple BBH events measured by LISA \citep{Maselli:2019mjd}.
With more detailed predictions from specific theories, parametrized constraints could be converted into constraints on a particular theory.

\paragraph{Merger}
%

The merger phase of a BBH inspiral is arguably when the nonlinear effects of gravity truly manifest themselves. This makes it a very challenging regime to model, requiring full nonlinear evolutions of the field equations. A model-independent self-consistency test, such as the IMR consistency test \citep{Ghosh_2016,Ghosh_2017} could be used in order to avoid having to model the merger in alternative theories of gravity. However, having theory-specific waveforms that include the merger is essential. It can provide more stringent constraints and it is necessary for interpreting them physically. It can also be used to quantitatively explore deviations from GR that do not affect other parts of the waveform --- e.g. new fields that are highly excited by nonlinear effects and decay rapidly --- and guide and calibrate parametrizations. 

Numerical simulations beyond GR have only been performed in a handful of cases, most notably for scalar nonminimally coupled to curvature invariants \citep{Benkel:2016rlz,Benkel:2016kcq,Okounkova:2017yby,Witek:2018dmd,Okounkova:2019dfo,Okounkova:2019zjf,Cayuso:2020lca,Silva:2020omi,East:2021bqk} and for Einstein--Maxwell-dilaton theory \citep{Hirschmann:2017psw}. Establishing whether the initial value problem is well-posed in alternative theories is particularly challenging \citep{Papallo:2017qvl,Cayuso:2017iqc,Sarbach:2019yso,Ripley:2019hxt,Kovacs:2020pns}. So far, known simulations have circumvented this problem by adopted Effective Field Theory inspired treatments: either  working perturbatively in the new coupling constants \citep{Benkel:2016rlz,Benkel:2016kcq,Okounkova:2017yby,Witek:2018dmd,Okounkova:2019dfo,Okounkova:2019zjf} or adopting a scheme similar to the Israel-Stewart treatment of viscous relativistic hydrodynamics, in which additional fields are introduced to render the system hyperbolic and then exponentially `damped' so that the evolution equation match the initial system asymptotically \citep{Cayuso:2017iqc}. Nonlinear evolution beyond GR is one of the major challenges in GW modelling.

\subsubsection{Test of GR with EMRIs (non-null tests)}


GW observations of EMRIs 
offer the opportunity to probe gravity in a mass range which is 
unique to LISA. In such systems a stellar mass object orbits around 
a more massive component, with typical mass ratios of the order 
of $q\sim 10^{-5} -10^{-7}$, leading to a large number of 
accumulated cycles before the merger, proportional to $\sim1/q$. The signal produced by the slow inspiral provides a detailed map of the spacetime that will be able to pinpoint deviations from GR predictions (if any) and requires detailed knowledge of the emitted waveform to avoid spurious systematics \citep{Pound:2015tma,Barack:2018yvs}. 

Despite progress within GR \citep{Pound:2019lzj}, tests of gravity with EMRIs have been so far limited to systematic calculations in modified theories, in which the different couplings to the gravity sector require, in general, a theory-by-theory analysis. These additional degrees of freedom, introduced as scalar, vector or tensor modes, activate extra emission channels that modify the binary phase evolution. These changes are expected to leave a detectable footprint in the emitted signal and be augmented by the large number of cycles followed by EMRIs. Calculations for non GR theories with non minimally coupled scalar fields have mainly investigated the changes in the emitted flux within the adiabatic approximation, for some orbital configurations (including generic orbits) around spinning BHs \citep{Pani:2011xj,Blazquez-Salcedo:2016enn,Cardoso:2011xi,Yunes:2011aa,Canizares:2012is,Fujita:2016yav}. These works have shown how, depending on the magnitude of the couplings, in some cases the accumulated GW phase can be large to produce hundreds of cycles of difference in the binary evolution compared to GR. The projected constraints inferred by LISA on the parameters of non-GR theories using approximate waveforms \citep{Yunes:2011aa,Canizares:2012is}, and modelling beyond the adiabatic approximation, i.e., taking into account self-force calculations \citep{Zimmerman:2015hua}, requires further work. 

Drastic simplifications in the EMRI modelling beyond GR have been 
recently proven to hold for a vast class of theories, for which no-hair theorems or separations of scales provide a decoupling of the metric and scalar perturbations \citep{Maselli:2020zgv}. This result allows one to describe the background spacetime as in GR, rendering all the modifications induced by the modified theory to be universally captured by the scalar field's charge only. It has already been used to show that the latter can be measured with unprecedented precision \citep{Maselli:2021men}. In this framework waveform modelling beyond GR can take advantage of all the efforts devoted so far to study the evolution of scalar charges around Kerr BHs \citep{Barack:2000zq,Detweiler:2002gi,Castillo:2018ibo,DiazRivera:2004ik,Warburton:2010eq,Warburton:2014bya,Gralla:2015rpa,Warburton:2011hp,Nasipak:2019hxh}. 

\subsubsection{Gravitational Memory and BMS symmetry}\label{subsec:memory}


GR predicts that the passage of a GW causes a permanent displacement in the relative position of 
two inertial GW detectors. This phenomenon is 
known as the displacement memory 
effect \citep{BMSAdd4,BMSAdd5,Christodoulou:1991cr,Thorne:1992sdb} (see \citealt{BMS5,BMS3b} 
for an account of historical references). Studies on the
detection of the displacement memory effects in actual
experiments have been presented 
in \citet{BMS9,BMS10,BMS11,BMS12}. 
More recently, prospects for  the memory effect detection
with LISA have been discussed in \citet{BMS13} using 
SMBH binaries undergoing coalescence
(See also Sect.~\ref{sec:testsofBHNature}).  In addition to the displacement memory 
effect, other and subdominant effects
such as the spin memory and center of mass memory 
effects can also take place \citep{BMSAdd3,BMS14,BMS15,BMS16}. 
Generalizations of the memory effect, not 
necessarily motivated or associated to a symmetry, 
which should in principle be measurable,
have also been discussed in \citet{BMS16,BMSAdd9}, and further memory effects with logarithmic branches have been inferred from graviton amplitude calculations \citep{BMSAdd1,BMSAdd2}.

Measurement of memory effects would not only act as a test of GR \citep{BMS14,BMS15,BMS16,Hou:2020tnd,Tahura:2020vsa,Tahura:2021hbk}, but it could also potentially shed light on puzzling infrared aspects of quantum field theories. 
Ideally isolated systems in GR can be
described by asymptotically flat spacetimes. At null
infinity, the symmetry group of asymptotically flat
spacetimes in GR is known
to be larger than the Poincare group of symmetries. 
It contains the original infinite-dimensional Bondi-Van der Burg-Metzner-Sachs (BMS) group which is a semi-direct product of the Lorentz group and of an infinite-dimensional group of ``angle dependent translations" called supertranslations \citep{BMS1,BMS2}. Further extensions of the BMS group have also been proposed \citep{BMS3c,BMS3d}. 
BMS supertranslations can be related to the gravitational
memory effect in that the relative positions of 
two inertial detectors before and after the passage of
a GW differ by a BMS supertranslation. 
In addition, perturbative quantum gravity admits remarkable infrared identities among its scattering amplitudes, the soft theorems, that have been demonstrated to be the Ward identities of BMS symmetries \citep{BMS4,BMS5}. The classical limit of the quantum soft theorems are equivalent after a Fourier transformation to the memory effects \citep{BMS4}. Measuring memory effects therefore directly probes the infrared structure of quantum gravity amplitudes. 


The detection of GW bursts with memory with the LISA instrument could be considered from SMBH binary mergers.  
Over a SMBH binary lifetime, memory undergoes a negligible growth prior to merger (corresponding to the slow time evolution of the binary's inspiral), rapid accumulation of power during coalescence, and eventual saturation to a constant value at ringdown. \citet{BMS13} study a simulation-suite of semi-analytic models for the SMBH  binary population with a signature of a memory signal from a SMBH  binary approximated in the time domain by a step-function centered at the moment of coalescence and assuming a simple power-law model to emulate any environmental interaction which could influence the
coalescence timing (e.g., final parsec problem \citealt{Begelman:1980vb}). SMBH could stall before reaching a regime where GW radiation can drive the
binary to coalesce). Considering SMBH  binaries at $z < 3$ with
masses in the range $(10^5 - 10^7)M_\odot$ and with mass ratios in the range $0.25 - 1$, LISA prospects could be SNR > 5 events occurring 0.3 - 2.8 times per year in the most optimistic environmental interaction model, and less than once per million years in the most pessimistic. As shown in~\citet{Sesana:2007sh,Klein:2015hvg} (and discussed
in the Astrophysics WG White Paper by \citealt{Amaro-Seoane:2022rxf}), most massive BBHs coalescences
whose inspiral signature yields LISA SNR > 5 lie beyond $z = 3$ for a 3-year LISA lifetime, meaning that the results of this study may be interpreted as lower limits on the number of LISA memory events.

\subsection{Burning Questions and Needed Developments}
\label{Sec:GR_burning}

Let us now summarize some important questions that ought to be further investigated in the context of tests of GR, without being however exhaustive. 
\begin{itemize}
    \item Even in theories where black holes deviate from Kerr, or strong field dynamics  deviates from GR, deviation can be very small once consistency and known viability constraints are imposed or once the dependence of any new charge on the mass or the spin are taken into account. Further work is needed to pin down theories and scenarios that could lead to deviations that are observable by LISA. 
    \item LISA inspiral observations will allow to put severe constraints on different non-GR predictions. This, however, will require a more accurate and complete development of GW waveforms including non-GR effects. 
    \item Numerical simulations with a systematic inclusion of non-GR effects for the merger and ringdowns phase have to be performed such as to get more adequate templates to be then used once data will be available. 
    \item Similarly more detailed studies of the EMRIs waveforms including non GR 
effects have to be performed in a systematic way.
    \item GW propagation tests are sensitive to potential couplings and decays of gravitons into other particles. Due to the larger oscillation baseline and lower frequency range LISA could improve substantially present LIGO/Virgo bounds. Present studies are preliminary and further analysis could be useful.
\end{itemize}

\section{Tests of the Nature of BHs}
\label{sec:testsofBHNature}



\subsection{The Kerr hypothesis}


In vacuum GR, the Carter--Robinson \citep{Carter:1971zc,Robinson:1975bv} uniqueness theorem, with later  refinements (see \citealt{Chrusciel:2012jk} for a review), establishes that  the Kerr geometry \citep{Kerr:1963ud} is  the  unique  physically acceptable equilibrium, asymptotically flat BH solution. This led to the more ambitious proposal that, regardless of the initial energy-matter content available in a gravitational collapse scenario, the dynamically formed equilibrium BHs belong to the Kerr family (in the absence of gauge charges) \citep{Ruffini:1971bza}. Accordingly, (near-)equilibrium astrophysical BH candidates are well described by the Kerr metric. This working proposal  is the \textit{Kerr hypothesis}.
Testing the Kerr hypothesis is an important cornerstone of  strong-gravity research in which GW science, and, in particular, LISA, are expected to give key contributions. 

Deviations from the Kerr hypothesis, that we shall  refer to as  \textit{non-Kerrness}, require either modified gravity (discussed in Section~\ref{sec-testsGR} and also below with a different twist) or non-vacuum GR (discussed in subsection \ref{subsec:matmod}). Moreover,  two approaches are, in principle, possible for studying non-Kerrness. The most explored one is theory dependent: to consider specific choices of matter contents or modified gravity models, compute the BH solutions (which, generically, will be non-Kerr), and finally explore the different phenomenology of a given model. The second one is theory agnostic: to consider parametrized deviations from the Kerr metric, regardless of the model they solve (if any). The latter has been fruitfully employed in studying BH phenomenology in stationary scenarios, e.g., \citet{Johannsen:2011dh,Cardoso:2014rha}, but its application in the study of dynamical properties is more limited, given the potential lack of an underlying theory.  

A substantial departure from the Kerr hypothesis is to admit deviations from, or even the absence of, a classical horizon. This leads to hypothetical \textit{exotic compact objects} (ECOs) with a  compactness comparable to that of (classical) BHs (see \citealt{Cardoso:2019rvt} for a review). One motivation for such a dramatic scenario is that, from Penrose's theorem \citep{Penrose:1964wq}, a classical BH (apparent) horizon implies the existence of spacetime singularities, under reasonable energy (and other) conditions. Thus, the absence of (or deviations from) a classical horizon could circumvent the singularity problem. Another motivation is that quantum corrections may be relevant at the horizon scale, even for small-curvature supermassive objects \citep{Almheiri:2012rt,Giddings:2006sj,Lunin:2001jy, Lunin:2002qf, Mathur:2005zp, Mathur:2008nj,Mathur:2009hf,Mayerson:2020tpn}. 

In this context it is also interesting that current LIGO/Virgo
GW observations (especially the recent GW190814; \citealt{Abbott:2020khf} and 
GW190521; \citealt{Abbott:2020tfl,Abbott:2020mjq}, respectively in the lower-mass and upper-mass gap forbidden for standard stellar-origin BHs) do not exclude the possibility that ECOs might co-exist along with BHs and NSs.
Models of ECOs are discussed in Sect.~\ref{Sec:Exotic_Compact_Objects}.

\subsection{Deviations from the Kerr hypothesis} 

\subsubsection{BHs in non-vacuum GR} 
\label{subsec:matmod}


There is a large class of models wherein (covariant) matter-energy is minimally coupled to Einstein's gravity. These models obey the EEP (cf. Sect.~\ref{subsec:ep}) and fall into the realm of GR.

Including minimally coupled matter fields, with standard kinetic terms and obeying some energy conditions (typically the dominant)  can be quite  restrictive for the admissible BH solutions. In many models it prevents the existence  of non-Kerr BHs.  This has been typically established  by model-specific no-hair theorems. Historically influential examples are the Bekenstein no-scalar and no-massive-vector hair theorems \citep{Bekenstein:1972ny} (see \citealt{Herdeiro:2015waa} for a review). Nonetheless, non-standard kinetic terms (e.g. Skyrme hair; \citealt{Luckock:1986tr}), negative energies (e.g. interacting real scalar hair; \citealt{Nucamendi:1995ex}), non-linear matter models (e.g. Yang--Mills hair; \citealt{Bizon:1990sr}) or symmetry non-inheritance between the geometry and matter fields (e.g. synchronised bosonic hair; \citealt{Herdeiro:2014goa,Herdeiro:2016tmi})  allow the existence of new families of BHs with hair\footnote{BH ``hair'' refers to new macroscopic degrees of freedom that are not associated to gauge symmetries, and cannot be computed by flux integrals at infinity.}, co-existing with the vacuum Kerr solution. 

The viability and relevance of any ``hairy'' BH model should be tested by dynamical considerations: besides demanding well posedness of the matter model, the non-Kerr BHs must have a dynamical formation mechanism and be sufficiently stable to play a  role in astrophysical processes. Asymptotically flat BHs with Yang-Mills hair, for instance, are known to be perturbatively unstable \citep{Zhou:1991nu} whereas BHs with Skyrme hair are perturbatively stable \citep{Heusler:1992av}. However, both these fields are best motivated by nuclear physics, in which case the corresponding BH hair is, likely,  astrophysically negligible (except, possibly, for small primordial BHs (PBHs)).

Potentially astrophysically relevant hairy BHs in GR occur in the presence of hypothetical (ultra-light) massive bosonic fields (such as the QCD axion, axion-like particles, dark photons, etc).  
These ultralight fields could be a significant component of 
the dark matter \citep{Arvanitaki:2009fg,Essig:2013lka,Marsh:2015xka,Hui:2016ltb} and are predicted in a multitude of scenarios beyond the standard model of particle physics \citep{Jaeckel:2010ni,Essig:2013lka,Hui:2016ltb,Irastorza:2018dyq}, including extra dimensions and string theories. They naturally interact very weekly and in a model-dependent fashion with baryonic matter, but their gravitational interaction is universal. 
The superradiant instability of Kerr BHs \citep{Brito:2015oca}, in the presence of (complex) ultralight bosonic fields \citep{East:2017ovw,Herdeiro:2017phl}, or mergers of self-gravitating lumps of such ultraligh bosons \citep{Sanchis-Gual:2020mzb} (known as bosonic stars - see Section~\ref{Sec:Exotic_Compact_Objects}) form BHs with synchronised bosonic hair. These BHs are themselves afflicted by superradiant instabilities \citep{Herdeiro:2014jaa,Ganchev:2017uuo}, 
but possibly on long timescales, even cosmologically long \citep{Degollado:2018ypf}, which can render them astrophysically relevant. Superradiance triggered by real ultralight bosonic fields, on the other hand, leads to other effects, such as a SGWB, continuous GW sources from isolated BHs, effects in compact binaries, BH mass-spin gaps, etc, all relevant for LISA science (cf. Section~\ref{Sec:Particle_low_mass}).

\subsubsection{ECOs: deviations from (or absence of) a classical horizon}
\label{Sec:Exotic_Compact_Objects}


ECOs \citep{Giudice:2016zpa} is a generic name for a class of hypothetical dark compact objects without a classical BH horizon that, nonetheless, can mimic the phenomenology of BHs at the classical level. They may be described by their compactness (i.e. the inverse of their --~possibly effective~-- radius in units of the total mass), reflectivity (as opposed to the perfect absorption by a classical BH horizon), and possible extra degrees of freedom related to additional fields \citep{Cardoso:2019rvt}. Their compactness should be comparable to that of BHs; they may be (albeit need not be) \textit{ultracompact}, $i.e.$  possess bound photon orbits, such as light rings. If they do, they could be further classified according to whether the typical light-crossing time of the object is longer or shorter than the instability time scale of circular null geodesics at the photon sphere, which in turn depends on the object compactness and internal composition \citep{Cardoso:2019rvt}.

Several models of ECOs have been conceived in order to overcome conceptual issues associated to BHs, such as their pathological inner structure and the information loss paradox. Under general conditions, Penrose's  theorem \citep{Penrose:1964wq} implies that an apparent horizon always hides a curvature singularity wherein Einstein's theory breaks down. Moreover, in the semi-classical approximation, BHs are thermodynamically unstable and have an entropy which is far in excess of a typical stellar progenitor \citep{Hawking:1976ra}. It has been argued that GWs may provide smoking guns for ECOs \citep{Cardoso:2016rao,Cardoso:2016oxy,Giudice:2016zpa,Barausse:2018vdb,Cardoso:2017cqb,Cardoso:2019rvt}.

ECOs fall into two classes: some models are solutions of concrete field theories coupled to gravity, with known dynamical properties; other models are \textit{ad hoc} proposals (to different extents) put forward to test phenomenological responses without a complete embedding in a concrete model.  In the former case their maximum compactness is constrained by the Buchdahl's theorem, when its hypotheses apply \citep{Cardoso:2019rvt}. In the latter case details about the dynamical formation of the ECOs are unknown. But, as a general principle, it has been argued that quantum effects in the near would-be horizon region could prevent the formation of a horizon in a variety of models and theories \citep{Giddings:2006sj,Lunin:2001jy, Lunin:2002qf,Mazur:2004fk, Mathur:2005zp, Mathur:2008nj,Mathur:2009hf,Mayerson:2020tpn}. 

Amongst the first class of ECOs one of the most studied examples corresponds to {\it bosonic stars}. These are self-gravitating solitons, composed of either scalar \citep{Kaup:1968zz,Ruffini:1969qy,Jetzer:1991jr} or vector \citep{Brito:2015pxa}, massive complex fields, minimally coupled to Einstein's gravity -- see also~\citet{Herdeiro:2017fhv,Herdeiro:2019mbz,Herdeiro:2020jzx} for comparisons. Bosonic stars\footnote{Often called simply \textit{boson stars} in the scalar case and \textit{Proca stars} in the  vector case.} arise in families of models with different classes of self-interactions of the bosonic fields, e.g., \citet{Colpi:1986ye,Schunck:2003kk,Kleihaus:2005me,Grandclement:2014msa,Minamitsuji:2018kof,Guerra:2019srj,Delgado:2020udb} and different field content, e.g., \citet{Alcubierre:2018ahf}; they may also be generalized to modified gravity, e.g., \citet{Herdeiro:2018wvd}. Bosonic stars circumvent Derrick  type no-soliton theorems \citep{Derrick:1964ww} due to a symmetry non-inheritance between matter and geometry, as the latter is static/stationary and the former includes a harmonic time dependence (but with a time-independent energy-momentum tensor). Some bosonic stars are dynamically robust \citep{Liebling:2012fv}, in particular perturbatively stable, with a known formation mechanism known as gravitational cooling \citep{Seidel:1993zk,DiGiovanni:2018bvo}. These can be evolved  in binaries yielding gravitational waveforms,  e.g., \citet{Liebling:2012fv,Palenzuela:2007dm,Palenzuela:2017kcg,Bezares:2017mzk,Sanchis-Gual:2018oui}, that can be used --~together with PN approximations \citep{Pacilio:2020jza}~-- as a basis to produce waveform approximants for GW searches. Their typical GW frequency depends crucially on the mass of the putative ultralight bosonic field and, depending on the range of the latter, the signal can fall in the frequency band of either LIGO/Virgo or LISA. Recently it was argued that one particular GW event, GW190521 \citep{Abbott:2020tfl}, is well mimicked by a very eccentric collision of spinning Proca stars \citep{CalderonBustillo:2020srq}.

Bosonic stars have a cousin family of solitons in the case of \textit{real} bosonic fields, called \textit{oscillatons} \citep{Seidel:1993zk}. They have a weak time-dependence and slow decay, but can be very long lived,  at least for spherical stars \citep{Page:2003rd}. Collisions of oscillatons and the corresponding waveforms have also been obtained, e.g., \citet{Clough:2018exo}.

Bosonic stars are the prototypical example of ECOs which are not meant to replace all BHs in the universe, but could in principle ``co-exist" with them and be exotic sources for LISA. They could also be especially interesting for the BH seed problem at large redshift. Indeed, just like ordinary NSs, bosonic stars have a maximum mass beyond which they are unstable against gravitational collapse and classically form an ordinary BH. Other models that share the same features are anisotropic stars \citep{1974ApJ...188..657B,Letelier:1980mxb,Bayin:1982vw} (see \citealt{Raposo:2018rjn} for a recent fully covariant model). Like bosonic stars, anistropic stars can evade Buchdahl's theorem due to their large anisotropies in the fluid.

A more ambitious first-principle model of ECO --~aiming instead at replacing the classical horizon completely~-- emerges in the \textit{fuzzball proposal} \citep{Lunin:2001jy, Lunin:2002qf, Mathur:2005zp, Mathur:2008nj}. In the latter the classical horizon is replaced by smooth horizonless  geometries with the same mass, charges, and angular momentum as the corresponding BH \citep{Myers:1997qi,Mathur:2005zp,Bena:2007kg,Balasubramanian:2008da,Bena:2013dka}. These geometries represent some of the microstates in the low-energy (super)gravity description. For special classes of extremal, charged, BHs \citep{Strominger:1996sh, Horowitz:1996ay, Maldacena:1997de} one can precisely count the microstates that account for the BH entropy, thus providing a regular, horizonless, microscopic description of a classical horizon. In the fuzzball paradigm, all properties of a BH geometry emerge 
in a coarse-grained description which ``averages'' over the large number of coherent superposition of microstates, or as a `collective behavior' of fuzzballs \citep{Bianchi:2017sds, Bianchi:2018kzy, Bena:2018mpb, Bena:2019azk, Bianchi:2020des}.  Crucially, in this model quantum gravity effects are not confined close to the BH singularity, rather the entire interior of the BH is ``filled'' by fluctuating geometries, regardless of its curvature.  It is worth noticing that, while being among the most motivated models for ECOs, fuzzballs anyway require beyond-GR physics confined at the horizon scale.

While microstate geometries emerge from a consistent low-energy truncation of string theory, other more phenomenological models sharing similar phenomenology have been proposed. For example, gravitational vacuum stars, or \textit{gravastars}, are dark energy stars whose interior spacetime is supported by a negative-pressure fluid which is compensated by a thin shell of an ultrarelativistic positive-pressure fluid. Gravastars are not endowed with an event horizon and have a regular interior. Their model has been conceived in order to overcome the surprisingly huge BH entropy and to provide a model for a thermodynamically stable dark compact object \citep{Mazur:2004fk}.
The negative pressure might arise as a hydrodynamical description of one-loop QFT effects in curved spacetime, so gravastars do not necessarily require exotic new physics \citep{Mottola:2006ew}. In these models, the Buchdahl limit is evaded both because the internal effective fluid is anisotropic \citep{Cattoen:2005he,Raposo:2018rjn} and because the negative pressure violates some of the energy conditions \citep{Mazur:2015kia}.
Gravastars can also be obtained as the BH-limit of constant-density stars, past the Buchdahl limit \citep{Mazur:2015kia,Camilo:2018goy}. In this regime such configurations were found to be dynamically stable \citep{Camilo:2018goy}.

Other models of ECOs include: {\it wormholes} \citep{Morris:1988cz,Visser:1995cc,Lemos:2003jb,Damour:2007ap}, {\it collapsed polymers} \citep{Brustein:2016msz,Brustein:2017kcj}, {\it nonlocal stars} in the context of infinite derivative gravity \citep{Buoninfante:2019swn}, {\it dark stars} \citep{Barcelo:2009tpa}, {\it naked singularities} and {\it superspinars} \citep{Gimon:2007ur}, {\it 2-2 holes} \citep{Holdom:2016nek}, and {\it quasi-BHs} \citep{Lemos:2003gx,Lemos:2008cv} (see \citealt{Carballo-Rubio:2018jzw,Cardoso:2019rvt} for some reviews on ECO models).
Finally, it is worth mentioning that some of the existing proposals to solve or circumvent the breakdown of unitarity in BH evaporation 
involve changes in the BH structure, without doing away with the horizon.
Some of the changes could involve ``soft'' modifications of the near-horizon region, such that the object still looks like a regular GR BH \citep{Giddings:2017mym,Giddings:2013kcj,Giddings:2019}, or drastic changes in the form of ``hard'' structures localized 
close to the horizon such as firewalls and other compact quantum objects \citep{Almheiri:2012rt,Kaplan:2018dqx,Giddings:2019}. A BH surrounded by some hard structure -- of quantum origin such as firewalls, or classical matter piled up close to the horizon -- behaves for many purposes as an ECO.

Despite the wealth of models, ECOs are not without challenges. In addition to the lack of plausible concrete formation mechanisms in many models, there are other generic problems. One issue is that spinning compact objects with an ergoregion but without an event horizon are prone to the ergoregion instability when spinning sufficiently fast \citep{Friedman:1978wla,10.1093/mnras/282.2.580}.
The endpoint of the instability could be a slowly spinning ECO \citep{Cardoso:2007az,Brito:2015oca} or dissipation within the object could lead to a stable remnant \citep{Maggio:2017ivp,Maggio:2018ivz}.
Another potential issue is that ultracompact ECOs which are topologically trivial have not one but at least a pair of light rings, one of which is stable, for physically reasonable matter sources \citep{Cunha:2017qtt}. Such stable light rings have been argued to source a spacetime instability at nonlinear level \citep{Keir:2014oka,Cardoso:2014sna}, whose timescale or endpoint, however, are unclear.  

\subsection{Observables and tests}

\subsubsection{Inspiral-based test with SMBH binaries, IMBH binaries, and EMRIs}


\paragraph{Non-gravitational emission channels by extra fundamental fields} 

An obvious difference between BHs and certain models of ECOs is that the latter could be charged under some gauge fields, as in the case of current fuzzball microstate solutions, quasi-BHs, and potentially other models that arise in extended theories of gravity. These fields might not be electromagnetic and can therefore avoid current bounds on the charge of astrophysical compact objects coming from charge neutralization and other effects \citep{Barausse:2014tra,Cardoso:2016olt}. In addition, their effective coupling might be suppressed, thus evading current constraints from the absence of dipole radiation in BBHs (see Sect.~\ref{sec-testsGR}). A detailed confrontation of given charged ECO models with current constraints on dipolar radiation remains to be done.

\paragraph{Multipolar structure \& Kerr bound}

The multipole moments of a Kerr BH satisfy an elegant relation \citep{Hansen:1974zz}\footnote{For a generic spacetime the multipole moments of order $\ell$ are rank-$\ell$ tensors, ${\cal M}_{\ell m}$ and ${\cal S}_{\ell m}$, which reduce to 
scalar quantities, ${\cal M}_\ell$ and ${\cal S}_\ell$, in the axisymmetric case, see e.g., \citet{Bianchi:2020bxa,Bianchi:2020miz} for the general definitions.} , 
\begin{equation}
 \mathcal{M}_\ell^{\rm BH}+{\rm i }  \mathcal{S}_\ell^{\rm BH}  
 =\mathcal{M}^{\ell+1}\left({\rm i } \chi\right)^\ell\,, \label{nohair}
\end{equation}
where $\mathcal{M}_\ell$ ($\mathcal{S}_\ell$) are the Geroch--Hansen mass (current) 
multipole moments \citep{Geroch:1970cd,Hansen:1974zz}, $\mathcal{M}={\cal M}_0$ is the mass, $\chi\equiv{\mathcal{J} }/{\mathcal{M}^2}$ the dimensionless spin, and 
$\mathcal{J}=\mathcal{S}_1$ the angular momentum. The multipole moments of the Kerr BH are non-trivial, but Eq.~\eqref{nohair} implies that they are completely determined by its mass and spin angular momentum. Thus, there is a multipolar structure, but not multipolar \textit{freedom} (unlike, say, in stars).

Furthermore, introducing the dimensionless quantities $\overline{{\cal M}}_\ell \equiv{\cal 
M}_\ell/{\cal M}^{\ell{+}1}$ and $\overline {\cal S}_\ell \equiv{\cal S}_\ell/{\cal M}^{\ell{+}1}$, the only nonvanishing moments of a Kerr BH are
\begin{equation}
 \overline{{\cal M}}_{2n}^{\rm BH}   = (-1)^n \chi^{2n} \quad, \quad
 \overline{{\cal S}}_{2n{+}1}^{\rm BH} = (-1)^n \chi^{2n{+}1} \label{momKerr}
\end{equation}
for $n=0,1,2,...$. The fact that ${\cal M}_\ell=0$ 
(${\cal S}_\ell=0$) when $\ell$ is odd (even) is a consequence of the equatorial symmetry of the Kerr metric, whereas the fact that all multipoles with $\ell\geq2$ are proportional to (powers of) the spin --~as well as their 
specific spin dependence~-- is a peculiarity of the Kerr metric.

Non-Kerr compact objects (BHs or ECOs) will have, in general, a different multipolar structure. Differences will be model dependent, but can be considerable in some cases, e.g. for boson stars \citep{Ryan:1996nk} and BHs with synchronised scalar hair \citep{Herdeiro:2014goa}. For ECOs, the tower of multipole moments is, in general, richer. The deformation of each multipole depends on the specific ECO's structure, and in general vanishes in the high-compactness limit, approaching the Kerr value \citep{Pani:2015tga,Glampedakis:2017cgd,Raposo:2018xkf,Raposo:2020yjy}. 
In particular, a smoking gun of the ``non-Kerrness" of an object would be the presence of moments that break the equatorial symmetry (e.g. the current quadrupole ${\cal S}_2$ or the mass octopole ${\cal M}_3$), or the axisymmetry (e.g. a generic mass quadrupole tensor ${\cal M}_{2m}$ with three independent components ($m=0,1,2$), as in the case of multipolar boson stars \citep{Herdeiro:2020kvf} and of fuzzball microstate geometries \citep{Bena:2020see,Bianchi:2020bxa,Bena:2020uup,Bianchi:2020miz}.

The multipolar structure of an object leaves a footprint in the GW signal emitted during the coalescence of a binary system, modifying the PN 
structure of the waveform at different orders. The lowest order contribution, entering at 2PN order is given by the intrinsic (typically spin-induced) quadrupole moment \citep{Barack:2006pq}. LISA can be able to detect deviations in the multipole 
 moments from supermassive binaries for comparable and unequal mass systems. So far proposed tests of the Kerr nature have been based on constraints of the spin-induced quadrupole $M_2$ \citep{Barack:2006pq,Krishnendu:2017shb}, spin-induced octopole $S_3$ \citep{Krishnendu:2019ebd}, and current quadrupole $S_2$ \citep{Fransen:2022jtw}.
 
 GW signals emitted by EMRIs will provide accurate measurements of the spin-induced quadrupole at the level of one part in $10^4$ \citep{Barack:2006pq,Babak:2017tow}, and of the equatorial symmetry breaking current quadrupole at the level of one part in $10^2$ \citep{Fransen:2022jtw}. A rather generic attempt to constrain the multipole moments of an axisymmetric and equatorially symmetric central object with EMRIs has been done in \citet{Ryan:1995wh,Ryan:1997hg} by mapping gauge-invariant geodesic quantities into multipole moments in a small-orbital velocity expansion. Constraining the radiative multipole moments of the entire binary system has been discussed in \citet{Kastha:2018bcr,Kastha:2019brk}; these constrain deviations from the GR expectation of the binary system without explicitly parametrizing the compact objects' multipole structure.

\paragraph{Tidal heating}

The compact objects in the binary produce a tidal field on each other which grows as the bodies approach their final plunge and merger. If the bodies dissipate some amount of radiation, these tides backreact on the orbit, transferring rotational energy from their spin into the orbit. This effect is known as tidal heating. For BHs, energy and angular momentum absorption by the horizon is responsible for tidal heating.
This effect is particularly significant for highly spinning BHs and mostly important in the latest stages of the inspiral. Tidal heating can contribute to thousands of radians of accumulated orbital phase for EMRIs in the LISA band \citep{Hughes:2001jr,Bernuzzi:2012ku,Taracchini:2013wfa,Harms:2014dqa,Datta:2019euh,Datta:2019epe,Maggio:2021uge}. If at least one binary member is an ECO instead of a BH, dissipation is in general smaller than in the BH case or even negligible, therefore significantly reducing the contribution of tidal heating to the GW phase. This would allow to distinguish between BBHs and binary involving other compact objects. For LISA binaries, constraints of the amount of dissipation would be stronger for highly spinning objects and for binaries with large mass ratios \citep{Maselli:2017cmm,Datta:2019epe,Maggio:2021uge}. For EMRIs in the LISA band, this effect could be used to put a very stringent upper bound on the reflectivity of ECOs, at the level of $0.01\%$ or better \citep{Datta:2019epe,Maggio:2021uge}. Absence of a horizon can also produce resonances that can be excited during EMRIs \citep{Pani:2010em,Macedo:2013jja,Cardoso:2019nis,Maggio:2021uge}.

\paragraph{Tidal deformability}

Tidal effects in compact binaries modify the dynamical evolution of the system, accelerating the coalescence. This modifies the orbital phase, and then in turn the GW emission \citep{poisson2014gravity,Hinderer:2018mrj}. The imprint on the waveform is encoded in a set of quantities which, as a first approximation, can be assumed to be constant during the coalescence \citep{Maselli:2012zq,Hinderer:2016eia,Steinhoff:2016rfi}: the {\rm tidal Love numbers} \citep{Hinderer:2007mb,Damour:2009vw,Binnington:2009bb}. These numbers can be thought of as the specific multipole moment induced by an external tidal field, in a way akin to the electric susceptibility in electrodynamics. The main contribution in the GW signal from a binary is given by the quadrupolar term $k_2$, connected to the tidal deformability $\lambda=\frac{2}{3}k_2 R^5$, or in its dimensionless form $\tilde{\lambda}=\frac{2}{3} k_2 {\cal C}^5$, where $R$ and ${\cal C}$ are the object radius and compactness, respectively.
 
The tidal Love numbers depend on the internal composition of the central object. So far, they have been used to constrain the properties of the nuclear equation of state through GW observations of binary NSs \citep{Abbott:2018exr}. For a fixed equation of state, i.e. composition, the Love numbers depend on the object compactness only.

The tidal Love numbers of a BH in GR are precisely zero. This was shown explicitly for Schwarzschild BHs, for both small \citep{Binnington:2009bb,Damour:2009vw} and  large \citep{Gurlebeck:2015xpa} tidal fields. The same result was shown to be valid for slowly rotating BHs up to the second (linear) order in the spin for axisymmetric (generic) tidal fields \citep{Pani:2015hfa,Pani:2015nua,Landry:2015zfa,Poisson:2014gka}.
Very recently, this result was extended to {\it any} tidal Love number of a Kerr BH with {\it arbitrary} spin (see \citealt{Chia:2020yla,LeTiec:2020spy,LeTiec:2020bos,Hui:2020xxx,Charalambous:2021kcz,Charalambous:2021mea} for literature on this topic).
 
For ECOs, the tidal Love numbers are generically different from zero. In analogy with the NS case they depend on the ECO's structure, and may be used to trace back the underlying properties of each model \citep{Pani:2015tga,Uchikata:2016qku,Porto:2016zng,Cardoso:2017cfl,Sennett:2017etc,Maselli:2017cmm,Johnson-McDaniel:2018uvs,Maselli:2018fay,Raposo:2018rjn,Giddings:2019ujs,Herdeiro:2020kba}. For nonrotating BH mimickers, featuring corrections at the horizon scale and that approach the BH compactness, the Love numbers vanish in the limit ${\cal C}_\textnormal{ECO}\rightarrow {\cal C}_\textnormal{BH}$, often logarithmically \citep{Cardoso:2017cfl}.
 
LISA will be able to measure the tidal Love numbers of BH mimickers \citep{Maselli:2017cmm}, which are otherwise unmeasurable by current and future ground based detectors \citep{Cardoso:2017cfl}. In the comparable-mass case, this measurement requires highly-spinning supermassive ECO binaries up to $10\,{\rm Gpc}$. LISA may also be able to perform model selection between different families of BH mimickers \citep{Maselli:2018fay}, although this will in general require detection of golden binaries (i.e.~binaries with a very large SNR) \citep{Addazi:2018uhd}. For a large class of slowly-rotating ECOs with compactness ${\cal C} \lesssim 0.3$,   LISA can measure the Love numbers with very good accuracy below 1\% \citep{Cardoso:2017cfl}. For (scalar) boson stars a recent study proposed a new data analysis strategy to consistently include several corrections (multipolar structure, tidal heating, tidal Love numbers) in the inspiral signal from boson star binaries, improving the accuracy on the measurement of the fundamental parameters of the theory by several orders of magnitude compared to the case in which the effects are considered independently \citep{Pacilio:2020jza}.

Finally, EMRI observations can set even more stringent constraints, since the measurement errors on the Love number scale as $q^{1/2}$, where $q\ll 1$ is the mass ratio of the binary \citep{Pani:2019cyc}. A simplistic Newtonian estimate (that should be corroborated by a more sophisticated modelling and data analysis) suggests that in this case the tidal Love number of the central object can be constrained at the level of one part in $10^5$ \citep{Pani:2019cyc}.

\paragraph{Integrability/Chaos} 

One particular probe of extreme gravity that is tailor-made for EMRI signals relates to chaos. For Hamiltonian systems, chaos refers to the non-integrability of the equations of motion, i.e. the non-existence of a smooth analytic function that interpolates between orbits, and has nothing to do with a system being non-deterministic \citep{Levin:2006zv}.

EMRIs in GR can be approximated, to zeroth-order, as geodesics of the Kerr spacetime, and the latter has enough symmetries to guarantee that geodesics are completely integrable and thus non-chaotic. Beyond the zeroth-order approximation, however, other effects, such as the spin of the small compact object, could break the integrability of the systems even within GR \citep{Zelenka:2019nyp}. In some modified theories, even the geodesic orbital motion might not be integrable \citep{LukesGerakopoulos:2010rc,Cardenas-Avendano:2018ocb}. This is also true for some models of ECOs, such as spinning scalar boson stars and non-Kerr BHs in GR \citep{Cunha:2016bjh}. In this sense, the presence of large chaotic features in the GWs emitted by EMRIs could signal a departure from the SEP, a violation of the Kerr hypothesis, or an environmental effect. 

Modifications to GR are expected to change the fundamental frequencies of the orbital motion of test particles, which will be then be imprinted on the GWs emitted by the system. A careful study of the evolution of these fundamental frequencies will allow us to understand the importance of chaos to GR and to the observations of GWs from EMRIs \citep{Gair:2007kr, LukesGerakopoulos:2010rc,Cardenas-Avendano:2018ocb,Destounis:2021mqv}.

\paragraph{Motion within ECOs}

If the ECO interior is made of weakly-interacting matter, a further discriminator of the absence of a horizon (or of a hard surface) would be the motion of 
test particles {\it within} the object and its peculiar GW signal, most notably as in the case of an EMRI moving inside a supermassive ECO. This motion can be driven by a combination of the self-gravity of the central object, accretion, and dynamical friction, etc. The study of geodesic motion inside solitonic boson stars was analyzed in e.g., \citet{Kesden:2004qx}. The effects of accretion and drag were included in \citet{Macedo:2013qea,Macedo:2013jja,Barausse:2014tra,Barausse:2014pra}. These effects are model independent to a certain extent, since they mostly depend on the density profile. For this reason they also share some similarities with environmental tests of dark matter (see Sect.~\ref{Sec:DM_and_PBHs}).
In general, they could be a smoking-gun signature for the existence of structures in supermassive ultracompact objects.

\subsubsection{Ringdown tests}


\paragraph{QNMs}

Similarly to what was discussed in Sect.~\ref{sec-testsGR}, measuring the ringdown modes in the post-merger signal of a binary coalescence provides a clean and robust way to test GR and the nature of the remnant. If the latter is a Kerr BH in GR, its (infinitely countable) QNM spectrum is entirely determined only in terms of its mass and spin. Thus, detecting several QNMs provides us with multiple independent null-hypothesis tests, and would allow us to perform GW spectroscopy \citep{Kokkotas:1999bd,Berti:2009kk}.
From a more theoretical perspective, the study of the QNMs of compact objects is crucial to assess their linear stability.

The ringdown waveform originates from the perturbed remnant object, and consists of a superposition of (complex) QNMs, whose amplitudes depend on the binary progenitors and on the underlying theory.
As previously discussed, the fundamental QNM frequency and damping time have been measured by LIGO/Virgo only for a few events, providing an independent measurement of the mass and spin of the remnant which is in agreement with what inferred from the inspiral-merger phase \citep{LIGOScientific:2019fpa,Abbott:2020jks}. Among the entire second GW transient catalogue \citep{Abbott:2020jks} the first GW event, GW150914, remains among those for which the fundamental QNM of the remnant has been measured with the highest precision (roughly $3\%$ and $7\%$ for the frequency and damping time, respectively).
More recently, the importance of overtones has attracted considerable attention, especially because they allow one to start the fitting of the ringdown signal closer to the peak of the signal, improving mass and spin measurements \citep{Isi:2019aib}. Overtones are particularly useful for tests of GR with equal-mass binaries (for which other angular modes can be suppressed) \citep{Bhagwat:2019dtm,Ota:2019bzl,Forteza:2020hbw}, but a detailed study for LISA remains to be done. 
Overall, tests of the no-hair theorem rely also on the ability to estimate the starting time of the ringdown when the signal is dominated by the QNMs of the remnant and on the modelling of higher modes \citep{Baibhav:2017jhs,Bhagwat:2017tkm,Brito:2018rfr,Giesler:2019uxc,Bhagwat:2019dtm,Ota:2019bzl,Forteza:2020hbw}.

The large SNR expected in LISA for ringdown signals of SMBH coalescences provides a unique opportunity to perform BH spectroscopy \citep{Dreyer:2003bv} and tests of the nature of the remnant. For a single "golden merger" up to redshift $z=10$ several QNMs can be measured with unprecedented precision \citep{Berti:2016lat}.

Besides introducing deformations in the QNM spectrum, if the remnant differs from a Kerr BH in GR, some further clear deviations in the prompt ringdown are: (i) possible presence of (or contamination from) other modes, e.g. fluid modes \citep{Pani:2009ss} in stars or extra degrees of freedom (e.g. scalar QNMs for boson stars; \citealt{Macedo:2013jja}), some of which --~being at low frequency~-- could be resonantly excited during the inspiral \citep{Pani:2010em,Macedo:2013jja,Cardoso:2019nis,Maggio:2021uge}; (ii) isospectrality breaking between modes that can be identified as even-parity and odd-parity in the zero-spin limit \citep{Maggio:2020jml}. This produces a characteristic “mode doublet” in the ringdown. A generic framework to study the ringdown of a dark compact object was recently proposed in \citet{Maggio:2020jml} by extending the BH membrane paradigm to ECOs.

\paragraph{Echoes}

GW echoes \citep{Cardoso:2016rao,Cardoso:2016oxy} in the post-merger signal of a compact binary coalescence might be a clear signature of near-horizon quantum structures \citep{Cardoso:2016rao,Cardoso:2016oxy,Abedi:2016hgu,Barcelo:2017lnx,Oshita:2018fqu,Wang:2019rcf}, ultracompact objects \citep{Cardoso:2016rao,Bueno:2017hyj}, exotic states of matter in ultracompact stars \citep{Ferrari:2000sr,Pani:2018flj,Buoninfante:2019swn}, 
and of modified theories of gravity \citep{Burgess:2018pmm,Buoninfante:2019teo,Delhom:2019btt} 
(see \citealt{Cardoso:2017cqb,Cardoso:2017njb,Cardoso:2019rvt,Abedi:2020ujo} for some recent reviews). 
Detecting echoes would give us the tantalizing prospect of probing the near-horizon structure of dark compact objects with the hope, in particular, to shed light on putative quantum properties of BHs \citep{Ikeda:2021uvc}.

If sufficiently compact, horizonless objects support quasi-bound modes trapped within their photon sphere \citep{Kokkotas:1999bd,Kokkotas:1995av,Cardoso:2016rao,Cardoso:2016oxy}. For ultracompact objects the prompt ringdown is identical to that of a BH, since the signal is initially due only to the perturbation of the photon sphere, whereas the BH horizon is reached in infinite coordinate time \citep{Cardoso:2016rao,Cardoso:2016oxy}. At late times, a modulated train of GW echoes appears as a result of multiple reflections of the GWs between the object interior and the photon sphere, leaking out to infinity at each reflection.
For the case of intermediate compactness, the prompt ringdown can show some differences with the BH case due to the interference with the first GW echoes \citep{Maggio:2020jml}.

The delay time between echoes is related to the compactness of the object through a logarithmic dependence, which allows for tests of Planckian corrections at the horizon scale \citep{Cardoso:2016oxy,Cardoso:2017cqb,Abedi:2020ujo, Oshita:2020dox}. The damping factor of subsequent echoes is related to the reflective properties of the compact object \citep{Price:2017cjr,Maggio:2018ivz,Cardoso:2019rvt,Maggio:2020jml}.

Several waveform templates for echo searches in LIGO/Virgo data have been developed, including: (i)~templates in time domain based on standard IMR templates with additional parameters \citep{Abedi:2016hgu,Nakano:2017fvh, Wang:2018gin}; (ii)~superposition of sine-Gaussians with free parameters \citep{Maselli:2017tfq}; (iii)~frequency-domain templates \footnote{Echo templates available at \noindent\url{http://www.darkgra.org}, \noindent\url{https://web.uniroma1.it/gmunu/}. }
based on the physical ECO parameters \citep{Mark:2017dnq, Testa:2018bzd, Maggio:2019zyv}. The former were developed for matched-filtered searches. In addition, unmodelled searches based on wavelets adapted from burst searches \citep{Tsang:2018uie,Tsang:2019zra} and on Fourier windows \citep{Conklin:2017lwb, Conklin:2019fcs} have been proposed. For a review on modelling and echo searches, see  \citet{Abedi:2020ujo}.

LIGO/Virgo O1 and O2 events triggered some controversial claims on hints of GW echoes detection. Independent searches found evidence for GW echoes in the O1-O2 events \citep{Abedi:2016hgu, Conklin:2017lwb, Abedi:2018npz}. However, a low statistical significance of such events has been claimed \citep{Westerweck:2017hus, Nielsen:2018lkf}, followed by more recent negative searches \citep{Tsang:2019zra, Uchikata:2019frs, Lo:2018sep}. Very recently, using a simplistic template, a dedicated search for echoes has been performed by the LIGO/Virgo Collaboration using the second GW transient catalogue (GWTC-2) \citep{Abbott:2020jks}, finding no evidence for echoes. This is consistent with independent studies using physically motivated templates, suggesting that that models  with almost perfect reflectivity can be excluded/detected with current instruments ($5\sigma$ confidence level with SNR in the ringdown of $\approx 10$), whereas probing values of 
the reflectivity smaller than $80\%$ at $3\sigma$ or more confidence level requires SNRs of $\mathcal{O}(100)$ in the post-merger phase \citep{Testa:2018bzd, Maggio:2019zyv}.
This makes LISA particularly well suited for echo searches.

\paragraph{GWs as messengers from the quantum world}
The high sensitivity of LISA or of other advanced detectors will also
allows to advance our understanding of quantum gravitational effects.
One of the most long-standing ideas in this context is the proposal that
the area of BH horizons is quantized in units of the Planck area
$A=\alpha\ell_p^2 N$,
where $N$ is an integer characterizing the BH quantum state, $\alpha$ is
an ${\cal O}(1)$ dimensionless coefficient and $\ell_p$ is Planck's
length \citep{Bekenstein:1974jk,Mukhanov:1986me,Bekenstein:1995ju}.
Transitions between states occur, surprisingly, at frequencies where
ground- and space-based detectors operate. Thus, quantum BHs may
have different tidal heating
properties \citep{Cardoso:2019apo,Agullo:2020hxe} or different tidal Love
numbers \citep{Brustein:2020tpg}, and the coalescence of two BHs
can lead to late-time echoes in the waveforms \citep{Cardoso:2019apo,Agullo:2020hxe}.

\subsection{Burning Questions and Needed Developments}
\label{Sec:BH_burning}

Several issues discussed in this section need further detailed studies, among which we list the following:
\begin{itemize}
    \item A better understanding of the consequences of superradiant instabilities triggered by  e.g. real ultralight bosonic fields on the SGWB, on continuous GW sources, and on binaries, which could be detected by LISA. 
    \item A comprehensive study of possible deviations of multipole moments for SMBH binaries is important to have a good tool at disposal for testing the Kerr nature of the BHs.
    \item First-principle models of (possibly stable) ultracompact objects should be developed to provide viable candidates for a BH mimicker. 
    \item In general for testing the different effects such as tidal heating, tidal deformability, multipole moments, resonances, and ringdown effects, and echoes there is need to develop quite accurate waveforms, with several parameters describing possible deviations, and to perform accurate statistical analysis (e.g. Bayesian model selection), especially in the case of a putative signature of new physics in the data. Clearly, this also requires to develop very accurate GR waveforms.
\end{itemize}

\section{Dark Matter and Primordial BHs}
\label{Sec:DM_and_PBHs}

\vspace{0.25cm}

The fundamental nature of dark matter (DM) is one of the major unresolved questions in physics and cosmology. Its presence is inferred from various observations, such as galaxy rotation curves, gravitational lensing, galaxy cluster dynamics and CMB data (see \citealt{Bertone:2004pz, Bertone:2016nfn, Carr:2016drx} for reviews). These observations constrain its behavior on larger scales: DM must broadly behave as non relativistic, collisionless matter, with an average density in galactic haloes of the order of $\sim 0.1 ~ {\rm M_\odot ~ pc^{-3}}$ ($1 ~ \rm{GeV ~ cm^{-3}}$), low velocity dispersions of order $\sim 100 ~{\rm km ~s^{-1}}$, and it must be interacting weakly, if at all, with baryonic matter and with itself. However, the physical properties of its constituents, in particular their individual masses and spins, remain very poorly constrained. 

\begin{figure}[h]
\begin{center}
\includegraphics[width=0.98\textwidth]{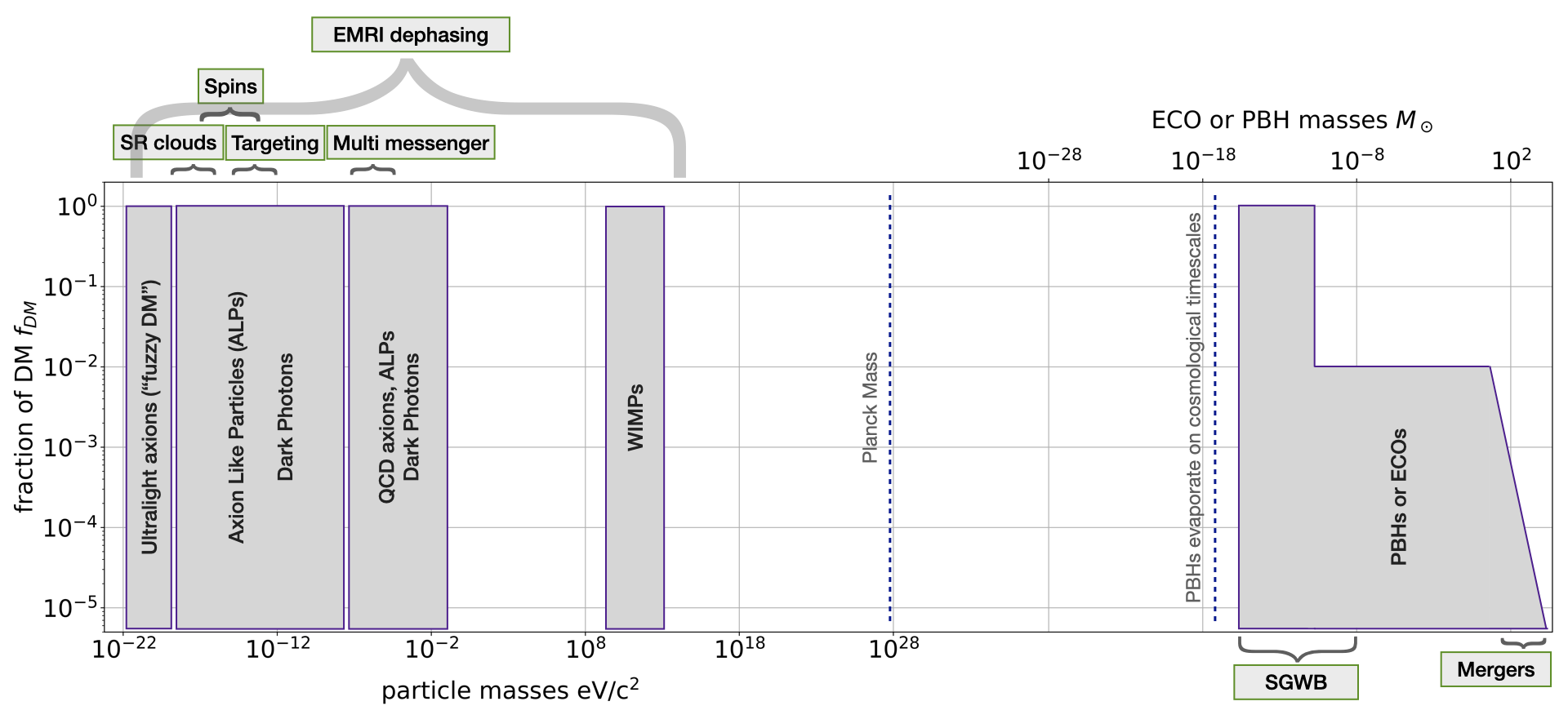}
\end{center}
\caption{Overview of key DM candidates \citep{Bertone:2019irm, Bertone:2018krk, Giudice:2017dde}. Currently excluded ranges of particle masses and DM fractions $f_{DM}$ are indicated in white. Parameter regions where LISA observations may provide constraints are gray coloured. We see that the masses of DM particle candidates remain currently very much unconstrained and that DM particles of any allowed mass can make up any fraction ($f_{DM} \leq 1$) of the DM. Macroscopic DM candidates such as PBHs and ECOs are constrained by current observations but may still constitute 100\% of the DM in the range $10^{-16} - 10^{-11} M_\odot$. Approximate regions where LISA can contribute to constraints are bracketed above and below the plot with a brief description - see the text for more details.}
\label{fig:LISA_DM}
\end{figure}

This Section is concerned with models where the DM consists of or is formed from (in the case of PBHs) some type of matter and not where the observed effects arise due to some modification to gravity, such as e.g., Modified Newtonian Dynamics (MOND) \citep{Sanders:2002pf}, which is the topic of Section \ref{sec-testsGR}. There are two broad categories of dark matter:
\begin{enumerate}
    \item Particle dark matter, e.g. WIMPs or axions, where the problem of identification lies in determining the mass, spin and fundamental interactions of the particle (or particles).
    \item Macroscopic objects, e.g. primordial BHs (PBHs) or exotic compact objects (ECOs), where the distributions of masses and spins, as well as the matter from which they formed, are key aspects to be understood.
\end{enumerate}
Nothing prevents DM from being a combination of several different components drawn from either or both of these categories. Moreover, some models (e.g. axions), are naturally hybrids, where the DM may primarily consist of unbound particles, but also form macroscopic gravitationally bound structures such as boson stars in overdense regions. A further aspect of DM on which observations may shed light is in identifying possible formation mechanisms, e.g. signatures of phase transitions, which often motivate particular DM models.

Several previous reviews have considered the potential for GWs to shed light on the nature of dark matter \citep{Bertone:2019irm, Bertone:2018krk, Giudice:2017dde}. Further, a complementary white paper by the LISA cosmology WG assesses the prospects of GW observations of DM from a cosmological angle. In this section we focus on the potential for LISA specifically to identify and constrain the particle nature of dark matter and its interactions. We summarise what is known to be possible, highlight promising areas which require further investigation, and detail the work that should be done prior to launch to enable us to correctly interpret the data that we will obtain. Fig. \ref{fig:LISA_DM} summarises the key candidates and regions which may be probed by LISA observations.

Before we proceed we note two caveats. First, effects described in this section may be degenerate with environmental effects described in Sect.~\ref{Sec:Astrophysics_Environmental_effects} and in the waveform modling white paper. Breaking this degeneracy will be a key factor in our ability to accurately identify DM signatures. Second, we have in most cases been optimistic in assuming that LISA measurements will be optimized for the searches we describe and that models can allow for sufficiently precise predictions to proceed. Turning these caveats around, they could both be regarded as a starting point for action items for future work in order to ensure LISA's science potential will be fully exploited. 

We now present a number of interesting particle candidates, or rather categories of candidates, that have been explored so far, without making the above qualifications necessarily explicit in each section. 
We start by reviewing particle candidates, separated into low mass (sub eV) and high mass cases. This roughly divides cases where wave-like and particle-like behaviour respectively would dominate on astrophysically significant scales. We then turn to macroscopic candidates such as PBHs, where we separate candidates that would clearly be of primordial origin (sub $M_\odot$) from those that could arise from stellar evolution. The last two sections review the potential for enhanced signatures from additional fundamental interactions.

\subsection{Low mass particles $m < eV$}
\label{Sec:Particle_low_mass}

In the past few years, the possibility that dark matter could be composed of ultralight bosons, with masses ranging anywhere between $10^{-22}\,{\rm eV}\lesssim 1$ eV, has become a popular hypothesis \citep{Ferreira:2020fam}. A non-exhaustive list of theoretical models predicting the existence of such particles includes the well-known QCD axion proposed to solve the strong CP problem of QCD \citep{Peccei:1977hh,Wilczek:1977pj}, axion-like particles arising in ``string axiverse'' scenarios \citep{Arvanitaki:2009fg}, dark photons \citep{Goodsell:2009xc,Nelson:2011sf} and ``fuzzy dark matter'' \citep{Hui:2016ltb}.

Quite remarkably, GW observatories such as LISA could be used to constrain or find evidence for the existence of these particles. This stems from the fact that rotating BHs can become unstable against the production of light bosonic particles, if such particles exist in nature \citep{Detweiler:1980uk,Damour:1976kh,Dolan:2007mj,Witek:2012tr,Pani:2012vp,Pani:2012bp,Brito:2013wya,Shlapentokh-Rothman:2013ysa,Moschidis:2016wew,East:2017mrj,Cardoso:2018tly,Frolov:2018ezx,Dolan:2018dqv,Brito:2020lup}, through a process known as BH superradiance \citep{zeldovich1,zeldovich2,Press:1972zz,Teukolsky:1974yv,Brito:2015oca}.  The instability spins the BH down, transferring up to a few percent of the BH's mass and angular momentum to the boson field,  forming a long-lived bosonic ``cloud'' outside the horizon \citep{Arvanitaki:2009fg,Arvanitaki:2010sy,Brito:2014wla,Arvanitaki:2014wva,Herdeiro:2014goa,Brito:2015oca,Arvanitaki:2016qwi,Baryakhtar:2017ngi,East:2017ovw,East:2018glu,Ficarra:2018rfu,Degollado:2018ypf}.  Superradiance is most effective when the boson's Compton wavelength is comparable to the BH's gravitational radius \citep{Dolan:2007mj,Witek:2012tr}, meaning that observations of astrophysical BHs in the supermassive to the stellar mass range, allow us to probe bosons with masses in the range $10^{-21}$~--~$10^{-10}$ eV.

For real bosonic fields, the cloud dissipates over long timescales through the emission of nearly-monochromatic GWs with a typical frequency $f\sim 2 c^2 m_b/h$, where $m_b$ is the mass of the field \citep{Arvanitaki:2009fg,Arvanitaki:2010sy,Yoshino:2013ofa,Arvanitaki:2014wva,Arvanitaki:2016qwi,Brito:2017zvb,Brito:2017wnc,Baryakhtar:2017ngi,East:2018glu,Siemonsen:2019ebd}. These signals could be observable individually or as a very strong SGWB \citep{Arvanitaki:2014wva,Arvanitaki:2016qwi,Baryakhtar:2017ngi,Brito:2017zvb,Brito:2017wnc,Tsukada:2018mbp,Isi:2018pzk,Ghosh:2018gaw,DAntonio:2018sff,Sun:2019mqb,Zhu:2020tht,Brito:2020lup,Tsukada:2020lgt}, and could thus be used to constrain the existence of light bosons in the absence of a detection. Current and future Earth-based detectors could detect GWs emitted by bosons in the range $m_b\sim 10^{-14}$--$10^{-11}$~eV, while LISA could be sensitive to bosons of mass $m_b\sim 10^{-19}$--$10^{-15}$~eV. It has also been proposed that LISA observations of BBHs with total masses in the range $\sim [100,6000]\,M_{\odot}$ will provide sufficient information to enable targeted searches for these monochromatic GW signals with future ground-based GW detectors for masses in the range $m_b\sim 10^{-14}$--$ 10^{-12}$~eV \citep{Ng:2020jqd}. 

Furthermore, since BHs affected by the instability would spin down in the process, accurate measurement of the spin of astrophysical BHs can be used to strongly constrain, or find evidence for, ultralight bosons \citep{Arvanitaki:2010sy,Pani:2012vp,Pani:2012bp,Brito:2013wya,Arvanitaki:2014wva,Arvanitaki:2016qwi,Baryakhtar:2017ngi,Brito:2017zvb,Cardoso:2018tly,Ng:2019jsx}. Requiring that the instability acts on timescales shorter than known astrophysical processes, such as accretion and mergers, current measurements of SMBHs spin using continuum fitting or the Iron K$\alpha$  (see e.g., \citealt{Jovanovic:2011tx}) can probe bosons in the mass range $m_b\sim 10^{-20}$--$10^{-17}$~eV, whereas similar measurements for stellar-mass BHs probe the mass range is $m_b\sim 10^{-12}$--$10^{-11}$~eV. LISA is expected to detect binaries and measure the spins of MBHs in the range $\sim 10^2$--$10^7\,M_{\odot}$ and will therefore be able to constrain the existence of light bosons in the intermediate range $m_b\sim 10^{-16}$--$10^{-13}$~eV \citep{Brito:2017zvb,Cardoso:2018tly,Brito:2020lup}. 

When the BH-boson cloud system is part of a binary, one may also expect new phenomena that could leave imprints in the GW signal emitted by the binary system. For example, the presence of a companion can result in resonant transitions between energy levels of the cloud \citep{Baumann:2018vus} or complete tidal disruption \citep{Cardoso:2020hca}, which could lead to a significant dephasing of the binary's GW signal. In particular, for EMRIs, resonant transitions could lead to long-lived floating orbits \citep{Ferreira:2017pth, Zhang:2018kib,Baumann:2019ztm}. The presence of a cloud would also be imprinted in the binary’s GW signal through its spin-induced multiple moment(s) and tidal Love number(s) \citep{Baumann:2018vus,Baumann:2019ztm,DeLuca:2021ite}. For eccentric orbits, resonances can occur over a larger range of frequencies, extending the possibility of detection \citep{Berti:2019wnn}. In the case of EMRIs, dynamical friction and the gravitational pull of the cloud may also leave sizable imprints on the GW waveform \citep{Macedo:2013qea,Ferreira:2017pth,Zhang:2019eid,Hannuksela:2018izj}. 

Ultralight scalars can also form self-gravitating structures. These
``dark matter stars'' are  candidates to describe (dark) haloes
in the central part of galaxies \citep{Hui:2016ltb}. In this context, it
is crucial to understand how such structures respond to external
excitations caused by compact objects (e.g. stars or BHs) living
in these environments. Stability, proper modes, excitation of
resonances, and depletion of the scalar through radiation are amongst
the important issues that must be addressed
here \citep{Annulli:2020ilw,Annulli:2020lyc}. 
There is also the potential for distinctive DM density enhancements due to accretion of DM onto BHs \citep{Hui:2019aqm, Clough:2019jpm, Bamber:2020bpu}, which is analogous to the case of dark matter spikes for high mass candidates (see Sect.~\ref{Sec:Particle_high_mass} below). However, such density enhancements are estimated to be several orders of magnitude weaker than those from a superradiant build up, and concentrated near the BH horizon. They are therefore below LISA sensitivity in the inspiral phase, but may have an impact on merger, something which is yet to be tested using NR simulations.
On the other hand, it is
also essential to know how the compact objects themselves are affected
by the surrounding dark matter. GW dephasing through
dynamical friction and scalar radiation deserves a proper and rigorous
exploration \citep{Hui:2016ltb,Annulli:2020ilw,Annulli:2020lyc}. In
particular, for sources in the LISA band, it was recently found that
scalar radiation from these structures affects the gravitational
waveform at leading -6 PN order with respect to the dominant quadrupolar
term \citep{Annulli:2020ilw,Annulli:2020lyc}. 

In summary, LISA has the potential to detect or constrain ultralight DM fields in regions of parameter space complementary to those covered by ground-based GW observations, although more work is needed to build waveform models that incorporate the effects of such fields and are sufficiently accurate to be used for data analysis purposes. It will also be important to explore in more detail BBHs with ultralight boson clouds with full NR simulations in order to understand how the presence of such boson clouds could be imprinted in the late stages of BBH mergers.

\subsection{High mass particles $m > eV$}
\label{Sec:Particle_high_mass}

We now discuss what we could learn with LISA about more massive cold DM particles. If one or both the components of a binary system are surrounded by a dense dark matter overdensity, the binary's inspiral will be driven by both dynamical friction \citep{Chandrasekhar:1943ys} and GW emission. The densities required to produce an effect observable by LISA are typically higher than the characteristic DM densities in most galaxies \citep{Barausse:2014tra,Cardoso:2019rou}.
There are however several  mechanisms by which the DM density can be enhanced to such higher levels, notably when a smaller seed BH grows in mass adiabatically in a DM halo to produce a DM ``spike'' \citep{Quinlan:1994ed,Gondolo:1999ef,Ullio:2001fb,Sadeghian:2013laa,Ferrer:2017xwm}. DM spikes around SMBHs at the center of galactic halos are prone to disruption by a number of processes, such as major mergers and gravitational interactions with stellar cusps \citep{Merritt:2002vj,Bertone:2005hw}. DM spikes around IMBHs might be more stable over cosmological timescales, and are arguably more promising targets for indirect and GW searches \citep{Bertone:2005xz}.
The density profile of spikes depends on the nature of DM, with warm, self-interacting, or self-annihilating DM candidates leading to shallower profiles with respect to cold, collisionless DM \citep{Gondolo:1999ef,Bertone:2005hw,Shapiro_2016,Hannuksela:2019vip,alvarez2021density}.

The faster rate of inspiral in the presence of DM can be distinguished from the slower inspiral in vacuum, thereby allowing LISA to infer the presence of DM around a binary \citep{Macedo:2013qea,Barausse:2014tra,Eda:2014kra}.
The prospects for detecting  DM spikes through their influence on the GWs emitted by IMRIs were studied in \citep{Eda:2014kra}, in which the dynamics of the IMRIs were treated in the leading Newtonian limit and the DM distribution was assumed to be static throughout the inspiral.
Fisher-matrix forecasts predict that the properties of the DM spike could be mapped precisely for the higher overdensities (comparable to the errors on the detector-frame chirp mass) and to a few tens of a percent for lower overdensities.

Conservation of energy means, however, that the energy dissipated through dynamical friction must be balanced by a corresponding change in the energy of the dark matter distribution.
For a range of different DM overdensities around different IMRIs, the energy dissipated when the DM spike is assumed to be static turns out to be a substantial fraction of the binding energy of the spike, rendering the assumption of staticity questionable. In these cases, it is important to jointly evolve both the IMRIs' orbits and their surrounding distributions of DM \citep{Kavanagh:2020cfn}. When this is done, the DM distribution is found to evolve non-trivially, significantly decreasing the amplitude of the dynamical friction force and hence the difference between an inspiral in vacuum and in a DM mini-spike \citep{Kavanagh:2020cfn}.
Yet for larger DM overdensities, the amplitude of the effect of the DM is seen to remain significant in the coupled evolution of the IMRI and DM spike. 

A Bayesian analysis of the detectability (signal-to-noise ratio), discoverability (discrimination against in-vacuum inspiral), and measurability (prospects for measuring the parameters) of dark matter environments shows that inspirals in presence of dark matter can be easily discriminated against in-vacuum inspirals.  In case of detection, the DM halo's density normalization can be distinguished from zero at high significance, with a $95\%$ credible interval of order 10\%. The DM halo's slope can also be measured, although the posterior exhibits degeneracy with the chirp mass and the mass ratio. Interestingly, unlike in-vacuum inspirals, the mass ratio can be measured even in the Newtonian limit \citep{Coogan:2021uqv}.

For EMRIs and IMRIs in vacuum, there is good evidence that gravitational waveforms that are accurate to first post-adiabatic (1PA) order will be sufficient to detect and perform parameter estimation  \citep{vandeMeent:2020xgc}. Such vacuum waveforms are under development and are expected to be ready for the beginning of the LISA mission.
By contrast, the waveform modelling for IMRIs and EMRIs with surrounding DM spikes coupled to their evolution is only just beginning: the calculations in \citet{Kavanagh:2020cfn}
assumed circular orbits and worked at leading Newtonian order. A more complete modelling of these systems will be necessary to produce waveforms that are suitable for describing the full range of possible orbits and DM distributions to cover the parameter space of these systems.
The possibility of finding a simpler effective description of IMRIs and EMRIs in DM spikes should be investigated in greater detail, in order to decrease the size of the parameter space needed to look for signatures of cold DM substructures.
At the other end of the scale, as in the light dark matter case, it may also be interesting to study whether there is any significant impact of these dark matter overdensities on the merger signal using NR simulations.

\subsection{PBHs $m < M_\odot$}
\label{Sec:PBH_low_mass}

PBHs might originate in the early universe from the collapse of large density perturbations from an enhancement of the scalar curvature perturbation power spectrum generated during inflation \citep{Ivanov:1994pa, GarciaBellido:1996qt, Ivanov:1997ia}.
PBHs may therefore span a wider range of masses compared to stellar BHs, which are expected to have a mass larger than the Chandrasekhar limit, around the solar mass, from stellar evolution. 

For masses smaller than about $10^{-17} M_\odot$, PBHs would have evaporated by now due to the emission of  Hawking radiation, and thus do not lead to late-time imprints  \citep{Sasaki:2018dmp}.
For  masses larger than this lower limit, while many observational bounds can be set on the PBH abundance in the Universe, these still leave some open windows for PBHs to constitute the totality of DM. A comprehensive review of the constraints on the PBH population can be found in \citet{Carr:2020gox}.

For PBHs with masses smaller than about the solar mass, the most stringent constraints on the PBH abundance arise from evaporation limits \citep{Barnacka:2012bm} and microlensing observations \citep{Niikura:2017zjd, Smyth:2019whb, Alcock:2000kd, Allsman:2000kg, Niikura:2019kqi, Oguri:2017ock}. Interestingly, there  exists an open window for PBHs with masses in the range $(10^{-16} - 10^{-11}) M_\odot$ to account for the totality of the dark matter in the universe (see \citealt{Carr:2020gox,Green:2020jor} for recent reviews). 
This range of masses is however notoriously difficult to probe with lensing observations. Indeed it was shown that both  femtolensing and microlensing searches are not able to constrain this mass window once the extended nature of the source as well as wave optics effects are properly considered \citep{Katz:2018zrn, Montero-Camacho:2019jte,Smyth:2019whb}. Also, bounds coming from the survival of white dwarfs (WDs) and NSs after asteroidal mass PBHs capture were shown to be unable to constrain the abundance in this range \citep{Montero-Camacho:2019jte}. 
Finally, the GWs from mergers of sub-solar mass PBHs, peaking at much higher frequencies compared to the ones testable by LISA, would not lead to an observable signal. 
However, LISA can search for SGWB signatures of the PBH formation in that interesting window, thus testing the possible nature of dark matter as asteroidal-mass PBHs, as we will describe in the following.

Curvature perturbations, responsible for the PBH production, would also lead to the emission
of a second-order induced SGWB due to the intrinsic non-linear nature of gravity \citep{Acquaviva:2002ud,Mollerach:2003nq}, analysed in details in \citet{Ananda:2006af,Baumann:2007zm,Bugaev:2009zh,Saito:2009jt,Garcia-Bellido:2016dkw,Ando:2017veq,Garcia-Bellido:2017aan, Espinosa:2018eve,Kohri:2018awv,Bartolo:2018qqn,Bartolo:2018rku,Bartolo:2018evs,Clesse:2018ogk,Wang:2019kaf,DeLuca:2019ufz, Inomata:2019yww,Yuan:2019fwv,Domenech:2019quo,Domenech:2020kqm,Pi:2020otn,Chang:2020iji,Fumagalli:2020nvq}.\footnote{Note that even perturbations that do not create PBHs contribute to the GWs produced, and it is the total of all these contributions that must be considered \citep{DeLuca:2019llr}.}
Since the emission mostly takes place when the length scale of the perturbation crosses the cosmological horizon, one can relate the peak of the GW frequency spectrum to the dominant PBH mass $M_\text{\tiny PBH}$ as \citep{Saito:2009jt}
\begin{equation}
\label{peak}
f_\star \simeq 3  \, \text{mHz} \left( \frac{M_\text{\tiny PBH}}{ 10^{-12}M_\odot}\right) ^{-1/2}.
\end{equation}
There are several current and future experiments searching for this SGWB in various frequency ranges.
In the ultra-low frequency range (around nHz), the null detections at Pulsar Timing Array experiments like PPTA \citep{Shannon:2015ect}, NANOGrav \citep{Arzoumanian:2018saf} and EPTA \citep{Lentati:2015qwp}, give rise to stringent constraints on the abundance of GWs (although NANOGrav shows some evidence for a background, as discussed further below). Future experiments like SKA \citep{2009IEEEP..97.1482D} (see also \citealt{Moore:2014lga}) will significantly improve the detection sensitivity. These constraints can also be translated into a bound on the amplitude of the comoving curvature perturbations at the corresponding scales \citep{Inomata:2018epa,Unal:2020mts}. 
Recently, the Earth-Moon system has been suggested as a detector of SGWB at $\mu$Hz frequencies \citep{Blas:2021mqw,Blas:2021mpc}, providing a complementary probe to LISA for signals in this range.
 
 In the intermediate frequency range (around mHz) the most relevant experiment is LISA. LISA will be able to probe the wavenumbers of the primordial curvature perturbation power spectrum on scales in the range $(10^{10} - 10^{15}) \, {\rm Mpc}^{-1}$ \citep{Garcia-Bellido:2016dkw}. Moreover, the production of  PBHs with masses around $M_\text{\tiny PBH} \sim {\cal O} \left( 10^{-15} - 10^{-8} \right) M_\odot$ would be associated 
 to GW signals with peak frequencies which fall within the LISA sensitivity band. LISA would be therefore able to search for DM candidates in the form of PBHs within this range of masses \citep{Bartolo:2018evs,Garcia-Bellido:2016dkw,Garcia-Bellido:2017aan}. Notice that, due to the exponential sensitivity of the PBH abundance to the amplitude of the curvature power spectrum, null detection of a SGWB at LISA would imply the PBH abundance in the corresponding mass range to be negligible. 
 
Physical SGWB spectra would typically have a white-noise ($\propto f^3$) behaviour at low frequencies \citep{Espinosa:2018eve,Cai:2019cdl}, a peak corresponding to the frequency provided in Eq.~\eqref{peak} and a high frequencies tail which goes like the squared curvature power spectrum, $\Omega_\text{\tiny GW} (f \gg f_\star) \sim \mathcal{P}_\zeta^2 (f)$. PBH formation from multi-field inflationary models \citep{Palma:2020ejf, Fumagalli:2020adf}, and more generally large particle production mechanisms involving a sharp feature, lead to oscillatory power spectra that result in a 10\% oscillatory modulation in the peak of the frequency profile of the SGWB \citep{Fumagalli:2020nvq,Braglia:2020taf}, thus representing a potentially observable characteristic of these models.

Recently the NANOGrav collaboration has published a 12.5 yrs analysis of pulsar timing data reporting strong evidence for a signal whose interpretation in terms of a stochastic common-spectrum process is strongly preferred against independent red-noise signals \citep{Arzoumanian:2020vkk}.
At the moment, the NANOGrav Collaboration does not yet claim a detection of a SGWB due to the absence of evidence for the characteristic quadrupole correlations. There are several models proposed so far to explain the  origin of this SGWB signal. One possibility is provided by the SGWB associated to PBH formation \citep{Vaskonen:2020lbd, DeLuca:2020agl, Kohri:2020qqd,Domenech:2020ers}. The relative PBH abundance in the mass range probed by NANOGrav would be subdominant \citep{Vaskonen:2020lbd}. However, the model proposed in \citet{DeLuca:2020agl}, where a flat power spectrum of the curvature perturbation \citep{Wands:1998yp, Leach:2000yw, Leach:2001zf, Biagetti:2018pjj} leads to a mass function peaked in the asteroidal mass range \citep{DeLuca:2020ioi}, a total abundance of DM in the forms of PBHs allowed by observational bounds and hinted by the lensing event candidate in the HSC data \citep{Kusenko:2020pcg,Sugiyama:2020roc}. This scenario predicts a flat SGWB spectrum which would be detectable and tested by LISA \citep{DeLuca:2020agl}.

As discussed in more detail in the complementary white paper of the cosmology WG, LISA will also be able to search for signatures of  primordial  non-Gaussianity \citep{Franciolini:2018vbk, Biagetti:2018pjj, Ezquiaga:2018gbw,Ezquiaga:2019ftu, Pattison:2021oen, Biagetti:2021eep}  at small scales. Primordial non-Gaussianity on small scales would have an impact on the shape of the SGWB, leading to potentially detectable signatures in the spectrum of frequencies of the observed monopole signal \citep{Cai:2018dig,Unal:2018yaa}. Possible small deformations smearing the GWs spectrum can also arise from similar effects \citep{Domcke:2020xmn}. Non-Gaussian signatures in the tensor three-point function would in principle be significant, but get washed out by   \citep{Bartolo:2018evs,Bartolo:2018rku} time-delay effects originating during the propagation of the GWs in the perturbed universe, see \citet{Bartolo:2018qqn,Margalit:2020sxp}. Non-Gaussian signatures may be detectable by searching for angular anisotropies in the SGWB which, in the absence of non-Gaussianity, are generally undetectable at LISA \citep{Bartolo:2019zvb} given its angular resolution of $\ell \lesssim 15$ \citep{Contaldi:2020rht, Baker:2019ync}. However, local scale-invariant non-Gaussianity, constrained by the Planck collaboration  to be $- 11.1 \leq f_\text{\tiny NL} \leq 9.3$ at 95\% \, C.L. \citep{Akrami:2019izv}, would correlate short and long scales and potentially lead to an enhancement of the SGWB, see \citet{Cai:2018dig,Unal:2018yaa,Cai:2019elf} for details, and possible large-scale anisotropies at detection \citep{Bartolo:2019zvb}. Now if PBHs constitute a large fraction of the DM, one expects a highly isotropic and Gaussian SGWB due to the strong constraints by CMB observations \citep{Akrami:2018odb} on the isocurvature modes in the DM density fluid, associated to the non-Gaussianity \citep{Young:2015kda}.
 On the other hand, the detection of a large amount of anisotropy in the signal would imply that only a small fraction of the DM can be  accounted by PBHs \citep{Bartolo:2019zvb}.
Moreover,  the propagation of GWs across disconnected regions of the universe may lead to large-scale anisotropies at detection, see e.g., \citet{Alba:2015cms,Contaldi:2016koz,Cusin:2017fwz,Cusin:2018avf,Jenkins:2018nty,Bartolo:2019oiq, Bartolo:2019yeu,Renzini:2019vmt, Bertacca:2019fnt}.

In this range of masses, the non detection of a SGWB will rule out the possibility of PBHs being the dark matter in the asteroidal mass range. However, this assumes the standard scenario where GWs are produced at PBH formation and the PBH abundance is exponentially sensitive to the size of the fluctuations. One should investigate whether non standard scenarios would change this conclusion, e.g. the potential role of non-Gaussianity or of other PBH formation mechanisms.

\subsection{PBHs $m > M_\odot$}
\label{Sec:PBH_high_mass}

For PBHs with masses larger than about a solar mass, LISA should be able to detect merger events of resolved sources, and unresolved signals in the form of a SGWB. 
The most relevant constraints come from CMB temperature and polarization anisotropies, which are impacted by the emission of ionizing radiation from PBHs accreting gas in the early universe at redshift between $z \sim (300 - 600)$, for which the PBH abundance is constrained to be below $f_\text{\tiny PBH}(M) \lesssim (M/10 M_\odot)^{-4}$ for masses smaller than $10^4 M_\odot$ \citep{Ali-Haimoud:2016mbv, Serpico:2020ehh, Carr:2020gox, Green:2020jor}.
At larger masses, constraints coming from CMB distortions become even  more stringent \citep{Nakama:2017xvq}. 

Late-time constraints can be set from the comparison of emitted EM signals from accreting PBHs with observations of galactic radio and X-ray isolated sources \citep{Gaggero:2016dpq,Manshanden:2018tze} and 
X-ray binaries \citep{Inoue:2017csr}, and from Dwarf Galaxy Heating observations \citep{Lu:2020bmd}. 
The totality of these bounds constrain the PBH abundance $f_\text{\tiny PBH}$ (the fraction of the dark matter composed of PBHs) to be below $f_\text{\tiny PBH} \lesssim 10^{-2}$ for masses larger than few tens of $M_\odot$.
A comparable constraint is set by the merger rates observed by the  LIGO/Virgo collaboration \citep{LIGOScientific:2018mvr}. 
Indeed, for masses around $30 M_\odot$, a PBH abundance larger than $f_\text{\tiny PBH} \thickapprox 10^{-3}$ would give an expected number of events larger than the one observed. This has been shown in \citet{Raidal:2018bbj,DeLuca:2020qqa, Hall:2020daa} using the GWTC-1 data and recently in \citet{Wong:2020yig} with the GWTC-2 catalog.
Even though PBHs are expected to follow a Poisson distribution at formation in the absence of non-Gaussianities \citep{Ali-Haimoud:2018dau,Desjacques:2018wuu,Ballesteros:2018swv,MoradinezhadDizgah:2019wjf}, they may start forming clusters even before the matter-radiation equality for large enough $f_\text{\tiny PBH}$ \citep{Inman:2019wvr}. Clustering has an impact on the PBH binary formation and subsequent disruption rates \citep{Raidal:2018bbj, Jedamzik:2020ypm, Jedamzik:2020omx, Trashorras:2020mwn}. While ineffective for abundances smaller than $f_\text{\tiny PBH} \lesssim 10^{-2}$, it is expected to reduce the merger rate prediction for larger abundances by an amount which is not enough to evade the LIGO/Virgo bound if one takes into account the fraction of PBHs in dense substructures \citep{Vaskonen:2019jpv, DeLuca:2020jug, Tkachev:2020uin}.
Overall, we stress that a complete description of PBH clustering up to very low redshift along with its impact on the GW signals coming from a PBH population still represents one of the main theoretical challenges, which deserves a thorough investigation.

Even if PBHs within this mass range can only account  for a small fraction of the dark matter in the universe, they could still both form binaries  observable by LISA and act as progenitors of SMBHs, observed at high redshift as large as $z \gtrsim 6$, through accretion of baryonic particles during the cosmological evolution \citep{Clesse:2015wea,Serpico:2020ehh}. The presence of a strong phase of accretion through the cosmological history before the reionization epoch has the effect of shifting 
the corresponding late-time mass function 
to larger masses relaxing the constraints \citep{DeLuca:2020fpg}, even though large uncertainties are present in modelling the accretion for massive compact objects. This is a key challenge to overcome in order to assess the PBH contribution to merging binaries observable by LISA.
Though difficult to conclusively associate with PBHs, the observation of events within this range of masses at high redshift by experiments like LISA would help in understanding the physical origin of SMBHs \citep{Volonteri_2010}.

Given that PBHs may potentially contribute to current GW data \citep{Franciolini:2021tla}, within this range of masses they need to be distinguished from astrophysical BHs formed after the collapse of stars in the late universe.
Various observables could however help in discriminating the two populations.
For instance, even though PBHs are expected to form with small spins \citep{Mirbabayi:2019uph, DeLuca:2019buf}, for PBH binaries in the mass range observable by LISA, the baryonic mass accretion leads to  spin growth, potentially increasing their spins for the binary components at the merger time depending on the accretion strength before the reionization epoch \citep{DeLuca:2020bjf}, which is still however affected by many uncertainties. It is of crucial importance to investigate in details the correlation between PBH masses and spins induced by baryonic accretion as a discriminant with the prediction for astrophysical BHs. 
Another possibility is to look for high redshift events (i.e. $z\gtrsim30$) where PBHs can contribute without any astrophysical contamination \citep{Koushiappas:2017kqm,Chen:2019irf,DeLuca:2021wjr}.

Furthermore, the massive BBH merger population currently observed at LIGO/Virgo is expected to give rise to a SGWB of unresolved sources with a tail at low frequencies which can be detected by LISA \citep{Clesse:2016ajp, Chen:2018rzo}.
The primordial scenario is expected to give a stronger contribution to the SGWB with respect to the astrophysical one, due to the additional contribution given by PBH mergers happening at higher redshift \citep{Raidal:2018bbj}. In particular, the PBHs merger rate is expected to increase with redshift, while the one of stellar BHs first increases and peaks at redshift  around $z \sim (1 - 2)$, and then rapidly decreases \citep{Chen:2018rzo}. Considering both observations of resolved  and unresolved signals may help to break the degeneracy within the two scenarios \citep{Mukherjee:2021ags,DeLuca:2021hde,Wang:2021djr,Mukherjee:2021itf}. 

LISA will also be able to detect the inspiral phase of binaries with masses of at least a few tens $M_\odot$ with a high precision. This mass range, falling in the large mass portion of the window observable by the LIGO/Virgo collaboration, can be filled by a PBH population. The presence of the known mass gap in the astrophysical BH spectrum  and the recent event GW190521 \citep{Abbott:2020tfl} make the primordial scenario
investigated in \citet{Clesse:2020ghq,DeLuca:2020sae,Kritos:2020wcl} potentially relevant (see however  \citealt{Abbott:2020mjq} for different scenarios). 

Finally, the detection of GWs from the coalescence of a BBH system including a compact object lighter than a solar mass could indicate primordial origin. Experiments like LISA, which target low frequency regimes, may detect GWs emitted from EMRIs of a SMBH with a light compact PBH, and GW constraints can be set on the PBH abundance from the expected PBH-SMBH merger rate \citep{Guo:2017njn,Kuhnel:2017bvu,Kuhnel:2018mlr}. A null-detection during a 5-year operation of the experiment would constrain the PBH abundance to values smaller than $f_\text{\tiny PBH} \lesssim 3 \times 10^{-4}$ for PBHs with masses in the range $(10^{-2} - 1) \, M_\odot$ \citep{Guo:2017njn}. 

Another interesting LISA observable with relevance to the DM PBH scenario arises from GW bursts from hyperbolic encounters of PBHs \citep{Garcia-Bellido:2017knh}. This would be particularly relevant for the clustered and broad mass spectrum PBH scenario, where the BHs cover a wide range of masses with a peak $\sim 1\,M_\odot$ \citep{Garcia-Bellido:2017fdg, Garcia-Bellido:2017mdw, Ezquiaga:2017fvi, Carr:2019kxo}, and have a clustered distribution which allows them to evade some, but not all, of the constraints discussed above, see \citet{Garcia-Bellido:2018leu,Garcia-Bellido:2019tvz} for a review.

To summarize, LISA would be sensitive to many different GW signals coming from a PBH population in this mass range. Given its sensitivity to BH mergers at high redshifts and large masses, even reaching the supermassive range, 
it will complement the reach of ground-based detectors to search for PBHs and inform models of PBH accretion.
On the theoretical side, a future goal is to sharpen the predictions for the impact of accretion and clustering on the evolution of PBH binaries and the generation of SMBHs from PBH seeds \citep{Serpico:2020ehh}, in order to differentiate between BHs with a primordial origin and those formed from stellar evolution.

\subsection{Exotic Compact Objects}

If Exotic Compact Objects (ECOs) exist, they could compose some fraction of the dark matter. ECOs have already been discussed in Sec. \ref{Sec:Exotic_Compact_Objects}, and for those with compactness similar to BHs, the opportunities for LISA to observe and constrain them will be similar to the PBH cases described in Sec. \ref{Sec:PBH_low_mass} and Sec. \ref{Sec:PBH_high_mass} above.

\subsection{Signatures from self-interactions}
\label{Sec:DM_self_interaction}

Self-interactions in DM were proposed as a way to solve tensions on small scales between N-body simulations and DM observations \citep{Spergel:1999mh}, but for hard interactions these are constrained to be of order $\sigma/m < 1 ~ {\rm cm^2 ~ g^{-1}} \sim 10^{-24} {\rm cm^2 ~ GeV^{-1}}$ to maintain consistency with structure formation and observations of the Bullet Cluster \citep{Randall:2007ph}. At the upper end of this range, they have a non-negligible effect on structures at typical DM halo densities and velocity dispersions (see \citealt{Tulin:2017ara} for a review). Smaller cross sections would  have an impact only  in more dense regions, e.g. in DM spikes around BHs, and so self-interacting dark matter (SIDM) could potentially have interesting consequences. However, we have yet to see a model with sufficient capture to realistically manifest these, at least where the SIDM composes the entirety of the DM. 

It has been proposed that a small amount of SIDM, composing less than 10\% of the DM, with cross section $\sigma/m \gg {\rm 1 ~ cm^2 /g}$, could have seeded the initial growth of SMBHs \citep{Pollack:2014rja}, which are a key LISA target. However, it seems unlikely that such a formation mechanism could be identified from LISA observations.  SIDM could also undergo dissipative collapse to form compact dark disks \citep{Fan:2013tia}, which could potentially be detectable via their consequent density enhancements in the same way as the superradiant clouds or DM spikes described above. 

In the case of light bosonic DM, self-interactions may play an important role in superradiant instabilities and, moreover, are an essential part of some models such as axions, where the massive potential is an approximation to a more general periodic function. Studies of superradiance have mostly treated self-interactions in the bosonic field as negligible, but in \citet{Yoshino:2012kn, Yoshino:2013ofa, Yoshino:2015nsa} attractive self-interactions were shown to destabilise the superradiant build up, resulting in a so called ``bosenova''. Such collapses have been conjectured to result in repeating instabilities, where the fall of the cloud back into the BH prior to saturation results in it being ``spun up'' again, and thus the superradiant growth resumes. This could weaken constraints that have been derived in the purely massive case, or lead to potentially observable GW signals from the bosenova explosions. The frequency of the bosenova signal would be $f \sim 1/\Delta t $ where $\Delta t$ is the duration of the bosenova. This is in the LISA band for SMBHs, but would be below detection sensitivity unless it occurs in our own or a nearby galaxy \citep{Yoshino:2012kn}. More work is required to simulate a complete build up and dissipation of the bosonic cloud, to further quantify the GW signals produced and to confirm that the superradiant build up is able to recover. Due to the long timescales involved, such simulations are very challenging, but feasible with current NR tools.

\subsection{Multi-messenger signatures}
\label{Sec:DM_multi_messenger}

If dark matter particles interact at all with standard model particles, the interaction must be weak, given the constraints that have been imposed by terrestrial direct detection experiments and indirect astrophysical observations (see e.g., \citealt{Kahlhoefer:2017dnp, Irastorza:2018dyq, Bertone:2004pz} for reviews). However, some dark matter candidates, such as the QCD axion, have a well defined coupling to standard model matter. In this case, coincident GW and EM signatures, for example, during SMBH mergers or in EMRIs, may provide distinctive signatures. This has been explored in the case of the QCD axion in \citet{Edwards:2019tzf}, on the basis of LISA observations of dephasing in the GW signal of a NS inspiral around an IMBH embedded within a DM spike. Such observations could identify axions in the mass range $10^{-7} - 10^{-5}$ eV by matching the signal to radio wave emission from axion-photon conversion. Similar multi-messenger effects could be relevant in other cases, such as where GWs from superradiant clouds or ECOs are detected, see e.g., \citet{Ikeda:2018nhb,Cardoso:2020nst,Blas:2020nbs,Blas:2020kaa,Amin:2020vja} for recent work in this direction.

\subsection{Burning Questions}
\label{Sec:DM_burning}

\begin{itemize}
    \item {The effects of dark matter on GW signals may often be degenerate with `environmental' astrophysical effects (cf.~Sect.~\ref{Sec:Astrophysics_Environmental_effects}). It will be absolutely crucial for the success of LISA to devote continual effort to disentangle these as much as possible.} 
    \item{LISA has the potential to detect or constrain the presence of both ultralight and heavier dark matter fields in regions of parameter space that are complementary to those covered by ground-based GW observations. However, the waveform modelling for IMRIs and EMRIs embedded in DM halos is but in its infancy. More work is needed to build waveform models that incorporate the effects of DM fields in mergers, covering a significant parameter space, and that are sufficiently accurate for data analysis purposes.}
    \item{Beyond this, it will be important to study BBH dynamics with ultralight boson clouds with full NR simulations in order to understand how the presence, and the dynamics, of such boson clouds could be imprinted in the late stages of BBH mergers.}
    \item{To properly interpret the observables related to PBHs and identify their origin, the impact of non-Gaussianity or novel PBH formation mechanisms in the small mass range, and clustering and accretion in the mass range above $M_\odot$ up to SMBHs, should be further characterised.}
\end{itemize}

\section{Tests of the $\Lambda$CDM model and dark energy}
\label{sec:LambdaCDM}

The past 20 years have seen the emergence of a standard model of the Universe based on a handful of parameters: the so-called $\Lambda$CDM model. This framework  can explain coherently a large number of observations but the nature of one of its main components, dark matter,  is still unknown, and the observed value of the cosmological constant $\Lambda$, a crucial element of the model, is still mysterious.  

In particular, many cosmological observations indicate that the Universe is currently undergoing 
accelerated expansion. To explain this observation assuming GR, the $\Lambda$CDM model requires that the energy density of the vacuum  is comparable to the total energy density of the Universe. However, quantum vacuum fluctuations of all  particles  contribute to the vacuum energy by an amount 
which is  expected to be many orders of magnitudes larger than its measured value. This mismatch is called the {\it old} cosmological constant problem  (see e.g., \citealt{Burgess:2013ara,Padilla:2015aaa}) and it is one of the most disconcerting puzzles in fundamental physics (see e.g., \citealt{Bousso:2007gp, Charmousis:2011bf, Kaloper:2013zca, Alberte:2016izw}).

This situation constitutes a theoretical motivation to consider alternatives to $\Lambda$CDM in which one assumes that for some unknown reason the value of $\Lambda$ is simply zero - dubbed the {\it new} cosmological constant problem -  and attempts to explain the accelerated expansion by different mechanisms. This has spurred the exploration of variations of the $\Lambda$CDM model in which the energy budget of the Universe is dominated by some dynamical dark energy, distinct from a constant $\Lambda$, or where the observed acceleration is explained by a modification of gravity (see \citealt{Clifton:2011jh,Amendola:2016saw} for a review). This   field of investigation has been rich of interesting theoretical developments and is currently energized by the prospect that a number of forthcoming galaxy surveys will test and tightly constrain these models in the near future (see e.g., \citealt{Frusciante:2019xia}).
A second theoretical motivation for considering 
modified gravity extensions of $\Lambda$CDM is the so-called {\it coincidence} problem \citep{Zlatev:1998tr}, the observation that the current vacuum energy density is comparable to the matter density, even though the former remains constant in time while the latter dilutes with the expansion. 

Quite apart from these theoretical considerations, several tensions have emerged in the $\Lambda$CDM model from recent observations. The most notable one is the discrepancy between the values of the Hubble constant measured through CMB \citep{Aghanim:2018eyx} and large-scale structure \citep{Philcox:2020vvt,DAmico:2020ods} observations on the one hand, and those measured by late-time probes \citep{Riess:2019cxk,Huang:2019yhh,Jee:2019hah,Pesce:2020xfe,Verde:2019ivm} on the other hand. 
Other milder tensions include the difference between the amplitude of matter fluctuations measured by Planck versus that inferred from weak lensing observations \citep{Heymans:2020gsg,Troster:2019ean}, as well as the lensing excess in the CMB temperature Planck data \citep{Ade:2013zuv,Adam:2015rua,Aghanim:2018eyx}.  
While it is possible that these tensions could be accounted for by some yet unknown systematics, several proposals have attempted to explain these discrepancies by new physics.

In this section we summarize how GW astronomy, in particular the LISA mission, can inform theoretical physics in the cosmology area, in particular by probing the fundamental physics underlying the $\Lambda$CDM model and its extensions involving dark energy and modified gravity. To this extent this Section is complementary to the Sections 2 and 4 of the white paper of the LISA cosmology WG, whose main goal is to develop the appropriate methods to probe cosmology with GW observations.
In the next subsection we present an overview of the dark energy and modified gravity models of interest for LISA. We then discuss the theoretical implications and the constraints  on GWs propagating in a homogeneous universe in Sect.~\ref{Sec5:Homcosmology}. We extend this analysis to an inhomogeneous universe in Sect.~\ref{Sec5:Inhomcosmology}, by discussing cosmological tests that involve the cross-correlation of GWs  with the large-scale structure of the Universe  and the use of the GW lensing to probe the distribution of structures. Future prospects are discussed in Sect.~\ref{Sec5:Future}.

\subsection{Dark energy and modified gravity} \label{Sec5:DE&MG}

Dark energy and modified gravity are extensions of the $\Lambda$CDM scenario. 
Although a strict distinction between the two does not exist \citep{Joyce:2016vqv}, the former usually refers to adding to the field equations the stress-energy tensor of a fluid responsible for the acceleration, leaving the Einstein tensor unchanged, while the latter refers to either changing the gravitational part of the field equations or adding a non-minimal coupling to matter. 
Since  at  low-energy GR is the {\it unique} Lorentz-invariant theory of a massless helicity-2 field \citep{Weinberg:1964kqu}, in most cases (see e.g., \citealt{Lin:2017oow} for an exception) one must add additional degrees of freedom to those present in GR.

\subsubsection{Effective field theory of dark energy}

In the simplest case, dark energy and modified gravity rely on the presence of a scalar field that spontaneously breaks time parametrizations, inducing a preferred slicing. Since in cosmology we are interested in studying fluctuations around  FLRW solutions, a particularly convenient way to parametrize the action in this case is to use the so-called Effective Field Theory of Dark Energy \citep{Creminelli:2008wc,Gubitosi:2012hu,Bloomfield:2012ff,Gleyzes:2013ooa}, which consists in writing the action in terms of all possible operators in this foliation \citep{Creminelli:2006xe,Cheung:2007st}. This allows one to describe a large class of models, including $f(R)$ theory (see \citealt{Sotiriou:2008rp} for a review), Horndeski theories \citep{Horndeski:1974wa,Deffayet:2009mn} and beyond \citep{Zumalacarregui:2013pma,Gleyzes:2014dya,Langlois:2015cwa, BenAchour:2016fzp}, and Lorentz violating theories \citep{Horava:2008ih},  in terms of a fewer number of parameters, corresponding to time-dependent functions. More generally, requirements such as locality, causality, unitarity and stability can be automatically enforced on the action so that the predicted signatures are consistent with well-established principles of physics. The generality of this approach has been successfully applied to the implementation of Einstein--Boltzmann solvers such as \texttt{EFTCAMB} \citep{Hu:2013twa,Raveri:2014cka} and \texttt{hi\_class} \citep{Zumalacarregui:2016pph} and the
derivation of observational constraints and forecasts in a model-independent fashion (see \citealt{Frusciante:2019xia} for a recent review). 

In these extensions of $\Lambda$CDM, we can identify two main classes of modifications to the propagation of GWs with observational implications. The first is a scale-independent additional friction term in the propagation equation, discussed in Sect.~\ref{Sec:GWmodprop}, which can be traced to the fact that the effective Planck mass that canonically normalizes the graviton can be time-dependent in models of modified gravity. In this case, the effect amounts simply  to a rescaling of the graviton amplitude between the time of emission and the one of observation.
The second class are scale-dependent modifications such as an anomalous speed of propagation, effects in the dispersion relation, GW decay, etc. These are thoroughly discussed in Secs.~\ref{Sec:Tests_with_GW_propagation} and \ref{sec:GW_propagation_test} and will not be treated here.

\subsubsection{Non-local model}

Another possibility, leading to scale-independent modifications that have been much investigated recently, is that the long-distance behavior of gravity is modified by quantum effects. The presence of infrared (IR) divergences in space-times of cosmological interest, such as de~Sitter \citep{Taylor:1989ua,Antoniadis:1986sb,Tsamis:1994ca}, suggests that, even without adding new degrees of freedom or modifying the fundamental action of the theory, the long-distance dynamics of gravity could be different from that derived from the classical Einstein-Hilbert action. When quantum effects enter into play the relevant quantity, rather than the fundamental action, is the quantum effective action. While the former, according to basic principles of quantum field theory, is local, the latter unavoidably develops non-local terms whenever the spectrum of the theory contains massless particles, such as the graviton. 

An extensive series of investigations 
(see \citealt{Belgacem:2017cqo,Belgacem:2020pdz} for reviews)
have eventually led to identify a unique candidate model
(the so-called `RT' model originally proposed in 
\citep{Maggiore:2013mea}). Physically, this model corresponds to assuming that IR effects generate a non-local mass term for the conformal mode of the metric.
At the conceptual level, this is  appealing because of various arguments that identify the conformal mode as the main candidate for producing strong IR quantum effects \citep{Antoniadis:1986sb,Antoniadis:1991fa}.
Tentative evidence for the generation of this specific  non-local term in the quantum effective action has also emerged from lattice gravity simulations \citep{Knorr:2018kog}. 
At the phenomenological level, the model passes Solar System tests \citep{Kehagias:2014sda}
and limits on the time variation of Newton's constant \citep{Belgacem:2018wtb},
it generates a dark energy dynamically \citep{Maggiore:2013mea}, and its cosmological perturbations, in the scalar sector, are well-behaved and quite close to those of $\Lambda$CDM \citep{Dirian:2014ara}. The model has been compared to CMB, BAO, SNe and structure formation data, and fits them at the same level as $\Lambda$CDM \citep{Dirian:2016puz,Belgacem:2017cqo}. A main surprise then came from the study of tensor perturbations \citep{Belgacem:2017cqo,Belgacem:2019lwx}: The model displays the phenomenon of modified GW propagation that will be discussed below, whose amplitude can be significant. At the redshifts relevant for LISA, the deviations with respect to GR can even be of order $80\%$ (depending on a free parameter of the model, related to the initial conditions). A complete and up-to-date review of the conceptual and phenomenological aspects of the model is given in \citet{Belgacem:2020pdz}.

Apart from its specific features, this model also provides an explicit example of a modification of GR that is very close to $\Lambda$CDM, at the percent level, for the background cosmological evolution and for the scalar perturbations (therefore complying with existing limits), while at the same time producing very large deviations from GR in the sector of tensor perturbations.
In the context of scalar-tensor theories described by the effective field theory of dark energy,  a similar case is discussed in \citet{Lombriser:2015sxa}.
This shows how the study of GWs on cosmological scales is a genuinely new window, accessible to LISA, that can produce significant surprises. 

\subsubsection{Early and interacting dark energy  }

Other interesting models to test with LISA are the so-called early dark energy \citep{Wetterich:2004pv} and interacting dark energy \citep{Wetterich:1994bg,Amendola:1999er} scenarios.
The former consists of a redshift-dependent dark energy component which gives a non-negligible cosmological contribution  at early times, in contrast to the more standard dark energy used to explain the late-time cosmic acceleration.
The presence of early dark energy in the cosmological energy budget affects the CMB anisotropies and polarization and the inferred constraints can change by  varying the epoch at which early dark energy starts to contribute \citep{Pettorino:2013ia}. 
In the context of LISA, early dark energy  can be tested by measuring the distance-redshift relation at high redshift (see Sect.~\ref{Sec:GWmodprop}). As such, LISA can be  complementary to other cosmological probes, such as the CMB, in helping to resolve the $H_0$-tension for which such early dark energy models are  candidate solutions \citep{Caprini:2016qxs}.

Interacting dark energy  is characterized by a direct coupling between  dark energy and dark matter, while ordinary matter remains uncoupled from dark energy and gravity is described by the ordinary Einstein-Hilbert  action (i.e., in the so-called  Einstein frame). It can be  motivated, for instance, by the fact that it can alleviate the  coincidence problem mentioned in the introduction \citep{Wetterich:1994bg,Amendola:1999er}. 
In this scenario, the propagation of GWs is modified  through a  change in the Hubble friction arising from an altered cosmic expansion history \citep{Caprini:2016qxs,Dalang:2019fma}. 
As for early dark energy, interacting dark energy  can be tested by LISA by accurately probing the background expansion history and the luminosity distance-redshift relation.  

\subsubsection{Screening} \label{sec:screening}

\noindent
Modifications of gravity that aim to explain the cosmic acceleration often entail extra light fields which, besides changing the cosmic expansion,  manifest themselves  at short distances, mediating a  {\it fifth force}. To pass Solar System and binary-pulsar tests, screening mechanisms are therefore necessary to hide, or strongly suppress, local deviations from GR (see e.g., \citealt{Joyce:2014kja}). Different types of screening mechanisms have been proposed over the years, all relying on non-linear physics becoming important close to matter sources. Here we discuss  only modified gravity models based on a scalar field. In chameleon and symmetron screening \citep{Khoury:2003rn, Hinterbichler:2010es}, non-trivial self-interactions and non-minimal couplings to matter dynamically suppress deviations from GR in high-density environments (see \citealt{Burrage:2017qrf} for a review), however these theories were shown to have a negligible effect on cosmological scales \citep{Wang:2012kj}. Self-interactions involving one or two derivatives of the field are at the origin of, respectively, $k$-mouflage \citep{Babichev:2009ee} and Vainshtein \citet{Vainshtein:1972sx, Babichev:2013usa} screening. These kinetic types of screening allow to reconcile theories such as Horndeski and DHOST, or massive (bi)gravity, with local tests of gravity. However, it is worth noticing that, until recently, kinetic screening had been thoroughly  tested in static/quasi-static configurations, while it was less studied in dynamical settings, e.g.~at GW generation \citep{deRham:2012fw,deRham:2012fg,Chu:2012kz,Dar:2018dra}.
Only very recently, numerical relativity simulations have been performed, which point at a partial breakdown of the screening in black-hole collapse \citep{terHaar:2020xxb,Bezares:2021yek} and in the late inspiral and merger of binary neutron stars \citep{Bezares:2021dma}. In more detail, stellar collapse seems (quite surprisingly) to produce a very low frequency signal potentially detectable by LISA, while waveforms from binary neutron stars seem to deviate from their GR counterparts at the quadrupole (but not dipole) multipole order.

In general, when screening is efficient, one can define three   {\it gravitational constants}. The one experienced by GWs, $G_{\rm gw}$;\footnote{This is proportional to the inverse square power of the effective Planck mass appearing in \eqref{eq:ratio_distance_Horndeski2} below.} the one felt by matter appearing in the Poisson equation, $G_{\rm dyn}$; the one felt by light intervening in null geodesics,  $G_{\rm light}$.
These three constants are generally different and independent from each other \citep{Lombriser:2015sxa, Tsujikawa:2019pih, Dalang:2019fma, Wolf:2019hun}, even when submitted to
the constraints coming from Solar System tests, such as lunar-laser ranging \citep{Williams:2004qba}, which tests $G_{\rm dyn}$, and the Cassini bound \citep{Bertotti:2003rm}, which tests a combination of $G_{\rm dyn}$ and $G_{\rm light}$. 
However, GW constraints can provide  independent relations among them.
For instance,  in  Horndeski theory  the combination of Solar System tests and the fact that $c_{\rm T}=1$ imposes that  $G_{\rm gw}$ varies very little with the redshift, implying that observational signatures in distance tests could be challenging  \citep{Dalang:2019fma, Dalang:2019rke, Wolf:2019hun,Lagos:2020mzy} (see Eq.~\eqref{eq:ratio_distance_Horndeski2} below).

This conclusion, however, is evaded in theories that extend Horndeski gravity, such as GLPV and DHOST theories, where additional parameters can lift the degeneracy.
In these theories, even when imposing $c_{\rm T} = 1$, the three constants remain independent \citep{Lombriser:2015sxa, Crisostomi:2017lbg, Langlois:2017dyl, Dima:2017pwp, Crisostomi:2019yfo} and there can be  effects in the luminosity distance. The recovered degeneracy can, however, be broken again if other constraints are imposed, such as those from the instability of GWs \citep{Creminelli:2019kjy}.
Another example where these constants remain independent is massive gravity \citep{deRham:2010kj} or bigravity \citep{Hassan:2011zd}, see \citet{Koyama:2011xz, Babichev:2013pfa}. 
Finally, non-local models do not rely on screening  and hence are not subject to these limitations.

\subsection{Homogeneous cosmology} \label{Sec5:Homcosmology}

In this subsection we consider GWs propagating on a homogeneous universe; we postpone the discussion of perturbations to the next subsection. 
We  assume a spatially-flat background space-time described by the Friedmann-Lema\^{i}tre-Robertson-Walker (FLRW) metric,
\be
\dd s^2 = - \dd t^2 + a^2(t) \delta_{ij} \dd x^i \dd x^j \;,
\ee
where $a(t)$ is the cosmological scale factor.
We describe GWs as perturbations of the spatial part of the above metric, $g_{ij} (t,\boldsymbol{x}) = a^2(t) \left[ \delta_{ij} + h_{ij} (t,\boldsymbol{x}) \right] $, using the traceless and transverse gauge $h_{ii} = \partial_i h_{ij} =0$. Moreover, we decompose $h_{ij}$ in its two polarisation modes, $h_{ij} = h_+ \epsilon^+_{ij} + h_\times \epsilon^\times_{ij} $, where $\epsilon_{ij}^\lambda$ are the polarization tensors. 

\subsubsection{Standard sirens} \label{Sec:standardsirens}
  
Standard sirens are GW sources at cosmological distances which provide a direct measurement of the luminosity distance. Standard sirens which come with an associated EM counterpart that enables one to determine  their redshift can be used to infer the luminosity distance-redshift relation, and thus trace the evolution of the Universe independently of other distance ladders.

For inspiraling binaries, at leading order in the PN approximation, the GW strain reads, for the two polarizations (see e.g., \citealt{Maggiore:1900zz}),
\begin{align}
h_+ &= \frac{2(1+\cos^2 \iota)}{d_{\rm L}(z)}(G\mathcal{M}_{\rm c})^{5/3}(\pi f)^{2/3}\cos \Phi\,,  \label{lum_dist1} \\
h_\times &= \frac{4\cos \iota}{d_{\rm L}(z)}(G\mathcal{M}_{\rm c})^{5/3}(\pi f)^{2/3}\sin \Phi\,,
\label{lum_dist2}
\end{align}
where $\Phi$ is the phase of the wave, $\mathcal{M}_{\rm c} \equiv (1+z) \mu^{3/5} m^{2/5}$  is the {\it redshifted} chirp mass ($\mu$ is the reduced mass and $m$ the sum of the  masses of the two bodies), $f$ is the observed GW frequency, $\iota$ is the inclination angle of the normal to the orbit with respect to the line of sight and $d_{\rm L}$ is the luminosity distance of the source, which for a spatially-flat universe reads $d_{\rm L}(z) \equiv (1+z) \int_0^z\frac{c}{H(z')}\,\dd z'$,
where $H \equiv \dot a /a$ is the Hubble parameter. One factor that hinders  the determination of the luminosity distance is  its degeneracy with the inclination angle. This degeneracy can be broken if higher order modes are included
%
or if  the binary has a precessing spin, in which case the characteristic modulation of the amplitude can disentangle the inclination angle. 

Standard sirens with LISA  will provide a formidable probe of the cosmic-expansion history, from low and intermediate redshifts of $z \lesssim 0.1$ and $0.1 \lesssim z \lesssim 1$ with stellar-mass BBHs and EMRIs, to deep redshifts of $z\lesssim 10$ for MBBHs.
 This will naturally place stringent constraints on a great amount of dark energy models (see e.g., \citealt{Tamanini:2016zlh, Belgacem:2019pkk}).
While the low-redshift data will be of particular interest for cosmic acceleration \citep{Lombriser:2015sxa}, the high-redshift data will provide invaluable constraints on early dark energy \citep{Caprini:2016qxs} in a regime that is not accessible to other observational methods such as  large-scale structure surveys.
 
For this scope, redshift measurements are of primary importance.  These can be obtained through a detection of an EM counterpart of the GW event.   
In the context of LISA only MBHBs are expected to produce counterparts at cosmological distances. There are non-standard scenarios to produce EM counterparts for SOBHBs and IMBHBs, which could be used as multi-band standard sirens.

When an EM counterpart is not available, one can still obtain the redshift information in a statistical way.
One method relies on the use of cross-correlation of each GW event with a galaxy catalogue, which provides a set of potential hosts inside the GW localization volume times its redshift uncertainty, computed from the GW distance measurements by taking into account the full prior range on all cosmological parameters \citep{Schutz:1986gp,MacLeod:2007jd,DelPozzo:2011yh,Chen:2017rfc,Fishbach:2018gjp,Gray:2019ksv,Soares-Santos:2019irc,Abbott:2019yzh,Palmese:2020aof,Vasylyev:2020hgb}.  
Combination of multiple events can significantly increase the constraining power of this method, as results from spurious galaxies will eventually average out. On the other hand, one has to properly take into account accurate modelling of the redshift uncertainties of galaxy catalogues, completeness issues---which can be relevant for sources at high redshift---and GW selection effects. The ideal candidates in this case are events with good localization and distance measurement, so that the potential hosts are in limited number (or even unique; \citealt{Borhanian:2020vyr}), at redshifts such that the available catalogues have a sufficient completeness \citep{DelPozzo:2017kme}.

Alternatively, one can exploit the full three-dimensional spatial cross-correlation between GW sources and the galaxy distribution \citep{Mukherjee:2020hyn,Bera:2020jhx}.
These techniques are the only possibility for LISA sources that are not expected to produce an EM counterpart, such as EMRIs,  SOBHBs, and IMBHBs. 
Although the cosmological potential of these sources is limited to low and intermediate redshift by the completeness of galaxy catalogues, they will anyway yield interesting constraints on $H_0$, reaching the percent level in the most optimistic scenarios \citep{MacLeod:2007jd,Kyutoku:2016zxn,DelPozzo:2017kme}.

As an alternative to the use of a galaxy catalogue, the redshift information can be obtained from the fact that the GW waveform gives the masses of the binaries redshifted by cosmic expansion. Information about the mass distribution of GW sources can then break this mass-redshift degeneracy and  allow a test of the luminosity distance-redshift relation. For stellar-mass BHs, this can be done by exploiting the presence of a mass scale in the  population \citep{Farr:2019twy,You:2020wju,Ezquiaga:2020tns}. For binary NSs, it is possible to jointly constrain the parameters of the NS mass function and the cosmological parameters  \citep{Taylor:2011fs,Taylor:2012db}.

\subsubsection{Luminosity distance}
\label{Sec:GWmodprop}

In GR, the observed gravitational strain is inversely proportional to the
luminosity distance to the GW source, see  \cref{lum_dist1,lum_dist2}. This can be understood by considering the free propagation of GWs, which in Fourier space is governed by  the equation
\be\label{4eqtensorsect}
{h}''_\lambda (\vk, \eta ) +2{\cal H} {h}'_\lambda  (\vk, \eta )+k^2 {h}_\lambda  (\vk, \eta ) =0\, ,
\ee
where $\lambda = +, \times$ is the polarization index (see above), a prime denotes the derivative with respect to conformal time $\eta \equiv \int \dd t/a(t)$ and
${\cal H}\equiv a'/a$ is the conformal Hubble parameter. This equation can be rewritten by defining   $\chi_\lambda \equiv a h_\lambda$, so that it  becomes
$ {\chi}''_\lambda+(k^2-a''/{a})  {\chi}_\lambda=0$.
For modes well inside the horizon, such as those considered for LISA,  the term $a''/a$  can be neglected and we obtain a frictionless wave equation for $\chi$.
For a spherical wave, the amplitude evolves as $\chi_\lambda\propto 1/r$, where $r$ denotes the comoving distance to the source, so that $h_\lambda \propto 1/(a r) = 1/d_{\rm L}$.

In alternative gravity models, however, the  strain is inversely proportional to a
``GW luminosity distance''\footnote{Strictly speaking, interferometers measure the strain of GWs, not their  luminosity, although the two are related. In modified gravity this relation is generally different from the GR one.} which is generally different from the one measured with EM signals. To distinguish between the two, in the following we denote  the former by  $d_{\rm L}^{\rm gw}$ and  the latter by $d_{\rm L}^{\rm em}$. Indeed, in non-GR theories the propagation equation (\ref{4eqtensorsect}) is generally modified. Here we consider only  scale-independent modifications (see Sects.~\ref{Sec:Tests_with_GW_propagation} and \ref{sec:GW_propagation_test} for scale-dependent ones) and thus one can have \citep{Saltas:2014dha,Gleyzes:2014rba,Lombriser:2015sxa,Nishizawa:2017nef,Arai:2017hxj,Belgacem:2017ihm,Amendola:2017ovw,Belgacem:2018lbp}
\be\label{prophmodgrav}
{h}''_\lambda (\vk, \eta ) +2{\cal H} [1-\delta(\eta)] {h}'_\lambda (\vk, \eta)+k^2 {h}_\lambda ( \vk, \eta) =0\, ,
\ee
for some function $\delta(\eta)$. In this case, to eliminate from the equation the  term proportional to $ {h}'_\lambda$, we
may introduce ${\chi}_\lambda\equiv \tilde{a}h_{\lambda}$ with $\tilde{a}'/{\tilde{a}}={\cal H}[1-\delta(\eta)]$. As before, $\chi_{\lambda}\propto 1/r$ for a spherical wave and hence $h_\lambda\propto 1/(\tilde{a}r)$. In other words, the GW luminosity distance is now related to the EM one by
\be\label{dLgwdLem}
d_{\rm L}^{\,\rm gw}(z)=d_{\rm L}^{\,\rm em}(z)\exp\left\{-\int_0^z \,\frac{\dd z'}{1+z'}\,\delta(z')\right\}\, .
\ee

Non-trivial background cosmological solutions such as self-accelerating ones \citep{Crisostomi:2017pjs, Lombriser:2015sxa, Lombriser:2016yzn} are of particular interest since they are responsible for the peculiar behavior of the friction term, which can change the ratio $d_{\rm L}^{\,\rm gw}/d_{\rm L}^{\,\rm em}$. For other works studying the damping of GWs see e.g., \citet{Deffayet:2007kf,Calabrese:2016bnu,Visinelli:2017bny,Amendola:2017ovw,Pardo:2018ipy,Frusciante:2021sio}. A deviation in the friction term  has potential signature in LISA \citep{Belgacem:2019pkk} or third-generation ground based standard sirens observations. GWs detectors are thus a complementary means with other cosmological surveys for constraining the friction term.  

In general, the reconstruction of a full function of redshift from the data is challenging and it is convenient to have a useful parametrization (analogous, for instance, to the $(w_0,w_a)$ parametrization of the dark energy equation of state). For modified GW propagation, a convenient parametrization can be given in terms of two parameters $(\Xi_0,n)$ \citep{Belgacem:2018lbp}, 
\be\label{eq:fit}
\frac{d_{\rm L}^{\,\rm gw}(z)}{d_{\rm L}^{\,\rm em}(z)}=\Xi_0 +\frac{1-\Xi_0}{(1+z)^n}\, .
\ee
This parametrization reproduces the fact that, as $z\ra 0$, $d_{\rm L}^{\,\rm gw}/d_{\rm L}^{\,\rm em}\ra 1$  since,  as the redshift of the source goes to zero, there can be no effect from modified propagation. In the limit of large redshifts, in \eq{eq:fit} $d_{\rm L}^{\,\rm gw}/d_{\rm L}^{\,\rm em}$ goes to a constant value $\Xi_0$. This is motivated by the fact that
in  typical  DE models
associated with the cosmic acceleration 
the deviations from GR only appear in the recent cosmological epoch, so $\delta(z)$ goes to zero at large redshift and, from \eq{dLgwdLem},
$d_{\rm L}^{\,\rm gw}(z)/d_{\rm L}^{\,\rm em}(z)$  saturates to a constant.\footnote{In a recent LISA Cosmology WG paper \citep{Belgacem:2019pkk}, the predictions of some of the best-studied modified gravity models (such as several examples of
Horndeski and DHOST theories, non-local infrared modifications of gravity, or  bigravity theories)  have been worked out, and it has been found that  all these models  predict  modified GW propagation of the form (\ref{prophmodgrav}), and the corresponding ratio $d_{\rm L}^{\,\rm gw}(z)/d_{\rm L}^{\,\rm em}(z)$ is well reproduced by the functional form (\ref{eq:fit}), except in bigravity, where the interaction between the two metric leads to oscillations in $d_{\rm L}^{\,\rm gw}(z)/d_{\rm L}^{\,\rm em}(z)$.}

For Horndeski and beyond-Horndeski theories the above discussion simplifies. In this case the damping in eq.~\eqref{prophmodgrav} comes from the time dependence of the effective Planck mass $M_*$ that canonically normalizes the graviton \citep{Gleyzes:2013ooa}. In particular, the quadratic action  for $h_\lambda$ reads,  using the conformal time, $S \propto \int \dd^3x \dd\eta (a M_*)^2 \left[  h_{\lambda}'{}^2 - (\partial_k h_{\lambda})^2 \right]$, which shows that in this case the quantity that satisfies a standard wave equation inside the horizon is $\chi_\lambda = a M_* h_\lambda $, with an extra $M_*(\eta)$ in the normalization. This implies that
\begin{equation}
\label{eq:ratio_distance_Horndeski2}
\frac{{\dgw} (z) }{\dem (z)} = \frac{M_*(0)}{M_*(z)} \ ,
\end{equation}
i.e., the ratio between the GW and EM luminosity distance simply reflects the ratio between the value of the effective Planck mass at the observer (at redshift $0$) and the one at the source at redshift $z$. If these are equal, for instance as a consequence of other constraints (see Sect.~\ref{sec:screening} above), then there is no effect \citep{Dalang:2019fma,Dalang:2019rke}. Another way to see this is by noting that  
the damping term in eq.~\eqref{prophmodgrav} is $\delta  = - \dd \ln M_*/\dd \ln a$   (see e.g., \citealt{Gleyzes:2014rba,Bellini:2014fua});  Eq.~\eqref{eq:ratio_distance_Horndeski2} then follows from Eq.~\eqref{dLgwdLem}.
 
In the past, studies of dark energy at the next generation of GW detectors have focused on the DE equation of state. However, 
while current EM observations show that in modified gravity theories deviations from the $\Lambda$CDM cannot be much larger than a few percent, modified GW propagation is a phenomenon accessible only to GW detectors, and therefore currently basically unconstrained.  For instance, 
there are explicit examples of phenomenologically viable modified gravity models that predict a very large effect. In particular, the RT non-local model mentioned above predicts a value of $\Xi_0$ that, depending on a free parameter of the model (related to the initial conditions) can be as large as $\Xi_0\simeq 1.8$ \citep{Belgacem:2017cqo,Belgacem:2019lwx}, corresponding to a $80\%$ deviation from the GR value $\Xi_0=1$.
Therefore, as observed in  \citet{Belgacem:2017ihm,Belgacem:2018lbp} modified GW propagation  provides  a complementary and  more promising observable than those inferred from EM observations.

\begin{figure}[t]
\begin{center}
\includegraphics[width=0.4\textwidth]{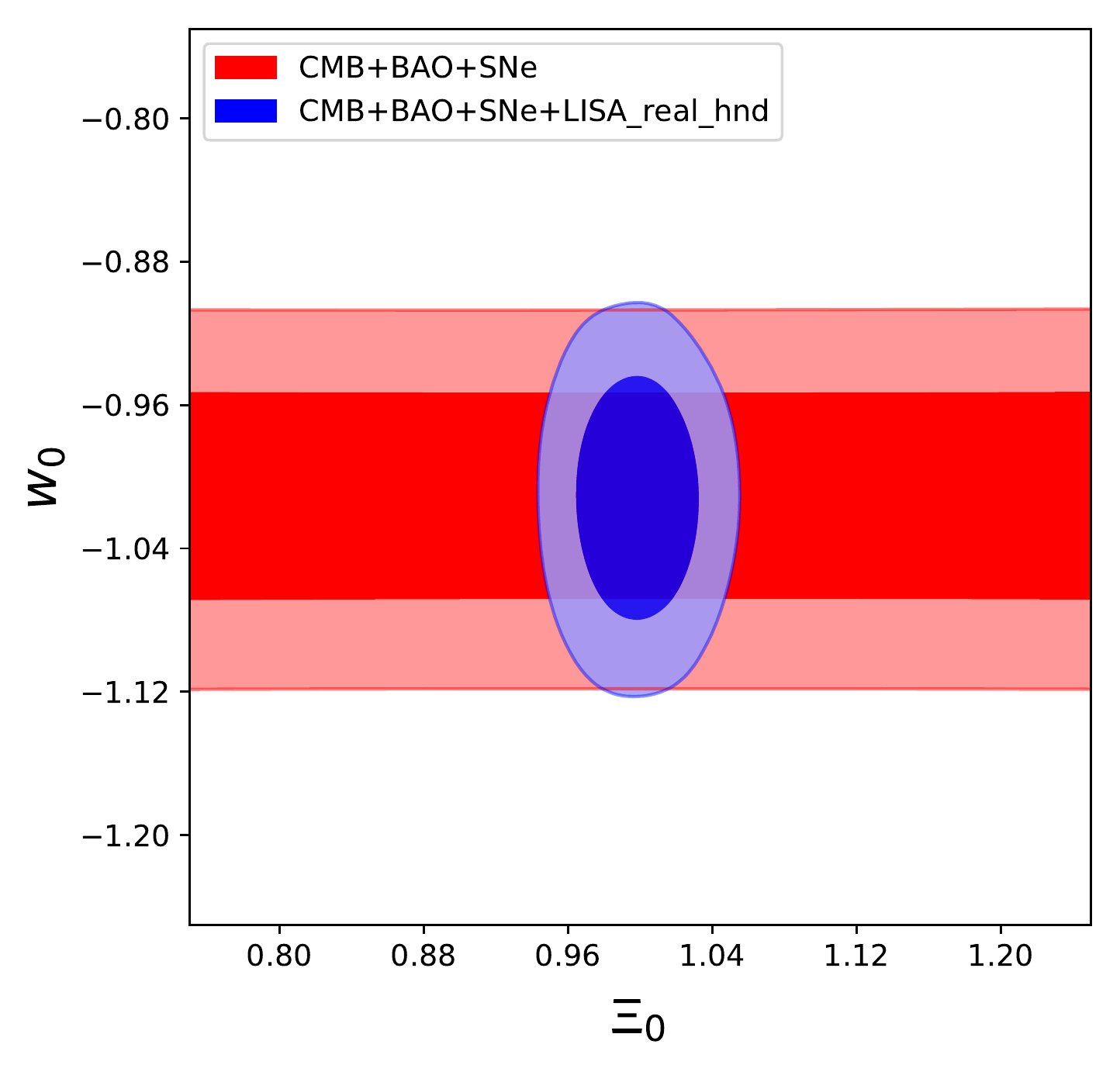}
\includegraphics[width=0.4\textwidth]{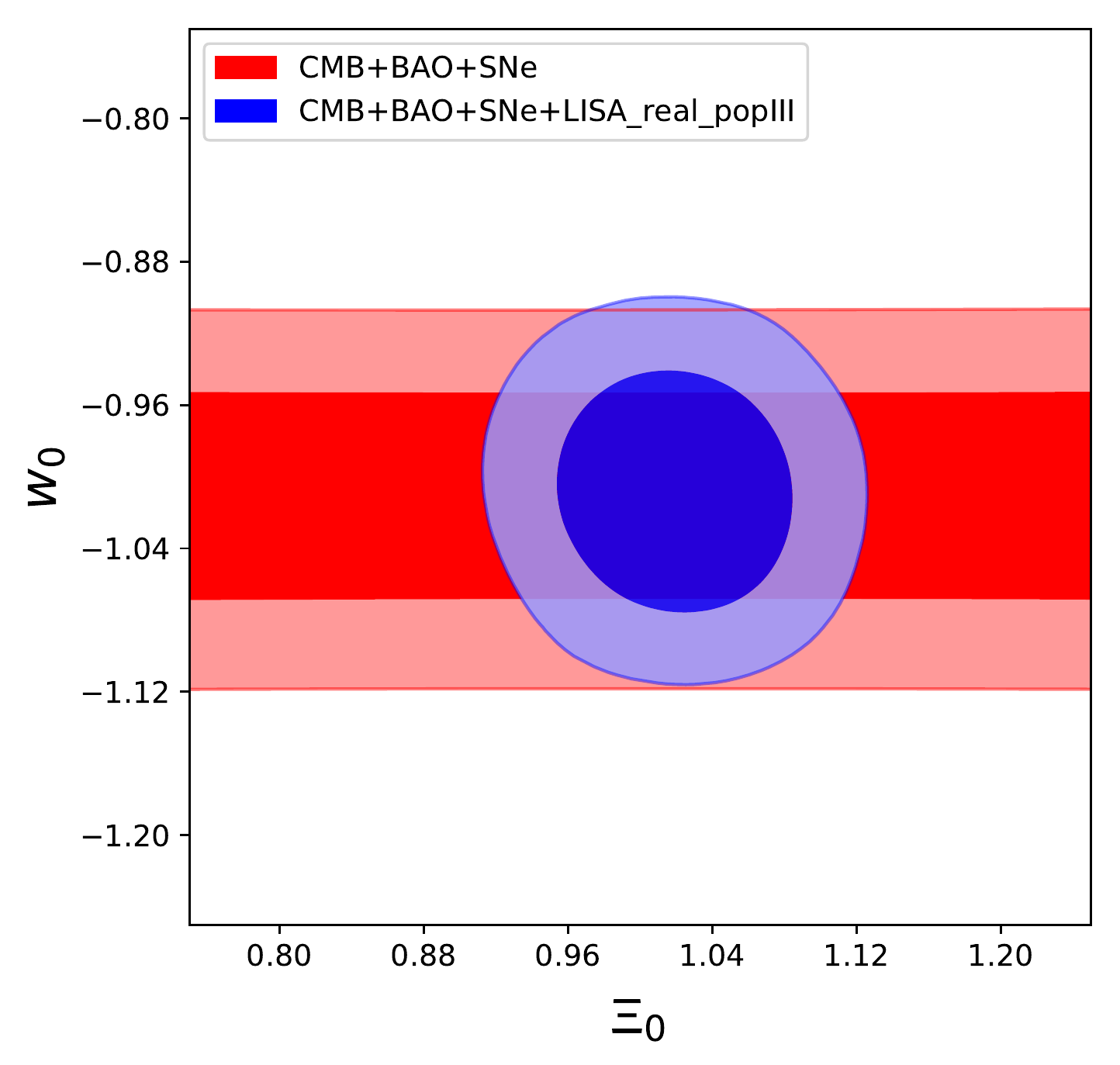}
\end{center}
\caption{The  $1\sigma$ and $2\sigma$
contours  of
 the  two-dimensional likelihood in the $(\Xi_0,w_0)$ plane, with the combined  contribution from 
CMB+BAO+SNe (red)  and the  combined contours
from  CMB+BAO+SNe+LISA standard sirens (blue).  Left: heavy seads and no-delay (``hnd'') scenario; right: ``pop~III'' seeds. From \citet{Belgacem:2019pkk}.}
\label{fig:xi0w0real}
\end{figure}

Constraints on a modified GW luminosity distance were placed after the multi-messenger event GW170817 \citep{Arai:2017hxj,Belgacem:2018lbp,Lagos:2019kds}.
The first limits on $\Xi_0$, using dark sirens from the O1-O2-O3a LIGO/Virgo runs have been recently presented in \citet{Finke:2021aom}, while constraints using the binary black hole mass distribution were obtained in \citet{Ezquiaga:2021ayr}.
Forecasts on the sensitivity of LISA to $\Xi_0$ (together with other parameters, such as the parameter $w_0$ that enters in the parametrization of the DE equation of state) have been  performed in \citet{Belgacem:2019pkk}, using  the coalescence of  SMBH binaries. These are  expected to produce powerful EM counterparts, since they are believed to merge in a gas rich environment, and are therefore potential standard candles for LISA. The results, using the astrophysical models of \citet{Barausse:2012fy,Klein:2015hvg},
which make a range of plausible assumptions
for the BH seeds and for the merger delay times,  and performing full MCMC exploration of the parameter spaces, confirm the strong potential of LISA for constraining, or detecting, modified GW propagation. As an example, Fig.~\ref{fig:xi0w0real}
shows the forecast in the $(\Xi_0,w_0)$ plane, for two different astrophysical scenarios.

To end this discussion, let us remark that in theories beyond GR that feature additional degrees of freedom, be it scalar, vector, or tensor, GWs are expected to exhibit new polarisations. In general, these additional polarisations will interact with the standard ones and such interactions may affect the amplitudes of the standard polarization modes, whose energy could {\it leak} into the new ones, thereby changing the measured gravitational distance $\dgw$. 
Over cosmological backgrounds GW mixings only appear when additional tensor modes are present \citep{Max:2017flc,Caldwell:2016sut,Jimenez:2019lrk} or cubic interactions are taken into account \citep{Creminelli:2018xsv,Creminelli:2019nok}. Over general space-times, quadratic mixing with scalars and vector modes are also possible \citep{Dalang:2020eaj,Ezquiaga:2020dao} (see also discussion in Sect.~\ref{Sec:Tests_with_GW_propagation}).
\subsubsection{Hubble constant}

The present value of the Hubble parameter $H_0$ is one of the fundamental parameters of the $\Lambda$CDM model. Over the last few years, a disagreement has persisted between local measurements obtained from the late-time distance ladder (such as the type Ia supernovae calibrated by cepheids \citep{Riess:2018byc}) and the  value inferred from early-time observables such as the CMB \citep{Aghanim:2018eyx} and the large-scale structure \citep{Philcox:2020vvt,DAmico:2020ods} (see e.g., \citealt{Shah:2021onj} for a recent review on the Hubble constant tension).
In the next decade, GWs will play a crucial role  in constraining $H_0$. The luminosity distance of a BBH and a BNS can be inferred and calibrated using an empirically constructed distance ladder at various distance scales. If combined with the known redshift of the host galaxy, the present rate of  $H_0$ can then be estimated \citep{Schutz:1986gp}, as it is done using type Ia supernovae via the local distance ladder \citep{Riess:2016jrr} 

The first standard siren-based $H_0$ measurement came from the  neutron-star merger event, GW170817, with the optical counterpart identification of the host galaxy, NGC 4993. While currently the observations of this merger set a loose constraint of $H_0= 70 ^{+12}_{-8}$ at 1$\sigma$ \citep{Abbott:2017xzu}, future detections of additional GW standard sirens,  obtained with ground-based second-generation detectors at full sensitivity, have the power to constrain the Hubble parameter to an accuracy of $\sim 1\%$ with an error that depends on the number of detected  events \citep{Nissanke:2013fka,Chen:2017rfc} and redshift distribution \citep{Dalal:2006qt}. 
Notably, for nearby sources, the dominant uncertainty comes from the error on the peculiar velocities when correcting the measurement of the observed redshift \citep{Boruah:2020fhl,Nicolaou:2019cip}.
A recent work by the LISA Cosmology WG \citep{Belgacem:2019pkk} forecast $H_0$ within the $\Lambda$CDM scenario using MBH and found the relative error to be 3.8\% in the optimistic case for the accuracy of redshift measurement and delensing and 7.7\% in the more realistic case (see also \citealt{Tamanini:2016zlh}). 

Furthermore, the rate of dark GW events without detection of EM counterpart, such as EMRIs, SOBHBs, or IMBHBs, should be much higher than those with EM counterpart \citep{Lyutikov:2016mgv}. To these dark sirens are also added all events whose EM counterpart will be difficult to identify, such as possible well-localised SMBHs at relatively low-redshift \citep{Petiteau:2011we}.
The strategy adopted by GW detectors then consists in estimating the Hubble parameter for each potential host galaxy of the GW, resulting in an average value of $H_0$ \citep{Schutz:1986gp,Chen:2017rfc,Fishbach:2018gjp}, as done for the GW170814
event \citep{Soares-Santos:2019irc}. In this context, the accurate sky location provided by LISA for the identification of the host galaxy of the GW event is definitely a major advantage \citep{Kyutoku:2016zxn} and the use of redshift catalogs from large-scale spectroscopic galaxy surveys such as the Dark Energy Spectroscopic Instrument (DESI) \citep{Aghamousa:2016zmz}, 4MOST \citep{deJong:2012nj} or \textit{Euclid} \citep{Amendola:2016saw} will therefore be required to improve the Hubble constant measurement.
Notably, LISA should be able to observe up to $\sim$100 stellar-mass BBHs within 100 Mpc, providing a few \% measurement of $H_0$ \citep{Kyutoku:2016ppx,DelPozzo:2017kme}.
This number however strongly depends on the performance of LISA at the high end of its frequency band, which is not guaranteed to provide the required sensitivity \citep{Moore:2019pke}.
A more promising population of dark sirens may be provided by EMRIs, which depending on the actual rate at which they will be detected, will yield constraints on $H_0$ at the few to one \% level \citep{Laghi:2021pqk}.

The current forecasts mentioned here above do not indicate that LISA will be able to contribute to solving the Hubble tension, which by the 2030s is expected to be largely solved by other observations. However they show that LISA will provide strong complementary constraints on $H_0$, which will be used to corroborate results obtained with other measurements and different techniques. This kind of cross-validation will be extremely important to reinforce our confidence on the actual value of the Hubble constant, especially if hints of deviations from $\Lambda$CDM will have been observed.

\subsection{Large-scale structure}\label{Sec5:Inhomcosmology}

The launch of LISA will be preceded by several large-scale structure surveys, such as DESI
and the Vera C.~Rubin Observatory
from ground, \textit{Euclid}
and Roman
from space, and SKA
in the radio band. These instruments will map a large part of the observable universe with unprecedented accuracy. Here we discuss how LISA observations can be combined with these surveys’ data and how the intervening large-scale structure will affect the observed GW signals.

\subsubsection{Cross-correlation with large-scale structure}

Another possible way to test cosmological models is represented by the cross correlation of GW detections with the large-scale structure (LSS) of the Universe.
This was pioneered initially to test dark energy and dark matter models and the distance-redshift relation in \citet{CameraGW, OguriGW, RaccanelliBHs}.
More recently, this probe became very popular and several different approaches have been suggested and investigated (see e.g., \citealt{Raccanelli2017GW, Scelfo:2018sny,Canas-Herrera:2019npr,Calore:2020bpd}), looking at both correlations with resolved sources and the SGWB \citep{Cusin:2018rsq, Jenkins2018PhRvD,Jenkins2019PhRvL, Bertacca2020PhRvD, 2020arXiv200710456Z, Mukherjee2020MNRAS, Alonso2020PhRvD}.

For the resolved-event case, the GW$\times$LSS correlation can be used for a variety of tests.
By measuring the effective bias of the hosts of compact-object mergers, one can get information about the presence and abundance of PBHs.
This happens because binary PBHs would preferentially merge in halos with low velocity dispersion, and hence halos with very little or absent star formation, which have a galaxy bias $b<1$. On the other hand, mergers of compact objects that are the endpoint of stellar evolution naturally happen for the vast majority within star forming-rich halos, which have larger galaxy bias values, $b>1$, where in the linear bias scheme, $b$ quantifies the mismatch between the distribution of matter and that of galaxies, $\delta_g=b\delta_m$.
Therefore, by measuring the amplitude of the angular cross correlation of galaxy maps with catalogues of compact-object mergers, which directly depends on the bias of the mergers' hosts, one can statistically probe the abundance of PBH mergers.
This information can also be used to discriminate between different astrophysical models (see e.g., \citealt{Scelfo2020})

When EM counterparts are available (or the redshift determination of BBHs is available through other methods), the GW$\times$LSS correlation can be used to test the distance-redshift relation \citep{OguriGW} and models of dark energy and modified gravity \citep{CameraGW,Raccanelli2017GW,RaccanelliVidotto2018, Mukherjee2020PhRvD}.

In the case of the SGWB, the cross correlation with the LSS will be fundamental to understand the observed signal.
GWs will experience projection effects that will need to be accounted for in order to observe the signal in the appropriate frame \citep{Bertacca2020PhRvD,Bellomo:2021mer}.
Moreover, the cross correlation with the LSS can allow us to disentangle the astrophysical SGWB from the primordial one coming from inflation.

\subsubsection{Gravitational lensing of GWs}

Similarly to EM radiation, GWs experience the gravitational potential of massive objects while traveling across the Universe. 
This opens the possibility of both probing the distribution of structures in the cosmos and the  underlying gravitational interactions. 
 LISA will offer a unique perspective since it will detect high-redshift GWs in a lower frequency band compared to ground-based detectors, increasing the lensing probabilities and the detectability of diffraction effects.

The probability that a GW is lensed depends on the redshift of the source and the distribution of lenses. For high-redshift GWs, intervening matter between the source and the observer can magnify the GW signal and introduce a systematic error in the determination of the luminosity distance. Correcting for this weak-lensing effect will be relevant when inferring cosmological and population parameters from GW standard sirens \citep{Holz:2002cn,Hirata:2010ba,Fleury:2016fda,Cusin:2020ezb}. 
A fraction of these GW events will pass close enough to the lens so that lensing will produce multiple images of the same event. 
LISA could detect a few strongly-lensed massive black-hole binaries during its mission \citep{Sereno:2010dr}. 
The detection of multiple lensing events could provide new means for cosmography with LISA \citep{Sereno:2011ty}. 

In the strong-lensing regime the time delay between the images increases linearly with the mass of the lens, reaching delays of months for a $10^{12}M_\odot$ galaxy lens. Each of the lensed images will have a different magnification and they will acquire a fixed phase shift depending on how many times the image has crossed a lens caustic \citep{Schneider:1992}. 
The precise measurement by LISA of the phase of long duration signals could be key in distinguishing different types of lensed images \citep{Ezquiaga:2020gdt}.

When time delays between lensed images become shorter than the signal's duration, interference effects among the images become relevant. This wave-optics regime is achieved when the wavelength of the GW is comparable to the Schwarzschild radius of the lens. Since LISA will observe lower frequencies, 
diffraction effects will be more relevant than for ground-based detectors \citep{Takahashi:2003ix}. 
In the diffraction limit the GW waveform can be distorted and one could use these features to identify the event as lensed \citep{Dai:2018enj}. 
In addition, these distortions may make the GW phase to appear to arrive earlier than an unlensed EM signal \citep{Takahashi:2016jom, Morita:2019sau}, although this is only an apparent superluminality \citep{Suyama:2020lbf,Ezquiaga:2020spg}. This effect has to be taken into account \citep{Ezquiaga:2020spg} when for example inferring constraints on the speed of gravity from the possible pre-merger modulated EM brightness of a super-massive BBH  \citep{Haiman:2017szj}. 
If these lensing features distorting the GW waveform are not properly identified, they could be misinterpreted as modified gravity, see e.g., \citet{Cusin:2019rmt,Ezquiaga:2020gdt,Toubiana:2020drf,Ezquiaga:2021ler}

\subsection{Burning Questions} \label{Sec5:Future}

We have reviewed several ways in which LISA will contribute to test the $\Lambda$CDM model, and to probe the fundamental physics nature of its extensions involving dark energy and modified gravity. There are many open directions. 
\begin{itemize}
    \item {}Screening is likely to play an important role, yet largely unexplored. In particular, it would be relevant to understand what happens when a GW propagates from, to, and through screened regions of the Universe.
The  considerations of Sect.~\ref{Sec:GWmodprop} were made in the context of an idealised homogeneous and isotropic FLRW background space-time and in the absence of extra polarization modes. It is thus natural to wonder which of its conclusions hold as GWs propagate through the actual inhomogeneous Universe or in the presence of extra modes. Theoretical analyses along these lines were initiated in \citet{Dalang:2019fma,Dalang:2019rke,Wolf:2019hun,Lagos:2020mzy} for Horndeski theories. 
\item{Another direction is to assess how other standard candles such as SMBH binaries and EMRIs will improve, using statistical methods, our measurements of $H_0$,  of the equation of state of dark energy and of the effects on the GW propagation.}
\item{Moreover, it would be relevant to explore, in terms of forecasts, how the synergy between LISA and large-scale structure surveys will help to improve the redshift estimate of GW sources and the inferred cosmological parameters.}
\item{Another possible synergy \citep{Piro:2021oaa} is  with the Athena observatory \citep{Nandra:2013jka}.
Measuring the X-ray signal and optical follow-up will make it possible to identify the host of an individual GW source. The distance-redshift relationship can then be measured to high accuracy ($1\%$) to probe the Hubble rate.} 
\item{Finally, we have discussed several effects of lensing of the GWs but their potential to constrain dark energy and modified gravity remains largely unexplored.}
\end{itemize}

\section{Model-Independent Tests }
\label{sec:model-indep-tests}


\vspace{0.25cm}

Although one can test a given model of gravity by comparing its predicted GW templates against data, a different approach is to perform model-agnostic tests. There are two main types of such tests. The first one is a consistency test of GR, where assuming that GR is correct one compares the GR waveform with GW data. The second one aims at constraining modified theories of gravity by performing a certain model-independent test and then mapping the outcome to specific modified gravity models. We will discuss these tests in some detail in what follows. 

\subsection{Consistency tests of GR}
Let us first discuss tests to check the consistency of GR predictions with the detected GW signals.
We begin with the residual test, which is the most generic, model-independent test that could, in principle, pick up arbitrary departures of the theory from the data. We will then describe a test looking for consistency in the intrinsic parameters of the source (e.g. the merger remnant's mass and spin) as determined from the early inspiral phase and the late IMR phase of the coalescence event. Finally, we discuss a generalization of the BH no-hair test using the entire IMR waveform, wherein one compares the intrinsic parameters of the source determined from the different multipole modes of the signal. 

It is worth noting that consistency tests are also possible by comparing the parameters obtained from the GW signal with those from other techniques in a multi-messenger scenario \citep{Baker:2019nct}. There are already systems observed in GW detectors (with LIGO/Virgo) and EM telescopes \citep{GBM:2017lvd}. This will become much more prevalent with LISA \citep{McGee:2018qwb,Wyithe:2002ep,Chen:2020wan,Korol:2017qcx,Mukherjee:2019wfw}. In particular, joint observations of GW systems \citep{Congedo:2018wfn,McGee:2018qwb,Edwards:2019tzf}, will provide independent estimates of the BH parameters assuming GR, which can be compared for consistency. 
\subsubsection{Residual Tests}

A Bayesian inference algorithm can be used to estimate the parameters of a signal/source in the data \citep{LIGOScientific:2019hgc}. Such an algorithm would essentially maximize the likelihood $P(d|h),$ where $d$ is the data and $h$ is the expected signal. The parameters $\lambda_\alpha,$ that maximize the likelihood, are deemed to be the best estimate of the signal parameters.  If GR gives the correct description of GWs, then the data would be consistent with the waveform predicted by GR. In that case the {\textit{residual}}, constructed by subtracting the best-fit waveform from the data, would be consistent with background noise. Failure of GR to accurately describe the data would lead to a residual that is statistically inconsistent with background noise. Such tests have been applied to the existing GW events by the LIGO/Virgo collaboration \citep{TheLIGOScientific:2016src,LIGOScientific:2019fpa}.

For the test to be effective, it is critical to characterize the statistical properties of the residual. The data from GW detectors is often contaminated by non-stationary and non-Gaussian background. Thus, one cannot simply ask if the residual is consistent with a Gaussian distribution but one must ask if it is consistent with the detector noise at times when no GW signals are known to be present. One can deploy statistical tools such as Anderson-Darling or Kolmogorov--Smirnov tests to compare the residual with the data at other times. Alternatively, one can use a transient detection algorithm, e.g. Bayeswave \citep{Cornish:2014kda}, to estimate the {\textit{coherent SNR}} in the residual data at a certain statistical significance and ask what is the probability that one gets an SNR as large as for data sets that contain only noise. 

By construction, the residual test is the most generic model-independent method that one can construct. This is because, the test does not require any non-GR waveforms, does not use additional parameters in the waveform to look for deviations (unlike, e.g., in parameterized tests of GR described in Sect.~\ref{sec:parameterized_tests1} below) and is sensitive to departures from GR outside the region where GR waveform has most of its support, (e.g. echoes). Its drawback, however, is that subtle departures from GR cannot be easily identified since the test is not phase coherent and the purely statistical nature will make it difficult to identify the origin of the failure of the theory. 

More work is needed to determine how to implement such a residual test in LISA. An approach similar to LIGO/Virgo's may not work because, unlike LIGO/Virgo currently, LISA will always have a background of sources that are on during the entire observation period. Therefore, removing only the loudest signal from the data will not necessarily lead to a residual that is consistent with noise, i.e.~there could be other signals hidden below the loudest one. Work has began to determine how to deal with multiple simultaneous sources in the data for parameter estimation reasons, but similar studies should be undertaken to generalize LIGO/Virgo's residual tests to LISA.

\subsubsection{Inspiral-merger-ringdown (IMR) tests}
\label{sec:IMRconsistency}
The orbital dynamics of a BBH essentially consists of three phases: (i) a long {\textit{adiabatic inspiral phase}} during which the companion objects slowly spiral in towards each other as they lose energy to GWs, (ii) a {\textit{rapid plunge-merger phase}} during which the system cannot be described by orbital dynamics as the horizons of the two BHs merge to form a single horizon, and (iii) a final {\textit{ringdown phase}} during which the merger remnant quickly settles down to its final stationary state. 
An IMR consistency test compares the parameters determined using the inspiral phase of the signal only to those obtained using the late-time merger-ringdown phase \citep{Ghosh:2016qgn,Ghosh:2017gfp}. If GR correctly describes both the inspiral and the plunge-merger regimes then the parameters determined from these two phases would be consistent with each other, within statistical uncertainties associated with the measurement; any discrepancy would then be a hint of a failure of the theory.

The IMR test currently implemented by the LIGO/Virgo collaboration \citep{TheLIGOScientific:2016src,LIGOScientific:2019fpa} translates the intrinsic parameters (the masses and spins) measured in the inspiral phase to the mass $M_f$ and the dimensionless spin magnitude $\chi_f$ of the merger remnant using a fitting formula derived from NR simulations. One then plots contours of the (two-dimensional) posterior probability density in the $M_f$-$\chi_f$ plane at different credible intervals. An IMR test therefore checks if the 90\% credible contours corresponding to the two different ways of measuring the parameters overlap with each other. Non-overlapping contours could indicate that one of the two phases is not consistent with the GR model. Large BH spins and small mass ratios are potentially more effective in such tests as they break the symmetries that could hide failure of GR in equal-mass systems with non-spinning BH companions. The IMR consistency test will strengthen its power when inspiral observations with LISA is combined with merger-ringdown observations with ground-based detectors through multiband GW observations \citep{Tso:2018pdv,Carson:2019rda,Carson:2019kkh}. 

One can go beyond the consistency check of GR and apply the IMR test to probe specific non-GR theories and spacetimes. Unlike the parameterized tests to be discussed in Sect.~\ref{sec:parameterized_tests1}, there is no simple known mapping to convert the information of the theory-agnostic IMR consistency test to bounds on specific theories or spacetimes, and hence the application needs to be done on a case-by-case basis. This has been demonstrated for a string-inspired quadratic gravity theory \citep{Carson:2020cqb} and generic BH spacetimes beyond Kerr \citep{Carson:2020iik}. The IMR tests with multiband GW observations will be able to probe such a theory and spacetimes more accurately than the existing GW and EM observations by a few orders of magnitude.

One problem that must be overcome to carry out such IMR tests with LISA is to develop a more thorough understanding of the mass and spin of the remnant in mergers of intermediate mass ratio systems and systems with non-negligible eccentricity and double-spin precession. Indeed, unlike LIGO/Virgo sources, many LISA binary sources are expected to be eccentric and spin-precessing. Waveform models for such systems are beginning to be developed, but detailed numerical simulations are currently lacking. We refer the interested reader to the waveform modeling white paper \citep{LISAWavWGWP} for further details. 

\subsubsection{Multipolar gravitational wave tests}

GWs at leading-order in a post-Minkowskian expansion is quadrupolar in GR. However, non-quadrupole (higher order) modes make appreciable contribution to the GWs from BBHs with large mass ratios and misaligned spins -- which was indeed the case for GW190412 \citep{LIGOScientific:2020stg} and GW190814 \citep{Abbott:2020khf}. The multipolar structure of the GWs is fully determined by the intrinsic parameters of a binary, e.g. masses and spin angular momenta of the companion BHs in a quasi-circular orbit. One can formulate multiple ways of testing the consistency of the observed GW signal with the expected multipolar structure of GWs from BBHs in GR. This is called a {\textit{“no-hair” test of BBHs}} \citep{Dhanpal:2018ufk,Islam:2019dmk} as it is similar to testing the “no-hair” theorems for isolated BHs through mutual consistency of the quasi-normal mode spectrum.   

The main idea of this test is to look for consistency of the source parameters determined independently from the quadrupole and higher order modes. We typically use GR waveforms when performing consistency tests, so we do not consider lower-order multipole modes, such as scalar dipole modes discussed in Sect.~\ref{sec:GR_test_MBH}. In spirit, this idea is similar to checking the consistency of cosmological parameters estimated from the low- and high multipoles of the CMB radiation. There are multiple ways in which the test could be implemented. For example, one could use different intrinsic parameters to describe the phase evolution of the quadrupole and higher-order modes and, as in the IMR consistency test described earlier, check to see if the parameters  determined from these modes are consistent with one another. Alternatively, one could introduce additional parameters to describe the amplitude of the higher modes while keeping the amplitude of the quadrupole mode as in GR. Such tests could reveal any modification from GR of the multipolar structure of the binary that is imprinted in the GWs observed at detectors.

An advance that would be useful in this context is the development of a mapping between such generic higher-mode deviations from GR and specific modified theories. Most studies of the dynamics of binary systems outside of GR have been only done to leading PN order in the inspiral. Going beyond leading order would be essential to find the higher-modes discussed above as predicted in a given theory. 

\subsection{Parametrized Tests}
\label{sec:parameterized_tests1}

We now discuss model-independent tests that introduce generic parameters capturing non-GR effects independent of the underlying gravitational theory. We first present a generic framework, followed by a few examples on how one can use such parameterized tests to probe specific non-GR theories and BH spacetimes beyond Kerr.

Many modified gravity theories have been proposed over the years, and sufficiently accurate gravitational waveform models are only available for a small subset of these theories, when their calculation is even possible at all \citep{Berti:2015itd}.  Therefore it is appealing to develop generic ``null'' tests of Einstein's theory, following the approach pursued with Solar System tests,  where the parametrized PN (PPN) framework proposed by Will and Nordtvedt \citep{Nordtvedt:1968qs,1971ApJ...163..611W,1972ApJ...177..757W,1972ApJ...177..775N} has provided a unifying scheme for tests of GR that has been in use for over 50 years.

\noindent
\subsubsection{Inspiral tests}
\label{sec:parameterized_insp}
In GW data analysis, a natural generalization of the PPN approach consists of verifying the PN structure of the waveform phase \citep{Arun:2006yw}. The idea is to decompose the Fourier-domain waveform model into a frequency-dependent amplitude and a frequency-dependent phase, and to then rewrite the phase (schematically, and ignoring logarithmic terms) as $\Psi_{\rm GR}(f) = \sum_{n=0}^{7} \alpha_{n} v(f)^{-5+n}$. In GR the PN coefficients $\alpha_{n}$ are known functions of the parameters of the binary (the individual masses $m_1$ and $m_2$ for non-spinning BH binaries of total mass $m=m_1+m_2$), and $v(f) = (\pi m f)^{1/3}$ is the orbital velocity. The idea is to treat all of these coefficients (or just a subset; \citealt{Arun:2006hn}) as independent, and find their best-fit values by comparing the above template waveform with the data. One can then check the consistency of the measured masses from each of the above coefficients. This procedure resembles binary pulsar tests in the parameterized post-Keplerian formalism \citep{Damour:1991rd,Stairs:2003eg}, but it has some limitations. There are known modified theories of gravity for which the Fourier phase does not have a leading-order term $\propto v^{-5}$: these include, for example, theories with dipole emission ($\propto v^{-7}$) and variability of the fundamental constants like the gravitational constant $G$ ($\propto v^{-13}$). Second, some modified gravity theories may modify the GW amplitude more than the phase: one example is gravitational birefringence \citep{Alexander:2007kv,Yunes:2010yf,Yagi:2017zhb}. 

One proposal to address these problems is the so-called ppE approach \citep{Yunes:2009ke}. In this framework, one extends the GR waveform model as follows:
\begin{equation}
\label{eq:PPE}
\tilde{h}(f) = \tilde{A}_{\rm GR}(f) \left[1 + \alpha_{\rm ppE}\, v(f)^{a}\right] e^{i \Psi_{\rm GR}(f) + i \beta_{\rm ppE}\, v(f)^{b}}\,.
\end{equation}
Here $\tilde{A}_{\rm GR}(f)$ and $\Psi_{\rm GR}(f)$ represent the most accurate GR models for the Fourier amplitude and phase. The quantities $(\alpha_{\rm ppE},\beta_{\rm ppE})$ are ppE constants that control the magnitude of deviations from GR, while $(a,b)$ are real numbers that determine the type of deviation that is being constrained. The ppE approach is similar in spirit to the parametric analysis performed by the LIGO/Virgo collaboration \citep{Abbott:2018lct,TheLIGOScientific:2016src}. Different variants and extensions of this idea have been proposed \citep{Cornish:2011ys,Chatziioannou:2012rf,Sampson:2013jpa,Sampson:2013lpa}. Constraints on ppE parameters can be mapped to constraints on specific extensions of GR \citep{Yunes:2013dva,Yunes:2016jcc}. Astrophysical bounds on ppE parameters have also been derived using observations of relativistic binary pulsars \citep{Yunes:2010qb}, and parametrized tests using the ppE formalism (in particular the amplitude corrections) can also be used to detect deviations from GR in the GW background emitted by BBHs \citep{Maselli:2016ekw,Saffer:2020xsw}.

In most data analysis applications (including recent tests of GR by the LIGO/Virgo collaboration \citep{Abbott:2018lct,TheLIGOScientific:2016src}), one considers variations of a single ppE term. This is often convenient, and it can generally be accurate enough because higher-order terms will not significantly alter the bounds coming from leading-order corrections (see e.g., \citealt{Chatziioannou:2012rf,Yunes:2016jcc}). However we can expect modifications appearing at different orders to be correlated, and in general it is important to take into account these correlations, because in any specific theory, all modifications of GR will depend on a single coupling constant (or on a finite number of coupling constants).

These parametrized tests of inspiral waveforms can be thought of as generic null tests of GR, although they can be translated into bounds on specific beyond-GR theories on a case-by-case basis (see Sect.~\ref{sec:mapping}). Interestingly, they can be applied to several source classes, including SOBHs, EMRIs and SMBHs. Each source class has distinct advantages and disadvantages: SMBHs are expected to have large SNR, EMRIs should compensate for their lower SNR with the large number ($\sim 10^5$) of observable inspiral cycles, and -- as we discuss below -- SOBHs can be used to test GR modifications at large negative PN orders through multiband observations in conjunction with 3G detectors (such as the Einstein Telescope and Cosmic Explorer) that should be operational by the time LISA flies. Each source class can produce tight constraints on different ppE modifications (and therefore on different modifications of GR), depending on astrophysical event rates and uncertain details on the future network of ground-based interferometers. One of the main challenges facing theorists, astrophysicists and experimentalists  is to figure out how this complex interplay between astrophysical event rates, beyond-GR waveform models, ground-based detector developments and data analysis methods will affect our ability to do fundamental physics with LISA.

Multiband observations of stellar-mass BH binaries can play an important role in constraining modified gravity via parametrized tests. This is because ``multibanding'' allows us to combine the information from the early inspiral dynamics using LISA with late inspiral, merger and ringdown observations using third-generation ground-based detectors. For example, it has been shown that the bounds on dipole GW emission from multibanding can be several orders of magnitude better than bounds coming from either of the bands \citep{Barausse:2016eii,Perkins:2020tra,Toubiana:2020vtf}. Single-parameter tests at positive PN orders are also expected to improve tremendously, as shown in \citet{Carson:2019rda,Carson:2019kkh,Gnocchi:2019jzp}. The impact of multibanding on multiparameter tests is equally profound:  \citet{Gupta:2020lxa} shows that multibanding observations may be the only way of carrying out accurate multiparameter tests of GR, where simultaneous measurement of most of the known PN coefficients could be achieved with percentage-level precision.

One important challenge of these parameterized inspiral tests is to extend the mapping between generic deviations and specific theories; we will present two examples below (in Sect.~\ref{sec:mapping}). Indeed, the more such mappings we possess, the more inferences about fundamental physics we will be able to make from parameterized inspiral tests. Another challenge is to determine how to handle parameterically modifications of GR that do not admit a simple PN expansion in the inspiral. Indeed, certain theories of gravity, such as those that predict the sudden activation of dipole radiation at a specific length scale, cannot be exactly modeled as described. The above parameterization may be approximate enough to still be useful, but this needs further study in the LISA context. 

\noindent
\subsubsection{Ringdown tests}
Any parametrization requires analytical control of the "unperturbed" GR waveform, therefore it is very difficult (if not impossible) to construct parametrizations of the nonlinear merger of compact binaries. However one can attempt to construct parametrized ringdown waveforms based on perturbative treatments of the ringdown in GR.

Such parametrizations are useful because a detailed, {\textit{ab-initio}} description of the BH ringdown phase beyond GR is very challenging, especially for rotating BHs. Despite years of efforts and significant advances in the 
field \citep{Barausse:2014tra,Glampedakis:2017cgd,Tattersall:2017erk,Franciolini:2018uyq,McManus:2019ulj,Glampedakis:2019dqh,Cardoso:2019mqo}, the vast majority of theoretical proposals to parametrize ringdown waveforms are currently plagued by certain approximations and assumptions that make them of limited utility in GW data analysis. Modified ringdown waveforms should be built upon the existence of BH solutions different from those predicted by GR (at least in theories where no-hair theorems do not apply), and in addition they must take into account that the perturbations will have different dynamics even for theories which admit the same solutions as GR \citep{Barausse:2008xv,Molina:2010fb,Tattersall:2017erk}.
These restrictions reflect the complexity of the problem, and they are of a varied nature. Some calculations were developed in specific modifications of GR, but most of them assume nonrotating BH backgrounds. Most follow-up works that attempt to relax the nonrotating approximation either deal only with scalar (rather than gravitational) perturbations, use the so-called geometric optics approximation, which is only valid in the eikonal (high-$\ell$) limit and in the absence of coupling between the metric and other degrees of freedom (but see \citealt{Silva:2019scu} for an attempt to include such a coupling with a small BH rotation), or work in the slow-rotation approximation \citep{Wagle:2021tam}. 

One of the most difficult open problems is the modelling of rotating BHs beyond GR (see also Sect.~\ref{sec2:BHbeyodGR}). This is important, because NR and recent GW observations indicate that the spin of the merger remnant is generally large \citep{Buonanno:2006ui,Berti:2007fi,Berti:2008af,Hofmann:2016yih,LIGOScientific:2018mvr}. 
Unfortunately most rotating BH solutions beyond GR are only known either perturbatively, 
as small-spin expansion around non-spinning backgrounds 
\citep{Pani:2011gy,Ayzenberg:2014aka,Maselli:2015tta,Cardoso:2018ptl,Cano:2019ore},
or they are given in the form of fully numerical solutions of the field equations \citep{Kleihaus:2011tg,Herdeiro:2016tmi,Cunha:2019dwb,Sullivan:2019vyi,Sullivan:2020zpf}, which makes the application of perturbation theory difficult.

The ``miraculous'' separability of the master equations for 
the perturbations in GR \citep{PhysRevLett.29.1114,1973ApJ...185..635T} is in general absent in modified theories of gravity, so the calculation of QNMs must rely on fits of time-domain waveforms computed from numerical simulations in selected classes of theories beyond GR, which are now possible (at least in the small-coupling limit and for selected theories \citep{Okounkova:2017yby,Witek:2018dmd,Okounkova:2019dfo,Okounkova:2019zep}) but technically challenging.

Rather than attacking the problem ``from the ground up'', various groups have considered a data-analysis-first approach, where one parametrizes deviations in the QNM frequencies and damping times \citep{Gossan:2011ha,Meidam:2014jpa,Carullo:2018sfu,Maselli:2019mjd}. In this framework, beyond-GR QNMs are described in terms of generic shifts from the corresponding Kerr QNM frequencies. Of course, once we set constraints on these shifts, we still need to map the frequency shifts to the new degrees of freedom of the underlying theory of gravity that produced them.

In conclusion, parametrized ringdown is an active research field which is still in its infancy.

\subsubsection{Mapping to non-GR Theories}
\label{sec:mapping}

Parametrized tests will show its power when bounds on generic, non-GR parameters can be mapped to those on theoretical constants in specific theories that represent violations of fundamental pillars in GR. In this subsection, we discuss two specific example theories, namely scalar-tensor theories and quadratic gravity. Mappings to other modified theories of gravity for both the phase and amplitude corrections can be found e.g. in Tables I and II of \citep{Tahura:2018zuq}.

One of the simple examples to which we can relate the parametrized tests of GR concerns scalar-tensor theories. In these theories the leading modification to GR is the presence of a dipolar radiation. Such a radiation is generated by the (dimensionless) scalar charge $\alpha_A$ of each body that violates the SEP. The parametrized dipolar test on the GW luminosity of $\dot{E}_{\mathrm{GW}} = \dot{E}_{\mathrm{GR}}\left(1+B\left(\frac{G m}{r c^2}\right)^{-1}\right)$ gives a bound on the parameter $B$ where $m$ is the total mass, $r$ is the binary separation and $B$ for scalar-tensor theories is $B_{\mathrm{ST}} = \frac{5}{96}\left(\alpha_{1}-\alpha_{2}\right)^{2}$ \citep{Barausse:2016eii}. Scalar charges are typically zero for BHs in scalar-tensor theories and $B_\mathrm{ST}=0$ for BBHs unless e.g. the scalar field is coupled to the Gauss--Bonnet invariant or some other forms of higher curvature invariant (see Sect.~\ref{sec2:BHbeyodGR}). However, $B_\mathrm{ST} \neq 0$ for EMRIs if the secondary objects are NSs.
Currently the best constraint on $B$ is given by the observation of binary pulsars, $|B|\lesssim 10^{-7}$ \citep{Yunes:2010qb,Shao:2017gwu}. 
This can be related with the ppE framework in Eq.~\eqref{eq:PPE} with $b_{\mathrm{ST}}=-7/3$ and $\beta = -\frac{3}{224}\eta^{2/5}B$. Such a correction enters at $-1$PN order to the waveform phase.

Another important example theory that can be tested within the ppE framework is quadratic gravity that introduces quadratic-curvature corrections (coupled to a scalar field) to the Einstein-Hilbert action. Under the small-coupling approximation, the leading-order modification to the Fourier phase $\Psi(f)$ takes on the ppE form \citep{Yunes:2009ke} in Eq.~\eqref{eq:PPE}, where $b_{{\mbox{\tiny dCS}}}=-1/3$ in dCS gravity (a 2PN correction) and $b_{{\mbox{\tiny EdGB}}} = -7/3$ in Einstein-dilaton Gauss--Bonnet gravity gravity (a $-1$PN correction).  The specific mapping to these two well-motivated theories can be found in \citet{Yagi:2011xp,Nair:2019iur} for EdGB, and in \citet{Yagi:2012vf,Nair:2019iur} for dCS.  From the data presented in first LIGO/Virgo catalogue, GWTC-1 \citep{LIGOScientific:2018mvr}, EdGB's coupling constant was constrained to ${\alpha}^{1/2}_{\mbox{\tiny EdGB}} \lesssim 5.6$ km \citep{Nair:2019iur} from single events and ${\alpha}^{1/2}_{\mbox{\tiny EdGB}} \lesssim 1.7$ km \citep{Yamada:2019zrb,Perkins:2021mhb} from combined multiple events, and the latter is better than previous bounds, such as those from BH low-mass x-ray binaries \citep{Yagi:2012gp}.  On the other hand, dCS gravity still remains unconstrained from the leading PN order correction to the waveform phase, although recent consistency constraints from the combination of LIGO/Virgo and NICER data require $\alpha_{\rm dCS}^{1/2} < 8.5$ km \citep{Silva:2020acr}. Work on other modified theories within the broad class of quadratic gravity models \citep{Yagi:2015oca} remains to be done.  

Apart from testing specific non-GR theories, one can use the parameterized tests to probe beyond-Kerr BH spacetime in a generic way. For example, it is possible to map the ppE constraints from the inspiral regime to constraints on parametric deviations from the GR BH metric \citep{Cardenas-Avendano:2019zxd,Carson:2020iik}. The modified ringdown frequencies can also be used to place constraints on deviations away from the GR BH metric \citep{Konoplya:2020hyk,Volkel:2020daa}. Similar to the IMR consistency tests discussed in Sect.~\ref{sec:IMRconsistency}, one can perform consistency checks between the above BH spacetime tests from the inspiral and ringdown. 
The two sets of constraints can be compared for discrepancies (which might be an indicator of new physics) or improving the overall constraints \citep{Carson:2020iik}. 

\subsection{Other Model-independent Tests}

There are other model-independent tests one can perform on GW observations with LISA (some of them have already been applied to the existing GW events by the LIGO/Virgo collaboration). In this section, we describe three such model-independent tests. First, one can look for additional GW (scalar and vector) polarization modes that are absent in GR. Next, one can study how the modified propagation of GWs affect the amplitude and phase of gravitational waveforms and how one can use multi-messenger observations or the arrival time difference of GWs at different detectors to probe such modified GW propagation. Lastly, one can use the SGWB of either astrophysical or cosmological (primordial) origin to test GR. 

\subsubsection{Polarization Tests}

A model-independent test of GR is based on the possibility of detecting additional polarizations, not present in GR. Metric theories of gravity can have up to six polarization modes -- two tensorial, two scalar and two vector modes. While in GR only the tensorial polarization modes are present, this is not the case for most of the viable alternative theories of gravity \citep{Will:2014kxa}. Such additional polarizations can be detected with the GW observations of inspiraling compact objects \citep{Chatziioannou:2012rf,Takeda:2018uai,Takeda:2019gwk,Pang:2020pfz}. Indeed, ground based detectors have already put an upper limit on such additional polarization modes in a theory-agnostic way \citep{Abbott:2017oio,Abbott:2018lct,LIGOScientific:2019fpa,Takeda:2020tjj,LIGOScientific:2020tif}, including the case with mixed polarizations of standard tensor modes with additional non-GR ones \citep{Chatziioannou:2021mij,Takeda:2021hgo}. 

LISA offers new possibilities for testing the non-GR polarizations with an increased sensitivity \citep{Tinto:2010hz}. For frequencies larger than roughly $6 \times 10^{−2}\,{\rm Hz}$, the sensitivity of  LISA for scalar-longitudinal and vector polarization modes can be up to ten times larger compared to the tensorial or scalar-transverse modes, while in the lower frequency, the sensitivity is of the same order. Therefore, it is expected that LISA will be able to assess the polarization of the detected GWs and thus put strong constraints on modified gravity.

The SGWBs (to be discussed in more detail in Sect.~\ref{sec:SGWB}) can also be a powerful test of non-GR polarizations \citep{Nishizawa:2009bf,Callister:2017ocg}. The sensitivities of LISA to circular-polarization modes in this context were addressed in \citet{Seto:2006hf,Seto:2006dz,Smith:2016jqs}. Such modes can appear for example in modifications of GR that lead to parity violation \citep{Alexander:2009tp}. 

Similar to the parameterized tests in Sect.~\ref{sec:parameterized_tests1}, we now discuss how GW polarization tests can be used to probe specific non-GR theories. Below, we focus on a few examples. 

In Horndeski theories, which form the most general class of non-degenerate scalar-tensor models leading to second-order equations of motion \citep{Horndeski:1974wa}, three degrees of freedom may propagate under the form of GWs. The first two are the usual plus and cross tensor polarisations (or equivalently the left- and right-handed helicities). Since the observation of GW170817/GRB~170817 \citep{TheLIGOScientific:2017qsa,Monitor:2017mdv}, these tensor modes are constrained to propagate at the speed of light (see Sect.~\ref{sec:GW_propagation_test}), which dramatically reduced the parameter space of viable Horndeski models \citep{Lombriser:2015sxa, Ezquiaga:2017ekz, Creminelli:2017sry, Sakstein:2017xjx, Langlois:2017dyl, Baker:2017hug}. Stringent constraints may also be obtained from birefrigence effects in lensed GWs \citep{Ezquiaga:2020dao}. However, even in such \textit{reduced} Horndeski theories, the third degree of freedom, namely scalar waves, may propagate at subluminal speeds.
While the tensor (plus and cross) modes produce shearing tidal forces that are transverse to the direction of propagation, scalar waves generally produce triaxial tidal forces. Specifically, in the rest frame of some inertial observer, there exists an orthonormal basis such that the curvature perturbation caused by a scalar wave reads \citep{Dalang:2020eaj}
\begin{equation}
(\delta R_{0i0j}^{\rm S})
=
\begin{pmatrix}
\mathcal{R}_1 & 0 & 0 \\
0 & \mathcal{R}_2 & 0 \\
0 & 0 & \mathcal{R}_3
\end{pmatrix}
\cos\Phi(t) \ ,
\end{equation}
where $\mathcal{R}_{1,2,3}$ are three amplitudes and $\Phi(t)$ is the phase of the scalar wave. Note that the third direction does not necessarily coincide with the direction of propagation of the wave. However, for luminal scalar waves, which only happens for conformally coupled quintessence, then $(\delta R_{0i0j}^{\rm S})$ reduces to a transverse ``breathing''  mode, i.e. $\mathcal{R}_1=\mathcal{R}_2$, and $\mathcal{R}_3=0$ in the direction of propagation.

Lorentz-violating gravity can have both scalar and vector modes. For example, Einstein-\AE ther theory has two vector modes and one scalar mode \citep{Foster:2006az,Foster:2007gr,Yagi:2013ava}, while Ho{\vr}ava gravity has one scalar mode with no vector modes \citep{Blas:2011zd,Yagi:2013ava}. GWs from compact binary inspirals have been derived in \citet{Hansen:2014ewa,Zhang:2019iim}, including the additional scalar and vector modes.

$f(T)$ gravity (with $T$ representing a torsion) is a popular modified gravity theory \citep{Ferraro:2006jd,Capozziello:2011et,Cai:2015emx} based on the teleparallel equivalent of general relativity (TEGR), see, e.g., \citet{Maluf:2013gaa} or \citet{Hayashi:1979qx} for New General Relativity, which is a variant of the teleparallel formulation. Some $f(T)$ gravity violates local Lorentz invariance \citep{Li:2010cg}. When a certain boundary term ($B = 2 \nabla_\mu T^\mu$ for a torsion vector $T^\mu$) is taken into account \citep{Bahamonde:2015zma} one can formulate a theory called $f(T,B)$ gravity which naturally links to $f(R)$ gravity. In the context of GWs these teleparallel theories are interesting models as they contain additional propagating degrees of freedom. While there is some controversy on the number of extra degrees of freedom in $f(T)$ gravity, see \citet{Li:2011rn,Ferraro:2018tpu,Ferraro:2018axk,Blagojevic:2020dyq}, there is agreement that the theory possesses at least one and up to three additional degrees of freedom. Consequently, these extra degrees of freedom should manifest themselves as extra polarizations. At first order in perturbation theory, extra polarizations only appear in $f(T,B)$  gravity and not in $f(T)$ gravity \citep{Farrugia:2018gyz}. Likewise, when considering New General Relativity it was found that up to six possible polarizations may exist, depending on the various parameter choices \citep{Hohmann:2018jso}.

Finally, we comment on probing extra dimensions through polarization tests. Within GR, the structure of the geodesic deviation equation in a general spacetime was first analyzed by \citet{Szekeres:1965ux} who decomposed the Riemann tensor into its canonical components, namely the transverse and traceless component, the longitudinal component, and the Newtonian-type tidal component.
 This analysis can be extended to a metric theory of gravity in any dimension $D$ \citep{Podolsky:2012he}. In such a case, the gravitational wave in (macroscopic) higher dimension ${D>4}$  is naturally identified as the transverse-traceless component with ${D(D-3)/2}$ independent polarization states. Higher-dimensional gravity also brings new effects. In particular, the presence of extra spatial dimensions into which the GWs could possibly propagate could be (in principle) observable as a violation of the traceless property of the gravitational wave measured by the detector located in our ${D=4}$ spacetime. Such an anomalous behavior could possibly serve as a sign of the existence of higher dimensions. Although this effect has already been described in \citet{Podolsky:2012he} and analyzed in the model of exact plane GWs (and their generalizations within the Kundt family of geometries) \citep{Podolsky:2013ola}, the specific representation of the violation of the transverse-traceless property in the triangular LISA configuration remains an open problem.

The development of polarization tests in the LISA context is still not mature enough for deployment in a data analysis pipeline. Few studies have been performed to determine how well LISA can constrain the extra polarizations of specific modified theories, given current constraints on the speed of tensor modes. The idea of the construction of null channels has been suggested for LIGO/Virgo \citep{Chatziioannou:2012rf}, but can also be extended for LISA in the ppE framework. The mapping of generic polarization tests to specific modified theories has also not yet been developed.  
 
\subsubsection{GW Propagation Tests}
\label{sec:GW_propagation_test}

As discussed in Secs.~\ref{Sec:Tests_with_GW_propagation} and~\ref{Sec:GWmodprop}, GW propagation over cosmological distances for generalised theories of gravity and other extensions of $\Lambda$CDM can be parametrised by the wave equation \citep{Saltas:2014dha,Gleyzes:2014rba,Lombriser:2015sxa,Nishizawa:2017nef,Belgacem:2017ihm,Lombriser:2018guo}
\begin{equation}
  h_{ij}'' + \left(3 + \frac{H'}{H} + \nu \right)h_{ij}' + \left[c_{\rm T}^2 \left(\frac{k}{aH}\right)^2 + \frac{\mu^2}{H^2} \right] h_{ij} = \frac{\Gamma}{H^2} \gamma_{ij} \,, \label{gwprop_eq:gw}
\end{equation}
where $h_{ij}$ is the linear traceless spatial tensor perturbation and primes indicate derivatives with respect to $\ln a$.
In addition to the Hubble friction, the damping term may receive a contribution from $\nu$, which can for instance parametrise an effective Planck mass evolution rate or the impact of extra dimensions.
The wave propagation may further be modified by a deviation in its speed $c_{\rm T}$ and the mass of the graviton $\mu$.
Finally, there can be a source term $\Gamma\gamma_{ij}$, which may for instance arise in bimetric theories or from anisotropic stress.
The modifications can generally be time and scale, or frequency, dependent, where GR/$\Lambda$CDM is recovered in the limit of $\nu=\mu=\Gamma=0$ and $c_{\rm T}=c$. 
The modifications in Eq.~(\ref{gwprop_eq:gw}) can also be cast into the ppE framework \citep{Yunes:2009ke} in Sect.~\ref{sec:parameterized_insp}, where the ppE amplitude parameter $\alpha$ can be expressed as an integral over $\nu$ while the ppE phase parameter $\beta$ may be expressed as an integral involving $c_T$ and $\mu$ \citep{Mirshekari:2011yq,Nishizawa:2017nef}.

Late-time constraints on the modified damping term have been forecasted in \citet{Lombriser:2015sxa,Belgacem:2017ihm,Amendola:2017ovw,Belgacem:2019pkk} based on standard sirens tests \citep{Schutz:1986gp,Holz:2005df}, 
and early-time modifications can be constrained by CMB B-modes \citep{Amendola:2014wma}.
So far there has not been much exploration of a possible frequency dependence of $\nu$ (see, however,  \citealt{Belgacem:2019pkk} for forecasts on oscillations in the GW amplitude).

The GW speed $c_T$ is constrained by a variety of measurements.
The detection of ultra high energy cosmic rays implies a strong constraint on gravitational Cherenkov radiation from a subluminal propagation of the waves as otherwise the radiation would decay away at a rate proportional to the square of their energy $\mathcal{O}(10^{11}~\textrm{GeV})$ before reaching us \citep{Moore:2001bv,Caves:1980jn,Kimura:2011qn}.
For galactic $\mathcal{O}(10~\textrm{kpc})$ or cosmological $\mathcal{O}(1~\textrm{Gpc})$ origin, the relative deviation in $c_{\rm T}$ is constrained to be smaller than $\mathcal{O}(10^{-15})$ or $\mathcal{O}(10^{-19})$, respectively.
This bound, however, only applies for subluminal propagation, redshifts of $z\lesssim0.1$, and modifications in the high-energy regime.
Another constraint on $c_T$ at the subpercent level can be inferred from the energy loss in binary pulsar systems \citep{Jimenez:2015bwa,Brax:2015dma}.
A stringent and prominent direct constraint on deviations of $c_T/c=1$ of $\lesssim\mathcal{O}(10^{-15})$ was obtained from the arrival times of the GW from the LIGO/Virgo event GW170817 
\citep{TheLIGOScientific:2017qsa,Monitor:2017mdv} and its EM counterparts (photons have been assumed to be massless given the stringent experimental upper bounds on their mass, though we note that some standard-model extensions imply Lorentz-violations
that naturally give rise to massive photons \citep{Collady:1997,Collady:1998,Bonetti:2017toa,Spallicci:2020diu}).
As anticipated the measurement left a strong impact across a wide range of cosmic acceleration models
\citep{Lombriser:2015sxa,McManus:2016kxu,Lombriser:2016yzn,Creminelli:2017sry,Sakstein:2017xjx,Ezquiaga:2017ekz,Baker:2017hug,Battye:2018ssx,deRham:2018red}.
Importantly, however, the constraint only applies to low redshifts of $z\lesssim0.01$ and the LIGO/Virgo frequency range \citep{Battye:2018ssx,deRham:2018red}.
In particular, it was argued in \citet{deRham:2018red} that UV completion terms for modified gravity theories naturally recover a luminal speed of gravity in the high-energy limit tested by GW170817 while allowing deviations at lower energies relevant to modifications that could drive cosmic acceleration.
Weak bounds on $c_T$ can be inferred without counterpart emissions from BH mergers \citep{Cornish:2017jml}, from the comparison of the GW arrival times between the terrestrial detectors \citep{Blas:2016qmn}, or the CMB B-mode power spectrum \citep{Raveri:2014eea} for early-time modifications.
Constraints on $c_{\rm T}$ have also been discussed as forecasts for potential arrival time comparisons with nearby supernovae emissions \citep{Nishizawa:2014zna,Saltas:2014dha,Lombriser:2015sxa,Brax:2015dma} (which are however very rare) and LISA eclipsing binary systems \citep{Bettoni:2016mij}. 

LISA will provide a twofold improvement over the current GW170817 bound.
With the detection of GWs from massive BBHs up to $z\sim8$ and their EM counterparts \citep{Tamanini:2016zlh}, $c_T$ will be tested across much larger distances, tightening the current constraint.
Furthermore, LISA tests the frequency range below the expected UV transition of modified gravity models \citep{deRham:2018red}. Besides providing a new test of GR at different energy scales, a measurement of the speed of GWs with LISA is thus of particular relevance to cosmic acceleration models.

Frequency dependent modifications in the velocity term of Eq.~(\ref{gwprop_eq:gw}) can also more generally be parametrised with the group velocity $(v_g/c) \simeq 1 + (\alpha - 1) A_{\alpha} E^{\alpha-2} + \mathcal{O}(A_{\alpha}^2)$, where $\alpha$ and $A_{\alpha}$ may parametrise quantum gravity effects, extra dimensions, or Lorentz invariance violations \citep{Mirshekari:2011yq}.
The parametrisation recovers for instance the effect of a massive graviton for $\alpha=0$ with $A_0 = m_g^2 c^4$
and $m_g=\mu h/c^2\neq0$, where the group velocity becomes $(v_g/c) \simeq 1 - (m_gc^2/E)^2$.
The combination of current LIGO/Virgo sources yields a constraint of $m_g\leq4.7\times10^{-23}~{\rm eV}/c^2$ \citep{LIGOScientific:2019fpa}.
More distant and more massive sources are generally more effective in constraining the dispersion relation and the lower energy range of LISA will be favourable for tightening the bounds on $m_g$ \citep{Will:1997bb,Berti:2004bd,Stavridis:2009mb,Yagi:2009zm,Berti:2011jz,Toubiana:2020vtf}.
Currently, the strongest bound on the mass of the graviton with $m_g\leq7\times10^{-32}~{\rm eV}/c^2$ is inferred from weak gravitational lensing \citep{Choudhury:2002pu}.

Finally, the presence of a source term $\Gamma\neq0$ further modifies the GW amplitude with an oscillatory correction of order $\Gamma\gamma_{ij}/(H k)$ \citep{Alexander:2009tp,Nishizawa:2017nef}.
The effect can be constrained with LISA standard sirens \citep{Belgacem:2019pkk}. 

\subsubsection{Stochastic Gravitational Wave Background Tests}
\label{sec:SGWB}

Astrophysical sources of the SGWB can come from a variety of phenomena \citep{Caprini:2019pxz,Schneider:2010ks}.  The first of these sources is the unresolved evolution of stellar BBH, BNS, and BWD systems \citep{Farmer:2003pa,Meacher:2015iua,Zhu:2011bd,Zhu:2012xw,TheLIGOScientific:2016wyq,Abbott:2017xzg,Zhao:2020iew}. Superradiant instabilities are another astrophysical source of the SGWB \citep{Yoshino:2013ofa,Brito:2017wnc,Brito:2017zvb,Barausse:2018vdb,Cardoso:2018tly,Huang:2018pbu}.  The presence of a population of BH-bosonic condensates can form a GW background, which depend on the formation rate, number density, and spin.  Superradiant instabilities occur when scalar field spinning around a BH transfer rotational energy and lead to the formation of a bosonic condensate. 
Other astrophysical SGWB sources that are less relevant for LISA include r-mode instabilities within NSs \citep{Andersson:1997xt,Friedman:1997uh,Ferrari:1998jf,Zhu:2011pt}, stellar core collapse \citep{Buonanno:2004tp,Crocker:2017agi,Mueller:2012sv,Ott:2012mr,Kuroda:2013rga}, and magnetars \citep{Regimbau:2005ey,Marassi:2010wj}.

Together, these sources make up a signal whose SNR is too small to be detected as individually-resolved detections, and thus combine to form a SGWB \citep{Allen:1997ad,Romano:2016dpx}.
The SGWB can be characterized by the spectral energy density as
\begin{equation}
	\label{eq:SGWB_Energy_Density}
	\Omega_{\rm GWB} = \frac{1}{\rho_c}\frac{d \rho_{\rm GW}}{d \ln f}\,,
\end{equation}
where $\rho_c=3H_0^2/(8\pi)$ is the critical density necessary for a closed universe with Hubble constant $H_0$, and $\rho_{\rm GW}$ is the energy density of all background GWs as a function of frequency.  This energy density may be approximated as a power law \citep{Thrane:2013oya} given by 
\begin{equation}
	\label{eq:SGWB_PowerLaw}
	\Omega_{\rm GWB} = \Omega_{\beta}\left( \frac{f}{f_*}\right)^\beta\,,
\end{equation}
where $\Omega_\beta$ is the GW spectral energy density at a reference frequency $f_*$ and the various factors of $\beta$ correspond to different sources for the SGWB. These can be given for binary systems ($\beta = 2/3$), r-mode instabilities ($\beta=2$), stellar core collapse ($\beta$ is model dependent), superradiant instabilities ($\beta=1-7$), and magnetars ($\beta = 3$) \citep{Kuroyanagi:2018csn}. One can use such SGWB spectrum to test GR in a theory agnostic way, in particular the amplitude corrections in the ppE framework \citep{Saffer:2020xsw}.

A SGWB can also originate from the primordial tensor perturbations generated in the early universe during inflation or via alternative mechanisms such as thermal production. Given a primordial tensor spectrum ${\cal P}_{\rm t}(k)$, one can compute the SGWB. The spectrum can be given directly by the model under study or, if allowed, in a parametrized form via the tensor spectral index $n_t$ and its running $\alpha_t$:
\begin{equation}\label{Ptk}
{\cal P}_{\rm t}(k)={\cal P}_{\rm t}(k_0)\,\exp\left[n_{\rm t}(k_0)\,\ln\frac{k}{k_0}+\frac{\alpha_{\rm t}(k_0)}{2}\left(\ln\frac{k}{k_0}\right)^2\right],\qquad {\cal P}_{\rm t}(k_0)=r(k_0)\,{\cal P}_{\rm s}(k_0)\,,
\end{equation}
where $n_t $, $\alpha_t$ and the tensor-to-scalar ratio $r$ are calculated at the pivot scale $k_0$, while ${\cal P}_{\rm s}(k_0)$ is the measured amplitude of the scalar perturbations. During inflation, the running term does not play an important role because $\alpha_{\rm t}\ll 1$ and $[\ln(k/k_0)]^2\ll 1$. However, at higher frequencies $[\ln(k/k_0)]^2$ increases and the running term can be large enough to affect the spectrum, an effect consistent with the above parametrization as long as $\alpha_{\rm t}$ is small. The relation between the primordial tensor spectrum and the SGWB observed today is
\begin{equation}\label{Omgw}
\Omega_\textsc{gw} =\frac{\pi^2f^2}{3a_0^2H_0^2}{\cal P}_{\rm t}(f)\,{\cal T}^2(f)\,,
\end{equation}
where the transfer function ${\cal T}$, which we do not write here, depends on the history of the universe and on the details of reheating \citep{Turner:1993vb,Watanabe:2006qe,Kuroyanagi:2014nba}. When the primordial tensor spectrum is blue-tilted, the amplitude of the SGWB at high frequencies can increase up to the sensitivity threshold of LISA and other interferometers. In general relativity with standard inflation, the tensor spectrum is red-tilted and therefore its associated SGWB is unobservable. However, models beyond the standard paradigm in the matter or gravity sector could have a blue tilt and a large-enough amplitude \citep{Bartolo:2016ami,Kuroyanagi:2018csn,Calcagni:2020}.

Methods for analyzing the LISA detection of an underlying SGWB signal have been developed in a source agnostic way \citep{Flauger:2020qyi,Caprini:2019pxz,Karnesis:2019mph,Pieroni:2020rob}. For example, a recent work \citep{Pieroni:2020rob} shows that the SGWB from stellar origin BH and NS mergers may be observable with LISA with an SNR of $\sim 50$. This acts as a foreground noise to other backgrounds. One can employ a principal component analysis to subtract this astrophysical foreground, which enables LISA to detect SGWB that is 10 times weaker than the foreground. Given these methods, and a better understanding of source modelling which may occur by launch, the prospects for detecting and analyzing a SGWB with LISA are promising.

A strategy to separate the astrophysical from the cosmological background in the context of LISA has been proposed in \citet{Boileau:2020rpg} using a Bayesian strategy based on an Adaptive Markov chain Monte-Carlo algorithm. This method has been later used in the case of a SGWB produced by a network of cosmic string loops \citep{Boileau:2021gbr}, updating a previous study where only the instrument noise was taken into account \citep{Auclair:2019wcv}. In particular, it has been shown that given the ability of LISA to simultaneously detect a large number of galactic double white dwarf binaries and a large number of compact binaries, a cosmic string tension (Newton's constant is denoted by $G$ and the string linear mass density by $\mu$) in the $G\mu\approx 10^{-16}$ (for loop distribution model; \citealt{Kibble:1984hp}) to $G\mu\approx 10^{-15}$
(for loop distribution model; \citealt{Ringeval:2005kr,Lorenz:2010sm,Blanco-Pillado:2013qja}) range or bigger could be measured by LISA, with the galactic foreground affecting this limit more than the astrophysical one.

\subsection{Burning Questions and Needed Developments}
\label{sec:model_indep_develop}

Throughout this section, we have described in passing a few of the many challenges we still face when developing and carrying out model-independent tests. We therefore end this section by presenting a brief summary of the main open problems.

\begin{itemize}

\item Residual tests, as well as other tests, need to be more carefully studied in the context of LISA, due to the expected abundance of sources at multiple frequencies. This is because the LISA data analysis challenge is a global one, in which multiple signals must be estimated simultaneously.

\item Regarding the IMR consistency tests (and for other tests), the challenge is to ascertain that the waveform models used in the measurement process are accurate enough so as not to cause systematics, especially at the large SNRs expected for BBH mergers in LISA.  This can affect both the extraction of parameters in the inspiral, as well as the inference of remnant masses and spins in the merger. It is therefore important to control modelling systematics to the required level. Work must begin now to address these issues to either demonstrate the readiness of the waveform models or to establish the required accuracy.

\item We currently lack complete IMR waveform model in non-GR theories, even with simple extension to GR, which prevents us from performing a full mapping between parametrized tests and specific theories. This is partially because we currently lack sufficient NR simulations performed in theories beyond GR. Furthermore, many of non-GR corrections to the inspiral portion have focused on the leading PN contribution, and therefore they neglect among other features, higher harmonics. Higher PN order terms may become important as the binary separation becomes smaller, especially at LISA SNRs.

\item Parameterized ringdown tests need further developments, such as including the BH spin without the slow-rotation approximation and finding more mappings to specific non-GR theories. This also requires that we construct  new rotating BH solutions in theories beyond GR at arbitrary rotation.

\item One needs to reveal how the complex interplay between astrophysical event rates, non-GR waveform models, ground-based detector developments and data analysis methods will affect our ability to extract fundamental physics information with LISA.

\item An interesting open problem regarding GW propagation is the mixture of propagation and source effects for extended theories of gravity and the role of screening mechanisms.
A PN expansion for the source emission in screened regimes can be performed using a scaling relation \citep{McManus:2017itv,Perkins:2018tir,Renevey:2020tvr}.
For some GR extensions the relevant modification in $\nu$ are in fact determined by the screened environments of emitter and observer rather than the cosmological background, whereas for $c_T$ screening effects may safely be neglected.

\item Tests that aim to constrain the existence of non-GR polarizations need further development, both from the standpoint of generic (null channel type) tests, as well as from the standpoint of specific examples in given modified theories. 

\item The tests described above, for the most part, require a specific or generic model for the signal, but LISA could have access to GWs that are not properly covered by such models. Efforts to develop methods to extract unexpected signals should also be pursued. 
\item More detailed analyses are needed to see how well one can remove the astrophysical background from stellar origin BH and neutron star mergers and how much its residuals affect tests of GR.

\end{itemize}

\section{Astrophysical and Waveform Systematics}
\label{Sec:Astrophysics_Environmental_effects}


\vspace{0.25cm}

The discovery of GWs has provided a powerful new tool to probe gravity in its most extreme regime and to search for signatures of new physics. To do so, however, requires the development of highly accurate waveform templates -- our ``filters'' to identify and interpret gravitational signals.
Before being able to claim a discovery of new physics we need to 
\begin{enumerate*}[label={(\roman*)}]
\item provide more accurate waveforms within GR to reduce systematic errors (and potential misinterpretation) due to modelling systematics;
\item understand astrophysical and environmental effects to avoid misinterpreting deviations from vacuum GR;
and
\item construct theory-specific waveforms for targeted tests of (classes of) beyond-GR theories.
\end{enumerate*}
A detailed account of the status and challenges of gravitational waveform modelling and
their systematics can be found in the LISA Waveform WG white paper \citep{XXX-cite-white-paper-WAVWG}.
Here we give a summary with focus on its relevance for the fundamental physics WG.

\subsection{Astrophysics: Environmental effects}

LISA sources, whether supermassive or stellar-mass, are likely to be surrounded by matter that could limit our ability to place constraints on fundamental physics. In the case of supermassive BBHs the impact of gaseous accretion raises questions regarding our ability to probe fundamental physics, especially in the case of EMRIs. Stellar mass BHs can also be sources for LISA and in this case if they are inside dense stellar clusters, then they could be in the form of hierarchical triplets. 

\subsubsection{Binary and disk interactions}

When a BBH is embedded in a magnetized gaseous disk, binary-disk interactions can induce change in the evolution of the GW phase. Typically the evolution of a binary disk-system is broadly composed of two phases: i) the pre-decoupling phase where the disk is coupled to the binary, and tracks the inspiral of the binary, because the GW timescale is very long compared to the effective disk viscous timescale which controls the inflow velocity of the disk; ii) the post-decoupling phase where the GW timescale becomes shorter than the  disk effective viscous timescale and the binary begins to run away from the disk which can no longer track the inspiral of the binary. 

\paragraph{Comparable mass binaries:} Equating the GW timescale with the disk viscous timescale \citep{2005ApJ...622L..93M}, one finds that the decoupling radius is given by \citep{2014PhRvD..90j4030G}
\begin{equation}
 \frac{a_d}{GM/c^2} \sim 140  \left(\frac{\alpha}{0.01}\right)^{-2/5}\left(\frac{H/R}{0.05}\right)^{-4/5}\left(\frac{\tilde \eta}{1}\right),
\end{equation}
where $M$ is the binary's total mass, $H/R$ is the disk aspect ratio, with $H$ the disk scale height (or thickness), $\alpha$ is the Shakura--Sunyaev viscous parameter \citep{1973A&A....24..337S}, and $\tilde \eta=4q/(q+1)^2$ is four times the symmetric mass ratio, with $q$ the binary mass ratio [see also \citealt{2002ApJ...567L...9A}). Notice that for extreme mass-ratio and intermediate mass-ratio inspirals the decoupling radius can be comparable to the gravitational radius of the binary, and thus affect its evolution if the disk is massive enough \citep{Kocsis:2011dr,Yunes:2011ws}. A good understanding of binary-disk interections is important to better understand the binary evolution into the LISA band. For a binary at the decoupling radius, the GW frequency is given by:
\begin{equation}
 f_{\rm GW} \sim 10^{-3} {\rm Hz} \left(\frac{a_d}{140GM/c^2}\right)^{-3/2}\left(\frac{M}{10^6M_\odot}\right)^{-1},
\end{equation}
which suggests that for typical values of the viscosity parameter, and if the disks are slim \citep{2013LRR....16....1A},  decoupling could occur inside the LISA band, and the effects of binary-disk could be imprinted on the GWs. For example, the binary residual eccentricity when it enters the LISA band is determined at binary-disk decoupling \citep{2011MNRAS.415.3033R}. For disk thickness, e.g., $H/R \sim 0.2$, for a comparable mass the decoupling radius becomes $a_d\sim 40 GM/c^2$ at $f_{\rm GW}\sim 0.01 \rm Hz$, i.e., well within the LISA band. The work of \citep{2011MNRAS.415.3033R} suggests that residual eccentricities of up to $10^{-2}$ in the LISA band are possible. How these effects could affect precision tests of GR has not been studied, yet.
Note however that at least for SOBHBs, environmental effects can be significant in gas-rich systems (such as AGN disks) \citep{Toubiana:2020drf,Caputo:2020irr}. Because these effects typically enter at low frequencies, they are expected to compete with (or even dominate over)  non-GR effects such as vacuum dipole emission).

\paragraph{EMRIs:}
EMRIs are thought to form in dense galactic nuclei where
a passing by stellar compact object (SCO), or a nascent SCO formed in the accretion disk
via fragmentation and/or coagulation of density enhancements
\citep{Milosavljevic:2004te,Goodman:2003sf,Levin:2006uc}, can be captured by the central
SMBH. It is expected that only EMRIs involving compact enough
objects, such as WDs, NSs, stellar-mass BHs or probably the Helium cores of giant stars, can survive to low redshifts ($z\lesssim 1$)
where EMRIs may be observed \citep{AmaroSeoane:2007aw,Barack:2004wc}.

A typical EMRI event will be emitting GWs with frequencies ranging between~$10^{-4}$
to $1\,\rm Hz$, which are expected to be in the LISA frequency band for many years,
allowing for accurate measurements of the parameters of the system. In particular, a GW event
detection with a SNR of~$\sim30$ will allow us to determine the mass of
the SMBH with a precision of~$\sim 10^{-4}$ and the distance to the source with a precision
of $\sim 3\%$ \citep{Gair:2010yu} (see also \citealt{Babak:2017tow}). Moreover, these events can be used to probe cosmological
parameters, test GR in the strong field regime (e.g., EMRI may provide a constraint
on deviations from the Kerr metric at a level of 0.01--1\%; \citealt{Gair:2017ynp,Babak:2017tow}), etc.
It is, therefore, necessary to assess whether accretion or other SCOs in the vicinity of
the EMRI can induce observable changes in the GW frequency. This is especially important for sources in active galaxies with BHs accreting near the Eddington limit, for which the disk mass can be comparable to the secondary object. 

Tidal interaction  of the SCO with the accretion disk may induce mutual angular momentum
exchange that modifies both the disk and the orbit of the SCO \citep{1997Icar..126..261W}.
The nature of this interaction depends basically on the ratio of their masses $q_{\rm
  CO,disk}\equiv M_{\rm SCO}/M_{\rm disk}$. If  $q_{\rm CO,disk}> 1$,  tidal torques
exerted by the SCO are strong enough  to carve a cavity in the disk. Radial inflow of the
gas disk then pushes the SCO inward causing the binary to harden on a time-scale comparable
to the viscous timescale  $t_{\rm visc}$ (Type-II migration) \citep{1997Icar..126..261W,
  Gould:1999ia}. Furthermore, if the mass of the SCO exceeds the mass of the nearby gas
disc, the migration slows down as the matter cannot remove angular momentum away from
the binary at a rate on which the gas flows \citep{Syer:1995hk,Ivanov:1998qk}. Analytic
work has predicted that the ``pile-up'' of gas disc around the orbit of the SCO mitigates
the slow down, and the hardening timescale is $t_{\rm hard}\approx q_{\rm 2, disk}^{\alpha}
t_{\rm visc}$ with a relatively shallow scaling of $\alpha\approx 0.5$ in a steady-state
limit \citep{Syer:1995hk,Ivanov:1998qk}. However, numerical simulations have found no evidence
for this pile-up, and instead suggest a simple scaling with $\alpha \ll 1$
\citep{Duffell:2014jma,2017A&A...598A..80D,2018ApJ...861..140K}. On the other hand, if the
$q_{\rm CO,disk}\lesssim 1$, the SCO is not massive enough to carve the cavity, it will
migrate inwards by exciting density waves in the disc \citep{1980ApJ...241..425G,Tanaka_2002}
(Type-I migration). However,  this mechanism is very sensitive to the temperature and opacity
of the disk \citep{JangCondell:2004jr,Menou:2003pu}. If the radiative cooling in the dense inner
regions surrounding the SCO is inefficient, torques exerted by the disc will induce outward
migration, instead of the usual inward migration found in locally isothermal disks, up to
regions when the disc opacity is low. As the accretion proceeds, these regions move inward,
and hence the SCO's migration will continue on a timescale much longer than the viscous
timescale \citep{Paardekooper:2006hr}. A slower migration mechanism has been recently
identified \citep{Kocsis:2012ui}. In this case, the gas piles-up significantly outside the orbit
of the SCO, but the viscosity increases to the point that in steady-state gas can reach the
SCO's Hill sphere, and is able to flow across its orbit (Type-1.5).

The effect of such interactions on the waveform might be degenerate with the effects due to modifications to GR \citep{Barausse:2014tra}. In such a scenario, the precision with which modifications to GR can be measured with LISA will be significantly affected, since, in order to be detectable, the effect induced by GR modifications should be larger than the \textit{environmental} effect. Effectively, this induces a lower bound on GR modifications observable with LISA. Development of waveforms beyond GR is in early stages, and the role of astrophysical systematics has not been studied in any detail. 

The inspiral part of the waveform will be largely unaffected by the astrophysical interactions as long as the BH resides in a thick disk environment, where densities are low. On the contrary, in thin-disk environments the waveform can be significantly perturbed by certain phenomena. Planetary migration can have a strong influence on the inspiral \citep{Yunes:2011ws,Kocsis:2011dr}, while dynamical friction \citep{Barausse:2007dy,Derdzinski:2018qzv,Derdzinski:2020wlw} and accretion \citep{Barausse:2007dy,Barausse:2014tra,Cardoso:2019rou} have a weaker but non-negligible effect. The effect of the disk's gravity can safely be neglected. All this can have a large enough effect to cause systematic errors when performing precision tests of gravity \citep{Barausse:2014tra}.

\paragraph{Ringdown:} A remarkable feature of the ringdown of GR BHs is the isospectrality of the polar and axial modes \citep{Yunes:2011ws,Kocsis:2011dr,Pani:2013ija,Wagle:2021tam}. Any deviation between the axial and polar modes could indicate new physics, but matter distribution around the BH also breaks the isospectrality \citep{Barausse:2014tra}, making it non-trivial to identify GR modifications this way. While a smoking-gun type test is not possible due to this degeneracy, the impact of typical matter distributions around a BH on the ringdown frequencies is very weak. Thus, the lower bounds introduced by the degeneracy is low enough for ringdown-based strong-fields tests of GR to be meaningful \citep{Barausse:2014tra}.

\paragraph{Plasma effects:} Even in the absence of a massive accretion disk around the binary, recent work suggests that for binaries embedded in a plasma environment, it may be challenging to perform fundamental physics tests that probe plasma-driven superradiant instabilities, EM emission from charged BH binaries and electromagnetically driven secondary modes in the GW ringdown signal \citep{Cardoso:2020nst}. The latter two effects can be relevant also in the case of massive fields propagating in vacuum, and provide an obstacle to tests of modified gravity with massive degrees of freedom. In addition, matter effects can change the tidal-deformability of a vacuum BBH \citep{Cardoso:2019upw}, which could potentially limit our ability to test for dark compact objects that are not BHs. 

\subsubsection{Interaction with stellar environments and hierarchical triplets}

In recent years interaction of BBHs with a third BH in stellar environments or near a SMBH have revealed new ways to think about LISA sources that exhibit effects which are absent in a standard isolated binary evolution. SMBH triplets can potentially induce binary eccentricities as high as 0.9 \citep{Bonetti:2018tpf} upon entrance in the LISA band, and could be as high as 0.1 at merger. Apart from eccentricity tidal effects induced on the binary by the companion BHs could contribute significantly to the binary waveform \citep{Yang:2017aht} by e.g. shifting the innermost stable orbit, and hence the onset of the plunge phase. BH binaries formed through such channels not only would require suitable templates for parameter estimation, but also necessitates an investigation on how eccentricity may affect tests of fundamental physics. 

In the case of stellar-mass BBHs, a non-zero graviton mass and dipole emission can be probed \citep{Toubiana:2020vtf} among other pieces of fundamental physics. However, if the binary is in a dense stellar environment or near a SMBH 3 body effects can become important \citep{Miller:2002pg,Blaes:2002cs,Wen:2002km,2013MNRAS.431.2155N,Yunes:2010sm,Randall:2018qna,Randall:2019sab}, e.g., through Kozai-Lidov mechanism \citep{1962AJ.....67..591K,LIDOV1962719} oscillations. The extent to which such 3-body effects can provide an obstacle to probing fundamental physics with stellar mass BH binaries has not been investigated, yet. 

\subsection{Waveform modelling and systematics for EMRIs}

The most promising avenue for GW modelling of EMRIs is perturbation theory (gravitational self-force), see e.g., \citet{Barack:2018yvs} for a recent review. 
The full calculation of the EMRI with perturbation theory is a formidable problem as it requires first-order metric perturbation for the gravitational  self-force, second-order metric perturbation for the dissipative part, the evolution of the inspiral and the final calculation of the waveform. The small parameter for the perturbation is the mass ratio $q$. According to \citet{Hughes:2016xwf} reliable templates will require that one keeps track of at least $O(q^2)$ terms. The state-of-the-art in the calculation of the self-force is the first-order calculation of \citet{vandeMeent:2017bcc} for generic orbits with spinning BHs, and the second-order calculation for circular orbits and non-spinning BHs \citep{Pound:2019lzj}, and including the spin of the secondary \citep{Akcay:2019bvk}. The state-of-the-art in fast calculation of EMRI GWs are the augmented kludge waveforms of \citet{Chua:2017ujo}, the two-timescale expansion approach \citep{Hinderer:2008dm,Miller:2020bft},  near-identity transformations \citep{vandeMeent:2018rms}, and the self-consistent evolution of \citet{Diener:2011cc}.

Apart from not yet having the full self-force calculation to second-order for generic primary BHs that include the spin of the secondary on generic orbits, potential systematics in the calculation involve the internal structure of the secondary \citep{Isoyama:2012in,Witzany:2019dii} and non-vacuum backgrounds \citep{Zimmerman:2014uja}, e.g., the case where the EMRI is embedded in a dark matter bosonic condensate. 

It is worth noting that several of the issues faced by EMRI modelling are also challenges in the case of intermediate mass-ratio inspirals.

\subsection{Waveform modelling in GR and systematics for comparable mass binaries}

Modelling the gravitational waveform of massive binaries covering their inspiral, merger and ringdown requires a combination of perturbative techniques and NR.
In particular, the inspiral is modelled by PN
theory, the late inspiral and merger where nonlinear effects of gravity become relevant require full NR,
and the ringing down of the final BH is described within black-hole perturbation theory.
These different pieces are combined into full IMR waveforms using effective-one-body (EoB) or phenomenological models.

\subsubsection{PN waveforms}

PN methods are analytic techniques based on a series expansion in weak fields and small velocities that describe the dynamics of a binary during the early inspiral phase. The accuracy of the description is controlled by the PN order, which identifies the relative order to which an expansion has been taken in small velocities and weak fields. While waveforms of non-spinning, quasi-circular binaries are currently known up to 4PN order \citep{Blanchet:2013haa}, those of eccentric, spinning or precessing binaries are currently only available up to 3PN and 3.5PN order, e.g., \citet{Moore:2019xkm,Tanay:2020gfb,Moore:2020rva}. Efforts to go beyond the current state of the art are under way, see, e.g., \citet{Foffa:2019yfl,Bini:2019nra,Bini:2020wpo}. Although for ground-based GW observatories going to high PN order will be beneficial, a recent analysis of requirements for LISA non-eccentric stellar mass BBHs finds that 3PN order is sufficient for all sources, while for ∼90\% of those sources, waveforms at PN order $\leq 2$ are sufficiently accurate for an unbiased recovery of the source parameters \citep{Mangiagli:2018kpu}. However, at this time it is unclear what PN accuracy is required for SMBH binaries. This will be crucial to determine as the high SNR ratio at which LISA is going to detect these binaries implies that the statistical errors are going to be smaller than the systematic errors in the theoretical waveforms. These systematic errors could hinder our ability to perform precision tests of fundamental physics with LISA.

\subsubsection{NR waveforms}
NR techniques are numerical methods for solving the equations of relativistic gravitation that are particularly suited in regimes where perturbation theory and PN methods fail (see e.g., \citealt{AlcubierreBook2008,BonaPalenzuelaBona2009,BaumgarteBook2010,GourgoulhonBook2012,ShibataBook2015} and references therein). In the case of BBH, NR is the only reliable method for capturing the late inspiral and merger of the binary from first principles. NR methods are used both in the case of GR and in modified gravity, on which we expand further below. 

Despite the tremendous strides made by NR since the breakthrough simulations of \citet{Pretorius:2005gq,Campanelli:2005dd,Baker:2005vv}, NR still faces several challenges: current methods become prohibitively expensive for high mass ratios ($q\gtrsim 10$) and near extremal BH spins. As a result, calibration of phenomenological waveforms such as Phenom or EOB (see next section) based on NR methods have systematic errors in the more extreme parts of the parameter space. Such systematic errors can pose challenges when performing precision tests of fundamental physics with LISA. Furthmore, it is known that in some cases numerically generated waveforms exhibit non-convergence at very high grid resolution \citep{Zlochower:2012fk}, which makes it challenging to assess the accuracy of the waveforms. Resolving these non-convergence problems may require new gauge conditions (see e.g., \citealt{Etienne:2014tia}) or next generation codes. Finally, comparisons in the context of the NRAR collaboration \citep{Hinder:2013oqa} showed that even when adopting the same methods at the analytic level, different NR codes (i.e. different implementations) do not entirely agree at the finite resolutions typically adopted to generate high-quality NR waveforms. Efforts like the NRAR collaboration and code comparisons are important to assess the accuracy requirements for LISA. Initial work toward assessing these requirements \citep{Ferguson:2020xnm} (see also \citealt{Purrer:2019jcp}) suggests that NR will require a substantial increase in accuracy compared to today's most accurate waveforms; for  example it is found that reaching with NR templates SNRs   above 1000,  finite-difference  NR  codes  would  have  to  efficiently scale to resolutions of at least $\Delta x< M/700$ ($M$ being the binary gravitational mass). Reaching resolutions for template indistinguishability  is  particularly  important  because residuals resulting  from  using  lower  resolution templates  can  be  comparable  to those  resulting  from ignoring higher modes entirely. 

\subsubsection{Phenom and EoB waveforms}

The phenomenological (Phenom) \citep{Ajith:2007qp,Ajith:2007kx,Ajith:2009bn,Santamaria:2010yb,Khan:2015jqa,Husa:2015iqa,Hannam:2013oca,Schmidt:2012rh, Schmidt:2014iyl,Khan:2019kot,Khan:2018fmp,Pratten:2020ceb,Garcia-Quiros:2020qpx} and EOB models  \citep{Buonanno:1998gg,Buonanno:2000ef,Barausse:2009aa, Barausse:2009xi, Barausse:2011ys,Taracchini:2013rva,Bohe:2016gbl,Damour:2001tu,Damour:2008qf,Nagar:2011fx,Balmelli:2013zna,Damour:2014sva,Damour:2007xr,Damour:2008gu,Damour:2002vi,Taracchini:2012ig,Damour:2012ky,Damour:2014yha,Pan:2010hz,Babak:2016tgq,Ossokine:2020kjp,Cotesta:2018fcv,Nagar:2020pcj,Nagar:2019wds,Chiaramello:2020ehz,Akcay:2018yyh,Nagar:2018zoe,Hinderer:2016eia,Steinhoff:2016rfi,Nagar:2018gnk,Nagar:2018plt,Babak:2016tgq} are the current state-of-the-art waveforms used in GW data analysis. Using two different waveform models is essential to assess the size of systematic uncertainties in the measurements. The Phenom and EOB modelling approaches are both based on combining information from analytical and NR. For binaries of BHs (and NS-BHs where tidal disruption does not occur \citep{Thompson:2020nei,Matas:2020wab}), they describe the entire signal from the early inspiral to the ringdown of the final remnant. The Phenom models provide a closed-form description of the frequency-domain signals, although see \citet{Estelles:2020osj} for a time-domain approach. They are focused primarily on efficiency and based on a modular assembly of combined theoretical and numerical insights into the main important features of the waveforms. The EOB models describe both the binary dynamics and waveforms in the time-domain. They are based on incorporating theoretical inputs from a variety of approximation methods into the structure of the model before introducing calibrations to NR. The efficiency of the EOB waveforms is then improved, e.g., by developing reduced-order frequency-domain models \citep{Field:2013cfa,Canizares:2014fya,Purrer:2014fza,Purrer:2015tud,Bohe:2016gbl,Lackey:2016krb} or using post-adiabatic approximations \citep{Gamba:2020ljo,Nagar:2018gnk}; see also \citet{Garcia-Quiros:2020qlt} for recent applications to Phenom models. 

There is ongoing rapid progress on improving the accuracy and physical realism of the waveform models within GR, as systematic differences are already starting to become noticeable for LIGO/Virgo/KAGRA detections such as \citet{LIGOScientific:2020stg}. For instance, the baseline Phenom model has recently been significantly upgraded for the dominant mode of aligned spin configurations \citep{Pratten:2020fqn}. Current models include several physical effects for circular orbits: a set of higher harmonic modes for aligned spins \citep{Garcia-Quiros:2020qpx,London:2017bcn,Nagar:2020pcj,Cotesta:2018fcv}, spin precession \citep{Ramos-Buades:2020noq,Khan:2018fmp,Schmidt:2014iyl,Hannam:2013oca}, and spin precession with higher harmonics \citep{Pratten:2020ceb,Khan:2019kot,Ossokine:2020kjp}. Models for nonprecessing binaries with mildly eccentric orbits have also recently been developed \citep{Chiaramello:2020ehz,Liu:2019jpg,Cao:2017ndf} and explored \citep{Ramos-Buades:2019uvh,Hinderer:2017jcs,Hinder:2017sxy}. 

The Phenom and EOB models have been widely tested against NR simulations mainly in the regime of comparable masses, i.e. mass ratios below four, and moderately high spins below eighty percent of the maximum, aside from a few more extreme cases. Ongoing efforts to improve the structure of the models aim to include more information from a mix of perturbative approaches involving PN calculations combined with results from scattering calculations within the post-Minkowski approximation and gravitational self-force results \citep{Antonelli:2019fmq,Antonelli:2019ytb,Khalil:2020mmr,Bini:2019nra,Bini:2020nsb}. Significant further work is required to turn these theoretical explorations into full, calibrated models, and deploy them for data analysis. While all of these developments mark important advances, waveforms for LISA will require models to account for yet more physics content and have higher accuracy than current waveform models. MBH binary mergers and intermediate mass ratio systems will generically involve eccentricity, arbitrary spins, substantially different masses, and SNR of up to several thousands. The resulting waveforms will have a very rich structure characterized by Fourier harmonics with several different frequencies. Further advances in both analytical and NR will be required as inputs for developing adequate Phenom and EOB models for these sources. 

For objects beyond BHs, Phenom and EOB models for the inspiral signals are also available. In this regime, a number of generic spin and tidal effects lead to characteristic GW signatures that depend on the objects' internal structure. Current models include the matter effects from spin-induced quadrupole moments and tidal deformability  \citep{Damour:2009wj,Bini:2012gu,Bernuzzi:2012ci,Bini:2014zxa,Bernuzzi:2014owa,Nagar:2018zoe,Nagar:2018gnk,Akcay:2018yyh,Nagar:2018plt,Flanagan:2007ix,Vines:2011ud,Damour:2012yf,Dietrich:2017aum,Dietrich:2018uni,Kawaguchi:2018gvj,Dietrich:2019kaq}, and the tidal excitation of fundamental quasi-normal modes \citep{Steinhoff:2016rfi,Hinderer:2016eia,Schmidt:2019wrl}. For the time-domain EOB models, the matter effects are based solely on theoretical results without any additional calibrations to NR. The frequency-domain models can either employ PN results for matter effects on top of the BH baseline, or, as has become standard for current data analysis, a calibrated tidal model based on NR simulations for binary NSs. In all of these models, matter signatures are described in a parametric form, where the characteristic coefficients such as tidal and rotational Love numbers, and quasi-normal mode frequencies encode the fundamental information on the matter. As these matter waveforms describe only the inspiral, they are tapered to zero in practical applications at the predicted merger frequency of double NS systems as determined from fits to NR simulations \citep{Bernuzzi:2015rla}, or when the fundamental mode resonance is reached. Future work remains on going beyond these restrictions and on enlarging the physics content of the models with matter phenomena. Another effect that has not yet been included are the objects' absorption coefficients. All of this remaining work presents no fundamental obstacles but will require time and effort to develop the full models.  Environmental effects such as from dynamical friction or energy-level transitions in clouds of ultralight fields are not yet included in the Phenom and EOB models. This is in part due to the fact that at present, studies of these phenomena are mainly exploratory. Developing a deeper understanding and more accurate descriptions of the backreaction and evolution of the system during an inspiral will be useful before including these phenomena in the full waveform models.   

As discussed in other parts of this white paper, there are many ways to test GR, but let us here focus on three broad classes: parameterized deviations of the waveforms from GR during all epochs, IMR consistency tests, and spectroscopy of the final remnant. For the Phenom models, where the signals are described by closed-form algebraic expressions, it is straightforward to add extra beyond-GR parameters \citep{Abbott:2018lct}. For the EOB model, a similar approach is used with a slightly different form of parameterized deformations. These are added to reduced-order models of frequency-domain EOB waveforms \citep{Sennett:2019bpc}. Current data analysis for these theory-agnostic tests constrain only one deviation parameter at a time, as it is unfeasible to obtain useful information on a very large number of extra parameters in the waveforms. Tests of the consistency of the inspiral and the ringdown with the EOB and Phenom  models \citep{Ghosh:2016qgn,LIGOScientific:2019fpa} rely on measuring the remnant properties \citep{Hughes:2004vw} and comparing with the final mass and spin predicted from the progenitor parameters by fits to NR data \citep{Haegel:2019uop,Jimenez-Forteza:2016oae,Bohe:2016gbl}. For spectroscopy using the ringdown of the remnant from a BBH merger, parameterized deformations of the quasi-normal modes, full waveform models can improve the constraints \citep{Brito:2018rfr}. Recent work has also started to develop foundations for theory-specific models as examples of non-GR waveforms. For instance, the EOB dynamics for nonspinning binaries have been calculated in scalar-tensor theories of gravity \citep{Julie:2017pkb,Julie:2017ucp}, Einstein-Maxwell-dilaton theories \citep{Khalil:2018aaj} and Einstein-scalar-Gauss--Bonnet gravity \citep{Julie:2019sab}. Significant further work and inputs from NR will be required to develop complete waveforms, and to incorporate the expected physical effects such as spins, mass ratio, and eccentricity. 

\subsection{Waveform modelling beyond GR and systematics for comparable mass binaries}

\subsubsection{NR waveforms}

Let us now discuss the current systematic errors of NR BBH merger waveforms beyond GR, and how to address these errors in future simulations. In particular, we will focus on recent work in beyond-GR effective field theories of gravity, including dynamical Chern--Simons (dCS) gravity, Einstein dilaton Gauss--Bonnet (EdGB), Einstein-scalar Gauss--Bonnet gravity (EsGB), Einstein dilaton Maxwell (EdM), and a flavor of a tensor-vector-scalar gravity as these are the only beyond-GR theories (aside from massive scalar-tensor gravity, which is identical to GR under ordinary initial and boundary conditions \citep{Berti:2015itd}) for which we have non-linear BBH evolutions \citep{Hirschmann:2017psw,Witek:2018dmd,Okounkova:2019dfo, Okounkova:2019zjf,Okounkova:2020rqw,East:2020hgw,Bozzola:2020mjx}.

For NR BBH simulations in GR, the greatest source of systematic error is finite numerical resolution (cf.\ \citealt{baumgarteShapiroBook}). To quantify numerical resolution errors, simulations are performed for a series of increasing numerical resolutions, checking for \textit{convergence} of the resulting gravitational waveforms. The error for a given waveform is reported as the mismatch between the waveform and the waveform from the next-lowest resolution simulation (see, e.g., the methods in \citealt{Boyle:2019kee} for technical details). For gravity theories which admit a well-posed initial value problem no approximation to the theory is necessary, and hence numerical accuracy is still the biggest source of systematic errors. For example, this is the case in the recent evolutions in EdM \citep{Hirschmann:2017psw}, which admits a well-posed initial value problem, and in EsGB \citep{East:2020hgw}, where a well-posed modified Harmonic formulation of the theory is adopted.

BBH simulations beyond-GR, such as in dCS gravity \citep{Okounkova:2019zjf} and Einstein dilaton Gauss--Bonnet gravity \citep{Okounkova:2020rqw} have numerical resolution errors comparable to those of GR (see the convergence analyses in \citealt{Witek:2018dmd, Okounkova:2018pql, Okounkova:2019zjf}), and thus the codes to generate these waveforms must improve in the same way as GR codes in order to be ready for LISA. 
However, in addition to numerical resolution errors (comparable to those of GR), present beyond-GR NR simulations have additional sources of systematic errors. In order to ensure a well-posed initial value problem (which may not exist in EdGB \citep{Papallo:2017qvl, Papallo:2017ddx, Ripley:2019hxt, Ripley:2019irj} and dCS \citep{Delsate:2014hba}), BBH mergers in quadratic gravity are performed in an \textit{order-reduction} scheme, where the spacetime metric is expanded about a GR BBH merger spacetime, and one computes the leading-order beyond-GR correction to the metric and gravitational waveform perturbatively. One then obtains the `full' gravitational waveform by combining the background GR metric with the leading-order correction, for a suitable choice of EdGB or dCS coupling constant. The order-reduction scheme operates under the assumption that corrections to GR are small, and that higher-order corrections (such as the second-order correction to the spacetime metric) are negligible. This assumption governs an \textit{instantaneous regime of validity}, where there is a maximum allowed value of the coupling constant that is allowed in order for the corrections to the spacetime metric to be smaller than the background metric (see \citealt{Okounkova:2019dfo} for technical details). Thus, the NR simulations performed in this scheme are not valid for all possible physical values of coupling constant. 

Additionally, the order-reduction scheme introduces systematic errors through \textit{secular dephasing}. Since the location of the BHs in the perturbative scheme is governed by GR (with no back-reaction from the beyond-GR fields), the phase of the BBH inspiral in the order-reduction scheme is different from that of the full theory, hence leading to \textit{secular dephasing}. This is a common feature of perturbative approximations to dynamical systems \citep{MR538168}, and in particular, PN approximations \citep{Will:2016pgm} and self-force BH perturbation theory \citep{Pound:2005fs} encounter these
challenges when applying perturbative approaches to long-duration inspirals. Hence, the present beyond-GR waveforms focus primarily on the merger (the phase that's crucial to numerically simulate for tests of GR \citep{Yunes:2016jcc}) and ringdown phases, ensuring that the beyond-GR simulation starts late enough so that the secular systematic errors are negligible \citep{Okounkova:2019zjf, Okounkova:2020rqw}. Finally, the extraction of GWs themselves is another potential source of systematic errors. For example, it is unclear whether the Newman--Penrose scalars that are commonly adopted in GR, have the same physical interpretation in modified gravity. 

Another modified theory where some work has been done is Moffat's tensor-vector-scalar theory \citep{Moffat:2014aja}. The theory admits a well-posed initial value problem, but for sufficiently small scales, it becomes mathematically equivalent to Einstein--Maxwell theory in vacuum, where BHs carry a ``gravitational" (not electric) charge, that is determined by the theory's coupling constant. First simulations in this context have been performed by \citet{Bozzola:2020mjx}. However, a systematic error in this case is whether the approximation of the theory as Einstein--Maxwell in vacuum holds true. Given that the theory admits a well-posed initial value problem, it is straightforward to move beyond this approximation.

Moving forward one would like to obtain NR beyond-GR mergers that are free of these systematic effects by being valid for all values of coupling constants and avoiding secular dephasing. In order to mitigate dephasing growth, one can `stitch' together results for simulations with different starting points of beyond-GR effects (hence different points of zero dephasing), as discussed for the case of EMRIs in \citet{Pound:2007th, Warburton:2011fk}. Similarly, one can apply the multiscale and dynamical renormalization group methods of \citet{Kunihiro:1995zt, Galley:2016zee, Hinderer:2008dm, Pound:2015wva} to get rid of secular dephasing effects. This is an active area of work. 
  
In order to produce beyond-GR BBH waveforms valid for all values of coupling parameter, one must move beyond perturbative methods and consider simulating BBHs in beyond-GR theories with well-posed initial value problems. Such approaches have been discussed using methods from fluid dynamics in \citet{Allwright:2018rut, Cayuso:2017iqc}. In particular, \citet{Witek:2020uzz} recently derived a 3+1 split for the (non-perturbative) equations of motion of EdGB in a step towards NR simulation. 

\subsubsection{PN waveforms}

Incomplete inspiral waveforms in theories beyond GR may give rise to systematic errors in tests of GR with GWs. The inspiral waveform is computed via the PN method. In most non-GR theories, only the leading PN correction has been computed, which can be mapped to the PPE framework (see Sect.~\ref{sec:parameterized_insp}). However, there are certain theories in which corrections to the waveform have been computed also to higher PN orders \citep{Yunes:2011aa,Sennett:2016klh,Zhang:2019iim,Shiralilou:2020gah,Shiralilou:2021mfl}. One such example is the Brans--Dicke theory, which is one of the simplest forms of scalar-tensor theories. The effect of higher PN corrections in this theory to constrain the theoretical coupling constant (Brans--Dicke parameter) has been studied in \citet{Yunes:2016jcc}. The authors showed that the bound on the Brans--Dicke parameter with GW150914 only changes by $\sim 10\%$  when one includes higher PN corrections or not. This shows that the higher-order corrections are not important when constraining Brans--Dicke theory, though there may be other non-GR theories where sub-leading PN corrections become important, and thus deriving such corrections is an important direction for future research (see Sect.~\ref{sec:model_indep_develop}).

\subsubsection{Effects of bosonic dark matter}

The effects of bosonic dark matter in the context of GR are discussed in Sect.~\ref{Sec:DM_and_PBHs}. Here we discuss effects related to modified gravity. 

Ultralight bosonic dark matter around spinning BHs can trigger superradiant instabilities, forming long-lived bosonic condensates outside the horizon which can alter the inspiral GW signal. A number of modified gravity theories with additional (scalar and/or vector) degrees of freedom exhibit phenomena such as dynamical scalarization (see e.g., \citealt{Benkel:2016kcq,Witek:2018dmd,Ripley:2019irj,East:2020hgw,Berti:2020kgk} for Einstein-Scalar-Gauss--Bonnet Theories and \citealt{Hirschmann:2017psw,Herdeiro:2018wub,Fernandes:2019rez} for Einstein-Scalar-Maxwell theories), where BHs in those theories spontaneously or dynamically acquire scalar hair. How BHs in modified gravity behave around ultra-light bosonic dark matter environments, and what new effects could arise as well as our ability to place constraints on deviations from GR and/or on dark matter has not yet been explored.

\subsubsection{EMRI waveforms}

With EMRI signals, we expect to be able to detect higher order-independent multipole moments of the central body \citep{Ryan:1995wh,Ryan:1997hg} and measure (if any) deviations from GR \citep{Barausse:2020rsu}. Most of the work in this direction, has so far considered BH spacetimes with a multipolar structure different from Kerr, such as bumpy BHs \citep{Collins:2004ex}, and constructed approximate ``kludge" waveforms \citep{Glampedakis:2002cb,Chua:2017ujo} generated considering geodesics in a perturbed Kerr spacetime with orbital parameters evolved using PN equations that assume radiation-reaction effects as in GR \citep{Glampedakis:2005cf,Barack:2006pq,Gair:2007kr,Apostolatos:2009vu,Moore:2017lxy}. 

This approach reproduces the orbit's main features but not to the precision required to determine that the inspiral is indeed an inspiral into a Kerr BH or not \citep{Sasaki:2003xr,Gair:2007kr}. Thus, further work on EMRI waveform modelling will require the consistent inclusion of radiation reaction not described as in GR, additional parameters related to the modification from GR, and environmental effects. Note that the presence of environmental effects may limit our ability to perform proper parameter estimation for some events \citep{Barausse:2014pra,Barausse:2014tra} if these are not properly taken into account. 

These enhanced waveforms will increase model degeneracies that will impact our ability to learn about fundamental physics in the strong-field regime of gravity. That is why it is paramount to fully understand possible discernible features, such as characteristic variations of the amplitude and the energy emission rate \citep{Suzuki:1999si} or the appearance of prolonged resonances \citep{Apostolatos:2009vu} in EMRI signals, as well as the development of novel data analysis techniques to extract information. 

\subsection{Burning Questions and Needed Developments}

In this chapter we have discussed how different systematics in the modelling of waveforms and induced by astrophysical environment can affect tests of GR. Below, we summarize the most important salient points. 

\begin{itemize}
    \item How does (residual) eccentricity induced by binary-disk or hierarchical triplet interactions limit our ability to perform fundamental physics tests with BH binaries?
    
    \item  Can accretion, plasma effects or other stellar compact objects in the vicinity of an EMRI induce observable changes in the GW frequency evolution during the inspiral and/or ringdown that can spoil fundamental physics tests?

    \item Self-force calculations for generic BBHs in vacuum or embedded in a background (e.g. dark matter boson cloud) at second-order are necessary for proper modelling of EMRIs, so that reliable waveforms are available to test for fundamental physics. 
    
    \item What PN accuracy is required for supermassive comparable-mass BH binaries so that we can perform precision tests of fundamental physics with LISA?
    
    \item Development of PN waveforms in beyond-GR theories may need to move beyond leading order corrections. The order to which one should go to constrain theories should be investigated. 
    
    \item What is the accuracy level that numerical relativity simulations must reach in order to produce high-fidelity waveforms for LISA?
    
    \item Numerical relativity must push through existing non-convergence problems to achieve high-fidelity waveforms near merger. Moreover, code efficiency and scaling must increase considerably to achieve resolutions necessary to supress systematic errors and to achieve template indistinguishability.
    
    \item Phenom and EOS models must be improved and calibrated against numerical relativity simulations through the entire parameter space of BBHs, and include eccentricity, arbitrary spins, and substantially different masses. Efforts must be made to include environmental effects such as dynamical friction or interactions with clouds of ultralight fields. Additionally, Phenom and EOB models must be extended to modified gravity.
    
    \item Numerical relativity efforts must expand to include more modified theories of gravity (especially those that admit a well-posed initial value problem). Efforts must be made to understand systematic errors of existing approaches that tackle ill-posed modified gravity theories. 
    
    \item A combination of the effects of bosonic dark matter and modified gravity must be considered in order to be able to understand how more complex deviations from standard general relativity can take place.
    
    \item Development of waveforms for EMRIs in modified gravity must take place. In particular, further work on EMRI waveform modelling requires the consistent inclusion of radiation reaction in beyond-GR theories, additional parameters related to the modification from GR, and any environmental effects.
    
    \item An important potential source of systematics that has not been included in this section (because no work has been done on this yet), but that is surely important to ensure the robustness of tests of GR are \textit{instrumental systematics.} These are uncertainties due to calibration errors, data gaps, or other issues with the instrument itself.

\end{itemize}

\section{Future Directions}

\vspace{0.25cm}

This white paper has been focused on the fundamental physics we can extract from LISA data. The white paper first discussed tests of the theoretical pillars of GR, such as the speed of propagation of GWs and the GW memory. The white paper then moved on to tests of the nature of BHs, discussing constraints on deviations from the Kerr hypothesis and other tests. Next, the white paper summarized tests of GR related to fundamental questions in cosmology, such as tests probing models that attempt to explain dark matter and others that attempt to explain dark energy (as an alternative to a cosmological constant). The white paper then concluded with a discussion of generic tests of GR, and the effects of systematics in all of these tests due to mismodelling and the influence of astrophysical environments. 

As we hope this white paper demonstrates, there is a tremendous amount of work yet to be done before LISA flies in order to fully realize and exploit LISA's scientific potential in this area. The most pressing questions have been discussed at the end of each of the sections summarized above, but let us here touch on a few of the key points.  Regarding tests of GR, the most burning questions all involve the development of more detailed models for the GWs emitted in specific modified theories, including not just for the inspiral phases, but also the merger and the ringdown for comparable-mass binaries and EMRIs. 

Much more work is also still needed in the context of LISA tests on the nature of BHs. Of particular note is the development of a systematic method to carry out Kerr hypothesis tests through the extraction of multipole moments of SMBHs in binaries.  Moreover, the development of accurate waveform models to study the effect of ultralight bosonic fields around SMBHs, as well as the inspiral, merger and ringdown of exotica is needed. Regarding the latter, further studies to determine whether all exotica are a priori possible or whether some theoretical restrictions should be imposed is also important to limit the number of possibilities to study. 

The extraction of fundamental physics information related to cosmology with LISA data will also require that we work hard to address several burning questions. Perhaps one of the most important ones is the construction of accurate waveform models that incorporate the effect of ultralight and heavier dark matter fields in the inspiral, merger and ringdown of compact objects of various mass ratios. Another important aspect that needs further studying is the effect of screening in the generation of propagating GWs, which would be particularly important when attempting to constrain modified gravity dark energy models. 

Let us now discuss the burning questions related to model-independent tests of GR. As discussed in Sect.~\ref{sec:model-indep-tests}, perhaps the most important action item here is the development of model-independent modifications to waveform models in all three phases of coalescence. The merger phase, in particular, is not well-understood in enough modified theories to allow for a model-independent parameterization. Another important action item is related to the extension of LIGO/Virgo tests \citep{LIGOScientific:2019hgc} to the LISA realm, such as the residual test and polarization tests. These tests will be different with LISA because LISA will be sensitive to a large number of sources that will be on during the entire observation, and because of the motion of LISA around the Sun. 

Perhaps one of the most important points to consider is to what extent tests of GR will be affected by waveform systematics or other systematics that could be induced by astrophysical environments. Waveform systematics will be particularly problematic for SMBH signals, because of their expected high SNRs. But this could be a problem even for intermediate mass-ratio systems, for which we may not have sufficiently accurate generic waveforms yet. Waveform systematics may also be a problem for EMRIs, although one expects second-order waveform calculations to be complete before LISA flies. However, astrophysical systematics may still be a particular problem for EMRIs, creating a floor to our ability to test GR. 

Given all of this, it is hopefully clear that a great amount of work is still needed to extract the most fundamental physics from LISA data and to ensure such inferences are robust. A close collaboration between the LISA fundamental physics WG, and other WGs, such as
the waveform modelling and the cosmology, is called for and will be paramount to
successfully overcome all these difficulties.
We remain nonetheless optimistic that through the infrastructure of the LISA Consortium these collaborations can be organized and structured, so that we can get the most science out of the data, when LISA flies.
\bigskip

\section*{Acknowledgments}
E.~Berti is supported by NSF Grants No. PHY-1912550 and AST-2006538, NASA ATP Grants No. 17-ATP17-0225 and 19-ATP19-0051, NSF-XSEDE Grant No. PHY-090003, and NSF Grant PHY-20043.
D.~Blas is supported by a `Ayuda Beatriz Galindo Senior' from the Spanish `Ministerio de Universidades', grant BG20/00228. IFAE is partially funded by the CERCA program of the Generalitat de Catalunya. The research leading of to these results has received funding from the Spanish Ministry of Science and Innovation (PID2020-115845GB-I00/AEI/10.13039/501100011033).
K.~Clough is supported by funding from the European Research Council (ERC) under the European Unions Horizon 2020 research and innovation programme (grant agreement No 693024).
A.~C{\'a}rdenas-Avenda{\~n}o acknowledges funding from the Fundaci\'on Universitaria Konrad Lorenz (Project 5INV1) and from Will and Kacie Snellings.
M.~Crisostomi and E.~Barausse are supported by the European Union's H2020 ERC Consolidator Grant ``GRavity from Astrophysical to Microscopic Scales'' (Grant No.  GRAMS-815673).
P.~Fleury received the support of a fellowship from ``la Caixa'' Foundation (ID 100010434). The fellowship code is LCF/BQ/PI19/11690018.
C.~Herdeiro thanks the support of the Center for Research and Development in Mathematics and Applications (CIDMA) through the Portuguese Foundation for Science and Technology (FCT - Funda\c c\~ao para a Ci\^encia e a Tecnologia), references UIDB/04106/2020 and UIDP/04106/2020, 
  the  projects PTDC/FIS-OUT/28407/2017, CERN/FIS-PAR/0027/2019,  PTDC/FIS-AST/3041/2020  and the European Union’s Horizon 2020 research and innovation (RISE) programme H2020-MSCA-RISE-2017 Grant No. FunFiCO-777740.  
P.~Pani and E.~Maggio acknowledge financial support provided under the European Union's H2020 ERC, Starting 
Grant agreement no.~DarkGRA--757480, and under the MIUR PRIN and FARE programmes (GW-NEXT, CUP:~B84I20000100001), and support from the Amaldi Research Center funded by the MIUR program "Dipartimento di Eccellenza" (CUP:~B81I18001170001).
N.Frusciante was supported by Funda\c{c}\~{a}o para a  Ci\^{e}ncia e a Tecnologia (FCT) through the research grants UIDB/04434/2020, UIDP/04434/2020, PTDC/FIS-OUT/29048/2017, CERN/FIS-PAR/0037/2019, the FCT project ``CosmoTests -- Cosmological tests of gravity theories beyond General Relativity" with ref.~number CEECIND/00017/2018 and the FCT project ``BEYLA --BEYond LAmbda" with ref. number PTDC/FIS-AST/0054/2021.
L.Lombriser~was supported by a Swiss National Science Foundation Professorship grant (No.~170547).
S.N. acknowledges support from the Alexander von Humboldt Foundation.
D.N.\ acknowledges support from the NSF Grant No.\ PHY-2011784.
R.B. acknowledges financial support from FCT – Fundação para a Ciência e a Tecnologia, I.P., under the Scientific Employment Stimulus - Individual Call - 2020.00470.CEECIND.
V. Paschalidis acknowledges support from NSF Grant PHY-1912619 and NASA Grant 80NSSC20K1542 to the University of Arizona.
B.S.S. is supported by NSF grants No. AST-2006384 and PHY-2012083.
C.F.S. is supported by contracts ESP2017-90084-P and PID2019-106515GB-I00/AEI/10.13039/501100011033 (Spanish Ministry of Science and Innovation) and 2017-SGR-1469 (AGAUR, Generalitat de Catalunya).
T. P. S. acknowledges partial support from the STFC Consolidated Grant No. ST/P000703/1.
M.~Ruiz acknowledges support from NASA Grant 80NSSC17K0070 to the
University of Illinois at Urbana-Champaign.
I.D. Saltas is supported by the Czech Science Foundation GAČR, Grant No. 21-16583M.
N. Stergioulas is supported by the ESA Prodex grant PEA:4000132310 "LISA Stochastic Signals Analysis Pipeline".
F.V.~acknowledges partial support from CNES.
K.Y. acknowledges support from NSF Grant PHY-1806776, NASA Grant 80NSSC20K0523, a Sloan Foundation Research Fellowship and the Owens Family Foundation. 
K.Y. would like to also acknowledge support by the COST Action GWverse CA16104 and JSPS KAKENHI Grants No. JP17H06358.
N.~Yunes acknowledges support from NASA Grants No. NNX16AB98G, 80NSSC17M0041 and 80NSSC18K1352, NSF Award No. 1759615, and the Simons Foundation through MPS Award Number 896696.
D.D. acknowledge financial support via an Emmy Noether Research Group funded by the German Research Foundation (DFG) under grant no. DO 1771/1-1.

\bibliographystyle{spbasic-FS}
\bibliography{sample-1, sample-2}

\begin{thebibliography}{1254}
\expandafter\ifx\csname url\endcsname\relax
 \def\url#1{\burl{#1}}\fi
\expandafter\ifx\csname urlprefix\endcsname\relax\def\urlprefix{URL }\fi
\providecommand{\bibinfo}[2]{#2}
\providecommand{\eprint}[2][]{\url{#2}}
\providecommand{\doi}[1]{\urlstyle{rm}\url{https://doi.org/#1}}

\bibitem[{Abbott et~al.(2016{\natexlab{a}})}]{TheLIGOScientific:2016wyq}
Abbott B, et~al. (2016{\natexlab{a}}) {GW150914: Implications for the
  stochastic gravitational wave background from binary black holes}. Phys Rev
  Lett 116(13):131102. \doi{10.1103/PhysRevLett.116.131102}.
  {\href{https://arxiv.org/abs/1602.03847}{{arXiv:1602.03847}}} {[gr-qc]}

\bibitem[{Abbott et~al.(2017{\natexlab{a}})}]{Abbott:2017xzu}
Abbott B, et~al. (2017{\natexlab{a}}) {A gravitational-wave standard siren
  measurement of the Hubble constant}. Nature 551(7678):85--88.
  \doi{10.1038/nature24471}.
  {\href{https://arxiv.org/abs/1710.05835}{{arXiv:1710.05835}}} {[astro-ph.CO]}

\bibitem[{Abbott et~al.(2017{\natexlab{b}})}]{Abbott:2017oio}
Abbott B, et~al. (2017{\natexlab{b}}) {GW170814: A Three-Detector Observation
  of Gravitational Waves from a Binary Black Hole Coalescence}. Phys Rev Lett
  119(14):141101. \doi{10.1103/PhysRevLett.119.141101}.
  {\href{https://arxiv.org/abs/1709.09660}{{arXiv:1709.09660}}} {[gr-qc]}

\bibitem[{Abbott et~al.(2017{\natexlab{c}})}]{TheLIGOScientific:2017qsa}
Abbott B, et~al. (2017{\natexlab{c}}) {GW170817: Observation of Gravitational
  Waves from a Binary Neutron Star Inspiral}. Phys Rev Lett 119(16):161101.
  \doi{10.1103/PhysRevLett.119.161101}.
  {\href{https://arxiv.org/abs/1710.05832}{{arXiv:1710.05832}}} {[gr-qc]}

\bibitem[{Abbott et~al.(2017{\natexlab{d}})}]{GBM:2017lvd}
Abbott B, et~al. (2017{\natexlab{d}}) {Multi-messenger Observations of a Binary
  Neutron Star Merger}. Astrophys J Lett 848(2):L12.
  \doi{10.3847/2041-8213/aa91c9}.
  {\href{https://arxiv.org/abs/1710.05833}{{arXiv:1710.05833}}} {[astro-ph.HE]}

\bibitem[{Abbott et~al.(2018{\natexlab{a}})}]{Abbott:2018exr}
Abbott B, et~al. (2018{\natexlab{a}}) {GW170817: Measurements of neutron star
  radii and equation of state}. Phys Rev Lett 121(16):161101.
  \doi{10.1103/PhysRevLett.121.161101}.
  {\href{https://arxiv.org/abs/1805.11581}{{arXiv:1805.11581}}} {[gr-qc]}

\bibitem[{Abbott et~al.(2019{\natexlab{a}})}]{Abbott:2019yzh}
Abbott B, et~al. (2019{\natexlab{a}}) {A gravitational-wave measurement of the
  Hubble constant following the second observing run of Advanced LIGO and
  Virgo} {\href{https://arxiv.org/abs/1908.06060}{{arXiv:1908.06060}}}
  {[astro-ph.CO]}

\bibitem[{Abbott et~al.(2019{\natexlab{b}})}]{LIGOScientific:2018mvr}
Abbott B, et~al. (2019{\natexlab{b}}) {GWTC-1: A Gravitational-Wave Transient
  Catalog of Compact Binary Mergers Observed by LIGO and Virgo during the First
  and Second Observing Runs}. Phys Rev X 9(3):031040.
  \doi{10.1103/PhysRevX.9.031040}.
  {\href{https://arxiv.org/abs/1811.12907}{{arXiv:1811.12907}}} {[astro-ph.HE]}

\bibitem[{Abbott et~al.(2019{\natexlab{c}})}]{Abbott:2018lct}
Abbott B, et~al. (2019{\natexlab{c}}) {Tests of General Relativity with
  GW170817}. Phys Rev Lett 123(1):011102. \doi{10.1103/PhysRevLett.123.011102}.
  {\href{https://arxiv.org/abs/1811.00364}{{arXiv:1811.00364}}} {[gr-qc]}

\bibitem[{Abbott et~al.(2019{\natexlab{d}})}]{LIGOScientific:2019fpa}
Abbott B, et~al. (2019{\natexlab{d}}) {Tests of General Relativity with the
  Binary Black Hole Signals from the LIGO-Virgo Catalog GWTC-1}. Phys Rev D
  100(10):104036. \doi{10.1103/PhysRevD.100.104036}.
  {\href{https://arxiv.org/abs/1903.04467}{{arXiv:1903.04467}}} {[gr-qc]}

\bibitem[{Abbott et~al.(2016{\natexlab{b}})}]{TheLIGOScientific:2016src}
Abbott BP, et~al. (2016{\natexlab{b}}) {Tests of general relativity with
  GW150914}. Phys Rev Lett 116(22):221101.
  \doi{10.1103/PhysRevLett.116.221101}.
  {\href{https://arxiv.org/abs/1602.03841}{{arXiv:1602.03841}}} {[gr-qc]}

\bibitem[{Abbott et~al.(2017{\natexlab{e}})}]{Monitor:2017mdv}
Abbott BP, et~al. (2017{\natexlab{e}}) {Gravitational Waves and Gamma-rays from
  a Binary Neutron Star Merger: GW170817 and GRB 170817A}. Astrophys J 848:L13.
  \doi{10.3847/2041-8213/aa920c}.
  {\href{https://arxiv.org/abs/1710.05834}{{arXiv:1710.05834}}} {[astro-ph.HE]}

\bibitem[{Abbott et~al.(2018{\natexlab{b}})}]{Abbott:2017xzg}
Abbott BP, et~al. (2018{\natexlab{b}}) {GW170817: Implications for the
  Stochastic Gravitational-Wave Background from Compact Binary Coalescences}.
  Phys Rev Lett 120(9):091101. \doi{10.1103/PhysRevLett.120.091101}.
  {\href{https://arxiv.org/abs/1710.05837}{{arXiv:1710.05837}}} {[gr-qc]}

\bibitem[{Abbott et~al.(2020{\natexlab{a}})}]{LIGOScientific:2019hgc}
Abbott BP, et~al. (2020{\natexlab{a}}) {A guide to LIGO\textendash{}Virgo
  detector noise and extraction of transient gravitational-wave signals}. Class
  Quant Grav 37(5):055002. \doi{10.1088/1361-6382/ab685e}.
  {\href{https://arxiv.org/abs/1908.11170}{{arXiv:1908.11170}}} {[gr-qc]}

\bibitem[{Abbott et~al.(2020{\natexlab{b}})}]{LIGOScientific:2020stg}
Abbott R, et~al. (2020{\natexlab{b}}) {GW190412: Observation of a
  Binary-Black-Hole Coalescence with Asymmetric Masses}. Phys Rev
  D102(4):043015. \doi{10.1103/PhysRevD.102.043015}.
  {\href{https://arxiv.org/abs/2004.08342}{{arXiv:2004.08342}}} {[astro-ph.HE]}

\bibitem[{Abbott et~al.(2020{\natexlab{c}})}]{Abbott:2020tfl}
Abbott R, et~al. (2020{\natexlab{c}}) {GW190521: A Binary Black Hole Merger
  with a Total Mass of 150\,\,M\ensuremath{\odot}}. Phys Rev Lett
  125(10):101102. \doi{10.1103/PhysRevLett.125.101102}.
  {\href{https://arxiv.org/abs/2009.01075}{{arXiv:2009.01075}}} {[gr-qc]}

\bibitem[{Abbott et~al.(2020{\natexlab{d}})}]{Abbott:2020khf}
Abbott R, et~al. (2020{\natexlab{d}}) {GW190814: Gravitational Waves from the
  Coalescence of a 23 Solar Mass Black Hole with a 2.6 Solar Mass Compact
  Object}. Astrophys J Lett 896(2):L44. \doi{10.3847/2041-8213/ab960f}.
  {\href{https://arxiv.org/abs/2006.12611}{{arXiv:2006.12611}}} {[astro-ph.HE]}

\bibitem[{Abbott et~al.(2020{\natexlab{e}})}]{Abbott:2020mjq}
Abbott R, et~al. (2020{\natexlab{e}}) {Properties and Astrophysical
  Implications of the 150 M$_\odot$ Binary Black Hole Merger GW190521}.
  Astrophys J 900(1):L13. \doi{10.3847/2041-8213/aba493}.
  {\href{https://arxiv.org/abs/2009.01190}{{arXiv:2009.01190}}} {[astro-ph.HE]}

\bibitem[{Abbott et~al.(2020{\natexlab{f}})}]{Abbott:2020jks}
Abbott R, et~al. (2020{\natexlab{f}}) {Tests of General Relativity with Binary
  Black Holes from the second LIGO-Virgo Gravitational-Wave Transient Catalog}
  {\href{https://arxiv.org/abs/2010.14529}{{arXiv:2010.14529}}} {[gr-qc]}

\bibitem[{Abbott et~al.(2021)}]{LIGOScientific:2020tif}
Abbott R, et~al. (2021) {Tests of general relativity with binary black holes
  from the second LIGO-Virgo gravitational-wave transient catalog}. Phys Rev D
  103(12):122002. \doi{10.1103/PhysRevD.103.122002}.
  {\href{https://arxiv.org/abs/2010.14529}{{arXiv:2010.14529}}} {[gr-qc]}

\bibitem[{Abedi and Afshordi(2018)}]{Abedi:2018npz}
Abedi J, Afshordi N (2018) {Echoes from the Abyss: A highly spinning black hole
  remnant for the binary neutron star merger GW170817}
  {\href{https://arxiv.org/abs/1803.10454}{{arXiv:1803.10454}}} {[gr-qc]}

\bibitem[{Abedi et~al.(2017)Abedi, Dykaar, and Afshordi}]{Abedi:2016hgu}
Abedi J, Dykaar H, Afshordi N (2017) {Echoes from the Abyss: Tentative evidence
  for Planck-scale structure at black hole horizons}. Phys Rev D96(8):082004.
  \doi{10.1103/PhysRevD.96.082004}.
  {\href{https://arxiv.org/abs/1612.00266}{{arXiv:1612.00266}}} {[gr-qc]}

\bibitem[{Abedi et~al.(2020)Abedi, Afshordi, Oshita, and Wang}]{Abedi:2020ujo}
Abedi J, Afshordi N, Oshita N, Wang Q (2020) {Quantum Black Holes in the Sky}.
  Universe 6(3):43. \doi{10.3390/universe6030043}.
  {\href{https://arxiv.org/abs/2001.09553}{{arXiv:2001.09553}}} {[gr-qc]}

\bibitem[{{Abramowicz} and {Fragile}(2013)}]{2013LRR....16....1A}
{Abramowicz} MA, {Fragile} PC (2013) {Foundations of Black Hole Accretion Disk
  Theory}. Living Reviews in Relativity 16(1):1. \doi{10.12942/lrr-2013-1}.
  {\href{https://arxiv.org/abs/1104.5499}{{arXiv:1104.5499}}} {[astro-ph.HE]}

\bibitem[{Acquaviva et~al.(2003)Acquaviva, Bartolo, Matarrese, and
  Riotto}]{Acquaviva:2002ud}
Acquaviva V, Bartolo N, Matarrese S, Riotto A (2003) {Second order cosmological
  perturbations from inflation}. Nucl Phys B 667:119--148.
  \doi{10.1016/S0550-3213(03)00550-9}.
  {\href{https://arxiv.org/abs/astro-ph/0209156}{{arXiv:astro-ph/0209156}}}

\bibitem[{Adam et~al.(2021)Adam, Figueras, Jacobson, and
  Wiseman}]{Adam:2021vsk}
Adam A, Figueras P, Jacobson T, Wiseman T (2021) {Rotating black holes in
  Einstein-aether theory}
  {\href{https://arxiv.org/abs/2108.00005}{{arXiv:2108.00005}}} {[gr-qc]}

\bibitem[{Adam et~al.(2016)}]{Adam:2015rua}
Adam R, et~al. (2016) {Planck 2015 results. I. Overview of products and
  scientific results}. Astron Astrophys 594:A1.
  \doi{10.1051/0004-6361/201527101}.
  {\href{https://arxiv.org/abs/1502.01582}{{arXiv:1502.01582}}} {[astro-ph.CO]}

\bibitem[{Addazi et~al.(2019)Addazi, Marciano, and Yunes}]{Addazi:2018uhd}
Addazi A, Marciano A, Yunes N (2019) {Can we probe Planckian corrections at the
  horizon scale with gravitational waves?} Phys Rev Lett 122(8):081301.
  \doi{10.1103/PhysRevLett.122.081301}.
  {\href{https://arxiv.org/abs/1810.10417}{{arXiv:1810.10417}}} {[gr-qc]}

\bibitem[{Ade et~al.(2014)}]{Ade:2013zuv}
Ade P, et~al. (2014) {Planck 2013 results. XVI. Cosmological parameters}.
  Astron Astrophys 571:A16. \doi{10.1051/0004-6361/201321591}.
  {\href{https://arxiv.org/abs/1303.5076}{{arXiv:1303.5076}}} {[astro-ph.CO]}

\bibitem[{Aghamousa et~al.(2016)}]{Aghamousa:2016zmz}
Aghamousa A, et~al. (2016) {The DESI Experiment Part I: Science,Targeting, and
  Survey Design} {\href{https://arxiv.org/abs/1611.00036}{{arXiv:1611.00036}}}
  {[astro-ph.IM]}

\bibitem[{Aghanim et~al.(2020)}]{Aghanim:2018eyx}
Aghanim N, et~al. (2020) {Planck 2018 results. VI. Cosmological parameters}.
  Astron Astrophys 641:A6. \doi{10.1051/0004-6361/201833910}.
  {\href{https://arxiv.org/abs/1807.06209}{{arXiv:1807.06209}}} {[astro-ph.CO]}

\bibitem[{Agullo et~al.(2020)Agullo, Cardoso, del Rio, Maggiore, and
  Pullin}]{Agullo:2020hxe}
Agullo I, Cardoso V, del Rio A, Maggiore M, Pullin J (2020) {Gravitational-wave
  signatures of quantum gravity}
  {\href{https://arxiv.org/abs/2007.13761}{{arXiv:2007.13761}}} {[gr-qc]}

\bibitem[{Ajith et~al.(2007)}]{Ajith:2007qp}
Ajith P, et~al. (2007) {Phenomenological template family for black-hole
  coalescence waveforms}. Class Quant Grav 24:S689--S700.
  \doi{10.1088/0264-9381/24/19/S31}.
  {\href{https://arxiv.org/abs/0704.3764}{{arXiv:0704.3764}}} {[gr-qc]}

\bibitem[{Ajith et~al.(2008)}]{Ajith:2007kx}
Ajith P, et~al. (2008) {A Template bank for gravitational waveforms from
  coalescing binary black holes. I. Non-spinning binaries}. Phys Rev D
  77:104017. \doi{10.1103/PhysRevD.77.104017}, [Erratum: Phys.Rev.D 79, 129901
  (2009)]. {\href{https://arxiv.org/abs/0710.2335}{{arXiv:0710.2335}}}
  {[gr-qc]}

\bibitem[{Ajith et~al.(2011)}]{Ajith:2009bn}
Ajith P, et~al. (2011) {Inspiral-merger-ringdown waveforms for black-hole
  binaries with non-precessing spins}. Phys Rev Lett 106:241101.
  \doi{10.1103/PhysRevLett.106.241101}.
  {\href{https://arxiv.org/abs/0909.2867}{{arXiv:0909.2867}}} {[gr-qc]}

\bibitem[{Akcay et~al.(2019)Akcay, Bernuzzi, Messina, Nagar, Ortiz, and
  Rettegno}]{Akcay:2018yyh}
Akcay S, Bernuzzi S, Messina F, Nagar A, Ortiz N, Rettegno P (2019)
  {Effective-one-body multipolar waveform for tidally interacting binary
  neutron stars up to merger}. Phys Rev D 99(4):044051.
  \doi{10.1103/PhysRevD.99.044051}.
  {\href{https://arxiv.org/abs/1812.02744}{{arXiv:1812.02744}}} {[gr-qc]}

\bibitem[{Akcay et~al.(2020)Akcay, Dolan, Kavanagh, Moxon, Warburton, and
  Wardell}]{Akcay:2019bvk}
Akcay S, Dolan SR, Kavanagh C, Moxon J, Warburton N, Wardell B (2020)
  {Dissipation in extreme-mass ratio binaries with a spinning secondary}. Phys
  Rev D 102(6):064013. \doi{10.1103/PhysRevD.102.064013}.
  {\href{https://arxiv.org/abs/1912.09461}{{arXiv:1912.09461}}} {[gr-qc]}

\bibitem[{Akrami et~al.(2018)}]{Akrami:2018odb}
Akrami Y, et~al. (2018) {Planck 2018 results. X. Constraints on inflation}
  {\href{https://arxiv.org/abs/1807.06211}{{arXiv:1807.06211}}} {[astro-ph.CO]}

\bibitem[{Akrami et~al.(2019)}]{Akrami:2019izv}
Akrami Y, et~al. (2019) {Planck 2018 results. IX. Constraints on primordial
  non-Gaussianity}
  {\href{https://arxiv.org/abs/1905.05697}{{arXiv:1905.05697}}} {[astro-ph.CO]}

\bibitem[{Alba and Maldacena(2016)}]{Alba:2015cms}
Alba V, Maldacena J (2016) {Primordial gravity wave background anisotropies}.
  JHEP 03:115. \doi{10.1007/JHEP03(2016)115}.
  {\href{https://arxiv.org/abs/1512.01531}{{arXiv:1512.01531}}} {[hep-th]}

\bibitem[{Alberte et~al.(2016)Alberte, Creminelli, Khmelnitsky, Pirtskhalava,
  and Trincherini}]{Alberte:2016izw}
Alberte L, Creminelli P, Khmelnitsky A, Pirtskhalava D, Trincherini E (2016)
  {Relaxing the Cosmological Constant: a Proof of Concept}. JHEP 12:022.
  \doi{10.1007/JHEP12(2016)022}.
  {\href{https://arxiv.org/abs/1608.05715}{{arXiv:1608.05715}}} {[hep-th]}

\bibitem[{Alcock et~al.(2001)}]{Alcock:2000kd}
Alcock C, et~al. (2001) {The MACHO project: microlensing detection efficiency}.
  Astrophys J Suppl 136:439--462. \doi{10.1086/322529}.
  {\href{https://arxiv.org/abs/astro-ph/0003392}{{arXiv:astro-ph/0003392}}}

\bibitem[{Alcubierre(2008)}]{AlcubierreBook2008}
Alcubierre M (2008) {Introduction to 3+1 Numerical Relativity}. Oxford
  University Press, Oxford, U.K.

\bibitem[{Alcubierre et~al.(2018)Alcubierre, Barranco, Bernal, Degollado,
  Diez-Tejedor, Megevand, Nunez, and Sarbach}]{Alcubierre:2018ahf}
Alcubierre M, Barranco J, Bernal A, Degollado JC, Diez-Tejedor A, Megevand M,
  Nunez D, Sarbach O (2018) {$\ell$-Boson stars}. Class Quant Grav
  35(19):19LT01. \doi{10.1088/1361-6382/aadcb6}.
  {\href{https://arxiv.org/abs/1805.11488}{{arXiv:1805.11488}}} {[gr-qc]}

\bibitem[{Alexander and Yunes(2009)}]{Alexander:2009tp}
Alexander S, Yunes N (2009) {Chern-Simons Modified General Relativity}. Phys
  Rept 480:1--55. \doi{10.1016/j.physrep.2009.07.002}.
  {\href{https://arxiv.org/abs/0907.2562}{{arXiv:0907.2562}}} {[hep-th]}

\bibitem[{Alexander et~al.(2008)Alexander, Finn, and Yunes}]{Alexander:2007kv}
Alexander S, Finn LS, Yunes N (2008) {A Gravitational-wave probe of effective
  quantum gravity}. Phys Rev D78:066005. \doi{10.1103/PhysRevD.78.066005}.
  {\href{https://arxiv.org/abs/0712.2542}{{arXiv:0712.2542}}} {[gr-qc]}

\bibitem[{Ali-Haïmoud(2018)}]{Ali-Haimoud:2018dau}
Ali-Haïmoud Y (2018) {Correlation Function of High-Threshold Regions and
  Application to the Initial Small-Scale Clustering of Primordial Black Holes}.
  Phys Rev Lett 121(8):081304. \doi{10.1103/PhysRevLett.121.081304}.
  {\href{https://arxiv.org/abs/1805.05912}{{arXiv:1805.05912}}} {[astro-ph.CO]}

\bibitem[{Ali-Haïmoud and Kamionkowski(2017)}]{Ali-Haimoud:2016mbv}
Ali-Haïmoud Y, Kamionkowski M (2017) {Cosmic microwave background limits on
  accreting primordial black holes}. Phys Rev D 95(4):043534.
  \doi{10.1103/PhysRevD.95.043534}.
  {\href{https://arxiv.org/abs/1612.05644}{{arXiv:1612.05644}}} {[astro-ph.CO]}

\bibitem[{Allen and Romano(1999)}]{Allen:1997ad}
Allen B, Romano JD (1999) {Detecting a stochastic background of gravitational
  radiation: Signal processing strategies and sensitivities}. Phys Rev D
  59:102001. \doi{10.1103/PhysRevD.59.102001}.
  {\href{https://arxiv.org/abs/gr-qc/9710117}{{arXiv:gr-qc/9710117}}}

\bibitem[{Allsman et~al.(2001)}]{Allsman:2000kg}
Allsman R, et~al. (2001) {MACHO project limits on black hole dark matter in the
  1-30 solar mass range}. Astrophys J Lett 550:L169. \doi{10.1086/319636}.
  {\href{https://arxiv.org/abs/astro-ph/0011506}{{arXiv:astro-ph/0011506}}}

\bibitem[{Allwright and Lehner(2019)}]{Allwright:2018rut}
Allwright G, Lehner L (2019) {Towards the nonlinear regime in extensions to GR:
  assessing possible options}. Class Quant Grav 36(8):084001.
  \doi{10.1088/1361-6382/ab0ee1}.
  {\href{https://arxiv.org/abs/1808.07897}{{arXiv:1808.07897}}} {[gr-qc]}

\bibitem[{Almheiri et~al.(2013)Almheiri, Marolf, Polchinski, and
  Sully}]{Almheiri:2012rt}
Almheiri A, Marolf D, Polchinski J, Sully J (2013) {Black Holes:
  Complementarity or Firewalls?} JHEP 02:062. \doi{10.1007/JHEP02(2013)062}.
  {\href{https://arxiv.org/abs/1207.3123}{{arXiv:1207.3123}}} {[hep-th]}

\bibitem[{{Alonso} et~al.(2020){Alonso}, {Cusin}, {Ferreira}, and
  {Pitrou}}]{Alonso2020PhRvD}
{Alonso} D, {Cusin} G, {Ferreira} PG, {Pitrou} C (2020) {Detecting the
  anisotropic astrophysical gravitational wave background in the presence of
  shot noise through cross-correlations}. \prd 102(2):023002.
  \doi{10.1103/PhysRevD.102.023002}.
  {\href{https://arxiv.org/abs/2002.02888}{{arXiv:2002.02888}}} {[astro-ph.CO]}

\bibitem[{Alvarez and Yu(2021)}]{alvarez2021density}
Alvarez G, Yu HB (2021) Density spikes near black holes in self-interacting
  dark matter halos and indirect detection constraints.
  {\href{https://arxiv.org/abs/2012.15050}{{arXiv:2012.15050}}} {[hep-ph]}

\bibitem[{Amaro-Seoane et~al.(2007)Amaro-Seoane, Gair, Freitag, Coleman~Miller,
  Mandel, Cutler, and Babak}]{AmaroSeoane:2007aw}
Amaro-Seoane P, Gair JR, Freitag M, Coleman~Miller M, Mandel I, Cutler CJ,
  Babak S (2007) {Astrophysics, detection and science applications of
  intermediate- and extreme mass-ratio inspirals}. Class Quant Grav
  24:R113--R169. \doi{10.1088/0264-9381/24/17/R01}.
  {\href{https://arxiv.org/abs/astro-ph/0703495}{{arXiv:astro-ph/0703495}}}

\bibitem[{Amaro-Seoane et~al.(2017)}]{LISA:2017pwj}
Amaro-Seoane P, et~al. (2017) {Laser Interferometer Space Antenna}
  {\href{https://arxiv.org/abs/1702.00786}{{arXiv:1702.00786}}} {[astro-ph.IM]}

\bibitem[{Amaro-Seoane et~al.(2022)}]{Amaro-Seoane:2022rxf}
Amaro-Seoane P, et~al. (2022) {Astrophysics with the Laser Interferometer Space
  Antenna} {\href{https://arxiv.org/abs/2203.06016}{{arXiv:2203.06016}}}
  {[gr-qc]}

\bibitem[{Amendola(2000)}]{Amendola:1999er}
Amendola L (2000) {Coupled quintessence}. Phys Rev D 62:043511.
  \doi{10.1103/PhysRevD.62.043511}.
  {\href{https://arxiv.org/abs/astro-ph/9908023}{{arXiv:astro-ph/9908023}}}

\bibitem[{Amendola et~al.(2014)Amendola, Ballesteros, and
  Pettorino}]{Amendola:2014wma}
Amendola L, Ballesteros G, Pettorino V (2014) {Effects of modified gravity on
  B-mode polarization}. Phys Rev D 90:043009. \doi{10.1103/PhysRevD.90.043009}.
  {\href{https://arxiv.org/abs/1405.7004}{{arXiv:1405.7004}}} {[astro-ph.CO]}

\bibitem[{Amendola et~al.(2018{\natexlab{a}})Amendola, Sawicki, Kunz, and
  Saltas}]{Amendola:2017ovw}
Amendola L, Sawicki I, Kunz M, Saltas ID (2018{\natexlab{a}}) {Direct detection
  of gravitational waves can measure the time variation of the Planck mass}.
  JCAP 1808:030. \doi{10.1088/1475-7516/2018/08/030}.
  {\href{https://arxiv.org/abs/1712.08623}{{arXiv:1712.08623}}} {[astro-ph.CO]}

\bibitem[{Amendola et~al.(2018{\natexlab{b}})}]{Amendola:2016saw}
Amendola L, et~al. (2018{\natexlab{b}}) {Cosmology and fundamental physics with
  the Euclid satellite}. Living Rev Rel 21(1):2.
  \doi{10.1007/s41114-017-0010-3}.
  {\href{https://arxiv.org/abs/1606.00180}{{arXiv:1606.00180}}} {[astro-ph.CO]}

\bibitem[{Amin and Mou(2020)}]{Amin:2020vja}
Amin MA, Mou ZG (2020) {Electromagnetic Bursts from Mergers of Oscillons in
  Axion-like Fields}
  {\href{https://arxiv.org/abs/2009.11337}{{arXiv:2009.11337}}} {[astro-ph.CO]}

\bibitem[{Ananda et~al.(2007)Ananda, Clarkson, and Wands}]{Ananda:2006af}
Ananda KN, Clarkson C, Wands D (2007) {The Cosmological gravitational wave
  background from primordial density perturbations}. Phys Rev D 75:123518.
  \doi{10.1103/PhysRevD.75.123518}.
  {\href{https://arxiv.org/abs/gr-qc/0612013}{{arXiv:gr-qc/0612013}}}

\bibitem[{Andersson(1998)}]{Andersson:1997xt}
Andersson N (1998) {A New class of unstable modes of rotating relativistic
  stars}. Astrophys J 502:708--713. \doi{10.1086/305919}.
  {\href{https://arxiv.org/abs/gr-qc/9706075}{{arXiv:gr-qc/9706075}}}

\bibitem[{Ando et~al.(2018)Ando, Inomata, Kawasaki, Mukaida, and
  Yanagida}]{Ando:2017veq}
Ando K, Inomata K, Kawasaki M, Mukaida K, Yanagida TT (2018) {Primordial black
  holes for the LIGO events in the axionlike curvaton model}. Phys Rev D
  97(12):123512. \doi{10.1103/PhysRevD.97.123512}.
  {\href{https://arxiv.org/abs/1711.08956}{{arXiv:1711.08956}}} {[astro-ph.CO]}

\bibitem[{Annulli et~al.(2020{\natexlab{a}})Annulli, Cardoso, and
  Vicente}]{Annulli:2020ilw}
Annulli L, Cardoso V, Vicente R (2020{\natexlab{a}}) {Stirred and shaken:
  dynamical behavior of boson stars and dark matter cores}
  {\href{https://arxiv.org/abs/2007.03700}{{arXiv:2007.03700}}} {[astro-ph.HE]}

\bibitem[{Annulli et~al.(2020{\natexlab{b}})Annulli, Cardoso, and
  Vicente}]{Annulli:2020lyc}
Annulli L, Cardoso V, Vicente R (2020{\natexlab{b}}) {The response of
  ultralight dark matter to supermassive black holes and binaries}
  {\href{https://arxiv.org/abs/2009.00012}{{arXiv:2009.00012}}} {[gr-qc]}

\bibitem[{Antonelli et~al.(2019)Antonelli, Buonanno, Steinhoff, van~de Meent,
  and Vines}]{Antonelli:2019ytb}
Antonelli A, Buonanno A, Steinhoff J, van~de Meent M, Vines J (2019)
  {Energetics of two-body Hamiltonians in post-Minkowskian gravity}. Phys Rev D
  99(10):104004. \doi{10.1103/PhysRevD.99.104004}.
  {\href{https://arxiv.org/abs/1901.07102}{{arXiv:1901.07102}}} {[gr-qc]}

\bibitem[{Antonelli et~al.(2020)Antonelli, van~de Meent, Buonanno, Steinhoff,
  and Vines}]{Antonelli:2019fmq}
Antonelli A, van~de Meent M, Buonanno A, Steinhoff J, Vines J (2020)
  {Quasicircular inspirals and plunges from nonspinning effective-one-body
  Hamiltonians with gravitational self-force information}. Phys Rev D
  101(2):024024. \doi{10.1103/PhysRevD.101.024024}.
  {\href{https://arxiv.org/abs/1907.11597}{{arXiv:1907.11597}}} {[gr-qc]}

\bibitem[{Antoniadis and Mottola(1991)}]{Antoniadis:1986sb}
Antoniadis I, Mottola E (1991) {Graviton Fluctuations in De Sitter Space}.
  JMathPhys 32:1037--1044. \doi{10.1063/1.529381}

\bibitem[{Antoniadis and Mottola(1992)}]{Antoniadis:1991fa}
Antoniadis I, Mottola E (1992) {4-D quantum gravity in the conformal sector}.
  PhysRev D45:2013--2025. \doi{10.1103/PhysRevD.45.2013}

\bibitem[{Antoniou et~al.(2018)Antoniou, Bakopoulos, and
  Kanti}]{Antoniou:2017acq}
Antoniou G, Bakopoulos A, Kanti P (2018) {Evasion of No-Hair Theorems and Novel
  Black-Hole Solutions in Gauss-Bonnet Theories}. Phys Rev Lett 120(13):131102.
  \doi{10.1103/PhysRevLett.120.131102}.
  {\href{https://arxiv.org/abs/1711.03390}{{arXiv:1711.03390}}} {[hep-th]}

\bibitem[{Antoniou et~al.(2020)Antoniou, Bordin, and
  Sotiriou}]{Antoniou:2020nax}
Antoniou G, Bordin L, Sotiriou TP (2020) {Compact object scalarization with
  general relativity as a cosmic attractor}
  {\href{https://arxiv.org/abs/2004.14985}{{arXiv:2004.14985}}} {[gr-qc]}

\bibitem[{Apostolatos et~al.(2009)Apostolatos, Lukes-Gerakopoulos, and
  Contopoulos}]{Apostolatos:2009vu}
Apostolatos TA, Lukes-Gerakopoulos G, Contopoulos G (2009) {How to Observe a
  Non-Kerr Spacetime Using Gravitational Waves}. Phys Rev Lett 103:111101.
  \doi{10.1103/PhysRevLett.103.111101}.
  {\href{https://arxiv.org/abs/0906.0093}{{arXiv:0906.0093}}} {[gr-qc]}

\bibitem[{Arai and Nishizawa(2018)}]{Arai:2017hxj}
Arai S, Nishizawa A (2018) {Generalized framework for testing gravity with
  gravitational-wave propagation. II. Constraints on Horndeski theory}. Phys
  Rev D97:104038. \doi{10.1103/PhysRevD.97.104038}.
  {\href{https://arxiv.org/abs/1711.03776}{{arXiv:1711.03776}}} {[gr-qc]}

\bibitem[{{Armitage} and {Natarajan}(2002)}]{2002ApJ...567L...9A}
{Armitage} PJ, {Natarajan} P (2002) {Accretion during the Merger of
  Supermassive Black Holes}. \apjl 567(1):L9--L12. \doi{10.1086/339770}.
  {\href{https://arxiv.org/abs/astro-ph/0201318}{{arXiv:astro-ph/0201318}}}
  {[astro-ph]}

\bibitem[{Arun et~al.(2006{\natexlab{a}})Arun, Iyer, Qusailah, and
  Sathyaprakash}]{Arun:2006hn}
Arun KG, Iyer BR, Qusailah MSS, Sathyaprakash BS (2006{\natexlab{a}}) {Probing
  the non-linear structure of general relativity with black hole binaries}.
  Phys Rev D74:024006. \doi{10.1103/PhysRevD.74.024006}.
  {\href{https://arxiv.org/abs/gr-qc/0604067}{{arXiv:gr-qc/0604067}}} {[gr-qc]}

\bibitem[{Arun et~al.(2006{\natexlab{b}})Arun, Iyer, Qusailah, and
  Sathyaprakash}]{Arun:2006yw}
Arun KG, Iyer BR, Qusailah MSS, Sathyaprakash BS (2006{\natexlab{b}}) {Testing
  post-Newtonian theory with gravitational wave observations}. Class Quant Grav
  23:L37--L43. \doi{10.1088/0264-9381/23/9/L01}.
  {\href{https://arxiv.org/abs/gr-qc/0604018}{{arXiv:gr-qc/0604018}}} {[gr-qc]}

\bibitem[{Arvanitaki and Dubovsky(2011)}]{Arvanitaki:2010sy}
Arvanitaki A, Dubovsky S (2011) {Exploring the String Axiverse with Precision
  Black Hole Physics}. Phys Rev D 83:044026. \doi{10.1103/PhysRevD.83.044026}.
  {\href{https://arxiv.org/abs/1004.3558}{{arXiv:1004.3558}}} {[hep-th]}

\bibitem[{Arvanitaki et~al.(2010)Arvanitaki, Dimopoulos, Dubovsky, Kaloper, and
  March-Russell}]{Arvanitaki:2009fg}
Arvanitaki A, Dimopoulos S, Dubovsky S, Kaloper N, March-Russell J (2010)
  {String Axiverse}. Phys Rev D 81:123530. \doi{10.1103/PhysRevD.81.123530}.
  {\href{https://arxiv.org/abs/0905.4720}{{arXiv:0905.4720}}} {[hep-th]}

\bibitem[{Arvanitaki et~al.(2015)Arvanitaki, Baryakhtar, and
  Huang}]{Arvanitaki:2014wva}
Arvanitaki A, Baryakhtar M, Huang X (2015) {Discovering the QCD Axion with
  Black Holes and Gravitational Waves}. Phys Rev D 91(8):084011.
  \doi{10.1103/PhysRevD.91.084011}.
  {\href{https://arxiv.org/abs/1411.2263}{{arXiv:1411.2263}}} {[hep-ph]}

\bibitem[{Arvanitaki et~al.(2017)Arvanitaki, Baryakhtar, Dimopoulos, Dubovsky,
  and Lasenby}]{Arvanitaki:2016qwi}
Arvanitaki A, Baryakhtar M, Dimopoulos S, Dubovsky S, Lasenby R (2017) {Black
  Hole Mergers and the QCD Axion at Advanced LIGO}. Phys Rev D 95(4):043001.
  \doi{10.1103/PhysRevD.95.043001}.
  {\href{https://arxiv.org/abs/1604.03958}{{arXiv:1604.03958}}} {[hep-ph]}

\bibitem[{Arzoumanian et~al.(2018)}]{Arzoumanian:2018saf}
Arzoumanian Z, et~al. (2018) {The NANOGrav 11-year Data Set: Pulsar-timing
  Constraints On The Stochastic Gravitational-wave Background}. Astrophys J
  859(1):47. \doi{10.3847/1538-4357/aabd3b}.
  {\href{https://arxiv.org/abs/1801.02617}{{arXiv:1801.02617}}} {[astro-ph.HE]}

\bibitem[{Arzoumanian et~al.(2020)}]{Arzoumanian:2020vkk}
Arzoumanian Z, et~al. (2020) {The NANOGrav 12.5-year Data Set: Search For An
  Isotropic Stochastic Gravitational-Wave Background}
  {\href{https://arxiv.org/abs/2009.04496}{{arXiv:2009.04496}}} {[astro-ph.HE]}

\bibitem[{Ashtekar et~al.(1989)Ashtekar, Balachandran, and
  Jo}]{Ashtekar:1988sw}
Ashtekar A, Balachandran A, Jo S (1989) {The \{CP\} Problem in Quantum
  Gravity}. Int J Mod Phys A 4:1493. \doi{10.1142/S0217751X89000649}

\bibitem[{Auclair et~al.(2020)}]{Auclair:2019wcv}
Auclair P, et~al. (2020) {Probing the gravitational wave background from cosmic
  strings with LISA}. JCAP 04:034. \doi{10.1088/1475-7516/2020/04/034}.
  {\href{https://arxiv.org/abs/1909.00819}{{arXiv:1909.00819}}} {[astro-ph.CO]}

\bibitem[{Ayzenberg and Yunes(2014)}]{Ayzenberg:2014aka}
Ayzenberg D, Yunes N (2014) {Slowly-Rotating Black Holes in
  Einstein-Dilaton-Gauss-Bonnet Gravity: Quadratic Order in Spin Solutions}.
  Phys Rev D 90:044066. \doi{10.1103/PhysRevD.90.044066}, [Erratum: Phys.Rev.D
  91, 069905 (2015)].
  {\href{https://arxiv.org/abs/1405.2133}{{arXiv:1405.2133}}} {[gr-qc]}

\bibitem[{Babak et~al.(2017{\natexlab{a}})Babak, Gair, Sesana, Barausse,
  Sopuerta, Berry, Berti, Amaro-Seoane, Petiteau, and Klein}]{Babak:2017tow}
Babak S, Gair J, Sesana A, Barausse E, Sopuerta CF, Berry CP, Berti E,
  Amaro-Seoane P, Petiteau A, Klein A (2017{\natexlab{a}}) {Science with the
  space-based interferometer LISA. V: Extreme mass-ratio inspirals}. Phys Rev D
  95(10):103012. \doi{10.1103/PhysRevD.95.103012}.
  {\href{https://arxiv.org/abs/1703.09722}{{arXiv:1703.09722}}} {[gr-qc]}

\bibitem[{Babak et~al.(2017{\natexlab{b}})Babak, Taracchini, and
  Buonanno}]{Babak:2016tgq}
Babak S, Taracchini A, Buonanno A (2017{\natexlab{b}}) {Validating the
  effective-one-body model of spinning, precessing binary black holes against
  numerical relativity}. Phys Rev D 95(2):024010.
  \doi{10.1103/PhysRevD.95.024010}.
  {\href{https://arxiv.org/abs/1607.05661}{{arXiv:1607.05661}}} {[gr-qc]}

\bibitem[{Babichev and Charmousis(2014)}]{Babichev:2013cya}
Babichev E, Charmousis C (2014) {Dressing a black hole with a time-dependent
  Galileon}. JHEP 08:106. \doi{10.1007/JHEP08(2014)106}.
  {\href{https://arxiv.org/abs/1312.3204}{{arXiv:1312.3204}}} {[gr-qc]}

\bibitem[{Babichev and Crisostomi(2013)}]{Babichev:2013pfa}
Babichev E, Crisostomi M (2013) {Restoring general relativity in massive
  bigravity theory}. Phys Rev D88(8):084002. \doi{10.1103/PhysRevD.88.084002}.
  {\href{https://arxiv.org/abs/1307.3640}{{arXiv:1307.3640}}} {[gr-qc]}

\bibitem[{Babichev and Deffayet(2013)}]{Babichev:2013usa}
Babichev E, Deffayet C (2013) {An introduction to the Vainshtein mechanism}.
  Class Quant Grav 30:184001. \doi{10.1088/0264-9381/30/18/184001}.
  {\href{https://arxiv.org/abs/1304.7240}{{arXiv:1304.7240}}} {[gr-qc]}

\bibitem[{Babichev et~al.(2009)Babichev, Deffayet, and Ziour}]{Babichev:2009ee}
Babichev E, Deffayet C, Ziour R (2009) {k-Mouflage gravity}. Int J Mod Phys
  D18:2147--2154. \doi{10.1142/S0218271809016107}.
  {\href{https://arxiv.org/abs/0905.2943}{{arXiv:0905.2943}}} {[hep-th]}

\bibitem[{Babichev et~al.(2017)Babichev, Charmousis, and
  Hassaine}]{Babichev:2017rti}
Babichev E, Charmousis C, Hassaine M (2017) {Black holes and solitons in an
  extended Proca theory}. JHEP 05:114. \doi{10.1007/JHEP05(2017)114}.
  {\href{https://arxiv.org/abs/1703.07676}{{arXiv:1703.07676}}} {[gr-qc]}

\bibitem[{Bahamonde et~al.(2015)Bahamonde, B\"ohmer, and
  Wright}]{Bahamonde:2015zma}
Bahamonde S, B\"ohmer CG, Wright M (2015) {Modified teleparallel theories of
  gravity}. Phys Rev D 92(10):104042. \doi{10.1103/PhysRevD.92.104042}.
  {\href{https://arxiv.org/abs/1508.05120}{{arXiv:1508.05120}}} {[gr-qc]}

\bibitem[{Baibhav et~al.(2018)Baibhav, Berti, Cardoso, and
  Khanna}]{Baibhav:2017jhs}
Baibhav V, Berti E, Cardoso V, Khanna G (2018) {Black Hole Spectroscopy:
  Systematic Errors and Ringdown Energy Estimates}. Phys Rev D 97(4):044048.
  \doi{10.1103/PhysRevD.97.044048}.
  {\href{https://arxiv.org/abs/1710.02156}{{arXiv:1710.02156}}} {[gr-qc]}

\bibitem[{Baker et~al.(2019{\natexlab{a}})}]{Baker:2019ync}
Baker J, et~al. (2019{\natexlab{a}}) {High angular resolution gravitational
  wave astronomy} {\href{https://arxiv.org/abs/1908.11410}{{arXiv:1908.11410}}}
  {[astro-ph.HE]}

\bibitem[{Baker et~al.(2019{\natexlab{b}})}]{Baker:2019nct}
Baker J, et~al. (2019{\natexlab{b}}) {Multimessenger science opportunities with
  mHz gravitational waves}
  {\href{https://arxiv.org/abs/1903.04417}{{arXiv:1903.04417}}} {[astro-ph.HE]}

\bibitem[{Baker et~al.(2006)Baker, Centrella, Choi, Koppitz, and van
  Meter}]{Baker:2005vv}
Baker JG, Centrella J, Choi DI, Koppitz M, van Meter J (2006) {Gravitational
  wave extraction from an inspiraling configuration of merging black holes}.
  Phys Rev Lett 96:111102. \doi{10.1103/PhysRevLett.96.111102}.
  {\href{https://arxiv.org/abs/gr-qc/0511103}{{arXiv:gr-qc/0511103}}}

\bibitem[{Baker et~al.(2017)Baker, Bellini, Ferreira, Lagos, Noller, and
  Sawicki}]{Baker:2017hug}
Baker T, Bellini E, Ferreira PG, Lagos M, Noller J, Sawicki I (2017) {Strong
  constraints on cosmological gravity from GW170817 and GRB 170817A}. Phys Rev
  Lett 119:251301. \doi{10.1103/PhysRevLett.119.251301}.
  {\href{https://arxiv.org/abs/1710.06394}{{arXiv:1710.06394}}} {[astro-ph.CO]}

\bibitem[{Balasubramanian et~al.(2008)Balasubramanian, de~Boer, El-Showk, and
  Messamah}]{Balasubramanian:2008da}
Balasubramanian V, de~Boer J, El-Showk S, Messamah I (2008) {Black Holes as
  Effective Geometries}. Class Quant Grav 25:214004.
  \doi{10.1088/0264-9381/25/21/214004}.
  {\href{https://arxiv.org/abs/0811.0263}{{arXiv:0811.0263}}} {[hep-th]}

\bibitem[{Ballesteros et~al.(2018)Ballesteros, Serpico, and
  Taoso}]{Ballesteros:2018swv}
Ballesteros G, Serpico PD, Taoso M (2018) {On the merger rate of primordial
  black holes: effects of nearest neighbours distribution and clustering}. JCAP
  10:043. \doi{10.1088/1475-7516/2018/10/043}.
  {\href{https://arxiv.org/abs/1807.02084}{{arXiv:1807.02084}}} {[astro-ph.CO]}

\bibitem[{Balmelli and Jetzer(2013)}]{Balmelli:2013zna}
Balmelli S, Jetzer P (2013) {Effective-one-body Hamiltonian with
  next-to-leading order spin-spin coupling for two nonprecessing black holes
  with aligned spins}. Phys Rev D 87(12):124036.
  \doi{10.1103/PhysRevD.87.124036}, [Erratum: Phys.Rev.D 90, 089905 (2014)].
  {\href{https://arxiv.org/abs/1305.5674}{{arXiv:1305.5674}}} {[gr-qc]}

\bibitem[{Bamber et~al.(2021)Bamber, Clough, Ferreira, Hui, and
  Lagos}]{Bamber:2020bpu}
Bamber J, Clough K, Ferreira PG, Hui L, Lagos M (2021) {Growth of accretion
  driven scalar hair around Kerr black holes}. Phys Rev D 103(4):044059.
  \doi{10.1103/PhysRevD.103.044059}.
  {\href{https://arxiv.org/abs/2011.07870}{{arXiv:2011.07870}}} {[gr-qc]}

\bibitem[{Barack and Burko(2000)}]{Barack:2000zq}
Barack L, Burko LM (2000) {Radiation reaction force on a particle plunging into
  a black hole}. Phys Rev D 62:084040. \doi{10.1103/PhysRevD.62.084040}.
  {\href{https://arxiv.org/abs/gr-qc/0007033}{{arXiv:gr-qc/0007033}}}

\bibitem[{Barack and Cutler(2004)}]{Barack:2004wc}
Barack L, Cutler C (2004) {Confusion noise from LISA capture sources}. Phys Rev
  D 70:122002. \doi{10.1103/PhysRevD.70.122002}.
  {\href{https://arxiv.org/abs/gr-qc/0409010}{{arXiv:gr-qc/0409010}}}

\bibitem[{Barack and Cutler(2007)}]{Barack:2006pq}
Barack L, Cutler C (2007) {Using LISA EMRI sources to test off-Kerr deviations
  in the geometry of massive black holes}. Phys Rev D 75:042003.
  \doi{10.1103/PhysRevD.75.042003}.
  {\href{https://arxiv.org/abs/gr-qc/0612029}{{arXiv:gr-qc/0612029}}}

\bibitem[{Barack and Pound(2019)}]{Barack:2018yvs}
Barack L, Pound A (2019) {Self-force and radiation reaction in general
  relativity}. Rept Prog Phys 82(1):016904. \doi{10.1088/1361-6633/aae552}.
  {\href{https://arxiv.org/abs/1805.10385}{{arXiv:1805.10385}}} {[gr-qc]}

\bibitem[{Barack et~al.(2019)}]{Barack:2018yly}
Barack L, et~al. (2019) {Black holes, gravitational waves and fundamental
  physics: a roadmap}. Class Quant Grav 36(14):143001.
  \doi{10.1088/1361-6382/ab0587}.
  {\href{https://arxiv.org/abs/1806.05195}{{arXiv:1806.05195}}} {[gr-qc]}

\bibitem[{Barausse(2012)}]{Barausse:2012fy}
Barausse E (2012) {The evolution of massive black holes and their spins in
  their galactic hosts}. Mon Not Roy Astron Soc 423:2533--2557.
  \doi{10.1111/j.1365-2966.2012.21057.x}.
  {\href{https://arxiv.org/abs/1201.5888}{{arXiv:1201.5888}}} {[astro-ph.CO]}

\bibitem[{Barausse and Buonanno(2010)}]{Barausse:2009xi}
Barausse E, Buonanno A (2010) {An Improved effective-one-body Hamiltonian for
  spinning black-hole binaries}. Phys Rev D 81:084024.
  \doi{10.1103/PhysRevD.81.084024}.
  {\href{https://arxiv.org/abs/0912.3517}{{arXiv:0912.3517}}} {[gr-qc]}

\bibitem[{Barausse and Buonanno(2011)}]{Barausse:2011ys}
Barausse E, Buonanno A (2011) {Extending the effective-one-body Hamiltonian of
  black-hole binaries to include next-to-next-to-leading spin-orbit couplings}.
  Phys Rev D 84:104027. \doi{10.1103/PhysRevD.84.104027}.
  {\href{https://arxiv.org/abs/1107.2904}{{arXiv:1107.2904}}} {[gr-qc]}

\bibitem[{Barausse and Rezzolla(2008)}]{Barausse:2007dy}
Barausse E, Rezzolla L (2008) {The Influence of the hydrodynamic drag from an
  accretion torus on extreme mass-ratio inspirals}. Phys Rev D 77:104027.
  \doi{10.1103/PhysRevD.77.104027}.
  {\href{https://arxiv.org/abs/0711.4558}{{arXiv:0711.4558}}} {[gr-qc]}

\bibitem[{Barausse and Sotiriou(2008)}]{Barausse:2008xv}
Barausse E, Sotiriou TP (2008) {Perturbed Kerr Black Holes can probe deviations
  from General Relativity}. Phys Rev Lett 101:099001.
  \doi{10.1103/PhysRevLett.101.099001}.
  {\href{https://arxiv.org/abs/0803.3433}{{arXiv:0803.3433}}} {[gr-qc]}

\bibitem[{Barausse et~al.(2008)Barausse, Sotiriou, and
  Miller}]{Barausse:2007pn}
Barausse E, Sotiriou TP, Miller JC (2008) {A No-go theorem for polytropic
  spheres in Palatini f(R) gravity}. Class Quant Grav 25:062001.
  \doi{10.1088/0264-9381/25/6/062001}.
  {\href{https://arxiv.org/abs/gr-qc/0703132}{{arXiv:gr-qc/0703132}}}

\bibitem[{Barausse et~al.(2009)Barausse, Racine, and
  Buonanno}]{Barausse:2009aa}
Barausse E, Racine E, Buonanno A (2009) {Hamiltonian of a spinning
  test-particle in curved spacetime}. Phys Rev D 80:104025.
  \doi{10.1103/PhysRevD.85.069904}, [Erratum: Phys.Rev.D 85, 069904 (2012)].
  {\href{https://arxiv.org/abs/0907.4745}{{arXiv:0907.4745}}} {[gr-qc]}

\bibitem[{Barausse et~al.(2011)Barausse, Jacobson, and
  Sotiriou}]{Barausse:2011pu}
Barausse E, Jacobson T, Sotiriou TP (2011) {Black holes in Einstein-aether and
  Horava-Lifshitz gravity}. Phys Rev D 83:124043.
  \doi{10.1103/PhysRevD.83.124043}.
  {\href{https://arxiv.org/abs/1104.2889}{{arXiv:1104.2889}}} {[gr-qc]}

\bibitem[{Barausse et~al.(2014)Barausse, Cardoso, and Pani}]{Barausse:2014tra}
Barausse E, Cardoso V, Pani P (2014) {Can environmental effects spoil precision
  gravitational-wave astrophysics?} Phys Rev D 89(10):104059.
  \doi{10.1103/PhysRevD.89.104059}.
  {\href{https://arxiv.org/abs/1404.7149}{{arXiv:1404.7149}}} {[gr-qc]}

\bibitem[{Barausse et~al.(2015)Barausse, Cardoso, and Pani}]{Barausse:2014pra}
Barausse E, Cardoso V, Pani P (2015) {Environmental Effects for
  Gravitational-wave Astrophysics}. J Phys Conf Ser 610(1):012044.
  \doi{10.1088/1742-6596/610/1/012044}.
  {\href{https://arxiv.org/abs/1404.7140}{{arXiv:1404.7140}}} {[astro-ph.CO]}

\bibitem[{Barausse et~al.(2016)Barausse, Yunes, and
  Chamberlain}]{Barausse:2016eii}
Barausse E, Yunes N, Chamberlain K (2016) {Theory-Agnostic Constraints on
  Black-Hole Dipole Radiation with Multiband Gravitational-Wave Astrophysics}.
  Phys Rev Lett 116(24):241104. \doi{10.1103/PhysRevLett.116.241104}.
  {\href{https://arxiv.org/abs/1603.04075}{{arXiv:1603.04075}}} {[gr-qc]}

\bibitem[{Barausse et~al.(2018)Barausse, Brito, Cardoso, Dvorkin, and
  Pani}]{Barausse:2018vdb}
Barausse E, Brito R, Cardoso V, Dvorkin I, Pani P (2018) {The stochastic
  gravitational-wave background in the absence of horizons}. Class Quant Grav
  35(20):20LT01. \doi{10.1088/1361-6382/aae1de}.
  {\href{https://arxiv.org/abs/1805.08229}{{arXiv:1805.08229}}} {[gr-qc]}

\bibitem[{Barausse et~al.(2020)}]{Barausse:2020rsu}
Barausse E, et~al. (2020) {Prospects for Fundamental Physics with LISA}. Gen
  Rel Grav 52(8):81. \doi{10.1007/s10714-020-02691-1}.
  {\href{https://arxiv.org/abs/2001.09793}{{arXiv:2001.09793}}} {[gr-qc]}

\bibitem[{Barcel\'o et~al.(2009)Barcel\'o, Liberati, Sonego, and
  Visser}]{Barcelo:2009tpa}
Barcel\'o C, Liberati S, Sonego S, Visser M (2009) {Black Stars, Not Holes}.
  Sci Am 301(4):38--45. \doi{10.1038/scientificamerican1009-38}

\bibitem[{Barcel\'o et~al.(2017)Barcel\'o, Carballo-Rubio, and
  Garay}]{Barcelo:2017lnx}
Barcel\'o C, Carballo-Rubio R, Garay LJ (2017) {Gravitational wave echoes from
  macroscopic quantum gravity effects}. JHEP 05:054.
  \doi{10.1007/JHEP05(2017)054}.
  {\href{https://arxiv.org/abs/1701.09156}{{arXiv:1701.09156}}} {[gr-qc]}

\bibitem[{Barnacka et~al.(2012)Barnacka, Glicenstein, and
  Moderski}]{Barnacka:2012bm}
Barnacka A, Glicenstein J, Moderski R (2012) {New constraints on primordial
  black holes abundance from femtolensing of gamma-ray bursts}. Phys Rev D
  86:043001. \doi{10.1103/PhysRevD.86.043001}.
  {\href{https://arxiv.org/abs/1204.2056}{{arXiv:1204.2056}}} {[astro-ph.CO]}

\bibitem[{Barnich and Troessaert(2010)}]{BMS3c}
Barnich G, Troessaert C (2010) {Aspects of the BMS/CFT correspondence}. JHEP
  2010(05):062. \doi{10.1007/978-3-030-04260-8}.
  {\href{https://arxiv.org/abs/1001.1541}{{arXiv:1001.1541}}} {[hep-th]}

\bibitem[{Barranco et~al.(2012)Barranco, Bernal, Degollado, Diez-Tejedor,
  Megevand, Alcubierre, Nunez, and Sarbach}]{Barranco:2012qs}
Barranco J, Bernal A, Degollado JC, Diez-Tejedor A, Megevand M, Alcubierre M,
  Nunez D, Sarbach O (2012) {Schwarzschild black holes can wear scalar wigs}.
  Phys Rev Lett 109:081102. \doi{10.1103/PhysRevLett.109.081102}.
  {\href{https://arxiv.org/abs/1207.2153}{{arXiv:1207.2153}}} {[gr-qc]}

\bibitem[{Barranco et~al.(2014)Barranco, Bernal, Degollado, Diez-Tejedor,
  Megevand, Alcubierre, N\'u\~nez, and Sarbach}]{Barranco:2013rua}
Barranco J, Bernal A, Degollado JC, Diez-Tejedor A, Megevand M, Alcubierre M,
  N\'u\~nez D, Sarbach O (2014) {Schwarzschild scalar wigs: spectral analysis
  and late time behavior}. Phys Rev D 89(8):083006.
  \doi{10.1103/PhysRevD.89.083006}.
  {\href{https://arxiv.org/abs/1312.5808}{{arXiv:1312.5808}}} {[gr-qc]}

\bibitem[{Bartolo et~al.(2018)Bartolo, Domcke, Figueroa, García-Bellido,
  Peloso, Pieroni, Ricciardone, Sakellariadou, Sorbo, and
  Tasinato}]{Bartolo:2018qqn}
Bartolo N, Domcke V, Figueroa DG, García-Bellido J, Peloso M, Pieroni M,
  Ricciardone A, Sakellariadou M, Sorbo L, Tasinato G (2018) {Probing
  non-Gaussian Stochastic Gravitational Wave Backgrounds with LISA}. JCAP
  11:034. \doi{10.1088/1475-7516/2018/11/034}.
  {\href{https://arxiv.org/abs/1806.02819}{{arXiv:1806.02819}}} {[astro-ph.CO]}

\bibitem[{Bartolo et~al.(2019{\natexlab{a}})Bartolo, Bertacca, Matarrese,
  Peloso, Ricciardone, Riotto, and Tasinato}]{Bartolo:2019oiq}
Bartolo N, Bertacca D, Matarrese S, Peloso M, Ricciardone A, Riotto A, Tasinato
  G (2019{\natexlab{a}}) {Anisotropies and non-Gaussianity of the Cosmological
  Gravitational Wave Background}. Phys Rev D 100(12):121501.
  \doi{10.1103/PhysRevD.100.121501}.
  {\href{https://arxiv.org/abs/1908.00527}{{arXiv:1908.00527}}} {[astro-ph.CO]}

\bibitem[{Bartolo et~al.(2019{\natexlab{b}})Bartolo, Bertacca, Matarrese,
  Peloso, Ricciardone, Riotto, and Tasinato}]{Bartolo:2019yeu}
Bartolo N, Bertacca D, Matarrese S, Peloso M, Ricciardone A, Riotto A, Tasinato
  G (2019{\natexlab{b}}) {Characterizing the Cosmological Gravitational Wave
  Background Anisotropies and non-Gaussianity}
  {\href{https://arxiv.org/abs/1912.09433}{{arXiv:1912.09433}}} {[astro-ph.CO]}

\bibitem[{Bartolo et~al.(2019{\natexlab{c}})Bartolo, De~Luca, Franciolini,
  Lewis, Peloso, and Riotto}]{Bartolo:2018evs}
Bartolo N, De~Luca V, Franciolini G, Lewis A, Peloso M, Riotto A
  (2019{\natexlab{c}}) {Primordial Black Hole Dark Matter: LISA Serendipity}.
  Phys Rev Lett 122(21):211301. \doi{10.1103/PhysRevLett.122.211301}.
  {\href{https://arxiv.org/abs/1810.12218}{{arXiv:1810.12218}}} {[astro-ph.CO]}

\bibitem[{Bartolo et~al.(2019{\natexlab{d}})Bartolo, De~Luca, Franciolini,
  Peloso, Racco, and Riotto}]{Bartolo:2018rku}
Bartolo N, De~Luca V, Franciolini G, Peloso M, Racco D, Riotto A
  (2019{\natexlab{d}}) {Testing primordial black holes as dark matter with
  LISA}. Phys Rev D 99(10):103521. \doi{10.1103/PhysRevD.99.103521}.
  {\href{https://arxiv.org/abs/1810.12224}{{arXiv:1810.12224}}} {[astro-ph.CO]}

\bibitem[{Bartolo et~al.(2020)Bartolo, Bertacca, De~Luca, Franciolini,
  Matarrese, Peloso, Ricciardone, Riotto, and Tasinato}]{Bartolo:2019zvb}
Bartolo N, Bertacca D, De~Luca V, Franciolini G, Matarrese S, Peloso M,
  Ricciardone A, Riotto A, Tasinato G (2020) {Gravitational wave anisotropies
  from primordial black holes}. JCAP 02:028.
  \doi{10.1088/1475-7516/2020/02/028}.
  {\href{https://arxiv.org/abs/1909.12619}{{arXiv:1909.12619}}} {[astro-ph.CO]}

\bibitem[{Bartolo et~al.(2016)}]{Bartolo:2016ami}
Bartolo N, et~al. (2016) {Science with the space-based interferometer LISA. IV:
  Probing inflation with gravitational waves}. JCAP 12:026.
  \doi{10.1088/1475-7516/2016/12/026}.
  {\href{https://arxiv.org/abs/1610.06481}{{arXiv:1610.06481}}} {[astro-ph.CO]}

\bibitem[{Baryakhtar et~al.(2017)Baryakhtar, Lasenby, and
  Teo}]{Baryakhtar:2017ngi}
Baryakhtar M, Lasenby R, Teo M (2017) {Black Hole Superradiance Signatures of
  Ultralight Vectors}. Phys Rev D 96(3):035019.
  \doi{10.1103/PhysRevD.96.035019}.
  {\href{https://arxiv.org/abs/1704.05081}{{arXiv:1704.05081}}} {[hep-ph]}

\bibitem[{Battye et~al.(2018)Battye, Pace, and Trinh}]{Battye:2018ssx}
Battye RA, Pace F, Trinh D (2018) {Gravitational wave constraints on dark
  sector models}. Phys Rev D 98(2):023504. \doi{10.1103/PhysRevD.98.023504}.
  {\href{https://arxiv.org/abs/1802.09447}{{arXiv:1802.09447}}} {[astro-ph.CO]}

\bibitem[{Baumann et~al.(2007)Baumann, Steinhardt, Takahashi, and
  Ichiki}]{Baumann:2007zm}
Baumann D, Steinhardt PJ, Takahashi K, Ichiki K (2007) {Gravitational Wave
  Spectrum Induced by Primordial Scalar Perturbations}. Phys Rev D 76:084019.
  \doi{10.1103/PhysRevD.76.084019}.
  {\href{https://arxiv.org/abs/hep-th/0703290}{{arXiv:hep-th/0703290}}}

\bibitem[{Baumann et~al.(2019)Baumann, Chia, and Porto}]{Baumann:2018vus}
Baumann D, Chia HS, Porto RA (2019) {Probing Ultralight Bosons with Binary
  Black Holes}. Phys Rev D 99(4):044001. \doi{10.1103/PhysRevD.99.044001}.
  {\href{https://arxiv.org/abs/1804.03208}{{arXiv:1804.03208}}} {[gr-qc]}

\bibitem[{Baumann et~al.(2020)Baumann, Chia, Porto, and
  Stout}]{Baumann:2019ztm}
Baumann D, Chia HS, Porto RA, Stout J (2020) {Gravitational Collider Physics}.
  Phys Rev D 101(8):083019. \doi{10.1103/PhysRevD.101.083019}.
  {\href{https://arxiv.org/abs/1912.04932}{{arXiv:1912.04932}}} {[gr-qc]}

\bibitem[{Baumgarte and Shapiro(2010{\natexlab{a}})}]{BaumgarteBook2010}
Baumgarte TW, Shapiro SL (2010{\natexlab{a}}) Numerical Relativity: Solving
  Einstein's Equations on the Computer. Cambridge university Press, Cambridge

\bibitem[{Baumgarte and Shapiro(2010{\natexlab{b}})}]{baumgarteShapiroBook}
Baumgarte TW, Shapiro SL (2010{\natexlab{b}}) Numerical Relativity: Solving
  Einstein's Equations on the Computer. Cambridge University Press, Cambridge,
  England

\bibitem[{Bayin(1982)}]{Bayin:1982vw}
Bayin SS (1982) {Anisotropic Fluid Spheres in General Relativity}. Phys Rev D
  26:1262. \doi{10.1103/PhysRevD.26.1262}

\bibitem[{Begelman et~al.(1980)Begelman, Blandford, and Rees}]{Begelman:1980vb}
Begelman M, Blandford R, Rees M (1980) {Massive black hole binaries in active
  galactic nuclei}. Nature 287:307--309. \doi{10.1038/287307a0}

\bibitem[{Bekenstein(1972)}]{Bekenstein:1972ny}
Bekenstein J (1972) {Transcendence of the law of baryon-number conservation in
  black hole physics}. Phys Rev Lett 28:452--455.
  \doi{10.1103/PhysRevLett.28.452}

\bibitem[{Bekenstein(1974)}]{Bekenstein:1974jk}
Bekenstein JD (1974) {The quantum mass spectrum of the Kerr black hole}. Lett
  Nuovo Cim 11:467. \doi{10.1007/BF02762768}

\bibitem[{Bekenstein(1996)}]{Bekenstein:1996pn}
Bekenstein JD (1996) {Black hole hair: 25 - years after}. In: {2nd
  International Sakharov Conference on Physics}. pp 216--219.
  {\href{https://arxiv.org/abs/gr-qc/9605059}{{arXiv:gr-qc/9605059}}}

\bibitem[{Bekenstein and Mukhanov(1995)}]{Bekenstein:1995ju}
Bekenstein JD, Mukhanov VF (1995) {Spectroscopy of the quantum black hole}.
  Phys Lett B 360:7--12. \doi{10.1016/0370-2693(95)01148-J}.
  {\href{https://arxiv.org/abs/gr-qc/9505012}{{arXiv:gr-qc/9505012}}}

\bibitem[{Belgacem et~al.(2018{\natexlab{a}})Belgacem, Dirian, Foffa, and
  Maggiore}]{Belgacem:2018lbp}
Belgacem E, Dirian Y, Foffa S, Maggiore M (2018{\natexlab{a}}) {Modified
  gravitational-wave propagation and standard sirens}. Phys Rev D98:023510.
  \doi{10.1103/PhysRevD.98.023510}.
  {\href{https://arxiv.org/abs/1805.08731}{{arXiv:1805.08731}}} {[gr-qc]}

\bibitem[{Belgacem et~al.(2018{\natexlab{b}})Belgacem, Dirian, Foffa, and
  Maggiore}]{Belgacem:2017cqo}
Belgacem E, Dirian Y, Foffa S, Maggiore M (2018{\natexlab{b}}) {Nonlocal
  gravity. Conceptual aspects and cosmological predictions}. JCAP 1803:002.
  \doi{10.1088/1475-7516/2018/03/002}.
  {\href{https://arxiv.org/abs/1712.07066}{{arXiv:1712.07066}}} {[hep-th]}

\bibitem[{Belgacem et~al.(2018{\natexlab{c}})Belgacem, Dirian, Foffa, and
  Maggiore}]{Belgacem:2017ihm}
Belgacem E, Dirian Y, Foffa S, Maggiore M (2018{\natexlab{c}}) {The
  gravitational-wave luminosity distance in modified gravity theories}. Phys
  Rev D97:104066. {\href{https://arxiv.org/abs/1712.08108}{{arXiv:1712.08108}}}
  {[astro-ph.CO]}

\bibitem[{Belgacem et~al.(2019{\natexlab{a}})Belgacem, Dirian, Finke, Foffa,
  and Maggiore}]{Belgacem:2019lwx}
Belgacem E, Dirian Y, Finke A, Foffa S, Maggiore M (2019{\natexlab{a}})
  {Nonlocal gravity and gravitational-wave observations}. JCAP 1911:022.
  \doi{10.1088/1475-7516/2019/11/022}.
  {\href{https://arxiv.org/abs/1907.02047}{{arXiv:1907.02047}}} {[astro-ph.CO]}

\bibitem[{Belgacem et~al.(2019{\natexlab{b}})Belgacem, Finke, Frassino, and
  Maggiore}]{Belgacem:2018wtb}
Belgacem E, Finke A, Frassino A, Maggiore M (2019{\natexlab{b}}) {Testing
  nonlocal gravity with Lunar Laser Ranging}. JCAP 1902:035.
  \doi{10.1088/1475-7516/2019/02/035}.
  {\href{https://arxiv.org/abs/1812.11181}{{arXiv:1812.11181}}} {[gr-qc]}

\bibitem[{Belgacem et~al.(2020)Belgacem, Dirian, Finke, Foffa, and
  Maggiore}]{Belgacem:2020pdz}
Belgacem E, Dirian Y, Finke A, Foffa S, Maggiore M (2020) {Gravity in the
  infrared and effective nonlocal models}. JCAP 04:010.
  \doi{10.1088/1475-7516/2020/04/010}.
  {\href{https://arxiv.org/abs/2001.07619}{{arXiv:2001.07619}}} {[astro-ph.CO]}

\bibitem[{Belgacem et~al.(2019{\natexlab{c}})}]{Belgacem:2019pkk}
Belgacem E, et~al. (2019{\natexlab{c}}) {Testing modified gravity at
  cosmological distances with LISA standard sirens}. JCAP 07:024.
  \doi{10.1088/1475-7516/2019/07/024}.
  {\href{https://arxiv.org/abs/1906.01593}{{arXiv:1906.01593}}} {[astro-ph.CO]}

\bibitem[{Bellini and Sawicki(2014)}]{Bellini:2014fua}
Bellini E, Sawicki I (2014) {Maximal freedom at minimum cost: linear
  large-scale structure in general modifications of gravity}. JCAP 1407:050.
  \doi{10.1088/1475-7516/2014/07/050}.
  {\href{https://arxiv.org/abs/1404.3713}{{arXiv:1404.3713}}} {[astro-ph.CO]}

\bibitem[{Bellomo et~al.(2021)Bellomo, Bertacca, Jenkins, Matarrese,
  Raccanelli, Regimbau, Ricciardone, and Sakellariadou}]{Bellomo:2021mer}
Bellomo N, Bertacca D, Jenkins AC, Matarrese S, Raccanelli A, Regimbau T,
  Ricciardone A, Sakellariadou M (2021) {CLASS\_GWB: robust modeling of the
  astrophysical gravitational wave background anisotropies}
  {\href{https://arxiv.org/abs/2110.15059}{{arXiv:2110.15059}}} {[gr-qc]}

\bibitem[{Beltran~Jimenez et~al.(2016)Beltran~Jimenez, Piazza, and
  Velten}]{Jimenez:2015bwa}
Beltran~Jimenez J, Piazza F, Velten H (2016) {Evading the Vainshtein Mechanism
  with Anomalous Gravitational Wave Speed: Constraints on Modified Gravity from
  Binary Pulsars}. Phys Rev Lett 116(6):061101.
  \doi{10.1103/PhysRevLett.116.061101}.
  {\href{https://arxiv.org/abs/1507.05047}{{arXiv:1507.05047}}} {[gr-qc]}

\bibitem[{Ben~Achour et~al.(2016)Ben~Achour, Crisostomi, Koyama, Langlois,
  Noui, and Tasinato}]{BenAchour:2016fzp}
Ben~Achour J, Crisostomi M, Koyama K, Langlois D, Noui K, Tasinato G (2016)
  {Degenerate higher order scalar-tensor theories beyond Horndeski up to cubic
  order}. JHEP 12:100. \doi{10.1007/JHEP12(2016)100}.
  {\href{https://arxiv.org/abs/1608.08135}{{arXiv:1608.08135}}} {[hep-th]}

\bibitem[{Bena and Mayerson(2020)}]{Bena:2020see}
Bena I, Mayerson DR (2020) {Multipole Ratios: A New Window into Black Holes}.
  Phys Rev Lett 125(22):22. \doi{10.1103/PhysRevLett.125.221602}.
  {\href{https://arxiv.org/abs/2006.10750}{{arXiv:2006.10750}}} {[hep-th]}

\bibitem[{Bena and Mayerson(2021)}]{Bena:2020uup}
Bena I, Mayerson DR (2021) {Black Holes Lessons from Multipole Ratios}. JHEP
  03:114. \doi{10.1007/JHEP03(2021)114}.
  {\href{https://arxiv.org/abs/2007.09152}{{arXiv:2007.09152}}} {[hep-th]}

\bibitem[{Bena and Warner(2008)}]{Bena:2007kg}
Bena I, Warner NP (2008) {Black holes, black rings and their microstates}. Lect
  Notes Phys 755:1--92. \doi{10.1007/978-3-540-79523-0_1}.
  {\href{https://arxiv.org/abs/hep-th/0701216}{{arXiv:hep-th/0701216}}}

\bibitem[{Bena and Warner(2013)}]{Bena:2013dka}
Bena I, Warner NP (2013) {Resolving the Structure of Black Holes:
  Philosophizing with a Hammer}
  {\href{https://arxiv.org/abs/1311.4538}{{arXiv:1311.4538}}} {[hep-th]}

\bibitem[{Bena et~al.(2019{\natexlab{a}})Bena, Heidmann, Monten, and
  Warner}]{Bena:2019azk}
Bena I, Heidmann P, Monten R, Warner NP (2019{\natexlab{a}}) {Thermal Decay
  without Information Loss in Horizonless Microstate Geometries}. SciPost Phys
  7(5):063. \doi{10.21468/SciPostPhys.7.5.063}.
  {\href{https://arxiv.org/abs/1905.05194}{{arXiv:1905.05194}}} {[hep-th]}

\bibitem[{Bena et~al.(2019{\natexlab{b}})Bena, Martinec, Walker, and
  Warner}]{Bena:2018mpb}
Bena I, Martinec EJ, Walker R, Warner NP (2019{\natexlab{b}}) {Early Scrambling
  and Capped BTZ Geometries}. JHEP 04:126. \doi{10.1007/JHEP04(2019)126}.
  {\href{https://arxiv.org/abs/1812.05110}{{arXiv:1812.05110}}} {[hep-th]}

\bibitem[{Bender and Orszag(1978)}]{MR538168}
Bender CM, Orszag SA (1978) Advanced mathematical methods for scientists and
  engineers. McGraw-Hill Book Co., New York. International Series in Pure and
  Applied Mathematics

\bibitem[{Benkel et~al.(2016)Benkel, Sotiriou, and Witek}]{Benkel:2016kcq}
Benkel R, Sotiriou TP, Witek H (2016) {Dynamical scalar hair formation around a
  Schwarzschild black hole}. Phys Rev D 94(12):121503.
  \doi{10.1103/PhysRevD.94.121503}.
  {\href{https://arxiv.org/abs/1612.08184}{{arXiv:1612.08184}}} {[gr-qc]}

\bibitem[{Benkel et~al.(2017)Benkel, Sotiriou, and Witek}]{Benkel:2016rlz}
Benkel R, Sotiriou TP, Witek H (2017) {Black hole hair formation in
  shift-symmetric generalised scalar-tensor gravity}. Class Quant Grav
  34(6):064001. \doi{10.1088/1361-6382/aa5ce7}.
  {\href{https://arxiv.org/abs/1610.09168}{{arXiv:1610.09168}}} {[gr-qc]}

\bibitem[{Bera et~al.(2020)Bera, Rana, More, and Bose}]{Bera:2020jhx}
Bera S, Rana D, More S, Bose S (2020) {Incompleteness be damned: Inference of
  $H_0$ from BBH-galaxy cross-correlations}
  {\href{https://arxiv.org/abs/2007.04271}{{arXiv:2007.04271}}} {[astro-ph.CO]}

\bibitem[{Berezhiani et~al.(2012)Berezhiani, Chkareuli, de~Rham, Gabadadze, and
  Tolley}]{Berezhiani:2011mt}
Berezhiani L, Chkareuli G, de~Rham C, Gabadadze G, Tolley A (2012) {On Black
  Holes in Massive Gravity}. Phys Rev D 85:044024.
  \doi{10.1103/PhysRevD.85.044024}.
  {\href{https://arxiv.org/abs/1111.3613}{{arXiv:1111.3613}}} {[hep-th]}

\bibitem[{Bernuzzi et~al.(2012{\natexlab{a}})Bernuzzi, Nagar, Thierfelder, and
  Brugmann}]{Bernuzzi:2012ci}
Bernuzzi S, Nagar A, Thierfelder M, Brugmann B (2012{\natexlab{a}}) {Tidal
  effects in binary neutron star coalescence}. Phys Rev D 86:044030.
  \doi{10.1103/PhysRevD.86.044030}.
  {\href{https://arxiv.org/abs/1205.3403}{{arXiv:1205.3403}}} {[gr-qc]}

\bibitem[{Bernuzzi et~al.(2012{\natexlab{b}})Bernuzzi, Nagar, and
  Zenginoglu}]{Bernuzzi:2012ku}
Bernuzzi S, Nagar A, Zenginoglu A (2012{\natexlab{b}}) {Horizon-absorption
  effects in coalescing black-hole binaries: An effective-one-body study of the
  non-spinning case}. Phys Rev D 86:104038. \doi{10.1103/PhysRevD.86.104038}.
  {\href{https://arxiv.org/abs/1207.0769}{{arXiv:1207.0769}}} {[gr-qc]}

\bibitem[{Bernuzzi et~al.(2015{\natexlab{a}})Bernuzzi, Dietrich, and
  Nagar}]{Bernuzzi:2015rla}
Bernuzzi S, Dietrich T, Nagar A (2015{\natexlab{a}}) {Modeling the complete
  gravitational wave spectrum of neutron star mergers}. Phys Rev Lett
  115(9):091101. \doi{10.1103/PhysRevLett.115.091101}.
  {\href{https://arxiv.org/abs/1504.01764}{{arXiv:1504.01764}}} {[gr-qc]}

\bibitem[{Bernuzzi et~al.(2015{\natexlab{b}})Bernuzzi, Nagar, Dietrich, and
  Damour}]{Bernuzzi:2014owa}
Bernuzzi S, Nagar A, Dietrich T, Damour T (2015{\natexlab{b}}) {Modeling the
  Dynamics of Tidally Interacting Binary Neutron Stars up to the Merger}. Phys
  Rev Lett 114(16):161103. \doi{10.1103/PhysRevLett.114.161103}.
  {\href{https://arxiv.org/abs/1412.4553}{{arXiv:1412.4553}}} {[gr-qc]}

\bibitem[{Bertacca et~al.(2020)Bertacca, Ricciardone, Bellomo, Jenkins,
  Matarrese, Raccanelli, Regimbau, and Sakellariadou}]{Bertacca:2019fnt}
Bertacca D, Ricciardone A, Bellomo N, Jenkins AC, Matarrese S, Raccanelli A,
  Regimbau T, Sakellariadou M (2020) {Projection effects on the observed
  angular spectrum of the astrophysical stochastic gravitational wave
  background}. Phys Rev D 101(10):103513. \doi{10.1103/PhysRevD.101.103513}.
  {\href{https://arxiv.org/abs/1909.11627}{{arXiv:1909.11627}}} {[astro-ph.CO]}

\bibitem[{{Bertacca} et~al.(2020){Bertacca}, {Ricciardone}, {Bellomo},
  {Jenkins}, {Matarrese}, {Raccanelli}, {Regimbau}, and
  {Sakellariadou}}]{Bertacca2020PhRvD}
{Bertacca} D, {Ricciardone} A, {Bellomo} N, {Jenkins} AC, {Matarrese} S,
  {Raccanelli} A, {Regimbau} T, {Sakellariadou} M (2020) {Projection effects on
  the observed angular spectrum of the astrophysical stochastic gravitational
  wave background}. \prd 101(10):103513. \doi{10.1103/PhysRevD.101.103513}.
  {\href{https://arxiv.org/abs/1909.11627}{{arXiv:1909.11627}}} {[astro-ph.CO]}

\bibitem[{Berti and Volonteri(2008)}]{Berti:2008af}
Berti E, Volonteri M (2008) {Cosmological black hole spin evolution by mergers
  and accretion}. Astrophys J 684:822--828. \doi{10.1086/590379}.
  {\href{https://arxiv.org/abs/0802.0025}{{arXiv:0802.0025}}} {[astro-ph]}

\bibitem[{Berti et~al.(2005)Berti, Buonanno, and Will}]{Berti:2004bd}
Berti E, Buonanno A, Will CM (2005) {Estimating spinning binary parameters and
  testing alternative theories of gravity with LISA}. Phys Rev D71:084025.
  \doi{10.1103/PhysRevD.71.084025}.
  {\href{https://arxiv.org/abs/gr-qc/0411129}{{arXiv:gr-qc/0411129}}} {[gr-qc]}

\bibitem[{Berti et~al.(2006)Berti, Cardoso, and Will}]{Berti:2005ys}
Berti E, Cardoso V, Will CM (2006) {On gravitational-wave spectroscopy of
  massive black holes with the space interferometer LISA}. Phys Rev D
  73:064030. \doi{10.1103/PhysRevD.73.064030}.
  {\href{https://arxiv.org/abs/gr-qc/0512160}{{arXiv:gr-qc/0512160}}}

\bibitem[{Berti et~al.(2007)Berti, Cardoso, Gonzalez, Sperhake, Hannam, Husa,
  and Bruegmann}]{Berti:2007fi}
Berti E, Cardoso V, Gonzalez JA, Sperhake U, Hannam M, Husa S, Bruegmann B
  (2007) {Inspiral, merger and ringdown of unequal mass black hole binaries: A
  Multipolar analysis}. Phys Rev D 76:064034. \doi{10.1103/PhysRevD.76.064034}.
  {\href{https://arxiv.org/abs/gr-qc/0703053}{{arXiv:gr-qc/0703053}}}

\bibitem[{Berti et~al.(2009)Berti, Cardoso, and Starinets}]{Berti:2009kk}
Berti E, Cardoso V, Starinets AO (2009) {Quasinormal modes of black holes and
  black branes}. Class Quant Grav 26:163001.
  \doi{10.1088/0264-9381/26/16/163001}.
  {\href{https://arxiv.org/abs/0905.2975}{{arXiv:0905.2975}}} {[gr-qc]}

\bibitem[{Berti et~al.(2011)Berti, Gair, and Sesana}]{Berti:2011jz}
Berti E, Gair J, Sesana A (2011) {Graviton mass bounds from space-based
  gravitational-wave observations of massive black hole populations}. Phys Rev
  D84:101501. \doi{10.1103/PhysRevD.84.101501}.
  {\href{https://arxiv.org/abs/1107.3528}{{arXiv:1107.3528}}} {[gr-qc]}

\bibitem[{Berti et~al.(2013)Berti, Cardoso, Gualtieri, Horbatsch, and
  Sperhake}]{Berti:2013gfa}
Berti E, Cardoso V, Gualtieri L, Horbatsch M, Sperhake U (2013) {Numerical
  simulations of single and binary black holes in scalar-tensor theories:
  circumventing the no-hair theorem}. Phys Rev D 87(12):124020.
  \doi{10.1103/PhysRevD.87.124020}.
  {\href{https://arxiv.org/abs/1304.2836}{{arXiv:1304.2836}}} {[gr-qc]}

\bibitem[{Berti et~al.(2016)Berti, Sesana, Barausse, Cardoso, and
  Belczynski}]{Berti:2016lat}
Berti E, Sesana A, Barausse E, Cardoso V, Belczynski K (2016) {Spectroscopy of
  Kerr black holes with Earth- and space-based interferometers}. Phys Rev Lett
  117(10):101102. \doi{10.1103/PhysRevLett.117.101102}.
  {\href{https://arxiv.org/abs/1605.09286}{{arXiv:1605.09286}}} {[gr-qc]}

\bibitem[{Berti et~al.(2018)Berti, Yagi, and Yunes}]{Berti:2018cxi}
Berti E, Yagi K, Yunes N (2018) {Extreme Gravity Tests with Gravitational Waves
  from Compact Binary Coalescences: (I) Inspiral-Merger}. Gen Rel Grav
  50(4):46. \doi{10.1007/s10714-018-2362-8}.
  {\href{https://arxiv.org/abs/1801.03208}{{arXiv:1801.03208}}} {[gr-qc]}

\bibitem[{Berti et~al.(2019)Berti, Brito, Macedo, Raposo, and
  Rosa}]{Berti:2019wnn}
Berti E, Brito R, Macedo CF, Raposo G, Rosa JL (2019) {Ultralight boson cloud
  depletion in binary systems}. Phys Rev D 99(10):104039.
  \doi{10.1103/PhysRevD.99.104039}.
  {\href{https://arxiv.org/abs/1904.03131}{{arXiv:1904.03131}}} {[gr-qc]}

\bibitem[{Berti et~al.(2020)Berti, Collodel, Kleihaus, and
  Kunz}]{Berti:2020kgk}
Berti E, Collodel LG, Kleihaus B, Kunz J (2020) {Spin-induced black-hole
  scalarization in Einstein-scalar-Gauss-Bonnet theory}
  {\href{https://arxiv.org/abs/2009.03905}{{arXiv:2009.03905}}} {[gr-qc]}

\bibitem[{Berti et~al.(2015)}]{Berti:2015itd}
Berti E, et~al. (2015) {Testing General Relativity with Present and Future
  Astrophysical Observations}. Class Quant Grav 32:243001.
  \doi{10.1088/0264-9381/32/24/243001}.
  {\href{https://arxiv.org/abs/1501.07274}{{arXiv:1501.07274}}} {[gr-qc]}

\bibitem[{Bertone and Hooper(2018)}]{Bertone:2016nfn}
Bertone G, Hooper D (2018) {History of dark matter}. Rev Mod Phys 90(4):045002.
  \doi{10.1103/RevModPhys.90.045002}.
  {\href{https://arxiv.org/abs/1605.04909}{{arXiv:1605.04909}}} {[astro-ph.CO]}

\bibitem[{Bertone and Merritt(2005)}]{Bertone:2005hw}
Bertone G, Merritt D (2005) {Time-dependent models for dark matter at the
  Galactic Center}. Phys Rev D 72:103502. \doi{10.1103/PhysRevD.72.103502}.
  {\href{https://arxiv.org/abs/astro-ph/0501555}{{arXiv:astro-ph/0501555}}}

\bibitem[{Bertone and Tait(2018)}]{Bertone:2018krk}
Bertone G, Tait M Tim (2018) {A new era in the search for dark matter}. Nature
  562(7725):51--56. \doi{10.1038/s41586-018-0542-z}.
  {\href{https://arxiv.org/abs/1810.01668}{{arXiv:1810.01668}}} {[astro-ph.CO]}

\bibitem[{Bertone et~al.(2005{\natexlab{a}})Bertone, Hooper, and
  Silk}]{Bertone:2004pz}
Bertone G, Hooper D, Silk J (2005{\natexlab{a}}) {Particle dark matter:
  Evidence, candidates and constraints}. Phys Rept 405:279--390.
  \doi{10.1016/j.physrep.2004.08.031}.
  {\href{https://arxiv.org/abs/hep-ph/0404175}{{arXiv:hep-ph/0404175}}}

\bibitem[{Bertone et~al.(2005{\natexlab{b}})Bertone, Zentner, and
  Silk}]{Bertone:2005xz}
Bertone G, Zentner AR, Silk J (2005{\natexlab{b}}) {A new signature of dark
  matter annihilations: gamma-rays from intermediate-mass black holes}. Phys
  Rev D 72:103517. \doi{10.1103/PhysRevD.72.103517}.
  {\href{https://arxiv.org/abs/astro-ph/0509565}{{arXiv:astro-ph/0509565}}}

\bibitem[{Bertone et~al.(2019)}]{Bertone:2019irm}
Bertone G, et~al. (2019) {Gravitational wave probes of dark matter: challenges
  and opportunities}
  {\href{https://arxiv.org/abs/1907.10610}{{arXiv:1907.10610}}} {[astro-ph.CO]}

\bibitem[{Bertotti et~al.(2003)Bertotti, Iess, and Tortora}]{Bertotti:2003rm}
Bertotti B, Iess L, Tortora P (2003) {A test of general relativity using radio
  links with the Cassini spacecraft}. Nature 425:374--376.
  \doi{10.1038/nature01997}

\bibitem[{Bettoni et~al.(2017)Bettoni, Ezquiaga, Hinterbichler, and
  Zumalac\'arregui}]{Bettoni:2016mij}
Bettoni D, Ezquiaga JM, Hinterbichler K, Zumalac\'arregui M (2017) {Speed of
  Gravitational Waves and the Fate of Scalar-Tensor Gravity}. Phys Rev D
  95(8):084029. \doi{10.1103/PhysRevD.95.084029}.
  {\href{https://arxiv.org/abs/1608.01982}{{arXiv:1608.01982}}} {[gr-qc]}

\bibitem[{Bezares et~al.(2017)Bezares, Palenzuela, and Bona}]{Bezares:2017mzk}
Bezares M, Palenzuela C, Bona C (2017) {Final fate of compact boson star
  mergers}. Phys Rev D 95(12):124005. \doi{10.1103/PhysRevD.95.124005}.
  {\href{https://arxiv.org/abs/1705.01071}{{arXiv:1705.01071}}} {[gr-qc]}

\bibitem[{Bezares et~al.(2021{\natexlab{a}})Bezares, Aguilera-Miret, ter Haar,
  Crisostomi, Palenzuela, and Barausse}]{Bezares:2021dma}
Bezares M, Aguilera-Miret R, ter Haar L, Crisostomi M, Palenzuela C, Barausse E
  (2021{\natexlab{a}}) {No evidence of kinetic screening in merging binary
  neutron stars} {\href{https://arxiv.org/abs/2107.05648}{{arXiv:2107.05648}}}
  {[gr-qc]}

\bibitem[{Bezares et~al.(2021{\natexlab{b}})Bezares, ter Haar, Crisostomi,
  Barausse, and Palenzuela}]{Bezares:2021yek}
Bezares M, ter Haar L, Crisostomi M, Barausse E, Palenzuela C
  (2021{\natexlab{b}}) {Kinetic screening in nonlinear stellar oscillations and
  gravitational collapse}. Phys Rev D 104(4):044022.
  \doi{10.1103/PhysRevD.104.044022}.
  {\href{https://arxiv.org/abs/2105.13992}{{arXiv:2105.13992}}} {[gr-qc]}

\bibitem[{Bhagwat et~al.(2018)Bhagwat, Okounkova, Ballmer, Brown, Giesler,
  Scheel, and Teukolsky}]{Bhagwat:2017tkm}
Bhagwat S, Okounkova M, Ballmer SW, Brown DA, Giesler M, Scheel MA, Teukolsky
  SA (2018) {On choosing the start time of binary black hole ringdowns}. Phys
  Rev D 97(10):104065. \doi{10.1103/PhysRevD.97.104065}.
  {\href{https://arxiv.org/abs/1711.00926}{{arXiv:1711.00926}}} {[gr-qc]}

\bibitem[{Bhagwat et~al.(2020)Bhagwat, Forteza, Pani, and
  Ferrari}]{Bhagwat:2019dtm}
Bhagwat S, Forteza XJ, Pani P, Ferrari V (2020) {Ringdown overtones, black hole
  spectroscopy, and no-hair theorem tests}. Phys Rev D 101(4):044033.
  \doi{10.1103/PhysRevD.101.044033}.
  {\href{https://arxiv.org/abs/1910.08708}{{arXiv:1910.08708}}} {[gr-qc]}

\bibitem[{Bhattacharyya et~al.(2016{\natexlab{a}})Bhattacharyya, Coates,
  Colombo, and Sotiriou}]{Bhattacharyya:2015uxt}
Bhattacharyya J, Coates A, Colombo M, Sotiriou TP (2016{\natexlab{a}})
  {Evolution and spherical collapse in Einstein-\AE{}ther theory and Ho\v{r}ava
  gravity}. Phys Rev D 93(6):064056. \doi{10.1103/PhysRevD.93.064056}.
  {\href{https://arxiv.org/abs/1512.04899}{{arXiv:1512.04899}}} {[gr-qc]}

\bibitem[{Bhattacharyya et~al.(2016{\natexlab{b}})Bhattacharyya, Colombo, and
  Sotiriou}]{Bhattacharyya:2015gwa}
Bhattacharyya J, Colombo M, Sotiriou TP (2016{\natexlab{b}}) {Causality and
  black holes in spacetimes with a preferred foliation}. Class Quant Grav
  33(23):235003. \doi{10.1088/0264-9381/33/23/235003}.
  {\href{https://arxiv.org/abs/1509.01558}{{arXiv:1509.01558}}} {[gr-qc]}

\bibitem[{Biagetti et~al.(2018)Biagetti, Franciolini, Kehagias, and
  Riotto}]{Biagetti:2018pjj}
Biagetti M, Franciolini G, Kehagias A, Riotto A (2018) {Primordial Black Holes
  from Inflation and Quantum Diffusion}. JCAP 07:032.
  \doi{10.1088/1475-7516/2018/07/032}.
  {\href{https://arxiv.org/abs/1804.07124}{{arXiv:1804.07124}}} {[astro-ph.CO]}

\bibitem[{Biagetti et~al.(2021)Biagetti, De~Luca, Franciolini, Kehagias, and
  Riotto}]{Biagetti:2021eep}
Biagetti M, De~Luca V, Franciolini G, Kehagias A, Riotto A (2021) {The
  Formation Probability of Primordial Black Holes}
  {\href{https://arxiv.org/abs/2105.07810}{{arXiv:2105.07810}}} {[astro-ph.CO]}

\bibitem[{Bianchi et~al.(2018)Bianchi, Consoli, and Morales}]{Bianchi:2017sds}
Bianchi M, Consoli D, Morales J (2018) {Probing Fuzzballs with Particles, Waves
  and Strings}. JHEP 06:157. \doi{10.1007/JHEP06(2018)157}.
  {\href{https://arxiv.org/abs/1711.10287}{{arXiv:1711.10287}}} {[hep-th]}

\bibitem[{Bianchi et~al.(2019)Bianchi, Consoli, Grillo, and
  Morales}]{Bianchi:2018kzy}
Bianchi M, Consoli D, Grillo A, Morales JF (2019) {The dark side of fuzzball
  geometries}. JHEP 05:126. \doi{10.1007/JHEP05(2019)126}.
  {\href{https://arxiv.org/abs/1811.02397}{{arXiv:1811.02397}}} {[hep-th]}

\bibitem[{Bianchi et~al.(2020{\natexlab{a}})Bianchi, Consoli, Grillo, Morales,
  Pani, and Raposo}]{Bianchi:2020bxa}
Bianchi M, Consoli D, Grillo A, Morales JF, Pani P, Raposo G
  (2020{\natexlab{a}}) {Distinguishing fuzzballs from black holes through their
  multipolar structure}. Phys Rev Lett 125(22):221601.
  \doi{10.1103/PhysRevLett.125.221601}.
  {\href{https://arxiv.org/abs/2007.01743}{{arXiv:2007.01743}}} {[hep-th]}

\bibitem[{Bianchi et~al.(2020{\natexlab{b}})Bianchi, Consoli, Grillo, Morales,
  Pani, and Raposo}]{Bianchi:2020miz}
Bianchi M, Consoli D, Grillo A, Morales JF, Pani P, Raposo G
  (2020{\natexlab{b}}) {The multipolar structure of fuzzballs}
  {\href{https://arxiv.org/abs/2008.01445}{{arXiv:2008.01445}}} {[hep-th]}

\bibitem[{Bianchi et~al.(2020{\natexlab{c}})Bianchi, Grillo, and
  Morales}]{Bianchi:2020des}
Bianchi M, Grillo A, Morales JF (2020{\natexlab{c}}) {Chaos at the rim of black
  hole and fuzzball shadows}. JHEP 05:078. \doi{10.1007/JHEP05(2020)078}.
  {\href{https://arxiv.org/abs/2002.05574}{{arXiv:2002.05574}}} {[hep-th]}

\bibitem[{Bini and Damour(2014)}]{Bini:2014zxa}
Bini D, Damour T (2014) {Gravitational self-force corrections to two-body tidal
  interactions and the effective one-body formalism}. Phys Rev D 90(12):124037.
  \doi{10.1103/PhysRevD.90.124037}.
  {\href{https://arxiv.org/abs/1409.6933}{{arXiv:1409.6933}}} {[gr-qc]}

\bibitem[{Bini et~al.(2012)Bini, Damour, and Faye}]{Bini:2012gu}
Bini D, Damour T, Faye G (2012) {Effective action approach to higher-order
  relativistic tidal interactions in binary systems and their effective one
  body description}. Phys Rev D 85:124034. \doi{10.1103/PhysRevD.85.124034}.
  {\href{https://arxiv.org/abs/1202.3565}{{arXiv:1202.3565}}} {[gr-qc]}

\bibitem[{Bini et~al.(2019)Bini, Damour, and Geralico}]{Bini:2019nra}
Bini D, Damour T, Geralico A (2019) {Novel approach to binary dynamics:
  application to the fifth post-Newtonian level}. Phys Rev Lett 123(23):231104.
  \doi{10.1103/PhysRevLett.123.231104}.
  {\href{https://arxiv.org/abs/1909.02375}{{arXiv:1909.02375}}} {[gr-qc]}

\bibitem[{Bini et~al.(2020{\natexlab{a}})Bini, Damour, and
  Geralico}]{Bini:2020wpo}
Bini D, Damour T, Geralico A (2020{\natexlab{a}}) {Binary dynamics at the fifth
  and fifth-and-a-half post-Newtonian orders}. Phys Rev D 102(2):024062.
  \doi{10.1103/PhysRevD.102.024062}.
  {\href{https://arxiv.org/abs/2003.11891}{{arXiv:2003.11891}}} {[gr-qc]}

\bibitem[{Bini et~al.(2020{\natexlab{b}})Bini, Damour, and
  Geralico}]{Bini:2020nsb}
Bini D, Damour T, Geralico A (2020{\natexlab{b}}) {Sixth post-Newtonian
  local-in-time dynamics of binary systems}. Phys Rev D 102(2):024061.
  \doi{10.1103/PhysRevD.102.024061}.
  {\href{https://arxiv.org/abs/2004.05407}{{arXiv:2004.05407}}} {[gr-qc]}

\bibitem[{Binnington and Poisson(2009)}]{Binnington:2009bb}
Binnington T, Poisson E (2009) {Relativistic theory of tidal Love numbers}.
  Phys Rev D 80:084018. \doi{10.1103/PhysRevD.80.084018}.
  {\href{https://arxiv.org/abs/0906.1366}{{arXiv:0906.1366}}} {[gr-qc]}

\bibitem[{Bizon(1990)}]{Bizon:1990sr}
Bizon P (1990) {Colored black holes}. Phys Rev Lett 64:2844--2847.
  \doi{10.1103/PhysRevLett.64.2844}

\bibitem[{Blaes et~al.(2002)Blaes, Lee, and Socrates}]{Blaes:2002cs}
Blaes O, Lee MH, Socrates A (2002) {The kozai mechanism and the evolution of
  binary supermassive black holes}. Astrophys J 578:775--786.
  \doi{10.1086/342655}.
  {\href{https://arxiv.org/abs/astro-ph/0203370}{{arXiv:astro-ph/0203370}}}

\bibitem[{Blagojevi\'c and Nester(2020)}]{Blagojevic:2020dyq}
Blagojevi\'c M, Nester JM (2020) {Local symmetries and physical degrees of
  freedom in $f(T)$ gravity: a Dirac Hamiltonian constraint analysis}. Phys Rev
  D 102(6):064025. \doi{10.1103/PhysRevD.102.064025}.
  {\href{https://arxiv.org/abs/2006.15303}{{arXiv:2006.15303}}} {[gr-qc]}

\bibitem[{Blanchet(2014)}]{Blanchet:2013haa}
Blanchet L (2014) {Gravitational Radiation from Post-Newtonian Sources and
  Inspiralling Compact Binaries}. Living Rev Rel 17:2.
  \doi{10.12942/lrr-2014-2}.
  {\href{https://arxiv.org/abs/1310.1528}{{arXiv:1310.1528}}} {[gr-qc]}

\bibitem[{Blanco-Pillado et~al.(2014)Blanco-Pillado, Olum, and
  Shlaer}]{Blanco-Pillado:2013qja}
Blanco-Pillado JJ, Olum KD, Shlaer B (2014) {The number of cosmic string
  loops}. Phys Rev D 89(2):023512. \doi{10.1103/PhysRevD.89.023512}.
  {\href{https://arxiv.org/abs/1309.6637}{{arXiv:1309.6637}}} {[astro-ph.CO]}

\bibitem[{Blas and Jenkins(2021{\natexlab{a}})}]{Blas:2021mqw}
Blas D, Jenkins AC (2021{\natexlab{a}}) {Bridging the $\mu$Hz gap in the
  gravitational-wave landscape with binary resonance}
  {\href{https://arxiv.org/abs/2107.04601}{{arXiv:2107.04601}}} {[astro-ph.CO]}

\bibitem[{Blas and Jenkins(2021{\natexlab{b}})}]{Blas:2021mpc}
Blas D, Jenkins AC (2021{\natexlab{b}}) {Detecting stochastic gravitational
  waves with binary resonance}
  {\href{https://arxiv.org/abs/2107.04063}{{arXiv:2107.04063}}} {[gr-qc]}

\bibitem[{Blas and Sanctuary(2011)}]{Blas:2011zd}
Blas D, Sanctuary H (2011) {Gravitational Radiation in Ho\v{r}ava Gravity}.
  Phys Rev D 84:064004. \doi{10.1103/PhysRevD.84.064004}.
  {\href{https://arxiv.org/abs/1105.5149}{{arXiv:1105.5149}}} {[gr-qc]}

\bibitem[{Blas and Sibiryakov(2011)}]{Blas:2011ni}
Blas D, Sibiryakov S (2011) {Horava gravity versus thermodynamics: The Black
  hole case}. Phys Rev D 84:124043. \doi{10.1103/PhysRevD.84.124043}.
  {\href{https://arxiv.org/abs/1110.2195}{{arXiv:1110.2195}}} {[hep-th]}

\bibitem[{Blas and Witte(2020{\natexlab{a}})}]{Blas:2020nbs}
Blas D, Witte SJ (2020{\natexlab{a}}) {Imprints of Axion Superradiance in the
  CMB}. Phys Rev D 102(10):103018. \doi{10.1103/PhysRevD.102.103018}.
  {\href{https://arxiv.org/abs/2009.10074}{{arXiv:2009.10074}}} {[astro-ph.CO]}

\bibitem[{Blas and Witte(2020{\natexlab{b}})}]{Blas:2020kaa}
Blas D, Witte SJ (2020{\natexlab{b}}) {Quenching Mechanisms of Photon
  Superradiance}. Phys Rev D 102:123018. \doi{10.1103/PhysRevD.102.123018}.
  {\href{https://arxiv.org/abs/2009.10075}{{arXiv:2009.10075}}} {[hep-ph]}

\bibitem[{Blas et~al.(2010)Blas, Pujolas, and Sibiryakov}]{Blas:2009qj}
Blas D, Pujolas O, Sibiryakov S (2010) {Consistent Extension of Horava
  Gravity}. Phys Rev Lett 104:181302. \doi{10.1103/PhysRevLett.104.181302}.
  {\href{https://arxiv.org/abs/0909.3525}{{arXiv:0909.3525}}} {[hep-th]}

\bibitem[{Blas et~al.(2016)Blas, Ivanov, Sawicki, and
  Sibiryakov}]{Blas:2016qmn}
Blas D, Ivanov MM, Sawicki I, Sibiryakov S (2016) {On constraining the speed of
  gravitational waves following GW150914}. JETP Lett 103(10):624--626.
  \doi{10.7868/S0370274X16100039}.
  {\href{https://arxiv.org/abs/1602.04188}{{arXiv:1602.04188}}} {[gr-qc]}

\bibitem[{Bl\'azquez-Salcedo et~al.(2018)Bl\'azquez-Salcedo, Doneva, Kunz, and
  Yazadjiev}]{Blazquez-Salcedo:2018jnn}
Bl\'azquez-Salcedo JL, Doneva DD, Kunz J, Yazadjiev SS (2018) {Radial
  perturbations of the scalarized Einstein-Gauss-Bonnet black holes}. Phys Rev
  D 98(8):084011. \doi{10.1103/PhysRevD.98.084011}.
  {\href{https://arxiv.org/abs/1805.05755}{{arXiv:1805.05755}}} {[gr-qc]}

\bibitem[{Bloomfield et~al.(2013)Bloomfield, Flanagan, Park, and
  Watson}]{Bloomfield:2012ff}
Bloomfield JK, Flanagan EE, Park M, Watson S (2013) {Dark energy or modified
  gravity? An effective field theory approach}. JCAP 1308:010.
  \doi{10.1088/1475-7516/2013/08/010}.
  {\href{https://arxiv.org/abs/1211.7054}{{arXiv:1211.7054}}} {[astro-ph.CO]}

\bibitem[{Blázquez-Salcedo et~al.(2016)Blázquez-Salcedo, Macedo, Cardoso,
  Ferrari, Gualtieri, Khoo, Kunz, and Pani}]{Blazquez-Salcedo:2016enn}
Blázquez-Salcedo JL, Macedo CFB, Cardoso V, Ferrari V, Gualtieri L, Khoo FS,
  Kunz J, Pani P (2016) {Perturbed black holes in Einstein-dilaton-Gauss-Bonnet
  gravity: Stability, ringdown, and gravitational-wave emission}. Phys Rev D
  94(10):104024. \doi{10.1103/PhysRevD.94.104024}.
  {\href{https://arxiv.org/abs/1609.01286}{{arXiv:1609.01286}}} {[gr-qc]}

\bibitem[{Boh\'e et~al.(2017)}]{Bohe:2016gbl}
Boh\'e A, et~al. (2017) {Improved effective-one-body model of spinning,
  nonprecessing binary black holes for the era of gravitational-wave
  astrophysics with advanced detectors}. Phys Rev D 95(4):044028.
  \doi{10.1103/PhysRevD.95.044028}.
  {\href{https://arxiv.org/abs/1611.03703}{{arXiv:1611.03703}}} {[gr-qc]}

\bibitem[{Boileau et~al.(2021)Boileau, Christensen, Meyer, and
  Cornish}]{Boileau:2020rpg}
Boileau G, Christensen N, Meyer R, Cornish NJ (2021) {Spectral separation of
  the stochastic gravitational-wave background for LISA: Observing both
  cosmological and astrophysical backgrounds}. Phys Rev D 103(10):103529.
  \doi{10.1103/PhysRevD.103.103529}.
  {\href{https://arxiv.org/abs/2011.05055}{{arXiv:2011.05055}}} {[gr-qc]}

\bibitem[{Boileau et~al.(2022)Boileau, Jenkins, Sakellariadou, Meyer, and
  Christensen}]{Boileau:2021gbr}
Boileau G, Jenkins AC, Sakellariadou M, Meyer R, Christensen N (2022) {Ability
  of LISA to detect a gravitational-wave background of cosmological origin: The
  cosmic string case}. Phys Rev D 105(2):023510.
  \doi{10.1103/PhysRevD.105.023510}.
  {\href{https://arxiv.org/abs/2109.06552}{{arXiv:2109.06552}}} {[gr-qc]}

\bibitem[{Bona et~al.(2009)Bona, Palenzuela-Luque, and
  Bona-Casas}]{BonaPalenzuelaBona2009}
Bona C, Palenzuela-Luque C, Bona-Casas C (2009) Elements of Numerical
  Relativity and Relativistic Hydrodynamics. Springer-Verlag Berlin Heidelberg,
  Heidelberg

\bibitem[{Bondi et~al.(1962)Bondi, van~der Burg, and Metzner}]{BMS1}
Bondi H, van~der Burg MGJ, Metzner AWK (1962) {Gravitational waves in general
  relativity. VII. waves from axi-symmetric isolated systems}. Proc R Soc Lond
  A 269:21. \doi{10.1098/rspa.1962.0161}

\bibitem[{Bonetti et~al.(2018)Bonetti, dos Santos~Filho, Helay\"el-Neto, and
  Spallicci}]{Bonetti:2017toa}
Bonetti L, dos Santos~Filho LR, Helay\"el-Neto JA, Spallicci ADAM (2018)
  {Photon sector analysis of Super and Lorentz symmetry breaking: effective
  photon mass, bi-refringence and dissipation}. Eur Phys J C 78(10):811.
  \doi{10.1140/epjc/s10052-018-6247-5}.
  {\href{https://arxiv.org/abs/1709.04995}{{arXiv:1709.04995}}} {[hep-th]}

\bibitem[{Bonetti et~al.(2019)Bonetti, Sesana, Haardt, Barausse, and
  Colpi}]{Bonetti:2018tpf}
Bonetti M, Sesana A, Haardt F, Barausse E, Colpi M (2019) {Post-Newtonian
  evolution of massive black hole triplets in galactic nuclei \textendash{} IV.
  Implications for LISA}. Mon Not Roy Astron Soc 486(3):4044--4060.
  \doi{10.1093/mnras/stz903}.
  {\href{https://arxiv.org/abs/1812.01011}{{arXiv:1812.01011}}} {[astro-ph.GA]}

\bibitem[{Borhanian et~al.(2020)Borhanian, Dhani, Gupta, Arun, and
  Sathyaprakash}]{Borhanian:2020vyr}
Borhanian S, Dhani A, Gupta A, Arun K, Sathyaprakash B (2020) {Dark Sirens to
  Resolve the Hubble-Lema\^itre Tension}
  {\href{https://arxiv.org/abs/2007.02883}{{arXiv:2007.02883}}} {[astro-ph.CO]}

\bibitem[{Boruah et~al.(2020)Boruah, Hudson, and Lavaux}]{Boruah:2020fhl}
Boruah SS, Hudson MJ, Lavaux G (2020) {Peculiar velocities in the local
  Universe: comparison of different models and the implications for $H_0$ and
  dark matter} {\href{https://arxiv.org/abs/2010.01119}{{arXiv:2010.01119}}}
  {[astro-ph.CO]}

\bibitem[{Bousso(2008)}]{Bousso:2007gp}
Bousso R (2008) {TASI Lectures on the Cosmological Constant}. Gen Rel Grav
  40:607--637. \doi{10.1007/s10714-007-0557-5}.
  {\href{https://arxiv.org/abs/0708.4231}{{arXiv:0708.4231}}} {[hep-th]}

\bibitem[{{Bowers} and {Liang}(1974)}]{1974ApJ...188..657B}
{Bowers} RL, {Liang} EPT (1974) {Anisotropic Spheres in General Relativity}.
  Astrophys J 188:657. \doi{10.1086/152760}

\bibitem[{Boyle et~al.(2019)}]{Boyle:2019kee}
Boyle M, et~al. (2019) {The SXS Collaboration catalog of binary black hole
  simulations}. Class Quant Grav 36(19):195006. \doi{10.1088/1361-6382/ab34e2}.
  {\href{https://arxiv.org/abs/1904.04831}{{arXiv:1904.04831}}} {[gr-qc]}

\bibitem[{Bozzola and Paschalidis(2021)}]{Bozzola:2020mjx}
Bozzola G, Paschalidis V (2021) {General Relativistic Simulations of the
  Quasicircular Inspiral and Merger of Charged Black Holes: GW150914 and
  Fundamental Physics Implications}. Phys Rev Lett 126(4):041103.
  \doi{10.1103/PhysRevLett.126.041103}.
  {\href{https://arxiv.org/abs/2006.15764}{{arXiv:2006.15764}}} {[gr-qc]}

\bibitem[{Braglia et~al.(2020)Braglia, Chen, and Hazra}]{Braglia:2020taf}
Braglia M, Chen X, Hazra DK (2020) {Probing Primordial Features with the
  Stochastic Gravitational Wave Background}
  {\href{https://arxiv.org/abs/2012.05821}{{arXiv:2012.05821}}} {[astro-ph.CO]}

\bibitem[{Brax et~al.(2016)Brax, Burrage, and Davis}]{Brax:2015dma}
Brax P, Burrage C, Davis AC (2016) {The Speed of Galileon Gravity}. JCAP
  03:004. \doi{10.1088/1475-7516/2016/03/004}.
  {\href{https://arxiv.org/abs/1510.03701}{{arXiv:1510.03701}}} {[gr-qc]}

\bibitem[{Brito et~al.(2013)Brito, Cardoso, and Pani}]{Brito:2013wya}
Brito R, Cardoso V, Pani P (2013) {Massive spin-2 fields on black hole
  spacetimes: Instability of the Schwarzschild and Kerr solutions and bounds on
  the graviton mass}. Phys Rev D 88(2):023514.
  \doi{10.1103/PhysRevD.88.023514}.
  {\href{https://arxiv.org/abs/1304.6725}{{arXiv:1304.6725}}} {[gr-qc]}

\bibitem[{Brito et~al.(2015{\natexlab{a}})Brito, Cardoso, and
  Pani}]{Brito:2014wla}
Brito R, Cardoso V, Pani P (2015{\natexlab{a}}) {Black holes as particle
  detectors: evolution of superradiant instabilities}. Class Quant Grav
  32(13):134001. \doi{10.1088/0264-9381/32/13/134001}.
  {\href{https://arxiv.org/abs/1411.0686}{{arXiv:1411.0686}}} {[gr-qc]}

\bibitem[{Brito et~al.(2015{\natexlab{b}})Brito, Cardoso, and
  Pani}]{Brito:2015oca}
Brito R, Cardoso V, Pani P (2015{\natexlab{b}}) {Superradiance}: {Energy
  Extraction, Black-Hole Bombs and Implications for Astrophysics and Particle
  Physics}, vol 906. Springer. \doi{10.1007/978-3-319-19000-6}.
  {\href{https://arxiv.org/abs/1501.06570}{{arXiv:1501.06570}}} {[gr-qc]}

\bibitem[{Brito et~al.(2016)Brito, Cardoso, Herdeiro, and Radu}]{Brito:2015pxa}
Brito R, Cardoso V, Herdeiro CAR, Radu E (2016) {Proca stars: Gravitating
  Bose\textendash{}Einstein condensates of massive spin 1 particles}. Phys Lett
  B 752:291--295. \doi{10.1016/j.physletb.2015.11.051}.
  {\href{https://arxiv.org/abs/1508.05395}{{arXiv:1508.05395}}} {[gr-qc]}

\bibitem[{Brito et~al.(2017{\natexlab{a}})Brito, Ghosh, Barausse, Berti,
  Cardoso, Dvorkin, Klein, and Pani}]{Brito:2017zvb}
Brito R, Ghosh S, Barausse E, Berti E, Cardoso V, Dvorkin I, Klein A, Pani P
  (2017{\natexlab{a}}) {Gravitational wave searches for ultralight bosons with
  LIGO and LISA}. Phys Rev D 96(6):064050. \doi{10.1103/PhysRevD.96.064050}.
  {\href{https://arxiv.org/abs/1706.06311}{{arXiv:1706.06311}}} {[gr-qc]}

\bibitem[{Brito et~al.(2017{\natexlab{b}})Brito, Ghosh, Barausse, Berti,
  Cardoso, Dvorkin, Klein, and Pani}]{Brito:2017wnc}
Brito R, Ghosh S, Barausse E, Berti E, Cardoso V, Dvorkin I, Klein A, Pani P
  (2017{\natexlab{b}}) {Stochastic and resolvable gravitational waves from
  ultralight bosons}. Phys Rev Lett 119(13):131101.
  \doi{10.1103/PhysRevLett.119.131101}.
  {\href{https://arxiv.org/abs/1706.05097}{{arXiv:1706.05097}}} {[gr-qc]}

\bibitem[{Brito et~al.(2018)Brito, Buonanno, and Raymond}]{Brito:2018rfr}
Brito R, Buonanno A, Raymond V (2018) {Black-hole Spectroscopy by Making Full
  Use of Gravitational-Wave Modeling}. Phys Rev D 98(8):084038.
  \doi{10.1103/PhysRevD.98.084038}.
  {\href{https://arxiv.org/abs/1805.00293}{{arXiv:1805.00293}}} {[gr-qc]}

\bibitem[{Brito et~al.(2020)Brito, Grillo, and Pani}]{Brito:2020lup}
Brito R, Grillo S, Pani P (2020) {Black Hole Superradiant Instability from
  Ultralight Spin-2 Fields}. Phys Rev Lett 124(21):211101.
  \doi{10.1103/PhysRevLett.124.211101}.
  {\href{https://arxiv.org/abs/2002.04055}{{arXiv:2002.04055}}} {[gr-qc]}

\bibitem[{Brustein and Medved(2017)}]{Brustein:2016msz}
Brustein R, Medved AJM (2017) {Black holes as collapsed polymers}. Fortsch Phys
  65(1):1600114. \doi{10.1002/prop.201600114}.
  {\href{https://arxiv.org/abs/1602.07706}{{arXiv:1602.07706}}} {[hep-th]}

\bibitem[{Brustein and Sherf(2020)}]{Brustein:2020tpg}
Brustein R, Sherf Y (2020) {Quantum Love}
  {\href{https://arxiv.org/abs/2008.02738}{{arXiv:2008.02738}}} {[gr-qc]}

\bibitem[{Brustein et~al.(2017)Brustein, Medved, and Yagi}]{Brustein:2017kcj}
Brustein R, Medved AJM, Yagi K (2017) {Discovering the interior of black
  holes}. Phys Rev D 96(12):124021. \doi{10.1103/PhysRevD.96.124021}.
  {\href{https://arxiv.org/abs/1701.07444}{{arXiv:1701.07444}}} {[gr-qc]}

\bibitem[{Bueno et~al.(2018)Bueno, Cano, Goelen, Hertog, and
  Vercnocke}]{Bueno:2017hyj}
Bueno P, Cano PA, Goelen F, Hertog T, Vercnocke B (2018) {Echoes of Kerr-like
  wormholes}. Phys Rev D 97(2):024040. \doi{10.1103/PhysRevD.97.024040}.
  {\href{https://arxiv.org/abs/1711.00391}{{arXiv:1711.00391}}} {[gr-qc]}

\bibitem[{Bugaev and Klimai(2010)}]{Bugaev:2009zh}
Bugaev E, Klimai P (2010) {Induced gravitational wave background and primordial
  black holes}. Phys Rev D 81:023517. \doi{10.1103/PhysRevD.81.023517}.
  {\href{https://arxiv.org/abs/0908.0664}{{arXiv:0908.0664}}} {[astro-ph.CO]}

\bibitem[{Bunting(1983)}]{buntingthesis}
Bunting GL (1983) Proof of the uniqueness conjecture for black holes. PhD
  thesis, University of New England

\bibitem[{Buonanno and Damour(1999)}]{Buonanno:1998gg}
Buonanno A, Damour T (1999) {Effective one-body approach to general
  relativistic two-body dynamics}. Phys Rev D 59:084006.
  \doi{10.1103/PhysRevD.59.084006}.
  {\href{https://arxiv.org/abs/gr-qc/9811091}{{arXiv:gr-qc/9811091}}}

\bibitem[{Buonanno and Damour(2000)}]{Buonanno:2000ef}
Buonanno A, Damour T (2000) {Transition from inspiral to plunge in binary black
  hole coalescences}. Phys Rev D 62:064015. \doi{10.1103/PhysRevD.62.064015}.
  {\href{https://arxiv.org/abs/gr-qc/0001013}{{arXiv:gr-qc/0001013}}}

\bibitem[{Buonanno et~al.(2005)Buonanno, Sigl, Raffelt, Janka, and
  Muller}]{Buonanno:2004tp}
Buonanno A, Sigl G, Raffelt GG, Janka HT, Muller E (2005) {Stochastic
  gravitational wave background from cosmological supernovae}. Phys Rev D
  72:084001. \doi{10.1103/PhysRevD.72.084001}.
  {\href{https://arxiv.org/abs/astro-ph/0412277}{{arXiv:astro-ph/0412277}}}

\bibitem[{Buonanno et~al.(2007)Buonanno, Cook, and Pretorius}]{Buonanno:2006ui}
Buonanno A, Cook GB, Pretorius F (2007) {Inspiral, merger and ring-down of
  equal-mass black-hole binaries}. Phys Rev D 75:124018.
  \doi{10.1103/PhysRevD.75.124018}.
  {\href{https://arxiv.org/abs/gr-qc/0610122}{{arXiv:gr-qc/0610122}}}

\bibitem[{Buoninfante and Mazumdar(2019)}]{Buoninfante:2019swn}
Buoninfante L, Mazumdar A (2019) {Nonlocal star as a blackhole mimicker}. Phys
  Rev D 100(2):024031. \doi{10.1103/PhysRevD.100.024031}.
  {\href{https://arxiv.org/abs/1903.01542}{{arXiv:1903.01542}}} {[gr-qc]}

\bibitem[{Buoninfante et~al.(2019)Buoninfante, Mazumdar, and
  Peng}]{Buoninfante:2019teo}
Buoninfante L, Mazumdar A, Peng J (2019) {Nonlocality amplifies echoes}. Phys
  Rev D 100(10):104059. \doi{10.1103/PhysRevD.100.104059}.
  {\href{https://arxiv.org/abs/1906.03624}{{arXiv:1906.03624}}} {[gr-qc]}

\bibitem[{Burgess(2015)}]{Burgess:2013ara}
Burgess C (2015) {The Cosmological Constant Problem: Why it's hard to get Dark
  Energy from Micro-physics}. In: {100e Ecole d'Ete de Physique: Post-Planck
  Cosmology}. pp 149--197. \doi{10.1093/acprof:oso/9780198728856.003.0004}.
  {\href{https://arxiv.org/abs/1309.4133}{{arXiv:1309.4133}}} {[hep-th]}

\bibitem[{Burgess et~al.(2018)Burgess, Plestid, and Rummel}]{Burgess:2018pmm}
Burgess CP, Plestid R, Rummel M (2018) {Effective Field Theory of Black Hole
  Echoes}. JHEP 09:113. \doi{10.1007/JHEP09(2018)113}.
  {\href{https://arxiv.org/abs/1808.00847}{{arXiv:1808.00847}}} {[gr-qc]}

\bibitem[{Burrage and Sakstein(2018)}]{Burrage:2017qrf}
Burrage C, Sakstein J (2018) {Tests of Chameleon Gravity}. Living Rev Rel
  21(1):1. \doi{10.1007/s41114-018-0011-x}.
  {\href{https://arxiv.org/abs/1709.09071}{{arXiv:1709.09071}}} {[astro-ph.CO]}

\bibitem[{Cai et~al.(2019{\natexlab{a}})Cai, Pi, and Sasaki}]{Cai:2018dig}
Cai Rg, Pi S, Sasaki M (2019{\natexlab{a}}) {Gravitational Waves Induced by
  non-Gaussian Scalar Perturbations}. Phys Rev Lett 122(20):201101.
  \doi{10.1103/PhysRevLett.122.201101}.
  {\href{https://arxiv.org/abs/1810.11000}{{arXiv:1810.11000}}} {[astro-ph.CO]}

\bibitem[{Cai et~al.(2019{\natexlab{b}})Cai, Pi, and Sasaki}]{Cai:2019cdl}
Cai RG, Pi S, Sasaki M (2019{\natexlab{b}}) {Universal infrared scaling of
  gravitational wave background spectra}
  {\href{https://arxiv.org/abs/1909.13728}{{arXiv:1909.13728}}} {[astro-ph.CO]}

\bibitem[{Cai et~al.(2019{\natexlab{c}})Cai, Pi, Wang, and Yang}]{Cai:2019elf}
Cai RG, Pi S, Wang SJ, Yang XY (2019{\natexlab{c}}) {Pulsar Timing Array
  Constraints on the Induced Gravitational Waves}. JCAP 10:059.
  \doi{10.1088/1475-7516/2019/10/059}.
  {\href{https://arxiv.org/abs/1907.06372}{{arXiv:1907.06372}}} {[astro-ph.CO]}

\bibitem[{Cai et~al.(2016)Cai, Capozziello, De~Laurentis, and
  Saridakis}]{Cai:2015emx}
Cai YF, Capozziello S, De~Laurentis M, Saridakis EN (2016) {f(T) teleparallel
  gravity and cosmology}. Rept Prog Phys 79(10):106901.
  \doi{10.1088/0034-4885/79/10/106901}.
  {\href{https://arxiv.org/abs/1511.07586}{{arXiv:1511.07586}}} {[gr-qc]}

\bibitem[{Calabrese et~al.(2016)Calabrese, Battaglia, and
  Spergel}]{Calabrese:2016bnu}
Calabrese E, Battaglia N, Spergel DN (2016) {Testing Gravity with Gravitational
  Wave Source Counts}. Class Quant Grav 33(16):165004.
  \doi{10.1088/0264-9381/33/16/165004}.
  {\href{https://arxiv.org/abs/1602.03883}{{arXiv:1602.03883}}} {[gr-qc]}

\bibitem[{Calcagni and Kuroyanagi(2021)}]{Calcagni:2020}
Calcagni G, Kuroyanagi S (2021) {Stochastic gravitational-wave background in
  quantum gravity}. JCAP 03:019. \doi{10.1088/1475-7516/2021/03/019}.
  {\href{https://arxiv.org/abs/2012.00170}{{arXiv:2012.00170}}} {[gr-qc]}

\bibitem[{Calder\'on~Bustillo et~al.(2020)Calder\'on~Bustillo, Sanchis-Gual,
  Torres-Forn\'e, Font, Vajpeyi, Smith, Herdeiro, Radu, and
  Leong}]{CalderonBustillo:2020srq}
Calder\'on~Bustillo J, Sanchis-Gual N, Torres-Forn\'e A, Font JA, Vajpeyi A,
  Smith R, Herdeiro C, Radu E, Leong SH (2020) {The (ultra) light in the dark:
  A potential vector boson of $8.7\times 10^{-13}$ eV from GW190521}
  {\href{https://arxiv.org/abs/2009.05376}{{arXiv:2009.05376}}} {[gr-qc]}

\bibitem[{Caldwell et~al.(2016)Caldwell, Devulder, and
  Maksimova}]{Caldwell:2016sut}
Caldwell R, Devulder C, Maksimova N (2016) {Gravitational
  wave\textendash{}Gauge field oscillations}. Phys Rev D 94(6):063005.
  \doi{10.1103/PhysRevD.94.063005}.
  {\href{https://arxiv.org/abs/1604.08939}{{arXiv:1604.08939}}} {[gr-qc]}

\bibitem[{Callister et~al.(2017)Callister, Biscoveanu, Christensen, Isi, Matas,
  Minazzoli, Regimbau, Sakellariadou, Tasson, and Thrane}]{Callister:2017ocg}
Callister T, Biscoveanu A, Christensen N, Isi M, Matas A, Minazzoli O, Regimbau
  T, Sakellariadou M, Tasson J, Thrane E (2017) {Polarization-based Tests of
  Gravity with the Stochastic Gravitational-Wave Background}. Phys Rev X
  7(4):041058. \doi{10.1103/PhysRevX.7.041058}.
  {\href{https://arxiv.org/abs/1704.08373}{{arXiv:1704.08373}}} {[gr-qc]}

\bibitem[{Calore et~al.(2020)Calore, Cuoco, Regimbau, Sachdev, and
  Serpico}]{Calore:2020bpd}
Calore F, Cuoco A, Regimbau T, Sachdev S, Serpico PD (2020) {Cross-correlating
  galaxy catalogs and gravitational waves: a tomographic approach}. Phys Rev
  Res 2:023314. \doi{10.1103/PhysRevResearch.2.023314}.
  {\href{https://arxiv.org/abs/2002.02466}{{arXiv:2002.02466}}} {[astro-ph.CO]}

\bibitem[{{Camera} and {Nishizawa}(2013)}]{CameraGW}
{Camera} S, {Nishizawa} A (2013) {Beyond Concordance Cosmology with
  Magnification of Gravitational-Wave Standard Sirens}. \prl 110(15):151103.
  \doi{10.1103/PhysRevLett.110.151103}.
  {\href{https://arxiv.org/abs/1303.5446}{{arXiv:1303.5446}}} {[astro-ph.CO]}

\bibitem[{Campanelli et~al.(2006)Campanelli, Lousto, Marronetti, and
  Zlochower}]{Campanelli:2005dd}
Campanelli M, Lousto CO, Marronetti P, Zlochower Y (2006) {Accurate evolutions
  of orbiting black-hole binaries without excision}. Phys Rev Lett 96:111101.
  \doi{10.1103/PhysRevLett.96.111101}.
  {\href{https://arxiv.org/abs/gr-qc/0511048}{{arXiv:gr-qc/0511048}}}

\bibitem[{Campbell et~al.(1992)Campbell, Kaloper, and Olive}]{Campbell:1991kz}
Campbell BA, Kaloper N, Olive KA (1992) {Classical hair for Kerr-Newman black
  holes in string gravity}. Phys Lett B 285:199--205.
  \doi{10.1016/0370-2693(92)91452-F}

\bibitem[{Campiglia and Laddha(2015)}]{BMS3d}
Campiglia M, Laddha A (2015) {New symmetries for the Gravitational S-matrix}.
  JHEP 2015(04):076. \doi{10.1007/978-3-030-04260-8}.
  {\href{https://arxiv.org/abs/1502.02318}{{arXiv:1502.02318}}} {[hep-th]}

\bibitem[{Canizares et~al.(2012)Canizares, Gair, and
  Sopuerta}]{Canizares:2012is}
Canizares P, Gair JR, Sopuerta CF (2012) {Testing Chern-Simons Modified Gravity
  with Gravitational-Wave Detections of Extreme-Mass-Ratio Binaries}. Phys Rev
  D 86:044010. \doi{10.1103/PhysRevD.86.044010}.
  {\href{https://arxiv.org/abs/1205.1253}{{arXiv:1205.1253}}} {[gr-qc]}

\bibitem[{Canizares et~al.(2015)Canizares, Field, Gair, Raymond, Smith, and
  Tiglio}]{Canizares:2014fya}
Canizares P, Field SE, Gair J, Raymond V, Smith R, Tiglio M (2015) {Accelerated
  gravitational-wave parameter estimation with reduced order modeling}. Phys
  Rev Lett 114(7):071104. \doi{10.1103/PhysRevLett.114.071104}.
  {\href{https://arxiv.org/abs/1404.6284}{{arXiv:1404.6284}}} {[gr-qc]}

\bibitem[{Cano and Ruip\'erez(2019)}]{Cano:2019ore}
Cano PA, Ruip\'erez A (2019) {Leading higher-derivative corrections to Kerr
  geometry}. JHEP 05:189. \doi{10.1007/JHEP05(2019)189}, [Erratum: JHEP 03, 187
  (2020)]. {\href{https://arxiv.org/abs/1901.01315}{{arXiv:1901.01315}}}
  {[gr-qc]}

\bibitem[{Cao and Han(2017)}]{Cao:2017ndf}
Cao Z, Han WB (2017) {Waveform model for an eccentric binary black hole based
  on the effective-one-body-numerical-relativity formalism}. Phys Rev D
  96(4):044028. \doi{10.1103/PhysRevD.96.044028}.
  {\href{https://arxiv.org/abs/1708.00166}{{arXiv:1708.00166}}} {[gr-qc]}

\bibitem[{Capozziello and De~Laurentis(2011)}]{Capozziello:2011et}
Capozziello S, De~Laurentis M (2011) {Extended Theories of Gravity}. Phys Rept
  509:167--321. \doi{10.1016/j.physrep.2011.09.003}.
  {\href{https://arxiv.org/abs/1108.6266}{{arXiv:1108.6266}}} {[gr-qc]}

\bibitem[{Caprini and Tamanini(2016)}]{Caprini:2016qxs}
Caprini C, Tamanini N (2016) {Constraining early and interacting dark energy
  with gravitational wave standard sirens: the potential of the eLISA mission}.
  JCAP 10:006. \doi{10.1088/1475-7516/2016/10/006}.
  {\href{https://arxiv.org/abs/1607.08755}{{arXiv:1607.08755}}} {[astro-ph.CO]}

\bibitem[{Caprini et~al.(2019)Caprini, Figueroa, Flauger, Nardini, Peloso,
  Pieroni, Ricciardone, and Tasinato}]{Caprini:2019pxz}
Caprini C, Figueroa D, Flauger R, Nardini G, Peloso M, Pieroni M, Ricciardone
  A, Tasinato G (2019) {Reconstructing the spectral shape of a stochastic
  gravitational wave background with LISA}. JCAP 12:017.
  \doi{10.1088/1475-7516/2019/11/017}.
  {\href{https://arxiv.org/abs/1906.09244}{{arXiv:1906.09244}}} {[astro-ph.CO]}

\bibitem[{Caputo et~al.(2020)Caputo, Sberna, Toubiana, Babak, Barausse, Marsat,
  and Pani}]{Caputo:2020irr}
Caputo A, Sberna L, Toubiana A, Babak S, Barausse E, Marsat S, Pani P (2020)
  {Gravitational-wave detection and parameter estimation for accreting
  black-hole binaries and their electromagnetic counterpart}. Astrophys J
  892(2):90. \doi{10.3847/1538-4357/ab7b66}.
  {\href{https://arxiv.org/abs/2001.03620}{{arXiv:2001.03620}}} {[astro-ph.HE]}

\bibitem[{Carballo-Rubio et~al.(2018)Carballo-Rubio, Di~Filippo, Liberati, and
  Visser}]{Carballo-Rubio:2018jzw}
Carballo-Rubio R, Di~Filippo F, Liberati S, Visser M (2018) {Phenomenological
  aspects of black holes beyond general relativity}. Phys Rev D 98(12):124009.
  \doi{10.1103/PhysRevD.98.124009}.
  {\href{https://arxiv.org/abs/1809.08238}{{arXiv:1809.08238}}} {[gr-qc]}

\bibitem[{Cardenas-Avendano et~al.(2020)Cardenas-Avendano, Nampalliwar, and
  Yunes}]{Cardenas-Avendano:2019zxd}
Cardenas-Avendano A, Nampalliwar S, Yunes N (2020) {Gravitational-wave versus
  X-ray tests of strong-field gravity}. Class Quant Grav 37(13):135008.
  \doi{10.1088/1361-6382/ab8f64}.
  {\href{https://arxiv.org/abs/1912.08062}{{arXiv:1912.08062}}} {[gr-qc]}

\bibitem[{Cardoso and Duque(2020)}]{Cardoso:2019upw}
Cardoso V, Duque F (2020) {Environmental effects in gravitational-wave physics:
  Tidal deformability of black holes immersed in matter}. Phys Rev D
  101(6):064028. \doi{10.1103/PhysRevD.101.064028}.
  {\href{https://arxiv.org/abs/1912.07616}{{arXiv:1912.07616}}} {[gr-qc]}

\bibitem[{Cardoso and Gualtieri(2009)}]{Cardoso:2009pk}
Cardoso V, Gualtieri L (2009) {Perturbations of Schwarzschild black holes in
  Dynamical Chern-Simons modified gravity}. Phys Rev D 80:064008.
  \doi{10.1103/PhysRevD.81.089903}, [Erratum: Phys.Rev.D 81, 089903 (2010)].
  {\href{https://arxiv.org/abs/0907.5008}{{arXiv:0907.5008}}} {[gr-qc]}

\bibitem[{Cardoso and Maselli(2019)}]{Cardoso:2019rou}
Cardoso V, Maselli A (2019) {Constraints on the astrophysical environment of
  binaries with gravitational-wave observations}
  {\href{https://arxiv.org/abs/1909.05870}{{arXiv:1909.05870}}} {[astro-ph.HE]}

\bibitem[{Cardoso and Pani(2017{\natexlab{a}})}]{Cardoso:2017cqb}
Cardoso V, Pani P (2017{\natexlab{a}}) {Tests for the existence of black holes
  through gravitational wave echoes}. Nature Astron 1(9):586--591.
  \doi{10.1038/s41550-017-0225-y}.
  {\href{https://arxiv.org/abs/1709.01525}{{arXiv:1709.01525}}} {[gr-qc]}

\bibitem[{Cardoso and Pani(2017{\natexlab{b}})}]{Cardoso:2017njb}
Cardoso V, Pani P (2017{\natexlab{b}}) {The observational evidence for
  horizons: from echoes to precision gravitational-wave physics}
  {\href{https://arxiv.org/abs/1707.03021}{{arXiv:1707.03021}}} {[gr-qc]}

\bibitem[{Cardoso and Pani(2019)}]{Cardoso:2019rvt}
Cardoso V, Pani P (2019) {Testing the nature of dark compact objects: a status
  report}. Living Rev Rel 22(1):4. \doi{10.1007/s41114-019-0020-4}.
  {\href{https://arxiv.org/abs/1904.05363}{{arXiv:1904.05363}}} {[gr-qc]}

\bibitem[{Cardoso et~al.(2008)Cardoso, Pani, Cadoni, and
  Cavaglia}]{Cardoso:2007az}
Cardoso V, Pani P, Cadoni M, Cavaglia M (2008) {Ergoregion instability of
  ultracompact astrophysical objects}. Phys Rev D77:124044.
  \doi{10.1103/PhysRevD.77.124044}.
  {\href{https://arxiv.org/abs/0709.0532}{{arXiv:0709.0532}}} {[gr-qc]}

\bibitem[{Cardoso et~al.(2011)Cardoso, Chakrabarti, Pani, Berti, and
  Gualtieri}]{Cardoso:2011xi}
Cardoso V, Chakrabarti S, Pani P, Berti E, Gualtieri L (2011) {Floating and
  sinking: The Imprint of massive scalars around rotating black holes}. Phys
  Rev Lett 107:241101. \doi{10.1103/PhysRevLett.107.241101}.
  {\href{https://arxiv.org/abs/1109.6021}{{arXiv:1109.6021}}} {[gr-qc]}

\bibitem[{Cardoso et~al.(2013{\natexlab{a}})Cardoso, Carucci, Pani, and
  Sotiriou}]{Cardoso:2013fwa}
Cardoso V, Carucci IP, Pani P, Sotiriou TP (2013{\natexlab{a}}) {Black holes
  with surrounding matter in scalar-tensor theories}. Phys Rev Lett 111:111101.
  \doi{10.1103/PhysRevLett.111.111101}.
  {\href{https://arxiv.org/abs/1308.6587}{{arXiv:1308.6587}}} {[gr-qc]}

\bibitem[{Cardoso et~al.(2013{\natexlab{b}})Cardoso, Carucci, Pani, and
  Sotiriou}]{Cardoso:2013opa}
Cardoso V, Carucci IP, Pani P, Sotiriou TP (2013{\natexlab{b}}) {Matter around
  Kerr black holes in scalar-tensor theories: scalarization and superradiant
  instability}. Phys Rev D 88:044056. \doi{10.1103/PhysRevD.88.044056}.
  {\href{https://arxiv.org/abs/1305.6936}{{arXiv:1305.6936}}} {[gr-qc]}

\bibitem[{Cardoso et~al.(2014{\natexlab{a}})Cardoso, Crispino, Macedo, Okawa,
  and Pani}]{Cardoso:2014sna}
Cardoso V, Crispino LCB, Macedo CFB, Okawa H, Pani P (2014{\natexlab{a}})
  {Light rings as observational evidence for event horizons: long-lived modes,
  ergoregions and nonlinear instabilities of ultracompact objects}. Phys Rev
  D90(4):044069. \doi{10.1103/PhysRevD.90.044069}.
  {\href{https://arxiv.org/abs/1406.5510}{{arXiv:1406.5510}}} {[gr-qc]}

\bibitem[{Cardoso et~al.(2014{\natexlab{b}})Cardoso, Pani, and
  Rico}]{Cardoso:2014rha}
Cardoso V, Pani P, Rico J (2014{\natexlab{b}}) {On generic parametrizations of
  spinning black-hole geometries}. Phys Rev D 89:064007.
  \doi{10.1103/PhysRevD.89.064007}.
  {\href{https://arxiv.org/abs/1401.0528}{{arXiv:1401.0528}}} {[gr-qc]}

\bibitem[{Cardoso et~al.(2016{\natexlab{a}})Cardoso, Franzin, and
  Pani}]{Cardoso:2016rao}
Cardoso V, Franzin E, Pani P (2016{\natexlab{a}}) {Is the gravitational-wave
  ringdown a probe of the event horizon?} Phys Rev Lett 116(17):171101.
  \doi{10.1103/PhysRevLett.116.171101}.
  {\href{https://arxiv.org/abs/1602.07309}{{arXiv:1602.07309}}} {[gr-qc]}

\bibitem[{Cardoso et~al.(2016{\natexlab{b}})Cardoso, Hopper, Macedo,
  Palenzuela, and Pani}]{Cardoso:2016oxy}
Cardoso V, Hopper S, Macedo CFB, Palenzuela C, Pani P (2016{\natexlab{b}})
  {Gravitational-wave signatures of exotic compact objects and of quantum
  corrections at the horizon scale}. Phys Rev D 94(8):084031.
  \doi{10.1103/PhysRevD.94.084031}.
  {\href{https://arxiv.org/abs/1608.08637}{{arXiv:1608.08637}}} {[gr-qc]}

\bibitem[{Cardoso et~al.(2016{\natexlab{c}})Cardoso, Macedo, Pani, and
  Ferrari}]{Cardoso:2016olt}
Cardoso V, Macedo CFB, Pani P, Ferrari V (2016{\natexlab{c}}) {Black holes and
  gravitational waves in models of minicharged dark matter}. JCAP 05:054.
  \doi{10.1088/1475-7516/2016/05/054}, [Erratum: JCAP 04, E01 (2020)].
  {\href{https://arxiv.org/abs/1604.07845}{{arXiv:1604.07845}}} {[hep-ph]}

\bibitem[{Cardoso et~al.(2017)Cardoso, Franzin, Maselli, Pani, and
  Raposo}]{Cardoso:2017cfl}
Cardoso V, Franzin E, Maselli A, Pani P, Raposo G (2017) {Testing strong-field
  gravity with tidal Love numbers}. Phys Rev D 95(8):084014.
  \doi{10.1103/PhysRevD.95.084014}, [Addendum: Phys.Rev.D 95, 089901 (2017)].
  {\href{https://arxiv.org/abs/1701.01116}{{arXiv:1701.01116}}} {[gr-qc]}

\bibitem[{Cardoso et~al.(2018{\natexlab{a}})Cardoso, Dias, Hartnett, Middleton,
  Pani, and Santos}]{Cardoso:2018tly}
Cardoso V, Dias OJ, Hartnett GS, Middleton M, Pani P, Santos JE
  (2018{\natexlab{a}}) {Constraining the mass of dark photons and axion-like
  particles through black-hole superradiance}. JCAP 03:043.
  \doi{10.1088/1475-7516/2018/03/043}.
  {\href{https://arxiv.org/abs/1801.01420}{{arXiv:1801.01420}}} {[gr-qc]}

\bibitem[{Cardoso et~al.(2018{\natexlab{b}})Cardoso, Kimura, Maselli, and
  Senatore}]{Cardoso:2018ptl}
Cardoso V, Kimura M, Maselli A, Senatore L (2018{\natexlab{b}}) {Black Holes in
  an Effective Field Theory Extension of General Relativity}. Phys Rev Lett
  121(25):251105. \doi{10.1103/PhysRevLett.121.251105}.
  {\href{https://arxiv.org/abs/1808.08962}{{arXiv:1808.08962}}} {[gr-qc]}

\bibitem[{Cardoso et~al.(2019{\natexlab{a}})Cardoso, Foit, and
  Kleban}]{Cardoso:2019apo}
Cardoso V, Foit VF, Kleban M (2019{\natexlab{a}}) {Gravitational wave echoes
  from black hole area quantization}. JCAP 08:006.
  \doi{10.1088/1475-7516/2019/08/006}.
  {\href{https://arxiv.org/abs/1902.10164}{{arXiv:1902.10164}}} {[hep-th]}

\bibitem[{Cardoso et~al.(2019{\natexlab{b}})Cardoso, Kimura, Maselli, Berti,
  Macedo, and McManus}]{Cardoso:2019mqo}
Cardoso V, Kimura M, Maselli A, Berti E, Macedo CF, McManus R
  (2019{\natexlab{b}}) {Parametrized black hole quasinormal ringdown: Decoupled
  equations for nonrotating black holes}. Phys Rev D 99(10):104077.
  \doi{10.1103/PhysRevD.99.104077}.
  {\href{https://arxiv.org/abs/1901.01265}{{arXiv:1901.01265}}} {[gr-qc]}

\bibitem[{Cardoso et~al.(2019{\natexlab{c}})Cardoso, del Rio, and
  Kimura}]{Cardoso:2019nis}
Cardoso V, del Rio A, Kimura M (2019{\natexlab{c}}) {Distinguishing black holes
  from horizonless objects through the excitation of resonances during
  inspiral}. Phys Rev D 100:084046. \doi{10.1103/PhysRevD.100.084046},
  [Erratum: Phys.Rev.D 101, 069902 (2020)].
  {\href{https://arxiv.org/abs/1907.01561}{{arXiv:1907.01561}}} {[gr-qc]}

\bibitem[{Cardoso et~al.(2020{\natexlab{a}})Cardoso, Duque, and
  Ikeda}]{Cardoso:2020hca}
Cardoso V, Duque F, Ikeda T (2020{\natexlab{a}}) {Tidal effects and disruption
  in superradiant clouds: a numerical investigation}. Phys Rev D 101(6):064054.
  \doi{10.1103/PhysRevD.101.064054}.
  {\href{https://arxiv.org/abs/2001.01729}{{arXiv:2001.01729}}} {[gr-qc]}

\bibitem[{Cardoso et~al.(2020{\natexlab{b}})Cardoso, Guo, Macedo, and
  Pani}]{Cardoso:2020nst}
Cardoso V, Guo Wd, Macedo CF, Pani P (2020{\natexlab{b}}) {The tune of the
  universe: the role of plasma in tests of strong-field gravity}
  {\href{https://arxiv.org/abs/2009.07287}{{arXiv:2009.07287}}} {[gr-qc]}

\bibitem[{Carr et~al.(2016)Carr, Kuhnel, and Sandstad}]{Carr:2016drx}
Carr B, Kuhnel F, Sandstad M (2016) {Primordial Black Holes as Dark Matter}.
  Phys Rev D 94(8):083504. \doi{10.1103/PhysRevD.94.083504}.
  {\href{https://arxiv.org/abs/1607.06077}{{arXiv:1607.06077}}} {[astro-ph.CO]}

\bibitem[{Carr et~al.(2020)Carr, Kohri, Sendouda, and Yokoyama}]{Carr:2020gox}
Carr B, Kohri K, Sendouda Y, Yokoyama J (2020) {Constraints on Primordial Black
  Holes} {\href{https://arxiv.org/abs/2002.12778}{{arXiv:2002.12778}}}
  {[astro-ph.CO]}

\bibitem[{Carr et~al.(2021)Carr, Clesse, García-Bellido, and
  Kühnel}]{Carr:2019kxo}
Carr B, Clesse S, García-Bellido J, Kühnel F (2021) {Cosmic conundra
  explained by thermal history and primordial black holes}. Phys Dark Univ
  31:100755. \doi{10.1016/j.dark.2020.100755}.
  {\href{https://arxiv.org/abs/1906.08217}{{arXiv:1906.08217}}} {[astro-ph.CO]}

\bibitem[{Carson and Yagi(2020{\natexlab{a}})}]{Carson:2019rda}
Carson Z, Yagi K (2020{\natexlab{a}}) {Multi-band gravitational wave tests of
  general relativity}. Class Quant Grav 37(2):02LT01.
  \doi{10.1088/1361-6382/ab5c9a}.
  {\href{https://arxiv.org/abs/1905.13155}{{arXiv:1905.13155}}} {[gr-qc]}

\bibitem[{Carson and Yagi(2020{\natexlab{b}})}]{Carson:2019kkh}
Carson Z, Yagi K (2020{\natexlab{b}}) {Parametrized and
  inspiral-merger-ringdown consistency tests of gravity with multiband
  gravitational wave observations}. Phys Rev D101(4):044047.
  \doi{10.1103/PhysRevD.101.044047}.
  {\href{https://arxiv.org/abs/1911.05258}{{arXiv:1911.05258}}} {[gr-qc]}

\bibitem[{Carson and Yagi(2020{\natexlab{c}})}]{Carson:2020iik}
Carson Z, Yagi K (2020{\natexlab{c}}) {Probing beyond-Kerr spacetimes with
  inspiral-ringdown corrections to gravitational waves}. Phys Rev D 101:084050.
  \doi{10.1103/PhysRevD.101.084050}.
  {\href{https://arxiv.org/abs/2003.02374}{{arXiv:2003.02374}}} {[gr-qc]}

\bibitem[{Carson and Yagi(2020{\natexlab{d}})}]{Carson:2020cqb}
Carson Z, Yagi K (2020{\natexlab{d}}) {Probing string-inspired gravity with the
  inspiral-merger-ringdown consistency tests of gravitational waves}. Class
  Quant Grav 37(21):215007. \doi{10.1088/1361-6382/aba221}.
  {\href{https://arxiv.org/abs/2002.08559}{{arXiv:2002.08559}}} {[gr-qc]}

\bibitem[{Carter(1971)}]{Carter:1971zc}
Carter B (1971) {Axisymmetric Black Hole Has Only Two Degrees of Freedom}. Phys
  Rev Lett 26:331--333. \doi{10.1103/PhysRevLett.26.331}

\bibitem[{Carullo et~al.(2018)}]{Carullo:2018sfu}
Carullo G, et~al. (2018) {Empirical tests of the black hole no-hair conjecture
  using gravitational-wave observations}. Phys Rev D 98(10):104020.
  \doi{10.1103/PhysRevD.98.104020}.
  {\href{https://arxiv.org/abs/1805.04760}{{arXiv:1805.04760}}} {[gr-qc]}

\bibitem[{Castillo et~al.(2018)Castillo, Vega, and Wardell}]{Castillo:2018ibo}
Castillo J, Vega I, Wardell B (2018) {Self-force on a scalar charge in a
  circular orbit about a Reissner-Nordström black hole}. Phys Rev D
  98(2):024024. \doi{10.1103/PhysRevD.98.024024}.
  {\href{https://arxiv.org/abs/1804.09224}{{arXiv:1804.09224}}} {[gr-qc]}

\bibitem[{Cattoen et~al.(2005)Cattoen, Faber, and Visser}]{Cattoen:2005he}
Cattoen C, Faber T, Visser M (2005) {Gravastars must have anisotropic
  pressures}. Class Quant Grav 22:4189--4202.
  \doi{10.1088/0264-9381/22/20/002}.
  {\href{https://arxiv.org/abs/gr-qc/0505137}{{arXiv:gr-qc/0505137}}}

\bibitem[{Caves(1980)}]{Caves:1980jn}
Caves C (1980) {GRAVITATIONAL RADIATION AND THE ULTIMATE SPEED IN ROSEN'S
  BIMETRIC THEORY OF GRAVITY}. Annals Phys 125:35--52.
  \doi{10.1016/0003-4916(80)90117-7}

\bibitem[{Cayuso et~al.(2017)Cayuso, Ortiz, and Lehner}]{Cayuso:2017iqc}
Cayuso J, Ortiz N, Lehner L (2017) {Fixing extensions to general relativity in
  the nonlinear regime}. Phys Rev D 96(8):084043.
  \doi{10.1103/PhysRevD.96.084043}.
  {\href{https://arxiv.org/abs/1706.07421}{{arXiv:1706.07421}}} {[gr-qc]}

\bibitem[{Cayuso and Lehner(2020)}]{Cayuso:2020lca}
Cayuso R, Lehner L (2020) {Nonlinear, noniterative treatment of EFT-motivated
  gravity}. Phys Rev D 102(8):084008. \doi{10.1103/PhysRevD.102.084008}.
  {\href{https://arxiv.org/abs/2005.13720}{{arXiv:2005.13720}}} {[gr-qc]}

\bibitem[{Chamberlain and Yunes(2017)}]{Chamberlain:2017fjl}
Chamberlain K, Yunes N (2017) {Theoretical Physics Implications of
  Gravitational Wave Observation with Future Detectors}. Phys Rev D
  96(8):084039. \doi{10.1103/PhysRevD.96.084039}.
  {\href{https://arxiv.org/abs/1704.08268}{{arXiv:1704.08268}}} {[gr-qc]}

\bibitem[{Chandrasekhar(1943)}]{Chandrasekhar:1943ys}
Chandrasekhar S (1943) {Dynamical Friction. I. General Considerations: the
  Coefficient of Dynamical Friction}. Astrophys J 97:255. \doi{10.1086/144517}

\bibitem[{Chang et~al.(2020)Chang, Wang, and Zhu}]{Chang:2020iji}
Chang Z, Wang S, Zhu QH (2020) {Gauge Invariant Second Order Gravitational
  Waves} {\href{https://arxiv.org/abs/2009.11994}{{arXiv:2009.11994}}}
  {[gr-qc]}

\bibitem[{Charalambous et~al.(2021{\natexlab{a}})Charalambous, Dubovsky, and
  Ivanov}]{Charalambous:2021kcz}
Charalambous P, Dubovsky S, Ivanov MM (2021{\natexlab{a}}) {Hidden Symmetry of
  Vanishing Love} {\href{https://arxiv.org/abs/2103.01234}{{arXiv:2103.01234}}}
  {[hep-th]}

\bibitem[{Charalambous et~al.(2021{\natexlab{b}})Charalambous, Dubovsky, and
  Ivanov}]{Charalambous:2021mea}
Charalambous P, Dubovsky S, Ivanov MM (2021{\natexlab{b}}) {On the Vanishing of
  Love Numbers for Kerr Black Holes}. JHEP 05:038.
  \doi{10.1007/JHEP05(2021)038}.
  {\href{https://arxiv.org/abs/2102.08917}{{arXiv:2102.08917}}} {[hep-th]}

\bibitem[{Charmousis et~al.(2012)Charmousis, Copeland, Padilla, and
  Saffin}]{Charmousis:2011bf}
Charmousis C, Copeland EJ, Padilla A, Saffin PM (2012) {General second order
  scalar-tensor theory, self tuning, and the Fab Four}. Phys Rev Lett
  108:051101. \doi{10.1103/PhysRevLett.108.051101}.
  {\href{https://arxiv.org/abs/1106.2000}{{arXiv:1106.2000}}} {[hep-th]}

\bibitem[{Chatziioannou et~al.(2012)Chatziioannou, Yunes, and
  Cornish}]{Chatziioannou:2012rf}
Chatziioannou K, Yunes N, Cornish N (2012) {Model-Independent Test of General
  Relativity: An Extended post-Einsteinian Framework with Complete Polarization
  Content}. Phys Rev D 86:022004. \doi{10.1103/PhysRevD.86.022004}, [Erratum:
  Phys.Rev.D 95, 129901 (2017)].
  {\href{https://arxiv.org/abs/1204.2585}{{arXiv:1204.2585}}} {[gr-qc]}

\bibitem[{Chatziioannou et~al.(2021)Chatziioannou, Isi, Haster, and
  Littenberg}]{Chatziioannou:2021mij}
Chatziioannou K, Isi M, Haster CJ, Littenberg TB (2021) {Morphology-independent
  test of the mixed polarization content of transient gravitational wave
  signals}. Phys Rev D 104(4):044005. \doi{10.1103/PhysRevD.104.044005}.
  {\href{https://arxiv.org/abs/2105.01521}{{arXiv:2105.01521}}} {[gr-qc]}

\bibitem[{Chen et~al.(2018)Chen, Fishbach, and Holz}]{Chen:2017rfc}
Chen HY, Fishbach M, Holz DE (2018) {A two per cent Hubble constant measurement
  from standard sirens within five years}. Nature 562:545.
  \doi{10.1038/s41586-018-0606-0}.
  {\href{https://arxiv.org/abs/1712.06531}{{arXiv:1712.06531}}} {[astro-ph.CO]}

\bibitem[{Chen et~al.(2020)Chen, Liu, and Wang}]{Chen:2020wan}
Chen WC, Liu DD, Wang B (2020) {Detectability of ultra-compact X-ray binaries
  as LISA sources}. Astrophys J 900(1):L8. \doi{10.3847/2041-8213/abae66}.
  {\href{https://arxiv.org/abs/2008.05143}{{arXiv:2008.05143}}} {[astro-ph.HE]}

\bibitem[{Chen and Huang(2020)}]{Chen:2019irf}
Chen ZC, Huang QG (2020) {Distinguishing Primordial Black Holes from
  Astrophysical Black Holes by Einstein Telescope and Cosmic Explorer}. JCAP
  08:039. \doi{10.1088/1475-7516/2020/08/039}.
  {\href{https://arxiv.org/abs/1904.02396}{{arXiv:1904.02396}}} {[astro-ph.CO]}

\bibitem[{Chen et~al.(2019)Chen, Huang, and Huang}]{Chen:2018rzo}
Chen ZC, Huang F, Huang QG (2019) {Stochastic Gravitational-wave Background
  from Binary Black Holes and Binary Neutron Stars and Implications for LISA}.
  Astrophys J 871(1):97. \doi{10.3847/1538-4357/aaf581}.
  {\href{https://arxiv.org/abs/1809.10360}{{arXiv:1809.10360}}} {[gr-qc]}

\bibitem[{Cheung et~al.(2008)Cheung, Creminelli, Fitzpatrick, Kaplan, and
  Senatore}]{Cheung:2007st}
Cheung C, Creminelli P, Fitzpatrick AL, Kaplan J, Senatore L (2008) {The
  Effective Field Theory of Inflation}. JHEP 03:014.
  \doi{10.1088/1126-6708/2008/03/014}.
  {\href{https://arxiv.org/abs/0709.0293}{{arXiv:0709.0293}}} {[hep-th]}

\bibitem[{Chia(2020)}]{Chia:2020yla}
Chia HS (2020) {Tidal Deformation and Dissipation of Rotating Black Holes}
  {\href{https://arxiv.org/abs/2010.07300}{{arXiv:2010.07300}}} {[gr-qc]}

\bibitem[{Chiaramello and Nagar(2020)}]{Chiaramello:2020ehz}
Chiaramello D, Nagar A (2020) {Faithful analytical effective-one-body waveform
  model for spin-aligned, moderately eccentric, coalescing black hole
  binaries}. Phys Rev D 101(10):101501. \doi{10.1103/PhysRevD.101.101501}.
  {\href{https://arxiv.org/abs/2001.11736}{{arXiv:2001.11736}}} {[gr-qc]}

\bibitem[{Choudhury et~al.(2004)Choudhury, Joshi, Mahajan, and
  McKellar}]{Choudhury:2002pu}
Choudhury S, Joshi GC, Mahajan S, McKellar BH (2004) {Probing large distance
  higher dimensional gravity from lensing data}. Astropart Phys 21:559--563.
  \doi{10.1016/j.astropartphys.2004.04.001}.
  {\href{https://arxiv.org/abs/hep-ph/0204161}{{arXiv:hep-ph/0204161}}}

\bibitem[{Christodoulou(1991)}]{Christodoulou:1991cr}
Christodoulou D (1991) {Nonlinear nature of gravitation and gravitational wave
  experiments}. Phys Rev Lett 67:1486--1489. \doi{10.1103/PhysRevLett.67.1486}

\bibitem[{Chrusciel et~al.(2012)Chrusciel, Lopes~Costa, and
  Heusler}]{Chrusciel:2012jk}
Chrusciel PT, Lopes~Costa J, Heusler M (2012) {Stationary Black Holes:
  Uniqueness and Beyond}. Living Rev Rel 15:7. \doi{10.12942/lrr-2012-7}.
  {\href{https://arxiv.org/abs/1205.6112}{{arXiv:1205.6112}}} {[gr-qc]}

\bibitem[{Chu and Trodden(2013)}]{Chu:2012kz}
Chu YZ, Trodden M (2013) {Retarded Green\textquoteright{}s function of a
  Vainshtein system and Galileon waves}. Phys Rev D 87(2):024011.
  \doi{10.1103/PhysRevD.87.024011}.
  {\href{https://arxiv.org/abs/1210.6651}{{arXiv:1210.6651}}} {[astro-ph.CO]}

\bibitem[{Chua et~al.(2017)Chua, Moore, and Gair}]{Chua:2017ujo}
Chua AJK, Moore CJ, Gair JR (2017) {Augmented kludge waveforms for detecting
  extreme-mass-ratio inspirals}. Phys Rev D 96(4):044005.
  \doi{10.1103/PhysRevD.96.044005}.
  {\href{https://arxiv.org/abs/1705.04259}{{arXiv:1705.04259}}} {[gr-qc]}

\bibitem[{Clesse and Garc\'ia-Bellido(2020)}]{Clesse:2020ghq}
Clesse S, Garc\'ia-Bellido J (2020) {GW190425, GW190521 and GW190814: Three
  candidate mergers of primordial black holes from the QCD epoch}
  {\href{https://arxiv.org/abs/2007.06481}{{arXiv:2007.06481}}} {[astro-ph.CO]}

\bibitem[{Clesse and García-Bellido(2015)}]{Clesse:2015wea}
Clesse S, García-Bellido J (2015) {Massive Primordial Black Holes from Hybrid
  Inflation as Dark Matter and the seeds of Galaxies}. Phys Rev D92(2):023524.
  \doi{10.1103/PhysRevD.92.023524}.
  {\href{https://arxiv.org/abs/1501.07565}{{arXiv:1501.07565}}} {[astro-ph.CO]}

\bibitem[{Clesse and García-Bellido(2017)}]{Clesse:2016ajp}
Clesse S, García-Bellido J (2017) {Detecting the gravitational wave background
  from primordial black hole dark matter}. Phys Dark Univ 18:105--114.
  \doi{10.1016/j.dark.2017.10.001}.
  {\href{https://arxiv.org/abs/1610.08479}{{arXiv:1610.08479}}} {[astro-ph.CO]}

\bibitem[{Clesse et~al.(2018)Clesse, García-Bellido, and
  Orani}]{Clesse:2018ogk}
Clesse S, García-Bellido J, Orani S (2018) {Detecting the Stochastic
  Gravitational Wave Background from Primordial Black Hole Formation}
  {\href{https://arxiv.org/abs/1812.11011}{{arXiv:1812.11011}}} {[astro-ph.CO]}

\bibitem[{Clifton et~al.(2012)Clifton, Ferreira, Padilla, and
  Skordis}]{Clifton:2011jh}
Clifton T, Ferreira PG, Padilla A, Skordis C (2012) {Modified Gravity and
  Cosmology}. Phys Rept 513:1--189. \doi{10.1016/j.physrep.2012.01.001}.
  {\href{https://arxiv.org/abs/1106.2476}{{arXiv:1106.2476}}} {[astro-ph.CO]}

\bibitem[{Clough et~al.(2018)Clough, Dietrich, and Niemeyer}]{Clough:2018exo}
Clough K, Dietrich T, Niemeyer JC (2018) {Axion star collisions with black
  holes and neutron stars in full 3D numerical relativity}. Phys Rev D
  98(8):083020. \doi{10.1103/PhysRevD.98.083020}.
  {\href{https://arxiv.org/abs/1808.04668}{{arXiv:1808.04668}}} {[gr-qc]}

\bibitem[{Clough et~al.(2019)Clough, Ferreira, and Lagos}]{Clough:2019jpm}
Clough K, Ferreira PG, Lagos M (2019) {Growth of massive scalar hair around a
  Schwarzschild black hole}. Phys Rev D 100(6):063014.
  \doi{10.1103/PhysRevD.100.063014}.
  {\href{https://arxiv.org/abs/1904.12783}{{arXiv:1904.12783}}} {[gr-qc]}

\bibitem[{Colladay and Kosteleck\'y(1997)}]{Collady:1997}
Colladay D, Kosteleck\'y VA (1997) Cptviolation and the standard model.
  Physical Review D 55(11):6760–6774. \doi{10.1103/physrevd.55.6760}

\bibitem[{Colladay and Kosteleck\'y(1998)}]{Collady:1998}
Colladay D, Kosteleck\'y VA (1998) Lorentz-violating extension of the standard
  model. Physical Review D 58(11). \doi{10.1103/physrevd.58.116002}

\bibitem[{Collins and Hughes(2004)}]{Collins:2004ex}
Collins NA, Hughes SA (2004) {Towards a formalism for mapping the space-times
  of massive compact objects: Bumpy black holes and their orbits}. Phys Rev D
  69:124022. \doi{10.1103/PhysRevD.69.124022}.
  {\href{https://arxiv.org/abs/gr-qc/0402063}{{arXiv:gr-qc/0402063}}}

\bibitem[{Collodel et~al.(2020{\natexlab{a}})Collodel, Doneva, and
  Yazadjiev}]{Collodel:2020gyp}
Collodel LG, Doneva DD, Yazadjiev SS (2020{\natexlab{a}}) {Rotating
  tensor-multi-scalar-$N=2$ black holes}
  {\href{https://arxiv.org/abs/2007.14143}{{arXiv:2007.14143}}} {[gr-qc]}

\bibitem[{Collodel et~al.(2020{\natexlab{b}})Collodel, Kleihaus, Kunz, and
  Berti}]{Collodel:2019kkx}
Collodel LG, Kleihaus B, Kunz J, Berti E (2020{\natexlab{b}}) {Spinning and
  excited black holes in Einstein-scalar-Gauss\textendash{}Bonnet theory}.
  Class Quant Grav 37(7):075018. \doi{10.1088/1361-6382/ab74f9}.
  {\href{https://arxiv.org/abs/1912.05382}{{arXiv:1912.05382}}} {[gr-qc]}

\bibitem[{Colpi et~al.(1986)Colpi, Shapiro, and Wasserman}]{Colpi:1986ye}
Colpi M, Shapiro S, Wasserman I (1986) {Boson Stars: Gravitational Equilibria
  of Selfinteracting Scalar Fields}. Phys Rev Lett 57:2485--2488.
  \doi{10.1103/PhysRevLett.57.2485}

\bibitem[{Comelli et~al.(2012)Comelli, Crisostomi, Nesti, and
  Pilo}]{Comelli:2011wq}
Comelli D, Crisostomi M, Nesti F, Pilo L (2012) {Spherically Symmetric
  Solutions in Ghost-Free Massive Gravity}. Phys Rev D 85:024044.
  \doi{10.1103/PhysRevD.85.024044}.
  {\href{https://arxiv.org/abs/1110.4967}{{arXiv:1110.4967}}} {[hep-th]}

\bibitem[{Comp\`ere(2019{\natexlab{a}})}]{BMS3b}
Comp\`ere G (2019{\natexlab{a}}) {Advanced Lectures on General Relativity}, vol
  952. Springer, Cham. \doi{10.1007/978-3-030-04260-8}

\bibitem[{Comp\`ere(2019{\natexlab{b}})}]{BMSAdd9}
Comp\`ere G (2019{\natexlab{b}}) {Infinite towers of supertranslation and
  superrotation memories}. Phys Rev Lett 123(2):021101.
  \doi{10.1103/PhysRevLett.123.021101}.
  {\href{https://arxiv.org/abs/1904.00280}{{arXiv:1904.00280}}} {[gr-qc]}

\bibitem[{Congedo and Taylor(2019)}]{Congedo:2018wfn}
Congedo G, Taylor A (2019) {Joint cosmological inference of standard sirens and
  gravitational wave weak lensing}. Phys Rev D 99(8):083526.
  \doi{10.1103/PhysRevD.99.083526}.
  {\href{https://arxiv.org/abs/1812.02730}{{arXiv:1812.02730}}} {[astro-ph.CO]}

\bibitem[{Conklin and Holdom(2019)}]{Conklin:2019fcs}
Conklin RS, Holdom B (2019) {Gravitational wave echo spectra}. Phys Rev D
  100(12):124030. \doi{10.1103/PhysRevD.100.124030}.
  {\href{https://arxiv.org/abs/1905.09370}{{arXiv:1905.09370}}} {[gr-qc]}

\bibitem[{Conklin et~al.(2018)Conklin, Holdom, and Ren}]{Conklin:2017lwb}
Conklin RS, Holdom B, Ren J (2018) {Gravitational wave echoes through new
  windows}. Phys Rev D 98(4):044021. \doi{10.1103/PhysRevD.98.044021}.
  {\href{https://arxiv.org/abs/1712.06517}{{arXiv:1712.06517}}} {[gr-qc]}

\bibitem[{Contaldi(2017)}]{Contaldi:2016koz}
Contaldi CR (2017) {Anisotropies of Gravitational Wave Backgrounds: A Line Of
  Sight Approach}. Phys Lett B 771:9--12. \doi{10.1016/j.physletb.2017.05.020}.
  {\href{https://arxiv.org/abs/1609.08168}{{arXiv:1609.08168}}} {[astro-ph.CO]}

\bibitem[{Contaldi et~al.(2020)Contaldi, Pieroni, Renzini, Cusin, Karnesis,
  Peloso, Ricciardone, and Tasinato}]{Contaldi:2020rht}
Contaldi CR, Pieroni M, Renzini AI, Cusin G, Karnesis N, Peloso M, Ricciardone
  A, Tasinato G (2020) {Maximum likelihood map-making with the Laser
  Interferometer Space Antenna}
  {\href{https://arxiv.org/abs/2006.03313}{{arXiv:2006.03313}}} {[astro-ph.CO]}

\bibitem[{Coogan et~al.(2021)Coogan, Bertone, Gaggero, Kavanagh, and
  Nichols}]{Coogan:2021uqv}
Coogan A, Bertone G, Gaggero D, Kavanagh BJ, Nichols DA (2021) {Measuring the
  dark matter environments of black hole binaries with gravitational waves}
  {\href{https://arxiv.org/abs/2108.04154}{{arXiv:2108.04154}}} {[gr-qc]}

\bibitem[{Cornish et~al.(2011)Cornish, Sampson, Yunes, and
  Pretorius}]{Cornish:2011ys}
Cornish N, Sampson L, Yunes N, Pretorius F (2011) {Gravitational Wave Tests of
  General Relativity with the Parameterized Post-Einsteinian Framework}. Phys
  Rev D84:062003. \doi{10.1103/PhysRevD.84.062003}.
  {\href{https://arxiv.org/abs/1105.2088}{{arXiv:1105.2088}}} {[gr-qc]}

\bibitem[{Cornish et~al.(2017)Cornish, Blas, and Nardini}]{Cornish:2017jml}
Cornish N, Blas D, Nardini G (2017) {Bounding the speed of gravity with
  gravitational wave observations}. Phys Rev Lett 119(16):161102.
  \doi{10.1103/PhysRevLett.119.161102}.
  {\href{https://arxiv.org/abs/1707.06101}{{arXiv:1707.06101}}} {[gr-qc]}

\bibitem[{Cornish and Littenberg(2015)}]{Cornish:2014kda}
Cornish NJ, Littenberg TB (2015) {BayesWave: Bayesian Inference for
  Gravitational Wave Bursts and Instrument Glitches}. Class Quant Grav
  32(13):135012. \doi{10.1088/0264-9381/32/13/135012}.
  {\href{https://arxiv.org/abs/1410.3835}{{arXiv:1410.3835}}} {[gr-qc]}

\bibitem[{Cotesta et~al.(2018)Cotesta, Buonanno, Boh\'e, Taracchini, Hinder,
  and Ossokine}]{Cotesta:2018fcv}
Cotesta R, Buonanno A, Boh\'e A, Taracchini A, Hinder I, Ossokine S (2018)
  {Enriching the Symphony of Gravitational Waves from Binary Black Holes by
  Tuning Higher Harmonics}. Phys Rev D 98(8):084028.
  \doi{10.1103/PhysRevD.98.084028}.
  {\href{https://arxiv.org/abs/1803.10701}{{arXiv:1803.10701}}} {[gr-qc]}

\bibitem[{Creminelli and Vernizzi(2017)}]{Creminelli:2017sry}
Creminelli P, Vernizzi F (2017) {Dark Energy after GW170817 and GRB170817A}.
  Phys Rev Lett 119:251302. \doi{10.1103/PhysRevLett.119.251302}.
  {\href{https://arxiv.org/abs/1710.05877}{{arXiv:1710.05877}}} {[astro-ph.CO]}

\bibitem[{Creminelli et~al.(2006)Creminelli, Luty, Nicolis, and
  Senatore}]{Creminelli:2006xe}
Creminelli P, Luty MA, Nicolis A, Senatore L (2006) {Starting the Universe:
  Stable Violation of the Null Energy Condition and Non-standard Cosmologies}.
  JHEP 12:080. \doi{10.1088/1126-6708/2006/12/080}.
  {\href{https://arxiv.org/abs/hep-th/0606090}{{arXiv:hep-th/0606090}}}

\bibitem[{Creminelli et~al.(2009)Creminelli, D'Amico, Norena, and
  Vernizzi}]{Creminelli:2008wc}
Creminelli P, D'Amico G, Norena J, Vernizzi F (2009) {The Effective Theory of
  Quintessence: the w\ensuremath{<}-1 Side Unveiled}. JCAP 02:018.
  \doi{10.1088/1475-7516/2009/02/018}.
  {\href{https://arxiv.org/abs/0811.0827}{{arXiv:0811.0827}}} {[astro-ph]}

\bibitem[{Creminelli et~al.(2018)Creminelli, Lewandowski, Tambalo, and
  Vernizzi}]{Creminelli:2018xsv}
Creminelli P, Lewandowski M, Tambalo G, Vernizzi F (2018) {Gravitational Wave
  Decay into Dark Energy}. JCAP 1812:025. \doi{10.1088/1475-7516/2018/12/025}.
  {\href{https://arxiv.org/abs/1809.03484}{{arXiv:1809.03484}}} {[astro-ph.CO]}

\bibitem[{Creminelli et~al.(2019)Creminelli, Tambalo, Vernizzi, and
  Yingcharoenrat}]{Creminelli:2019nok}
Creminelli P, Tambalo G, Vernizzi F, Yingcharoenrat V (2019) {Resonant Decay of
  Gravitational Waves into Dark Energy}. JCAP 10:072.
  \doi{10.1088/1475-7516/2019/10/072}.
  {\href{https://arxiv.org/abs/1906.07015}{{arXiv:1906.07015}}} {[gr-qc]}

\bibitem[{Creminelli et~al.(2020)Creminelli, Tambalo, Vernizzi, and
  Yingcharoenrat}]{Creminelli:2019kjy}
Creminelli P, Tambalo G, Vernizzi F, Yingcharoenrat V (2020) {Dark-Energy
  Instabilities induced by Gravitational Waves}. JCAP 2005:002.
  \doi{10.1088/1475-7516/2020/05/002}.
  {\href{https://arxiv.org/abs/1910.14035}{{arXiv:1910.14035}}} {[gr-qc]}

\bibitem[{Crisostomi and Koyama(2018{\natexlab{a}})}]{Crisostomi:2017pjs}
Crisostomi M, Koyama K (2018{\natexlab{a}}) {Self-accelerating universe in
  scalar-tensor theories after GW170817}. Phys Rev D97(8):084004.
  \doi{10.1103/PhysRevD.97.084004}.
  {\href{https://arxiv.org/abs/1712.06556}{{arXiv:1712.06556}}} {[astro-ph.CO]}

\bibitem[{Crisostomi and Koyama(2018{\natexlab{b}})}]{Crisostomi:2017lbg}
Crisostomi M, Koyama K (2018{\natexlab{b}}) {Vainshtein mechanism after
  GW170817}. Phys Rev D97(2):021301. \doi{10.1103/PhysRevD.97.021301}.
  {\href{https://arxiv.org/abs/1711.06661}{{arXiv:1711.06661}}} {[astro-ph.CO]}

\bibitem[{Crisostomi et~al.(2016)Crisostomi, Koyama, and
  Tasinato}]{Crisostomi:2016czh}
Crisostomi M, Koyama K, Tasinato G (2016) {Extended Scalar-Tensor Theories of
  Gravity}. JCAP 1604:044. \doi{10.1088/1475-7516/2016/04/044}.
  {\href{https://arxiv.org/abs/1602.03119}{{arXiv:1602.03119}}} {[hep-th]}

\bibitem[{Crisostomi et~al.(2019)Crisostomi, Lewandowski, and
  Vernizzi}]{Crisostomi:2019yfo}
Crisostomi M, Lewandowski M, Vernizzi F (2019) {Vainshtein regime in
  scalar-tensor gravity: Constraints on degenerate higher-order scalar-tensor
  theories}. Phys Rev D100(2):024025. \doi{10.1103/PhysRevD.100.024025}.
  {\href{https://arxiv.org/abs/1903.11591}{{arXiv:1903.11591}}} {[gr-qc]}

\bibitem[{Crocker et~al.(2017)Crocker, Prestegard, Mandic, Regimbau, Olive, and
  Vangioni}]{Crocker:2017agi}
Crocker K, Prestegard T, Mandic V, Regimbau T, Olive K, Vangioni E (2017)
  {Systematic study of the stochastic gravitational-wave background due to
  stellar core collapse}. Phys Rev D 95(6):063015.
  \doi{10.1103/PhysRevD.95.063015}.
  {\href{https://arxiv.org/abs/1701.02638}{{arXiv:1701.02638}}} {[astro-ph.CO]}

\bibitem[{Cunha et~al.(2019)Cunha, Herdeiro, and Radu}]{Cunha:2019dwb}
Cunha PV, Herdeiro CA, Radu E (2019) {Spontaneously Scalarized Kerr Black Holes
  in Extended Scalar-Tensor--Gauss-Bonnet Gravity}. Phys Rev Lett
  123(1):011101. \doi{10.1103/PhysRevLett.123.011101}.
  {\href{https://arxiv.org/abs/1904.09997}{{arXiv:1904.09997}}} {[gr-qc]}

\bibitem[{Cunha et~al.(2016)Cunha, Grover, Herdeiro, Radu, Runarsson, and
  Wittig}]{Cunha:2016bjh}
Cunha PVP, Grover J, Herdeiro C, Radu E, Runarsson H, Wittig A (2016) {Chaotic
  lensing around boson stars and Kerr black holes with scalar hair}. Phys Rev D
  94(10):104023. \doi{10.1103/PhysRevD.94.104023}.
  {\href{https://arxiv.org/abs/1609.01340}{{arXiv:1609.01340}}} {[gr-qc]}

\bibitem[{Cunha et~al.(2017)Cunha, Berti, and Herdeiro}]{Cunha:2017qtt}
Cunha PVP, Berti E, Herdeiro CAR (2017) {Light-Ring Stability for Ultracompact
  Objects}. Phys Rev Lett 119(25):251102. \doi{10.1103/PhysRevLett.119.251102}.
  {\href{https://arxiv.org/abs/1708.04211}{{arXiv:1708.04211}}} {[gr-qc]}

\bibitem[{Cusin and Lagos(2020)}]{Cusin:2019rmt}
Cusin G, Lagos M (2020) {Gravitational wave propagation beyond geometric
  optics}. Phys Rev D 101(4):044041. \doi{10.1103/PhysRevD.101.044041}.
  {\href{https://arxiv.org/abs/1910.13326}{{arXiv:1910.13326}}} {[gr-qc]}

\bibitem[{Cusin and Tamanini(2021)}]{Cusin:2020ezb}
Cusin G, Tamanini N (2021) {Characterization of lensing selection effects for
  LISA massive black hole binary mergers}. Mon Not Roy Astron Soc
  504(3):3610--3618. \doi{10.1093/mnras/stab1130}.
  {\href{https://arxiv.org/abs/2011.15109}{{arXiv:2011.15109}}} {[astro-ph.CO]}

\bibitem[{Cusin et~al.(2017)Cusin, Pitrou, and Uzan}]{Cusin:2017fwz}
Cusin G, Pitrou C, Uzan JP (2017) {Anisotropy of the astrophysical
  gravitational wave background: Analytic expression of the angular power
  spectrum and correlation with cosmological observations}. Phys Rev D
  96(10):103019. \doi{10.1103/PhysRevD.96.103019}.
  {\href{https://arxiv.org/abs/1704.06184}{{arXiv:1704.06184}}} {[astro-ph.CO]}

\bibitem[{Cusin et~al.(2018)Cusin, Dvorkin, Pitrou, and Uzan}]{Cusin:2018rsq}
Cusin G, Dvorkin I, Pitrou C, Uzan JP (2018) {First predictions of the angular
  power spectrum of the astrophysical gravitational wave background}. Phys Rev
  Lett 120:231101. \doi{10.1103/PhysRevLett.120.231101}.
  {\href{https://arxiv.org/abs/1803.03236}{{arXiv:1803.03236}}} {[astro-ph.CO]}

\bibitem[{Cusin et~al.(2019)Cusin, Durrer, and Ferreira}]{Cusin:2018avf}
Cusin G, Durrer R, Ferreira PG (2019) {Polarization of a stochastic
  gravitational wave background through diffusion by massive structures}. Phys
  Rev D 99(2):023534. \doi{10.1103/PhysRevD.99.023534}.
  {\href{https://arxiv.org/abs/1807.10620}{{arXiv:1807.10620}}} {[astro-ph.CO]}

\bibitem[{Cárdenas-Avendaño et~al.(2018)Cárdenas-Avendaño, Gutierrez,
  Pachón, and Yunes}]{Cardenas-Avendano:2018ocb}
Cárdenas-Avendaño A, Gutierrez AF, Pachón LA, Yunes N (2018) {The exact
  dynamical Chern--Simons metric for a spinning black hole possesses a fourth
  constant of motion: A dynamical-systems-based conjecture}. Class Quant Grav
  35(16):165010. \doi{10.1088/1361-6382/aad06f}.
  {\href{https://arxiv.org/abs/1804.04002}{{arXiv:1804.04002}}} {[gr-qc]}

\bibitem[{Dai et~al.(2018)Dai, Li, Zackay, Mao, and Lu}]{Dai:2018enj}
Dai L, Li SS, Zackay B, Mao S, Lu Y (2018) {Detecting Lensing-Induced
  Diffraction in Astrophysical Gravitational Waves}. Phys Rev D 98(10):104029.
  \doi{10.1103/PhysRevD.98.104029}.
  {\href{https://arxiv.org/abs/1810.00003}{{arXiv:1810.00003}}} {[gr-qc]}

\bibitem[{Dalal et~al.(2006)Dalal, Holz, Hughes, and Jain}]{Dalal:2006qt}
Dalal N, Holz DE, Hughes SA, Jain B (2006) {Short grb and binary black hole
  standard sirens as a probe of dark energy}. Phys Rev D 74:063006.
  \doi{10.1103/PhysRevD.74.063006}.
  {\href{https://arxiv.org/abs/astro-ph/0601275}{{arXiv:astro-ph/0601275}}}

\bibitem[{Dalang and Lombriser(2019)}]{Dalang:2019fma}
Dalang C, Lombriser L (2019) {Limitations on Standard Sirens tests of gravity
  from screening}. JCAP 10:013. \doi{10.1088/1475-7516/2019/10/013}.
  {\href{https://arxiv.org/abs/1906.12333}{{arXiv:1906.12333}}} {[astro-ph.CO]}

\bibitem[{Dalang et~al.(2020)Dalang, Fleury, and Lombriser}]{Dalang:2019rke}
Dalang C, Fleury P, Lombriser L (2020) {Horndeski gravity and standard sirens}.
  Phys Rev D 102(4):044036. \doi{10.1103/PhysRevD.102.044036}.
  {\href{https://arxiv.org/abs/1912.06117}{{arXiv:1912.06117}}} {[gr-qc]}

\bibitem[{Dalang et~al.(2021)Dalang, Fleury, and Lombriser}]{Dalang:2020eaj}
Dalang C, Fleury P, Lombriser L (2021) {Scalar and tensor gravitational waves}.
  Phys Rev D 103(6):064075. \doi{10.1103/PhysRevD.103.064075}.
  {\href{https://arxiv.org/abs/2009.11827}{{arXiv:2009.11827}}} {[gr-qc]}

\bibitem[{D'Amico et~al.(2020)D'Amico, Senatore, Zhang, and
  Zheng}]{DAmico:2020ods}
D'Amico G, Senatore L, Zhang P, Zheng H (2020) {The Hubble Tension in Light of
  the Full-Shape Analysis of Large-Scale Structure Data}
  {\href{https://arxiv.org/abs/2006.12420}{{arXiv:2006.12420}}} {[astro-ph.CO]}

\bibitem[{Damour(2001)}]{Damour:2001tu}
Damour T (2001) {Coalescence of two spinning black holes: an effective one-body
  approach}. Phys Rev D 64:124013. \doi{10.1103/PhysRevD.64.124013}.
  {\href{https://arxiv.org/abs/gr-qc/0103018}{{arXiv:gr-qc/0103018}}}

\bibitem[{Damour and Nagar(2007)}]{Damour:2007xr}
Damour T, Nagar A (2007) {Faithful effective-one-body waveforms of
  small-mass-ratio coalescing black-hole binaries}. Phys Rev D 76:064028.
  \doi{10.1103/PhysRevD.76.064028}.
  {\href{https://arxiv.org/abs/0705.2519}{{arXiv:0705.2519}}} {[gr-qc]}

\bibitem[{Damour and Nagar(2009)}]{Damour:2009vw}
Damour T, Nagar A (2009) {Relativistic tidal properties of neutron stars}. Phys
  Rev D 80:084035. \doi{10.1103/PhysRevD.80.084035}.
  {\href{https://arxiv.org/abs/0906.0096}{{arXiv:0906.0096}}} {[gr-qc]}

\bibitem[{Damour and Nagar(2010)}]{Damour:2009wj}
Damour T, Nagar A (2010) {Effective One Body description of tidal effects in
  inspiralling compact binaries}. Phys Rev D 81:084016.
  \doi{10.1103/PhysRevD.81.084016}.
  {\href{https://arxiv.org/abs/0911.5041}{{arXiv:0911.5041}}} {[gr-qc]}

\bibitem[{Damour and Nagar(2014{\natexlab{a}})}]{Damour:2014yha}
Damour T, Nagar A (2014{\natexlab{a}}) {A new analytic representation of the
  ringdown waveform of coalescing spinning black hole binaries}. Phys Rev D
  90(2):024054. \doi{10.1103/PhysRevD.90.024054}.
  {\href{https://arxiv.org/abs/1406.0401}{{arXiv:1406.0401}}} {[gr-qc]}

\bibitem[{Damour and Nagar(2014{\natexlab{b}})}]{Damour:2014sva}
Damour T, Nagar A (2014{\natexlab{b}}) {New effective-one-body description of
  coalescing nonprecessing spinning black-hole binaries}. Phys Rev D
  90(4):044018. \doi{10.1103/PhysRevD.90.044018}.
  {\href{https://arxiv.org/abs/1406.6913}{{arXiv:1406.6913}}} {[gr-qc]}

\bibitem[{Damour and Solodukhin(2007)}]{Damour:2007ap}
Damour T, Solodukhin SN (2007) {Wormholes as black hole foils}. Phys Rev D
  76:024016. \doi{10.1103/PhysRevD.76.024016}.
  {\href{https://arxiv.org/abs/0704.2667}{{arXiv:0704.2667}}} {[gr-qc]}

\bibitem[{Damour and Taylor(1992)}]{Damour:1991rd}
Damour T, Taylor JH (1992) {Strong field tests of relativistic gravity and
  binary pulsars}. Phys Rev D45:1840--1868. \doi{10.1103/PhysRevD.45.1840}

\bibitem[{Damour et~al.(1976)Damour, Deruelle, and Ruffini}]{Damour:1976kh}
Damour T, Deruelle N, Ruffini R (1976) {On Quantum Resonances in Stationary
  Geometries}. Lett Nuovo Cim 15:257--262. \doi{10.1007/BF02725534}

\bibitem[{Damour et~al.(2003)Damour, Iyer, Jaranowski, and
  Sathyaprakash}]{Damour:2002vi}
Damour T, Iyer BR, Jaranowski P, Sathyaprakash BS (2003) {Gravitational waves
  from black hole binary inspiral and merger: The Span of third postNewtonian
  effective one-body templates}. Phys Rev D 67:064028.
  \doi{10.1103/PhysRevD.67.064028}.
  {\href{https://arxiv.org/abs/gr-qc/0211041}{{arXiv:gr-qc/0211041}}}

\bibitem[{Damour et~al.(2008)Damour, Jaranowski, and Schaefer}]{Damour:2008qf}
Damour T, Jaranowski P, Schaefer G (2008) {Effective one body approach to the
  dynamics of two spinning black holes with next-to-leading order spin-orbit
  coupling}. Phys Rev D 78:024009. \doi{10.1103/PhysRevD.78.024009}.
  {\href{https://arxiv.org/abs/0803.0915}{{arXiv:0803.0915}}} {[gr-qc]}

\bibitem[{Damour et~al.(2009)Damour, Iyer, and Nagar}]{Damour:2008gu}
Damour T, Iyer BR, Nagar A (2009) {Improved resummation of post-Newtonian
  multipolar waveforms from circularized compact binaries}. Phys Rev D
  79:064004. \doi{10.1103/PhysRevD.79.064004}.
  {\href{https://arxiv.org/abs/0811.2069}{{arXiv:0811.2069}}} {[gr-qc]}

\bibitem[{Damour et~al.(2012)Damour, Nagar, and Villain}]{Damour:2012yf}
Damour T, Nagar A, Villain L (2012) {Measurability of the tidal polarizability
  of neutron stars in late-inspiral gravitational-wave signals}. Phys Rev D
  85:123007. \doi{10.1103/PhysRevD.85.123007}.
  {\href{https://arxiv.org/abs/1203.4352}{{arXiv:1203.4352}}} {[gr-qc]}

\bibitem[{Damour et~al.(2013)Damour, Nagar, and Bernuzzi}]{Damour:2012ky}
Damour T, Nagar A, Bernuzzi S (2013) {Improved effective-one-body description
  of coalescing nonspinning black-hole binaries and its numerical-relativity
  completion}. Phys Rev D 87(8):084035. \doi{10.1103/PhysRevD.87.084035}.
  {\href{https://arxiv.org/abs/1212.4357}{{arXiv:1212.4357}}} {[gr-qc]}

\bibitem[{D'Antonio et~al.(2018)}]{DAntonio:2018sff}
D'Antonio S, et~al. (2018) {Semicoherent analysis method to search for
  continuous gravitational waves emitted by ultralight boson clouds around
  spinning black holes}. Phys Rev D 98(10):103017.
  \doi{10.1103/PhysRevD.98.103017}.
  {\href{https://arxiv.org/abs/1809.07202}{{arXiv:1809.07202}}} {[gr-qc]}

\bibitem[{Dar et~al.(2019)Dar, De~Rham, Deskins, Giblin, and
  Tolley}]{Dar:2018dra}
Dar F, De~Rham C, Deskins JT, Giblin JT, Tolley AJ (2019) {Scalar Gravitational
  Radiation from Binaries: Vainshtein Mechanism in Time-dependent Systems}.
  Class Quant Grav 36(2):025008. \doi{10.1088/1361-6382/aaf5e8}.
  {\href{https://arxiv.org/abs/1808.02165}{{arXiv:1808.02165}}} {[hep-th]}

\bibitem[{Datta and Bose(2019)}]{Datta:2019euh}
Datta S, Bose S (2019) {Probing the nature of central objects in
  extreme-mass-ratio inspirals with gravitational waves}. Phys Rev D
  99(8):084001. \doi{10.1103/PhysRevD.99.084001}.
  {\href{https://arxiv.org/abs/1902.01723}{{arXiv:1902.01723}}} {[gr-qc]}

\bibitem[{Datta et~al.(2020)Datta, Brito, Bose, Pani, and
  Hughes}]{Datta:2019epe}
Datta S, Brito R, Bose S, Pani P, Hughes SA (2020) {Tidal heating as a
  discriminator for horizons in extreme mass ratio inspirals}. Phys Rev D
  101(4):044004. \doi{10.1103/PhysRevD.101.044004}.
  {\href{https://arxiv.org/abs/1910.07841}{{arXiv:1910.07841}}} {[gr-qc]}

\bibitem[{De~Luca and Pani(2021)}]{DeLuca:2021ite}
De~Luca V, Pani P (2021) {Tidal deformability of dressed black holes and tests
  of ultralight bosons in extended mass ranges}
  {\href{https://arxiv.org/abs/2106.14428}{{arXiv:2106.14428}}} {[gr-qc]}

\bibitem[{De~Luca et~al.(2019{\natexlab{a}})De~Luca, Desjacques, Franciolini,
  Malhotra, and Riotto}]{DeLuca:2019buf}
De~Luca V, Desjacques V, Franciolini G, Malhotra A, Riotto A
  (2019{\natexlab{a}}) {The initial spin probability distribution of primordial
  black holes}. JCAP 05:018. \doi{10.1088/1475-7516/2019/05/018}.
  {\href{https://arxiv.org/abs/1903.01179}{{arXiv:1903.01179}}} {[astro-ph.CO]}

\bibitem[{De~Luca et~al.(2019{\natexlab{b}})De~Luca, Desjacques, Franciolini,
  and Riotto}]{DeLuca:2019llr}
De~Luca V, Desjacques V, Franciolini G, Riotto A (2019{\natexlab{b}})
  {Gravitational Waves from Peaks}. JCAP 09:059.
  \doi{10.1088/1475-7516/2019/09/059}.
  {\href{https://arxiv.org/abs/1905.13459}{{arXiv:1905.13459}}} {[astro-ph.CO]}

\bibitem[{De~Luca et~al.(2020{\natexlab{a}})De~Luca, Desjacques, Franciolini,
  Pani, and Riotto}]{DeLuca:2020sae}
De~Luca V, Desjacques V, Franciolini G, Pani P, Riotto A (2020{\natexlab{a}})
  {The GW190521 Mass Gap Event and the Primordial Black Hole Scenario}
  {\href{https://arxiv.org/abs/2009.01728}{{arXiv:2009.01728}}} {[astro-ph.CO]}

\bibitem[{De~Luca et~al.(2020{\natexlab{b}})De~Luca, Desjacques, Franciolini,
  and Riotto}]{DeLuca:2020jug}
De~Luca V, Desjacques V, Franciolini G, Riotto A (2020{\natexlab{b}}) {The
  Clustering Evolution of Primordial Black Holes}
  {\href{https://arxiv.org/abs/2009.04731}{{arXiv:2009.04731}}} {[astro-ph.CO]}

\bibitem[{De~Luca et~al.(2020{\natexlab{c}})De~Luca, Franciolini, Kehagias, and
  Riotto}]{DeLuca:2019ufz}
De~Luca V, Franciolini G, Kehagias A, Riotto A (2020{\natexlab{c}}) {On the
  Gauge Invariance of Cosmological Gravitational Waves}. JCAP 03:014.
  \doi{10.1088/1475-7516/2020/03/014}.
  {\href{https://arxiv.org/abs/1911.09689}{{arXiv:1911.09689}}} {[gr-qc]}

\bibitem[{De~Luca et~al.(2020{\natexlab{d}})De~Luca, Franciolini, Pani, and
  Riotto}]{DeLuca:2020fpg}
De~Luca V, Franciolini G, Pani P, Riotto A (2020{\natexlab{d}}) {Constraints on
  Primordial Black Holes: the Importance of Accretion}. Phys Rev D
  102(4):043505. \doi{10.1103/PhysRevD.102.043505}.
  {\href{https://arxiv.org/abs/2003.12589}{{arXiv:2003.12589}}} {[astro-ph.CO]}

\bibitem[{De~Luca et~al.(2020{\natexlab{e}})De~Luca, Franciolini, Pani, and
  Riotto}]{DeLuca:2020qqa}
De~Luca V, Franciolini G, Pani P, Riotto A (2020{\natexlab{e}}) {Primordial
  Black Holes Confront LIGO/Virgo data: Current situation}. JCAP 06:044.
  \doi{10.1088/1475-7516/2020/06/044}.
  {\href{https://arxiv.org/abs/2005.05641}{{arXiv:2005.05641}}} {[astro-ph.CO]}

\bibitem[{De~Luca et~al.(2020{\natexlab{f}})De~Luca, Franciolini, Pani, and
  Riotto}]{DeLuca:2020bjf}
De~Luca V, Franciolini G, Pani P, Riotto A (2020{\natexlab{f}}) {The evolution
  of primordial black holes and their final observable spins}. JCAP 04:052.
  \doi{10.1088/1475-7516/2020/04/052}.
  {\href{https://arxiv.org/abs/2003.02778}{{arXiv:2003.02778}}} {[astro-ph.CO]}

\bibitem[{De~Luca et~al.(2020{\natexlab{g}})De~Luca, Franciolini, and
  Riotto}]{DeLuca:2020agl}
De~Luca V, Franciolini G, Riotto A (2020{\natexlab{g}}) {NANOGrav Hints to
  Primordial Black Holes as Dark Matter}
  {\href{https://arxiv.org/abs/2009.08268}{{arXiv:2009.08268}}} {[astro-ph.CO]}

\bibitem[{De~Luca et~al.(2020{\natexlab{h}})De~Luca, Franciolini, and
  Riotto}]{DeLuca:2020ioi}
De~Luca V, Franciolini G, Riotto A (2020{\natexlab{h}}) {On the Primordial
  Black Hole Mass Function for Broad Spectra}. Phys Lett B 807:135550.
  \doi{10.1016/j.physletb.2020.135550}.
  {\href{https://arxiv.org/abs/2001.04371}{{arXiv:2001.04371}}} {[astro-ph.CO]}

\bibitem[{De~Luca et~al.(2021{\natexlab{a}})De~Luca, Franciolini, Pani, and
  Riotto}]{DeLuca:2021wjr}
De~Luca V, Franciolini G, Pani P, Riotto A (2021{\natexlab{a}}) {Bayesian
  Evidence for Both Astrophysical and Primordial Black Holes: Mapping the
  GWTC-2 Catalog to Third-Generation Detectors}. JCAP 05:003.
  \doi{10.1088/1475-7516/2021/05/003}.
  {\href{https://arxiv.org/abs/2102.03809}{{arXiv:2102.03809}}} {[astro-ph.CO]}

\bibitem[{De~Luca et~al.(2021{\natexlab{b}})De~Luca, Franciolini, Pani, and
  Riotto}]{DeLuca:2021hde}
De~Luca V, Franciolini G, Pani P, Riotto A (2021{\natexlab{b}}) {The Minimum
  Testable Abundance of Primordial Black Holes at Future Gravitational-Wave
  Detectors} {\href{https://arxiv.org/abs/2106.13769}{{arXiv:2106.13769}}}
  {[astro-ph.CO]}

\bibitem[{Deffayet and Menou(2007)}]{Deffayet:2007kf}
Deffayet C, Menou K (2007) {Probing Gravity with Spacetime Sirens}. Astrophys J
  Lett 668:L143--L146. \doi{10.1086/522931}.
  {\href{https://arxiv.org/abs/0709.0003}{{arXiv:0709.0003}}} {[astro-ph]}

\bibitem[{Deffayet et~al.(2009)Deffayet, Deser, and
  Esposito-Farese}]{Deffayet:2009mn}
Deffayet C, Deser S, Esposito-Farese G (2009) {Generalized Galileons: All
  scalar models whose curved background extensions maintain second-order field
  equations and stress-tensors}. Phys Rev D 80:064015.
  \doi{10.1103/PhysRevD.80.064015}.
  {\href{https://arxiv.org/abs/0906.1967}{{arXiv:0906.1967}}} {[gr-qc]}

\bibitem[{Deffayet et~al.(2011)Deffayet, Gao, Steer, and
  Zahariade}]{Deffayet:2011gz}
Deffayet C, Gao X, Steer D, Zahariade G (2011) {From k-essence to generalised
  Galileons}. PhysRev D84:064039. \doi{10.1103/PhysRevD.84.064039}.
  {\href{https://arxiv.org/abs/1103.3260}{{arXiv:1103.3260}}} {[hep-th]}

\bibitem[{Degollado et~al.(2018)Degollado, Herdeiro, and
  Radu}]{Degollado:2018ypf}
Degollado JC, Herdeiro CA, Radu E (2018) {Effective stability against
  superradiance of Kerr black holes with synchronised hair}. Phys Lett B
  781:651--655. \doi{10.1016/j.physletb.2018.04.052}.
  {\href{https://arxiv.org/abs/1802.07266}{{arXiv:1802.07266}}} {[gr-qc]}

\bibitem[{Del~Pozzo(2012)}]{DelPozzo:2011yh}
Del~Pozzo W (2012) {Inference of the cosmological parameters from gravitational
  waves: application to second generation interferometers}. Phys Rev
  D86:043011. \doi{10.1103/PhysRevD.86.043011}.
  {\href{https://arxiv.org/abs/1108.1317}{{arXiv:1108.1317}}} {[astro-ph.CO]}

\bibitem[{Del~Pozzo et~al.(2018)Del~Pozzo, Sesana, and
  Klein}]{DelPozzo:2017kme}
Del~Pozzo W, Sesana A, Klein A (2018) {Stellar binary black holes in the LISA
  band: a new class of standard sirens}. Mon Not Roy Astron Soc
  475(3):3485--3492. \doi{10.1093/mnras/sty057}.
  {\href{https://arxiv.org/abs/1703.01300}{{arXiv:1703.01300}}} {[astro-ph.CO]}

\bibitem[{Delgado et~al.(2019)Delgado, Herdeiro, and Radu}]{Delgado:2019prc}
Delgado JF, Herdeiro CA, Radu E (2019) {Kerr black holes with synchronised
  scalar hair and higher azimuthal harmonic index}. Phys Lett B 792:436--444.
  \doi{10.1016/j.physletb.2019.04.009}.
  {\href{https://arxiv.org/abs/1903.01488}{{arXiv:1903.01488}}} {[gr-qc]}

\bibitem[{Delgado et~al.(2020{\natexlab{a}})Delgado, Herdeiro, and
  Radu}]{Delgado:2020udb}
Delgado JF, Herdeiro CA, Radu E (2020{\natexlab{a}}) {Rotating Axion Boson
  Stars}. JCAP 06:037. \doi{10.1088/1475-7516/2020/06/037}.
  {\href{https://arxiv.org/abs/2005.05982}{{arXiv:2005.05982}}} {[gr-qc]}

\bibitem[{Delgado et~al.(2020{\natexlab{b}})Delgado, Herdeiro, and
  Radu}]{Delgado:2020rev}
Delgado JF, Herdeiro CA, Radu E (2020{\natexlab{b}}) {Spinning black holes in
  shift-symmetric Horndeski theory}. JHEP 04:180.
  \doi{10.1007/JHEP04(2020)180}.
  {\href{https://arxiv.org/abs/2002.05012}{{arXiv:2002.05012}}} {[gr-qc]}

\bibitem[{Delhom et~al.(2019)Delhom, Macedo, Olmo, and
  Crispino}]{Delhom:2019btt}
Delhom A, Macedo CFB, Olmo GJ, Crispino LCB (2019) {Absorption by black hole
  remnants in metric-affine gravity}. Phys Rev D 100(2):024016.
  \doi{10.1103/PhysRevD.100.024016}.
  {\href{https://arxiv.org/abs/1906.06411}{{arXiv:1906.06411}}} {[gr-qc]}

\bibitem[{Delsate et~al.(2015)Delsate, Hilditch, and Witek}]{Delsate:2014hba}
Delsate T, Hilditch D, Witek H (2015) {Initial value formulation of dynamical
  Chern-Simons gravity}. Phys Rev D91(2):024027.
  \doi{10.1103/PhysRevD.91.024027}.
  {\href{https://arxiv.org/abs/1407.6727}{{arXiv:1407.6727}}} {[gr-qc]}

\bibitem[{Delsate et~al.(2018)Delsate, Herdeiro, and Radu}]{Delsate:2018ome}
Delsate T, Herdeiro C, Radu E (2018) {Non-perturbative spinning black holes in
  dynamical Chern-Simons gravity}
  {\href{https://arxiv.org/abs/1806.06700}{{arXiv:1806.06700}}} {[gr-qc]}

\bibitem[{Derdzinski et~al.(2021)Derdzinski, D'Orazio, Duffell, Haiman, and
  MacFadyen}]{Derdzinski:2020wlw}
Derdzinski A, D'Orazio D, Duffell P, Haiman Z, MacFadyen A (2021) {Evolution of
  gas disc\textendash{}embedded intermediate mass ratio inspirals in the $LISA$
  band}. Mon Not Roy Astron Soc 501(3):3540--3557.
  \doi{10.1093/mnras/staa3976}.
  {\href{https://arxiv.org/abs/2005.11333}{{arXiv:2005.11333}}} {[astro-ph.HE]}

\bibitem[{Derdzinski et~al.(2019)Derdzinski, D'Orazio, Duffell, Haiman, and
  MacFadyen}]{Derdzinski:2018qzv}
Derdzinski AM, D'Orazio D, Duffell P, Haiman Z, MacFadyen A (2019) {Probing gas
  disc physics with LISA: simulations of an intermediate mass ratio inspiral in
  an accretion disc}. Mon Not Roy Astron Soc 486(2):2754--2765.
  \doi{10.1093/mnras/stz1026}, [Erratum: Mon.Not.Roy.Astron.Soc. 489,
  4860--4861 (2019)].
  {\href{https://arxiv.org/abs/1810.03623}{{arXiv:1810.03623}}} {[astro-ph.HE]}

\bibitem[{Derrick(1964)}]{Derrick:1964ww}
Derrick G (1964) {Comments on nonlinear wave equations as models for elementary
  particles}. J Math Phys 5:1252--1254. \doi{10.1063/1.1704233}

\bibitem[{Desjacques and Riotto(2018)}]{Desjacques:2018wuu}
Desjacques V, Riotto A (2018) {Spatial clustering of primordial black holes}.
  Phys Rev D 98(12):123533. \doi{10.1103/PhysRevD.98.123533}.
  {\href{https://arxiv.org/abs/1806.10414}{{arXiv:1806.10414}}} {[astro-ph.CO]}

\bibitem[{Destounis et~al.(2021)Destounis, Suvorov, and
  Kokkotas}]{Destounis:2021mqv}
Destounis K, Suvorov AG, Kokkotas KD (2021) {Gravitational-wave glitches in
  chaotic extreme-mass-ratio inspirals}. Phys Rev Lett 126(14):141102.
  \doi{10.1103/PhysRevLett.126.141102}.
  {\href{https://arxiv.org/abs/2103.05643}{{arXiv:2103.05643}}} {[gr-qc]}

\bibitem[{Detweiler(1980{\natexlab{a}})}]{Detweiler:1980gk}
Detweiler SL (1980{\natexlab{a}}) {BLACK HOLES AND GRAVITATIONAL WAVES. III.
  THE RESONANT FREQUENCIES OF ROTATING HOLES}. Astrophys J 239:292--295.
  \doi{10.1086/158109}

\bibitem[{Detweiler(1980{\natexlab{b}})}]{Detweiler:1980uk}
Detweiler SL (1980{\natexlab{b}}) {KLEIN-GORDON EQUATION AND ROTATING BLACK
  HOLES}. Phys Rev D 22:2323--2326. \doi{10.1103/PhysRevD.22.2323}

\bibitem[{Detweiler et~al.(2003)Detweiler, Messaritaki, and
  Whiting}]{Detweiler:2002gi}
Detweiler SL, Messaritaki E, Whiting BF (2003) {Selfforce of a scalar field for
  circular orbits about a Schwarzschild black hole}. Phys Rev D 67:104016.
  \doi{10.1103/PhysRevD.67.104016}.
  {\href{https://arxiv.org/abs/gr-qc/0205079}{{arXiv:gr-qc/0205079}}}

\bibitem[{{Dewdney} et~al.(2009){Dewdney}, {Hall}, {Schilizzi}, and
  {Lazio}}]{2009IEEEP..97.1482D}
{Dewdney} PE, {Hall} PJ, {Schilizzi} RT, {Lazio} TJLW (2009) {The Square
  Kilometre Array}. IEEE Proceedings 97(8):1482--1496.
  \doi{10.1109/JPROC.2009.2021005}

\bibitem[{Dhanpal et~al.(2019)Dhanpal, Ghosh, Mehta, Ajith, and
  Sathyaprakash}]{Dhanpal:2018ufk}
Dhanpal S, Ghosh A, Mehta AK, Ajith P, Sathyaprakash BS (2019) {A no-hair test
  for binary black holes}. Phys Rev D99(10):104056.
  \doi{10.1103/PhysRevD.99.104056}.
  {\href{https://arxiv.org/abs/1804.03297}{{arXiv:1804.03297}}} {[gr-qc]}

\bibitem[{Di~Giovanni et~al.(2018)Di~Giovanni, Sanchis-Gual, Herdeiro, and
  Font}]{DiGiovanni:2018bvo}
Di~Giovanni F, Sanchis-Gual N, Herdeiro CA, Font JA (2018) {Dynamical formation
  of Proca stars and quasistationary solitonic objects}. Phys Rev D
  98(6):064044. \doi{10.1103/PhysRevD.98.064044}.
  {\href{https://arxiv.org/abs/1803.04802}{{arXiv:1803.04802}}} {[gr-qc]}

\bibitem[{Diaz-Rivera et~al.(2004)Diaz-Rivera, Messaritaki, Whiting, and
  Detweiler}]{DiazRivera:2004ik}
Diaz-Rivera LM, Messaritaki E, Whiting BF, Detweiler SL (2004) {Scalar field
  self-force effects on orbits about a Schwarzschild black hole}. Phys Rev D
  70:124018. \doi{10.1103/PhysRevD.70.124018}.
  {\href{https://arxiv.org/abs/gr-qc/0410011}{{arXiv:gr-qc/0410011}}}

\bibitem[{Diener et~al.(2012)Diener, Vega, Wardell, and
  Detweiler}]{Diener:2011cc}
Diener P, Vega I, Wardell B, Detweiler S (2012) {Self-consistent orbital
  evolution of a particle around a Schwarzschild black hole}. Phys Rev Lett
  108:191102. \doi{10.1103/PhysRevLett.108.191102}.
  {\href{https://arxiv.org/abs/1112.4821}{{arXiv:1112.4821}}} {[gr-qc]}

\bibitem[{Dietrich et~al.(2017)Dietrich, Bernuzzi, and
  Tichy}]{Dietrich:2017aum}
Dietrich T, Bernuzzi S, Tichy W (2017) {Closed-form tidal approximants for
  binary neutron star gravitational waveforms constructed from high-resolution
  numerical relativity simulations}. Phys Rev D 96(12):121501.
  \doi{10.1103/PhysRevD.96.121501}.
  {\href{https://arxiv.org/abs/1706.02969}{{arXiv:1706.02969}}} {[gr-qc]}

\bibitem[{Dietrich et~al.(2019{\natexlab{a}})Dietrich, Samajdar, Khan,
  Johnson-McDaniel, Dudi, and Tichy}]{Dietrich:2019kaq}
Dietrich T, Samajdar A, Khan S, Johnson-McDaniel NK, Dudi R, Tichy W
  (2019{\natexlab{a}}) {Improving the NRTidal model for binary neutron star
  systems}. Phys Rev D 100(4):044003. \doi{10.1103/PhysRevD.100.044003}.
  {\href{https://arxiv.org/abs/1905.06011}{{arXiv:1905.06011}}} {[gr-qc]}

\bibitem[{Dietrich et~al.(2019{\natexlab{b}})}]{Dietrich:2018uni}
Dietrich T, et~al. (2019{\natexlab{b}}) {Matter imprints in waveform models for
  neutron star binaries: Tidal and self-spin effects}. Phys Rev D 99(2):024029.
  \doi{10.1103/PhysRevD.99.024029}.
  {\href{https://arxiv.org/abs/1804.02235}{{arXiv:1804.02235}}} {[gr-qc]}

\bibitem[{Dima and Vernizzi(2018)}]{Dima:2017pwp}
Dima A, Vernizzi F (2018) {Vainshtein Screening in Scalar-Tensor Theories
  before and after GW170817: Constraints on Theories beyond Horndeski}. Phys
  Rev D97(10):101302. \doi{10.1103/PhysRevD.97.101302}.
  {\href{https://arxiv.org/abs/1712.04731}{{arXiv:1712.04731}}} {[gr-qc]}

\bibitem[{Dima et~al.(2020)Dima, Barausse, Franchini, and
  Sotiriou}]{Dima:2020yac}
Dima A, Barausse E, Franchini N, Sotiriou TP (2020) {Spin-induced black hole
  spontaneous scalarization}
  {\href{https://arxiv.org/abs/2006.03095}{{arXiv:2006.03095}}} {[gr-qc]}

\bibitem[{Dirian et~al.(2014)Dirian, Foffa, Khosravi, Kunz, and
  Maggiore}]{Dirian:2014ara}
Dirian Y, Foffa S, Khosravi N, Kunz M, Maggiore M (2014) {Cosmological
  perturbations and structure formation in nonlocal infrared modifications of
  general relativity}. JCAP 1406:033. \doi{10.1088/1475-7516/2014/06/033}.
  {\href{https://arxiv.org/abs/1403.6068}{{arXiv:1403.6068}}} {[astro-ph.CO]}

\bibitem[{Dirian et~al.(2016)Dirian, Foffa, Kunz, Maggiore, and
  Pettorino}]{Dirian:2016puz}
Dirian Y, Foffa S, Kunz M, Maggiore M, Pettorino V (2016) {Non-local gravity
  and comparison with observational datasets. II. Updated results and Bayesian
  model comparison with $\Lambda$CDM}. JCAP 1605:068.
  \doi{10.1088/1475-7516/2016/05/068}.
  {\href{https://arxiv.org/abs/1602.03558}{{arXiv:1602.03558}}} {[astro-ph.CO]}

\bibitem[{Dolan(2007)}]{Dolan:2007mj}
Dolan SR (2007) {Instability of the massive Klein-Gordon field on the Kerr
  spacetime}. Phys Rev D 76:084001. \doi{10.1103/PhysRevD.76.084001}.
  {\href{https://arxiv.org/abs/0705.2880}{{arXiv:0705.2880}}} {[gr-qc]}

\bibitem[{Dolan(2018)}]{Dolan:2018dqv}
Dolan SR (2018) {Instability of the Proca field on Kerr spacetime}. Phys Rev D
  98(10):104006. \doi{10.1103/PhysRevD.98.104006}.
  {\href{https://arxiv.org/abs/1806.01604}{{arXiv:1806.01604}}} {[gr-qc]}

\bibitem[{Domcke et~al.(2020)Domcke, Jinno, and Rubira}]{Domcke:2020xmn}
Domcke V, Jinno R, Rubira H (2020) {Deformation of the gravitational wave
  spectrum by density perturbations}. JCAP 06:046.
  \doi{10.1088/1475-7516/2020/06/046}.
  {\href{https://arxiv.org/abs/2002.11083}{{arXiv:2002.11083}}} {[astro-ph.CO]}

\bibitem[{Dom\`enech and Pi(2020)}]{Domenech:2020ers}
Dom\`enech G, Pi S (2020) {NANOGrav Hints on Planet-Mass Primordial Black
  Holes} {\href{https://arxiv.org/abs/2010.03976}{{arXiv:2010.03976}}}
  {[astro-ph.CO]}

\bibitem[{Domènech(2019)}]{Domenech:2019quo}
Domènech G (2019) {Induced gravitational waves in a general cosmological
  background} \doi{10.1142/S0218271820500285}.
  {\href{https://arxiv.org/abs/1912.05583}{{arXiv:1912.05583}}} {[gr-qc]}

\bibitem[{Domènech et~al.(2020)Domènech, Pi, and Sasaki}]{Domenech:2020kqm}
Domènech G, Pi S, Sasaki M (2020) {Induced gravitational waves as a probe of
  thermal history of the universe}
  {\href{https://arxiv.org/abs/2005.12314}{{arXiv:2005.12314}}} {[gr-qc]}

\bibitem[{Doneva and Yazadjiev(2020)}]{Doneva:2020dji}
Doneva D, Yazadjiev S (2020) {No-hair theorems for non-canonical
  self-gravitating static multiple scalar fields}
  {\href{https://arxiv.org/abs/2008.01965}{{arXiv:2008.01965}}} {[gr-qc]}

\bibitem[{Doneva and Yazadjiev(2018)}]{Doneva:2017bvd}
Doneva DD, Yazadjiev SS (2018) {New Gauss-Bonnet Black Holes with
  Curvature-Induced Scalarization in Extended Scalar-Tensor Theories}. Phys Rev
  Lett 120(13):131103. \doi{10.1103/PhysRevLett.120.131103}.
  {\href{https://arxiv.org/abs/1711.01187}{{arXiv:1711.01187}}} {[gr-qc]}

\bibitem[{Doneva et~al.(2020{\natexlab{a}})Doneva, Collodel, Kr\"uger, and
  Yazadjiev}]{Doneva:2020nbb}
Doneva DD, Collodel LG, Kr\"uger CJ, Yazadjiev SS (2020{\natexlab{a}}) {Black
  hole scalarization induced by the spin: 2+1 time evolution}. Phys Rev D
  102(10):104027. \doi{10.1103/PhysRevD.102.104027}.
  {\href{https://arxiv.org/abs/2008.07391}{{arXiv:2008.07391}}} {[gr-qc]}

\bibitem[{Doneva et~al.(2020{\natexlab{b}})Doneva, Collodel, Kr\"uger, and
  Yazadjiev}]{Doneva:2020kfv}
Doneva DD, Collodel LG, Kr\"uger CJ, Yazadjiev SS (2020{\natexlab{b}})
  {Spin-induced scalarization of Kerr black holes with a massive scalar field}.
  Eur Phys J C 80(12):1205. \doi{10.1140/epjc/s10052-020-08765-3}.
  {\href{https://arxiv.org/abs/2009.03774}{{arXiv:2009.03774}}} {[gr-qc]}

\bibitem[{Dreyer et~al.(2004)Dreyer, Kelly, Krishnan, Finn, Garrison, and
  Lopez-Aleman}]{Dreyer:2003bv}
Dreyer O, Kelly BJ, Krishnan B, Finn LS, Garrison D, Lopez-Aleman R (2004)
  {Black hole spectroscopy: Testing general relativity through gravitational
  wave observations}. Class Quant Grav 21:787--804.
  \doi{10.1088/0264-9381/21/4/003}.
  {\href{https://arxiv.org/abs/gr-qc/0309007}{{arXiv:gr-qc/0309007}}}

\bibitem[{Duffell et~al.(2014)Duffell, Haiman, MacFadyen, D'Orazio, and
  Farris}]{Duffell:2014jma}
Duffell PC, Haiman Z, MacFadyen AI, D'Orazio DJ, Farris BD (2014) {The
  Migration of Gap-Opening Planets is not Locked to Viscous Disk Evolution}.
  Astrophys J Lett 792:L10. \doi{10.1088/2041-8205/792/1/L10}.
  {\href{https://arxiv.org/abs/1405.3711}{{arXiv:1405.3711}}} {[astro-ph.EP]}

\bibitem[{{D{\"u}rmann} and {Kley}(2017)}]{2017A&A...598A..80D}
{D{\"u}rmann} C, {Kley} W (2017) {The accretion of migrating giant planets}.
  \aap 598:A80. \doi{10.1051/0004-6361/201629074}.
  {\href{https://arxiv.org/abs/1611.01070}{{arXiv:1611.01070}}} {[astro-ph.EP]}

\bibitem[{East(2017)}]{East:2017mrj}
East WE (2017) {Superradiant instability of massive vector fields around
  spinning black holes in the relativistic regime}. Phys Rev D 96(2):024004.
  \doi{10.1103/PhysRevD.96.024004}.
  {\href{https://arxiv.org/abs/1705.01544}{{arXiv:1705.01544}}} {[gr-qc]}

\bibitem[{East(2018)}]{East:2018glu}
East WE (2018) {Massive Boson Superradiant Instability of Black Holes:
  Nonlinear Growth, Saturation, and Gravitational Radiation}. Phys Rev Lett
  121(13):131104. \doi{10.1103/PhysRevLett.121.131104}.
  {\href{https://arxiv.org/abs/1807.00043}{{arXiv:1807.00043}}} {[gr-qc]}

\bibitem[{East and Pretorius(2017)}]{East:2017ovw}
East WE, Pretorius F (2017) {Superradiant Instability and Backreaction of
  Massive Vector Fields around Kerr Black Holes}. Phys Rev Lett 119(4):041101.
  \doi{10.1103/PhysRevLett.119.041101}.
  {\href{https://arxiv.org/abs/1704.04791}{{arXiv:1704.04791}}} {[gr-qc]}

\bibitem[{East and Ripley(2021{\natexlab{a}})}]{East:2021bqk}
East WE, Ripley JL (2021{\natexlab{a}}) {Dynamics of spontaneous black hole
  scalarization and mergers in Einstein-scalar-Gauss-Bonnet gravity}
  {\href{https://arxiv.org/abs/2105.08571}{{arXiv:2105.08571}}} {[gr-qc]}

\bibitem[{East and Ripley(2021{\natexlab{b}})}]{East:2020hgw}
East WE, Ripley JL (2021{\natexlab{b}}) {Evolution of
  Einstein-scalar-Gauss-Bonnet gravity using a modified harmonic formulation}.
  Phys Rev D 103(4):044040. \doi{10.1103/PhysRevD.103.044040}.
  {\href{https://arxiv.org/abs/2011.03547}{{arXiv:2011.03547}}} {[gr-qc]}

\bibitem[{Eda et~al.(2015)Eda, Itoh, Kuroyanagi, and Silk}]{Eda:2014kra}
Eda K, Itoh Y, Kuroyanagi S, Silk J (2015) {Gravitational waves as a probe of
  dark matter minispikes}. Phys Rev D 91(4):044045.
  \doi{10.1103/PhysRevD.91.044045}.
  {\href{https://arxiv.org/abs/1408.3534}{{arXiv:1408.3534}}} {[gr-qc]}

\bibitem[{Edwards et~al.(2020)Edwards, Chianese, Kavanagh, Nissanke, and
  Weniger}]{Edwards:2019tzf}
Edwards TD, Chianese M, Kavanagh BJ, Nissanke SM, Weniger C (2020) {Unique
  Multimessenger Signal of QCD Axion Dark Matter}. Phys Rev Lett
  124(16):161101. \doi{10.1103/PhysRevLett.124.161101}.
  {\href{https://arxiv.org/abs/1905.04686}{{arXiv:1905.04686}}} {[hep-ph]}

\bibitem[{Eling and Jacobson(2006)}]{Eling:2006ec}
Eling C, Jacobson T (2006) {Black Holes in Einstein-Aether Theory}. Class Quant
  Grav 23:5643--5660. \doi{10.1088/0264-9381/23/18/009}, [Erratum:
  Class.Quant.Grav. 27, 049802 (2010)].
  {\href{https://arxiv.org/abs/gr-qc/0604088}{{arXiv:gr-qc/0604088}}}

\bibitem[{Emir~Gümrükçüoğlu et~al.(2018)Emir~Gümrükçüoğlu, Saravani,
  and Sotiriou}]{Gumrukcuoglu:2017ijh}
Emir~Gümrükçüoğlu A, Saravani M, Sotiriou TP (2018) {Hořava gravity after
  GW170817}. Phys Rev D97(2):024032. \doi{10.1103/PhysRevD.97.024032}.
  {\href{https://arxiv.org/abs/1711.08845}{{arXiv:1711.08845}}} {[gr-qc]}

\bibitem[{Espinosa et~al.(2018)Espinosa, Racco, and Riotto}]{Espinosa:2018eve}
Espinosa JR, Racco D, Riotto A (2018) {A Cosmological Signature of the SM Higgs
  Instability: Gravitational Waves}. JCAP 09:012.
  \doi{10.1088/1475-7516/2018/09/012}.
  {\href{https://arxiv.org/abs/1804.07732}{{arXiv:1804.07732}}} {[hep-ph]}

\bibitem[{Essig et~al.(2013)}]{Essig:2013lka}
Essig R, et~al. (2013) {Working Group Report: New Light Weakly Coupled
  Particles}. In: {Community Summer Study 2013}: {Snowmass on the Mississippi}.
  {\href{https://arxiv.org/abs/1311.0029}{{arXiv:1311.0029}}} {[hep-ph]}

\bibitem[{Estell\'es et~al.(2021)Estell\'es, Ramos-Buades, Husa,
  Garc\'\i{}a-Quir\'os, Colleoni, Haegel, and Jaume}]{Estelles:2020osj}
Estell\'es H, Ramos-Buades A, Husa S, Garc\'\i{}a-Quir\'os C, Colleoni M,
  Haegel L, Jaume R (2021) {Phenomenological time domain model for dominant
  quadrupole gravitational wave signal of coalescing binary black holes}. Phys
  Rev D 103(12):124060. \doi{10.1103/PhysRevD.103.124060}.
  {\href{https://arxiv.org/abs/2004.08302}{{arXiv:2004.08302}}} {[gr-qc]}

\bibitem[{Etienne et~al.(2014)Etienne, Baker, Paschalidis, Kelly, and
  Shapiro}]{Etienne:2014tia}
Etienne ZB, Baker JG, Paschalidis V, Kelly BJ, Shapiro SL (2014) {Improved
  Moving Puncture Gauge Conditions for Compact Binary Evolutions}. Phys Rev D
  90(6):064032. \doi{10.1103/PhysRevD.90.064032}.
  {\href{https://arxiv.org/abs/1404.6523}{{arXiv:1404.6523}}} {[astro-ph.HE]}

\bibitem[{Ezquiaga(2021)}]{Ezquiaga:2021ayr}
Ezquiaga JM (2021) {Hearing gravity from the cosmos: GWTC-2 probes general
  relativity at cosmological scales}
  {\href{https://arxiv.org/abs/2104.05139}{{arXiv:2104.05139}}} {[astro-ph.CO]}

\bibitem[{Ezquiaga and García-Bellido(2018)}]{Ezquiaga:2018gbw}
Ezquiaga JM, García-Bellido J (2018) {Quantum diffusion beyond slow-roll:
  implications for primordial black-hole production}. JCAP 1808:018.
  \doi{10.1088/1475-7516/2018/08/018}.
  {\href{https://arxiv.org/abs/1805.06731}{{arXiv:1805.06731}}} {[astro-ph.CO]}

\bibitem[{Ezquiaga and Holz(2020)}]{Ezquiaga:2020tns}
Ezquiaga JM, Holz DE (2020) {Jumping the gap: searching for LIGO's biggest
  black holes} {\href{https://arxiv.org/abs/2006.02211}{{arXiv:2006.02211}}}
  {[astro-ph.HE]}

\bibitem[{Ezquiaga and Zumalac\'arregui(2017)}]{Ezquiaga:2017ekz}
Ezquiaga JM, Zumalac\'arregui M (2017) {Dark Energy After GW170817: Dead Ends
  and the Road Ahead}. Phys Rev Lett 119:251304.
  \doi{10.1103/PhysRevLett.119.251304}.
  {\href{https://arxiv.org/abs/1710.05901}{{arXiv:1710.05901}}} {[astro-ph.CO]}

\bibitem[{Ezquiaga and Zumalac\'arregui(2020)}]{Ezquiaga:2020dao}
Ezquiaga JM, Zumalac\'arregui M (2020) {Gravitational wave lensing beyond
  general relativity: birefringence, echoes and shadows}
  {\href{https://arxiv.org/abs/2009.12187}{{arXiv:2009.12187}}} {[gr-qc]}

\bibitem[{Ezquiaga et~al.(2018)Ezquiaga, Garc\'ia-Bellido, and
  Ruiz~Morales}]{Ezquiaga:2017fvi}
Ezquiaga JM, Garc\'ia-Bellido J, Ruiz~Morales E (2018) {Primordial Black Hole
  production in Critical Higgs Inflation}. Phys Lett B776:345--349.
  \doi{10.1016/j.physletb.2017.11.039}.
  {\href{https://arxiv.org/abs/1705.04861}{{arXiv:1705.04861}}} {[astro-ph.CO]}

\bibitem[{Ezquiaga et~al.(2020{\natexlab{a}})Ezquiaga, García-Bellido, and
  Vennin}]{Ezquiaga:2019ftu}
Ezquiaga JM, García-Bellido J, Vennin V (2020{\natexlab{a}}) {The exponential
  tail of inflationary fluctuations: consequences for primordial black holes}.
  JCAP 2003:029. \doi{10.1088/1475-7516/2020/03/029}.
  {\href{https://arxiv.org/abs/1912.05399}{{arXiv:1912.05399}}} {[astro-ph.CO]}

\bibitem[{Ezquiaga et~al.(2020{\natexlab{b}})Ezquiaga, Holz, Hu, Lagos, and
  Wald}]{Ezquiaga:2020gdt}
Ezquiaga JM, Holz DE, Hu W, Lagos M, Wald RM (2020{\natexlab{b}}) {Phase
  effects from strong gravitational lensing of gravitational waves}
  {\href{https://arxiv.org/abs/2008.12814}{{arXiv:2008.12814}}} {[gr-qc]}

\bibitem[{Ezquiaga et~al.(2020{\natexlab{c}})Ezquiaga, Hu, and
  Lagos}]{Ezquiaga:2020spg}
Ezquiaga JM, Hu W, Lagos M (2020{\natexlab{c}}) {Apparent Superluminality of
  Lensed Gravitational Waves}. Phys Rev D 102(2):023531.
  \doi{10.1103/PhysRevD.102.023531}.
  {\href{https://arxiv.org/abs/2005.10702}{{arXiv:2005.10702}}} {[astro-ph.CO]}

\bibitem[{Ezquiaga et~al.(2021)Ezquiaga, Hu, Lagos, and Lin}]{Ezquiaga:2021ler}
Ezquiaga JM, Hu W, Lagos M, Lin MX (2021) {Gravitational wave propagation
  beyond general relativity: waveform distortions and echoes}
  {\href{https://arxiv.org/abs/2108.10872}{{arXiv:2108.10872}}} {[astro-ph.CO]}

\bibitem[{Fan et~al.(2013)Fan, Katz, Randall, and Reece}]{Fan:2013tia}
Fan J, Katz A, Randall L, Reece M (2013) {Dark-Disk Universe}. Phys Rev Lett
  110(21):211302. \doi{10.1103/PhysRevLett.110.211302}.
  {\href{https://arxiv.org/abs/1303.3271}{{arXiv:1303.3271}}} {[hep-ph]}

\bibitem[{Farmer and Phinney(2003)}]{Farmer:2003pa}
Farmer AJ, Phinney E (2003) {The gravitational wave background from
  cosmological compact binaries}. Mon Not Roy Astron Soc 346:1197.
  \doi{10.1111/j.1365-2966.2003.07176.x}.
  {\href{https://arxiv.org/abs/astro-ph/0304393}{{arXiv:astro-ph/0304393}}}

\bibitem[{Farr et~al.(2019)Farr, Fishbach, Ye, and Holz}]{Farr:2019twy}
Farr WM, Fishbach M, Ye J, Holz D (2019) {A Future Percent-Level Measurement of
  the Hubble Expansion at Redshift 0.8 With Advanced LIGO}. Astrophys J Lett
  883(2):L42. \doi{10.3847/2041-8213/ab4284}.
  {\href{https://arxiv.org/abs/1908.09084}{{arXiv:1908.09084}}} {[astro-ph.CO]}

\bibitem[{Farrugia et~al.(2018)Farrugia, Levi~Said, Gakis, and
  Saridakis}]{Farrugia:2018gyz}
Farrugia G, Levi~Said J, Gakis V, Saridakis EN (2018) {Gravitational Waves in
  Modified Teleparallel Theories}. Phys Rev D 97(12):124064.
  \doi{10.1103/PhysRevD.97.124064}.
  {\href{https://arxiv.org/abs/1804.07365}{{arXiv:1804.07365}}} {[gr-qc]}

\bibitem[{Favata(2009{\natexlab{a}})}]{BMS10}
Favata M (2009{\natexlab{a}}) {Nonlinear gravitational-wave memory from binary
  black hole mergers}. Astrophys J Lett 696:L159.
  \doi{10.1088/0004-637X/696/2/L159}.
  {\href{https://arxiv.org/abs/0902.3660}{{arXiv:0902.3660}}} {[astro-ph.SR]}

\bibitem[{Favata(2009{\natexlab{b}})}]{BMS9}
Favata M (2009{\natexlab{b}}) {Post-Newtonian corrections to the
  gravitational-wave memory for quasi-circular, inspiralling compact binaries}.
  Phys Rev D 80:024002. \doi{10.1103/PhysRevD.80.024002}.
  {\href{https://arxiv.org/abs/0812.0069}{{arXiv:0812.0069}}} {[gr-qc]}

\bibitem[{Favata(2010)}]{BMS11}
Favata M (2010) {The gravitational-wave memory effect}. Class Quant Grav
  27:084036. \doi{10.1088/0264-9381/27/8/084036}.
  {\href{https://arxiv.org/abs/1003.3486}{{arXiv:1003.3486}}} {[gr-qc]}

\bibitem[{Favata(2011)}]{BMS12}
Favata M (2011) {The Gravitational-wave memory from eccentric binaries}. Phys
  Rev D 84:124013. \doi{10.1103/PhysRevD.84.124013}.
  {\href{https://arxiv.org/abs/1108.3121}{{arXiv:1108.3121}}} {[gr-qc]}

\bibitem[{Ferguson et~al.(2020)Ferguson, Jani, Laguna, and
  Shoemaker}]{Ferguson:2020xnm}
Ferguson D, Jani K, Laguna P, Shoemaker D (2020) {Assessing the Readiness of
  Numerical Relativity for LISA and 3G Detectors}
  {\href{https://arxiv.org/abs/2006.04272}{{arXiv:2006.04272}}} {[gr-qc]}

\bibitem[{Fernandes et~al.(2019)Fernandes, Herdeiro, Pombo, Radu, and
  Sanchis-Gual}]{Fernandes:2019rez}
Fernandes PGS, Herdeiro CAR, Pombo AM, Radu E, Sanchis-Gual N (2019)
  {Spontaneous Scalarisation of Charged Black Holes: Coupling Dependence and
  Dynamical Features}. Class Quant Grav 36(13):134002.
  \doi{10.1088/1361-6382/ab23a1}, [Erratum: Class.Quant.Grav. 37, 049501
  (2020)]. {\href{https://arxiv.org/abs/1902.05079}{{arXiv:1902.05079}}}
  {[gr-qc]}

\bibitem[{Ferrari and Kokkotas(2000)}]{Ferrari:2000sr}
Ferrari V, Kokkotas KD (2000) {Scattering of particles by neutron stars: Time
  evolutions for axial perturbations}. Phys Rev D 62:107504.
  \doi{10.1103/PhysRevD.62.107504}.
  {\href{https://arxiv.org/abs/gr-qc/0008057}{{arXiv:gr-qc/0008057}}}

\bibitem[{Ferrari et~al.(1999)Ferrari, Matarrese, and
  Schneider}]{Ferrari:1998jf}
Ferrari V, Matarrese S, Schneider R (1999) {Stochastic background of
  gravitational waves generated by a cosmological population of young, rapidly
  rotating neutron stars}. Mon Not Roy Astron Soc 303:258.
  \doi{10.1046/j.1365-8711.1999.02207.x}.
  {\href{https://arxiv.org/abs/astro-ph/9806357}{{arXiv:astro-ph/9806357}}}

\bibitem[{Ferraro and Fiorini(2007)}]{Ferraro:2006jd}
Ferraro R, Fiorini F (2007) {Modified teleparallel gravity: Inflation without
  inflaton}. Phys Rev D 75:084031. \doi{10.1103/PhysRevD.75.084031}.
  {\href{https://arxiv.org/abs/gr-qc/0610067}{{arXiv:gr-qc/0610067}}}

\bibitem[{Ferraro and Guzm\'an(2018{\natexlab{a}})}]{Ferraro:2018tpu}
Ferraro R, Guzm\'an MJ (2018{\natexlab{a}}) {Hamiltonian formalism for f(T)
  gravity}. Phys Rev D 97(10):104028. \doi{10.1103/PhysRevD.97.104028}.
  {\href{https://arxiv.org/abs/1802.02130}{{arXiv:1802.02130}}} {[gr-qc]}

\bibitem[{Ferraro and Guzm\'an(2018{\natexlab{b}})}]{Ferraro:2018axk}
Ferraro R, Guzm\'an MJ (2018{\natexlab{b}}) {Quest for the extra degree of
  freedom in $f(T)$ gravity}. Phys Rev D 98(12):124037.
  \doi{10.1103/PhysRevD.98.124037}.
  {\href{https://arxiv.org/abs/1810.07171}{{arXiv:1810.07171}}} {[gr-qc]}

\bibitem[{Ferreira(2020)}]{Ferreira:2020fam}
Ferreira EG (2020) {Ultra-Light Dark Matter}
  {\href{https://arxiv.org/abs/2005.03254}{{arXiv:2005.03254}}} {[astro-ph.CO]}

\bibitem[{Ferreira et~al.(2017)Ferreira, Macedo, and
  Cardoso}]{Ferreira:2017pth}
Ferreira MC, Macedo CFB, Cardoso V (2017) {Orbital fingerprints of ultralight
  scalar fields around black holes}. Phys Rev D 96(8):083017.
  \doi{10.1103/PhysRevD.96.083017}.
  {\href{https://arxiv.org/abs/1710.00830}{{arXiv:1710.00830}}} {[gr-qc]}

\bibitem[{Ferrer et~al.(2017)Ferrer, da~Rosa, and Will}]{Ferrer:2017xwm}
Ferrer F, da~Rosa AM, Will CM (2017) {Dark matter spikes in the vicinity of
  Kerr black holes}. Phys Rev D 96(8):083014. \doi{10.1103/PhysRevD.96.083014}.
  {\href{https://arxiv.org/abs/1707.06302}{{arXiv:1707.06302}}} {[astro-ph.CO]}

\bibitem[{Ficarra et~al.(2019)Ficarra, Pani, and Witek}]{Ficarra:2018rfu}
Ficarra G, Pani P, Witek H (2019) {Impact of multiple modes on the black-hole
  superradiant instability}. Phys Rev D 99(10):104019.
  \doi{10.1103/PhysRevD.99.104019}.
  {\href{https://arxiv.org/abs/1812.02758}{{arXiv:1812.02758}}} {[gr-qc]}

\bibitem[{Field et~al.(2014)Field, Galley, Hesthaven, Kaye, and
  Tiglio}]{Field:2013cfa}
Field SE, Galley CR, Hesthaven JS, Kaye J, Tiglio M (2014) {Fast prediction and
  evaluation of gravitational waveforms using surrogate models}. Phys Rev X
  4(3):031006. \doi{10.1103/PhysRevX.4.031006}.
  {\href{https://arxiv.org/abs/1308.3565}{{arXiv:1308.3565}}} {[gr-qc]}

\bibitem[{Finke et~al.(2021)Finke, Foffa, Iacovelli, Maggiore, and
  Mancarella}]{Finke:2021aom}
Finke A, Foffa S, Iacovelli F, Maggiore M, Mancarella M (2021) {Cosmology with
  LIGO/Virgo dark sirens: Hubble parameter and modified gravitational wave
  propagation} {\href{https://arxiv.org/abs/2101.12660}{{arXiv:2101.12660}}}
  {[astro-ph.CO]}

\bibitem[{Fishbach et~al.(2019)}]{Fishbach:2018gjp}
Fishbach M, et~al. (2019) {A Standard Siren Measurement of the Hubble Constant
  from GW170817 without the Electromagnetic Counterpart}. Astrophys J Lett
  871:L13. \doi{10.3847/2041-8213/aaf96e}.
  {\href{https://arxiv.org/abs/1807.05667}{{arXiv:1807.05667}}} {[astro-ph.CO]}

\bibitem[{Flanagan(2004)}]{Flanagan:2003rb}
Flanagan EE (2004) {Palatini form of 1/R gravity}. Phys Rev Lett 92:071101.
  \doi{10.1103/PhysRevLett.92.071101}.
  {\href{https://arxiv.org/abs/astro-ph/0308111}{{arXiv:astro-ph/0308111}}}

\bibitem[{Flanagan and Hinderer(2008)}]{Flanagan:2007ix}
Flanagan EE, Hinderer T (2008) {Constraining neutron star tidal Love numbers
  with gravitational wave detectors}. Phys Rev D77:021502.
  \doi{10.1103/PhysRevD.77.021502}.
  {\href{https://arxiv.org/abs/0709.1915}{{arXiv:0709.1915}}} {[astro-ph]}

\bibitem[{Flanagan et~al.(2019)Flanagan, Grant, Harte, and Nichols}]{BMS16}
Flanagan EE, Grant AM, Harte AI, Nichols DA (2019) {Persistent gravitational
  wave observables: general framework}. Phys Rev D 99:084044.
  \doi{10.1103/PhysRevD.99.084044}.
  {\href{https://arxiv.org/abs/1901.00021}{{arXiv:1901.00021}}} {[gr-qc]}

\bibitem[{Flauger et~al.(2021)Flauger, Karnesis, Nardini, Pieroni, Ricciardone,
  and Torrado}]{Flauger:2020qyi}
Flauger R, Karnesis N, Nardini G, Pieroni M, Ricciardone A, Torrado J (2021)
  {Improved reconstruction of a stochastic gravitational wave background with
  LISA}. JCAP 01:059. \doi{10.1088/1475-7516/2021/01/059}.
  {\href{https://arxiv.org/abs/2009.11845}{{arXiv:2009.11845}}} {[astro-ph.CO]}

\bibitem[{Fleury et~al.(2017)Fleury, Clarkson, and Maartens}]{Fleury:2016fda}
Fleury P, Clarkson C, Maartens R (2017) {How does the cosmic large-scale
  structure bias the Hubble diagram?} JCAP 03:062.
  \doi{10.1088/1475-7516/2017/03/062}.
  {\href{https://arxiv.org/abs/1612.03726}{{arXiv:1612.03726}}} {[astro-ph.CO]}

\bibitem[{Foffa et~al.(2019)Foffa, Porto, Rothstein, and
  Sturani}]{Foffa:2019yfl}
Foffa S, Porto RA, Rothstein I, Sturani R (2019) {Conservative dynamics of
  binary systems to fourth Post-Newtonian order in the EFT approach II:
  Renormalized Lagrangian}. Phys Rev D 100(2):024048.
  \doi{10.1103/PhysRevD.100.024048}.
  {\href{https://arxiv.org/abs/1903.05118}{{arXiv:1903.05118}}} {[gr-qc]}

\bibitem[{Foster(2006)}]{Foster:2006az}
Foster BZ (2006) {Radiation damping in Einstein-aether theory}. Phys Rev D
  73:104012. \doi{10.1103/PhysRevD.75.129904}, [Erratum: Phys.Rev.D 75, 129904
  (2007)]. {\href{https://arxiv.org/abs/gr-qc/0602004}{{arXiv:gr-qc/0602004}}}

\bibitem[{Foster(2007)}]{Foster:2007gr}
Foster BZ (2007) {Strong field effects on binary systems in Einstein-aether
  theory}. Phys Rev D 76:084033. \doi{10.1103/PhysRevD.76.084033}.
  {\href{https://arxiv.org/abs/0706.0704}{{arXiv:0706.0704}}} {[gr-qc]}

\bibitem[{Franchini and Sotiriou(2020)}]{Franchini:2019npi}
Franchini N, Sotiriou TP (2020) {Cosmology with subdominant Horndeski scalar
  field}. Phys Rev D 101(6):064068. \doi{10.1103/PhysRevD.101.064068}.
  {\href{https://arxiv.org/abs/1903.05427}{{arXiv:1903.05427}}} {[gr-qc]}

\bibitem[{Franciolini et~al.(2018)Franciolini, Kehagias, Matarrese, and
  Riotto}]{Franciolini:2018vbk}
Franciolini G, Kehagias A, Matarrese S, Riotto A (2018) {Primordial Black Holes
  from Inflation and non-Gaussianity}. JCAP 03:016.
  \doi{10.1088/1475-7516/2018/03/016}.
  {\href{https://arxiv.org/abs/1801.09415}{{arXiv:1801.09415}}} {[astro-ph.CO]}

\bibitem[{Franciolini et~al.(2019)Franciolini, Hui, Penco, Santoni, and
  Trincherini}]{Franciolini:2018uyq}
Franciolini G, Hui L, Penco R, Santoni L, Trincherini E (2019) {Effective Field
  Theory of Black Hole Quasinormal Modes in Scalar-Tensor Theories}. JHEP
  02:127. \doi{10.1007/JHEP02(2019)127}.
  {\href{https://arxiv.org/abs/1810.07706}{{arXiv:1810.07706}}} {[hep-th]}

\bibitem[{Franciolini et~al.(2021)Franciolini, Baibhav, De~Luca, Ng, Wong,
  Berti, Pani, Riotto, and Vitale}]{Franciolini:2021tla}
Franciolini G, Baibhav V, De~Luca V, Ng KKY, Wong KWK, Berti E, Pani P, Riotto
  A, Vitale S (2021) {Quantifying the evidence for primordial black holes in
  LIGO/Virgo gravitational-wave data}
  {\href{https://arxiv.org/abs/2105.03349}{{arXiv:2105.03349}}} {[gr-qc]}

\bibitem[{Fransen and Mayerson(2022)}]{Fransen:2022jtw}
Fransen K, Mayerson DR (2022) {On Detecting Equatorial Symmetry Breaking with
  LISA} {\href{https://arxiv.org/abs/2201.03569}{{arXiv:2201.03569}}} {[gr-qc]}

\bibitem[{Friedman(1978)}]{Friedman:1978wla}
Friedman JL (1978) {Generic instability of rotating relativistic stars}. Commun
  Math Phys 62(3):247--278. \doi{10.1007/BF01202527}

\bibitem[{Friedman and Morsink(1998)}]{Friedman:1997uh}
Friedman JL, Morsink SM (1998) {Axial instability of rotating relativistic
  stars}. Astrophys J 502:714--720. \doi{10.1086/305920}.
  {\href{https://arxiv.org/abs/gr-qc/9706073}{{arXiv:gr-qc/9706073}}}

\bibitem[{Frolov et~al.(2018)Frolov, Krtou\v~s, Kubiz\v~nák, and
  Santos}]{Frolov:2018ezx}
Frolov VP, Krtou\v~s P, Kubiz\v~nák D, Santos JE (2018) {Massive Vector Fields
  in Rotating Black-Hole Spacetimes: Separability and Quasinormal Modes}. Phys
  Rev Lett 120:231103. \doi{10.1103/PhysRevLett.120.231103}.
  {\href{https://arxiv.org/abs/1804.00030}{{arXiv:1804.00030}}} {[hep-th]}

\bibitem[{Frusciante(2021)}]{Frusciante:2021sio}
Frusciante N (2021) {Signatures of $f(Q)$-gravity in cosmology}. Phys Rev D
  103(4):044021. \doi{10.1103/PhysRevD.103.044021}.
  {\href{https://arxiv.org/abs/2101.09242}{{arXiv:2101.09242}}} {[astro-ph.CO]}

\bibitem[{Frusciante and Perenon(2020)}]{Frusciante:2019xia}
Frusciante N, Perenon L (2020) {Effective field theory of dark energy: A
  review}. Phys Rept 857:1--63. \doi{10.1016/j.physrep.2020.02.004}.
  {\href{https://arxiv.org/abs/1907.03150}{{arXiv:1907.03150}}} {[astro-ph.CO]}

\bibitem[{Fujita and Cardoso(2017)}]{Fujita:2016yav}
Fujita R, Cardoso V (2017) {Ultralight scalars and resonances in black-hole
  physics}. Phys Rev D 95(4):044016. \doi{10.1103/PhysRevD.95.044016}.
  {\href{https://arxiv.org/abs/1612.00978}{{arXiv:1612.00978}}} {[gr-qc]}

\bibitem[{Fumagalli et~al.(2020{\natexlab{a}})Fumagalli, Renaux-Petel, Ronayne,
  and Witkowski}]{Fumagalli:2020adf}
Fumagalli J, Renaux-Petel S, Ronayne JW, Witkowski LT (2020{\natexlab{a}})
  {Turning in the landscape: a new mechanism for generating Primordial Black
  Holes} {\href{https://arxiv.org/abs/2004.08369}{{arXiv:2004.08369}}}
  {[hep-th]}

\bibitem[{Fumagalli et~al.(2020{\natexlab{b}})Fumagalli, Renaux-Petel, and
  Witkowski}]{Fumagalli:2020nvq}
Fumagalli J, Renaux-Petel S, Witkowski LT (2020{\natexlab{b}}) {Oscillations in
  the stochastic gravitational wave background from sharp features and particle
  production during inflation}
  {\href{https://arxiv.org/abs/2012.02761}{{arXiv:2012.02761}}} {[astro-ph.CO]}

\bibitem[{Gaggero et~al.(2017)Gaggero, Bertone, Calore, Connors, Lovell,
  Markoff, and Storm}]{Gaggero:2016dpq}
Gaggero D, Bertone G, Calore F, Connors RMT, Lovell M, Markoff S, Storm E
  (2017) {Searching for Primordial Black Holes in the radio and X-ray sky}.
  Phys Rev Lett 118(24):241101. \doi{10.1103/PhysRevLett.118.241101}.
  {\href{https://arxiv.org/abs/1612.00457}{{arXiv:1612.00457}}} {[astro-ph.HE]}

\bibitem[{Gair et~al.(2008)Gair, Li, and Mandel}]{Gair:2007kr}
Gair JR, Li C, Mandel I (2008) {Observable Properties of Orbits in Exact Bumpy
  Spacetimes}. Phys Rev D 77:024035. \doi{10.1103/PhysRevD.77.024035}.
  {\href{https://arxiv.org/abs/0708.0628}{{arXiv:0708.0628}}} {[gr-qc]}

\bibitem[{Gair et~al.(2010)Gair, Tang, and Volonteri}]{Gair:2010yu}
Gair JR, Tang C, Volonteri M (2010) {LISA extreme-mass-ratio inspiral events as
  probes of the black hole mass function}. Phys Rev D 81:104014.
  \doi{10.1103/PhysRevD.81.104014}.
  {\href{https://arxiv.org/abs/1004.1921}{{arXiv:1004.1921}}} {[astro-ph.GA]}

\bibitem[{Gair et~al.(2017)Gair, Babak, Sesana, Amaro-Seoane, Barausse, Berry,
  Berti, and Sopuerta}]{Gair:2017ynp}
Gair JR, Babak S, Sesana A, Amaro-Seoane P, Barausse E, Berry CPL, Berti E,
  Sopuerta C (2017) {Prospects for observing extreme-mass-ratio inspirals with
  LISA}. J Phys Conf Ser 840(1):012021. \doi{10.1088/1742-6596/840/1/012021}.
  {\href{https://arxiv.org/abs/1704.00009}{{arXiv:1704.00009}}} {[astro-ph.GA]}

\bibitem[{Galley and Rothstein(2017)}]{Galley:2016zee}
Galley CR, Rothstein IZ (2017) {Deriving analytic solutions for compact binary
  inspirals without recourse to adiabatic approximations}. Phys Rev
  D95(10):104054. \doi{10.1103/PhysRevD.95.104054}.
  {\href{https://arxiv.org/abs/1609.08268}{{arXiv:1609.08268}}} {[gr-qc]}

\bibitem[{Gamba et~al.(2020)Gamba, Bernuzzi, and Nagar}]{Gamba:2020ljo}
Gamba R, Bernuzzi S, Nagar A (2020) {Fast, faithful, frequency-domain
  effective-one-body waveforms for compact binary coalescences}
  {\href{https://arxiv.org/abs/2012.00027}{{arXiv:2012.00027}}} {[gr-qc]}

\bibitem[{Ganchev and Santos(2018)}]{Ganchev:2017uuo}
Ganchev B, Santos JE (2018) {Scalar Hairy Black Holes in Four Dimensions are
  Unstable}. Phys Rev Lett 120(17):171101.
  \doi{10.1103/PhysRevLett.120.171101}.
  {\href{https://arxiv.org/abs/1711.08464}{{arXiv:1711.08464}}} {[gr-qc]}

\bibitem[{Garc\'ia-Bellido and Ruiz~Morales(2017)}]{Garcia-Bellido:2017mdw}
Garc\'ia-Bellido J, Ruiz~Morales E (2017) {Primordial black holes from single
  field models of inflation}. Phys Dark Univ 18:47--54.
  \doi{10.1016/j.dark.2017.09.007}.
  {\href{https://arxiv.org/abs/1702.03901}{{arXiv:1702.03901}}} {[astro-ph.CO]}

\bibitem[{Garcia-Bellido et~al.(1996)Garcia-Bellido, Linde, and
  Wands}]{GarciaBellido:1996qt}
Garcia-Bellido J, Linde AD, Wands D (1996) {Density perturbations and black
  hole formation in hybrid inflation}. Phys Rev D 54:6040--6058.
  \doi{10.1103/PhysRevD.54.6040}.
  {\href{https://arxiv.org/abs/astro-ph/9605094}{{arXiv:astro-ph/9605094}}}

\bibitem[{Garc\'ia-Bellido et~al.(2016)Garc\'ia-Bellido, Peloso, and
  Unal}]{Garcia-Bellido:2016dkw}
Garc\'ia-Bellido J, Peloso M, Unal C (2016) {Gravitational waves at
  interferometer scales and primordial black holes in axion inflation}. JCAP
  1612:031. \doi{10.1088/1475-7516/2016/12/031}.
  {\href{https://arxiv.org/abs/1610.03763}{{arXiv:1610.03763}}} {[astro-ph.CO]}

\bibitem[{Garc\'ia-Bellido et~al.(2017)Garc\'ia-Bellido, Peloso, and
  Unal}]{Garcia-Bellido:2017aan}
Garc\'ia-Bellido J, Peloso M, Unal C (2017) {Gravitational Wave signatures of
  inflationary models from Primordial Black Hole Dark Matter}. JCAP 1709:013.
  \doi{10.1088/1475-7516/2017/09/013}.
  {\href{https://arxiv.org/abs/1707.02441}{{arXiv:1707.02441}}} {[astro-ph.CO]}

\bibitem[{Garc\'\i{}a-Quir\'os et~al.(2020)Garc\'\i{}a-Quir\'os, Colleoni,
  Husa, Estell\'es, Pratten, Ramos-Buades, Mateu-Lucena, and
  Jaume}]{Garcia-Quiros:2020qpx}
Garc\'\i{}a-Quir\'os C, Colleoni M, Husa S, Estell\'es H, Pratten G,
  Ramos-Buades A, Mateu-Lucena M, Jaume R (2020) {Multimode frequency-domain
  model for the gravitational wave signal from nonprecessing black-hole
  binaries}. Phys Rev D 102(6):064002. \doi{10.1103/PhysRevD.102.064002}.
  {\href{https://arxiv.org/abs/2001.10914}{{arXiv:2001.10914}}} {[gr-qc]}

\bibitem[{Garc\'\i{}a-Quir\'os et~al.(2021)Garc\'\i{}a-Quir\'os, Husa,
  Mateu-Lucena, and Borchers}]{Garcia-Quiros:2020qlt}
Garc\'\i{}a-Quir\'os C, Husa S, Mateu-Lucena M, Borchers A (2021) {Accelerating
  the evaluation of inspiral\textendash{}merger\textendash{}ringdown waveforms
  with adapted grids}. Class Quant Grav 38(1):015006.
  \doi{10.1088/1361-6382/abc36e}.
  {\href{https://arxiv.org/abs/2001.10897}{{arXiv:2001.10897}}} {[gr-qc]}

\bibitem[{García-Bellido(2017)}]{Garcia-Bellido:2017fdg}
García-Bellido J (2017) {Massive Primordial Black Holes as Dark Matter and
  their detection with Gravitational Waves}. J Phys Conf Ser 840(1):012032.
  \doi{10.1088/1742-6596/840/1/012032}.
  {\href{https://arxiv.org/abs/1702.08275}{{arXiv:1702.08275}}} {[astro-ph.CO]}

\bibitem[{García-Bellido(2018)}]{Garcia-Bellido:2018leu}
García-Bellido J (2018) {Primordial Black Holes}. PoS EDSU2018:042.
  \doi{10.22323/1.335.0042}

\bibitem[{García-Bellido(2019)}]{Garcia-Bellido:2019tvz}
García-Bellido J (2019) {Primordial black holes and the origin of the
  matter–antimatter asymmetry}. Phil Trans Roy Soc Lond A377(2161):20190091.
  \doi{10.1098/rsta.2019.0091}

\bibitem[{García-Bellido and Nesseris(2018)}]{Garcia-Bellido:2017knh}
García-Bellido J, Nesseris S (2018) {Gravitational wave energy emission and
  detection rates of Primordial Black Hole hyperbolic encounters}. Phys Dark
  Univ 21:61--69. \doi{10.1016/j.dark.2018.06.001}.
  {\href{https://arxiv.org/abs/1711.09702}{{arXiv:1711.09702}}} {[astro-ph.HE]}

\bibitem[{Geroch(1970)}]{Geroch:1970cd}
Geroch RP (1970) {Multipole moments. II. Curved space}. J Math Phys
  11:2580--2588. \doi{10.1063/1.1665427}

\bibitem[{Ghosh et~al.(2016{\natexlab{a}})Ghosh, Ghosh, Johnson-McDaniel,
  Mishra, Ajith, Del~Pozzo, Nichols, Chen, Nielsen, Berry, and
  et~al.}]{Ghosh_2016}
Ghosh A, Ghosh A, Johnson-McDaniel NK, Mishra CK, Ajith P, Del~Pozzo W, Nichols
  DA, Chen Y, Nielsen AB, Berry CP, et~al (2016{\natexlab{a}}) Testing general
  relativity using golden black-hole binaries. Physical Review D 94(2).
  \doi{10.1103/physrevd.94.021101}

\bibitem[{Ghosh et~al.(2017)Ghosh, Johnson-McDaniel, Ghosh, Mishra, Ajith,
  Pozzo, Berry, Nielsen, and London}]{Ghosh_2017}
Ghosh A, Johnson-McDaniel NK, Ghosh A, Mishra CK, Ajith P, Pozzo WD, Berry CPL,
  Nielsen AB, London L (2017) Testing general relativity using gravitational
  wave signals from the inspiral, merger and ringdown of binary black holes.
  Classical and Quantum Gravity 35(1):014002. \doi{10.1088/1361-6382/aa972e}

\bibitem[{Ghosh et~al.(2018)Ghosh, Johnson-Mcdaniel, Ghosh, Mishra, Ajith,
  Del~Pozzo, Berry, Nielsen, and London}]{Ghosh:2017gfp}
Ghosh A, Johnson-Mcdaniel NK, Ghosh A, Mishra CK, Ajith P, Del~Pozzo W, Berry
  CPL, Nielsen AB, London L (2018) {Testing general relativity using
  gravitational wave signals from the inspiral, merger and ringdown of binary
  black holes}. Class Quant Grav 35(1):014002. \doi{10.1088/1361-6382/aa972e}.
  {\href{https://arxiv.org/abs/1704.06784}{{arXiv:1704.06784}}} {[gr-qc]}

\bibitem[{Ghosh et~al.(2016{\natexlab{b}})}]{Ghosh:2016qgn}
Ghosh A, et~al. (2016{\natexlab{b}}) {Testing general relativity using golden
  black-hole binaries}. Phys Rev D94(2):021101.
  \doi{10.1103/PhysRevD.94.021101}.
  {\href{https://arxiv.org/abs/1602.02453}{{arXiv:1602.02453}}} {[gr-qc]}

\bibitem[{Ghosh et~al.(2019)Ghosh, Berti, Brito, and Richartz}]{Ghosh:2018gaw}
Ghosh S, Berti E, Brito R, Richartz M (2019) {Follow-up signals from
  superradiant instabilities of black hole merger remnants}. Phys Rev D
  99(10):104030. \doi{10.1103/PhysRevD.99.104030}.
  {\href{https://arxiv.org/abs/1812.01620}{{arXiv:1812.01620}}} {[gr-qc]}

\bibitem[{Giddings(2006)}]{Giddings:2006sj}
Giddings SB (2006) {Black hole information, unitarity, and nonlocality}. Phys
  Rev D 74:106005. \doi{10.1103/PhysRevD.74.106005}.
  {\href{https://arxiv.org/abs/hep-th/0605196}{{arXiv:hep-th/0605196}}}

\bibitem[{Giddings(2013)}]{Giddings:2013kcj}
Giddings SB (2013) {Nonviolent information transfer from black holes: A field
  theory parametrization}. Phys Rev D 88(2):024018.
  \doi{10.1103/PhysRevD.88.024018}.
  {\href{https://arxiv.org/abs/1302.2613}{{arXiv:1302.2613}}} {[hep-th]}

\bibitem[{Giddings(2017)}]{Giddings:2017mym}
Giddings SB (2017) {Nonviolent unitarization: basic postulates to soft quantum
  structure of black holes}. JHEP 12:047. \doi{10.1007/JHEP12(2017)047}.
  {\href{https://arxiv.org/abs/1701.08765}{{arXiv:1701.08765}}} {[hep-th]}

\bibitem[{Giddings et~al.(2019{\natexlab{a}})Giddings, Koren, and
  Treviño}]{Giddings:2019}
Giddings SB, Koren S, Treviño G (2019{\natexlab{a}}) {Exploring strong-field
  deviations from general relativity via gravitational waves}
  {\href{https://arxiv.org/abs/1904.04258}{{arXiv:1904.04258}}} {[gr-qc]}

\bibitem[{Giddings et~al.(2019{\natexlab{b}})Giddings, Koren, and
  Treviño}]{Giddings:2019ujs}
Giddings SB, Koren S, Treviño G (2019{\natexlab{b}}) {Exploring strong-field
  deviations from general relativity via gravitational waves}. Phys Rev D
  100(4):044005. \doi{10.1103/PhysRevD.100.044005}.
  {\href{https://arxiv.org/abs/1904.04258}{{arXiv:1904.04258}}} {[gr-qc]}

\bibitem[{Giesler et~al.(2019)Giesler, Isi, Scheel, and
  Teukolsky}]{Giesler:2019uxc}
Giesler M, Isi M, Scheel MA, Teukolsky S (2019) {Black Hole Ringdown: The
  Importance of Overtones}. Phys Rev X 9(4):041060.
  \doi{10.1103/PhysRevX.9.041060}.
  {\href{https://arxiv.org/abs/1903.08284}{{arXiv:1903.08284}}} {[gr-qc]}

\bibitem[{Gimon and Horava(2009)}]{Gimon:2007ur}
Gimon EG, Horava P (2009) {Astrophysical violations of the Kerr bound as a
  possible signature of string theory}. Phys Lett B 672:299--302.
  \doi{10.1016/j.physletb.2009.01.026}.
  {\href{https://arxiv.org/abs/0706.2873}{{arXiv:0706.2873}}} {[hep-th]}

\bibitem[{Giudice(2017)}]{Giudice:2017dde}
Giudice GF (2017) {Hunting for dark particles with gravitational waves}. EPJ
  Web Conf 164:02004. \doi{10.1051/epjconf/201716402004}

\bibitem[{Giudice et~al.(2016)Giudice, McCullough, and
  Urbano}]{Giudice:2016zpa}
Giudice GF, McCullough M, Urbano A (2016) {Hunting for Dark Particles with
  Gravitational Waves}. JCAP 10:001. \doi{10.1088/1475-7516/2016/10/001}.
  {\href{https://arxiv.org/abs/1605.01209}{{arXiv:1605.01209}}} {[hep-ph]}

\bibitem[{Glampedakis and Babak(2006)}]{Glampedakis:2005cf}
Glampedakis K, Babak S (2006) {Mapping spacetimes with LISA: Inspiral of a
  test-body in a `quasi-Kerr' field}. Class Quant Grav 23:4167--4188.
  \doi{10.1088/0264-9381/23/12/013}.
  {\href{https://arxiv.org/abs/gr-qc/0510057}{{arXiv:gr-qc/0510057}}}

\bibitem[{Glampedakis and Pappas(2018)}]{Glampedakis:2017cgd}
Glampedakis K, Pappas G (2018) {How well can ultracompact bodies imitate black
  hole ringdowns?} Phys Rev D 97(4):041502. \doi{10.1103/PhysRevD.97.041502}.
  {\href{https://arxiv.org/abs/1710.02136}{{arXiv:1710.02136}}} {[gr-qc]}

\bibitem[{Glampedakis and Silva(2019)}]{Glampedakis:2019dqh}
Glampedakis K, Silva HO (2019) {Eikonal quasinormal modes of black holes beyond
  General Relativity}. Phys Rev D 100(4):044040.
  \doi{10.1103/PhysRevD.100.044040}.
  {\href{https://arxiv.org/abs/1906.05455}{{arXiv:1906.05455}}} {[gr-qc]}

\bibitem[{Glampedakis et~al.(2002)Glampedakis, Hughes, and
  Kennefick}]{Glampedakis:2002cb}
Glampedakis K, Hughes SA, Kennefick D (2002) {Approximating the inspiral of
  test bodies into Kerr black holes}. Phys Rev D 66:064005.
  \doi{10.1103/PhysRevD.66.064005}.
  {\href{https://arxiv.org/abs/gr-qc/0205033}{{arXiv:gr-qc/0205033}}}

\bibitem[{Glampedakis et~al.(2017)Glampedakis, Pappas, Silva, and
  Berti}]{Glampedakis:2017dvb}
Glampedakis K, Pappas G, Silva HO, Berti E (2017) {Post-Kerr black hole
  spectroscopy}. Phys Rev D 96(6):064054. \doi{10.1103/PhysRevD.96.064054}.
  {\href{https://arxiv.org/abs/1706.07658}{{arXiv:1706.07658}}} {[gr-qc]}

\bibitem[{Gleyzes et~al.(2013)Gleyzes, Langlois, Piazza, and
  Vernizzi}]{Gleyzes:2013ooa}
Gleyzes J, Langlois D, Piazza F, Vernizzi F (2013) {Essential Building Blocks
  of Dark Energy}. JCAP 08:025. \doi{10.1088/1475-7516/2013/08/025}.
  {\href{https://arxiv.org/abs/1304.4840}{{arXiv:1304.4840}}} {[hep-th]}

\bibitem[{Gleyzes et~al.(2014)Gleyzes, Langlois, and
  Vernizzi}]{Gleyzes:2014rba}
Gleyzes J, Langlois D, Vernizzi F (2014) {A unifying description of dark
  energy}. Int J Mod Phys D23:1443010. \doi{10.1142/S021827181443010X}.
  {\href{https://arxiv.org/abs/1411.3712}{{arXiv:1411.3712}}} {[hep-th]}

\bibitem[{Gleyzes et~al.(2015)Gleyzes, Langlois, Piazza, and
  Vernizzi}]{Gleyzes:2014dya}
Gleyzes J, Langlois D, Piazza F, Vernizzi F (2015) {Healthy theories beyond
  Horndeski}. Phys Rev Lett 114(21):211101.
  \doi{10.1103/PhysRevLett.114.211101}.
  {\href{https://arxiv.org/abs/1404.6495}{{arXiv:1404.6495}}} {[hep-th]}

\bibitem[{Gnocchi et~al.(2019)Gnocchi, Maselli, Abdelsalhin, Giacobbo, and
  Mapelli}]{Gnocchi:2019jzp}
Gnocchi G, Maselli A, Abdelsalhin T, Giacobbo N, Mapelli M (2019) {Bounding
  alternative theories of gravity with multiband GW observations}. Phys Rev D
  100(6):064024. \doi{10.1103/PhysRevD.100.064024}.
  {\href{https://arxiv.org/abs/1905.13460}{{arXiv:1905.13460}}} {[gr-qc]}

\bibitem[{{Gold} et~al.(2014){Gold}, {Paschalidis}, {Ruiz}, {Shapiro},
  {Etienne}, and {Pfeiffer}}]{2014PhRvD..90j4030G}
{Gold} R, {Paschalidis} V, {Ruiz} M, {Shapiro} SL, {Etienne} ZB, {Pfeiffer} HP
  (2014) {Accretion disks around binary black holes of unequal mass: General
  relativistic MHD simulations of postdecoupling and merger}. \prd
  90(10):104030. \doi{10.1103/PhysRevD.90.104030}.
  {\href{https://arxiv.org/abs/1410.1543}{{arXiv:1410.1543}}} {[astro-ph.GA]}

\bibitem[{{Goldreich} and {Tremaine}(1980)}]{1980ApJ...241..425G}
{Goldreich} P, {Tremaine} S (1980) {Disk-satellite interactions.} \apj
  241:425--441. \doi{10.1086/158356}

\bibitem[{Gondolo and Silk(1999)}]{Gondolo:1999ef}
Gondolo P, Silk J (1999) {Dark matter annihilation at the galactic center}.
  Phys Rev Lett 83:1719--1722. \doi{10.1103/PhysRevLett.83.1719}.
  {\href{https://arxiv.org/abs/astro-ph/9906391}{{arXiv:astro-ph/9906391}}}

\bibitem[{Goodman and Tan(2004)}]{Goodman:2003sf}
Goodman J, Tan JC (2004) {Supermassive stars in quasar disks}. Astrophys J
  608:108--118. \doi{10.1086/386360}.
  {\href{https://arxiv.org/abs/astro-ph/0307361}{{arXiv:astro-ph/0307361}}}

\bibitem[{Goodsell et~al.(2009)Goodsell, Jaeckel, Redondo, and
  Ringwald}]{Goodsell:2009xc}
Goodsell M, Jaeckel J, Redondo J, Ringwald A (2009) {Naturally Light Hidden
  Photons in LARGE Volume String Compactifications}. JHEP 11:027.
  \doi{10.1088/1126-6708/2009/11/027}.
  {\href{https://arxiv.org/abs/0909.0515}{{arXiv:0909.0515}}} {[hep-ph]}

\bibitem[{Gossan et~al.(2012)Gossan, Veitch, and Sathyaprakash}]{Gossan:2011ha}
Gossan S, Veitch J, Sathyaprakash B (2012) {Bayesian model selection for
  testing the no-hair theorem with black hole ringdowns}. Phys Rev D 85:124056.
  \doi{10.1103/PhysRevD.85.124056}.
  {\href{https://arxiv.org/abs/1111.5819}{{arXiv:1111.5819}}} {[gr-qc]}

\bibitem[{Gould and Rix(2000)}]{Gould:1999ia}
Gould A, Rix HW (2000) {Binary black hole mergers from planet-like migrations}.
  Astrophys J Lett 532:L29. \doi{10.1086/312562}.
  {\href{https://arxiv.org/abs/astro-ph/9912111}{{arXiv:astro-ph/9912111}}}

\bibitem[{Gourgoulhon(2012)}]{GourgoulhonBook2012}
Gourgoulhon E (2012) 3+1 Formalism in General Relativity. Springer-Verlag
  Berlin Heidelberg, Heidelberg

\bibitem[{Gralla et~al.(2015)Gralla, Porfyriadis, and
  Warburton}]{Gralla:2015rpa}
Gralla SE, Porfyriadis AP, Warburton N (2015) {Particle on the Innermost Stable
  Circular Orbit of a Rapidly Spinning Black Hole}. Phys Rev D 92(6):064029.
  \doi{10.1103/PhysRevD.92.064029}.
  {\href{https://arxiv.org/abs/1506.08496}{{arXiv:1506.08496}}} {[gr-qc]}

\bibitem[{Grandclement et~al.(2014)Grandclement, Som\'e, and
  Gourgoulhon}]{Grandclement:2014msa}
Grandclement P, Som\'e C, Gourgoulhon E (2014) {Models of rotating boson stars
  and geodesics around them: new type of orbits}. Phys Rev D 90(2):024068.
  \doi{10.1103/PhysRevD.90.024068}.
  {\href{https://arxiv.org/abs/1405.4837}{{arXiv:1405.4837}}} {[gr-qc]}

\bibitem[{Gray et~al.(2020)}]{Gray:2019ksv}
Gray R, et~al. (2020) {Cosmological Inference using Gravitational Wave Standard
  Sirens: A Mock Data Challenge}. Phys Rev D 101:122001.
  \doi{10.1103/PhysRevD.101.122001}.
  {\href{https://arxiv.org/abs/1908.06050}{{arXiv:1908.06050}}} {[gr-qc]}

\bibitem[{Green and Kavanagh(2021)}]{Green:2020jor}
Green AM, Kavanagh BJ (2021) {Primordial Black Holes as a dark matter
  candidate}. J Phys G 48(4):043001. \doi{10.1088/1361-6471/abc534}.
  {\href{https://arxiv.org/abs/2007.10722}{{arXiv:2007.10722}}} {[astro-ph.CO]}

\bibitem[{Gubitosi et~al.(2013)Gubitosi, Piazza, and
  Vernizzi}]{Gubitosi:2012hu}
Gubitosi G, Piazza F, Vernizzi F (2013) {The Effective Field Theory of Dark
  Energy}. JCAP 1302:032. \doi{10.1088/1475-7516/2013/02/032},
  [JCAP1302,032(2013)].
  {\href{https://arxiv.org/abs/1210.0201}{{arXiv:1210.0201}}} {[hep-th]}

\bibitem[{Guerra et~al.(2019)Guerra, Macedo, and Pani}]{Guerra:2019srj}
Guerra D, Macedo CF, Pani P (2019) {Axion boson stars}. JCAP 09(09):061.
  \doi{10.1088/1475-7516/2019/09/061}, [Erratum: JCAP 06, E01 (2020)].
  {\href{https://arxiv.org/abs/1909.05515}{{arXiv:1909.05515}}} {[gr-qc]}

\bibitem[{Guo et~al.(2019)Guo, Shu, and Zhao}]{Guo:2017njn}
Guo HK, Shu J, Zhao Y (2019) {Using LISA-like Gravitational Wave Detectors to
  Search for Primordial Black Holes}. Phys Rev D 99(2):023001.
  \doi{10.1103/PhysRevD.99.023001}.
  {\href{https://arxiv.org/abs/1709.03500}{{arXiv:1709.03500}}} {[astro-ph.CO]}

\bibitem[{Gupta et~al.(2020)Gupta, Datta, Kastha, Borhanian, Arun, and
  Sathyaprakash}]{Gupta:2020lxa}
Gupta A, Datta S, Kastha S, Borhanian S, Arun K, Sathyaprakash B (2020)
  {Multiparameter tests of general relativity using multiband
  gravitational-wave observations}
  {\href{https://arxiv.org/abs/2005.09607}{{arXiv:2005.09607}}} {[gr-qc]}

\bibitem[{Gürlebeck(2015)}]{Gurlebeck:2015xpa}
Gürlebeck N (2015) {No-hair theorem for Black Holes in Astrophysical
  Environments}. Phys Rev Lett 114(15):151102.
  \doi{10.1103/PhysRevLett.114.151102}.
  {\href{https://arxiv.org/abs/1503.03240}{{arXiv:1503.03240}}} {[gr-qc]}

\bibitem[{ter Haar et~al.(2021)ter Haar, Bezares, Crisostomi, Barausse, and
  Palenzuela}]{terHaar:2020xxb}
ter Haar L, Bezares M, Crisostomi M, Barausse E, Palenzuela C (2021) {Dynamics
  of Screening in Modified Gravity}. Phys Rev Lett 126:091102.
  \doi{10.1103/PhysRevLett.126.091102}.
  {\href{https://arxiv.org/abs/2009.03354}{{arXiv:2009.03354}}} {[gr-qc]}

\bibitem[{Haegel and Husa(2020)}]{Haegel:2019uop}
Haegel L, Husa S (2020) {Predicting the properties of black-hole merger
  remnants with deep neural networks}. Class Quant Grav 37(13):135005.
  \doi{10.1088/1361-6382/ab905c}.
  {\href{https://arxiv.org/abs/1911.01496}{{arXiv:1911.01496}}} {[gr-qc]}

\bibitem[{Haiman(2017)}]{Haiman:2017szj}
Haiman Z (2017) {Electromagnetic chirp of a compact binary black hole: A phase
  template for the gravitational wave inspiral}. Phys Rev D 96(2):023004.
  \doi{10.1103/PhysRevD.96.023004}.
  {\href{https://arxiv.org/abs/1705.06765}{{arXiv:1705.06765}}} {[astro-ph.HE]}

\bibitem[{Hall et~al.(2020)Hall, Gow, and Byrnes}]{Hall:2020daa}
Hall A, Gow AD, Byrnes CT (2020) {Bayesian analysis of LIGO-Virgo mergers:
  Primordial vs. astrophysical black hole populations}
  {\href{https://arxiv.org/abs/2008.13704}{{arXiv:2008.13704}}} {[astro-ph.CO]}

\bibitem[{Hannam et~al.(2014)Hannam, Schmidt, Boh\'e, Haegel, Husa, Ohme,
  Pratten, and P\"urrer}]{Hannam:2013oca}
Hannam M, Schmidt P, Boh\'e A, Haegel L, Husa S, Ohme F, Pratten G, P\"urrer M
  (2014) {Simple Model of Complete Precessing Black-Hole-Binary Gravitational
  Waveforms}. Phys Rev Lett 113(15):151101.
  \doi{10.1103/PhysRevLett.113.151101}.
  {\href{https://arxiv.org/abs/1308.3271}{{arXiv:1308.3271}}} {[gr-qc]}

\bibitem[{Hannuksela et~al.(2019)Hannuksela, Wong, Brito, Berti, and
  Li}]{Hannuksela:2018izj}
Hannuksela OA, Wong KW, Brito R, Berti E, Li TG (2019) {Probing the existence
  of ultralight bosons with a single gravitational-wave measurement}. Nature
  Astron 3(5):447--451. \doi{10.1038/s41550-019-0712-4}.
  {\href{https://arxiv.org/abs/1804.09659}{{arXiv:1804.09659}}} {[astro-ph.HE]}

\bibitem[{Hannuksela et~al.(2020)Hannuksela, Ng, and Li}]{Hannuksela:2019vip}
Hannuksela OA, Ng KCY, Li TGF (2020) {Extreme dark matter tests with extreme
  mass ratio inspirals}. Phys Rev D 102(10):103022.
  \doi{10.1103/PhysRevD.102.103022}.
  {\href{https://arxiv.org/abs/1906.11845}{{arXiv:1906.11845}}} {[astro-ph.CO]}

\bibitem[{Hansen et~al.(2015)Hansen, Yunes, and Yagi}]{Hansen:2014ewa}
Hansen D, Yunes N, Yagi K (2015) {Projected Constraints on Lorentz-Violating
  Gravity with Gravitational Waves}. Phys Rev D91(8):082003.
  \doi{10.1103/PhysRevD.91.082003}.
  {\href{https://arxiv.org/abs/1412.4132}{{arXiv:1412.4132}}} {[gr-qc]}

\bibitem[{Hansen(1974)}]{Hansen:1974zz}
Hansen R (1974) {Multipole moments of stationary space-times}. J Math Phys
  15:46--52. \doi{10.1063/1.1666501}

\bibitem[{Harms et~al.(2014)Harms, Bernuzzi, Nagar, and
  Zenginoglu}]{Harms:2014dqa}
Harms E, Bernuzzi S, Nagar A, Zenginoglu A (2014) {A new gravitational wave
  generation algorithm for particle perturbations of the Kerr spacetime}. Class
  Quant Grav 31(24):245004. \doi{10.1088/0264-9381/31/24/245004}.
  {\href{https://arxiv.org/abs/1406.5983}{{arXiv:1406.5983}}} {[gr-qc]}

\bibitem[{Hassan and Rosen(2012)}]{Hassan:2011zd}
Hassan SF, Rosen RA (2012) {Bimetric Gravity from Ghost-free Massive Gravity}.
  JHEP 02:126. \doi{10.1007/JHEP02(2012)126}.
  {\href{https://arxiv.org/abs/1109.3515}{{arXiv:1109.3515}}} {[hep-th]}

\bibitem[{Hawking(1972)}]{Hawking:1972qk}
Hawking S (1972) {Black holes in the Brans-Dicke theory of gravitation}. Commun
  Math Phys 25:167--171. \doi{10.1007/BF01877518}

\bibitem[{Hawking(1976)}]{Hawking:1976ra}
Hawking SW (1976) {Breakdown of Predictability in Gravitational Collapse}. Phys
  Rev D 14:2460--2473. \doi{10.1103/PhysRevD.14.2460}

\bibitem[{Hayashi and Shirafuji(1979)}]{Hayashi:1979qx}
Hayashi K, Shirafuji T (1979) {New General Relativity}. Phys Rev D
  19:3524--3553. \doi{10.1103/PhysRevD.19.3524}, [Addendum: Phys.Rev.D 24,
  3312--3314 (1982)]

\bibitem[{Healy and Lousto(2020)}]{Healy:2020vre}
Healy J, Lousto CO (2020) {The Third RIT binary black hole simulations catalog}
  {\href{https://arxiv.org/abs/2007.07910}{{arXiv:2007.07910}}} {[gr-qc]}

\bibitem[{Healy et~al.(2012)Healy, Bode, Haas, Pazos, Laguna, Shoemaker, and
  Yunes}]{Healy:2011ef}
Healy J, Bode T, Haas R, Pazos E, Laguna P, Shoemaker D, Yunes N (2012) {Late
  Inspiral and Merger of Binary Black Holes in Scalar-Tensor Theories of
  Gravity}. Class Quant Grav 29:232002. \doi{10.1088/0264-9381/29/23/232002}.
  {\href{https://arxiv.org/abs/1112.3928}{{arXiv:1112.3928}}} {[gr-qc]}

\bibitem[{Heisenberg et~al.(2017)Heisenberg, Kase, Minamitsuji, and
  Tsujikawa}]{Heisenberg:2017xda}
Heisenberg L, Kase R, Minamitsuji M, Tsujikawa S (2017) {Hairy black-hole
  solutions in generalized Proca theories}. Phys Rev D 96(8):084049.
  \doi{10.1103/PhysRevD.96.084049}.
  {\href{https://arxiv.org/abs/1705.09662}{{arXiv:1705.09662}}} {[gr-qc]}

\bibitem[{Herdeiro and Radu(2014{\natexlab{a}})}]{Herdeiro:2014jaa}
Herdeiro C, Radu E (2014{\natexlab{a}}) {Ergosurfaces for Kerr black holes with
  scalar hair}. Phys Rev D 89(12):124018. \doi{10.1103/PhysRevD.89.124018}.
  {\href{https://arxiv.org/abs/1406.1225}{{arXiv:1406.1225}}} {[gr-qc]}

\bibitem[{Herdeiro et~al.(2016)Herdeiro, Radu, and
  R\'unarsson}]{Herdeiro:2016tmi}
Herdeiro C, Radu E, R\'unarsson H (2016) {Kerr black holes with Proca hair}.
  Class Quant Grav 33(15):154001. \doi{10.1088/0264-9381/33/15/154001}.
  {\href{https://arxiv.org/abs/1603.02687}{{arXiv:1603.02687}}} {[gr-qc]}

\bibitem[{Herdeiro et~al.(2018{\natexlab{a}})Herdeiro, Kunz, Radu, and
  Subagyo}]{Herdeiro:2017oyt}
Herdeiro C, Kunz J, Radu E, Subagyo B (2018{\natexlab{a}}) {Probing the
  universality of synchronised hair around rotating black holes with Q-clouds}.
  Phys Lett B 779:151--159. \doi{10.1016/j.physletb.2018.01.083}.
  {\href{https://arxiv.org/abs/1712.04286}{{arXiv:1712.04286}}} {[gr-qc]}

\bibitem[{Herdeiro et~al.(2019)Herdeiro, Perapechka, Radu, and
  Shnir}]{Herdeiro:2019mbz}
Herdeiro C, Perapechka I, Radu E, Shnir Y (2019) {Asymptotically flat spinning
  scalar, Dirac and Proca stars}. Phys Lett B 797:134845.
  \doi{10.1016/j.physletb.2019.134845}.
  {\href{https://arxiv.org/abs/1906.05386}{{arXiv:1906.05386}}} {[gr-qc]}

\bibitem[{Herdeiro and Radu(2015)}]{Herdeiro:2015waa}
Herdeiro CA, Radu E (2015) {Asymptotically flat black holes with scalar hair: a
  review}. Int J Mod Phys D 24(09):1542014. \doi{10.1142/S0218271815420146}.
  {\href{https://arxiv.org/abs/1504.08209}{{arXiv:1504.08209}}} {[gr-qc]}

\bibitem[{Herdeiro and Radu(2018)}]{Herdeiro:2018wvd}
Herdeiro CA, Radu E (2018) {Spinning boson stars and hairy black holes with
  nonminimal coupling}. Int J Mod Phys D 27(11):1843009.
  \doi{10.1142/S0218271818430095}.
  {\href{https://arxiv.org/abs/1803.08149}{{arXiv:1803.08149}}} {[gr-qc]}

\bibitem[{Herdeiro et~al.(2018{\natexlab{b}})Herdeiro, Radu, Sanchis-Gual, and
  Font}]{Herdeiro:2018wub}
Herdeiro CA, Radu E, Sanchis-Gual N, Font JA (2018{\natexlab{b}}) {Spontaneous
  Scalarization of Charged Black Holes}. Phys Rev Lett 121(10):101102.
  \doi{10.1103/PhysRevLett.121.101102}.
  {\href{https://arxiv.org/abs/1806.05190}{{arXiv:1806.05190}}} {[gr-qc]}

\bibitem[{Herdeiro et~al.(2020{\natexlab{a}})Herdeiro, Radu, Silva, Sotiriou,
  and Yunes}]{Herdeiro:2020wei}
Herdeiro CA, Radu E, Silva HO, Sotiriou TP, Yunes N (2020{\natexlab{a}})
  {Spin-induced scalarized black holes}
  {\href{https://arxiv.org/abs/2009.03904}{{arXiv:2009.03904}}} {[gr-qc]}

\bibitem[{Herdeiro and Radu(2014{\natexlab{b}})}]{Herdeiro:2014goa}
Herdeiro CAR, Radu E (2014{\natexlab{b}}) {Kerr black holes with scalar hair}.
  Phys Rev Lett 112:221101. \doi{10.1103/PhysRevLett.112.221101}.
  {\href{https://arxiv.org/abs/1403.2757}{{arXiv:1403.2757}}} {[gr-qc]}

\bibitem[{Herdeiro and Radu(2017)}]{Herdeiro:2017phl}
Herdeiro CAR, Radu E (2017) {Dynamical Formation of Kerr Black Holes with
  Synchronized Hair: An Analytic Model}. Phys Rev Lett 119(26):261101.
  \doi{10.1103/PhysRevLett.119.261101}.
  {\href{https://arxiv.org/abs/1706.06597}{{arXiv:1706.06597}}} {[gr-qc]}

\bibitem[{Herdeiro and Radu(2020)}]{Herdeiro:2020jzx}
Herdeiro CAR, Radu E (2020) {Asymptotically flat, spherical, self-interacting
  scalar, Dirac and Proca stars}. Symmetry 12(12):2032.
  \doi{10.3390/sym12122032}.
  {\href{https://arxiv.org/abs/2012.03595}{{arXiv:2012.03595}}} {[gr-qc]}

\bibitem[{Herdeiro et~al.(2015)Herdeiro, Radu, and
  R\'unarsson}]{Herdeiro:2015tia}
Herdeiro CAR, Radu E, R\'unarsson H (2015) {Kerr black holes with
  self-interacting scalar hair: hairier but not heavier}. Phys Rev D
  92(8):084059. \doi{10.1103/PhysRevD.92.084059}.
  {\href{https://arxiv.org/abs/1509.02923}{{arXiv:1509.02923}}} {[gr-qc]}

\bibitem[{Herdeiro et~al.(2017)Herdeiro, Pombo, and Radu}]{Herdeiro:2017fhv}
Herdeiro CAR, Pombo AM, Radu E (2017) {Asymptotically flat scalar, Dirac and
  Proca stars: discrete vs. continuous families of solutions}. Phys Lett B
  773:654--662. \doi{10.1016/j.physletb.2017.09.036}.
  {\href{https://arxiv.org/abs/1708.05674}{{arXiv:1708.05674}}} {[gr-qc]}

\bibitem[{Herdeiro et~al.(2020{\natexlab{b}})Herdeiro, Panotopoulos, and
  Radu}]{Herdeiro:2020kba}
Herdeiro CAR, Panotopoulos G, Radu E (2020{\natexlab{b}}) {Tidal Love numbers
  of Proca stars}. JCAP 08:029. \doi{10.1088/1475-7516/2020/08/029}.
  {\href{https://arxiv.org/abs/2006.11083}{{arXiv:2006.11083}}} {[gr-qc]}

\bibitem[{Herdeiro et~al.(2021)Herdeiro, Kunz, Perapechka, Radu, and
  Shnir}]{Herdeiro:2020kvf}
Herdeiro CAR, Kunz J, Perapechka I, Radu E, Shnir Y (2021) {Multipolar boson
  stars: macroscopic Bose-Einstein condensates akin to hydrogen orbitals}. Phys
  Lett B 812:136027. \doi{10.1016/j.physletb.2020.136027}.
  {\href{https://arxiv.org/abs/2008.10608}{{arXiv:2008.10608}}} {[gr-qc]}

\bibitem[{Ca\~nas Herrera et~al.(2020)Ca\~nas Herrera, Contigiani, and
  Vardanyan}]{Canas-Herrera:2019npr}
Ca\~nas Herrera G, Contigiani O, Vardanyan V (2020) {Cross-correlation of the
  astrophysical gravitational-wave background with galaxy clustering}. Phys Rev
  D 102(4):043513. \doi{10.1103/PhysRevD.102.043513}.
  {\href{https://arxiv.org/abs/1910.08353}{{arXiv:1910.08353}}} {[astro-ph.CO]}

\bibitem[{Heusler(1996)}]{heusler_1996}
Heusler M (1996) Black Hole Uniqueness Theorems. Cambridge Lecture Notes in
  Physics, Cambridge University Press

\bibitem[{Heusler et~al.(1992)Heusler, Droz, and Straumann}]{Heusler:1992av}
Heusler M, Droz S, Straumann N (1992) {Linear stability of Einstein Skyrme
  black holes}. Phys Lett B 285:21--26. \doi{10.1016/0370-2693(92)91294-J}

\bibitem[{Heymans and other(2020)}]{Heymans:2020gsg}
Heymans C, other (2020) {KiDS-1000 Cosmology: Multi-probe weak gravitational
  lensing and spectroscopic galaxy clustering constraints}
  {\href{https://arxiv.org/abs/2007.15632}{{arXiv:2007.15632}}} {[astro-ph.CO]}

\bibitem[{Hinder et~al.(2018)Hinder, Kidder, and Pfeiffer}]{Hinder:2017sxy}
Hinder I, Kidder LE, Pfeiffer HP (2018) {Eccentric binary black hole
  inspiral-merger-ringdown gravitational waveform model from numerical
  relativity and post-Newtonian theory}. Phys Rev D 98(4):044015.
  \doi{10.1103/PhysRevD.98.044015}.
  {\href{https://arxiv.org/abs/1709.02007}{{arXiv:1709.02007}}} {[gr-qc]}

\bibitem[{Hinder et~al.(2014)}]{Hinder:2013oqa}
Hinder I, et~al. (2014) {Error-analysis and comparison to analytical models of
  numerical waveforms produced by the NRAR Collaboration}. Class Quant Grav
  31:025012. \doi{10.1088/0264-9381/31/2/025012}.
  {\href{https://arxiv.org/abs/1307.5307}{{arXiv:1307.5307}}} {[gr-qc]}

\bibitem[{Hinderer(2008)}]{Hinderer:2007mb}
Hinderer T (2008) {Tidal Love numbers of neutron stars}. Astrophys J
  677:1216--1220. \doi{10.1086/533487}.
  {\href{https://arxiv.org/abs/0711.2420}{{arXiv:0711.2420}}} {[astro-ph]}

\bibitem[{Hinderer and Babak(2017)}]{Hinderer:2017jcs}
Hinderer T, Babak S (2017) {Foundations of an effective-one-body model for
  coalescing binaries on eccentric orbits}. Phys Rev D 96(10):104048.
  \doi{10.1103/PhysRevD.96.104048}.
  {\href{https://arxiv.org/abs/1707.08426}{{arXiv:1707.08426}}} {[gr-qc]}

\bibitem[{Hinderer and Flanagan(2008)}]{Hinderer:2008dm}
Hinderer T, Flanagan EE (2008) {Two timescale analysis of extreme mass ratio
  inspirals in Kerr. I. Orbital Motion}. Phys Rev D78:064028.
  \doi{10.1103/PhysRevD.78.064028}.
  {\href{https://arxiv.org/abs/0805.3337}{{arXiv:0805.3337}}} {[gr-qc]}

\bibitem[{Hinderer et~al.(2018)Hinderer, Rezzolla, and
  Baiotti}]{Hinderer:2018mrj}
Hinderer T, Rezzolla L, Baiotti L (2018) {Gravitational Waves from Merging
  Binary Neutron-Star Systems}, vol 457, pp 575--635.
  \doi{10.1007/978-3-319-97616-7\_10}

\bibitem[{Hinderer et~al.(2016)}]{Hinderer:2016eia}
Hinderer T, et~al. (2016) {Effects of neutron-star dynamic tides on
  gravitational waveforms within the effective-one-body approach}. Phys Rev
  Lett 116(18):181101. \doi{10.1103/PhysRevLett.116.181101}.
  {\href{https://arxiv.org/abs/1602.00599}{{arXiv:1602.00599}}} {[gr-qc]}

\bibitem[{Hinterbichler and Khoury(2010)}]{Hinterbichler:2010es}
Hinterbichler K, Khoury J (2010) {Symmetron Fields: Screening Long-Range Forces
  Through Local Symmetry Restoration}. Phys Rev Lett 104:231301.
  \doi{10.1103/PhysRevLett.104.231301}.
  {\href{https://arxiv.org/abs/1001.4525}{{arXiv:1001.4525}}} {[hep-th]}

\bibitem[{Hirata et~al.(2010)Hirata, Holz, and Cutler}]{Hirata:2010ba}
Hirata CM, Holz DE, Cutler C (2010) {Reducing the weak lensing noise for the
  gravitational wave Hubble diagram using the non-Gaussianity of the
  magnification distribution}. Phys Rev D 81:124046.
  \doi{10.1103/PhysRevD.81.124046}.
  {\href{https://arxiv.org/abs/1004.3988}{{arXiv:1004.3988}}} {[astro-ph.CO]}

\bibitem[{Hirschmann et~al.(2018)Hirschmann, Lehner, Liebling, and
  Palenzuela}]{Hirschmann:2017psw}
Hirschmann EW, Lehner L, Liebling SL, Palenzuela C (2018) {Black Hole Dynamics
  in Einstein-Maxwell-Dilaton Theory}. Phys Rev D 97(6):064032.
  \doi{10.1103/PhysRevD.97.064032}.
  {\href{https://arxiv.org/abs/1706.09875}{{arXiv:1706.09875}}} {[gr-qc]}

\bibitem[{Hod(2020)}]{Hod:2020jjy}
Hod S (2020) {Onset of spontaneous scalarization in spinning Gauss-Bonnet black
  holes} {\href{https://arxiv.org/abs/2006.09399}{{arXiv:2006.09399}}}
  {[gr-qc]}

\bibitem[{Hofmann et~al.(2016)Hofmann, Barausse, and
  Rezzolla}]{Hofmann:2016yih}
Hofmann F, Barausse E, Rezzolla L (2016) {The final spin from binary black
  holes in quasi-circular orbits}. Astrophys J Lett 825(2):L19.
  \doi{10.3847/2041-8205/825/2/L19}.
  {\href{https://arxiv.org/abs/1605.01938}{{arXiv:1605.01938}}} {[gr-qc]}

\bibitem[{Hohmann et~al.(2018)Hohmann, Kr\v{s}\v{s}\'ak, Pfeifer, and
  Ualikhanova}]{Hohmann:2018jso}
Hohmann M, Kr\v{s}\v{s}\'ak M, Pfeifer C, Ualikhanova U (2018) {Propagation of
  gravitational waves in teleparallel gravity theories}. Phys Rev D
  98(12):124004. \doi{10.1103/PhysRevD.98.124004}.
  {\href{https://arxiv.org/abs/1807.04580}{{arXiv:1807.04580}}} {[gr-qc]}

\bibitem[{Holdom and Ren(2017)}]{Holdom:2016nek}
Holdom B, Ren J (2017) {Not quite a black hole}. Phys Rev D 95(8):084034.
  \doi{10.1103/PhysRevD.95.084034}.
  {\href{https://arxiv.org/abs/1612.04889}{{arXiv:1612.04889}}} {[gr-qc]}

\bibitem[{Holz and Hughes(2005)}]{Holz:2005df}
Holz DE, Hughes SA (2005) {Using gravitational-wave standard sirens}. Astrophys
  J 629:15--22. \doi{10.1086/431341}.
  {\href{https://arxiv.org/abs/astro-ph/0504616}{{arXiv:astro-ph/0504616}}}

\bibitem[{Horava(2009{\natexlab{a}})}]{Horava:2008ih}
Horava P (2009{\natexlab{a}}) {Membranes at Quantum Criticality}. JHEP 03:020.
  \doi{10.1088/1126-6708/2009/03/020}.
  {\href{https://arxiv.org/abs/0812.4287}{{arXiv:0812.4287}}} {[hep-th]}

\bibitem[{Horava(2009{\natexlab{b}})}]{Horava:2009uw}
Horava P (2009{\natexlab{b}}) {Quantum Gravity at a Lifshitz Point}. Phys Rev D
  79:084008. \doi{10.1103/PhysRevD.79.084008}.
  {\href{https://arxiv.org/abs/0901.3775}{{arXiv:0901.3775}}} {[hep-th]}

\bibitem[{Horndeski(1974)}]{Horndeski:1974wa}
Horndeski GW (1974) {Second-order scalar-tensor field equations in a
  four-dimensional space}. Int J Theor Phys 10:363--384.
  \doi{10.1007/BF01807638}

\bibitem[{Horowitz et~al.(1996)Horowitz, Maldacena, and
  Strominger}]{Horowitz:1996ay}
Horowitz GT, Maldacena JM, Strominger A (1996) {Nonextremal black hole
  microstates and U duality}. Phys Lett B 383:151--159.
  \doi{10.1016/0370-2693(96)00738-1}.
  {\href{https://arxiv.org/abs/hep-th/9603109}{{arXiv:hep-th/9603109}}}

\bibitem[{Hou and Zhu(2020)}]{Hou:2020tnd}
Hou S, Zhu ZH (2020) {Gravitational memory effects and Bondi-Metzner-Sachs
  symmetries in scalar-tensor theories}
  {\href{https://arxiv.org/abs/2005.01310}{{arXiv:2005.01310}}} {[gr-qc]}

\bibitem[{Hu et~al.(2014)Hu, Raveri, Frusciante, and Silvestri}]{Hu:2013twa}
Hu B, Raveri M, Frusciante N, Silvestri A (2014) {Effective Field Theory of
  Cosmic Acceleration: an implementation in CAMB}. Phys Rev D 89(10):103530.
  \doi{10.1103/PhysRevD.89.103530}.
  {\href{https://arxiv.org/abs/1312.5742}{{arXiv:1312.5742}}} {[astro-ph.CO]}

\bibitem[{Huang et~al.(2019{\natexlab{a}})Huang, Riess, Yuan, Macri, Zakamska,
  Casertano, Whitelock, Hoffmann, Filippenko, and Scolnic}]{Huang:2019yhh}
Huang CD, Riess AG, Yuan W, Macri LM, Zakamska NL, Casertano S, Whitelock PA,
  Hoffmann SL, Filippenko AV, Scolnic D (2019{\natexlab{a}}) {Hubble Space
  Telescope Observations of Mira Variables in the Type Ia Supernova Host NGC
  1559: An Alternative Candle to Measure the Hubble Constant}
  \doi{10.3847/1538-4357/ab5dbd}.
  {\href{https://arxiv.org/abs/1908.10883}{{arXiv:1908.10883}}} {[astro-ph.CO]}

\bibitem[{Huang et~al.(2019{\natexlab{b}})Huang, Johnson, Sagunski,
  Sakellariadou, and Zhang}]{Huang:2018pbu}
Huang J, Johnson MC, Sagunski L, Sakellariadou M, Zhang J (2019{\natexlab{b}})
  {Prospects for axion searches with Advanced LIGO through binary mergers}.
  Phys Rev D 99(6):063013. \doi{10.1103/PhysRevD.99.063013}.
  {\href{https://arxiv.org/abs/1807.02133}{{arXiv:1807.02133}}} {[hep-ph]}

\bibitem[{Hughes and Holz(2003)}]{Holz:2002cn}
Hughes S, Holz D (2003) {Cosmology with coalescing massive black holes}. Class
  Quant Grav 20:S65--S72. \doi{10.1088/0264-9381/20/10/308}.
  {\href{https://arxiv.org/abs/astro-ph/0212218}{{arXiv:astro-ph/0212218}}}

\bibitem[{Hughes(2001)}]{Hughes:2001jr}
Hughes SA (2001) {Evolution of circular, nonequatorial orbits of Kerr black
  holes due to gravitational wave emission. II. Inspiral trajectories and
  gravitational wave forms}. Phys Rev D 64:064004.
  \doi{10.1103/PhysRevD.64.064004}, [Erratum: Phys.Rev.D 88, 109902 (2013)].
  {\href{https://arxiv.org/abs/gr-qc/0104041}{{arXiv:gr-qc/0104041}}}

\bibitem[{Hughes(2016)}]{Hughes:2016xwf}
Hughes SA (2016) {Adiabatic and post-adiabatic approaches to extreme mass ratio
  inspiral}. In: {14th Marcel Grossmann Meeting on Recent Developments in
  Theoretical and Experimental General Relativity, Astrophysics, and
  Relativistic Field Theories}. \doi{10.1142/9789813226609_0208}.
  {\href{https://arxiv.org/abs/1601.02042}{{arXiv:1601.02042}}} {[gr-qc]}

\bibitem[{Hughes and Menou(2005)}]{Hughes:2004vw}
Hughes SA, Menou K (2005) {Golden binaries for LISA: Robust probes of
  strong-field gravity}. Astrophys J 623:689--699. \doi{10.1086/428826}.
  {\href{https://arxiv.org/abs/astro-ph/0410148}{{arXiv:astro-ph/0410148}}}

\bibitem[{Hui and Nicolis(2013)}]{Hui:2012qt}
Hui L, Nicolis A (2013) {No-Hair Theorem for the Galileon}. Phys Rev Lett
  110:241104. \doi{10.1103/PhysRevLett.110.241104}.
  {\href{https://arxiv.org/abs/1202.1296}{{arXiv:1202.1296}}} {[hep-th]}

\bibitem[{Hui et~al.(2017)Hui, Ostriker, Tremaine, and Witten}]{Hui:2016ltb}
Hui L, Ostriker JP, Tremaine S, Witten E (2017) {Ultralight scalars as
  cosmological dark matter}. Phys Rev D 95(4):043541.
  \doi{10.1103/PhysRevD.95.043541}.
  {\href{https://arxiv.org/abs/1610.08297}{{arXiv:1610.08297}}} {[astro-ph.CO]}

\bibitem[{Hui et~al.(2019)Hui, Kabat, Li, Santoni, and Wong}]{Hui:2019aqm}
Hui L, Kabat D, Li X, Santoni L, Wong SS (2019) {Black Hole Hair from Scalar
  Dark Matter}. JCAP 06:038. \doi{10.1088/1475-7516/2019/06/038}.
  {\href{https://arxiv.org/abs/1904.12803}{{arXiv:1904.12803}}} {[gr-qc]}

\bibitem[{Hui et~al.(2021)Hui, Joyce, Penco, Santoni, and
  Solomon}]{Hui:2020xxx}
Hui L, Joyce A, Penco R, Santoni L, Solomon AR (2021) {Static response and Love
  numbers of Schwarzschild black holes}. JCAP 04:052.
  \doi{10.1088/1475-7516/2021/04/052}.
  {\href{https://arxiv.org/abs/2010.00593}{{arXiv:2010.00593}}} {[hep-th]}

\bibitem[{Husa et~al.(2016)Husa, Khan, Hannam, Pürrer, Ohme, Jiménez~Forteza,
  and Bohé}]{Husa:2015iqa}
Husa S, Khan S, Hannam M, Pürrer M, Ohme F, Jiménez~Forteza X, Bohé A (2016)
  {Frequency-domain gravitational waves from nonprecessing black-hole binaries.
  I. New numerical waveforms and anatomy of the signal}. Phys Rev
  D93(4):044006. \doi{10.1103/PhysRevD.93.044006}.
  {\href{https://arxiv.org/abs/1508.07250}{{arXiv:1508.07250}}} {[gr-qc]}

\bibitem[{Ikeda et~al.(2019)Ikeda, Brito, and Cardoso}]{Ikeda:2018nhb}
Ikeda T, Brito R, Cardoso V (2019) {Blasts of Light from Axions}. Phys Rev Lett
  122(8):081101. \doi{10.1103/PhysRevLett.122.081101}.
  {\href{https://arxiv.org/abs/1811.04950}{{arXiv:1811.04950}}} {[gr-qc]}

\bibitem[{Ikeda et~al.(2021)Ikeda, Bianchi, Consoli, Grillo, Morales, Pani, and
  Raposo}]{Ikeda:2021uvc}
Ikeda T, Bianchi M, Consoli D, Grillo A, Morales JF, Pani P, Raposo G (2021)
  {Black-hole microstate spectroscopy: ringdown, quasinormal modes, and echoes}
  {\href{https://arxiv.org/abs/2103.10960}{{arXiv:2103.10960}}} {[gr-qc]}

\bibitem[{Inman and Ali-Ha\"\i{}moud(2019)}]{Inman:2019wvr}
Inman D, Ali-Ha\"\i{}moud Y (2019) {Early structure formation in primordial
  black hole cosmologies}. Phys Rev D 100(8):083528.
  \doi{10.1103/PhysRevD.100.083528}.
  {\href{https://arxiv.org/abs/1907.08129}{{arXiv:1907.08129}}} {[astro-ph.CO]}

\bibitem[{Inomata and Nakama(2019)}]{Inomata:2018epa}
Inomata K, Nakama T (2019) {Gravitational waves induced by scalar perturbations
  as probes of the small-scale primordial spectrum}. Phys Rev D 99(4):043511.
  \doi{10.1103/PhysRevD.99.043511}.
  {\href{https://arxiv.org/abs/1812.00674}{{arXiv:1812.00674}}} {[astro-ph.CO]}

\bibitem[{Inomata and Terada(2020)}]{Inomata:2019yww}
Inomata K, Terada T (2020) {Gauge Independence of Induced Gravitational Waves}.
  Phys Rev D 101(2):023523. \doi{10.1103/PhysRevD.101.023523}.
  {\href{https://arxiv.org/abs/1912.00785}{{arXiv:1912.00785}}} {[gr-qc]}

\bibitem[{Inoue and Kusenko(2017)}]{Inoue:2017csr}
Inoue Y, Kusenko A (2017) {New X-ray bound on density of primordial black
  holes}. JCAP 1710:034. \doi{10.1088/1475-7516/2017/10/034}.
  {\href{https://arxiv.org/abs/1705.00791}{{arXiv:1705.00791}}} {[astro-ph.CO]}

\bibitem[{Irastorza and Redondo(2018)}]{Irastorza:2018dyq}
Irastorza IG, Redondo J (2018) {New experimental approaches in the search for
  axion-like particles}. Prog Part Nucl Phys 102:89--159.
  \doi{10.1016/j.ppnp.2018.05.003}.
  {\href{https://arxiv.org/abs/1801.08127}{{arXiv:1801.08127}}} {[hep-ph]}

\bibitem[{Isi et~al.(2019{\natexlab{a}})Isi, Giesler, Farr, Scheel, and
  Teukolsky}]{Isi:2019aib}
Isi M, Giesler M, Farr WM, Scheel MA, Teukolsky SA (2019{\natexlab{a}})
  {Testing the no-hair theorem with GW150914}. Phys Rev Lett 123(11):111102.
  \doi{10.1103/PhysRevLett.123.111102}.
  {\href{https://arxiv.org/abs/1905.00869}{{arXiv:1905.00869}}} {[gr-qc]}

\bibitem[{Isi et~al.(2019{\natexlab{b}})Isi, Sun, Brito, and
  Melatos}]{Isi:2018pzk}
Isi M, Sun L, Brito R, Melatos A (2019{\natexlab{b}}) {Directed searches for
  gravitational waves from ultralight bosons}. Phys Rev D 99(8):084042.
  \doi{10.1103/PhysRevD.99.084042}, [Erratum: Phys.Rev.D 102, 049901 (2020)].
  {\href{https://arxiv.org/abs/1810.03812}{{arXiv:1810.03812}}} {[gr-qc]}

\bibitem[{Islam et~al.(2020)Islam, Mehta, Ghosh, Varma, Ajith, and
  Sathyaprakash}]{Islam:2019dmk}
Islam T, Mehta AK, Ghosh A, Varma V, Ajith P, Sathyaprakash BS (2020) {Testing
  the no-hair nature of binary black holes using the consistency of multipolar
  gravitational radiation}. Phys Rev D101(2):024032.
  \doi{10.1103/PhysRevD.101.024032}.
  {\href{https://arxiv.org/abs/1910.14259}{{arXiv:1910.14259}}} {[gr-qc]}

\bibitem[{Islo et~al.(2019)Islo, Simon, Burke-Spolaor, and Siemens}]{BMS13}
Islo K, Simon J, Burke-Spolaor S, Siemens X (2019) {Prospects for Memory
  Detection with Low-Frequency Gravitational Wave Detectors}
  {\href{https://arxiv.org/abs/1906.11936}{{arXiv:1906.11936}}} {[astro-ph.HE]}

\bibitem[{Isoyama and Poisson(2012)}]{Isoyama:2012in}
Isoyama S, Poisson E (2012) {Self-force as probe of internal structure}. Class
  Quant Grav 29:155012. \doi{10.1088/0264-9381/29/15/155012}.
  {\href{https://arxiv.org/abs/1205.1236}{{arXiv:1205.1236}}} {[gr-qc]}

\bibitem[{Ivanov(1998)}]{Ivanov:1997ia}
Ivanov P (1998) {Nonlinear metric perturbations and production of primordial
  black holes}. Phys Rev D 57:7145--7154. \doi{10.1103/PhysRevD.57.7145}.
  {\href{https://arxiv.org/abs/astro-ph/9708224}{{arXiv:astro-ph/9708224}}}

\bibitem[{Ivanov et~al.(1994)Ivanov, Naselsky, and Novikov}]{Ivanov:1994pa}
Ivanov P, Naselsky P, Novikov I (1994) {Inflation and primordial black holes as
  dark matter}. Phys Rev D 50:7173--7178. \doi{10.1103/PhysRevD.50.7173}

\bibitem[{Ivanov et~al.(1999)Ivanov, Papaloizou, and Polnarev}]{Ivanov:1998qk}
Ivanov PB, Papaloizou JCB, Polnarev AG (1999) {The evolution of a supermassive
  binary caused by an accretion disc}. Mon Not Roy Astron Soc 307:79.
  \doi{10.1046/j.1365-8711.1999.02623.x}.
  {\href{https://arxiv.org/abs/astro-ph/9812198}{{arXiv:astro-ph/9812198}}}

\bibitem[{Jackiw and Pi(2003)}]{Jackiw:2003pm}
Jackiw R, Pi S (2003) {Chern-Simons modification of general relativity}. Phys
  Rev D 68:104012. \doi{10.1103/PhysRevD.68.104012}.
  {\href{https://arxiv.org/abs/gr-qc/0308071}{{arXiv:gr-qc/0308071}}}

\bibitem[{Jacobson(1999)}]{Jacobson:1999vr}
Jacobson T (1999) {Primordial black hole evolution in tensor scalar cosmology}.
  Phys Rev Lett 83:2699--2702. \doi{10.1103/PhysRevLett.83.2699}.
  {\href{https://arxiv.org/abs/astro-ph/9905303}{{arXiv:astro-ph/9905303}}}

\bibitem[{Jacobson(2007)}]{Jacobson:2008aj}
Jacobson T (2007) {Einstein-aether gravity: A Status report}. PoS QG-PH:020.
  \doi{10.22323/1.043.0020}.
  {\href{https://arxiv.org/abs/0801.1547}{{arXiv:0801.1547}}} {[gr-qc]}

\bibitem[{Jacobson and Mattingly(2001)}]{Jacobson:2000xp}
Jacobson T, Mattingly D (2001) {Gravity with a dynamical preferred frame}. Phys
  Rev D 64:024028. \doi{10.1103/PhysRevD.64.024028}.
  {\href{https://arxiv.org/abs/gr-qc/0007031}{{arXiv:gr-qc/0007031}}}

\bibitem[{Jaeckel and Ringwald(2010)}]{Jaeckel:2010ni}
Jaeckel J, Ringwald A (2010) {The Low-Energy Frontier of Particle Physics}. Ann
  Rev Nucl Part Sci 60:405--437. \doi{10.1146/annurev.nucl.012809.104433}.
  {\href{https://arxiv.org/abs/1002.0329}{{arXiv:1002.0329}}} {[hep-ph]}

\bibitem[{Jang-Condell and Sasselov(2005)}]{JangCondell:2004jr}
Jang-Condell H, Sasselov DD (2005) {Type I migration in a non-isothermal
  protoplanetary disk}. Astrophys J 619:1123--1131. \doi{10.1086/426577}.
  {\href{https://arxiv.org/abs/astro-ph/0410550}{{arXiv:astro-ph/0410550}}}

\bibitem[{Jani et~al.(2016)Jani, Healy, Clark, London, Laguna, and
  Shoemaker}]{Jani:2016wkt}
Jani K, Healy J, Clark JA, London L, Laguna P, Shoemaker D (2016) {Georgia Tech
  Catalog of Gravitational Waveforms}. Class Quant Grav 33(20):204001.
  \doi{10.1088/0264-9381/33/20/204001}.
  {\href{https://arxiv.org/abs/1605.03204}{{arXiv:1605.03204}}} {[gr-qc]}

\bibitem[{Jedamzik(2020{\natexlab{a}})}]{Jedamzik:2020omx}
Jedamzik K (2020{\natexlab{a}}) {Evidence for primordial black hole dark matter
  from LIGO/Virgo merger rates}
  {\href{https://arxiv.org/abs/2007.03565}{{arXiv:2007.03565}}} {[astro-ph.CO]}

\bibitem[{Jedamzik(2020{\natexlab{b}})}]{Jedamzik:2020ypm}
Jedamzik K (2020{\natexlab{b}}) {Primordial Black Hole Dark Matter and the
  LIGO/Virgo observations}. JCAP 09:022. \doi{10.1088/1475-7516/2020/09/022}.
  {\href{https://arxiv.org/abs/2006.11172}{{arXiv:2006.11172}}} {[astro-ph.CO]}

\bibitem[{Jee et~al.(2019)Jee, Suyu, Komatsu, Fassnacht, Hilbert, and
  Koopmans}]{Jee:2019hah}
Jee I, Suyu S, Komatsu E, Fassnacht CD, Hilbert S, Koopmans LV (2019) {A
  measurement of the Hubble constant from angular diameter distances to two
  gravitational lenses} \doi{10.1126/science.aat7371}.
  {\href{https://arxiv.org/abs/1909.06712}{{arXiv:1909.06712}}} {[astro-ph.CO]}

\bibitem[{Jenkins and Sakellariadou(2018)}]{Jenkins:2018nty}
Jenkins AC, Sakellariadou M (2018) {Anisotropies in the stochastic
  gravitational-wave background: Formalism and the cosmic string case}. Phys
  Rev D 98(6):063509. \doi{10.1103/PhysRevD.98.063509}.
  {\href{https://arxiv.org/abs/1802.06046}{{arXiv:1802.06046}}} {[astro-ph.CO]}

\bibitem[{{Jenkins} and {Sakellariadou}(2018)}]{Jenkins2018PhRvD}
{Jenkins} AC, {Sakellariadou} M (2018) {Anisotropies in the stochastic
  gravitational-wave background: Formalism and the cosmic string case}. \prd
  98(6):063509. \doi{10.1103/PhysRevD.98.063509}.
  {\href{https://arxiv.org/abs/1802.06046}{{arXiv:1802.06046}}} {[astro-ph.CO]}

\bibitem[{{Jenkins} et~al.(2019){Jenkins}, {O'Shaughnessy}, {Sakellariadou},
  and {Wysocki}}]{Jenkins2019PhRvL}
{Jenkins} AC, {O'Shaughnessy} R, {Sakellariadou} M, {Wysocki} D (2019)
  {Anisotropies in the Astrophysical Gravitational-Wave Background: The Impact
  of Black Hole Distributions}. \prl 122(11):111101.
  \doi{10.1103/PhysRevLett.122.111101}.
  {\href{https://arxiv.org/abs/1810.13435}{{arXiv:1810.13435}}} {[astro-ph.CO]}

\bibitem[{Jetzer(1992)}]{Jetzer:1991jr}
Jetzer P (1992) {Boson stars}. Phys Rept 220:163--227.
  \doi{10.1016/0370-1573(92)90123-H}

\bibitem[{Jim\'enez et~al.(2020)Jim\'enez, Ezquiaga, and
  Heisenberg}]{Jimenez:2019lrk}
Jim\'enez JB, Ezquiaga JM, Heisenberg L (2020) {Probing cosmological fields
  with gravitational wave oscillations}. JCAP 04:027.
  \doi{10.1088/1475-7516/2020/04/027}.
  {\href{https://arxiv.org/abs/1912.06104}{{arXiv:1912.06104}}} {[astro-ph.CO]}

\bibitem[{Jim\'enez-Forteza et~al.(2017)Jim\'enez-Forteza, Keitel, Husa,
  Hannam, Khan, and P\"urrer}]{Jimenez-Forteza:2016oae}
Jim\'enez-Forteza X, Keitel D, Husa S, Hannam M, Khan S, P\"urrer M (2017)
  {Hierarchical data-driven approach to fitting numerical relativity data for
  nonprecessing binary black holes with an application to final spin and
  radiated energy}. Phys Rev D 95(6):064024. \doi{10.1103/PhysRevD.95.064024}.
  {\href{https://arxiv.org/abs/1611.00332}{{arXiv:1611.00332}}} {[gr-qc]}

\bibitem[{Jim\'enez~Forteza et~al.(2020)Jim\'enez~Forteza, Bhagwat, Pani, and
  Ferrari}]{Forteza:2020hbw}
Jim\'enez~Forteza X, Bhagwat S, Pani P, Ferrari V (2020) {Spectroscopy of
  binary black hole ringdown using overtones and angular modes}. Phys Rev D
  102(4):044053. \doi{10.1103/PhysRevD.102.044053}.
  {\href{https://arxiv.org/abs/2005.03260}{{arXiv:2005.03260}}} {[gr-qc]}

\bibitem[{Johannsen and Psaltis(2011)}]{Johannsen:2011dh}
Johannsen T, Psaltis D (2011) {A Metric for Rapidly Spinning Black Holes
  Suitable for Strong-Field Tests of the No-Hair Theorem}. Phys Rev D
  83:124015. \doi{10.1103/PhysRevD.83.124015}.
  {\href{https://arxiv.org/abs/1105.3191}{{arXiv:1105.3191}}} {[gr-qc]}

\bibitem[{Johnson-Mcdaniel et~al.(2018)Johnson-Mcdaniel, Mukherjee, Kashyap,
  Ajith, Del~Pozzo, and Vitale}]{Johnson-McDaniel:2018uvs}
Johnson-Mcdaniel NK, Mukherjee A, Kashyap R, Ajith P, Del~Pozzo W, Vitale S
  (2018) {Constraining black hole mimickers with gravitational wave
  observations} {\href{https://arxiv.org/abs/1804.08026}{{arXiv:1804.08026}}}
  {[gr-qc]}

\bibitem[{de~Jong et~al.(2012)}]{deJong:2012nj}
de~Jong RS, et~al. (2012) {4MOST - 4-metre Multi-Object Spectroscopic
  Telescope}. Proc SPIE Int Soc Opt Eng 8446:84460T. \doi{10.1117/12.926239}.
  {\href{https://arxiv.org/abs/1206.6885}{{arXiv:1206.6885}}} {[astro-ph.IM]}

\bibitem[{Jovanovic(2012)}]{Jovanovic:2011tx}
Jovanovic P (2012) {The broad Fe K$\alpha$ line and supermassive black holes}.
  New Astron Rev 56:37--48. \doi{10.1016/j.newar.2011.11.002}.
  {\href{https://arxiv.org/abs/1112.0172}{{arXiv:1112.0172}}} {[astro-ph.CO]}

\bibitem[{Joyce et~al.(2015)Joyce, Jain, Khoury, and Trodden}]{Joyce:2014kja}
Joyce A, Jain B, Khoury J, Trodden M (2015) {Beyond the Cosmological Standard
  Model}. Phys Rept 568:1--98. \doi{10.1016/j.physrep.2014.12.002}.
  {\href{https://arxiv.org/abs/1407.0059}{{arXiv:1407.0059}}} {[astro-ph.CO]}

\bibitem[{Joyce et~al.(2016)Joyce, Lombriser, and Schmidt}]{Joyce:2016vqv}
Joyce A, Lombriser L, Schmidt F (2016) {Dark Energy Versus Modified Gravity}.
  Ann Rev Nucl Part Sci 66:95--122. \doi{10.1146/annurev-nucl-102115-044553}.
  {\href{https://arxiv.org/abs/1601.06133}{{arXiv:1601.06133}}} {[astro-ph.CO]}

\bibitem[{Juli\'e(2018)}]{Julie:2017ucp}
Juli\'e FL (2018) {Reducing the two-body problem in scalar-tensor theories to
  the motion of a test particle : a scalar-tensor effective-one-body approach}.
  Phys Rev D 97(2):024047. \doi{10.1103/PhysRevD.97.024047}.
  {\href{https://arxiv.org/abs/1709.09742}{{arXiv:1709.09742}}} {[gr-qc]}

\bibitem[{Juli\'e and Berti(2019)}]{Julie:2019sab}
Juli\'e FL, Berti E (2019) {Post-Newtonian dynamics and black hole
  thermodynamics in Einstein-scalar-Gauss-Bonnet gravity}. Phys Rev D
  100(10):104061. \doi{10.1103/PhysRevD.100.104061}.
  {\href{https://arxiv.org/abs/1909.05258}{{arXiv:1909.05258}}} {[gr-qc]}

\bibitem[{Juli\'e and Deruelle(2017)}]{Julie:2017pkb}
Juli\'e FL, Deruelle N (2017) {Two-body problem in Scalar-Tensor theories as a
  deformation of General Relativity : an Effective-One-Body approach}. Phys Rev
  D 95(12):124054. \doi{10.1103/PhysRevD.95.124054}.
  {\href{https://arxiv.org/abs/1703.05360}{{arXiv:1703.05360}}} {[gr-qc]}

\bibitem[{Kahlhoefer(2017)}]{Kahlhoefer:2017dnp}
Kahlhoefer F (2017) {Review of LHC Dark Matter Searches}. Int J Mod Phys A
  32(13):1730006. \doi{10.1142/S0217751X1730006X}.
  {\href{https://arxiv.org/abs/1702.02430}{{arXiv:1702.02430}}} {[hep-ph]}

\bibitem[{Kaloper and Padilla(2014)}]{Kaloper:2013zca}
Kaloper N, Padilla A (2014) {Sequestering the Standard Model Vacuum Energy}.
  Phys Rev Lett 112(9):091304. \doi{10.1103/PhysRevLett.112.091304}.
  {\href{https://arxiv.org/abs/1309.6562}{{arXiv:1309.6562}}} {[hep-th]}

\bibitem[{Kamaretsos et~al.(2012)Kamaretsos, Hannam, Husa, and
  Sathyaprakash}]{Kamaretsos:2011um}
Kamaretsos I, Hannam M, Husa S, Sathyaprakash B (2012) {Black-hole hair loss:
  learning about binary progenitors from ringdown signals}. Phys Rev D
  85:024018. \doi{10.1103/PhysRevD.85.024018}.
  {\href{https://arxiv.org/abs/1107.0854}{{arXiv:1107.0854}}} {[gr-qc]}

\bibitem[{{Kanagawa} et~al.(2018){Kanagawa}, {Tanaka}, and
  {Szuszkiewicz}}]{2018ApJ...861..140K}
{Kanagawa} KD, {Tanaka} H, {Szuszkiewicz} E (2018) {Radial Migration of
  Gap-opening Planets in Protoplanetary Disks. I. The Case of a Single Planet}.
  \apj 861(2):140. \doi{10.3847/1538-4357/aac8d9}.
  {\href{https://arxiv.org/abs/1805.11101}{{arXiv:1805.11101}}} {[astro-ph.EP]}

\bibitem[{Kanti et~al.(1996)Kanti, Mavromatos, Rizos, Tamvakis, and
  Winstanley}]{Kanti:1995vq}
Kanti P, Mavromatos N, Rizos J, Tamvakis K, Winstanley E (1996) {Dilatonic
  black holes in higher curvature string gravity}. Phys Rev D 54:5049--5058.
  \doi{10.1103/PhysRevD.54.5049}.
  {\href{https://arxiv.org/abs/hep-th/9511071}{{arXiv:hep-th/9511071}}}

\bibitem[{Kaplan and Rajendran(2019)}]{Kaplan:2018dqx}
Kaplan DE, Rajendran S (2019) {Firewalls in General Relativity}. Phys Rev D
  99(4):044033. \doi{10.1103/PhysRevD.99.044033}.
  {\href{https://arxiv.org/abs/1812.00536}{{arXiv:1812.00536}}} {[hep-th]}

\bibitem[{Karnesis et~al.(2019)Karnesis, Lilley, and
  Petiteau}]{Karnesis:2019mph}
Karnesis N, Lilley M, Petiteau A (2019) {Assessing the detectability of a
  Stochastic Gravitational Wave Background with LISA, using an excess of power
  approach} {\href{https://arxiv.org/abs/1906.09027}{{arXiv:1906.09027}}}
  {[astro-ph.IM]}

\bibitem[{Kase et~al.(2018)Kase, Minamitsuji, and Tsujikawa}]{Kase:2018owh}
Kase R, Minamitsuji M, Tsujikawa S (2018) {Black holes in quartic-order
  beyond-generalized Proca theories}. Phys Lett B 782:541--550.
  \doi{10.1016/j.physletb.2018.05.078}.
  {\href{https://arxiv.org/abs/1803.06335}{{arXiv:1803.06335}}} {[gr-qc]}

\bibitem[{Kastha et~al.(2018)Kastha, Gupta, Arun, Sathyaprakash, and Van
  Den~Broeck}]{Kastha:2018bcr}
Kastha S, Gupta A, Arun K, Sathyaprakash B, Van Den~Broeck C (2018) {Testing
  the multipole structure of compact binaries using gravitational wave
  observations}. Phys Rev D 98(12):124033. \doi{10.1103/PhysRevD.98.124033}.
  {\href{https://arxiv.org/abs/1809.10465}{{arXiv:1809.10465}}} {[gr-qc]}

\bibitem[{Kastha et~al.(2019)Kastha, Gupta, Arun, Sathyaprakash, and Van
  Den~Broeck}]{Kastha:2019brk}
Kastha S, Gupta A, Arun K, Sathyaprakash B, Van Den~Broeck C (2019) {Testing
  the multipole structure and conservative dynamics of compact binaries using
  gravitational wave observations: The spinning case}. Phys Rev D
  100(4):044007. \doi{10.1103/PhysRevD.100.044007}.
  {\href{https://arxiv.org/abs/1905.07277}{{arXiv:1905.07277}}} {[gr-qc]}

\bibitem[{Katz et~al.(2018)Katz, Kopp, Sibiryakov, and Xue}]{Katz:2018zrn}
Katz A, Kopp J, Sibiryakov S, Xue W (2018) {Femtolensing by Dark Matter
  Revisited}. JCAP 12:005. \doi{10.1088/1475-7516/2018/12/005}.
  {\href{https://arxiv.org/abs/1807.11495}{{arXiv:1807.11495}}} {[astro-ph.CO]}

\bibitem[{Kaup(1968)}]{Kaup:1968zz}
Kaup DJ (1968) {Klein-Gordon Geon}. Phys Rev 172:1331--1342.
  \doi{10.1103/PhysRev.172.1331}

\bibitem[{Kavanagh et~al.(2020)Kavanagh, Nichols, Bertone, and
  Gaggero}]{Kavanagh:2020cfn}
Kavanagh BJ, Nichols DA, Bertone G, Gaggero D (2020) {Detecting dark matter
  around black holes with gravitational waves: Effects of dark-matter dynamics
  on the gravitational waveform}
  {\href{https://arxiv.org/abs/2002.12811}{{arXiv:2002.12811}}} {[gr-qc]}

\bibitem[{Kawaguchi et~al.(2018)Kawaguchi, Kiuchi, Kyutoku, Sekiguchi, Shibata,
  and Taniguchi}]{Kawaguchi:2018gvj}
Kawaguchi K, Kiuchi K, Kyutoku K, Sekiguchi Y, Shibata M, Taniguchi K (2018)
  {Frequency-domain gravitational waveform models for inspiraling binary
  neutron stars}. Phys Rev D 97(4):044044. \doi{10.1103/PhysRevD.97.044044}.
  {\href{https://arxiv.org/abs/1802.06518}{{arXiv:1802.06518}}} {[gr-qc]}

\bibitem[{Kehagias and Maggiore(2014)}]{Kehagias:2014sda}
Kehagias A, Maggiore M (2014) {Spherically symmetric static solutions in a
  non-local infrared modification of General Relativity}. JHEP 1408:029.
  \doi{10.1007/JHEP08(2014)029}.
  {\href{https://arxiv.org/abs/1401.8289}{{arXiv:1401.8289}}} {[hep-th]}

\bibitem[{Keir(2016)}]{Keir:2014oka}
Keir J (2016) {Slowly decaying waves on spherically symmetric spacetimes and
  ultracompact neutron stars}. Class Quant Grav 33(13):135009.
  \doi{10.1088/0264-9381/33/13/135009}.
  {\href{https://arxiv.org/abs/1404.7036}{{arXiv:1404.7036}}} {[gr-qc]}

\bibitem[{Kerr(1963)}]{Kerr:1963ud}
Kerr RP (1963) {Gravitational field of a spinning mass as an example of
  algebraically special metrics}. Phys Rev Lett 11:237--238.
  \doi{10.1103/PhysRevLett.11.237}

\bibitem[{Kesden et~al.(2005)Kesden, Gair, and Kamionkowski}]{Kesden:2004qx}
Kesden M, Gair J, Kamionkowski M (2005) {Gravitational-wave signature of an
  inspiral into a supermassive horizonless object}. Phys Rev D71:044015.
  \doi{10.1103/PhysRevD.71.044015}.
  {\href{https://arxiv.org/abs/astro-ph/0411478}{{arXiv:astro-ph/0411478}}}
  {[astro-ph]}

\bibitem[{Khalil et~al.(2018)Khalil, Sennett, Steinhoff, Vines, and
  Buonanno}]{Khalil:2018aaj}
Khalil M, Sennett N, Steinhoff J, Vines J, Buonanno A (2018) {Hairy binary
  black holes in Einstein-Maxwell-dilaton theory and their effective-one-body
  description}. Phys Rev D 98(10):104010. \doi{10.1103/PhysRevD.98.104010}.
  {\href{https://arxiv.org/abs/1809.03109}{{arXiv:1809.03109}}} {[gr-qc]}

\bibitem[{Khalil et~al.(2019)Khalil, Sennett, Steinhoff, and
  Buonanno}]{Khalil:2019wyy}
Khalil M, Sennett N, Steinhoff J, Buonanno A (2019) {Theory-agnostic framework
  for dynamical scalarization of compact binaries}. Phys Rev D 100(12):124013.
  \doi{10.1103/PhysRevD.100.124013}.
  {\href{https://arxiv.org/abs/1906.08161}{{arXiv:1906.08161}}} {[gr-qc]}

\bibitem[{Khalil et~al.(2020)Khalil, Steinhoff, Vines, and
  Buonanno}]{Khalil:2020mmr}
Khalil M, Steinhoff J, Vines J, Buonanno A (2020) {Fourth post-Newtonian
  effective-one-body Hamiltonians with generic spins}. Phys Rev D
  101(10):104034. \doi{10.1103/PhysRevD.101.104034}.
  {\href{https://arxiv.org/abs/2003.04469}{{arXiv:2003.04469}}} {[gr-qc]}

\bibitem[{Khan et~al.(2016)Khan, Husa, Hannam, Ohme, Pürrer, Jiménez~Forteza,
  and Bohé}]{Khan:2015jqa}
Khan S, Husa S, Hannam M, Ohme F, Pürrer M, Jiménez~Forteza X, Bohé A (2016)
  {Frequency-domain gravitational waves from nonprecessing black-hole binaries.
  II. A phenomenological model for the advanced detector era}. Phys Rev
  D93(4):044007. \doi{10.1103/PhysRevD.93.044007}.
  {\href{https://arxiv.org/abs/1508.07253}{{arXiv:1508.07253}}} {[gr-qc]}

\bibitem[{Khan et~al.(2019)Khan, Chatziioannou, Hannam, and
  Ohme}]{Khan:2018fmp}
Khan S, Chatziioannou K, Hannam M, Ohme F (2019) {Phenomenological model for
  the gravitational-wave signal from precessing binary black holes with
  two-spin effects}. Phys Rev D 100(2):024059.
  \doi{10.1103/PhysRevD.100.024059}.
  {\href{https://arxiv.org/abs/1809.10113}{{arXiv:1809.10113}}} {[gr-qc]}

\bibitem[{Khan et~al.(2020)Khan, Ohme, Chatziioannou, and
  Hannam}]{Khan:2019kot}
Khan S, Ohme F, Chatziioannou K, Hannam M (2020) {Including higher order
  multipoles in gravitational-wave models for precessing binary black holes}.
  Phys Rev D 101(2):024056. \doi{10.1103/PhysRevD.101.024056}.
  {\href{https://arxiv.org/abs/1911.06050}{{arXiv:1911.06050}}} {[gr-qc]}

\bibitem[{Khoury and Weltman(2004)}]{Khoury:2003rn}
Khoury J, Weltman A (2004) {Chameleon cosmology}. Phys Rev D69:044026.
  \doi{10.1103/PhysRevD.69.044026}.
  {\href{https://arxiv.org/abs/astro-ph/0309411}{{arXiv:astro-ph/0309411}}}
  {[astro-ph]}

\bibitem[{Kibble(1985)}]{Kibble:1984hp}
Kibble TWB (1985) {Evolution of a system of cosmic strings}. Nucl Phys B
  252:227. \doi{10.1016/0550-3213(85)90596-6}, [Erratum: Nucl.Phys.B 261, 750
  (1985)]

\bibitem[{Kimura and Yamamoto(2012)}]{Kimura:2011qn}
Kimura R, Yamamoto K (2012) {Constraints on general second-order scalar-tensor
  models from gravitational Cherenkov radiation}. JCAP 07:050.
  \doi{10.1088/1475-7516/2012/07/050}.
  {\href{https://arxiv.org/abs/1112.4284}{{arXiv:1112.4284}}} {[astro-ph.CO]}

\bibitem[{Kleihaus et~al.(2005)Kleihaus, Kunz, and List}]{Kleihaus:2005me}
Kleihaus B, Kunz J, List M (2005) {Rotating boson stars and Q-balls}. Phys Rev
  D 72:064002. \doi{10.1103/PhysRevD.72.064002}.
  {\href{https://arxiv.org/abs/gr-qc/0505143}{{arXiv:gr-qc/0505143}}}

\bibitem[{Kleihaus et~al.(2011)Kleihaus, Kunz, and Radu}]{Kleihaus:2011tg}
Kleihaus B, Kunz J, Radu E (2011) {Rotating Black Holes in Dilatonic
  Einstein-Gauss-Bonnet Theory}. Phys Rev Lett 106:151104.
  \doi{10.1103/PhysRevLett.106.151104}.
  {\href{https://arxiv.org/abs/1101.2868}{{arXiv:1101.2868}}} {[gr-qc]}

\bibitem[{Kleihaus et~al.(2015)Kleihaus, Kunz, and
  Yazadjiev}]{Kleihaus:2015iea}
Kleihaus B, Kunz J, Yazadjiev S (2015) {Scalarized Hairy Black Holes}. Phys
  Lett B 744:406--412. \doi{10.1016/j.physletb.2015.04.014}.
  {\href{https://arxiv.org/abs/1503.01672}{{arXiv:1503.01672}}} {[gr-qc]}

\bibitem[{Klein et~al.(2016)}]{Klein:2015hvg}
Klein A, et~al. (2016) {Science with the space-based interferometer eLISA:
  Supermassive black hole binaries}. Phys Rev D 93(2):024003.
  \doi{10.1103/PhysRevD.93.024003}.
  {\href{https://arxiv.org/abs/1511.05581}{{arXiv:1511.05581}}} {[gr-qc]}

\bibitem[{Knorr and Saueressig(2018)}]{Knorr:2018kog}
Knorr B, Saueressig F (2018) {Towards reconstructing the quantum effective
  action of gravity}. Phys Rev Lett 121:161304.
  \doi{10.1103/PhysRevLett.121.161304}.
  {\href{https://arxiv.org/abs/1804.03846}{{arXiv:1804.03846}}} {[hep-th]}

\bibitem[{Kobayashi(2019)}]{Kobayashi:2019hrl}
Kobayashi T (2019) {Horndeski theory and beyond: a review}. Rept Prog Phys
  82(8):086901. \doi{10.1088/1361-6633/ab2429}.
  {\href{https://arxiv.org/abs/1901.07183}{{arXiv:1901.07183}}} {[gr-qc]}

\bibitem[{Kocsis et~al.(2011)Kocsis, Yunes, and Loeb}]{Kocsis:2011dr}
Kocsis B, Yunes N, Loeb A (2011) {Observable Signatures of EMRI Black Hole
  Binaries Embedded in Thin Accretion Disks}. Phys Rev D 84:024032.
  \doi{10.1103/PhysRevD.86.049907}.
  {\href{https://arxiv.org/abs/1104.2322}{{arXiv:1104.2322}}} {[astro-ph.GA]}

\bibitem[{Kocsis et~al.(2012)Kocsis, Haiman, and Loeb}]{Kocsis:2012ui}
Kocsis B, Haiman Z, Loeb A (2012) {Gas pile-up, gap overflow, and Type 1.5
  migration in circumbinary disks: application to supermassive black hole
  binaries}. Mon Not Roy Astron Soc 427:2680--2700.
  \doi{10.1111/j.1365-2966.2012.22118.x}.
  {\href{https://arxiv.org/abs/1205.5268}{{arXiv:1205.5268}}} {[astro-ph.HE]}

\bibitem[{Kohri and Terada(2018)}]{Kohri:2018awv}
Kohri K, Terada T (2018) {Semianalytic calculation of gravitational wave
  spectrum nonlinearly induced from primordial curvature perturbations}. Phys
  Rev D 97(12):123532. \doi{10.1103/PhysRevD.97.123532}.
  {\href{https://arxiv.org/abs/1804.08577}{{arXiv:1804.08577}}} {[gr-qc]}

\bibitem[{Kohri and Terada(2020)}]{Kohri:2020qqd}
Kohri K, Terada T (2020) {Possible Solar-Mass Primordial Black Holes for
  NANOGrav Hint of Gravitational Waves}
  {\href{https://arxiv.org/abs/2009.11853}{{arXiv:2009.11853}}} {[astro-ph.CO]}

\bibitem[{Kokkotas(1995)}]{Kokkotas:1995av}
Kokkotas KD (1995) {Pulsating relativistic stars}. In: {Relativistic
  gravitation and gravitational radiation. Proceedings, School of Physics, Les
  Houches, France, September 26-October 6, 1995}. pp 89--102.
  {\href{https://arxiv.org/abs/gr-qc/9603024}{{arXiv:gr-qc/9603024}}} {[gr-qc]}

\bibitem[{Kokkotas and Schmidt(1999)}]{Kokkotas:1999bd}
Kokkotas KD, Schmidt BG (1999) {Quasinormal modes of stars and black holes}.
  Living Rev Rel 2:2. \doi{10.12942/lrr-1999-2}.
  {\href{https://arxiv.org/abs/gr-qc/9909058}{{arXiv:gr-qc/9909058}}} {[gr-qc]}

\bibitem[{Konoplya and Zhidenko(2020)}]{Konoplya:2020hyk}
Konoplya R, Zhidenko A (2020) {General parametrization of black holes: The only
  parameters that matter}. Phys Rev D 101(12):124004.
  \doi{10.1103/PhysRevD.101.124004}.
  {\href{https://arxiv.org/abs/2001.06100}{{arXiv:2001.06100}}} {[gr-qc]}

\bibitem[{Korol et~al.(2017)Korol, Rossi, Groot, Nelemans, Toonen, and
  Brown}]{Korol:2017qcx}
Korol V, Rossi EM, Groot PJ, Nelemans G, Toonen S, Brown AG (2017) {Prospects
  for detection of detached double white dwarf binaries with Gaia, LSST and
  LISA}. Mon Not Roy Astron Soc 470(2):1894--1910. \doi{10.1093/mnras/stx1285}.
  {\href{https://arxiv.org/abs/1703.02555}{{arXiv:1703.02555}}} {[astro-ph.HE]}

\bibitem[{Kosteleck\'y and Mewes(2016)}]{Kostelecky:2016kfm}
Kosteleck\'y VA, Mewes M (2016) {Testing local Lorentz invariance with
  gravitational waves}. Phys Lett B 757:510--514.
  \doi{10.1016/j.physletb.2016.04.040}.
  {\href{https://arxiv.org/abs/1602.04782}{{arXiv:1602.04782}}} {[gr-qc]}

\bibitem[{Koushiappas and Loeb(2017)}]{Koushiappas:2017kqm}
Koushiappas SM, Loeb A (2017) {Maximum redshift of gravitational wave merger
  events}. Phys Rev Lett 119(22):221104. \doi{10.1103/PhysRevLett.119.221104}.
  {\href{https://arxiv.org/abs/1708.07380}{{arXiv:1708.07380}}} {[astro-ph.CO]}

\bibitem[{Kov\'acs and Reall(2020)}]{Kovacs:2020pns}
Kov\'acs AD, Reall HS (2020) {Well-Posed Formulation of Scalar-Tensor Effective
  Field Theory}. Phys Rev Lett 124(22):221101.
  \doi{10.1103/PhysRevLett.124.221101}.
  {\href{https://arxiv.org/abs/2003.04327}{{arXiv:2003.04327}}} {[gr-qc]}

\bibitem[{Koyama et~al.(2011)Koyama, Niz, and Tasinato}]{Koyama:2011xz}
Koyama K, Niz G, Tasinato G (2011) {Analytic solutions in non-linear massive
  gravity}. Phys Rev Lett 107:131101. \doi{10.1103/PhysRevLett.107.131101}.
  {\href{https://arxiv.org/abs/1103.4708}{{arXiv:1103.4708}}} {[hep-th]}

\bibitem[{{Kozai}(1962)}]{1962AJ.....67..591K}
{Kozai} Y (1962) {Secular perturbations of asteroids with high inclination and
  eccentricity}. \aj 67:591--598. \doi{10.1086/108790}

\bibitem[{Krishnendu and Yelikar(2019)}]{Krishnendu:2019ebd}
Krishnendu N, Yelikar AB (2019) {Testing the Kerr Nature of Intermediate-Mass
  and Supermassive Black Hole Binaries Using Spin-Induced Multipole Moment
  Measurements} {\href{https://arxiv.org/abs/1904.12712}{{arXiv:1904.12712}}}
  {[gr-qc]}

\bibitem[{Krishnendu et~al.(2017)Krishnendu, Arun, and
  Mishra}]{Krishnendu:2017shb}
Krishnendu N, Arun K, Mishra CK (2017) {Testing the binary black hole nature of
  a compact binary coalescence}. Phys Rev Lett 119(9):091101.
  \doi{10.1103/PhysRevLett.119.091101}.
  {\href{https://arxiv.org/abs/1701.06318}{{arXiv:1701.06318}}} {[gr-qc]}

\bibitem[{Kritos et~al.(2020)Kritos, De~Luca, Franciolini, Kehagias, and
  Riotto}]{Kritos:2020wcl}
Kritos K, De~Luca V, Franciolini G, Kehagias A, Riotto A (2020) {The
  Astro-Primordial Black Hole Merger Rates: a Reappraisal}
  {\href{https://arxiv.org/abs/2012.03585}{{arXiv:2012.03585}}} {[gr-qc]}

\bibitem[{K\"uhnel et~al.(2017)K\"uhnel, Starkman, and Freese}]{Kuhnel:2017bvu}
K\"uhnel F, Starkman GD, Freese K (2017) {Primordial Black-Hole and Macroscopic
  Dark-Matter Constraints with LISA}
  {\href{https://arxiv.org/abs/1705.10361}{{arXiv:1705.10361}}} {[gr-qc]}

\bibitem[{Kuhnel et~al.(2020)Kuhnel, Matas, Starkman, and
  Freese}]{Kuhnel:2018mlr}
Kuhnel F, Matas A, Starkman GD, Freese K (2020) {Waves from the Centre: Probing
  PBH and other Macroscopic Dark Matter with LISA}. Eur Phys J C 80(7):627.
  \doi{10.1140/epjc/s10052-020-8183-4}.
  {\href{https://arxiv.org/abs/1811.06387}{{arXiv:1811.06387}}} {[gr-qc]}

\bibitem[{Kunihiro(1995)}]{Kunihiro:1995zt}
Kunihiro T (1995) {A Geometrical formulation of the renormalization group
  method for global analysis}. Prog Theor Phys 94:503--514.
  \doi{10.1143/PTP.94.503}, [Erratum: Prog. Theor. Phys.95,835(1996)].
  {\href{https://arxiv.org/abs/hep-th/9505166}{{arXiv:hep-th/9505166}}}
  {[hep-th]}

\bibitem[{Kuroda et~al.(2014)Kuroda, Takiwaki, and Kotake}]{Kuroda:2013rga}
Kuroda T, Takiwaki T, Kotake K (2014) {Gravitational Wave Signatures from
  Low-mode Spiral Instabilities in Rapidly Rotating Supernova Cores}. Phys Rev
  D 89(4):044011. \doi{10.1103/PhysRevD.89.044011}.
  {\href{https://arxiv.org/abs/1304.4372}{{arXiv:1304.4372}}} {[astro-ph.HE]}

\bibitem[{Kuroyanagi et~al.(2015)Kuroyanagi, Takahashi, and
  Yokoyama}]{Kuroyanagi:2014nba}
Kuroyanagi S, Takahashi T, Yokoyama S (2015) {Blue-tilted Tensor Spectrum and
  Thermal History of the Universe}. JCAP 02:003.
  \doi{10.1088/1475-7516/2015/02/003}.
  {\href{https://arxiv.org/abs/1407.4785}{{arXiv:1407.4785}}} {[astro-ph.CO]}

\bibitem[{Kuroyanagi et~al.(2018)Kuroyanagi, Chiba, and
  Takahashi}]{Kuroyanagi:2018csn}
Kuroyanagi S, Chiba T, Takahashi T (2018) {Probing the Universe through the
  Stochastic Gravitational Wave Background}. JCAP 11:038.
  \doi{10.1088/1475-7516/2018/11/038}.
  {\href{https://arxiv.org/abs/1807.00786}{{arXiv:1807.00786}}} {[astro-ph.CO]}

\bibitem[{Kusenko et~al.(2020)Kusenko, Sasaki, Sugiyama, Takada, Takhistov, and
  Vitagliano}]{Kusenko:2020pcg}
Kusenko A, Sasaki M, Sugiyama S, Takada M, Takhistov V, Vitagliano E (2020)
  {Exploring Primordial Black Holes from Multiverse with Optical Telescopes}
  {\href{https://arxiv.org/abs/2001.09160}{{arXiv:2001.09160}}} {[astro-ph.CO]}

\bibitem[{Kyutoku and Seto(2016)}]{Kyutoku:2016ppx}
Kyutoku K, Seto N (2016) {Concise estimate of the expected number of detections
  for stellar-mass binary black holes by eLISA}. Mon Not Roy Astron Soc
  462(2):2177--2183. \doi{10.1093/mnras/stw1767}.
  {\href{https://arxiv.org/abs/1606.02298}{{arXiv:1606.02298}}} {[astro-ph.HE]}

\bibitem[{Kyutoku and Seto(2017)}]{Kyutoku:2016zxn}
Kyutoku K, Seto N (2017) {Gravitational-wave cosmography with LISA and the
  Hubble tension}. Phys Rev D 95(8):083525. \doi{10.1103/PhysRevD.95.083525}.
  {\href{https://arxiv.org/abs/1609.07142}{{arXiv:1609.07142}}} {[astro-ph.CO]}

\bibitem[{Lackey et~al.(2017)Lackey, Bernuzzi, Galley, Meidam, and Van
  Den~Broeck}]{Lackey:2016krb}
Lackey BD, Bernuzzi S, Galley CR, Meidam J, Van Den~Broeck C (2017)
  {Effective-one-body waveforms for binary neutron stars using surrogate
  models}. Phys Rev D 95(10):104036. \doi{10.1103/PhysRevD.95.104036}.
  {\href{https://arxiv.org/abs/1610.04742}{{arXiv:1610.04742}}} {[gr-qc]}

\bibitem[{Laddha and Sen(2019)}]{BMSAdd2}
Laddha A, Sen A (2019) {Observational Signature of the Logarithmic Terms in the
  Soft Graviton Theorem}. Phys Rev D 100(2):024009.
  \doi{10.1103/PhysRevD.100.024009}.
  {\href{https://arxiv.org/abs/1806.01872}{{arXiv:1806.01872}}} {[hep-th]}

\bibitem[{Laghi et~al.(2021)Laghi, Tamanini, Del~Pozzo, Sesana, Gair, and
  Babak}]{Laghi:2021pqk}
Laghi D, Tamanini N, Del~Pozzo W, Sesana A, Gair J, Babak S (2021)
  {Gravitational wave cosmology with extreme mass-ratio inspirals}
  {\href{https://arxiv.org/abs/2102.01708}{{arXiv:2102.01708}}} {[astro-ph.CO]}

\bibitem[{Lagos and Zhu(2020)}]{Lagos:2020mzy}
Lagos M, Zhu H (2020) {Gravitational couplings in Chameleon models}. JCAP
  06:061. \doi{10.1088/1475-7516/2020/06/061}.
  {\href{https://arxiv.org/abs/2003.01038}{{arXiv:2003.01038}}} {[gr-qc]}

\bibitem[{Lagos et~al.(2019)Lagos, Fishbach, Landry, and Holz}]{Lagos:2019kds}
Lagos M, Fishbach M, Landry P, Holz DE (2019) {Standard sirens with a running
  Planck mass}. Phys Rev D 99(8):083504. \doi{10.1103/PhysRevD.99.083504}.
  {\href{https://arxiv.org/abs/1901.03321}{{arXiv:1901.03321}}} {[astro-ph.CO]}

\bibitem[{Landry and Poisson(2015)}]{Landry:2015zfa}
Landry P, Poisson E (2015) {Tidal deformation of a slowly rotating material
  body. External metric}. Phys Rev D 91:104018.
  \doi{10.1103/PhysRevD.91.104018}.
  {\href{https://arxiv.org/abs/1503.07366}{{arXiv:1503.07366}}} {[gr-qc]}

\bibitem[{Langlois and Noui(2016)}]{Langlois:2015cwa}
Langlois D, Noui K (2016) {Degenerate higher derivative theories beyond
  Horndeski: evading the Ostrogradski instability}. JCAP 1602:034.
  \doi{10.1088/1475-7516/2016/02/034}.
  {\href{https://arxiv.org/abs/1510.06930}{{arXiv:1510.06930}}} {[gr-qc]}

\bibitem[{Langlois et~al.(2018)Langlois, Saito, Yamauchi, and
  Noui}]{Langlois:2017dyl}
Langlois D, Saito R, Yamauchi D, Noui K (2018) {Scalar-tensor theories and
  modified gravity in the wake of GW170817}. Phys Rev D97(6):061501.
  \doi{10.1103/PhysRevD.97.061501}.
  {\href{https://arxiv.org/abs/1711.07403}{{arXiv:1711.07403}}} {[gr-qc]}

\bibitem[{Le~Tiec and Casals(2020)}]{LeTiec:2020spy}
Le~Tiec A, Casals M (2020) {Spinning Black Holes Fall in Love}
  {\href{https://arxiv.org/abs/2007.00214}{{arXiv:2007.00214}}} {[gr-qc]}

\bibitem[{Le~Tiec et~al.(2020)Le~Tiec, Casals, and Franzin}]{LeTiec:2020bos}
Le~Tiec A, Casals M, Franzin E (2020) {Tidal Love Numbers of Kerr Black Holes}
  {\href{https://arxiv.org/abs/2010.15795}{{arXiv:2010.15795}}} {[gr-qc]}

\bibitem[{Leach and Liddle(2001)}]{Leach:2000yw}
Leach SM, Liddle AR (2001) {Inflationary perturbations near horizon crossing}.
  Phys Rev D 63:043508. \doi{10.1103/PhysRevD.63.043508}.
  {\href{https://arxiv.org/abs/astro-ph/0010082}{{arXiv:astro-ph/0010082}}}

\bibitem[{Leach et~al.(2001)Leach, Sasaki, Wands, and Liddle}]{Leach:2001zf}
Leach SM, Sasaki M, Wands D, Liddle AR (2001) {Enhancement of superhorizon
  scale inflationary curvature perturbations}. Phys Rev D 64:023512.
  \doi{10.1103/PhysRevD.64.023512}.
  {\href{https://arxiv.org/abs/astro-ph/0101406}{{arXiv:astro-ph/0101406}}}

\bibitem[{Lemos and Weinberg(2004)}]{Lemos:2003gx}
Lemos JPS, Weinberg EJ (2004) {Quasiblack holes from extremal charged dust}.
  Phys Rev D 69:104004. \doi{10.1103/PhysRevD.69.104004}.
  {\href{https://arxiv.org/abs/gr-qc/0311051}{{arXiv:gr-qc/0311051}}}

\bibitem[{Lemos and Zaslavskii(2008)}]{Lemos:2008cv}
Lemos JPS, Zaslavskii OB (2008) {Black hole mimickers: Regular versus singular
  behavior}. Phys Rev D78:024040. \doi{10.1103/PhysRevD.78.024040}.
  {\href{https://arxiv.org/abs/0806.0845}{{arXiv:0806.0845}}} {[gr-qc]}

\bibitem[{Lemos et~al.(2003)Lemos, Lobo, and Quinet~de Oliveira}]{Lemos:2003jb}
Lemos JPS, Lobo FSN, Quinet~de Oliveira S (2003) {Morris-Thorne wormholes with
  a cosmological constant}. Phys Rev D68:064004.
  \doi{10.1103/PhysRevD.68.064004}.
  {\href{https://arxiv.org/abs/gr-qc/0302049}{{arXiv:gr-qc/0302049}}} {[gr-qc]}

\bibitem[{Lentati et~al.(2015)}]{Lentati:2015qwp}
Lentati L, et~al. (2015) {European Pulsar Timing Array Limits On An Isotropic
  Stochastic Gravitational-Wave Background}. Mon Not Roy Astron Soc
  453(3):2576--2598. \doi{10.1093/mnras/stv1538}.
  {\href{https://arxiv.org/abs/1504.03692}{{arXiv:1504.03692}}} {[astro-ph.CO]}

\bibitem[{Letelier(1980)}]{Letelier:1980mxb}
Letelier PS (1980) {Anisotropic fluids with two-perfect-fluid components}. Phys
  Rev D 22(4):807. \doi{10.1103/PhysRevD.22.807}

\bibitem[{Levin(2006)}]{Levin:2006zv}
Levin J (2006) {Chaos and Order in Models of Black Hole Pairs}. Phys Rev D
  74:124027. \doi{10.1103/PhysRevD.74.124027}.
  {\href{https://arxiv.org/abs/gr-qc/0612003}{{arXiv:gr-qc/0612003}}}

\bibitem[{Levin(2007)}]{Levin:2006uc}
Levin Y (2007) {Starbursts near supermassive black holes: young stars in the
  Galactic Center, and gravitational waves in LISA band}. Mon Not Roy Astron
  Soc 374:515--524. \doi{10.1111/j.1365-2966.2006.11155.x}.
  {\href{https://arxiv.org/abs/astro-ph/0603583}{{arXiv:astro-ph/0603583}}}

\bibitem[{Li et~al.(2011{\natexlab{a}})Li, Sotiriou, and Barrow}]{Li:2010cg}
Li B, Sotiriou TP, Barrow JD (2011{\natexlab{a}}) {$f(T)$ gravity and local
  Lorentz invariance}. Phys Rev D 83:064035. \doi{10.1103/PhysRevD.83.064035}.
  {\href{https://arxiv.org/abs/1010.1041}{{arXiv:1010.1041}}} {[gr-qc]}

\bibitem[{Li et~al.(2011{\natexlab{b}})Li, Miao, and Miao}]{Li:2011rn}
Li M, Miao RX, Miao YG (2011{\natexlab{b}}) {Degrees of freedom of $f(T)$
  gravity}. JHEP 07:108. \doi{10.1007/JHEP07(2011)108}.
  {\href{https://arxiv.org/abs/1105.5934}{{arXiv:1105.5934}}} {[hep-th]}

\bibitem[{Lidov(1962)}]{LIDOV1962719}
Lidov ML (1962) The evolution of orbits of artificial satellites of planets
  under the action of gravitational perturbations of external bodies. Planetary
  and Space Science 9(10):719 -- 759. \doi{10.1016/0032-0633(62)90129-0}

\bibitem[{Liebling and Palenzuela(2017)}]{Liebling:2012fv}
Liebling SL, Palenzuela C (2017) {Dynamical Boson Stars}. Living Rev Rel
  20(1):5. \doi{10.12942/lrr-2012-6}.
  {\href{https://arxiv.org/abs/1202.5809}{{arXiv:1202.5809}}} {[gr-qc]}

\bibitem[{Lin and Mukohyama(2017)}]{Lin:2017oow}
Lin C, Mukohyama S (2017) {A Class of Minimally Modified Gravity Theories}.
  JCAP 10:033. \doi{10.1088/1475-7516/2017/10/033}.
  {\href{https://arxiv.org/abs/1708.03757}{{arXiv:1708.03757}}} {[gr-qc]}

\bibitem[{LISA-Waveform-Working-Group(2022)}]{LISAWavWGWP}
LISA-Waveform-Working-Group (2022) {The LISA Waveform Working Group White
  Paper.} In prep.

\bibitem[{Liu et~al.(2020)Liu, Cao, and Shao}]{Liu:2019jpg}
Liu X, Cao Z, Shao L (2020) {Validating the Effective-One-Body
  Numerical-Relativity Waveform Models for Spin-aligned Binary Black Holes
  along Eccentric Orbits}. Phys Rev D 101(4):044049.
  \doi{10.1103/PhysRevD.101.044049}.
  {\href{https://arxiv.org/abs/1910.00784}{{arXiv:1910.00784}}} {[gr-qc]}

\bibitem[{Lo et~al.(2019)Lo, Li, and Weinstein}]{Lo:2018sep}
Lo RKL, Li TGF, Weinstein AJ (2019) {Template-based Gravitational-Wave Echoes
  Search Using Bayesian Model Selection}. Phys Rev D 99(8):084052.
  \doi{10.1103/PhysRevD.99.084052}.
  {\href{https://arxiv.org/abs/1811.07431}{{arXiv:1811.07431}}} {[gr-qc]}

\bibitem[{Lombriser(2018)}]{Lombriser:2018guo}
Lombriser L (2018) {Parametrizations for tests of gravity}. Int J Mod Phys D
  27(15):1848002. \doi{10.1142/S0218271818480024}.
  {\href{https://arxiv.org/abs/1908.07892}{{arXiv:1908.07892}}} {[astro-ph.CO]}

\bibitem[{Lombriser and Lima(2017)}]{Lombriser:2016yzn}
Lombriser L, Lima NA (2017) {Challenges to Self-Acceleration in Modified
  Gravity from Gravitational Waves and Large-Scale Structure}. Phys Lett B
  765:382--385. \doi{10.1016/j.physletb.2016.12.048}.
  {\href{https://arxiv.org/abs/1602.07670}{{arXiv:1602.07670}}} {[astro-ph.CO]}

\bibitem[{Lombriser and Taylor(2016)}]{Lombriser:2015sxa}
Lombriser L, Taylor A (2016) {Breaking a Dark Degeneracy with Gravitational
  Waves}. JCAP 1603:031. \doi{10.1088/1475-7516/2016/03/031}.
  {\href{https://arxiv.org/abs/1509.08458}{{arXiv:1509.08458}}} {[astro-ph.CO]}

\bibitem[{London et~al.(2018)London, Khan, Fauchon-Jones, Garc\'\i{}a, Hannam,
  Husa, Jim\'enez-Forteza, Kalaghatgi, Ohme, and Pannarale}]{London:2017bcn}
London L, Khan S, Fauchon-Jones E, Garc\'\i{}a C, Hannam M, Husa S,
  Jim\'enez-Forteza X, Kalaghatgi C, Ohme F, Pannarale F (2018) {First
  higher-multipole model of gravitational waves from spinning and coalescing
  black-hole binaries}. Phys Rev Lett 120(16):161102.
  \doi{10.1103/PhysRevLett.120.161102}.
  {\href{https://arxiv.org/abs/1708.00404}{{arXiv:1708.00404}}} {[gr-qc]}

\bibitem[{Lorenz et~al.(2010)Lorenz, Ringeval, and
  Sakellariadou}]{Lorenz:2010sm}
Lorenz L, Ringeval C, Sakellariadou M (2010) {Cosmic string loop distribution
  on all length scales and at any redshift}. JCAP 10:003.
  \doi{10.1088/1475-7516/2010/10/003}.
  {\href{https://arxiv.org/abs/1006.0931}{{arXiv:1006.0931}}} {[astro-ph.CO]}

\bibitem[{Lovelock(1972)}]{Lovelock:1972vz}
Lovelock D (1972) {The four-dimensionality of space and the einstein tensor}. J
  Math Phys 13:874--876. \doi{10.1063/1.1666069}

\bibitem[{Lu et~al.(2020)Lu, Takhistov, Gelmini, Hayashi, Inoue, and
  Kusenko}]{Lu:2020bmd}
Lu P, Takhistov V, Gelmini GB, Hayashi K, Inoue Y, Kusenko A (2020)
  {Constraining Primordial Black Holes with Dwarf Galaxy Heating}
  {\href{https://arxiv.org/abs/2007.02213}{{arXiv:2007.02213}}} {[astro-ph.CO]}

\bibitem[{Luckock and Moss(1986)}]{Luckock:1986tr}
Luckock H, Moss I (1986) {BLACK HOLES HAVE SKYRMION HAIR}. Phys Lett B
  176:341--345. \doi{10.1016/0370-2693(86)90175-9}

\bibitem[{{Ludvigsen}(1989)}]{BMSAdd4}
{Ludvigsen} M (1989) {Geodesic deviation at null infinity and the physical
  effects of very long wave gravitational radiation}. General Relativity and
  Gravitation 21(12):1205--1212. \doi{10.1007/BF00763308}

\bibitem[{Lukes-Gerakopoulos et~al.(2010)Lukes-Gerakopoulos, Apostolatos, and
  Contopoulos}]{LukesGerakopoulos:2010rc}
Lukes-Gerakopoulos G, Apostolatos TA, Contopoulos G (2010) {Observable
  signature of a background deviating from the Kerr metric}. Phys Rev D
  81:124005. \doi{10.1103/PhysRevD.81.124005}.
  {\href{https://arxiv.org/abs/1003.3120}{{arXiv:1003.3120}}} {[gr-qc]}

\bibitem[{Lunin and Mathur(2002{\natexlab{a}})}]{Lunin:2001jy}
Lunin O, Mathur SD (2002{\natexlab{a}}) {AdS / CFT duality and the black hole
  information paradox}. Nucl Phys B 623:342--394.
  \doi{10.1016/S0550-3213(01)00620-4}.
  {\href{https://arxiv.org/abs/hep-th/0109154}{{arXiv:hep-th/0109154}}}

\bibitem[{Lunin and Mathur(2002{\natexlab{b}})}]{Lunin:2002qf}
Lunin O, Mathur SD (2002{\natexlab{b}}) {Statistical interpretation of
  Bekenstein entropy for systems with a stretched horizon}. Phys Rev Lett
  88:211303. \doi{10.1103/PhysRevLett.88.211303}.
  {\href{https://arxiv.org/abs/hep-th/0202072}{{arXiv:hep-th/0202072}}}

\bibitem[{Lyutikov(2016)}]{Lyutikov:2016mgv}
Lyutikov M (2016) {Fermi GBM signal contemporaneous with GW150914 - an unlikely
  association} {\href{https://arxiv.org/abs/1602.07352}{{arXiv:1602.07352}}}
  {[astro-ph.HE]}

\bibitem[{Macedo et~al.(2013{\natexlab{a}})Macedo, Pani, Cardoso, and
  Crispino}]{Macedo:2013qea}
Macedo CF, Pani P, Cardoso V, Crispino LC (2013{\natexlab{a}}) {Into the lair:
  gravitational-wave signatures of dark matter}. Astrophys J 774:48.
  \doi{10.1088/0004-637X/774/1/48}.
  {\href{https://arxiv.org/abs/1302.2646}{{arXiv:1302.2646}}} {[gr-qc]}

\bibitem[{Macedo et~al.(2013{\natexlab{b}})Macedo, Pani, Cardoso, and
  Crispino}]{Macedo:2013jja}
Macedo CF, Pani P, Cardoso V, Crispino LCB (2013{\natexlab{b}}) {Astrophysical
  signatures of boson stars: quasinormal modes and inspiral resonances}. Phys
  Rev D 88(6):064046. \doi{10.1103/PhysRevD.88.064046}.
  {\href{https://arxiv.org/abs/1307.4812}{{arXiv:1307.4812}}} {[gr-qc]}

\bibitem[{Macedo et~al.(2019)Macedo, Sakstein, Berti, Gualtieri, Silva, and
  Sotiriou}]{Macedo:2019sem}
Macedo CF, Sakstein J, Berti E, Gualtieri L, Silva HO, Sotiriou TP (2019)
  {Self-interactions and Spontaneous Black Hole Scalarization}. Phys Rev D
  99(10):104041. \doi{10.1103/PhysRevD.99.104041}.
  {\href{https://arxiv.org/abs/1903.06784}{{arXiv:1903.06784}}} {[gr-qc]}

\bibitem[{MacLeod and Hogan(2008)}]{MacLeod:2007jd}
MacLeod CL, Hogan CJ (2008) {Precision of Hubble constant derived using black
  hole binary absolute distances and statistical redshift information}. Phys
  Rev D 77:043512. \doi{10.1103/PhysRevD.77.043512}.
  {\href{https://arxiv.org/abs/0712.0618}{{arXiv:0712.0618}}} {[astro-ph]}

\bibitem[{Maggio et~al.(2017)Maggio, Pani, and Ferrari}]{Maggio:2017ivp}
Maggio E, Pani P, Ferrari V (2017) {Exotic Compact Objects and How to Quench
  their Ergoregion Instability}. Phys Rev D 96(10):104047.
  \doi{10.1103/PhysRevD.96.104047}.
  {\href{https://arxiv.org/abs/1703.03696}{{arXiv:1703.03696}}} {[gr-qc]}

\bibitem[{Maggio et~al.(2019{\natexlab{a}})Maggio, Cardoso, Dolan, and
  Pani}]{Maggio:2018ivz}
Maggio E, Cardoso V, Dolan SR, Pani P (2019{\natexlab{a}}) {Ergoregion
  instability of exotic compact objects: electromagnetic and gravitational
  perturbations and the role of absorption}. Phys Rev D 99(6):064007.
  \doi{10.1103/PhysRevD.99.064007}.
  {\href{https://arxiv.org/abs/1807.08840}{{arXiv:1807.08840}}} {[gr-qc]}

\bibitem[{Maggio et~al.(2019{\natexlab{b}})Maggio, Testa, Bhagwat, and
  Pani}]{Maggio:2019zyv}
Maggio E, Testa A, Bhagwat S, Pani P (2019{\natexlab{b}}) {Analytical model for
  gravitational-wave echoes from spinning remnants}. Phys Rev D 100(6):064056.
  \doi{10.1103/PhysRevD.100.064056}.
  {\href{https://arxiv.org/abs/1907.03091}{{arXiv:1907.03091}}} {[gr-qc]}

\bibitem[{Maggio et~al.(2020)Maggio, Buoninfante, Mazumdar, and
  Pani}]{Maggio:2020jml}
Maggio E, Buoninfante L, Mazumdar A, Pani P (2020) {How does a dark compact
  object ringdown?} Phys Rev D 102(6):064053.
  \doi{10.1103/PhysRevD.102.064053}.
  {\href{https://arxiv.org/abs/2006.14628}{{arXiv:2006.14628}}} {[gr-qc]}

\bibitem[{Maggio et~al.(2021)Maggio, van~de Meent, and Pani}]{Maggio:2021uge}
Maggio E, van~de Meent M, Pani P (2021) {Extreme mass-ratio inspirals around a
  spinning horizonless compact object}
  {\href{https://arxiv.org/abs/2106.07195}{{arXiv:2106.07195}}} {[gr-qc]}

\bibitem[{Maggiore(2007)}]{Maggiore:1900zz}
Maggiore M (2007) {Gravitational Waves. Vol. 1: Theory and Experiments}. Oxford
  Master Series in Physics, Oxford University Press.
  \doi{10.1093/acprof:oso/9780198570745.001.0001}

\bibitem[{Maggiore(2014)}]{Maggiore:2013mea}
Maggiore M (2014) {Phantom dark energy from nonlocal infrared modifications of
  general relativity}. PhysRev D89:043008.
  {\href{https://arxiv.org/abs/1307.3898}{{arXiv:1307.3898}}} {[hep-th]}

\bibitem[{Maldacena et~al.(1997)Maldacena, Strominger, and
  Witten}]{Maldacena:1997de}
Maldacena JM, Strominger A, Witten E (1997) {Black hole entropy in M theory}.
  JHEP 12:002. \doi{10.1088/1126-6708/1997/12/002}.
  {\href{https://arxiv.org/abs/hep-th/9711053}{{arXiv:hep-th/9711053}}}

\bibitem[{Maluf(2013)}]{Maluf:2013gaa}
Maluf J (2013) {The teleparallel equivalent of general relativity}. Annalen
  Phys 525:339--357. \doi{10.1002/andp.201200272}.
  {\href{https://arxiv.org/abs/1303.3897}{{arXiv:1303.3897}}} {[gr-qc]}

\bibitem[{Mangiagli et~al.(2019)Mangiagli, Klein, Sesana, Barausse, and
  Colpi}]{Mangiagli:2018kpu}
Mangiagli A, Klein A, Sesana A, Barausse E, Colpi M (2019) {Post-Newtonian
  phase accuracy requirements for stellar black hole binaries with LISA}. Phys
  Rev D 99(6):064056. \doi{10.1103/PhysRevD.99.064056}.
  {\href{https://arxiv.org/abs/1811.01805}{{arXiv:1811.01805}}} {[gr-qc]}

\bibitem[{Manshanden et~al.(2019)Manshanden, Gaggero, Bertone, Connors, and
  Ricotti}]{Manshanden:2018tze}
Manshanden J, Gaggero D, Bertone G, Connors RM, Ricotti M (2019)
  {Multi-wavelength astronomical searches for primordial black holes}. JCAP
  06:026. \doi{10.1088/1475-7516/2019/06/026}.
  {\href{https://arxiv.org/abs/1812.07967}{{arXiv:1812.07967}}} {[astro-ph.HE]}

\bibitem[{Marassi et~al.(2011)Marassi, Ciolfi, Schneider, Stella, and
  Ferrari}]{Marassi:2010wj}
Marassi S, Ciolfi R, Schneider R, Stella L, Ferrari V (2011) {Stochastic
  background of gravitational waves emitted by magnetars}. Mon Not Roy Astron
  Soc 411:2549. \doi{10.1111/j.1365-2966.2010.17861.x}.
  {\href{https://arxiv.org/abs/1009.1240}{{arXiv:1009.1240}}} {[astro-ph.CO]}

\bibitem[{Margalit et~al.(2020)Margalit, Contaldi, and
  Pieroni}]{Margalit:2020sxp}
Margalit A, Contaldi CR, Pieroni M (2020) {Phase Decoherence of Gravitational
  Wave Backgrounds}
  {\href{https://arxiv.org/abs/2004.01727}{{arXiv:2004.01727}}} {[astro-ph.CO]}

\bibitem[{Mark et~al.(2017)Mark, Zimmerman, Du, and Chen}]{Mark:2017dnq}
Mark Z, Zimmerman A, Du SM, Chen Y (2017) {A recipe for echoes from exotic
  compact objects}. Phys Rev D96(8):084002. \doi{10.1103/PhysRevD.96.084002}.
  {\href{https://arxiv.org/abs/1706.06155}{{arXiv:1706.06155}}} {[gr-qc]}

\bibitem[{Marsh(2016)}]{Marsh:2015xka}
Marsh DJE (2016) {Axion Cosmology}. Phys Rept 643:1--79.
  \doi{10.1016/j.physrep.2016.06.005}.
  {\href{https://arxiv.org/abs/1510.07633}{{arXiv:1510.07633}}} {[astro-ph.CO]}

\bibitem[{Maselli et~al.(2012)Maselli, Gualtieri, Pannarale, and
  Ferrari}]{Maselli:2012zq}
Maselli A, Gualtieri L, Pannarale F, Ferrari V (2012) {On the validity of the
  adiabatic approximation in compact binary inspirals}. Phys Rev D 86:044032.
  \doi{10.1103/PhysRevD.86.044032}.
  {\href{https://arxiv.org/abs/1205.7006}{{arXiv:1205.7006}}} {[gr-qc]}

\bibitem[{Maselli et~al.(2015)Maselli, Pani, Gualtieri, and
  Ferrari}]{Maselli:2015tta}
Maselli A, Pani P, Gualtieri L, Ferrari V (2015) {Rotating black holes in
  Einstein-Dilaton-Gauss-Bonnet gravity with finite coupling}. Phys Rev D
  92(8):083014. \doi{10.1103/PhysRevD.92.083014}.
  {\href{https://arxiv.org/abs/1507.00680}{{arXiv:1507.00680}}} {[gr-qc]}

\bibitem[{Maselli et~al.(2016)Maselli, Marassi, Ferrari, Kokkotas, and
  Schneider}]{Maselli:2016ekw}
Maselli A, Marassi S, Ferrari V, Kokkotas K, Schneider R (2016) {Constraining
  Modified Theories of Gravity with Gravitational-Wave Stochastic Backgrounds}.
  Phys Rev Lett 117(9):091102. \doi{10.1103/PhysRevLett.117.091102}.
  {\href{https://arxiv.org/abs/1606.04996}{{arXiv:1606.04996}}} {[gr-qc]}

\bibitem[{Maselli et~al.(2017)Maselli, V\"olkel, and
  Kokkotas}]{Maselli:2017tfq}
Maselli A, V\"olkel SH, Kokkotas KD (2017) {Parameter estimation of
  gravitational wave echoes from exotic compact objects}. Phys Rev D
  96(6):064045. \doi{10.1103/PhysRevD.96.064045}.
  {\href{https://arxiv.org/abs/1708.02217}{{arXiv:1708.02217}}} {[gr-qc]}

\bibitem[{Maselli et~al.(2018)Maselli, Pani, Cardoso, Abdelsalhin, Gualtieri,
  and Ferrari}]{Maselli:2017cmm}
Maselli A, Pani P, Cardoso V, Abdelsalhin T, Gualtieri L, Ferrari V (2018)
  {Probing Planckian corrections at the horizon scale with LISA binaries}. Phys
  Rev Lett 120(8):081101. \doi{10.1103/PhysRevLett.120.081101}.
  {\href{https://arxiv.org/abs/1703.10612}{{arXiv:1703.10612}}} {[gr-qc]}

\bibitem[{Maselli et~al.(2019)Maselli, Pani, Cardoso, Abdelsalhin, Gualtieri,
  and Ferrari}]{Maselli:2018fay}
Maselli A, Pani P, Cardoso V, Abdelsalhin T, Gualtieri L, Ferrari V (2019)
  {From micro to macro and back: probing near-horizon quantum structures with
  gravitational waves}. Class Quant Grav 36(16):167001.
  \doi{10.1088/1361-6382/ab30ff}.
  {\href{https://arxiv.org/abs/1811.03689}{{arXiv:1811.03689}}} {[gr-qc]}

\bibitem[{Maselli et~al.(2020{\natexlab{a}})Maselli, Franchini, Gualtieri, and
  Sotiriou}]{Maselli:2020zgv}
Maselli A, Franchini N, Gualtieri L, Sotiriou TP (2020{\natexlab{a}})
  {Detecting scalar fields with Extreme Mass Ratio Inspirals}. Phys Rev Lett
  125(14):141101. \doi{10.1103/PhysRevLett.125.141101}.
  {\href{https://arxiv.org/abs/2004.11895}{{arXiv:2004.11895}}} {[gr-qc]}

\bibitem[{Maselli et~al.(2020{\natexlab{b}})Maselli, Pani, Gualtieri, and
  Berti}]{Maselli:2019mjd}
Maselli A, Pani P, Gualtieri L, Berti E (2020{\natexlab{b}}) {Parametrized
  ringdown spin expansion coefficients: a data-analysis framework for
  black-hole spectroscopy with multiple events}. Phys Rev D 101(2):024043.
  \doi{10.1103/PhysRevD.101.024043}.
  {\href{https://arxiv.org/abs/1910.12893}{{arXiv:1910.12893}}} {[gr-qc]}

\bibitem[{Maselli et~al.(2021)Maselli, Franchini, Gualtieri, Sotiriou,
  Barsanti, and Pani}]{Maselli:2021men}
Maselli A, Franchini N, Gualtieri L, Sotiriou TP, Barsanti S, Pani P (2021)
  {Detecting new fundamental fields with LISA}
  {\href{https://arxiv.org/abs/2106.11325}{{arXiv:2106.11325}}} {[gr-qc]}

\bibitem[{Matas et~al.(2020)}]{Matas:2020wab}
Matas A, et~al. (2020) {Aligned-spin neutron-star\textendash{}black-hole
  waveform model based on the effective-one-body approach and
  numerical-relativity simulations}. Phys Rev D 102(4):043023.
  \doi{10.1103/PhysRevD.102.043023}.
  {\href{https://arxiv.org/abs/2004.10001}{{arXiv:2004.10001}}} {[gr-qc]}

\bibitem[{Mathur(2005)}]{Mathur:2005zp}
Mathur SD (2005) {The Fuzzball proposal for black holes: An Elementary review}.
  Fortsch Phys 53:793--827. \doi{10.1002/prop.200410203}.
  {\href{https://arxiv.org/abs/hep-th/0502050}{{arXiv:hep-th/0502050}}}

\bibitem[{Mathur(2008)}]{Mathur:2008nj}
Mathur SD (2008) {Fuzzballs and the information paradox: A Summary and
  conjectures} {\href{https://arxiv.org/abs/0810.4525}{{arXiv:0810.4525}}}
  {[hep-th]}

\bibitem[{Mathur(2009)}]{Mathur:2009hf}
Mathur SD (2009) {The Information paradox: A Pedagogical introduction}. Class
  Quant Grav 26:224001. \doi{10.1088/0264-9381/26/22/224001}.
  {\href{https://arxiv.org/abs/0909.1038}{{arXiv:0909.1038}}} {[hep-th]}

\bibitem[{Max et~al.(2017)Max, Platscher, and Smirnov}]{Max:2017flc}
Max K, Platscher M, Smirnov J (2017) {Gravitational Wave Oscillations in
  Bigravity}. Phys Rev Lett 119(11):111101.
  \doi{10.1103/PhysRevLett.119.111101}.
  {\href{https://arxiv.org/abs/1703.07785}{{arXiv:1703.07785}}} {[gr-qc]}

\bibitem[{Mayerson(2020)}]{Mayerson:2020tpn}
Mayerson DR (2020) {Fuzzballs and Observations}. Gen Rel Grav 52(12):115.
  \doi{10.1007/s10714-020-02769-w}.
  {\href{https://arxiv.org/abs/2010.09736}{{arXiv:2010.09736}}} {[hep-th]}

\bibitem[{Mazur(1982)}]{Mazur:1982db}
Mazur PO (1982) {Proof of uniqueness of the Kerr-Newman black hole solution}. J
  Phys A 15:3173--3180. \doi{10.1088/0305-4470/15/10/021}

\bibitem[{Mazur and Mottola(2004)}]{Mazur:2004fk}
Mazur PO, Mottola E (2004) {Gravitational vacuum condensate stars}. Proc Nat
  Acad Sci 101:9545--9550. \doi{10.1073/pnas.0402717101}.
  {\href{https://arxiv.org/abs/gr-qc/0407075}{{arXiv:gr-qc/0407075}}} {[gr-qc]}

\bibitem[{Mazur and Mottola(2015)}]{Mazur:2015kia}
Mazur PO, Mottola E (2015) {Surface tension and negative pressure interior of a
  non-singular \textquoteleft{}black hole\textquoteright{}}. Class Quant Grav
  32(21):215024. \doi{10.1088/0264-9381/32/21/215024}.
  {\href{https://arxiv.org/abs/1501.03806}{{arXiv:1501.03806}}} {[gr-qc]}

\bibitem[{McGee et~al.(2020)McGee, Sesana, and Vecchio}]{McGee:2018qwb}
McGee S, Sesana A, Vecchio A (2020) {Linking gravitational waves and X-ray
  phenomena with joint LISA and Athena observations}. Nature Astron
  4(1):26--31. \doi{10.1038/s41550-019-0969-7}.
  {\href{https://arxiv.org/abs/1811.00050}{{arXiv:1811.00050}}} {[astro-ph.HE]}

\bibitem[{McManus et~al.(2016)McManus, Lombriser, and
  Pe\~narrubia}]{McManus:2016kxu}
McManus R, Lombriser L, Pe\~narrubia J (2016) {Finding Horndeski theories with
  Einstein gravity limits}. JCAP 11:006. \doi{10.1088/1475-7516/2016/11/006}.
  {\href{https://arxiv.org/abs/1606.03282}{{arXiv:1606.03282}}} {[gr-qc]}

\bibitem[{McManus et~al.(2017)McManus, Lombriser, and
  Pe\~narrubia}]{McManus:2017itv}
McManus R, Lombriser L, Pe\~narrubia J (2017) {Parameterised Post-Newtonian
  Expansion in Screened Regions}. JCAP 12:031.
  \doi{10.1088/1475-7516/2017/12/031}.
  {\href{https://arxiv.org/abs/1705.05324}{{arXiv:1705.05324}}} {[gr-qc]}

\bibitem[{McManus et~al.(2019)McManus, Berti, Macedo, Kimura, Maselli, and
  Cardoso}]{McManus:2019ulj}
McManus R, Berti E, Macedo CF, Kimura M, Maselli A, Cardoso V (2019)
  {Parametrized black hole quasinormal ringdown. II. Coupled equations and
  quadratic corrections for nonrotating black holes}. Phys Rev D 100(4):044061.
  \doi{10.1103/PhysRevD.100.044061}.
  {\href{https://arxiv.org/abs/1906.05155}{{arXiv:1906.05155}}} {[gr-qc]}

\bibitem[{Meacher et~al.(2015)Meacher, Coughlin, Morris, Regimbau, Christensen,
  Kandhasamy, Mandic, Romano, and Thrane}]{Meacher:2015iua}
Meacher D, Coughlin M, Morris S, Regimbau T, Christensen N, Kandhasamy S,
  Mandic V, Romano JD, Thrane E (2015) {Mock data and science challenge for
  detecting an astrophysical stochastic gravitational-wave background with
  Advanced LIGO and Advanced Virgo}. Phys Rev D 92(6):063002.
  \doi{10.1103/PhysRevD.92.063002}.
  {\href{https://arxiv.org/abs/1506.06744}{{arXiv:1506.06744}}} {[astro-ph.HE]}

\bibitem[{van~de Meent(2018)}]{vandeMeent:2017bcc}
van~de Meent M (2018) {Gravitational self-force on generic bound geodesics in
  Kerr spacetime}. Phys Rev D 97(10):104033. \doi{10.1103/PhysRevD.97.104033}.
  {\href{https://arxiv.org/abs/1711.09607}{{arXiv:1711.09607}}} {[gr-qc]}

\bibitem[{van~de Meent and Pfeiffer(2020)}]{vandeMeent:2020xgc}
van~de Meent M, Pfeiffer HP (2020) {Intermediate mass-ratio black hole
  binaries: Applicability of small mass-ratio perturbation theory}
  {\href{https://arxiv.org/abs/2006.12036}{{arXiv:2006.12036}}} {[gr-qc]}

\bibitem[{Meidam et~al.(2014)Meidam, Agathos, Van Den~Broeck, Veitch, and
  Sathyaprakash}]{Meidam:2014jpa}
Meidam J, Agathos M, Van Den~Broeck C, Veitch J, Sathyaprakash B (2014)
  {Testing the no-hair theorem with black hole ringdowns using TIGER}. Phys Rev
  D 90(6):064009. \doi{10.1103/PhysRevD.90.064009}.
  {\href{https://arxiv.org/abs/1406.3201}{{arXiv:1406.3201}}} {[gr-qc]}

\bibitem[{Menou and Goodman(2004)}]{Menou:2003pu}
Menou K, Goodman J (2004) {Low - mass proto-planet migration in T-Tauri
  alpha-disks}. Astrophys J 606:520--531. \doi{10.1086/382947}.
  {\href{https://arxiv.org/abs/astro-ph/0310169}{{arXiv:astro-ph/0310169}}}

\bibitem[{Merritt et~al.(2002)Merritt, Milosavljevic, Verde, and
  Jimenez}]{Merritt:2002vj}
Merritt D, Milosavljevic M, Verde L, Jimenez R (2002) {Dark matter spikes and
  annihilation radiation from the galactic center}. Phys Rev Lett 88:191301.
  \doi{10.1103/PhysRevLett.88.191301}.
  {\href{https://arxiv.org/abs/astro-ph/0201376}{{arXiv:astro-ph/0201376}}}

\bibitem[{Metsaev and Tseytlin(1987)}]{Metsaev:1986yb}
Metsaev R, Tseytlin AA (1987) {Curvature Cubed Terms in String Theory Effective
  Actions}. Phys Lett B 185:52--58. \doi{10.1016/0370-2693(87)91527-9}

\bibitem[{Miller and Pound(2021)}]{Miller:2020bft}
Miller J, Pound A (2021) {Two-timescale evolution of extreme-mass-ratio
  inspirals: waveform generation scheme for quasicircular orbits in
  Schwarzschild spacetime}. Phys Rev D 103(6):064048.
  \doi{10.1103/PhysRevD.103.064048}.
  {\href{https://arxiv.org/abs/2006.11263}{{arXiv:2006.11263}}} {[gr-qc]}

\bibitem[{Miller and Hamilton(2002)}]{Miller:2002pg}
Miller MC, Hamilton DP (2002) {Four-body effects in globular cluster black hole
  coalescence}. Astrophys J 576:894. \doi{10.1086/341788}.
  {\href{https://arxiv.org/abs/astro-ph/0202298}{{arXiv:astro-ph/0202298}}}

\bibitem[{Milosavljevic and Loeb(2004)}]{Milosavljevic:2004te}
Milosavljevic M, Loeb A (2004) {The Link between warm molecular disks in maser
  nuclei and star formation near the black hole at the Galactic Center}.
  Astrophys J Lett 604:L45. \doi{10.1086/383467}.
  {\href{https://arxiv.org/abs/astro-ph/0401221}{{arXiv:astro-ph/0401221}}}

\bibitem[{{Milosavljevi{\'c}} and {Phinney}(2005)}]{2005ApJ...622L..93M}
{Milosavljevi{\'c}} M, {Phinney} ES (2005) {The Afterglow of Massive Black Hole
  Coalescence}. \apjl 622(2):L93--L96. \doi{10.1086/429618}.
  {\href{https://arxiv.org/abs/astro-ph/0410343}{{arXiv:astro-ph/0410343}}}
  {[astro-ph]}

\bibitem[{Minamitsuji(2017)}]{Minamitsuji:2017aan}
Minamitsuji M (2017) {Black holes in the generalized Proca theory}. Gen Rel
  Grav 49(7):86. \doi{10.1007/s10714-017-2250-7}

\bibitem[{Minamitsuji(2018)}]{Minamitsuji:2018kof}
Minamitsuji M (2018) {Vector boson star solutions with a quartic order
  self-interaction}. Phys Rev D 97(10):104023.
  \doi{10.1103/PhysRevD.97.104023}.
  {\href{https://arxiv.org/abs/1805.09867}{{arXiv:1805.09867}}} {[gr-qc]}

\bibitem[{Mirbabayi et~al.(2020)Mirbabayi, Gruzinov, and
  Nore\~na}]{Mirbabayi:2019uph}
Mirbabayi M, Gruzinov A, Nore\~na J (2020) {Spin of Primordial Black Holes}.
  JCAP 03:017. \doi{10.1088/1475-7516/2020/03/017}.
  {\href{https://arxiv.org/abs/1901.05963}{{arXiv:1901.05963}}} {[astro-ph.CO]}

\bibitem[{Mirshekari et~al.(2012)Mirshekari, Yunes, and
  Will}]{Mirshekari:2011yq}
Mirshekari S, Yunes N, Will CM (2012) {Constraining Generic Lorentz Violation
  and the Speed of the Graviton with Gravitational Waves}. Phys Rev D
  85:024041. \doi{10.1103/PhysRevD.85.024041}.
  {\href{https://arxiv.org/abs/1110.2720}{{arXiv:1110.2720}}} {[gr-qc]}

\bibitem[{Moffat(2015)}]{Moffat:2014aja}
Moffat JW (2015) {Black Holes in Modified Gravity (MOG)}. Eur Phys J C
  75(4):175. \doi{10.1140/epjc/s10052-015-3405-x}.
  {\href{https://arxiv.org/abs/1412.5424}{{arXiv:1412.5424}}} {[gr-qc]}

\bibitem[{Molina et~al.(2010)Molina, Pani, Cardoso, and
  Gualtieri}]{Molina:2010fb}
Molina C, Pani P, Cardoso V, Gualtieri L (2010) {Gravitational signature of
  Schwarzschild black holes in dynamical Chern-Simons gravity}. Phys Rev D
  81:124021. \doi{10.1103/PhysRevD.81.124021}.
  {\href{https://arxiv.org/abs/1004.4007}{{arXiv:1004.4007}}} {[gr-qc]}

\bibitem[{Mollerach et~al.(2004)Mollerach, Harari, and
  Matarrese}]{Mollerach:2003nq}
Mollerach S, Harari D, Matarrese S (2004) {CMB polarization from secondary
  vector and tensor modes}. Phys Rev D 69:063002.
  \doi{10.1103/PhysRevD.69.063002}.
  {\href{https://arxiv.org/abs/astro-ph/0310711}{{arXiv:astro-ph/0310711}}}

\bibitem[{Montero-Camacho et~al.(2019)Montero-Camacho, Fang, Vasquez, Silva,
  and Hirata}]{Montero-Camacho:2019jte}
Montero-Camacho P, Fang X, Vasquez G, Silva M, Hirata CM (2019) {Revisiting
  constraints on asteroid-mass primordial black holes as dark matter
  candidates}. JCAP 08:031. \doi{10.1088/1475-7516/2019/08/031}.
  {\href{https://arxiv.org/abs/1906.05950}{{arXiv:1906.05950}}} {[astro-ph.CO]}

\bibitem[{Moore and Yunes(2019)}]{Moore:2019xkm}
Moore B, Yunes N (2019) {A 3PN Fourier Domain Waveform for Non-Spinning
  Binaries with Moderate Eccentricity}. Class Quant Grav 36(18):185003.
  \doi{10.1088/1361-6382/ab3778}.
  {\href{https://arxiv.org/abs/1903.05203}{{arXiv:1903.05203}}} {[gr-qc]}

\bibitem[{Moore and Yunes(2020)}]{Moore:2020rva}
Moore B, Yunes N (2020) {Constraining Gravity with Eccentric Gravitational
  Waves: Projected Upper Bounds and Model Selection}. Class Quant Grav
  37(16):165006. \doi{10.1088/1361-6382/ab8bb6}.
  {\href{https://arxiv.org/abs/2002.05775}{{arXiv:2002.05775}}} {[gr-qc]}

\bibitem[{Moore et~al.(2015)Moore, Cole, and Berry}]{Moore:2014lga}
Moore C, Cole R, Berry C (2015) {Gravitational-wave sensitivity curves}. Class
  Quant Grav 32(1):015014. \doi{10.1088/0264-9381/32/1/015014}.
  {\href{https://arxiv.org/abs/1408.0740}{{arXiv:1408.0740}}} {[gr-qc]}

\bibitem[{Moore et~al.(2017)Moore, Chua, and Gair}]{Moore:2017lxy}
Moore CJ, Chua AJK, Gair JR (2017) {Gravitational waves from extreme mass ratio
  inspirals around bumpy black holes}. Class Quant Grav 34(19):195009.
  \doi{10.1088/1361-6382/aa85fa}.
  {\href{https://arxiv.org/abs/1707.00712}{{arXiv:1707.00712}}} {[gr-qc]}

\bibitem[{Moore et~al.(2019)Moore, Gerosa, and Klein}]{Moore:2019pke}
Moore CJ, Gerosa D, Klein A (2019) {Are stellar-mass black-hole binaries too
  quiet for LISA?} Mon Not Roy Astron Soc 488(1):L94--L98.
  \doi{10.1093/mnrasl/slz104}.
  {\href{https://arxiv.org/abs/1905.11998}{{arXiv:1905.11998}}} {[astro-ph.HE]}

\bibitem[{Moore and Nelson(2001)}]{Moore:2001bv}
Moore GD, Nelson AE (2001) {Lower bound on the propagation speed of gravity
  from gravitational Cherenkov radiation}. JHEP 09:023.
  \doi{10.1088/1126-6708/2001/09/023}.
  {\href{https://arxiv.org/abs/hep-ph/0106220}{{arXiv:hep-ph/0106220}}}

\bibitem[{Moradinezhad~Dizgah et~al.(2019)Moradinezhad~Dizgah, Franciolini, and
  Riotto}]{MoradinezhadDizgah:2019wjf}
Moradinezhad~Dizgah A, Franciolini G, Riotto A (2019) {Primordial Black Holes
  from Broad Spectra: Abundance and Clustering}. JCAP 11:001.
  \doi{10.1088/1475-7516/2019/11/001}.
  {\href{https://arxiv.org/abs/1906.08978}{{arXiv:1906.08978}}} {[astro-ph.CO]}

\bibitem[{Morita and Soda(2019)}]{Morita:2019sau}
Morita T, Soda J (2019) {Arrival Time Differences of Lensed Massive
  Gravitational Waves}
  {\href{https://arxiv.org/abs/1911.07435}{{arXiv:1911.07435}}} {[gr-qc]}

\bibitem[{Morris and Thorne(1988)}]{Morris:1988cz}
Morris MS, Thorne KS (1988) {Wormholes in space-time and their use for
  interstellar travel: A tool for teaching general relativity}. Am J Phys
  56:395--412. \doi{10.1119/1.15620}

\bibitem[{Moschidis(2016)}]{Moschidis:2016wew}
Moschidis G (2016) {Superradiant instabilities for short-range non-negative
  potentials on Kerr spacetimes and applications}
  {\href{https://arxiv.org/abs/1608.02041}{{arXiv:1608.02041}}} {[math.AP]}

\bibitem[{Mottola and Vaulin(2006)}]{Mottola:2006ew}
Mottola E, Vaulin R (2006) {Macroscopic Effects of the Quantum Trace Anomaly}.
  Phys Rev D 74:064004. \doi{10.1103/PhysRevD.74.064004}.
  {\href{https://arxiv.org/abs/gr-qc/0604051}{{arXiv:gr-qc/0604051}}}

\bibitem[{Mueller et~al.(2013)Mueller, Janka, and Marek}]{Mueller:2012sv}
Mueller B, Janka HT, Marek A (2013) {A New Multi-Dimensional General
  Relativistic Neutrino Hydrodynamics Code of Core-Collapse Supernovae III.
  Gravitational Wave Signals from Supernova Explosion Models}. Astrophys J
  766:43. \doi{10.1088/0004-637X/766/1/43}.
  {\href{https://arxiv.org/abs/1210.6984}{{arXiv:1210.6984}}} {[astro-ph.SR]}

\bibitem[{Mukhanov(1986)}]{Mukhanov:1986me}
Mukhanov VF (1986) {ARE BLACK HOLES QUANTIZED?} JETP Lett 44:63--66

\bibitem[{Mukherjee and Silk(2021)}]{Mukherjee:2021ags}
Mukherjee S, Silk J (2021) {Can we distinguish astrophysical from primordial
  black holes via the stochastic gravitational wave background?}
  \doi{10.1093/mnras/stab1932}.
  {\href{https://arxiv.org/abs/2105.11139}{{arXiv:2105.11139}}} {[gr-qc]}

\bibitem[{Mukherjee et~al.(2020)Mukherjee, Wandelt, Nissanke, and
  Silvestri}]{Mukherjee:2020hyn}
Mukherjee S, Wandelt BD, Nissanke SM, Silvestri A (2020) {Accurate and
  precision Cosmology with redshift unknown gravitational wave sources}
  {\href{https://arxiv.org/abs/2007.02943}{{arXiv:2007.02943}}} {[astro-ph.CO]}

\bibitem[{{Mukherjee} et~al.(2020){Mukherjee}, {Wandelt}, and
  {Silk}}]{Mukherjee2020PhRvD}
{Mukherjee} S, {Wandelt} BD, {Silk} J (2020) {Multimessenger tests of gravity
  with weakly lensed gravitational waves}. \prd 101(10):103509.
  \doi{10.1103/PhysRevD.101.103509}.
  {\href{https://arxiv.org/abs/1908.08950}{{arXiv:1908.08950}}} {[astro-ph.CO]}

\bibitem[{Mukherjee et~al.(2020)Mukherjee, Wandelt, and
  Silk}]{Mukherjee:2019wfw}
Mukherjee S, Wandelt BD, Silk J (2020) {Multimessenger tests of gravity with
  weakly lensed gravitational waves}. Phys Rev D 101(10):103509.
  \doi{10.1103/PhysRevD.101.103509}.
  {\href{https://arxiv.org/abs/1908.08950}{{arXiv:1908.08950}}} {[astro-ph.CO]}

\bibitem[{{Mukherjee} et~al.(2020){Mukherjee}, {Wandelt}, and
  {Silk}}]{Mukherjee2020MNRAS}
{Mukherjee} S, {Wandelt} BD, {Silk} J (2020) {Probing the theory of gravity
  with gravitational lensing of gravitational waves and galaxy surveys}. \mnras
  494(2):1956--1970. \doi{10.1093/mnras/staa827}.
  {\href{https://arxiv.org/abs/1908.08951}{{arXiv:1908.08951}}} {[astro-ph.CO]}

\bibitem[{Mukherjee et~al.(2021)Mukherjee, Meinema, and
  Silk}]{Mukherjee:2021itf}
Mukherjee S, Meinema MSP, Silk J (2021) {Prospects of discovering sub-solar
  primordial black holes using the stochastic gravitational wave background
  from third-generation detectors}
  {\href{https://arxiv.org/abs/2107.02181}{{arXiv:2107.02181}}} {[astro-ph.CO]}

\bibitem[{Myers(1997)}]{Myers:1997qi}
Myers RC (1997) {Pure states don't wear black}. Gen Rel Grav 29:1217--1222.
  \doi{10.1023/A:1018855611972}.
  {\href{https://arxiv.org/abs/gr-qc/9705065}{{arXiv:gr-qc/9705065}}}

\bibitem[{Nagar(2011)}]{Nagar:2011fx}
Nagar A (2011) {Effective one body Hamiltonian of two spinning black-holes with
  next-to-next-to-leading order spin-orbit coupling}. Phys Rev D 84:084028.
  \doi{10.1103/PhysRevD.84.084028}, [Erratum: Phys.Rev.D 88, 089901 (2013)].
  {\href{https://arxiv.org/abs/1106.4349}{{arXiv:1106.4349}}} {[gr-qc]}

\bibitem[{Nagar and Rettegno(2019)}]{Nagar:2018gnk}
Nagar A, Rettegno P (2019) {Efficient effective one body time-domain
  gravitational waveforms}. Phys Rev D 99(2):021501.
  \doi{10.1103/PhysRevD.99.021501}.
  {\href{https://arxiv.org/abs/1805.03891}{{arXiv:1805.03891}}} {[gr-qc]}

\bibitem[{Nagar et~al.(2019)Nagar, Messina, Rettegno, Bini, Damour, Geralico,
  Akcay, and Bernuzzi}]{Nagar:2018plt}
Nagar A, Messina F, Rettegno P, Bini D, Damour T, Geralico A, Akcay S, Bernuzzi
  S (2019) {Nonlinear-in-spin effects in effective-one-body waveform models of
  spin-aligned, inspiralling, neutron star binaries}. Phys Rev D 99(4):044007.
  \doi{10.1103/PhysRevD.99.044007}.
  {\href{https://arxiv.org/abs/1812.07923}{{arXiv:1812.07923}}} {[gr-qc]}

\bibitem[{Nagar et~al.(2020{\natexlab{a}})Nagar, Pratten, Riemenschneider, and
  Gamba}]{Nagar:2019wds}
Nagar A, Pratten G, Riemenschneider G, Gamba R (2020{\natexlab{a}}) {Multipolar
  effective one body model for nonspinning black hole binaries}. Phys Rev D
  101(2):024041. \doi{10.1103/PhysRevD.101.024041}.
  {\href{https://arxiv.org/abs/1904.09550}{{arXiv:1904.09550}}} {[gr-qc]}

\bibitem[{Nagar et~al.(2020{\natexlab{b}})Nagar, Riemenschneider, Pratten,
  Rettegno, and Messina}]{Nagar:2020pcj}
Nagar A, Riemenschneider G, Pratten G, Rettegno P, Messina F
  (2020{\natexlab{b}}) {Multipolar effective one body waveform model for
  spin-aligned black hole binaries}. Phys Rev D 102(2):024077.
  \doi{10.1103/PhysRevD.102.024077}.
  {\href{https://arxiv.org/abs/2001.09082}{{arXiv:2001.09082}}} {[gr-qc]}

\bibitem[{Nagar et~al.(2018)}]{Nagar:2018zoe}
Nagar A, et~al. (2018) {Time-domain effective-one-body gravitational waveforms
  for coalescing compact binaries with nonprecessing spins, tides and self-spin
  effects}. Phys Rev D 98(10):104052. \doi{10.1103/PhysRevD.98.104052}.
  {\href{https://arxiv.org/abs/1806.01772}{{arXiv:1806.01772}}} {[gr-qc]}

\bibitem[{Nair et~al.(2019)Nair, Perkins, Silva, and Yunes}]{Nair:2019iur}
Nair R, Perkins S, Silva HO, Yunes N (2019) {Fundamental Physics Implications
  for Higher-Curvature Theories from Binary Black Hole Signals in the
  LIGO-Virgo Catalog GWTC-1}. Phys Rev Lett 123(19):191101.
  \doi{10.1103/PhysRevLett.123.191101}.
  {\href{https://arxiv.org/abs/1905.00870}{{arXiv:1905.00870}}} {[gr-qc]}

\bibitem[{Nakama et~al.(2018)Nakama, Carr, and Silk}]{Nakama:2017xvq}
Nakama T, Carr B, Silk J (2018) {Limits on primordial black holes from $\mu$
  distortions in cosmic microwave background}. Phys Rev D 97(4):043525.
  \doi{10.1103/PhysRevD.97.043525}.
  {\href{https://arxiv.org/abs/1710.06945}{{arXiv:1710.06945}}} {[astro-ph.CO]}

\bibitem[{Nakano et~al.(2017)Nakano, Sago, Tagoshi, and
  Tanaka}]{Nakano:2017fvh}
Nakano H, Sago N, Tagoshi H, Tanaka T (2017) {Black hole ringdown echoes and
  howls}. PTEP 2017(7):071E01. \doi{10.1093/ptep/ptx093}.
  {\href{https://arxiv.org/abs/1704.07175}{{arXiv:1704.07175}}} {[gr-qc]}

\bibitem[{Nandra et~al.(2013)}]{Nandra:2013jka}
Nandra K, et~al. (2013) {The Hot and Energetic Universe: A White Paper
  presenting the science theme motivating the Athena+ mission}
  {\href{https://arxiv.org/abs/1306.2307}{{arXiv:1306.2307}}} {[astro-ph.HE]}

\bibitem[{{Naoz} et~al.(2013){Naoz}, {Farr}, {Lithwick}, {Rasio}, and
  {Teyssandier}}]{2013MNRAS.431.2155N}
{Naoz} S, {Farr} WM, {Lithwick} Y, {Rasio} FA, {Teyssandier} J (2013) {Secular
  dynamics in hierarchical three-body systems}. \mnras 431(3):2155--2171.
  \doi{10.1093/mnras/stt302}.
  {\href{https://arxiv.org/abs/1107.2414}{{arXiv:1107.2414}}} {[astro-ph.EP]}

\bibitem[{Nasipak et~al.(2019)Nasipak, Osburn, and Evans}]{Nasipak:2019hxh}
Nasipak Z, Osburn T, Evans CR (2019) {Repeated faint quasinormal bursts in
  extreme-mass-ratio inspiral waveforms: Evidence from frequency-domain scalar
  self-force calculations on generic Kerr orbits}. Phys Rev D 100(6):064008.
  \doi{10.1103/PhysRevD.100.064008}.
  {\href{https://arxiv.org/abs/1905.13237}{{arXiv:1905.13237}}} {[gr-qc]}

\bibitem[{Nelson and Scholtz(2011)}]{Nelson:2011sf}
Nelson AE, Scholtz J (2011) {Dark Light, Dark Matter and the Misalignment
  Mechanism}. Phys Rev D 84:103501. \doi{10.1103/PhysRevD.84.103501}.
  {\href{https://arxiv.org/abs/1105.2812}{{arXiv:1105.2812}}} {[hep-ph]}

\bibitem[{Ng et~al.(2019)Ng, Hannuksela, Vitale, and Li}]{Ng:2019jsx}
Ng KK, Hannuksela OA, Vitale S, Li TG (2019) {Searching for ultralight bosons
  within spin measurements of a population of binary black hole mergers}
  {\href{https://arxiv.org/abs/1908.02312}{{arXiv:1908.02312}}} {[gr-qc]}

\bibitem[{Ng et~al.(2020)Ng, Isi, Haster, and Vitale}]{Ng:2020jqd}
Ng KK, Isi M, Haster CJ, Vitale S (2020) {Multiband gravitational-wave searches
  for ultralight bosons}
  {\href{https://arxiv.org/abs/2007.12793}{{arXiv:2007.12793}}} {[gr-qc]}

\bibitem[{Nichols(2017)}]{BMS14}
Nichols DA (2017) {Spin memory effect for compact binaries in the
  post-Newtonian approximation}. Phys Rev D 95:084048.
  \doi{10.1103/PhysRevD.95.084048}.
  {\href{https://arxiv.org/abs/1702.03300}{{arXiv:1702.03300}}} {[gr-qc]}

\bibitem[{Nichols(2018)}]{BMS15}
Nichols DA (2018) {Center-of-mass angular momentum and memory effect in
  asymptotically flat spacetimes}. Phys Rev D 98:064032.
  \doi{10.1103/PhysRevD.98.064032}.
  {\href{https://arxiv.org/abs/1807.08767}{{arXiv:1807.08767}}} {[gr-qc]}

\bibitem[{Nicolaou et~al.(2020)Nicolaou, Lahav, Lemos, Hartley, and
  Braden}]{Nicolaou:2019cip}
Nicolaou C, Lahav O, Lemos P, Hartley W, Braden J (2020) {The Impact of
  Peculiar Velocities on the Estimation of the Hubble Constant from
  Gravitational Wave Standard Sirens}. Mon Not Roy Astron Soc 495(1):90--97.
  \doi{10.1093/mnras/staa1120}.
  {\href{https://arxiv.org/abs/1909.09609}{{arXiv:1909.09609}}} {[astro-ph.CO]}

\bibitem[{Nielsen et~al.(2019)Nielsen, Capano, Birnholtz, and
  Westerweck}]{Nielsen:2018lkf}
Nielsen AB, Capano CD, Birnholtz O, Westerweck J (2019) {Parameter estimation
  and statistical significance of echoes following black hole signals in the
  first Advanced LIGO observing run}. Phys Rev D 99(10):104012.
  \doi{10.1103/PhysRevD.99.104012}.
  {\href{https://arxiv.org/abs/1811.04904}{{arXiv:1811.04904}}} {[gr-qc]}

\bibitem[{Niikura et~al.(2019{\natexlab{a}})Niikura, Takada, Yokoyama, Sumi,
  and Masaki}]{Niikura:2019kqi}
Niikura H, Takada M, Yokoyama S, Sumi T, Masaki S (2019{\natexlab{a}})
  {Constraints on Earth-mass primordial black holes from OGLE 5-year
  microlensing events}. Phys Rev D 99(8):083503.
  \doi{10.1103/PhysRevD.99.083503}.
  {\href{https://arxiv.org/abs/1901.07120}{{arXiv:1901.07120}}} {[astro-ph.CO]}

\bibitem[{Niikura et~al.(2019{\natexlab{b}})}]{Niikura:2017zjd}
Niikura H, et~al. (2019{\natexlab{b}}) {Microlensing constraints on primordial
  black holes with Subaru/HSC Andromeda observations}. Nature Astron
  3(6):524--534. \doi{10.1038/s41550-019-0723-1}.
  {\href{https://arxiv.org/abs/1701.02151}{{arXiv:1701.02151}}} {[astro-ph.CO]}

\bibitem[{Nishizawa(2018)}]{Nishizawa:2017nef}
Nishizawa A (2018) {Generalized framework for testing gravity with
  gravitational-wave propagation. I. Formulation}. Phys Rev D97:104037.
  \doi{10.1103/PhysRevD.97.104037}.
  {\href{https://arxiv.org/abs/1710.04825}{{arXiv:1710.04825}}} {[gr-qc]}

\bibitem[{Nishizawa and Nakamura(2014)}]{Nishizawa:2014zna}
Nishizawa A, Nakamura T (2014) {Measuring Speed of Gravitational Waves by
  Observations of Photons and Neutrinos from Compact Binary Mergers and
  Supernovae}. Phys Rev D 90(4):044048. \doi{10.1103/PhysRevD.90.044048}.
  {\href{https://arxiv.org/abs/1406.5544}{{arXiv:1406.5544}}} {[gr-qc]}

\bibitem[{Nishizawa et~al.(2009)Nishizawa, Taruya, Hayama, Kawamura, and
  Sakagami}]{Nishizawa:2009bf}
Nishizawa A, Taruya A, Hayama K, Kawamura S, Sakagami Ma (2009) {Probing
  non-tensorial polarizations of stochastic gravitational-wave backgrounds with
  ground-based laser interferometers}. Phys Rev D 79:082002.
  \doi{10.1103/PhysRevD.79.082002}.
  {\href{https://arxiv.org/abs/0903.0528}{{arXiv:0903.0528}}} {[astro-ph.CO]}

\bibitem[{Nissanke et~al.(2013)Nissanke, Holz, Dalal, Hughes, Sievers, and
  Hirata}]{Nissanke:2013fka}
Nissanke S, Holz DE, Dalal N, Hughes SA, Sievers JL, Hirata CM (2013)
  {Determining the Hubble constant from gravitational wave observations of
  merging compact binaries}
  {\href{https://arxiv.org/abs/1307.2638}{{arXiv:1307.2638}}} {[astro-ph.CO]}

\bibitem[{Noller et~al.(2020)Noller, Santoni, Trincherini, and
  Trombetta}]{Noller:2019chl}
Noller J, Santoni L, Trincherini E, Trombetta LG (2020) {Black Hole Ringdown as
  a Probe for Dark Energy}. Phys Rev D 101:084049.
  \doi{10.1103/PhysRevD.101.084049}.
  {\href{https://arxiv.org/abs/1911.11671}{{arXiv:1911.11671}}} {[gr-qc]}

\bibitem[{{Nordtvedt} and {Will}(1972)}]{1972ApJ...177..775N}
{Nordtvedt} J Kenneth, {Will} CM (1972) {Conservation Laws and Preferred Frames
  in Relativistic Gravity. II. Experimental Evidence to Rule Out
  Preferred-Frame Theories of Gravity}. \apj 177:775. \doi{10.1086/151755}

\bibitem[{Nordtvedt(1968)}]{Nordtvedt:1968qs}
Nordtvedt K (1968) {Equivalence Principle for Massive Bodies. 2. Theory}. Phys
  Rev 169:1017--1025. \doi{10.1103/PhysRev.169.1017}

\bibitem[{Nucamendi and Salgado(2003)}]{Nucamendi:1995ex}
Nucamendi U, Salgado M (2003) {Scalar hairy black holes and solitons in
  asymptotically flat space-times}. Phys Rev D 68:044026.
  \doi{10.1103/PhysRevD.68.044026}.
  {\href{https://arxiv.org/abs/gr-qc/0301062}{{arXiv:gr-qc/0301062}}}

\bibitem[{{Oguri}(2016)}]{OguriGW}
{Oguri} M (2016) {Measuring the distance-redshift relation with the
  cross-correlation of gravitational wave standard sirens and galaxies}. \prd
  93(8):083511. \doi{10.1103/PhysRevD.93.083511}.
  {\href{https://arxiv.org/abs/1603.02356}{{arXiv:1603.02356}}} {[astro-ph.CO]}

\bibitem[{Oguri et~al.(2018)Oguri, Diego, Kaiser, Kelly, and
  Broadhurst}]{Oguri:2017ock}
Oguri M, Diego JM, Kaiser N, Kelly PL, Broadhurst T (2018) {Understanding
  caustic crossings in giant arcs: characteristic scales, event rates, and
  constraints on compact dark matter}. Phys Rev D 97(2):023518.
  \doi{10.1103/PhysRevD.97.023518}.
  {\href{https://arxiv.org/abs/1710.00148}{{arXiv:1710.00148}}} {[astro-ph.CO]}

\bibitem[{Okounkova(2019)}]{Okounkova:2019zep}
Okounkova M (2019) {Stability of Rotating Black Holes in Einstein Dilaton
  Gauss-Bonnet Gravity}. Phys Rev D 100(12):124054.
  \doi{10.1103/PhysRevD.100.124054}.
  {\href{https://arxiv.org/abs/1909.12251}{{arXiv:1909.12251}}} {[gr-qc]}

\bibitem[{Okounkova(2020)}]{Okounkova:2020rqw}
Okounkova M (2020) {Numerical relativity simulation of GW150914 in Einstein
  dilaton Gauss-Bonnet gravity}
  {\href{https://arxiv.org/abs/2001.03571}{{arXiv:2001.03571}}} {[gr-qc]}

\bibitem[{Okounkova et~al.(2017)Okounkova, Stein, Scheel, and
  Hemberger}]{Okounkova:2017yby}
Okounkova M, Stein LC, Scheel MA, Hemberger DA (2017) {Numerical binary black
  hole mergers in dynamical Chern-Simons gravity: Scalar field}. Phys Rev D
  96(4):044020. \doi{10.1103/PhysRevD.96.044020}.
  {\href{https://arxiv.org/abs/1705.07924}{{arXiv:1705.07924}}} {[gr-qc]}

\bibitem[{Okounkova et~al.(2019{\natexlab{a}})Okounkova, Scheel, and
  Teukolsky}]{Okounkova:2018pql}
Okounkova M, Scheel MA, Teukolsky SA (2019{\natexlab{a}}) {Evolving Metric
  Perturbations in dynamical Chern-Simons Gravity}. Phys Rev D 99(4):044019.
  \doi{10.1103/PhysRevD.99.044019}.
  {\href{https://arxiv.org/abs/1811.10713}{{arXiv:1811.10713}}} {[gr-qc]}

\bibitem[{Okounkova et~al.(2019{\natexlab{b}})Okounkova, Stein, Scheel, and
  Teukolsky}]{Okounkova:2019dfo}
Okounkova M, Stein LC, Scheel MA, Teukolsky SA (2019{\natexlab{b}}) {Numerical
  binary black hole collisions in dynamical Chern-Simons gravity}. Phys Rev D
  100(10):104026. \doi{10.1103/PhysRevD.100.104026}.
  {\href{https://arxiv.org/abs/1906.08789}{{arXiv:1906.08789}}} {[gr-qc]}

\bibitem[{Okounkova et~al.(2020)Okounkova, Stein, Moxon, Scheel, and
  Teukolsky}]{Okounkova:2019zjf}
Okounkova M, Stein LC, Moxon J, Scheel MA, Teukolsky SA (2020) {Numerical
  relativity simulation of GW150914 beyond general relativity}. Phys Rev D
  101(10):104016. \doi{10.1103/PhysRevD.101.104016}.
  {\href{https://arxiv.org/abs/1911.02588}{{arXiv:1911.02588}}} {[gr-qc]}

\bibitem[{Oost et~al.(2018)Oost, Mukohyama, and Wang}]{Oost:2018tcv}
Oost J, Mukohyama S, Wang A (2018) {Constraints on Einstein-aether theory after
  GW170817}. Phys Rev D97(12):124023. \doi{10.1103/PhysRevD.97.124023}.
  {\href{https://arxiv.org/abs/1802.04303}{{arXiv:1802.04303}}} {[gr-qc]}

\bibitem[{Oshita and Afshordi(2019)}]{Oshita:2018fqu}
Oshita N, Afshordi N (2019) {Probing microstructure of black hole spacetimes
  with gravitational wave echoes}. Phys Rev D 99(4):044002.
  \doi{10.1103/PhysRevD.99.044002}.
  {\href{https://arxiv.org/abs/1807.10287}{{arXiv:1807.10287}}} {[gr-qc]}

\bibitem[{Oshita et~al.(2020)Oshita, Tsuna, and Afshordi}]{Oshita:2020dox}
Oshita N, Tsuna D, Afshordi N (2020) {Quantum Black Hole Seismology I: Echoes,
  Ergospheres, and Spectra}. Phys Rev D 102(2):024045.
  \doi{10.1103/PhysRevD.102.024045}.
  {\href{https://arxiv.org/abs/2001.11642}{{arXiv:2001.11642}}} {[gr-qc]}

\bibitem[{Ossokine et~al.(2020)}]{Ossokine:2020kjp}
Ossokine S, et~al. (2020) {Multipolar Effective-One-Body Waveforms for
  Precessing Binary Black Holes: Construction and Validation}. Phys Rev D
  102(4):044055. \doi{10.1103/PhysRevD.102.044055}.
  {\href{https://arxiv.org/abs/2004.09442}{{arXiv:2004.09442}}} {[gr-qc]}

\bibitem[{Ota and Chirenti(2020)}]{Ota:2019bzl}
Ota I, Chirenti C (2020) {Overtones or higher harmonics? Prospects for testing
  the no-hair theorem with gravitational wave detections}. Phys Rev D
  101(10):104005. \doi{10.1103/PhysRevD.101.104005}.
  {\href{https://arxiv.org/abs/1911.00440}{{arXiv:1911.00440}}} {[gr-qc]}

\bibitem[{Ott et~al.(2013)Ott, Abdikamalov, M\"osta, Haas, Drasco, O'Connor,
  Reisswig, Meakin, and Schnetter}]{Ott:2012mr}
Ott CD, Abdikamalov E, M\"osta P, Haas R, Drasco S, O'Connor EP, Reisswig C,
  Meakin CA, Schnetter E (2013) {General-Relativistic Simulations of
  Three-Dimensional Core-Collapse Supernovae}. Astrophys J 768:115.
  \doi{10.1088/0004-637X/768/2/115}.
  {\href{https://arxiv.org/abs/1210.6674}{{arXiv:1210.6674}}} {[astro-ph.HE]}

\bibitem[{Paardekooper and Mellema(2006)}]{Paardekooper:2006hr}
Paardekooper SJ, Mellema G (2006) {Halting Type I planet migration in
  non-isothermal disks}. Astron Astrophys 459:L17.
  \doi{10.1051/0004-6361:20066304}.
  {\href{https://arxiv.org/abs/astro-ph/0608658}{{arXiv:astro-ph/0608658}}}

\bibitem[{Pacilio et~al.(2020)Pacilio, Vaglio, Maselli, and
  Pani}]{Pacilio:2020jza}
Pacilio C, Vaglio M, Maselli A, Pani P (2020) {Gravitational-wave detectors as
  particle-physics laboratories: Constraining scalar interactions with
  boson-star binaries}
  {\href{https://arxiv.org/abs/2007.05264}{{arXiv:2007.05264}}} {[gr-qc]}

\bibitem[{Padilla(2015)}]{Padilla:2015aaa}
Padilla A (2015) {Lectures on the Cosmological Constant Problem}
  {\href{https://arxiv.org/abs/1502.05296}{{arXiv:1502.05296}}} {[hep-th]}

\bibitem[{Page(2004)}]{Page:2003rd}
Page DN (2004) {Classical and quantum decay of oscillatons: Oscillating
  selfgravitating real scalar field solitons}. Phys Rev D 70:023002.
  \doi{10.1103/PhysRevD.70.023002}.
  {\href{https://arxiv.org/abs/gr-qc/0310006}{{arXiv:gr-qc/0310006}}}

\bibitem[{Palenzuela et~al.(2008)Palenzuela, Lehner, and
  Liebling}]{Palenzuela:2007dm}
Palenzuela C, Lehner L, Liebling SL (2008) {Orbital Dynamics of Binary Boson
  Star Systems}. Phys Rev D 77:044036. \doi{10.1103/PhysRevD.77.044036}.
  {\href{https://arxiv.org/abs/0706.2435}{{arXiv:0706.2435}}} {[gr-qc]}

\bibitem[{Palenzuela et~al.(2017)Palenzuela, Pani, Bezares, Cardoso, Lehner,
  and Liebling}]{Palenzuela:2017kcg}
Palenzuela C, Pani P, Bezares M, Cardoso V, Lehner L, Liebling S (2017)
  {Gravitational Wave Signatures of Highly Compact Boson Star Binaries}. Phys
  Rev D 96(10):104058. \doi{10.1103/PhysRevD.96.104058}.
  {\href{https://arxiv.org/abs/1710.09432}{{arXiv:1710.09432}}} {[gr-qc]}

\bibitem[{Palma et~al.(2020)Palma, Sypsas, and Zenteno}]{Palma:2020ejf}
Palma GA, Sypsas S, Zenteno C (2020) {Seeding primordial black holes in
  multifield inflation}. Phys Rev Lett 125(12):121301.
  \doi{10.1103/PhysRevLett.125.121301}.
  {\href{https://arxiv.org/abs/2004.06106}{{arXiv:2004.06106}}} {[astro-ph.CO]}

\bibitem[{Palmese et~al.(2020)}]{Palmese:2020aof}
Palmese A, et~al. (2020) {A statistical standard siren measurement of the
  Hubble constant from the LIGO/Virgo gravitational wave compact object merger
  GW190814 and Dark Energy Survey galaxies}
  {\href{https://arxiv.org/abs/2006.14961}{{arXiv:2006.14961}}} {[astro-ph.CO]}

\bibitem[{Pan et~al.(2011)Pan, Buonanno, Fujita, Racine, and
  Tagoshi}]{Pan:2010hz}
Pan Y, Buonanno A, Fujita R, Racine E, Tagoshi H (2011) {Post-Newtonian
  factorized multipolar waveforms for spinning, non-precessing black-hole
  binaries}. Phys Rev D 83:064003. \doi{10.1103/PhysRevD.83.064003}, [Erratum:
  Phys.Rev.D 87, 109901 (2013)].
  {\href{https://arxiv.org/abs/1006.0431}{{arXiv:1006.0431}}} {[gr-qc]}

\bibitem[{Pang et~al.(2020)Pang, Lo, Wong, Li, and Van
  Den~Broeck}]{Pang:2020pfz}
Pang PT, Lo RK, Wong IC, Li TG, Van Den~Broeck C (2020) {Generic searches for
  alternative gravitational wave polarizations with networks of interferometric
  detectors}. Phys Rev D 101(10):104055. \doi{10.1103/PhysRevD.101.104055}.
  {\href{https://arxiv.org/abs/2003.07375}{{arXiv:2003.07375}}} {[gr-qc]}

\bibitem[{Pani(2015)}]{Pani:2015tga}
Pani P (2015) {I-Love-Q relations for gravastars and the approach to the
  black-hole limit}. Phys Rev D 92(12):124030.
  \doi{10.1103/PhysRevD.95.049902}, [Erratum: Phys.Rev.D 95, 049902 (2017)].
  {\href{https://arxiv.org/abs/1506.06050}{{arXiv:1506.06050}}} {[gr-qc]}

\bibitem[{Pani and Ferrari(2018)}]{Pani:2018flj}
Pani P, Ferrari V (2018) {On gravitational-wave echoes from neutron-star binary
  coalescences}. Class Quant Grav 35(15):15LT01.
  \doi{10.1088/1361-6382/aacb8f}.
  {\href{https://arxiv.org/abs/1804.01444}{{arXiv:1804.01444}}} {[gr-qc]}

\bibitem[{Pani and Maselli(2019)}]{Pani:2019cyc}
Pani P, Maselli A (2019) {Love in Extrema Ratio}. Int J Mod Phys D
  28(14):1944001. \doi{10.1142/S0218271819440012}.
  {\href{https://arxiv.org/abs/1905.03947}{{arXiv:1905.03947}}} {[gr-qc]}

\bibitem[{Pani and Sotiriou(2012)}]{Pani:2012qd}
Pani P, Sotiriou TP (2012) {Surface singularities in Eddington-inspired
  Born-Infeld gravity}. Phys Rev Lett 109:251102.
  \doi{10.1103/PhysRevLett.109.251102}.
  {\href{https://arxiv.org/abs/1209.2972}{{arXiv:1209.2972}}} {[gr-qc]}

\bibitem[{Pani et~al.(2009)Pani, Berti, Cardoso, Chen, and Norte}]{Pani:2009ss}
Pani P, Berti E, Cardoso V, Chen Y, Norte R (2009) {Gravitational wave
  signatures of the absence of an event horizon. I. Nonradial oscillations of a
  thin-shell gravastar}. Phys Rev D 80:124047.
  \doi{10.1103/PhysRevD.80.124047}.
  {\href{https://arxiv.org/abs/0909.0287}{{arXiv:0909.0287}}} {[gr-qc]}

\bibitem[{Pani et~al.(2010)Pani, Berti, Cardoso, Chen, and Norte}]{Pani:2010em}
Pani P, Berti E, Cardoso V, Chen Y, Norte R (2010) {Gravitational-wave
  signatures of the absence of an event horizon. II. Extreme mass ratio
  inspirals in the spacetime of a thin-shell gravastar}. Phys Rev D81:084011.
  \doi{10.1103/PhysRevD.81.084011}.
  {\href{https://arxiv.org/abs/1001.3031}{{arXiv:1001.3031}}} {[gr-qc]}

\bibitem[{Pani et~al.(2011{\natexlab{a}})Pani, Cardoso, and
  Gualtieri}]{Pani:2011xj}
Pani P, Cardoso V, Gualtieri L (2011{\natexlab{a}}) {Gravitational waves from
  extreme mass-ratio inspirals in Dynamical Chern-Simons gravity}. Phys Rev D
  83:104048. \doi{10.1103/PhysRevD.83.104048}.
  {\href{https://arxiv.org/abs/1104.1183}{{arXiv:1104.1183}}} {[gr-qc]}

\bibitem[{Pani et~al.(2011{\natexlab{b}})Pani, Macedo, Crispino, and
  Cardoso}]{Pani:2011gy}
Pani P, Macedo CF, Crispino LC, Cardoso V (2011{\natexlab{b}}) {Slowly rotating
  black holes in alternative theories of gravity}. Phys Rev D 84:087501.
  \doi{10.1103/PhysRevD.84.087501}.
  {\href{https://arxiv.org/abs/1109.3996}{{arXiv:1109.3996}}} {[gr-qc]}

\bibitem[{Pani et~al.(2012{\natexlab{a}})Pani, Cardoso, Gualtieri, Berti, and
  Ishibashi}]{Pani:2012vp}
Pani P, Cardoso V, Gualtieri L, Berti E, Ishibashi A (2012{\natexlab{a}})
  {Black hole bombs and photon mass bounds}. Phys Rev Lett 109:131102.
  \doi{10.1103/PhysRevLett.109.131102}.
  {\href{https://arxiv.org/abs/1209.0465}{{arXiv:1209.0465}}} {[gr-qc]}

\bibitem[{Pani et~al.(2012{\natexlab{b}})Pani, Cardoso, Gualtieri, Berti, and
  Ishibashi}]{Pani:2012bp}
Pani P, Cardoso V, Gualtieri L, Berti E, Ishibashi A (2012{\natexlab{b}})
  {Perturbations of slowly rotating black holes: massive vector fields in the
  Kerr metric}. Phys Rev D 86:104017. \doi{10.1103/PhysRevD.86.104017}.
  {\href{https://arxiv.org/abs/1209.0773}{{arXiv:1209.0773}}} {[gr-qc]}

\bibitem[{Pani et~al.(2013{\natexlab{a}})Pani, Berti, and
  Gualtieri}]{Pani:2013ija}
Pani P, Berti E, Gualtieri L (2013{\natexlab{a}}) {Gravitoelectromagnetic
  Perturbations of Kerr-Newman Black Holes: Stability and Isospectrality in the
  Slow-Rotation Limit}. Phys Rev Lett 110(24):241103.
  \doi{10.1103/PhysRevLett.110.241103}.
  {\href{https://arxiv.org/abs/1304.1160}{{arXiv:1304.1160}}} {[gr-qc]}

\bibitem[{Pani et~al.(2013{\natexlab{b}})Pani, Sotiriou, and
  Vernieri}]{Pani:2013qfa}
Pani P, Sotiriou TP, Vernieri D (2013{\natexlab{b}}) {Gravity with Auxiliary
  Fields}. Phys Rev D 88(12):121502. \doi{10.1103/PhysRevD.88.121502}.
  {\href{https://arxiv.org/abs/1306.1835}{{arXiv:1306.1835}}} {[gr-qc]}

\bibitem[{Pani et~al.(2015{\natexlab{a}})Pani, Gualtieri, and
  Ferrari}]{Pani:2015nua}
Pani P, Gualtieri L, Ferrari V (2015{\natexlab{a}}) {Tidal Love numbers of a
  slowly spinning neutron star}. Phys Rev D 92(12):124003.
  \doi{10.1103/PhysRevD.92.124003}.
  {\href{https://arxiv.org/abs/1509.02171}{{arXiv:1509.02171}}} {[gr-qc]}

\bibitem[{Pani et~al.(2015{\natexlab{b}})Pani, Gualtieri, Maselli, and
  Ferrari}]{Pani:2015hfa}
Pani P, Gualtieri L, Maselli A, Ferrari V (2015{\natexlab{b}}) {Tidal
  deformations of a spinning compact object}. Phys Rev D 92(2):024010.
  \doi{10.1103/PhysRevD.92.024010}.
  {\href{https://arxiv.org/abs/1503.07365}{{arXiv:1503.07365}}} {[gr-qc]}

\bibitem[{Papallo(2017)}]{Papallo:2017ddx}
Papallo G (2017) {On the hyperbolicity of the most general Horndeski theory}.
  Phys Rev D96(12):124036. \doi{10.1103/PhysRevD.96.124036}.
  {\href{https://arxiv.org/abs/1710.10155}{{arXiv:1710.10155}}} {[gr-qc]}

\bibitem[{Papallo and Reall(2017)}]{Papallo:2017qvl}
Papallo G, Reall HS (2017) {On the local well-posedness of Lovelock and
  Horndeski theories}. Phys Rev D 96(4):044019.
  \doi{10.1103/PhysRevD.96.044019}.
  {\href{https://arxiv.org/abs/1705.04370}{{arXiv:1705.04370}}} {[gr-qc]}

\bibitem[{Pardo et~al.(2018)Pardo, Fishbach, Holz, and Spergel}]{Pardo:2018ipy}
Pardo K, Fishbach M, Holz DE, Spergel DN (2018) {Limits on the number of
  spacetime dimensions from GW170817}. JCAP 07:048.
  \doi{10.1088/1475-7516/2018/07/048}.
  {\href{https://arxiv.org/abs/1801.08160}{{arXiv:1801.08160}}} {[gr-qc]}

\bibitem[{Pasterski et~al.(2016)Pasterski, Strominger, and Zhiboedov}]{BMSAdd3}
Pasterski S, Strominger A, Zhiboedov A (2016) {New Gravitational Memories}.
  JHEP 12:053. \doi{10.1007/JHEP12(2016)053}.
  {\href{https://arxiv.org/abs/1502.06120}{{arXiv:1502.06120}}} {[hep-th]}

\bibitem[{Pattison et~al.(2021)Pattison, Vennin, Wands, and
  Assadullahi}]{Pattison:2021oen}
Pattison C, Vennin V, Wands D, Assadullahi H (2021) {Ultra-slow-roll inflation
  with quantum diffusion}
  {\href{https://arxiv.org/abs/2101.05741}{{arXiv:2101.05741}}} {[astro-ph.CO]}

\bibitem[{{Payne}(1983)}]{BMSAdd5}
{Payne} PN (1983) {Smarr's zero-frequency-limit calculation}. Phys Rev D
  28(8):1894--1897. \doi{10.1103/PhysRevD.28.1894}

\bibitem[{Peccei and Quinn(1977)}]{Peccei:1977hh}
Peccei R, Quinn HR (1977) {CP Conservation in the Presence of Instantons}. Phys
  Rev Lett 38:1440--1443. \doi{10.1103/PhysRevLett.38.1440}

\bibitem[{Penrose(1965)}]{Penrose:1964wq}
Penrose R (1965) {Gravitational collapse and space-time singularities}. Phys
  Rev Lett 14:57--59. \doi{10.1103/PhysRevLett.14.57}

\bibitem[{Perkins and Yunes(2019)}]{Perkins:2018tir}
Perkins S, Yunes N (2019) {Probing Screening and the Graviton Mass with
  Gravitational Waves}. Class Quant Grav 36(5):055013.
  \doi{10.1088/1361-6382/aafce6}.
  {\href{https://arxiv.org/abs/1811.02533}{{arXiv:1811.02533}}} {[gr-qc]}

\bibitem[{Perkins et~al.(2021{\natexlab{a}})Perkins, Nair, Silva, and
  Yunes}]{Perkins:2021mhb}
Perkins SE, Nair R, Silva HO, Yunes N (2021{\natexlab{a}}) {Improved
  gravitational-wave constraints on higher-order curvature theories of gravity}
  {\href{https://arxiv.org/abs/2104.11189}{{arXiv:2104.11189}}} {[gr-qc]}

\bibitem[{Perkins et~al.(2021{\natexlab{b}})Perkins, Yunes, and
  Berti}]{Perkins:2020tra}
Perkins SE, Yunes N, Berti E (2021{\natexlab{b}}) {Probing Fundamental Physics
  with Gravitational Waves: The Next Generation}. Phys Rev D 103(4):044024.
  \doi{10.1103/PhysRevD.103.044024}.
  {\href{https://arxiv.org/abs/2010.09010}{{arXiv:2010.09010}}} {[gr-qc]}

\bibitem[{Pesce et~al.(2020)}]{Pesce:2020xfe}
Pesce D, et~al. (2020) {The Megamaser Cosmology Project. XIII. Combined Hubble
  constant constraints}. Astrophys J Lett 891(1):L1.
  \doi{10.3847/2041-8213/ab75f0}.
  {\href{https://arxiv.org/abs/2001.09213}{{arXiv:2001.09213}}} {[astro-ph.CO]}

\bibitem[{Petiteau et~al.(2011)Petiteau, Babak, and Sesana}]{Petiteau:2011we}
Petiteau A, Babak S, Sesana A (2011) {Constraining the dark energy equation of
  state using LISA observations of spinning Massive Black Hole binaries}.
  Astrophys J 732:82. \doi{10.1088/0004-637X/732/2/82}.
  {\href{https://arxiv.org/abs/1102.0769}{{arXiv:1102.0769}}} {[astro-ph.CO]}

\bibitem[{Pettorino et~al.(2013)Pettorino, Amendola, and
  Wetterich}]{Pettorino:2013ia}
Pettorino V, Amendola L, Wetterich C (2013) {How early is early dark energy?}
  Phys Rev D 87:083009. \doi{10.1103/PhysRevD.87.083009}.
  {\href{https://arxiv.org/abs/1301.5279}{{arXiv:1301.5279}}} {[astro-ph.CO]}

\bibitem[{Philcox et~al.(2020)Philcox, Ivanov, Simonovi\'c, and
  Zaldarriaga}]{Philcox:2020vvt}
Philcox OH, Ivanov MM, Simonovi\'c M, Zaldarriaga M (2020) {Combining
  Full-Shape and BAO Analyses of Galaxy Power Spectra: A 1.6\% CMB-independent
  constraint on H0}. JCAP 05:032. \doi{10.1088/1475-7516/2020/05/032}.
  {\href{https://arxiv.org/abs/2002.04035}{{arXiv:2002.04035}}} {[astro-ph.CO]}

\bibitem[{Pi and Sasaki(2020)}]{Pi:2020otn}
Pi S, Sasaki M (2020) {Gravitational Waves Induced by Scalar Perturbations with
  a Lognormal Peak}
  {\href{https://arxiv.org/abs/2005.12306}{{arXiv:2005.12306}}} {[gr-qc]}

\bibitem[{Pieroni and Barausse(2020)}]{Pieroni:2020rob}
Pieroni M, Barausse E (2020) {Foreground cleaning and template-free stochastic
  background extraction for LISA}. JCAP 07:021.
  \doi{10.1088/1475-7516/2020/07/021}, [Erratum: JCAP 09, E01 (2020)].
  {\href{https://arxiv.org/abs/2004.01135}{{arXiv:2004.01135}}} {[astro-ph.CO]}

\bibitem[{Piro et~al.(2021)}]{Piro:2021oaa}
Piro L, et~al. (2021) {Multi-messenger-Athena Synergy White Paper}
  {\href{https://arxiv.org/abs/2110.15677}{{arXiv:2110.15677}}} {[astro-ph.HE]}

\bibitem[{Podolsky and Svarc(2012)}]{Podolsky:2012he}
Podolsky J, Svarc R (2012) {Interpreting spacetimes of any dimension using
  geodesic deviation}. Phys Rev D 85:044057. \doi{10.1103/PhysRevD.85.044057}.
  {\href{https://arxiv.org/abs/1201.4790}{{arXiv:1201.4790}}} {[gr-qc]}

\bibitem[{Podolsk\'y and \v{S}varc(2013)}]{Podolsky:2013ola}
Podolsk\'y J, \v{S}varc R (2013) {Physical interpretation of Kundt spacetimes
  using geodesic deviation}. Class Quant Grav 30:205016.
  \doi{10.1088/0264-9381/30/20/205016}.
  {\href{https://arxiv.org/abs/1306.6554}{{arXiv:1306.6554}}} {[gr-qc]}

\bibitem[{Poisson(2015)}]{Poisson:2014gka}
Poisson E (2015) {Tidal deformation of a slowly rotating black hole}. Phys Rev
  D 91(4):044004. \doi{10.1103/PhysRevD.91.044004}.
  {\href{https://arxiv.org/abs/1411.4711}{{arXiv:1411.4711}}} {[gr-qc]}

\bibitem[{Poisson and Will(2014)}]{poisson2014gravity}
Poisson E, Will C (2014) Gravity: Newtonian, Post-Newtonian, Relativistic.
  Cambridge University Press

\bibitem[{Pollack et~al.(2015)Pollack, Spergel, and
  Steinhardt}]{Pollack:2014rja}
Pollack J, Spergel DN, Steinhardt PJ (2015) {Supermassive Black Holes from
  Ultra-Strongly Self-Interacting Dark Matter}. Astrophys J 804(2):131.
  \doi{10.1088/0004-637X/804/2/131}.
  {\href{https://arxiv.org/abs/1501.00017}{{arXiv:1501.00017}}} {[astro-ph.CO]}

\bibitem[{Porto(2016)}]{Porto:2016zng}
Porto RA (2016) {The Tune of Love and the Nature(ness) of Spacetime}. Fortsch
  Phys 64(10):723--729. \doi{10.1002/prop.201600064}.
  {\href{https://arxiv.org/abs/1606.08895}{{arXiv:1606.08895}}} {[gr-qc]}

\bibitem[{Posada and Chirenti(2019)}]{Camilo:2018goy}
Posada C, Chirenti C (2019) {On the radial stability of ultra compact
  Schwarzschild stars beyond the Buchdahl limit}. Class Quant Grav 36:065004.
  \doi{10.1088/1361-6382/ab0526}.
  {\href{https://arxiv.org/abs/1811.09589}{{arXiv:1811.09589}}} {[gr-qc]}

\bibitem[{Pound(2015{\natexlab{a}})}]{Pound:2015tma}
Pound A (2015{\natexlab{a}}) {Motion of small objects in curved spacetimes: An
  introduction to gravitational self-force}. Fund Theor Phys 179:399--486.
  \doi{10.1007/978-3-319-18335-0_13}.
  {\href{https://arxiv.org/abs/1506.06245}{{arXiv:1506.06245}}} {[gr-qc]}

\bibitem[{Pound(2015{\natexlab{b}})}]{Pound:2015wva}
Pound A (2015{\natexlab{b}}) {Second-order perturbation theory: problems on
  large scales}. Phys Rev D92(10):104047. \doi{10.1103/PhysRevD.92.104047}.
  {\href{https://arxiv.org/abs/1510.05172}{{arXiv:1510.05172}}} {[gr-qc]}

\bibitem[{Pound and Poisson(2008)}]{Pound:2007th}
Pound A, Poisson E (2008) {Osculating orbits in Schwarzschild spacetime, with
  an application to extreme mass-ratio inspirals}. Phys Rev D77:044013.
  \doi{10.1103/PhysRevD.77.044013}.
  {\href{https://arxiv.org/abs/0708.3033}{{arXiv:0708.3033}}} {[gr-qc]}

\bibitem[{Pound et~al.(2005)Pound, Poisson, and Nickel}]{Pound:2005fs}
Pound A, Poisson E, Nickel BG (2005) {Limitations of the adiabatic
  approximation to the gravitational self-force}. Phys Rev D72:124001.
  \doi{10.1103/PhysRevD.72.124001}.
  {\href{https://arxiv.org/abs/gr-qc/0509122}{{arXiv:gr-qc/0509122}}} {[gr-qc]}

\bibitem[{Pound et~al.(2020)Pound, Wardell, Warburton, and
  Miller}]{Pound:2019lzj}
Pound A, Wardell B, Warburton N, Miller J (2020) {Second-Order Self-Force
  Calculation of Gravitational Binding Energy in Compact Binaries}. Phys Rev
  Lett 124(2):021101. \doi{10.1103/PhysRevLett.124.021101}.
  {\href{https://arxiv.org/abs/1908.07419}{{arXiv:1908.07419}}} {[gr-qc]}

\bibitem[{Pratten et~al.(2020)Pratten, Husa, Garcia-Quiros, Colleoni,
  Ramos-Buades, Estelles, and Jaume}]{Pratten:2020fqn}
Pratten G, Husa S, Garcia-Quiros C, Colleoni M, Ramos-Buades A, Estelles H,
  Jaume R (2020) {Setting the cornerstone for a family of models for
  gravitational waves from compact binaries: The dominant harmonic for
  nonprecessing quasicircular black holes}. Phys Rev D 102(6):064001.
  \doi{10.1103/PhysRevD.102.064001}.
  {\href{https://arxiv.org/abs/2001.11412}{{arXiv:2001.11412}}} {[gr-qc]}

\bibitem[{Pratten et~al.(2021)}]{Pratten:2020ceb}
Pratten G, et~al. (2021) {Computationally efficient models for the dominant and
  subdominant harmonic modes of precessing binary black holes}. Phys Rev D
  103(10):104056. \doi{10.1103/PhysRevD.103.104056}.
  {\href{https://arxiv.org/abs/2004.06503}{{arXiv:2004.06503}}} {[gr-qc]}

\bibitem[{Press and Teukolsky(1972)}]{Press:1972zz}
Press WH, Teukolsky SA (1972) {Floating Orbits, Superradiant Scattering and the
  Black-hole Bomb}. Nature 238:211--212. \doi{10.1038/238211a0}

\bibitem[{Pretorius(2005)}]{Pretorius:2005gq}
Pretorius F (2005) {Evolution of binary black hole spacetimes}. Phys Rev Lett
  95:121101. \doi{10.1103/PhysRevLett.95.121101}.
  {\href{https://arxiv.org/abs/gr-qc/0507014}{{arXiv:gr-qc/0507014}}}

\bibitem[{Price and Khanna(2017)}]{Price:2017cjr}
Price RH, Khanna G (2017) {Gravitational wave sources: reflections and echoes}.
  Class Quant Grav 34(22):225005. \doi{10.1088/1361-6382/aa8f29}.
  {\href{https://arxiv.org/abs/1702.04833}{{arXiv:1702.04833}}} {[gr-qc]}

\bibitem[{P\"urrer(2014)}]{Purrer:2014fza}
P\"urrer M (2014) {Frequency domain reduced order models for gravitational
  waves from aligned-spin compact binaries}. Class Quant Grav 31(19):195010.
  \doi{10.1088/0264-9381/31/19/195010}.
  {\href{https://arxiv.org/abs/1402.4146}{{arXiv:1402.4146}}} {[gr-qc]}

\bibitem[{P\"urrer(2016)}]{Purrer:2015tud}
P\"urrer M (2016) {Frequency domain reduced order model of aligned-spin
  effective-one-body waveforms with generic mass-ratios and spins}. Phys Rev D
  93(6):064041. \doi{10.1103/PhysRevD.93.064041}.
  {\href{https://arxiv.org/abs/1512.02248}{{arXiv:1512.02248}}} {[gr-qc]}

\bibitem[{P\"urrer and Haster(2020)}]{Purrer:2019jcp}
P\"urrer M, Haster CJ (2020) {Gravitational waveform accuracy requirements for
  future ground-based detectors}. Phys Rev Res 2(2):023151.
  \doi{10.1103/PhysRevResearch.2.023151}.
  {\href{https://arxiv.org/abs/1912.10055}{{arXiv:1912.10055}}} {[gr-qc]}

\bibitem[{Quinlan et~al.(1995)Quinlan, Hernquist, and
  Sigurdsson}]{Quinlan:1994ed}
Quinlan GD, Hernquist L, Sigurdsson S (1995) {Models of Galaxies with Central
  Black Holes: Adiabatic Growth in Spherical Galaxies}. Astrophys J
  440:554--564. \doi{10.1086/175295}.
  {\href{https://arxiv.org/abs/astro-ph/9407005}{{arXiv:astro-ph/9407005}}}

\bibitem[{{Raccanelli}(2017)}]{Raccanelli2017GW}
{Raccanelli} A (2017) {Gravitational wave astronomy with radio galaxy surveys}.
  \mnras 469(1):656--670. \doi{10.1093/mnras/stx835}.
  {\href{https://arxiv.org/abs/1609.09377}{{arXiv:1609.09377}}} {[astro-ph.CO]}

\bibitem[{{Raccanelli} et~al.(2016){Raccanelli}, {Kovetz}, {Bird}, {Cholis},
  and {Mu{\~n}oz}}]{RaccanelliBHs}
{Raccanelli} A, {Kovetz} ED, {Bird} S, {Cholis} I, {Mu{\~n}oz} JB (2016)
  {Determining the progenitors of merging black-hole binaries}. \prd
  94(2):023516. \doi{10.1103/PhysRevD.94.023516}.
  {\href{https://arxiv.org/abs/1605.01405}{{arXiv:1605.01405}}} {[astro-ph.CO]}

\bibitem[{{Raccanelli} et~al.(2018){Raccanelli}, {Vidotto}, and
  {Verde}}]{RaccanelliVidotto2018}
{Raccanelli} A, {Vidotto} F, {Verde} L (2018) {Effects of primordial black
  holes quantum gravity decay on galaxy clustering}. \jcap 2018(8):003.
  \doi{10.1088/1475-7516/2018/08/003}.
  {\href{https://arxiv.org/abs/1708.02588}{{arXiv:1708.02588}}} {[astro-ph.CO]}

\bibitem[{Rahman and Sen(2019)}]{Rahman:2018fgy}
Rahman M, Sen AA (2019) {Astrophysical Signatures of Black holes in Generalized
  Proca Theories}. Phys Rev D 99(2):024052. \doi{10.1103/PhysRevD.99.024052}.
  {\href{https://arxiv.org/abs/1810.09200}{{arXiv:1810.09200}}} {[gr-qc]}

\bibitem[{Raidal et~al.(2019)Raidal, Spethmann, Vaskonen, and
  Veerm\"ae}]{Raidal:2018bbj}
Raidal M, Spethmann C, Vaskonen V, Veerm\"ae H (2019) {Formation and Evolution
  of Primordial Black Hole Binaries in the Early Universe}. JCAP 02:018.
  \doi{10.1088/1475-7516/2019/02/018}.
  {\href{https://arxiv.org/abs/1812.01930}{{arXiv:1812.01930}}} {[astro-ph.CO]}

\bibitem[{Ramos and Barausse(2019)}]{Ramos:2018oku}
Ramos O, Barausse E (2019) {Constraints on Ho\v{r}ava gravity from binary black
  hole observations}. Phys Rev D 99(2):024034.
  \doi{10.1103/PhysRevD.99.024034}.
  {\href{https://arxiv.org/abs/1811.07786}{{arXiv:1811.07786}}} {[gr-qc]}

\bibitem[{Ramos-Buades et~al.(2020{\natexlab{a}})Ramos-Buades, Husa, Pratten,
  Estell\'es, Garc\'\i{}a-Quir\'os, Mateu-Lucena, Colleoni, and
  Jaume}]{Ramos-Buades:2019uvh}
Ramos-Buades A, Husa S, Pratten G, Estell\'es H, Garc\'\i{}a-Quir\'os C,
  Mateu-Lucena M, Colleoni M, Jaume R (2020{\natexlab{a}}) {First survey of
  spinning eccentric black hole mergers: Numerical relativity simulations,
  hybrid waveforms, and parameter estimation}. Phys Rev D 101(8):083015.
  \doi{10.1103/PhysRevD.101.083015}.
  {\href{https://arxiv.org/abs/1909.11011}{{arXiv:1909.11011}}} {[gr-qc]}

\bibitem[{Ramos-Buades et~al.(2020{\natexlab{b}})Ramos-Buades, Schmidt,
  Pratten, and Husa}]{Ramos-Buades:2020noq}
Ramos-Buades A, Schmidt P, Pratten G, Husa S (2020{\natexlab{b}}) {Validity of
  common modeling approximations for precessing binary black holes with
  higher-order modes}. Phys Rev D 101(10):103014.
  \doi{10.1103/PhysRevD.101.103014}.
  {\href{https://arxiv.org/abs/2001.10936}{{arXiv:2001.10936}}} {[gr-qc]}

\bibitem[{Randall and Xianyu(2018)}]{Randall:2018qna}
Randall L, Xianyu ZZ (2018) {An Analytical Portrait of Binary Mergers in
  Hierarchical Triple Systems}. Astrophys J 864(2):134.
  \doi{10.3847/1538-4357/aad7fe}.
  {\href{https://arxiv.org/abs/1802.05718}{{arXiv:1802.05718}}} {[gr-qc]}

\bibitem[{Randall and Xianyu(2019)}]{Randall:2019sab}
Randall L, Xianyu ZZ (2019) {Observing Eccentricity Oscillations of Binary
  Black Holes in LISA}
  {\href{https://arxiv.org/abs/1902.08604}{{arXiv:1902.08604}}} {[astro-ph.HE]}

\bibitem[{Randall et~al.(2008)Randall, Markevitch, Clowe, Gonzalez, and
  Bradac}]{Randall:2007ph}
Randall SW, Markevitch M, Clowe D, Gonzalez AH, Bradac M (2008) {Constraints on
  the Self-Interaction Cross-Section of Dark Matter from Numerical Simulations
  of the Merging Galaxy Cluster 1E 0657-56}. Astrophys J 679:1173--1180.
  \doi{10.1086/587859}.
  {\href{https://arxiv.org/abs/0704.0261}{{arXiv:0704.0261}}} {[astro-ph]}

\bibitem[{Raposo and Pani(2020)}]{Raposo:2020yjy}
Raposo G, Pani P (2020) {Axisymmetric deformations of neutron stars and
  gravitational-wave astronomy}
  {\href{https://arxiv.org/abs/2002.02555}{{arXiv:2002.02555}}} {[gr-qc]}

\bibitem[{Raposo et~al.(2019{\natexlab{a}})Raposo, Pani, Bezares, Palenzuela,
  and Cardoso}]{Raposo:2018rjn}
Raposo G, Pani P, Bezares M, Palenzuela C, Cardoso V (2019{\natexlab{a}})
  {Anisotropic stars as ultracompact objects in General Relativity}. Phys Rev D
  99(10):104072. \doi{10.1103/PhysRevD.99.104072}.
  {\href{https://arxiv.org/abs/1811.07917}{{arXiv:1811.07917}}} {[gr-qc]}

\bibitem[{Raposo et~al.(2019{\natexlab{b}})Raposo, Pani, and
  Emparan}]{Raposo:2018xkf}
Raposo G, Pani P, Emparan R (2019{\natexlab{b}}) {Exotic compact objects with
  soft hair}. Phys Rev D 99(10):104050. \doi{10.1103/PhysRevD.99.104050}.
  {\href{https://arxiv.org/abs/1812.07615}{{arXiv:1812.07615}}} {[gr-qc]}

\bibitem[{Raveri et~al.(2014)Raveri, Hu, Frusciante, and
  Silvestri}]{Raveri:2014cka}
Raveri M, Hu B, Frusciante N, Silvestri A (2014) {Effective Field Theory of
  Cosmic Acceleration: constraining dark energy with CMB data}. Phys Rev D
  90(4):043513. \doi{10.1103/PhysRevD.90.043513}.
  {\href{https://arxiv.org/abs/1405.1022}{{arXiv:1405.1022}}} {[astro-ph.CO]}

\bibitem[{Raveri et~al.(2015)Raveri, Baccigalupi, Silvestri, and
  Zhou}]{Raveri:2014eea}
Raveri M, Baccigalupi C, Silvestri A, Zhou SY (2015) {Measuring the speed of
  cosmological gravitational waves}. Phys Rev D 91(6):061501.
  \doi{10.1103/PhysRevD.91.061501}.
  {\href{https://arxiv.org/abs/1405.7974}{{arXiv:1405.7974}}} {[astro-ph.CO]}

\bibitem[{Regimbau and de~Freitas~Pacheco(2006)}]{Regimbau:2005ey}
Regimbau T, de~Freitas~Pacheco JA (2006) {Gravitational wave background from
  magnetars}. Astron Astrophys 447:1. \doi{10.1051/0004-6361:20053702}.
  {\href{https://arxiv.org/abs/astro-ph/0509880}{{arXiv:astro-ph/0509880}}}

\bibitem[{Renevey et~al.(2020)Renevey, Kennedy, and
  Lombriser}]{Renevey:2020tvr}
Renevey C, Kennedy J, Lombriser L (2020) {Parameterised post-Newtonian
  formalism for the effective field theory of dark energy via screened
  reconstructed Horndeski theories}
  {\href{https://arxiv.org/abs/2006.09910}{{arXiv:2006.09910}}} {[gr-qc]}

\bibitem[{Renzini and Contaldi(2019)}]{Renzini:2019vmt}
Renzini A, Contaldi C (2019) {Improved limits on a stochastic
  gravitational-wave background and its anisotropies from Advanced LIGO O1 and
  O2 runs}. Phys Rev D 100(6):063527. \doi{10.1103/PhysRevD.100.063527}.
  {\href{https://arxiv.org/abs/1907.10329}{{arXiv:1907.10329}}} {[gr-qc]}

\bibitem[{de~Rham and Melville(2018)}]{deRham:2018red}
de~Rham C, Melville S (2018) {Gravitational Rainbows: LIGO and Dark Energy at
  its Cutoff}. Phys Rev Lett 121(22):221101.
  \doi{10.1103/PhysRevLett.121.221101}.
  {\href{https://arxiv.org/abs/1806.09417}{{arXiv:1806.09417}}} {[hep-th]}

\bibitem[{de~Rham et~al.(2011)de~Rham, Gabadadze, and Tolley}]{deRham:2010kj}
de~Rham C, Gabadadze G, Tolley AJ (2011) {Resummation of Massive Gravity}. Phys
  Rev Lett 106:231101. \doi{10.1103/PhysRevLett.106.231101}.
  {\href{https://arxiv.org/abs/1011.1232}{{arXiv:1011.1232}}} {[hep-th]}

\bibitem[{de~Rham et~al.(2013{\natexlab{a}})de~Rham, Matas, and
  Tolley}]{deRham:2012fg}
de~Rham C, Matas A, Tolley AJ (2013{\natexlab{a}}) {Galileon Radiation from
  Binary Systems}. Phys Rev D 87(6):064024. \doi{10.1103/PhysRevD.87.064024}.
  {\href{https://arxiv.org/abs/1212.5212}{{arXiv:1212.5212}}} {[hep-th]}

\bibitem[{de~Rham et~al.(2013{\natexlab{b}})de~Rham, Tolley, and
  Wesley}]{deRham:2012fw}
de~Rham C, Tolley AJ, Wesley DH (2013{\natexlab{b}}) {Vainshtein Mechanism in
  Binary Pulsars}. Phys Rev D 87(4):044025. \doi{10.1103/PhysRevD.87.044025}.
  {\href{https://arxiv.org/abs/1208.0580}{{arXiv:1208.0580}}} {[gr-qc]}

\bibitem[{Riess et~al.(2019)Riess, Casertano, Yuan, Macri, and
  Scolnic}]{Riess:2019cxk}
Riess AG, Casertano S, Yuan W, Macri LM, Scolnic D (2019) {Large Magellanic
  Cloud Cepheid Standards Provide a 1
  Hubble Constant and Stronger Evidence for Physics beyond $\Lambda$CDM}.
  Astrophys J 876(1):85. \doi{10.3847/1538-4357/ab1422}.
  {\href{https://arxiv.org/abs/1903.07603}{{arXiv:1903.07603}}} {[astro-ph.CO]}

\bibitem[{Riess et~al.(2016)}]{Riess:2016jrr}
Riess AG, et~al. (2016) {A 2.4\% Determination of the Local Value of the Hubble
  Constant}. Astrophys J 826(1):56. \doi{10.3847/0004-637X/826/1/56}.
  {\href{https://arxiv.org/abs/1604.01424}{{arXiv:1604.01424}}} {[astro-ph.CO]}

\bibitem[{Riess et~al.(2018)}]{Riess:2018byc}
Riess AG, et~al. (2018) {Milky Way Cepheid Standards for Measuring Cosmic
  Distances and Application to Gaia DR2: Implications for the Hubble Constant}.
  Astrophys J 861(2):126. \doi{10.3847/1538-4357/aac82e}.
  {\href{https://arxiv.org/abs/1804.10655}{{arXiv:1804.10655}}} {[astro-ph.CO]}

\bibitem[{Ringeval et~al.(2007)Ringeval, Sakellariadou, and
  Bouchet}]{Ringeval:2005kr}
Ringeval C, Sakellariadou M, Bouchet F (2007) {Cosmological evolution of cosmic
  string loops}. JCAP 02:023. \doi{10.1088/1475-7516/2007/02/023}.
  {\href{https://arxiv.org/abs/astro-ph/0511646}{{arXiv:astro-ph/0511646}}}

\bibitem[{Ripley and Pretorius(2019{\natexlab{a}})}]{Ripley:2019irj}
Ripley JL, Pretorius F (2019{\natexlab{a}}) {Gravitational collapse in Einstein
  dilaton-Gauss–Bonnet gravity}. Class Quant Grav 36(13):134001.
  \doi{10.1088/1361-6382/ab2416}.
  {\href{https://arxiv.org/abs/1903.07543}{{arXiv:1903.07543}}} {[gr-qc]}

\bibitem[{Ripley and Pretorius(2019{\natexlab{b}})}]{Ripley:2019hxt}
Ripley JL, Pretorius F (2019{\natexlab{b}}) {Hyperbolicity in Spherical
  Gravitational Collapse in a Horndeski Theory}. Phys Rev D 99(8):084014.
  \doi{10.1103/PhysRevD.99.084014}.
  {\href{https://arxiv.org/abs/1902.01468}{{arXiv:1902.01468}}} {[gr-qc]}

\bibitem[{Robinson(1975)}]{Robinson:1975bv}
Robinson D (1975) {Uniqueness of the Kerr black hole}. Phys Rev Lett
  34:905--906. \doi{10.1103/PhysRevLett.34.905}

\bibitem[{{Roedig} et~al.(2011){Roedig}, {Dotti}, {Sesana}, {Cuadra}, and
  {Colpi}}]{2011MNRAS.415.3033R}
{Roedig} C, {Dotti} M, {Sesana} A, {Cuadra} J, {Colpi} M (2011) {Limiting
  eccentricity of subparsec massive black hole binaries surrounded by
  self-gravitating gas discs}. \mnras 415(4):3033--3041.
  \doi{10.1111/j.1365-2966.2011.18927.x}.
  {\href{https://arxiv.org/abs/1104.3868}{{arXiv:1104.3868}}} {[astro-ph.CO]}

\bibitem[{Romano and Cornish(2017)}]{Romano:2016dpx}
Romano JD, Cornish NJ (2017) {Detection methods for stochastic
  gravitational-wave backgrounds: a unified treatment}. Living Rev Rel 20(1):2.
  \doi{10.1007/s41114-017-0004-1}.
  {\href{https://arxiv.org/abs/1608.06889}{{arXiv:1608.06889}}} {[gr-qc]}

\bibitem[{Rosen(2017)}]{Rosen:2017dvn}
Rosen RA (2017) {Non-Singular Black Holes in Massive Gravity: Time-Dependent
  Solutions}. JHEP 10:206. \doi{10.1007/JHEP10(2017)206}.
  {\href{https://arxiv.org/abs/1702.06543}{{arXiv:1702.06543}}} {[hep-th]}

\bibitem[{Ruffini and Bonazzola(1969)}]{Ruffini:1969qy}
Ruffini R, Bonazzola S (1969) {Systems of selfgravitating particles in general
  relativity and the concept of an equation of state}. Phys Rev 187:1767--1783.
  \doi{10.1103/PhysRev.187.1767}

\bibitem[{Ruffini and Wheeler(1971)}]{Ruffini:1971bza}
Ruffini R, Wheeler JA (1971) {Introducing the black hole}. Phys Today 24(1):30.
  \doi{10.1063/1.3022513}

\bibitem[{Ryan(1995)}]{Ryan:1995wh}
Ryan F (1995) {Gravitational waves from the inspiral of a compact object into a
  massive, axisymmetric body with arbitrary multipole moments}. Phys Rev D
  52:5707--5718. \doi{10.1103/PhysRevD.52.5707}

\bibitem[{Ryan(1997{\natexlab{a}})}]{Ryan:1997hg}
Ryan FD (1997{\natexlab{a}}) {Accuracy of estimating the multipole moments of a
  massive body from the gravitational waves of a binary inspiral}. Phys Rev D
  56:1845--1855. \doi{10.1103/PhysRevD.56.1845}

\bibitem[{Ryan(1997{\natexlab{b}})}]{Ryan:1996nk}
Ryan FD (1997{\natexlab{b}}) {Spinning boson stars with large selfinteraction}.
  Phys Rev D 55:6081--6091. \doi{10.1103/PhysRevD.55.6081}

\bibitem[{Sachs(1962)}]{BMS2}
Sachs RK (1962) {Gravitational Waves in General Relativity. VIII. Waves in
  asymptotically flat space-time}. Proc R Soc Lond A 270:103.
  \doi{10.1098/rspa.1962.0206}

\bibitem[{Sadeghian et~al.(2013)Sadeghian, Ferrer, and
  Will}]{Sadeghian:2013laa}
Sadeghian L, Ferrer F, Will CM (2013) {Dark matter distributions around massive
  black holes: A general relativistic analysis}. Phys Rev D 88(6):063522.
  \doi{10.1103/PhysRevD.88.063522}.
  {\href{https://arxiv.org/abs/1305.2619}{{arXiv:1305.2619}}} {[astro-ph.GA]}

\bibitem[{Saffer and Yagi(2020)}]{Saffer:2020xsw}
Saffer A, Yagi K (2020) {Parameter Estimation for Tests of General Relativity
  with the Astrophysical Stochastic Gravitational Wave Background}. Phys Rev
  D102(2):024001. \doi{10.1103/PhysRevD.102.024001}.
  {\href{https://arxiv.org/abs/2003.11128}{{arXiv:2003.11128}}} {[gr-qc]}

\bibitem[{Sahoo and Sen(2019)}]{BMSAdd1}
Sahoo B, Sen A (2019) {Classical and Quantum Results on Logarithmic Terms in
  the Soft Theorem in Four Dimensions}. JHEP 02:086.
  \doi{10.1007/JHEP02(2019)086}.
  {\href{https://arxiv.org/abs/1808.03288}{{arXiv:1808.03288}}} {[hep-th]}

\bibitem[{Saito and Yokoyama(2010)}]{Saito:2009jt}
Saito R, Yokoyama J (2010) {Gravitational-Wave Constraints on the Abundance of
  Primordial Black Holes}. Prog Theor Phys 123:867--886.
  \doi{10.1143/PTP.126.351}, [Erratum: Prog.Theor.Phys. 126, 351--352 (2011)].
  {\href{https://arxiv.org/abs/0912.5317}{{arXiv:0912.5317}}} {[astro-ph.CO]}

\bibitem[{Sakstein and Jain(2017)}]{Sakstein:2017xjx}
Sakstein J, Jain B (2017) {Implications of the Neutron Star Merger GW170817 for
  Cosmological Scalar-Tensor Theories}. Phys Rev Lett 119:251303.
  \doi{10.1103/PhysRevLett.119.251303}.
  {\href{https://arxiv.org/abs/1710.05893}{{arXiv:1710.05893}}} {[astro-ph.CO]}

\bibitem[{Saltas et~al.(2014)Saltas, Sawicki, Amendola, and
  Kunz}]{Saltas:2014dha}
Saltas ID, Sawicki I, Amendola L, Kunz M (2014) {Anisotropic Stress as a
  Signature of Nonstandard Propagation of Gravitational Waves}. Phys Rev Lett
  113:191101. \doi{10.1103/PhysRevLett.113.191101}.
  {\href{https://arxiv.org/abs/1406.7139}{{arXiv:1406.7139}}} {[astro-ph.CO]}

\bibitem[{Sampson et~al.(2013)Sampson, Cornish, and Yunes}]{Sampson:2013lpa}
Sampson L, Cornish N, Yunes N (2013) {Gravitational Wave Tests of Strong Field
  General Relativity with Binary Inspirals: Realistic Injections and Optimal
  Model Selection}. Phys Rev D87(10):102001. \doi{10.1103/PhysRevD.87.102001}.
  {\href{https://arxiv.org/abs/1303.1185}{{arXiv:1303.1185}}} {[gr-qc]}

\bibitem[{Sampson et~al.(2014)Sampson, Cornish, and Yunes}]{Sampson:2013jpa}
Sampson L, Cornish N, Yunes N (2014) {Mismodeling in gravitational-wave
  astronomy: The trouble with templates}. Phys Rev D89(6):064037.
  \doi{10.1103/PhysRevD.89.064037}.
  {\href{https://arxiv.org/abs/1311.4898}{{arXiv:1311.4898}}} {[gr-qc]}

\bibitem[{Sanchis-Gual et~al.(2019)Sanchis-Gual, Herdeiro, Font, Radu, and
  Di~Giovanni}]{Sanchis-Gual:2018oui}
Sanchis-Gual N, Herdeiro C, Font JA, Radu E, Di~Giovanni F (2019) {Head-on
  collisions and orbital mergers of Proca stars}. Phys Rev D 99(2):024017.
  \doi{10.1103/PhysRevD.99.024017}.
  {\href{https://arxiv.org/abs/1806.07779}{{arXiv:1806.07779}}} {[gr-qc]}

\bibitem[{Sanchis-Gual et~al.(2020)Sanchis-Gual, Zilh\~ao, Herdeiro,
  Di~Giovanni, Font, and Radu}]{Sanchis-Gual:2020mzb}
Sanchis-Gual N, Zilh\~ao M, Herdeiro C, Di~Giovanni F, Font JA, Radu E (2020)
  {Synchronized gravitational atoms from mergers of bosonic stars}. Phys Rev D
  102(10):101504. \doi{10.1103/PhysRevD.102.101504}.
  {\href{https://arxiv.org/abs/2007.11584}{{arXiv:2007.11584}}} {[gr-qc]}

\bibitem[{Sanders and McGaugh(2002)}]{Sanders:2002pf}
Sanders RH, McGaugh SS (2002) {Modified Newtonian dynamics as an alternative to
  dark matter}. Ann Rev Astron Astrophys 40:263--317.
  \doi{10.1146/annurev.astro.40.060401.093923}.
  {\href{https://arxiv.org/abs/astro-ph/0204521}{{arXiv:astro-ph/0204521}}}

\bibitem[{Santamaria et~al.(2010)}]{Santamaria:2010yb}
Santamaria L, et~al. (2010) {Matching post-Newtonian and numerical relativity
  waveforms: systematic errors and a new phenomenological model for
  non-precessing black hole binaries}. Phys Rev D 82:064016.
  \doi{10.1103/PhysRevD.82.064016}.
  {\href{https://arxiv.org/abs/1005.3306}{{arXiv:1005.3306}}} {[gr-qc]}

\bibitem[{Santos et~al.(2020)Santos, Benone, Crispino, Herdeiro, and
  Radu}]{Santos:2020pmh}
Santos NM, Benone CL, Crispino LC, Herdeiro CA, Radu E (2020) {Black holes with
  synchronised Proca hair: linear clouds and fundamental non-linear solutions}.
  JHEP 07:010. \doi{10.1007/JHEP07(2020)010}.
  {\href{https://arxiv.org/abs/2004.09536}{{arXiv:2004.09536}}} {[gr-qc]}

\bibitem[{Saravani and Sotiriou(2019)}]{Saravani:2019xwx}
Saravani M, Sotiriou TP (2019) {Classification of shift-symmetric Horndeski
  theories and hairy black holes}. Phys Rev D 99(12):124004.
  \doi{10.1103/PhysRevD.99.124004}.
  {\href{https://arxiv.org/abs/1903.02055}{{arXiv:1903.02055}}} {[gr-qc]}

\bibitem[{Sarbach et~al.(2019)Sarbach, Barausse, and
  Preciado-L\'opez}]{Sarbach:2019yso}
Sarbach O, Barausse E, Preciado-L\'opez JA (2019) {Well-posed Cauchy
  formulation for Einstein-\ae{}ther theory}. Class Quant Grav 36(16):165007.
  \doi{10.1088/1361-6382/ab2e13}.
  {\href{https://arxiv.org/abs/1902.05130}{{arXiv:1902.05130}}} {[gr-qc]}

\bibitem[{Sasaki and Tagoshi(2003)}]{Sasaki:2003xr}
Sasaki M, Tagoshi H (2003) {Analytic black hole perturbation approach to
  gravitational radiation}. Living Rev Rel 6:6. \doi{10.12942/lrr-2003-6}.
  {\href{https://arxiv.org/abs/gr-qc/0306120}{{arXiv:gr-qc/0306120}}}

\bibitem[{Sasaki et~al.(2018)Sasaki, Suyama, Tanaka, and
  Yokoyama}]{Sasaki:2018dmp}
Sasaki M, Suyama T, Tanaka T, Yokoyama S (2018) {Primordial black
  holes-perspectives in gravitational wave astronomy}. Class Quant Grav
  35(6):063001. \doi{10.1088/1361-6382/aaa7b4}.
  {\href{https://arxiv.org/abs/1801.05235}{{arXiv:1801.05235}}} {[astro-ph.CO]}

\bibitem[{Scelfo et~al.(2018)Scelfo, Bellomo, Raccanelli, Matarrese, and
  Verde}]{Scelfo:2018sny}
Scelfo G, Bellomo N, Raccanelli A, Matarrese S, Verde L (2018)
  {GW\textbackslash{}times\$LSS: chasing the progenitors of merging binary
  black holes}. JCAP 09:039. \doi{10.1088/1475-7516/2018/09/039}.
  {\href{https://arxiv.org/abs/1809.03528}{{arXiv:1809.03528}}} {[astro-ph.CO]}

\bibitem[{{Scelfo} et~al.(2020){Scelfo}, {Boco}, {Lapi}, and
  {Viel}}]{Scelfo2020}
{Scelfo} G, {Boco} L, {Lapi} A, {Viel} M (2020) {Exploring
  galaxies-gravitational waves cross-correlations as an astrophysical probe}.
  arXiv e-prints arXiv:2007.08534.
  {\href{https://arxiv.org/abs/2007.08534}{{arXiv:2007.08534}}} {[astro-ph.CO]}

\bibitem[{Schmidt and Hinderer(2019)}]{Schmidt:2019wrl}
Schmidt P, Hinderer T (2019) {Frequency domain model of $f$-mode dynamic tides
  in gravitational waveforms from compact binary inspirals}. Phys Rev D
  100(2):021501. \doi{10.1103/PhysRevD.100.021501}.
  {\href{https://arxiv.org/abs/1905.00818}{{arXiv:1905.00818}}} {[gr-qc]}

\bibitem[{Schmidt et~al.(2012)Schmidt, Hannam, and Husa}]{Schmidt:2012rh}
Schmidt P, Hannam M, Husa S (2012) {Towards models of gravitational waveforms
  from generic binaries: A simple approximate mapping between precessing and
  non-precessing inspiral signals}. Phys Rev D 86:104063.
  \doi{10.1103/PhysRevD.86.104063}.
  {\href{https://arxiv.org/abs/1207.3088}{{arXiv:1207.3088}}} {[gr-qc]}

\bibitem[{Schmidt et~al.(2015)Schmidt, Ohme, and Hannam}]{Schmidt:2014iyl}
Schmidt P, Ohme F, Hannam M (2015) {Towards models of gravitational waveforms
  from generic binaries II: Modelling precession effects with a single
  effective precession parameter}. Phys Rev D 91(2):024043.
  \doi{10.1103/PhysRevD.91.024043}.
  {\href{https://arxiv.org/abs/1408.1810}{{arXiv:1408.1810}}} {[gr-qc]}

\bibitem[{Schneider et~al.(1992)Schneider, Ehlers, and Falco}]{Schneider:1992}
Schneider P, Ehlers J, Falco E (1992) { Gravitational Lenses}. Springer-Verlag
  Berlin Heidelberg. \doi{10.1007/978-3-662-03758-4}

\bibitem[{Schneider et~al.(2010)Schneider, Marassi, and
  Ferrari}]{Schneider:2010ks}
Schneider R, Marassi S, Ferrari V (2010) {Stochastic backgrounds of
  gravitational waves from extragalactic sources}. Class Quant Grav 27:194007.
  \doi{10.1088/0264-9381/27/19/194007}.
  {\href{https://arxiv.org/abs/1005.0977}{{arXiv:1005.0977}}} {[astro-ph.CO]}

\bibitem[{Schunck and Mielke(2003)}]{Schunck:2003kk}
Schunck FE, Mielke EW (2003) {General relativistic boson stars}. Class Quant
  Grav 20:R301--R356. \doi{10.1088/0264-9381/20/20/201}.
  {\href{https://arxiv.org/abs/0801.0307}{{arXiv:0801.0307}}} {[astro-ph]}

\bibitem[{Schutz(1986)}]{Schutz:1986gp}
Schutz BF (1986) {Determining the Hubble Constant from Gravitational Wave
  Observations}. Nature 323:310--311. \doi{10.1038/323310a0}

\bibitem[{Seidel and Suen(1994)}]{Seidel:1993zk}
Seidel E, Suen WM (1994) {Formation of solitonic stars through gravitational
  cooling}. Phys Rev Lett 72:2516--2519. \doi{10.1103/PhysRevLett.72.2516}.
  {\href{https://arxiv.org/abs/gr-qc/9309015}{{arXiv:gr-qc/9309015}}}

\bibitem[{Sennett et~al.(2016)Sennett, Marsat, and Buonanno}]{Sennett:2016klh}
Sennett N, Marsat S, Buonanno A (2016) {Gravitational waveforms in
  scalar-tensor gravity at 2PN relative order}. Phys Rev D 94(8):084003.
  \doi{10.1103/PhysRevD.94.084003}.
  {\href{https://arxiv.org/abs/1607.01420}{{arXiv:1607.01420}}} {[gr-qc]}

\bibitem[{Sennett et~al.(2017)Sennett, Hinderer, Steinhoff, Buonanno, and
  Ossokine}]{Sennett:2017etc}
Sennett N, Hinderer T, Steinhoff J, Buonanno A, Ossokine S (2017)
  {Distinguishing Boson Stars from Black Holes and Neutron Stars from Tidal
  Interactions in Inspiraling Binary Systems}. Phys Rev D 96(2):024002.
  \doi{10.1103/PhysRevD.96.024002}.
  {\href{https://arxiv.org/abs/1704.08651}{{arXiv:1704.08651}}} {[gr-qc]}

\bibitem[{Sennett et~al.(2020)Sennett, Brito, Buonanno, Gorbenko, and
  Senatore}]{Sennett:2019bpc}
Sennett N, Brito R, Buonanno A, Gorbenko V, Senatore L (2020)
  {Gravitational-Wave Constraints on an Effective Field-Theory Extension of
  General Relativity}. Phys Rev D 102(4):044056.
  \doi{10.1103/PhysRevD.102.044056}.
  {\href{https://arxiv.org/abs/1912.09917}{{arXiv:1912.09917}}} {[gr-qc]}

\bibitem[{Sereno et~al.(2010)Sereno, Sesana, Bleuler, Jetzer, Volonteri, and
  Begelman}]{Sereno:2010dr}
Sereno M, Sesana A, Bleuler A, Jetzer P, Volonteri M, Begelman M (2010) {Strong
  lensing of gravitational waves as seen by LISA}. Phys Rev Lett 105:251101.
  \doi{10.1103/PhysRevLett.105.251101}.
  {\href{https://arxiv.org/abs/1011.5238}{{arXiv:1011.5238}}} {[astro-ph.CO]}

\bibitem[{Sereno et~al.(2011)Sereno, Jetzer, Sesana, and
  Volonteri}]{Sereno:2011ty}
Sereno M, Jetzer P, Sesana A, Volonteri M (2011) {Cosmography with strong
  lensing of LISA gravitational wave sources}. Mon Not Roy Astron Soc 415:2773.
  \doi{10.1111/j.1365-2966.2011.18895.x}.
  {\href{https://arxiv.org/abs/1104.1977}{{arXiv:1104.1977}}} {[astro-ph.CO]}

\bibitem[{Serpico et~al.(2020)Serpico, Poulin, Inman, and
  Kohri}]{Serpico:2020ehh}
Serpico PD, Poulin V, Inman D, Kohri K (2020) {Cosmic microwave background
  bounds on primordial black holes including dark matter halo accretion}. Phys
  Rev Res 2(2):023204. \doi{10.1103/PhysRevResearch.2.023204}.
  {\href{https://arxiv.org/abs/2002.10771}{{arXiv:2002.10771}}} {[astro-ph.CO]}

\bibitem[{Sesana(2016)}]{Sesana:2016ljz}
Sesana A (2016) {Prospects for Multiband Gravitational-Wave Astronomy after
  GW150914}. Phys Rev Lett 116(23):231102.
  \doi{10.1103/PhysRevLett.116.231102}.
  {\href{https://arxiv.org/abs/1602.06951}{{arXiv:1602.06951}}} {[gr-qc]}

\bibitem[{Sesana et~al.(2007)Sesana, Volonteri, and Haardt}]{Sesana:2007sh}
Sesana A, Volonteri M, Haardt F (2007) {The imprint of massive black hole
  formation models on the LISA data stream}. Mon Not Roy Astron Soc
  377:1711--1716. \doi{10.1111/j.1365-2966.2007.11734.x}.
  {\href{https://arxiv.org/abs/astro-ph/0701556}{{arXiv:astro-ph/0701556}}}

\bibitem[{Seto(2006)}]{Seto:2006hf}
Seto N (2006) {Prospects for direct detection of circular polarization of
  gravitational-wave background}. Phys Rev Lett 97:151101.
  \doi{10.1103/PhysRevLett.97.151101}.
  {\href{https://arxiv.org/abs/astro-ph/0609504}{{arXiv:astro-ph/0609504}}}

\bibitem[{Seto(2007)}]{Seto:2006dz}
Seto N (2007) {Quest for circular polarization of gravitational wave background
  and orbits of laser interferometers in space}. Phys Rev D 75:061302.
  \doi{10.1103/PhysRevD.75.061302}.
  {\href{https://arxiv.org/abs/astro-ph/0609633}{{arXiv:astro-ph/0609633}}}

\bibitem[{Shah et~al.(2021)Shah, Lemos, and Lahav}]{Shah:2021onj}
Shah P, Lemos P, Lahav O (2021) {A buyer's guide to the Hubble Constant}
  {\href{https://arxiv.org/abs/2109.01161}{{arXiv:2109.01161}}} {[astro-ph.CO]}

\bibitem[{{Shakura} and {Sunyaev}(1973)}]{1973A&A....24..337S}
{Shakura} NI, {Sunyaev} RA (1973) {Reprint of 1973A\&A....24..337S. Black holes
  in binary systems. Observational appearance.} \aap 500:33--51

\bibitem[{Shannon et~al.(2015)}]{Shannon:2015ect}
Shannon R, et~al. (2015) {Gravitational waves from binary supermassive black
  holes missing in pulsar observations}. Science 349(6255):1522--1525.
  \doi{10.1126/science.aab1910}.
  {\href{https://arxiv.org/abs/1509.07320}{{arXiv:1509.07320}}} {[astro-ph.CO]}

\bibitem[{Shao(2020)}]{Shao:2020shv}
Shao L (2020) {Combined search for anisotropic birefringence in the
  gravitational-wave transient catalog GWTC-1}. Phys Rev D 101:104019.
  \doi{10.1103/PhysRevD.101.104019}.
  {\href{https://arxiv.org/abs/2002.01185}{{arXiv:2002.01185}}} {[hep-ph]}

\bibitem[{Shao et~al.(2017)Shao, Sennett, Buonanno, Kramer, and
  Wex}]{Shao:2017gwu}
Shao L, Sennett N, Buonanno A, Kramer M, Wex N (2017) {Constraining
  nonperturbative strong-field effects in scalar-tensor gravity by combining
  pulsar timing and laser-interferometer gravitational-wave detectors}. Phys
  Rev X 7(4):041025. \doi{10.1103/PhysRevX.7.041025}.
  {\href{https://arxiv.org/abs/1704.07561}{{arXiv:1704.07561}}} {[gr-qc]}

\bibitem[{Shapiro and Shelton(2016)}]{Shapiro_2016}
Shapiro SL, Shelton J (2016) Weak annihilation cusp inside the dark matter
  spike about a black hole. Physical Review D 93(12).
  \doi{10.1103/physrevd.93.123510}

\bibitem[{Shibata(2015)}]{ShibataBook2015}
Shibata M (2015) Numerical Relativity. World Scientific, Singapore

\bibitem[{Shiralilou et~al.(2021{\natexlab{a}})Shiralilou, Hinderer, Nissanke,
  Ortiz, and Witek}]{Shiralilou:2020gah}
Shiralilou B, Hinderer T, Nissanke S, Ortiz N, Witek H (2021{\natexlab{a}})
  {Nonlinear curvature effects in gravitational waves from inspiralling black
  hole binaries}. Phys Rev D 103(12):L121503.
  \doi{10.1103/PhysRevD.103.L121503}.
  {\href{https://arxiv.org/abs/2012.09162}{{arXiv:2012.09162}}} {[gr-qc]}

\bibitem[{Shiralilou et~al.(2021{\natexlab{b}})Shiralilou, Hinderer, Nissanke,
  Ortiz, and Witek}]{Shiralilou:2021mfl}
Shiralilou B, Hinderer T, Nissanke S, Ortiz N, Witek H (2021{\natexlab{b}})
  {Post-Newtonian Gravitational and Scalar Waves in Scalar-Gauss-Bonnet
  Gravity} {\href{https://arxiv.org/abs/2105.13972}{{arXiv:2105.13972}}}
  {[gr-qc]}

\bibitem[{Shlapentokh-Rothman(2014)}]{Shlapentokh-Rothman:2013ysa}
Shlapentokh-Rothman Y (2014) {Exponentially growing finite energy solutions for
  the Klein-Gordon equation on sub-extremal Kerr spacetimes}. Commun Math Phys
  329:859--891. \doi{10.1007/s00220-014-2033-x}.
  {\href{https://arxiv.org/abs/1302.3448}{{arXiv:1302.3448}}} {[gr-qc]}

\bibitem[{Siemonsen and East(2020)}]{Siemonsen:2019ebd}
Siemonsen N, East WE (2020) {Gravitational wave signatures of ultralight vector
  bosons from black hole superradiance}. Phys Rev D 101(2):024019.
  \doi{10.1103/PhysRevD.101.024019}.
  {\href{https://arxiv.org/abs/1910.09476}{{arXiv:1910.09476}}} {[gr-qc]}

\bibitem[{Silva and Glampedakis(2020)}]{Silva:2019scu}
Silva HO, Glampedakis K (2020) {Eikonal quasinormal modes of black holes beyond
  general relativity. II. Generalized scalar-tensor perturbations}. Phys Rev
  D101(4):044051. \doi{10.1103/PhysRevD.101.044051}.
  {\href{https://arxiv.org/abs/1912.09286}{{arXiv:1912.09286}}} {[gr-qc]}

\bibitem[{Silva et~al.(2018)Silva, Sakstein, Gualtieri, Sotiriou, and
  Berti}]{Silva:2017uqg}
Silva HO, Sakstein J, Gualtieri L, Sotiriou TP, Berti E (2018) {Spontaneous
  scalarization of black holes and compact stars from a Gauss-Bonnet coupling}.
  Phys Rev Lett 120(13):131104. \doi{10.1103/PhysRevLett.120.131104}.
  {\href{https://arxiv.org/abs/1711.02080}{{arXiv:1711.02080}}} {[gr-qc]}

\bibitem[{Silva et~al.(2019)Silva, Macedo, Sotiriou, Gualtieri, Sakstein, and
  Berti}]{Silva:2018qhn}
Silva HO, Macedo CF, Sotiriou TP, Gualtieri L, Sakstein J, Berti E (2019)
  {Stability of scalarized black hole solutions in scalar-Gauss-Bonnet
  gravity}. Phys Rev D 99(6):064011. \doi{10.1103/PhysRevD.99.064011}.
  {\href{https://arxiv.org/abs/1812.05590}{{arXiv:1812.05590}}} {[gr-qc]}

\bibitem[{Silva et~al.(2021{\natexlab{a}})Silva, Holgado,
  C\'ardenas-Avenda\~no, and Yunes}]{Silva:2020acr}
Silva HO, Holgado AM, C\'ardenas-Avenda\~no A, Yunes N (2021{\natexlab{a}})
  {Astrophysical and theoretical physics implications from multimessenger
  neutron star observations}. Phys Rev Lett 126(18):181101.
  \doi{10.1103/PhysRevLett.126.181101}.
  {\href{https://arxiv.org/abs/2004.01253}{{arXiv:2004.01253}}} {[gr-qc]}

\bibitem[{Silva et~al.(2021{\natexlab{b}})Silva, Witek, Elley, and
  Yunes}]{Silva:2020omi}
Silva HO, Witek H, Elley M, Yunes N (2021{\natexlab{b}}) {Dynamical
  Descalarization in Binary Black Hole Mergers}. Phys Rev Lett 127(3):031101.
  \doi{10.1103/PhysRevLett.127.031101}.
  {\href{https://arxiv.org/abs/2012.10436}{{arXiv:2012.10436}}} {[gr-qc]}

\bibitem[{Smith and Caldwell(2017)}]{Smith:2016jqs}
Smith TL, Caldwell R (2017) {Sensitivity to a Frequency-Dependent Circular
  Polarization in an Isotropic Stochastic Gravitational Wave Background}. Phys
  Rev D 95(4):044036. \doi{10.1103/PhysRevD.95.044036}.
  {\href{https://arxiv.org/abs/1609.05901}{{arXiv:1609.05901}}} {[gr-qc]}

\bibitem[{Smyth et~al.(2020)Smyth, Profumo, English, Jeltema, McKinnon, and
  Guhathakurta}]{Smyth:2019whb}
Smyth N, Profumo S, English S, Jeltema T, McKinnon K, Guhathakurta P (2020)
  {Updated Constraints on Asteroid-Mass Primordial Black Holes as Dark Matter}.
  Phys Rev D 101(6):063005. \doi{10.1103/PhysRevD.101.063005}.
  {\href{https://arxiv.org/abs/1910.01285}{{arXiv:1910.01285}}} {[astro-ph.CO]}

\bibitem[{Soares-Santos et~al.(2019)}]{Soares-Santos:2019irc}
Soares-Santos M, et~al. (2019) {First Measurement of the Hubble Constant from a
  Dark Standard Siren using the Dark Energy Survey Galaxies and the LIGO/Virgo
  Binary--Black-hole Merger GW170814}. Astrophys J Lett 876:L7.
  \doi{10.3847/2041-8213/ab14f1}.
  {\href{https://arxiv.org/abs/1901.01540}{{arXiv:1901.01540}}} {[astro-ph.CO]}

\bibitem[{Sotiriou(2011)}]{Sotiriou:2010wn}
Sotiriou TP (2011) {Horava-Lifshitz gravity: a status report}. J Phys Conf Ser
  283:012034. \doi{10.1088/1742-6596/283/1/012034}.
  {\href{https://arxiv.org/abs/1010.3218}{{arXiv:1010.3218}}} {[hep-th]}

\bibitem[{Sotiriou(2015{\natexlab{a}})}]{Sotiriou:2015pka}
Sotiriou TP (2015{\natexlab{a}}) {Black Holes and Scalar Fields}. Class Quant
  Grav 32(21):214002. \doi{10.1088/0264-9381/32/21/214002}.
  {\href{https://arxiv.org/abs/1505.00248}{{arXiv:1505.00248}}} {[gr-qc]}

\bibitem[{Sotiriou(2015{\natexlab{b}})}]{Sotiriou:2015lxa}
Sotiriou TP (2015{\natexlab{b}}) {Gravity and Scalar Fields}. Lect Notes Phys
  892:3--24. \doi{10.1007/978-3-319-10070-8_1}.
  {\href{https://arxiv.org/abs/1404.2955}{{arXiv:1404.2955}}} {[gr-qc]}

\bibitem[{Sotiriou(2018)}]{Sotiriou:2017obf}
Sotiriou TP (2018) {Detecting Lorentz Violations with Gravitational Waves from
  Black Hole Binaries}. Phys Rev Lett 120(4):041104.
  \doi{10.1103/PhysRevLett.120.041104}.
  {\href{https://arxiv.org/abs/1709.00940}{{arXiv:1709.00940}}} {[gr-qc]}

\bibitem[{Sotiriou and Faraoni(2010)}]{Sotiriou:2008rp}
Sotiriou TP, Faraoni V (2010) {f(R) Theories Of Gravity}. Rev Mod Phys
  82:451--497. \doi{10.1103/RevModPhys.82.451}.
  {\href{https://arxiv.org/abs/0805.1726}{{arXiv:0805.1726}}} {[gr-qc]}

\bibitem[{Sotiriou and Faraoni(2012)}]{Sotiriou:2011dz}
Sotiriou TP, Faraoni V (2012) {Black holes in scalar-tensor gravity}. Phys Rev
  Lett 108:081103. \doi{10.1103/PhysRevLett.108.081103}.
  {\href{https://arxiv.org/abs/1109.6324}{{arXiv:1109.6324}}} {[gr-qc]}

\bibitem[{Sotiriou and Zhou(2014{\natexlab{a}})}]{Sotiriou:2013qea}
Sotiriou TP, Zhou SY (2014{\natexlab{a}}) {Black hole hair in generalized
  scalar-tensor gravity}. Phys Rev Lett 112:251102.
  \doi{10.1103/PhysRevLett.112.251102}.
  {\href{https://arxiv.org/abs/1312.3622}{{arXiv:1312.3622}}} {[gr-qc]}

\bibitem[{Sotiriou and Zhou(2014{\natexlab{b}})}]{Sotiriou:2014pfa}
Sotiriou TP, Zhou SY (2014{\natexlab{b}}) {Black hole hair in generalized
  scalar-tensor gravity: An explicit example}. Phys Rev D 90:124063.
  \doi{10.1103/PhysRevD.90.124063}.
  {\href{https://arxiv.org/abs/1408.1698}{{arXiv:1408.1698}}} {[gr-qc]}

\bibitem[{Spallicci et~al.(2021)Spallicci, Helay\"el-Neto, L\'opez-Corredoira,
  and Capozziello}]{Spallicci:2020diu}
Spallicci ADAM, Helay\"el-Neto JA, L\'opez-Corredoira M, Capozziello S (2021)
  {Cosmology and the massive photon frequency shift in the Standard-Model
  Extension}. Eur Phys J C 81(1):4. \doi{10.1140/epjc/s10052-020-08703-3}.
  {\href{https://arxiv.org/abs/2011.12608}{{arXiv:2011.12608}}} {[astro-ph.CO]}

\bibitem[{Spergel and Steinhardt(2000)}]{Spergel:1999mh}
Spergel DN, Steinhardt PJ (2000) {Observational evidence for selfinteracting
  cold dark matter}. Phys Rev Lett 84:3760--3763.
  \doi{10.1103/PhysRevLett.84.3760}.
  {\href{https://arxiv.org/abs/astro-ph/9909386}{{arXiv:astro-ph/9909386}}}

\bibitem[{Stairs(2003)}]{Stairs:2003eg}
Stairs IH (2003) {Testing general relativity with pulsar timing}. Living Rev
  Rel 6:5. \doi{10.12942/lrr-2003-5}.
  {\href{https://arxiv.org/abs/astro-ph/0307536}{{arXiv:astro-ph/0307536}}}
  {[astro-ph]}

\bibitem[{Stavridis and Will(2009)}]{Stavridis:2009mb}
Stavridis A, Will CM (2009) {Bounding the mass of the graviton with
  gravitational waves: Effect of spin precessions in massive black hole
  binaries}. Phys Rev D80:044002. \doi{10.1103/PhysRevD.80.044002}.
  {\href{https://arxiv.org/abs/0906.3602}{{arXiv:0906.3602}}} {[gr-qc]}

\bibitem[{Stein(2014)}]{Stein:2014xba}
Stein LC (2014) {Rapidly rotating black holes in dynamical Chern-Simons
  gravity: Decoupling limit solutions and breakdown}. Phys Rev D 90(4):044061.
  \doi{10.1103/PhysRevD.90.044061}.
  {\href{https://arxiv.org/abs/1407.2350}{{arXiv:1407.2350}}} {[gr-qc]}

\bibitem[{Steinhoff et~al.(2016)Steinhoff, Hinderer, Buonanno, and
  Taracchini}]{Steinhoff:2016rfi}
Steinhoff J, Hinderer T, Buonanno A, Taracchini A (2016) {Dynamical Tides in
  General Relativity: Effective Action and Effective-One-Body Hamiltonian}.
  Phys Rev D 94(10):104028. \doi{10.1103/PhysRevD.94.104028}.
  {\href{https://arxiv.org/abs/1608.01907}{{arXiv:1608.01907}}} {[gr-qc]}

\bibitem[{Strominger(2018)}]{BMS5}
Strominger A (2018) {Lectures on the infrared structure of gravity and gauge
  theory}. Princeton University Press.
  {\href{https://arxiv.org/abs/1703.05448}{{arXiv:1703.05448}}} {[hep-th]}

\bibitem[{Strominger and Vafa(1996)}]{Strominger:1996sh}
Strominger A, Vafa C (1996) {Microscopic origin of the Bekenstein-Hawking
  entropy}. Phys Lett B 379:99--104. \doi{10.1016/0370-2693(96)00345-0}.
  {\href{https://arxiv.org/abs/hep-th/9601029}{{arXiv:hep-th/9601029}}}

\bibitem[{Strominger and Zhiboedov(2016)}]{BMS4}
Strominger A, Zhiboedov A (2016) {Gravitational Memory, BMS Supertranslations
  and Soft Theorems}. JHEP 01:086. \doi{10.1007/JHEP01(2016)086}.
  {\href{https://arxiv.org/abs/1411.5745}{{arXiv:1411.5745}}} {[hep-th]}

\bibitem[{Sugiyama et~al.(2020)Sugiyama, Takhistov, Vitagliano, Kusenko,
  Sasaki, and Takada}]{Sugiyama:2020roc}
Sugiyama S, Takhistov V, Vitagliano E, Kusenko A, Sasaki M, Takada M (2020)
  {Testing Stochastic Gravitational Wave Signals from Primordial Black Holes
  with Optical Telescopes}
  {\href{https://arxiv.org/abs/2010.02189}{{arXiv:2010.02189}}} {[astro-ph.CO]}

\bibitem[{Sullivan et~al.(2020)Sullivan, Yunes, and
  Sotiriou}]{Sullivan:2019vyi}
Sullivan A, Yunes N, Sotiriou TP (2020) {Numerical black hole solutions in
  modified gravity theories: Spherical symmetry case}. Phys Rev D
  101(4):044024. \doi{10.1103/PhysRevD.101.044024}.
  {\href{https://arxiv.org/abs/1903.02624}{{arXiv:1903.02624}}} {[gr-qc]}

\bibitem[{Sullivan et~al.(2021)Sullivan, Yunes, and
  Sotiriou}]{Sullivan:2020zpf}
Sullivan A, Yunes N, Sotiriou TP (2021) {Numerical black hole solutions in
  modified gravity theories: Axial symmetry case}. Phys Rev D 103(12):124058.
  \doi{10.1103/PhysRevD.103.124058}.
  {\href{https://arxiv.org/abs/2009.10614}{{arXiv:2009.10614}}} {[gr-qc]}

\bibitem[{Sun et~al.(2020)Sun, Brito, and Isi}]{Sun:2019mqb}
Sun L, Brito R, Isi M (2020) {Search for ultralight bosons in Cygnus X-1 with
  Advanced LIGO}. Phys Rev D 101(6):063020. \doi{10.1103/PhysRevD.101.063020}.
  {\href{https://arxiv.org/abs/1909.11267}{{arXiv:1909.11267}}} {[gr-qc]}

\bibitem[{Suyama(2020)}]{Suyama:2020lbf}
Suyama T (2020) {On arrival time difference between lensed gravitational waves
  and light}. Astrophys J 896(1):46. \doi{10.3847/1538-4357/ab8d3f}.
  {\href{https://arxiv.org/abs/2003.11748}{{arXiv:2003.11748}}} {[gr-qc]}

\bibitem[{Suzuki and Maeda(2000)}]{Suzuki:1999si}
Suzuki S, Maeda Ki (2000) {Signature of chaos in gravitational waves from a
  spinning particle}. Phys Rev D 61:024005. \doi{10.1103/PhysRevD.61.024005}.
  {\href{https://arxiv.org/abs/gr-qc/9910064}{{arXiv:gr-qc/9910064}}}

\bibitem[{Syer and Clarke(1995)}]{Syer:1995hk}
Syer D, Clarke CJ (1995) {Satellites in discs: regulating the accretion
  luminosity}. Mon Not Roy Astron Soc 277:758. \doi{10.1093/mnras/277.3.758}.
  {\href{https://arxiv.org/abs/astro-ph/9505021}{{arXiv:astro-ph/9505021}}}

\bibitem[{Szekeres(1965)}]{Szekeres:1965ux}
Szekeres P (1965) {The Gravitational compass}. J Math Phys 6:1387--1391.
  \doi{10.1063/1.1704788}

\bibitem[{Tahura and Yagi(2018)}]{Tahura:2018zuq}
Tahura S, Yagi K (2018) {Parameterized Post-Einsteinian Gravitational Waveforms
  in Various Modified Theories of Gravity}. Phys Rev D98(8):084042.
  \doi{10.1103/PhysRevD.101.109902, 10.1103/PhysRevD.98.084042}, [Erratum:
  Phys. Rev.D101,no.10,109902(2020)].
  {\href{https://arxiv.org/abs/1809.00259}{{arXiv:1809.00259}}} {[gr-qc]}

\bibitem[{Tahura et~al.(2020)Tahura, Nichols, Saffer, Stein, and
  Yagi}]{Tahura:2020vsa}
Tahura S, Nichols DA, Saffer A, Stein LC, Yagi K (2020) {Brans-Dicke theory in
  Bondi-Sachs form: Asymptotically flat solutions, asymptotic symmetries and
  gravitational-wave memory effects}
  {\href{https://arxiv.org/abs/2007.13799}{{arXiv:2007.13799}}} {[gr-qc]}

\bibitem[{Tahura et~al.(2021)Tahura, Nichols, and Yagi}]{Tahura:2021hbk}
Tahura S, Nichols DA, Yagi K (2021) {Gravitational-wave memory effects in
  Brans-Dicke theory: Waveforms and effects in the post-Newtonian
  approximation} {\href{https://arxiv.org/abs/2107.02208}{{arXiv:2107.02208}}}
  {[gr-qc]}

\bibitem[{Takahashi(2017)}]{Takahashi:2016jom}
Takahashi R (2017) {Arrival time differences between gravitational waves and
  electromagnetic signals due to gravitational lensing}. Astrophys J
  835(1):103. \doi{10.3847/1538-4357/835/1/103}.
  {\href{https://arxiv.org/abs/1606.00458}{{arXiv:1606.00458}}} {[astro-ph.CO]}

\bibitem[{Takahashi and Nakamura(2003)}]{Takahashi:2003ix}
Takahashi R, Nakamura T (2003) {Wave effects in gravitational lensing of
  gravitational waves from chirping binaries}. Astrophys J 595:1039--1051.
  \doi{10.1086/377430}.
  {\href{https://arxiv.org/abs/astro-ph/0305055}{{arXiv:astro-ph/0305055}}}

\bibitem[{Takeda et~al.(2018)Takeda, Nishizawa, Michimura, Nagano, Komori,
  Ando, and Hayama}]{Takeda:2018uai}
Takeda H, Nishizawa A, Michimura Y, Nagano K, Komori K, Ando M, Hayama K (2018)
  {Polarization test of gravitational waves from compact binary coalescences}.
  Phys Rev D 98(2):022008. \doi{10.1103/PhysRevD.98.022008}.
  {\href{https://arxiv.org/abs/1806.02182}{{arXiv:1806.02182}}} {[gr-qc]}

\bibitem[{Takeda et~al.(2019)Takeda, Nishizawa, Nagano, Michimura, Komori,
  Ando, and Hayama}]{Takeda:2019gwk}
Takeda H, Nishizawa A, Nagano K, Michimura Y, Komori K, Ando M, Hayama K (2019)
  {Prospects for gravitational-wave polarization tests from compact binary
  mergers with future ground-based detectors}. Phys Rev D 100(4):042001.
  \doi{10.1103/PhysRevD.100.042001}.
  {\href{https://arxiv.org/abs/1904.09989}{{arXiv:1904.09989}}} {[gr-qc]}

\bibitem[{Takeda et~al.(2021{\natexlab{a}})Takeda, Morisaki, and
  Nishizawa}]{Takeda:2020tjj}
Takeda H, Morisaki S, Nishizawa A (2021{\natexlab{a}}) {Pure polarization test
  of GW170814 and GW170817 using waveforms consistent with modified theories of
  gravity}. Phys Rev D 103(6):064037. \doi{10.1103/PhysRevD.103.064037}.
  {\href{https://arxiv.org/abs/2010.14538}{{arXiv:2010.14538}}} {[gr-qc]}

\bibitem[{Takeda et~al.(2021{\natexlab{b}})Takeda, Morisaki, and
  Nishizawa}]{Takeda:2021hgo}
Takeda H, Morisaki S, Nishizawa A (2021{\natexlab{b}}) {Scalar-tensor mixed
  polarization search of gravitational waves}
  {\href{https://arxiv.org/abs/2105.00253}{{arXiv:2105.00253}}} {[gr-qc]}

\bibitem[{Tamanini et~al.(2016)Tamanini, Caprini, Barausse, Sesana, Klein, and
  Petiteau}]{Tamanini:2016zlh}
Tamanini N, Caprini C, Barausse E, Sesana A, Klein A, Petiteau A (2016)
  {Science with the space-based interferometer eLISA. III: Probing the
  expansion of the Universe using gravitational wave standard sirens}. JCAP
  04:002. \doi{10.1088/1475-7516/2016/04/002}.
  {\href{https://arxiv.org/abs/1601.07112}{{arXiv:1601.07112}}} {[astro-ph.CO]}

\bibitem[{Tanaka et~al.(2002)Tanaka, Takeuchi, and Ward}]{Tanaka_2002}
Tanaka H, Takeuchi T, Ward WR (2002) Three-dimensional interaction between a
  planet and an isothermal gaseous disk. i. corotation and lindblad torques and
  planet migration. The Astrophysical Journal 565(2):1257--1274.
  \doi{10.1086/324713}, \urlprefix\url{10.1086/324713}

\bibitem[{Tanay et~al.(2021)Tanay, Stein, and G\'alvez~Ghersi}]{Tanay:2020gfb}
Tanay S, Stein LC, G\'alvez~Ghersi JT (2021) {Integrability of eccentric,
  spinning black hole binaries up to second post-Newtonian order}. Phys Rev D
  103(6):064066. \doi{10.1103/PhysRevD.103.064066}.
  {\href{https://arxiv.org/abs/2012.06586}{{arXiv:2012.06586}}} {[gr-qc]}

\bibitem[{Taracchini et~al.(2012)Taracchini, Pan, Buonanno, Barausse, Boyle,
  Chu, Lovelace, Pfeiffer, and Scheel}]{Taracchini:2012ig}
Taracchini A, Pan Y, Buonanno A, Barausse E, Boyle M, Chu T, Lovelace G,
  Pfeiffer HP, Scheel MA (2012) {Prototype effective-one-body model for
  nonprecessing spinning inspiral-merger-ringdown waveforms}. Phys Rev D
  86:024011. \doi{10.1103/PhysRevD.86.024011}.
  {\href{https://arxiv.org/abs/1202.0790}{{arXiv:1202.0790}}} {[gr-qc]}

\bibitem[{Taracchini et~al.(2013)Taracchini, Buonanno, Hughes, and
  Khanna}]{Taracchini:2013wfa}
Taracchini A, Buonanno A, Hughes SA, Khanna G (2013) {Modeling the
  horizon-absorbed gravitational flux for equatorial-circular orbits in Kerr
  spacetime}. Phys Rev D 88:044001. \doi{10.1103/PhysRevD.88.044001}, [Erratum:
  Phys.Rev.D 88, 109903 (2013)].
  {\href{https://arxiv.org/abs/1305.2184}{{arXiv:1305.2184}}} {[gr-qc]}

\bibitem[{Taracchini et~al.(2014)}]{Taracchini:2013rva}
Taracchini A, et~al. (2014) {Effective-one-body model for black-hole binaries
  with generic mass ratios and spins}. Phys Rev D 89(6):061502.
  \doi{10.1103/PhysRevD.89.061502}.
  {\href{https://arxiv.org/abs/1311.2544}{{arXiv:1311.2544}}} {[gr-qc]}

\bibitem[{Tattersall and Ferreira(2018)}]{Tattersall:2018nve}
Tattersall OJ, Ferreira PG (2018) {Quasinormal modes of black holes in
  Horndeski gravity}. Phys Rev D 97(10):104047.
  \doi{10.1103/PhysRevD.97.104047}.
  {\href{https://arxiv.org/abs/1804.08950}{{arXiv:1804.08950}}} {[gr-qc]}

\bibitem[{Tattersall et~al.(2018)Tattersall, Ferreira, and
  Lagos}]{Tattersall:2017erk}
Tattersall OJ, Ferreira PG, Lagos M (2018) {General theories of linear
  gravitational perturbations to a Schwarzschild Black Hole}. Phys Rev D
  97(4):044021. \doi{10.1103/PhysRevD.97.044021}.
  {\href{https://arxiv.org/abs/1711.01992}{{arXiv:1711.01992}}} {[gr-qc]}

\bibitem[{Taylor and Gair(2012)}]{Taylor:2012db}
Taylor SR, Gair JR (2012) {Cosmology with the lights off: standard sirens in
  the Einstein Telescope era}. Phys Rev D86:023502.
  \doi{10.1103/PhysRevD.86.023502}.
  {\href{https://arxiv.org/abs/1204.6739}{{arXiv:1204.6739}}} {[astro-ph.CO]}

\bibitem[{Taylor et~al.(2012)Taylor, Gair, and Mandel}]{Taylor:2011fs}
Taylor SR, Gair JR, Mandel I (2012) {Hubble without the Hubble: Cosmology using
  advanced gravitational-wave detectors alone}. Phys Rev D85:023535.
  \doi{10.1103/PhysRevD.85.023535}.
  {\href{https://arxiv.org/abs/1108.5161}{{arXiv:1108.5161}}} {[gr-qc]}

\bibitem[{Taylor and Veneziano(1990)}]{Taylor:1989ua}
Taylor T, Veneziano G (1990) {Quantum Gravity at Large Distances and the
  Cosmological Constant}. NuclPhys B345:210--230.
  \doi{10.1016/0550-3213(90)90615-K}

\bibitem[{Testa and Pani(2018)}]{Testa:2018bzd}
Testa A, Pani P (2018) {Analytical template for gravitational-wave echoes:
  signal characterization and prospects of detection with current and future
  interferometers}. Phys Rev D 98(4):044018. \doi{10.1103/PhysRevD.98.044018}.
  {\href{https://arxiv.org/abs/1806.04253}{{arXiv:1806.04253}}} {[gr-qc]}

\bibitem[{Teukolsky and Press(1974)}]{Teukolsky:1974yv}
Teukolsky S, Press W (1974) {Perturbations of a rotating black hole. III -
  Interaction of the hole with gravitational and electromagnet ic radiation}.
  Astrophys J 193:443--461. \doi{10.1086/153180}

\bibitem[{Teukolsky(1972)}]{PhysRevLett.29.1114}
Teukolsky SA (1972) Rotating black holes: Separable wave equations for
  gravitational and electromagnetic perturbations. Phys Rev Lett 29:1114--1118.
  \doi{10.1103/PhysRevLett.29.1114}

\bibitem[{{Teukolsky}(1973)}]{1973ApJ...185..635T}
{Teukolsky} SA (1973) {Perturbations of a Rotating Black Hole. I. Fundamental
  Equations for Gravitational, Electromagnetic, and Neutrino-Field
  Perturbations}. The Astroph J 185:635--648. \doi{10.1086/152444}

\bibitem[{Thompson et~al.(2020)Thompson, Fauchon-Jones, Khan, Nitoglia,
  Pannarale, Dietrich, and Hannam}]{Thompson:2020nei}
Thompson JE, Fauchon-Jones E, Khan S, Nitoglia E, Pannarale F, Dietrich T,
  Hannam M (2020) {Modeling the gravitational wave signature of neutron star
  black hole coalescences}. Phys Rev D 101(12):124059.
  \doi{10.1103/PhysRevD.101.124059}.
  {\href{https://arxiv.org/abs/2002.08383}{{arXiv:2002.08383}}} {[gr-qc]}

\bibitem[{Thorne(1992)}]{Thorne:1992sdb}
Thorne KS (1992) {Gravitational-wave bursts with memory: The Christodoulou
  effect}. Phys Rev D45(2):520--524. \doi{10.1103/PhysRevD.45.520}

\bibitem[{Thrane and Romano(2013)}]{Thrane:2013oya}
Thrane E, Romano JD (2013) {Sensitivity curves for searches for
  gravitational-wave backgrounds}. Phys Rev D 88(12):124032.
  \doi{10.1103/PhysRevD.88.124032}.
  {\href{https://arxiv.org/abs/1310.5300}{{arXiv:1310.5300}}} {[astro-ph.IM]}

\bibitem[{Tinto and da~Silva~Alves(2010)}]{Tinto:2010hz}
Tinto M, da~Silva~Alves ME (2010) {LISA Sensitivities to Gravitational Waves
  from Relativistic Metric Theories of Gravity}. Phys Rev D 82:122003.
  \doi{10.1103/PhysRevD.82.122003}.
  {\href{https://arxiv.org/abs/1010.1302}{{arXiv:1010.1302}}} {[gr-qc]}

\bibitem[{Tkachev et~al.(2020)Tkachev, Pilipenko, and Yepes}]{Tkachev:2020uin}
Tkachev M, Pilipenko S, Yepes G (2020) {Dark Matter Simulations with Primordial
  Black Holes in the Early Universe}
  {\href{https://arxiv.org/abs/2009.07813}{{arXiv:2009.07813}}} {[astro-ph.CO]}

\bibitem[{Toubiana et~al.(2020)Toubiana, Marsat, Barausse, Babak, and
  Baker}]{Toubiana:2020vtf}
Toubiana A, Marsat S, Barausse E, Babak S, Baker J (2020) {Tests of general
  relativity with stellar-mass black hole binaries observed by LISA}. Phys Rev
  D 101(10):104038. \doi{10.1103/PhysRevD.101.104038}.
  {\href{https://arxiv.org/abs/2004.03626}{{arXiv:2004.03626}}} {[gr-qc]}

\bibitem[{Toubiana et~al.(2021)}]{Toubiana:2020drf}
Toubiana A, et~al. (2021) {Detectable environmental effects in GW190521-like
  black-hole binaries with LISA}. Phys Rev Lett 126(10):101105.
  \doi{10.1103/PhysRevLett.126.101105}.
  {\href{https://arxiv.org/abs/2010.06056}{{arXiv:2010.06056}}} {[astro-ph.HE]}

\bibitem[{Trashorras et~al.(2021)Trashorras, García-Bellido, and
  Nesseris}]{Trashorras:2020mwn}
Trashorras M, García-Bellido J, Nesseris S (2021) {The clustering dynamics of
  primordial black boles in $ N $-body simulations}. Universe 7(1):18.
  \doi{10.3390/universe7010018}.
  {\href{https://arxiv.org/abs/2006.15018}{{arXiv:2006.15018}}} {[astro-ph.CO]}

\bibitem[{Tr\"oster et~al.(2020)}]{Troster:2019ean}
Tr\"oster T, et~al. (2020) {Cosmology from large-scale structure: Constraining
  $\Lambda$CDM with BOSS}. Astron Astrophys 633:L10.
  \doi{10.1051/0004-6361/201936772}.
  {\href{https://arxiv.org/abs/1909.11006}{{arXiv:1909.11006}}} {[astro-ph.CO]}

\bibitem[{Tsamis and Woodard(1995)}]{Tsamis:1994ca}
Tsamis NC, Woodard RP (1995) {Strong infrared effects in quantum gravity}.
  Annals Phys 238:1--82. \doi{10.1006/aphy.1995.1015}

\bibitem[{Tsang et~al.(2018)Tsang, Rollier, Ghosh, Samajdar, Agathos,
  Chatziioannou, Cardoso, Khanna, and Van Den~Broeck}]{Tsang:2018uie}
Tsang KW, Rollier M, Ghosh A, Samajdar A, Agathos M, Chatziioannou K, Cardoso
  V, Khanna G, Van Den~Broeck C (2018) {A morphology-independent data analysis
  method for detecting and characterizing gravitational wave echoes}. Phys Rev
  D 98(2):024023. \doi{10.1103/PhysRevD.98.024023}.
  {\href{https://arxiv.org/abs/1804.04877}{{arXiv:1804.04877}}} {[gr-qc]}

\bibitem[{Tsang et~al.(2020)Tsang, Ghosh, Samajdar, Chatziioannou,
  Mastrogiovanni, Agathos, and Van Den~Broeck}]{Tsang:2019zra}
Tsang KW, Ghosh A, Samajdar A, Chatziioannou K, Mastrogiovanni S, Agathos M,
  Van Den~Broeck C (2020) {A morphology-independent search for gravitational
  wave echoes in data from the first and second observing runs of Advanced LIGO
  and Advanced Virgo}. Phys Rev D 101(6):064012.
  \doi{10.1103/PhysRevD.101.064012}.
  {\href{https://arxiv.org/abs/1906.11168}{{arXiv:1906.11168}}} {[gr-qc]}

\bibitem[{Tso et~al.(2019)Tso, Gerosa, and Chen}]{Tso:2018pdv}
Tso R, Gerosa D, Chen Y (2019) {Optimizing LIGO with LISA forewarnings to
  improve black-hole spectroscopy}. Phys Rev D 99(12):124043.
  \doi{10.1103/PhysRevD.99.124043}.
  {\href{https://arxiv.org/abs/1807.00075}{{arXiv:1807.00075}}} {[gr-qc]}

\bibitem[{Tsujikawa(2019)}]{Tsujikawa:2019pih}
Tsujikawa S (2019) {Lunar Laser Ranging constraints on nonminimally coupled
  dark energy and standard sirens}. Phys Rev D100(4):043510.
  \doi{10.1103/PhysRevD.100.043510}.
  {\href{https://arxiv.org/abs/1903.07092}{{arXiv:1903.07092}}} {[gr-qc]}

\bibitem[{Tsukada et~al.(2019)Tsukada, Callister, Matas, and
  Meyers}]{Tsukada:2018mbp}
Tsukada L, Callister T, Matas A, Meyers P (2019) {First search for a stochastic
  gravitational-wave background from ultralight bosons}. Phys Rev D
  99(10):103015. \doi{10.1103/PhysRevD.99.103015}.
  {\href{https://arxiv.org/abs/1812.09622}{{arXiv:1812.09622}}} {[astro-ph.HE]}

\bibitem[{Tsukada et~al.(2020)Tsukada, Brito, East, and
  Siemonsen}]{Tsukada:2020lgt}
Tsukada L, Brito R, East WE, Siemonsen N (2020) {Modeling and searching for a
  stochastic gravitational-wave background from ultralight vector bosons}
  {\href{https://arxiv.org/abs/2011.06995}{{arXiv:2011.06995}}} {[astro-ph.HE]}

\bibitem[{Tulin and Yu(2018)}]{Tulin:2017ara}
Tulin S, Yu HB (2018) {Dark Matter Self-interactions and Small Scale
  Structure}. Phys Rept 730:1--57. \doi{10.1016/j.physrep.2017.11.004}.
  {\href{https://arxiv.org/abs/1705.02358}{{arXiv:1705.02358}}} {[hep-ph]}

\bibitem[{Turner et~al.(1993)Turner, White, and Lidsey}]{Turner:1993vb}
Turner MS, White MJ, Lidsey JE (1993) {Tensor perturbations in inflationary
  models as a probe of cosmology}. Phys Rev D 48:4613--4622.
  \doi{10.1103/PhysRevD.48.4613}.
  {\href{https://arxiv.org/abs/astro-ph/9306029}{{arXiv:astro-ph/9306029}}}

\bibitem[{Uchikata et~al.(2016)Uchikata, Yoshida, and Pani}]{Uchikata:2016qku}
Uchikata N, Yoshida S, Pani P (2016) {Tidal deformability and I-Love-Q
  relations for gravastars with polytropic thin shells}. Phys Rev D
  94(6):064015. \doi{10.1103/PhysRevD.94.064015}.
  {\href{https://arxiv.org/abs/1607.03593}{{arXiv:1607.03593}}} {[gr-qc]}

\bibitem[{Uchikata et~al.(2019)Uchikata, Nakano, Narikawa, Sago, Tagoshi, and
  Tanaka}]{Uchikata:2019frs}
Uchikata N, Nakano H, Narikawa T, Sago N, Tagoshi H, Tanaka T (2019) {Searching
  for black hole echoes from the LIGO-Virgo Catalog GWTC-1}
  {\href{https://arxiv.org/abs/1906.00838}{{arXiv:1906.00838}}} {[gr-qc]}

\bibitem[{Ullio et~al.(2001)Ullio, Zhao, and Kamionkowski}]{Ullio:2001fb}
Ullio P, Zhao H, Kamionkowski M (2001) {A Dark matter spike at the galactic
  center?} Phys Rev D 64:043504. \doi{10.1103/PhysRevD.64.043504}.
  {\href{https://arxiv.org/abs/astro-ph/0101481}{{arXiv:astro-ph/0101481}}}

\bibitem[{Unal(2019)}]{Unal:2018yaa}
Unal C (2019) {Imprints of Primordial Non-Gaussianity on Gravitational Wave
  Spectrum}. Phys Rev D 99(4):041301. \doi{10.1103/PhysRevD.99.041301}.
  {\href{https://arxiv.org/abs/1811.09151}{{arXiv:1811.09151}}} {[astro-ph.CO]}

\bibitem[{\"Unal et~al.(2021)\"Unal, Kovetz, and Patil}]{Unal:2020mts}
\"Unal C, Kovetz ED, Patil SP (2021) {Multimessenger probes of inflationary
  fluctuations and primordial black holes}. Phys Rev D 103(6):063519.
  \doi{10.1103/PhysRevD.103.063519}.
  {\href{https://arxiv.org/abs/2008.11184}{{arXiv:2008.11184}}} {[astro-ph.CO]}

\bibitem[{Vainshtein(1972)}]{Vainshtein:1972sx}
Vainshtein AI (1972) {To the problem of nonvanishing gravitation mass}. Phys
  Lett 39B:393--394. \doi{10.1016/0370-2693(72)90147-5}

\bibitem[{Van De~Meent and Warburton(2018)}]{vandeMeent:2018rms}
Van De~Meent M, Warburton N (2018) {Fast Self-forced Inspirals}. Class Quant
  Grav 35(14):144003. \doi{10.1088/1361-6382/aac8ce}.
  {\href{https://arxiv.org/abs/1802.05281}{{arXiv:1802.05281}}} {[gr-qc]}

\bibitem[{Vaskonen and Veerm\"ae(2020{\natexlab{a}})}]{Vaskonen:2020lbd}
Vaskonen V, Veerm\"ae H (2020{\natexlab{a}}) {Did NANOGrav see a signal from
  primordial black hole formation?}
  {\href{https://arxiv.org/abs/2009.07832}{{arXiv:2009.07832}}} {[astro-ph.CO]}

\bibitem[{Vaskonen and Veerm\"ae(2020{\natexlab{b}})}]{Vaskonen:2019jpv}
Vaskonen V, Veerm\"ae H (2020{\natexlab{b}}) {Lower bound on the primordial
  black hole merger rate}. Phys Rev D 101(4):043015.
  \doi{10.1103/PhysRevD.101.043015}.
  {\href{https://arxiv.org/abs/1908.09752}{{arXiv:1908.09752}}} {[astro-ph.CO]}

\bibitem[{Vasylyev and Filippenko(2020)}]{Vasylyev:2020hgb}
Vasylyev S, Filippenko A (2020) {A Measurement of the Hubble Constant using
  Gravitational Waves from the Neutron-Star Black-Hole Candidate GW190814}
  {\href{https://arxiv.org/abs/2007.11148}{{arXiv:2007.11148}}} {[astro-ph.CO]}

\bibitem[{Verde et~al.(2019)Verde, Treu, and Riess}]{Verde:2019ivm}
Verde L, Treu T, Riess A (2019) {Tensions between the Early and the Late
  Universe}. \doi{10.1038/s41550-019-0902-0}.
  {\href{https://arxiv.org/abs/1907.10625}{{arXiv:1907.10625}}} {[astro-ph.CO]}

\bibitem[{Vines et~al.(2011)Vines, Flanagan, and Hinderer}]{Vines:2011ud}
Vines J, Flanagan EE, Hinderer T (2011) {Post-1-Newtonian tidal effects in the
  gravitational waveform from binary inspirals}. Phys Rev D 83:084051.
  \doi{10.1103/PhysRevD.83.084051}.
  {\href{https://arxiv.org/abs/1101.1673}{{arXiv:1101.1673}}} {[gr-qc]}

\bibitem[{Visinelli et~al.(2018)Visinelli, Bolis, and
  Vagnozzi}]{Visinelli:2017bny}
Visinelli L, Bolis N, Vagnozzi S (2018) {Brane-world extra dimensions in light
  of GW170817}. Phys Rev D 97(6):064039. \doi{10.1103/PhysRevD.97.064039}.
  {\href{https://arxiv.org/abs/1711.06628}{{arXiv:1711.06628}}} {[gr-qc]}

\bibitem[{Visser(1995)}]{Visser:1995cc}
Visser M (1995) {Lorentzian wormholes: From Einstein to Hawking}

\bibitem[{V\"olkel and Barausse(2020)}]{Volkel:2020daa}
V\"olkel SH, Barausse E (2020) {Bayesian Metric Reconstruction with
  Gravitational Wave Observations}
  {\href{https://arxiv.org/abs/2007.02986}{{arXiv:2007.02986}}} {[gr-qc]}

\bibitem[{Volonteri(2010)}]{Volonteri_2010}
Volonteri M (2010) Formation of supermassive black holes. The Astronomy and
  Astrophysics Review 18(3):279–315. \doi{10.1007/s00159-010-0029-x}

\bibitem[{Wagle et~al.(2021)Wagle, Yunes, and Silva}]{Wagle:2021tam}
Wagle PK, Yunes N, Silva HO (2021) {Quasinormal modes of slowly-rotating black
  holes in dynamical Chern-Simons gravity}
  {\href{https://arxiv.org/abs/2103.09913}{{arXiv:2103.09913}}} {[gr-qc]}

\bibitem[{Wands(1999)}]{Wands:1998yp}
Wands D (1999) {Duality invariance of cosmological perturbation spectra}. Phys
  Rev D 60:023507. \doi{10.1103/PhysRevD.60.023507}.
  {\href{https://arxiv.org/abs/gr-qc/9809062}{{arXiv:gr-qc/9809062}}}

\bibitem[{Wang et~al.(2012)Wang, Hui, and Khoury}]{Wang:2012kj}
Wang J, Hui L, Khoury J (2012) {No-Go Theorems for Generalized Chameleon Field
  Theories}. Phys Rev Lett 109:241301. \doi{10.1103/PhysRevLett.109.241301}.
  {\href{https://arxiv.org/abs/1208.4612}{{arXiv:1208.4612}}} {[astro-ph.CO]}

\bibitem[{Wang and Afshordi(2018)}]{Wang:2018gin}
Wang Q, Afshordi N (2018) {Black hole echology: The observer\textquoteright{}s
  manual}. Phys Rev D 97(12):124044. \doi{10.1103/PhysRevD.97.124044}.
  {\href{https://arxiv.org/abs/1803.02845}{{arXiv:1803.02845}}} {[gr-qc]}

\bibitem[{Wang et~al.(2019{\natexlab{a}})Wang, Oshita, and
  Afshordi}]{Wang:2019rcf}
Wang Q, Oshita N, Afshordi N (2019{\natexlab{a}}) {Echoes from Quantum Black
  Holes} {\href{https://arxiv.org/abs/1905.00446}{{arXiv:1905.00446}}}
  {[gr-qc]}

\bibitem[{Wang and Kohri(2021)}]{Wang:2021djr}
Wang S, Kohri K (2021) {Probing Primordial Black Holes with Angular Power
  Spectrum for Anisotropies in Stochastic Gravitational-Wave Background}
  {\href{https://arxiv.org/abs/2107.01935}{{arXiv:2107.01935}}} {[gr-qc]}

\bibitem[{Wang et~al.(2019{\natexlab{b}})Wang, Terada, and
  Kohri}]{Wang:2019kaf}
Wang S, Terada T, Kohri K (2019{\natexlab{b}}) {Prospective constraints on the
  primordial black hole abundance from the stochastic gravitational-wave
  backgrounds produced by coalescing events and curvature perturbations}. Phys
  Rev D 99(10):103531. \doi{10.1103/PhysRevD.99.103531}, [Erratum: Phys.Rev.D
  101, 069901 (2020)].
  {\href{https://arxiv.org/abs/1903.05924}{{arXiv:1903.05924}}} {[astro-ph.CO]}

\bibitem[{Warburton(2015)}]{Warburton:2014bya}
Warburton N (2015) {Self force on a scalar charge in Kerr spacetime: inclined
  circular orbits}. Phys Rev D 91(2):024045. \doi{10.1103/PhysRevD.91.024045}.
  {\href{https://arxiv.org/abs/1408.2885}{{arXiv:1408.2885}}} {[gr-qc]}

\bibitem[{Warburton and Barack(2010)}]{Warburton:2010eq}
Warburton N, Barack L (2010) {Self force on a scalar charge in Kerr spacetime:
  circular equatorial orbits}. Phys Rev D 81:084039.
  \doi{10.1103/PhysRevD.81.084039}.
  {\href{https://arxiv.org/abs/1003.1860}{{arXiv:1003.1860}}} {[gr-qc]}

\bibitem[{Warburton and Barack(2011)}]{Warburton:2011hp}
Warburton N, Barack L (2011) {Self force on a scalar charge in Kerr spacetime:
  eccentric equatorial orbits}. Phys Rev D 83:124038.
  \doi{10.1103/PhysRevD.83.124038}.
  {\href{https://arxiv.org/abs/1103.0287}{{arXiv:1103.0287}}} {[gr-qc]}

\bibitem[{Warburton et~al.(2012)Warburton, Akcay, Barack, Gair, and
  Sago}]{Warburton:2011fk}
Warburton N, Akcay S, Barack L, Gair JR, Sago N (2012) {Evolution of inspiral
  orbits around a Schwarzschild black hole}. Phys Rev D85:061501.
  \doi{10.1103/PhysRevD.85.061501}.
  {\href{https://arxiv.org/abs/1111.6908}{{arXiv:1111.6908}}} {[gr-qc]}

\bibitem[{{Ward}(1997)}]{1997Icar..126..261W}
{Ward} WR (1997) {Protoplanet Migration by Nebula Tides}. \icarus
  126(2):261--281. \doi{10.1006/icar.1996.5647}

\bibitem[{Watanabe and Komatsu(2006)}]{Watanabe:2006qe}
Watanabe Y, Komatsu E (2006) {Improved Calculation of the Primordial
  Gravitational Wave Spectrum in the Standard Model}. Phys Rev D 73:123515.
  \doi{10.1103/PhysRevD.73.123515}.
  {\href{https://arxiv.org/abs/astro-ph/0604176}{{arXiv:astro-ph/0604176}}}

\bibitem[{Weinberg(1964)}]{Weinberg:1964kqu}
Weinberg S (1964) {Derivation of gauge invariance and the equivalence principle
  from Lorentz invariance of the S- matrix}. Phys Lett 9(4):357--359.
  \doi{10.1016/0031-9163(64)90396-8}

\bibitem[{Wen(2003)}]{Wen:2002km}
Wen L (2003) {On the eccentricity distribution of coalescing black hole
  binaries driven by the Kozai mechanism in globular clusters}. Astrophys J
  598:419--430. \doi{10.1086/378794}.
  {\href{https://arxiv.org/abs/astro-ph/0211492}{{arXiv:astro-ph/0211492}}}

\bibitem[{Westerweck et~al.(2018)Westerweck, Nielsen, Fischer-Birnholtz,
  Cabero, Capano, Dent, Krishnan, Meadors, and Nitz}]{Westerweck:2017hus}
Westerweck J, Nielsen A, Fischer-Birnholtz O, Cabero M, Capano C, Dent T,
  Krishnan B, Meadors G, Nitz AH (2018) {Low significance of evidence for black
  hole echoes in gravitational wave data}. Phys Rev D 97(12):124037.
  \doi{10.1103/PhysRevD.97.124037}.
  {\href{https://arxiv.org/abs/1712.09966}{{arXiv:1712.09966}}} {[gr-qc]}

\bibitem[{Wetterich(1995)}]{Wetterich:1994bg}
Wetterich C (1995) {The Cosmon model for an asymptotically vanishing time
  dependent cosmological 'constant'}. Astron Astrophys 301:321--328.
  {\href{https://arxiv.org/abs/hep-th/9408025}{{arXiv:hep-th/9408025}}}

\bibitem[{Wetterich(2004)}]{Wetterich:2004pv}
Wetterich C (2004) {Phenomenological parameterization of quintessence}. Phys
  Lett B 594:17--22. \doi{10.1016/j.physletb.2004.05.008}.
  {\href{https://arxiv.org/abs/astro-ph/0403289}{{arXiv:astro-ph/0403289}}}

\bibitem[{Wilczek(1978)}]{Wilczek:1977pj}
Wilczek F (1978) {Problem of Strong $P$ and $T$ Invariance in the Presence of
  Instantons}. Phys Rev Lett 40:279--282. \doi{10.1103/PhysRevLett.40.279}

\bibitem[{Will(1993)}]{Will:1993ns}
Will C (1993) {Theory and experiment in gravitational physics}

\bibitem[{{Will}(1971)}]{1971ApJ...163..611W}
{Will} CM (1971) {Theoretical Frameworks for Testing Relativistic Gravity. II.
  Parametrized Post-Newtonian Hydrodynamics, and the Nordtvedt Effect}. \apj
  163:611. \doi{10.1086/150804}

\bibitem[{Will(1998)}]{Will:1997bb}
Will CM (1998) {Bounding the mass of the graviton using gravitational wave
  observations of inspiralling compact binaries}. Phys Rev D 57:2061--2068.
  \doi{10.1103/PhysRevD.57.2061}.
  {\href{https://arxiv.org/abs/gr-qc/9709011}{{arXiv:gr-qc/9709011}}} {[gr-qc]}

\bibitem[{Will(2014)}]{Will:2014kxa}
Will CM (2014) {The Confrontation between General Relativity and Experiment}.
  Living Rev Rel 17:4. \doi{10.12942/lrr-2014-4}.
  {\href{https://arxiv.org/abs/1403.7377}{{arXiv:1403.7377}}} {[gr-qc]}

\bibitem[{Will and Maitra(2017)}]{Will:2016pgm}
Will CM, Maitra M (2017) {Relativistic orbits around spinning supermassive
  black holes. Secular evolution to 4.5 post-Newtonian order}. Phys Rev
  D95(6):064003. \doi{10.1103/PhysRevD.95.064003}.
  {\href{https://arxiv.org/abs/1611.06931}{{arXiv:1611.06931}}} {[gr-qc]}

\bibitem[{{Will} and {Nordtvedt}(1972)}]{1972ApJ...177..757W}
{Will} CM, {Nordtvedt} J Kenneth (1972) {Conservation Laws and Preferred Frames
  in Relativistic Gravity. I. Preferred-Frame Theories and an Extended PPN
  Formalism}. \apj 177:757. \doi{10.1086/151754}

\bibitem[{Williams et~al.(2004)Williams, Turyshev, and
  Boggs}]{Williams:2004qba}
Williams JG, Turyshev SG, Boggs DH (2004) {Progress in lunar laser ranging
  tests of relativistic gravity}. Phys Rev Lett 93:261101.
  \doi{10.1103/PhysRevLett.93.261101}.
  {\href{https://arxiv.org/abs/gr-qc/0411113}{{arXiv:gr-qc/0411113}}}

\bibitem[{Witek et~al.(2013)Witek, Cardoso, Ishibashi, and
  Sperhake}]{Witek:2012tr}
Witek H, Cardoso V, Ishibashi A, Sperhake U (2013) {Superradiant instabilities
  in astrophysical systems}. Phys Rev D 87(4):043513.
  \doi{10.1103/PhysRevD.87.043513}.
  {\href{https://arxiv.org/abs/1212.0551}{{arXiv:1212.0551}}} {[gr-qc]}

\bibitem[{Witek et~al.(2019)Witek, Gualtieri, Pani, and
  Sotiriou}]{Witek:2018dmd}
Witek H, Gualtieri L, Pani P, Sotiriou TP (2019) {Black holes and binary
  mergers in scalar Gauss-Bonnet gravity: scalar field dynamics}. Phys Rev D
  99(6):064035. \doi{10.1103/PhysRevD.99.064035}.
  {\href{https://arxiv.org/abs/1810.05177}{{arXiv:1810.05177}}} {[gr-qc]}

\bibitem[{Witek et~al.(2020)Witek, Gualtieri, and Pani}]{Witek:2020uzz}
Witek H, Gualtieri L, Pani P (2020) {Towards numerical relativity in scalar
  Gauss-Bonnet gravity: $3+1$ decomposition beyond the small-coupling limit}.
  Phys Rev D 101(12):124055. \doi{10.1103/PhysRevD.101.124055}.
  {\href{https://arxiv.org/abs/2004.00009}{{arXiv:2004.00009}}} {[gr-qc]}

\bibitem[{Witzany(2019)}]{Witzany:2019dii}
Witzany V (2019) {Spin-perturbed orbits near black holes}
  {\href{https://arxiv.org/abs/1903.03649}{{arXiv:1903.03649}}} {[gr-qc]}

\bibitem[{Wolf and Lagos(2020)}]{Wolf:2019hun}
Wolf WJ, Lagos M (2020) {Standard Sirens as a Novel Probe of Dark Energy}. Phys
  Rev Lett 124(6):061101. \doi{10.1103/PhysRevLett.124.061101}.
  {\href{https://arxiv.org/abs/1910.10580}{{arXiv:1910.10580}}} {[gr-qc]}

\bibitem[{Wong et~al.(2020)Wong, Franciolini, De~Luca, Baibhav, Berti, Pani,
  and Riotto}]{Wong:2020yig}
Wong K, Franciolini G, De~Luca V, Baibhav V, Berti E, Pani P, Riotto A (2020)
  {Constraining the primordial black hole scenario with Bayesian inference and
  machine learning: the GWTC-2 gravitational wave catalog}
  {\href{https://arxiv.org/abs/2011.01865}{{arXiv:2011.01865}}} {[gr-qc]}

\bibitem[{Wyithe and Loeb(2003)}]{Wyithe:2002ep}
Wyithe JB, Loeb A (2003) {Low - frequency gravitational waves from massive
  black hole binaries: Predictions for LISA and pulsar timing arrays}.
  Astrophys J 590:691--706. \doi{10.1086/375187}.
  {\href{https://arxiv.org/abs/astro-ph/0211556}{{arXiv:astro-ph/0211556}}}

\bibitem[{Yagi(2012)}]{Yagi:2012gp}
Yagi K (2012) {A New constraint on scalar Gauss-Bonnet gravity and a possible
  explanation for the excess of the orbital decay rate in a low-mass X-ray
  binary}. Phys Rev D86:081504. \doi{10.1103/PhysRevD.86.081504}.
  {\href{https://arxiv.org/abs/1204.4524}{{arXiv:1204.4524}}} {[gr-qc]}

\bibitem[{Yagi and Stein(2016)}]{Yagi:2016jml}
Yagi K, Stein LC (2016) {Black Hole Based Tests of General Relativity}. Class
  Quant Grav 33(5):054001. \doi{10.1088/0264-9381/33/5/054001}.
  {\href{https://arxiv.org/abs/1602.02413}{{arXiv:1602.02413}}} {[gr-qc]}

\bibitem[{Yagi and Tanaka(2010)}]{Yagi:2009zm}
Yagi K, Tanaka T (2010) {Constraining alternative theories of gravity by
  gravitational waves from precessing eccentric compact binaries with LISA}.
  Phys Rev D81:064008. \doi{10.1103/PhysRevD.81.109902,
  10.1103/PhysRevD.81.064008}, [Erratum: Phys. Rev.D81,109902(2010)].
  {\href{https://arxiv.org/abs/0906.4269}{{arXiv:0906.4269}}} {[gr-qc]}

\bibitem[{Yagi and Yang(2018)}]{Yagi:2017zhb}
Yagi K, Yang H (2018) {Probing Gravitational Parity Violation with
  Gravitational Waves from Stellar-mass Black Hole Binaries}. Phys Rev
  D97(10):104018. \doi{10.1103/PhysRevD.97.104018}.
  {\href{https://arxiv.org/abs/1712.00682}{{arXiv:1712.00682}}} {[gr-qc]}

\bibitem[{Yagi et~al.(2012{\natexlab{a}})Yagi, Stein, Yunes, and
  Tanaka}]{Yagi:2011xp}
Yagi K, Stein LC, Yunes N, Tanaka T (2012{\natexlab{a}}) {Post-Newtonian,
  Quasi-Circular Binary Inspirals in Quadratic Modified Gravity}. Phys Rev D
  85:064022. \doi{10.1103/PhysRevD.85.064022}, [Erratum: Phys.Rev.D 93, 029902
  (2016)]. {\href{https://arxiv.org/abs/1110.5950}{{arXiv:1110.5950}}}
  {[gr-qc]}

\bibitem[{Yagi et~al.(2012{\natexlab{b}})Yagi, Yunes, and Tanaka}]{Yagi:2012vf}
Yagi K, Yunes N, Tanaka T (2012{\natexlab{b}}) {Gravitational Waves from
  Quasi-Circular Black Hole Binaries in Dynamical Chern-Simons Gravity}. Phys
  Rev Lett 109:251105. \doi{10.1103/PhysRevLett.116.169902}, [Erratum:
  Phys.Rev.Lett. 116, 169902 (2016), Erratum: Phys.Rev.Lett. 124, 029901
  (2020)]. {\href{https://arxiv.org/abs/1208.5102}{{arXiv:1208.5102}}}
  {[gr-qc]}

\bibitem[{Yagi et~al.(2012{\natexlab{c}})Yagi, Yunes, and Tanaka}]{Yagi:2012ya}
Yagi K, Yunes N, Tanaka T (2012{\natexlab{c}}) {Slowly Rotating Black Holes in
  Dynamical Chern-Simons Gravity: Deformation Quadratic in the Spin}. Phys Rev
  D 86:044037. \doi{10.1103/PhysRevD.86.044037}, [Erratum: Phys.Rev.D 89,
  049902 (2014)]. {\href{https://arxiv.org/abs/1206.6130}{{arXiv:1206.6130}}}
  {[gr-qc]}

\bibitem[{Yagi et~al.(2014)Yagi, Blas, Barausse, and Yunes}]{Yagi:2013ava}
Yagi K, Blas D, Barausse E, Yunes N (2014) {Constraints on Einstein-\AE{}ther
  theory and Ho\v{r}ava gravity from binary pulsar observations}. Phys Rev D
  89(8):084067. \doi{10.1103/PhysRevD.89.084067}, [Erratum: Phys.Rev.D 90,
  069902 (2014), Erratum: Phys.Rev.D 90, 069901 (2014)].
  {\href{https://arxiv.org/abs/1311.7144}{{arXiv:1311.7144}}} {[gr-qc]}

\bibitem[{Yagi et~al.(2016)Yagi, Stein, and Yunes}]{Yagi:2015oca}
Yagi K, Stein LC, Yunes N (2016) {Challenging the Presence of Scalar Charge and
  Dipolar Radiation in Binary Pulsars}. Phys Rev D 93(2):024010.
  \doi{10.1103/PhysRevD.93.024010}.
  {\href{https://arxiv.org/abs/1510.02152}{{arXiv:1510.02152}}} {[gr-qc]}

\bibitem[{Yamada et~al.(2019)Yamada, Narikawa, and Tanaka}]{Yamada:2019zrb}
Yamada K, Narikawa T, Tanaka T (2019) {Testing massive-field modifications of
  gravity via gravitational waves}. PTEP 2019(10):103E01.
  \doi{10.1093/ptep/ptz103}.
  {\href{https://arxiv.org/abs/1905.11859}{{arXiv:1905.11859}}} {[gr-qc]}

\bibitem[{Yang and Casals(2017)}]{Yang:2017aht}
Yang H, Casals M (2017) {General Relativistic Dynamics of an Extreme Mass-Ratio
  Binary interacting with an External Body}. Phys Rev D 96(8):083015.
  \doi{10.1103/PhysRevD.96.083015}.
  {\href{https://arxiv.org/abs/1704.02022}{{arXiv:1704.02022}}} {[gr-qc]}

\bibitem[{Yoshida and Eriguchi(1996)}]{10.1093/mnras/282.2.580}
Yoshida S, Eriguchi Y (1996) {Ergoregion instability revisited — a new and
  general method for numerical analysis of stability}. Monthly Notices of the
  Royal Astronomical Society 282(2):580--586. \doi{10.1093/mnras/282.2.580}

\bibitem[{Yoshino and Kodama(2012)}]{Yoshino:2012kn}
Yoshino H, Kodama H (2012) {Bosenova collapse of axion cloud around a rotating
  black hole}. Prog Theor Phys 128:153--190. \doi{10.1143/PTP.128.153}.
  {\href{https://arxiv.org/abs/1203.5070}{{arXiv:1203.5070}}} {[gr-qc]}

\bibitem[{Yoshino and Kodama(2014)}]{Yoshino:2013ofa}
Yoshino H, Kodama H (2014) {Gravitational radiation from an axion cloud around
  a black hole: Superradiant phase}. PTEP 2014:043E02.
  \doi{10.1093/ptep/ptu029}.
  {\href{https://arxiv.org/abs/1312.2326}{{arXiv:1312.2326}}} {[gr-qc]}

\bibitem[{Yoshino and Kodama(2015)}]{Yoshino:2015nsa}
Yoshino H, Kodama H (2015) {The bosenova and axiverse}. Class Quant Grav
  32(21):214001. \doi{10.1088/0264-9381/32/21/214001}.
  {\href{https://arxiv.org/abs/1505.00714}{{arXiv:1505.00714}}} {[gr-qc]}

\bibitem[{You et~al.(2020)You, Zhu, Ashton, Thrane, and Zhu}]{You:2020wju}
You ZQ, Zhu XJ, Ashton G, Thrane E, Zhu ZH (2020) {Standard-siren cosmology
  using gravitational waves from binary black holes}
  {\href{https://arxiv.org/abs/2004.00036}{{arXiv:2004.00036}}} {[astro-ph.CO]}

\bibitem[{Young and Byrnes(2015)}]{Young:2015kda}
Young S, Byrnes CT (2015) {Signatures of non-gaussianity in the isocurvature
  modes of primordial black hole dark matter}. JCAP 04:034.
  \doi{10.1088/1475-7516/2015/04/034}.
  {\href{https://arxiv.org/abs/1503.01505}{{arXiv:1503.01505}}} {[astro-ph.CO]}

\bibitem[{Yuan et~al.(2020)Yuan, Chen, and Huang}]{Yuan:2019fwv}
Yuan C, Chen ZC, Huang QG (2020) {Scalar induced gravitational waves in
  different gauges}. Phys Rev D 101(6):063018.
  \doi{10.1103/PhysRevD.101.063018}.
  {\href{https://arxiv.org/abs/1912.00885}{{arXiv:1912.00885}}} {[astro-ph.CO]}

\bibitem[{Yunes and Hughes(2010)}]{Yunes:2010qb}
Yunes N, Hughes SA (2010) {Binary Pulsar Constraints on the Parameterized
  post-Einsteinian Framework}. Phys Rev D 82:082002.
  \doi{10.1103/PhysRevD.82.082002}.
  {\href{https://arxiv.org/abs/1007.1995}{{arXiv:1007.1995}}} {[gr-qc]}

\bibitem[{Yunes and Pretorius(2009{\natexlab{a}})}]{Yunes:2009hc}
Yunes N, Pretorius F (2009{\natexlab{a}}) {Dynamical Chern-Simons Modified
  Gravity. I. Spinning Black Holes in the Slow-Rotation Approximation}. Phys
  Rev D 79:084043. \doi{10.1103/PhysRevD.79.084043}.
  {\href{https://arxiv.org/abs/0902.4669}{{arXiv:0902.4669}}} {[gr-qc]}

\bibitem[{Yunes and Pretorius(2009{\natexlab{b}})}]{Yunes:2009ke}
Yunes N, Pretorius F (2009{\natexlab{b}}) {Fundamental Theoretical Bias in
  Gravitational Wave Astrophysics and the Parameterized Post-Einsteinian
  Framework}. Phys Rev D 80:122003. \doi{10.1103/PhysRevD.80.122003}.
  {\href{https://arxiv.org/abs/0909.3328}{{arXiv:0909.3328}}} {[gr-qc]}

\bibitem[{Yunes and Siemens(2013)}]{Yunes:2013dva}
Yunes N, Siemens X (2013) {Gravitational-Wave Tests of General Relativity with
  Ground-Based Detectors and Pulsar Timing-Arrays}. Living Rev Rel 16:9.
  \doi{10.12942/lrr-2013-9}.
  {\href{https://arxiv.org/abs/1304.3473}{{arXiv:1304.3473}}} {[gr-qc]}

\bibitem[{Yunes and Stein(2011)}]{Yunes:2011we}
Yunes N, Stein LC (2011) {Non-Spinning Black Holes in Alternative Theories of
  Gravity}. Phys Rev D 83:104002. \doi{10.1103/PhysRevD.83.104002}.
  {\href{https://arxiv.org/abs/1101.2921}{{arXiv:1101.2921}}} {[gr-qc]}

\bibitem[{Yunes et~al.(2010)Yunes, O'Shaughnessy, Owen, and
  Alexander}]{Yunes:2010yf}
Yunes N, O'Shaughnessy R, Owen BJ, Alexander S (2010) {Testing gravitational
  parity violation with coincident gravitational waves and short gamma-ray
  bursts}. Phys Rev D82:064017. \doi{10.1103/PhysRevD.82.064017}.
  {\href{https://arxiv.org/abs/1005.3310}{{arXiv:1005.3310}}} {[gr-qc]}

\bibitem[{Yunes et~al.(2011{\natexlab{a}})Yunes, Coleman~Miller, and
  Thornburg}]{Yunes:2010sm}
Yunes N, Coleman~Miller M, Thornburg J (2011{\natexlab{a}}) {The Effect of
  Massive Perturbers on Extreme Mass-Ratio Inspiral Waveforms}. Phys Rev D
  83:044030. \doi{10.1103/PhysRevD.83.044030}.
  {\href{https://arxiv.org/abs/1010.1721}{{arXiv:1010.1721}}} {[astro-ph.GA]}

\bibitem[{Yunes et~al.(2011{\natexlab{b}})Yunes, Kocsis, Loeb, and
  Haiman}]{Yunes:2011ws}
Yunes N, Kocsis B, Loeb A, Haiman Z (2011{\natexlab{b}}) {Imprint of Accretion
  Disk-Induced Migration on Gravitational Waves from Extreme Mass Ratio
  Inspirals}. Phys Rev Lett 107:171103. \doi{10.1103/PhysRevLett.107.171103}.
  {\href{https://arxiv.org/abs/1103.4609}{{arXiv:1103.4609}}} {[astro-ph.CO]}

\bibitem[{Yunes et~al.(2012)Yunes, Pani, and Cardoso}]{Yunes:2011aa}
Yunes N, Pani P, Cardoso V (2012) {Gravitational Waves from Quasicircular
  Extreme Mass-Ratio Inspirals as Probes of Scalar-Tensor Theories}. Phys Rev D
  85:102003. \doi{10.1103/PhysRevD.85.102003}.
  {\href{https://arxiv.org/abs/1112.3351}{{arXiv:1112.3351}}} {[gr-qc]}

\bibitem[{Yunes et~al.(2016)Yunes, Yagi, and Pretorius}]{Yunes:2016jcc}
Yunes N, Yagi K, Pretorius F (2016) {Theoretical Physics Implications of the
  Binary Black-Hole Mergers GW150914 and GW151226}. Phys Rev D 94(8):084002.
  \doi{10.1103/PhysRevD.94.084002}.
  {\href{https://arxiv.org/abs/1603.08955}{{arXiv:1603.08955}}} {[gr-qc]}

\bibitem[{Zel'dovich(1971)}]{zeldovich1}
Zel'dovich YB (1971) Generation of waves by a rotating body. Pis'ma Zh Eksp
  Teor Fiz 14:270 [JETP Lett. {\bf14}, 180 (1971)]

\bibitem[{Zel'dovich(1972)}]{zeldovich2}
Zel'dovich YB (1972) Amplification of cylindrical electromagnetic waves
  reflected from a rotating body. Zh Eksp Teor Fiz 62:2076 [Sov.Phys. JETP {\bf
  35}, 1085 (1972)]

\bibitem[{Zelenka et~al.(2020)Zelenka, Lukes-Gerakopoulos, Witzany, and
  Kop\'a\v{c}ek}]{Zelenka:2019nyp}
Zelenka O, Lukes-Gerakopoulos G, Witzany V, Kop\'a\v{c}ek O (2020) {Growth of
  resonances and chaos for a spinning test particle in the Schwarzschild
  background}. Phys Rev D 101(2):024037. \doi{10.1103/PhysRevD.101.024037}.
  {\href{https://arxiv.org/abs/1911.00414}{{arXiv:1911.00414}}} {[gr-qc]}

\bibitem[{Zhang et~al.(2020)Zhang, Zhao, Wang, Wang, Yagi, Yunes, Zhao, and
  Zhu}]{Zhang:2019iim}
Zhang C, Zhao X, Wang A, Wang B, Yagi K, Yunes N, Zhao W, Zhu T (2020)
  {Gravitational waves from the quasicircular inspiral of compact binaries in
  Einstein-aether theory}. Phys Rev D101(4):044002.
  \doi{10.1103/PhysRevD.101.044002}.
  {\href{https://arxiv.org/abs/1911.10278}{{arXiv:1911.10278}}} {[gr-qc]}

\bibitem[{Zhang and Yang(2019)}]{Zhang:2018kib}
Zhang J, Yang H (2019) {Gravitational floating orbits around hairy black
  holes}. Phys Rev D 99(6):064018. \doi{10.1103/PhysRevD.99.064018}.
  {\href{https://arxiv.org/abs/1808.02905}{{arXiv:1808.02905}}} {[gr-qc]}

\bibitem[{Zhang and Yang(2020)}]{Zhang:2019eid}
Zhang J, Yang H (2020) {Dynamic Signatures of Black Hole Binaries with
  Superradiant Clouds}. Phys Rev D 101(4):043020.
  \doi{10.1103/PhysRevD.101.043020}.
  {\href{https://arxiv.org/abs/1907.13582}{{arXiv:1907.13582}}} {[gr-qc]}

\bibitem[{Zhao and Lu(2020)}]{Zhao:2020iew}
Zhao Y, Lu Y (2020) {Stochastic Gravitational Wave Background and Eccentric
  Stellar Compact Binaries}
  {\href{https://arxiv.org/abs/2009.01436}{{arXiv:2009.01436}}} {[astro-ph.HE]}

\bibitem[{Zhou and Straumann(1991)}]{Zhou:1991nu}
Zhou Zh, Straumann N (1991) {Nonlinear perturbations of Einstein Yang-Mills
  solitons and nonAbelian black holes}. Nucl Phys B 360:180--196.
  \doi{10.1016/0550-3213(91)90439-5}

\bibitem[{Zhu et~al.(2020)Zhu, Baryakhtar, Papa, Tsuna, Kawanaka, and
  Eggenstein}]{Zhu:2020tht}
Zhu SJ, Baryakhtar M, Papa MA, Tsuna D, Kawanaka N, Eggenstein HB (2020)
  {Characterizing the continuous gravitational-wave signal from boson clouds
  around Galactic isolated black holes}
  {\href{https://arxiv.org/abs/2003.03359}{{arXiv:2003.03359}}} {[gr-qc]}

\bibitem[{Zhu et~al.(2011{\natexlab{a}})Zhu, Fan, and Zhu}]{Zhu:2011pt}
Zhu XJ, Fan XL, Zhu ZH (2011{\natexlab{a}}) {Stochastic Gravitational Wave
  Background from Neutron Star r-mode Instability Revisited}. Astrophys J
  729:59. \doi{10.1088/0004-637X/729/1/59}.
  {\href{https://arxiv.org/abs/1102.2786}{{arXiv:1102.2786}}} {[astro-ph.CO]}

\bibitem[{Zhu et~al.(2011{\natexlab{b}})Zhu, Howell, Regimbau, Blair, and
  Zhu}]{Zhu:2011bd}
Zhu XJ, Howell E, Regimbau T, Blair D, Zhu ZH (2011{\natexlab{b}}) {Stochastic
  Gravitational Wave Background from Coalescing Binary Black Holes}. Astrophys
  J 739:86. \doi{10.1088/0004-637X/739/2/86}.
  {\href{https://arxiv.org/abs/1104.3565}{{arXiv:1104.3565}}} {[gr-qc]}

\bibitem[{Zhu et~al.(2013)Zhu, Howell, Blair, and Zhu}]{Zhu:2012xw}
Zhu XJ, Howell EJ, Blair DG, Zhu ZH (2013) {On the gravitational wave
  background from compact binary coalescences in the band of ground-based
  interferometers}. Mon Not Roy Astron Soc 431(1):882--899.
  \doi{10.1093/mnras/stt207}.
  {\href{https://arxiv.org/abs/1209.0595}{{arXiv:1209.0595}}} {[gr-qc]}

\bibitem[{Zimmerman(2015)}]{Zimmerman:2015hua}
Zimmerman P (2015) {Gravitational self-force in scalar-tensor gravity}. Phys
  Rev D 92(6):064051. \doi{10.1103/PhysRevD.92.064051}.
  {\href{https://arxiv.org/abs/1507.04076}{{arXiv:1507.04076}}} {[gr-qc]}

\bibitem[{Zimmerman and Poisson(2014)}]{Zimmerman:2014uja}
Zimmerman P, Poisson E (2014) {Gravitational self-force in nonvacuum
  spacetimes}. Phys Rev D 90(8):084030. \doi{10.1103/PhysRevD.90.084030}.
  {\href{https://arxiv.org/abs/1406.5111}{{arXiv:1406.5111}}} {[gr-qc]}

\bibitem[{{Ziyan Yang} et~al.(2020){Ziyan Yang}, {Mandic}, {Scarlata}, and
  {Banagiri}}]{2020arXiv200710456Z}
{Ziyan Yang} K, {Mandic} V, {Scarlata} C, {Banagiri} S (2020) {Searching for
  Cross-Correlation Between Stochastic Gravitational Wave Background and Galaxy
  Number Counts}. arXiv e-prints arXiv:2007.10456.
  {\href{https://arxiv.org/abs/2007.10456}{{arXiv:2007.10456}}} {[astro-ph.CO]}

\bibitem[{Zlatev et~al.(1999)Zlatev, Wang, and Steinhardt}]{Zlatev:1998tr}
Zlatev I, Wang LM, Steinhardt PJ (1999) {Quintessence, cosmic coincidence, and
  the cosmological constant}. Phys Rev Lett 82:896--899.
  \doi{10.1103/PhysRevLett.82.896}.
  {\href{https://arxiv.org/abs/astro-ph/9807002}{{arXiv:astro-ph/9807002}}}

\bibitem[{Zlochower et~al.(2012)Zlochower, Ponce, and
  Lousto}]{Zlochower:2012fk}
Zlochower Y, Ponce M, Lousto CO (2012) {Accuracy Issues for Numerical
  Waveforms}. Phys Rev D 86:104056. \doi{10.1103/PhysRevD.86.104056}.
  {\href{https://arxiv.org/abs/1208.5494}{{arXiv:1208.5494}}} {[gr-qc]}

\bibitem[{Zumalac{\'a}rregui and
  Garc{\'\i}a-Bellido(2014)}]{Zumalacarregui:2013pma}
Zumalac{\'a}rregui M, Garc{\'\i}a-Bellido J (2014) {Transforming gravity: from
  derivative couplings to matter to second-order scalar-tensor theories beyond
  the Horndeski Lagrangian}. PhysRev D89(6):064046.
  \doi{10.1103/PhysRevD.89.064046}.
  {\href{https://arxiv.org/abs/1308.4685}{{arXiv:1308.4685}}} {[gr-qc]}

\bibitem[{Zumalac\'arregui et~al.(2017)Zumalac\'arregui, Bellini, Sawicki,
  Lesgourgues, and Ferreira}]{Zumalacarregui:2016pph}
Zumalac\'arregui M, Bellini E, Sawicki I, Lesgourgues J, Ferreira PG (2017)
  {hi\_class: Horndeski in the Cosmic Linear Anisotropy Solving System}. JCAP
  08:019. \doi{10.1088/1475-7516/2017/08/019}.
  {\href{https://arxiv.org/abs/1605.06102}{{arXiv:1605.06102}}} {[astro-ph.CO]}

\end{thebibliography}

\end{document}